\documentclass[iop,apjl,numberedappendix,appendixfloats,floatfix]{emulateapj}
\usepackage{lineno}
\usepackage{color}
\usepackage{amsmath}
\usepackage{url}
\usepackage[figuresright]{rotating}
\usepackage{yfonts}
\usepackage[T1]{fontenc}

\newcommand{\solar}{$_{\odot}$}
\newcommand{\tco}{$^{12}$CO}
\newcommand{\ttco}{$^{13}$CO}
\newcommand{\ceto}{C$^{18}$O}
\newcommand{\hcop}{HCO$^+$}

\newcommand{\joz}{$J$=1$\rightarrow$0}
\newcommand{\jto}{$J$=2$\rightarrow$1}

\newcommand{\kms}{\,km\,s$^{-1}$}
\newcommand{\degree}{$^{\circ}$}
\newcommand{\fdeg}{$^{\circ}$\hspace{-1mm}.}

\newcommand{\tex}{$T_{\rm ex}$}

\newcommand{\td}{$T_{\rm dust}$}
\newcommand{\vlsr}{$V_{\rm LSR}$}
\newcommand{\sigv}{$\sigma_V$}

\newcommand{\htwo}{H$_2$}

\newcommand{\nhtwo}{$N_{\rm H_2}$}
\newcommand{\nco}{$N_{\rm CO}$}
\newcommand{\ntco}{$N_{\rm ^{12}CO}$}

\newcommand{\ico}{$I_{\rm CO}$}
\newcommand{\itco}{$I_{\rm ^{12}CO}$}
\newcommand{\ittco}{$I_{\rm ^{13}CO}$}
\newcommand{\iceto}{$I_{\rm C^{18}O}$}

\newcommand{\xco}{$X_{\rm CO}$}

\def\lapp{\ifmmode\stackrel{<}{_{\sim}}\else$\stackrel{<}{_{\sim}}$\fi}
\def\gapp{\ifmmode\stackrel{>}{_{\sim}}\else$\stackrel{>}{_{\sim}}$\fi}
\newcommand{\lbv}{($l$,$b$,$V$)}
\newcommand{\lb}{($l$,$b$)}
\newcommand{\lv}{($l$,$V$)}
\newcommand{\tnt}{\tex,$N$,$\tau$}
\newcommand{\uv}{$u_0$,$v_0$}
\newcommand{\uvw}{$u_0$,$v_0$,$w_0$}
\newcommand{\xy}{($x$,$y$)}
\newcommand{\xyz}{($x$,$y$,$z$)}
\newcommand{\ld}{($l$,$d$)}
\newcommand{\lbd}{($l$,$b$,$d$)}

\shorttitle{ThrUMMS III: 3D Architecture of the Southern Milky Way}
\shortauthors{Barnes et al.}

\begin{document}

\title{The Three-mm Ultimate Mopra Milky Way Survey. III. \\
    Data Release 6, An Atlas of Physical Conditions, Global Mass Conversion Laws, and \\
    3D Physical Architecture of the Molecular ISM in the Fourth Quadrant
    }

\author{Peter J.\ Barnes$^{1}$, Dylan G.\ H.\ Barnes$^{2}$, Audra K.\ Hern\'andez$^{3}$, Sebastian Lopez$^{4,5}$, Erik Muller$^{6}$}
\affiliation{$^{1}$Space Science Institute, 4765 Walnut St, Suite B, Boulder, CO 80301, USA}
%\altaffiltext{2}{School of Science and Technology, University of New England, NSW 2351, Australia}
%\altaffiltext{3}{Eastside High School, 1201 SE 43rd St, Gainesville, FL 32641, USA}
\affiliation{$^{2}$Dept of Aerospace, Physics \& Space Sciences, Florida Institute of Technology, 150 W.\ University Blvd, Melbourne, FL 32901, USA}
\affiliation{$^{3}$University of Wisconsin, 500 Lincoln Drive, Madison, WI 53706, USA}
\affiliation{$^{4}$Department of Astronomy, Ohio State University, 140 W.\ 18th Ave., Columbus, OH 43210, USA}
\affiliation{$^{5}$Center for Cosmology \& AstroParticle Physics, Ohio State University, 191 W.\ Woodruff Ave., Columbus, OH 43210, USA}
\affiliation{$^{6}$School of Engineering, University of Sydney, NSW 2006, Australia}

\email{pbarnes@spacescience.org}

\begin{abstract}
We present Data Release 6 of ThrUMMS, consisting of complete data cubes and various moments of line emission (\tco, \ttco, \ceto) from molecular clouds, across 60\degree$\times$2\degree\ of the Fourth Quadrant (4Q) of the Milky Way at a resolution of 72$''$ in \lb\ and 0.09\kms\ in \vlsr.  From LTE radiative transfer %50 words
analysis of the data cubes, we compute cubes and moments of the lines' opacity, excitation temperature, and column density \ntco.  Combining \itco\ and \ntco\ data, we derive a global mass conversion law $N$=$N_{0}I^{p}$, where $N_{0}$$\approx$10$^{18}$mol/m$^{2}$ and $p$=2 at this resolution.  We argue that the standard linear $N$=$XI$ is only approximately %100 words
valid: $p$$\sim$1.5--1.0 at coarser resolutions or in atypical locations, such as Galactic Center clouds.  The velocity dispersion distributions are very different between \itco\ and \ntco, the former preferentially tracing more diffuse molecular gas.

\hspace{3mm}We re-evaluated Galactic rotation parameters for the 4Q, defining a new ``BGT'' model, and deprojected the %149 words
\lv\ data onto \ld\ and \xy\ grids using standard kinematic procedures.  To automate distance disambiguation inside the solar circle, we developed a simple $\zeta^{+}$ discriminator function and applied it to our deprojections.  We discovered two previously unrecognised features of the molecular cloud population: widespread ripples in the midplane of wavelength 4\,kpc %201 words
and amplitude 50\,pc, potentially generated by the last perigalactic passage of the Sgr dwarf; and three distant, massive molecular structures, the Far Ara clouds, two of which exhibit an exceptional velocity gradient, possibly lying in the far end of the Galaxy's Bar or a gas-rich dwarf galaxy $\sim$20--300\,kpc beyond the disk. %254
\end{abstract}


%%%%%%%%%%
%%   Section 1  %%
%%%%%%%%%%
\section{Introduction}\label{intro}

\vspace{1mm}Our home Galaxy, the Milky Way, continues to provide our best observational opportunity in terms of sensitivity and resolution to study a range of fundamental astrophysical processes that occur in most disk galaxies throughout the Universe.  Among these are the origin and structure of spiral arms, and the star formation that occurs from the gaseous interstellar medium in the disk, with its several distinct phases and rich panorama of observables.  In particular, cold molecular clouds are the specific location where stars form, and their properties and distribution in the Galactic disk are of fundamental importance in deciphering the physical processes of star formation, and how they are related to spiral structure and the evolution of the Galaxy.

\vspace{1mm}Since the last major review of this field \citep{hd15}, molecular cloud properties have continued to be explored by a range of increasingly sophisticated efforts.  A particular innovation was the development of powerful new digital backends for mm-wave receivers, exemplified by the commissioning of the MOPS spectrometer on the Mopra\footnote{Operation of the Mopra radio telescope during 2012--15 was made possible by funding from the National Astronomical Observatory of Japan (NAOJ), the University of New South Wales, the University of Adelaide, and the Commonwealth of Australia through CSIRO/Australia Telescope National Facility (ATNF).} 
antenna during the period 2006--10 \citep{wmf06,b11}.  These have enabled simultaneous multi-species observations and wide-field surveys that vastly sped up data acquisition and simplified cross-calibration, such as CHaMP \citep{b11,b16,b18}, ThrUMMS \citep[][hereafter Papers I and II]{b15,q15}, %h19
and MALT90 \citep{jrb13}, all at Mopra; SEDIGISM \citep{s17,dc20} at APEX; FUGIN \citep{um17} at Nobeyama; and several others.

\vspace{1mm}One of the earliest multi-species wide-field surveys, ThrUMMS' main goals were to provide the first parsec-scale mapping survey of the physical conditions in the bulk of the molecular gas lying in the Fourth Galactic Quadrant (hereafter 4Q), and we now present the next full ThrUMMS Data Release, DR6.  This includes a standard set of data products together with some second-generation data analyses, focused on a subset of key topics in Galactic astrophysics, namely mass conversion laws and kinematic distances.  We anticipate these publicly available products and results will be widely useful to the Galactic ISM community, and provide important legacy value for a number of follow-up studies.

\vspace{1mm}In this paper, we describe results of compiling the overall properties of the ThrUMMS molecular cloud population in the 4Q, as traced by the \joz\ emission from the three ``iso-CO'' species \tco, \ttco, and \ceto\ (because of the hyperfine structure in the $N$=1$\rightarrow$0 line of CN, we deferred moment analysis of this line).  In the main part of the paper, we give brief summaries of our results, with full details presented in extensive Appendix material.  We start with a description of the processing and product design (\S\ref{design}), from which we explore the global view of structural and physical features of the iso-CO emission, both projected on the sky ($l$,$b$) and as $P$$V$ maps.  We then present results of %a 3D SCIMES analysis of the \ttco\ cubes in \S\ref{scimes}, comprising a comprehensive catalogue of {\color{red}6789} molecular clouds and their properties; a preliminary version of this cloud catalogue was given in \cite{h19} but is updated here.  Finally we present 
a radiative transfer analysis of the iso-CO species' emission (\S\ref{radxfer}), which yields a complete atlas of the \tex, $\tau$, and $N_{\rm CO}$ distribution (\S\ref{atlas}) plus determination of a global mass conversion law for CO across the 4Q (\S\ref{convlaw}).  Finally we perform a detailed kinematic analysis of the data in order to improve prior range-finding techniques for the southern Milky Way, including a height-based statistical technique for discriminating between near and far kinematic distances, and revealing previously unrecognised features of the Galaxy's molecular layer (\S\ref{3d}).  Our conclusions appear in \S\ref{concl}.

%%%%%%%%%%
%%   Section 2  %%
%%%%%%%%%%
\section{Data Release 6}\label{design}

%%%%%%%%%%%
%%   Section 2.1  %%
%%%%%%%%%%%
\subsection{Processing History}\label{history}

\vspace{1mm}Paper I presented the ThrUMMS DR3 data available at that time, covering about 65\% of the intended 4Q survey area.  During 2014--15 (i.e., while Paper I was being published), we completed observations of the planned 60\degree$\times$2\degree\ coverage with Mopra (i.e., 360\degree\ $>$ $l$ $>$ 300\degree\ and |$b$| $<$ 1\degree), including re-observing as many fields as possible that were affected by bad weather or (rarely) hardware problems.  As a result, processing of all the ThrUMMS raw data files produces cubes of even higher quality and wider coverage than available in DR3.  So after Paper I, in 2016 Feb we released DR4, containing all \ttco\ data cubes across the 4Q, on the U.\,Florida ThrUMMS web pages (operational until 2021, but now defunct).  At the same time, equivalent DR4-quality \tco\ and \ceto\ data cubes over selected areas were also made available to researchers on request, pending future releases.

\vspace{1mm}DR5 was an expanded version of this, containing all cubes for all four spectral lines across the 4Q, originally made available on a National Astronomical Observatory of Japan web page during 2016--17, but this also became unavailable.  Therefore, in 2017 May--July, we reprocessed all the raw \tco\ data again, and in 2018 Sep--2019 Jan we did the same for the \ceto\ and CN data, but with more selective editing of bad data within files, in order to minimise their impact on the final products.  Details of the processing itself, including the standard software pipeline and our improvements thereto, are given in Paper I and summarised here and below.

\vspace{1mm}As part of this processing, we also produced a standard set of moment maps from each of the iso-CO data cubes using the {\em smooth-and-mask} (SAM) algorithm \citep{rbv90,b15}, which helped us diagnose and address most of the artifacts in the raw data, resulting in the best possible pipeline-produced products with the completed observations.  In late 2022, we re-established ThrUMMS data access at SSI\footnote{See gemelli.spacescience.org/$\sim$pbarnes/research/thrumms}.  We now make the latest cubes and moment maps available as DR6, and this access is mirrored at IPAC\footnote{See irsa.ipac.caltech.edu/data/ThrUMMS/overview.html, Di-gital Object Identifier doi.org/10.26131/IRSA628}.  Here we also present results from further processing on: the radiative transfer analysis, application of those results to constructing new physical property atlases, and deriving a new mass conversion law (performed during 2019 June--2020 March and 2023 March--Sep); and exploring the 3D structure of these data (performed during 2023 Oct--2025 Feb).

% Figure 1
\begin{figure*}[ht]
			\centerline{\includegraphics[angle=0,scale=0.052]{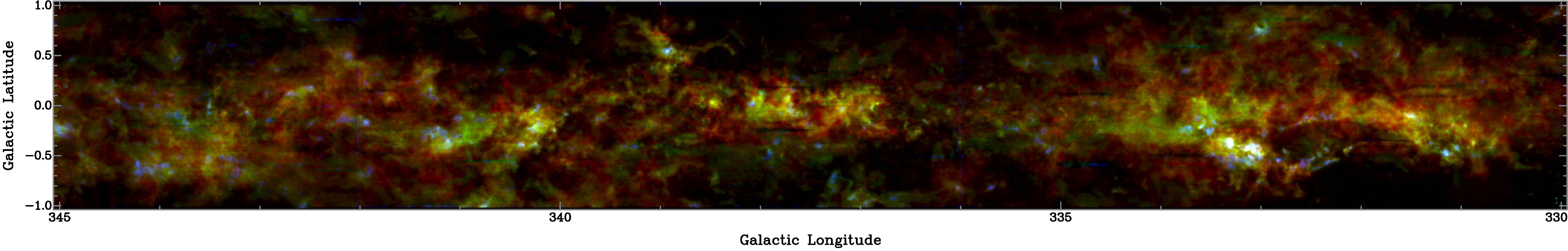}\hspace{1mm}}
\vspace{-3.5mm}\centerline{\includegraphics[angle=0,scale=0.052]{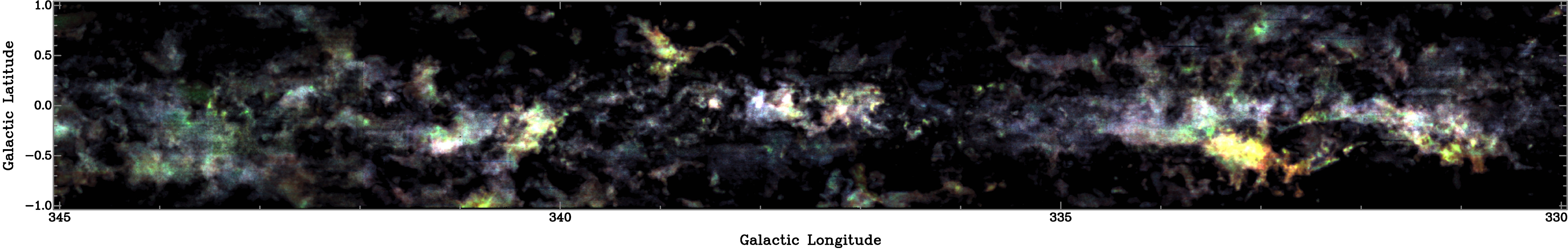}\hspace{1mm}}
\vspace{-53mm}
{\color{red}\hspace{10mm}{\tco}}
{\color{green}\hspace{1mm}{\ttco}}
{\color{cyan}\hspace{1mm}{\ceto}}

\vspace{22mm}
{\color{red}\hspace{10mm}{\tex}}
{\color{green}\hspace{1mm}{$N_{\rm co}$}}
{\color{cyan}\hspace{1mm}{$\tau_{\rm co}$}}
\vspace{25mm}
\caption{\footnotesize Sample 15\degree$\times$2\degree\ area ($\tfrac{1}{4}$) of ThrUMMS DR6; the full mosaics are given in Appendix \ref{fullmos}. 
{\em Top.}  Pseudocolour image of integrated intensity (zeroth moment) in the three iso-CO \joz\ lines, as labelled.  Note the wide variations in line ratios across the Galactic Plane, evident from the varying colours in this image.  The data maximum, saturated colour, median uncertainty, and black levels in this image are respectively at 1137, 265, 1.91, --5\,K\kms\ (\tco); 229, 70, 0.75, --2.4\,K\kms\ (\ttco); and 33, 9.6, 0.37, 0.0\,K\kms\ (\ceto). \\
{\em Bottom.}  Same area as top panel, but with colour-coded physical parameter solutions (as labelled) %(red = mean \tex, green = \nco, blue = mean $\tau_{\rm ^{12}CO}$) 
from the analysis in \S\ref{radxfer}, which makes manifest the radiative transfer implied in the iso-CO line ratios.  The data maximum, saturated colour, median uncertainty, and black levels in this image are respectively at 3.16, 0.85, 0.054, --0.03\,K (mean \tex), 257, 44, 1.43, \& --0.4$\times$10$^{24}$\,molec\,m$^{-2}$ (\nco), and 5.53, 3.67, 0.0093, --0.11 (mean $\tau_{\rm ^{12}CO}$).  Note that the \tex\ and $\tau$ scales are artificially low, since the averaging has been done over channels with no solution (taken as 0), putting the intensity in each on a relative scale only. $$ $$ % nevertheless, the values are proportional to the contribution from non-zero \tex\ or $\tau$ channels.  %% Nerr0 corresp. to ZMerr0 = 2.68468 Msun/pc2
\label{samplemos}}
\vspace{-9mm}
\end{figure*}

%%%%%%%%%%%
%%   Section 2.2  %%
%%%%%%%%%%%
\subsection{Observed Data Cubes\label{lines}}

\vspace{1mm}The structure of DR6 is similar to the previous Data Releases \citep{b15}.  The raw data were assembled into data cubes covering limited areas, necessary to facilitate the handling of the huge data volume and processing through the \textsc{livedata/gridzilla} package \citep{b01}.  Each cube of size 6\degree(longitude)$\times$ 2\degree(latitude), with velocity as the third dimension, constitutes one ``Sector'' of the 4Q, with ten Sectors making up the 60\degree\ longitude coverage of the survey.  The maps are generated from the raw data with an effective angular resolution of 72$''$, effectively beam-sampled compared to the Mopra telescope's intrinsic resolution of 33$''$ (or to a more typical 37$''$ beam with a small smoothing kernel, in standard observing modes that implement Nyquist sampling).  The beam sampling, with a long-standing history in radio-frequency surveys, was done to enable a full 4Q survey in only $\sim$1200 hrs of clock time.  Each of the ten ($l$,$b$) Sector maps therefore has a ``spatial dynamic range'' (SDR), defined as the field of view divided by half the angular resolution, of 600$\times$200 = 120,000.  For the survey as a whole, the SDR = 1.2$\times$10$^6$ in each species.

\vspace{1mm}The velocity coverage is given by the 4096 channels available in each spectral ``Zoom'' window, multiplied by the 33\,kHz ($\approx$\,0.09\kms) channel width ($\approx$ velocity resolution), but truncated to a \vlsr\ range selected to encompass all significant emission visible in the Columbia-CfA survey \citep{dht01}.  This is typically from a minimum of --100 to --200\kms, depending on the Sector, to a maximum around +50\kms.  The cubes in the pre-DR6 files were binned by 4 channels in order to limit file sizes and improve the spectral S/N; this is also the default for DR6, but the full velocity resolution data are now also available upon request.  Therefore, the spectral dynamic range (SpDR) in the binned data is 450--750, or 1800--3000 in the full spectral resolution cubes, both depending on the Sector.

\vspace{1mm}The rms noise in the data cubes was computed at each pixel from channels that were emission-free across the whole Sector being analysed.  This yielded full rms maps across all pixels in each Sector.

\vspace{1mm}The total information content in the whole survey is then SDR $\times$ SpDR $\times$ 4 species = 1.2$\times$10$^{10}$ independently observed voxels (i.e., not including derived quantities).

%%%%%%%%%%%
%%   Section 2.3  %%
%%%%%%%%%%%
\subsection{Moment Maps and Mosaics}\label{mmm}

\vspace{1mm}In our implementation of the SAM algorithm (see Paper I), we routinely compute the zeroth (integrated intensity \ico), first (intensity-weighted mean velocity field \vlsr), \& second (velocity dispersion \sigv) $V$-moments; the peak \& mean brightness; and the rms spectral noise as above, for each pixel across 30 (Sector $\times$ species) data cubes.  We also compute the formal error maps for the 0th--2nd moments.  These are all 2D ($l$,$b$) projections of information across all \vlsr\ in the full 3D ($l$,$b$,$V$) cubes, with the benefit of SAM's noise suppression characteristics, and give high-quality renderings of each moment.  We can also compute similar moments projecting across $b$, such as a standard longitude-velocity ($lV$) diagram for the zeroth moment, and so on.

\vspace{1mm}The 10 Sector moments of a single type (e.g., \ttco\ $V_{\rm LSR}$) can also be mosaicked into a single 60\degree$\times$2\degree\ image and analysed as a unit.  In such mosaics, the merging of data at overlapping Sector boundaries or ``seams'' (i.e., near $l$=306\degree, 312\degree, etc.), already optimised by the noise-based masking of the SAM algorithm, is further minimised by the per-pixel rms$^{-2}$ weighting in creating the mosaics, efficiently eliminating many artifacts due to end-of-scan sampling irregularities during observing.  Especially with the bright (high S/N) \tco\ and \ttco\ emission, this leaves only a small number of suspect, low-S/N features at the seams, which are statistically negligible.  A few seams still have evident artifacts, but these are easily allowed for by eye.  A selection of the mosaicked moment maps is included in DR6 and presented in {\color{red}Appendix \ref{fullmos}}; samples are shown in {\color{red}Figure \ref{samplemos}}.

\vspace{1mm}For some of these mosaics, we use an RGB colour composite to overlay different species or physical parameters in one image: e.g., in the top panel of Fig.\,\ref{samplemos}, we use red for \tco, green for \ttco, and blue for \ceto.  The contrast and brightness in each colour channel are chosen to maximise the total colour contrast across each image, in order to make as clear as possible to casual inspection, various changes in line or parameter ratio across different emission regions.  For example, in Figure \ref{samplemos}, the colours vary widely from strong red to orange to yellow to green, with a number of more compact clumps showing cyan or strong blue colours.  This doesn't mean there are widespread areas where \ittco\ $>$ \itco\ or \iceto\ $>$ \ittco\ (in fact, there are relatively few voxels or pixels where either condition occurs, especially at higher S/N); rather, it means that the ratios span a range of values which the colours track.  A similar approach was chosen to render the physical parameter (\tex\ = red, \nco\ = green, $\tau$ = blue) overlays, described next.

%%%%%%%%%%%
%%   Section 2.4  %%
%%%%%%%%%%%
\subsection{Derived Data\label{physcubes}}

\vspace{1mm}Because our radiative transfer analysis (see \S\ref{radxfer}) operates on each voxel of the observed data, each of the solutions \tex, \nco, $\tau$ of this analysis are also obtained as full ($l$,$b$,$V$) cubes (see Appendix \ref{fullmos}).  As such, they can also be subject to moment analysis, and this gives a particular advantage for downstream physical analysis, compared to moments of observed emission lines: the opportunity to obtain mass- or column-weighted properties of an emission region, rather than emission-weighted properties.

\vspace{1mm}This is important because observed emission-line intensities \ico\ are a non-linear function (at mm wavelengths, at least) of the excitation temperature and optical depth in the lines' emitting areas \citep{b15,b18,p19,p21}.  Determining \nco\ from \ico\ without knowing \tex\ or $\tau$ can lead to biases towards areas with either high \tex\ or high $\tau$, potentially undersampling low-excitation or low-opacity regions that may nevertheless contribute to overall cloud physics. This has been a vexing problem for decades, with the potential biases being largely unaddressed through the widespread use of a single conversion factor (\xco) from \ico\ to \nco\ \citep[e.g., see][]{bwl13}.  Instead, $N$ is more likely a power-law of \ico, with index $p$ varying somewhat by locale.  Studies to date have found $p$ generally in the range $\sim$1.3--2 \citep{b15,b18}, but we reassess this here as well (see \S\ref{convlaw}).  The advantage of having explicit 3D solutions for \tex\ and $\tau$ by radiative transfer methods, and being able to calculate \nco\ without further assumptions, not only reduces the difficulties inherent in a simple $I$-weighted or $X$ factor approach, but can reveal new insights into the physics of molecular clouds.

%%%%%%%%%%
%%   Section 3  %%
%%%%%%%%%%
\section{Radiative Transfer Analysis}\label{radxfer}
%{\color{red}delicious!}

%%%%%%%%%%%
%%   Section 3.1  %%
%%%%%%%%%%%
\subsection{The Method}\label{method}

\vspace{1mm}Our radiative transfer calculations and procedures have been fully described in Paper I and \cite{b18}, and the interested reader is referred to these works for all details.  Our full results for this work are presented in {\color{red}Appendix \ref{rta}}.  We discuss some interesting highlights next, and include a number of moment maps from the derived quantities' data cubes in DR6 and Appendix \ref{fullmos}, with a sample in Figure \ref{samplemos}.

\vspace{1mm}In brief, the method takes the radiative transfer equation
\begin{equation}
	T_{{\rm mb,}i} = [S_{\nu}(T_{\rm ex})-S_{\nu}(T_{\rm bg})]\,(1-e^{-\tau_{i}})~			%% ONE
\end{equation}
%\begin{equation} % THREE
%	\frac{T_{\rm 13}}{T_{\rm 12}} = \frac{1-e^{-\tau_{\rm 13}}}{1-e^{-R_{\rm 13}\tau_{\rm 13}}}   ,
%\end{equation}
%\begin{displaymath}
%	\hspace{-55.5mm}X_{\rm CO} = N_{H_2}/I_{\rm CO}\vspace{-2mm}
%\end{displaymath}
%\begin{equation} % FOUR
%	= 1.8\times10^{24}~{\rm H}_2\,{\rm molecules\,m}^{-2}/({\rm K\,km\,s}^{-1})
%\end{equation}
for each species $i$ = \tco, \ttco, \ceto\ and solves for the common \tex\ (and, incidentally, the implied abundance ratio $R_{18}$ $\equiv$ [\ttco]/[\ceto]), assuming $\tau_{12}$/$\tau_{13}$ = $R_{13}$ $\equiv$ [\tco]/[\ttco] is fixed to a single value everywhere.  Then, having derived $\tau_{i}$ and $T_{\rm ex}$ at each voxel, we compute the CO column density via
\begin{equation}
	N_{\rm CO} = \frac{3h}{8\pi^3\mu^2}~\frac{Q(T_{\rm ex})e^{E_u/kT_{\rm ex}}}{J_u(e^{h\nu/kT_{\rm ex}}-1)}~\int\tau{\rm d}V  ,		%% TWO
\end{equation}
where %the quantities are all as described by \cite{b18}. %
$\mu$ is the CO dipole moment, $Q$ is the rotational partition function, $E_u$ and $J_u$ are the energy and quantum number of the upper level of the transition at frequency $\nu$, and the integral is over the velocity range of either a single channel or the whole emission for a cloud (depending on the computational objective).  %Here, instead of using a constant $X_{\rm CO}$ factor to convert the observed $I_{\rm CO}$ into a molecular gas mass, as is sometimes done in the literature despite \citet{dht01}'s caveat against doing so, we can use the ($\tau$,$T_{\rm ex}$) estimates above (\S\ref{tauratios}) to compute a more realistic CO column density.  This can then be compared with other estimates of molecular gas mass (e.g., SED fits to the Hi-GAL data) to derive CO abundance maps.

\vspace{1mm}One must still include an important caveat to this treatment: it is not the best line analysis that can be conceived, since it relies on simplifying assumptions which make the determination of physical solutions tractable, but not necessarily ideal.  These are: a plane-parallel geometry, local thermodynamic equilibrium (LTE), and a fixed gas-phase abundance ratio $R_{13}$ = 60.  In contrast, there are a number of multi-line, non-LTE studies of particular regions based on existing packages, which indubitably do a better job of recovering the gas conditions from the data than our code, but they tend to be of smaller areas (usually $<$1\degree\ across) and are much more computationally intensive than our approach \citep[e.g.,][]{rgg21}.  Thus, these methods are still prohibitive to apply to such a large dataset.

% Figure 2
\begin{figure*}[ht]
\hspace{0mm}\includegraphics[angle=-90,scale=0.169]{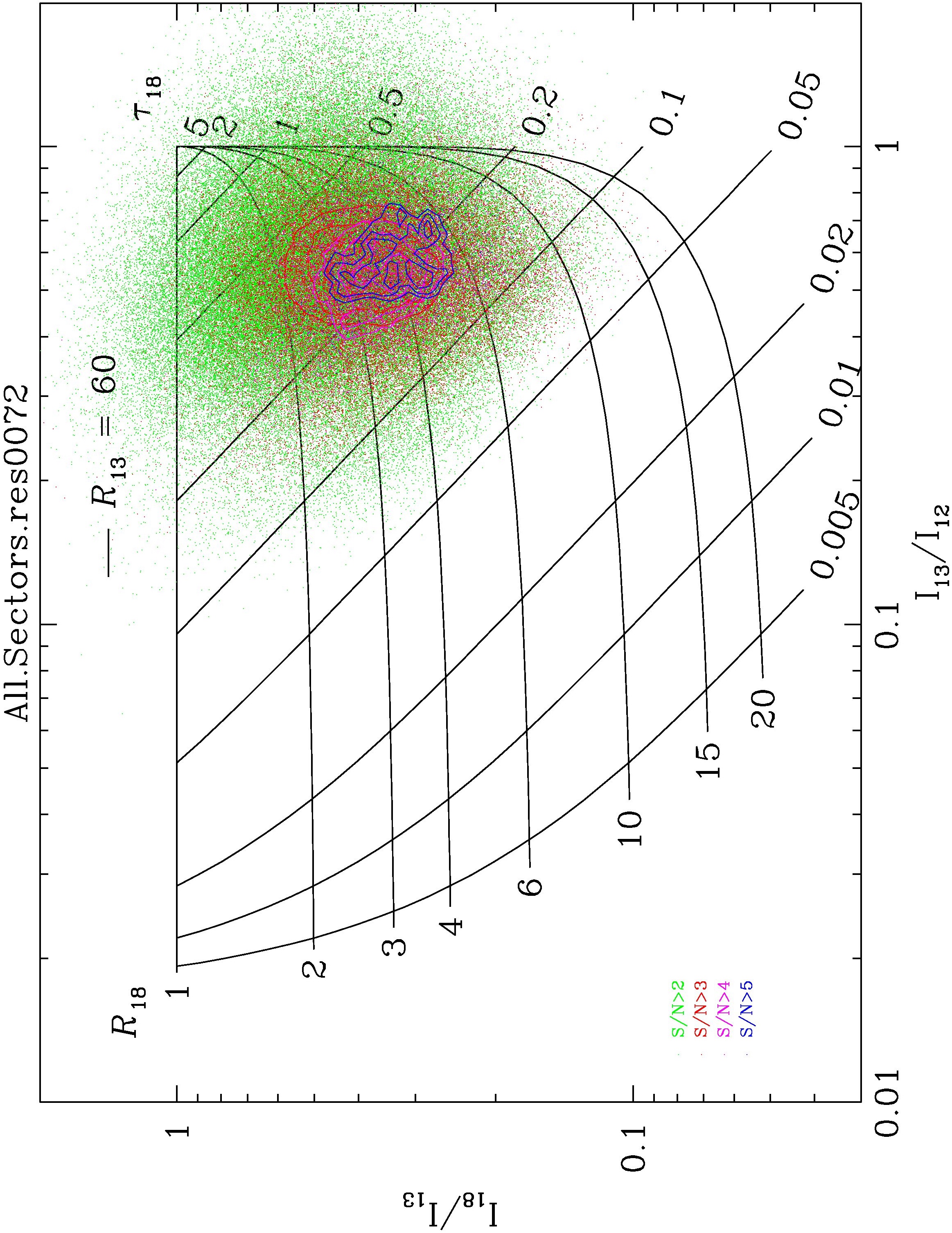} % NOT Iratios-All-sn1.jpg

\vspace{-2mm}
\caption{\footnotesize Iso-CO line intensity ratio-ratio diagram across all ThrUMMS DR6 data.  Each point is one \lbv\ data voxel where all three iso-CO lines are above the colour-coded S/N ratio, from among 360,000 voxels with S/N$>$0.  Contours of pixel density, at 67\%, 78\%, and 89\% of the peak density, are also shown for each of the top three S/N levels, to highlight their near-concentricality around ratios $I_{13}$/$I_{12}$ = 0.6$\pm$0.2 and $I_{18}$/$I_{13}$ = 0.35$\pm$0.10.  Also shown is the grid of radiative transfer solutions from \S\ref{radxfer} for any ratio-ratio point.  If we use a different value for $R_{13}$ than that shown here (say, 40), it changes the underlying grid (and the solutions for all the data points) very little, compressing only the 3 leftmost contours of $\tau_{18}$ in the top left corner of the diagram slightly to the right, far from any data. $$ $$
\label{bigiratios}}
\vspace{-10mm}
\end{figure*}

% Figure 2, formerly (= small version of usetabe Fig A3)
%\begin{figure}[t]
%\hspace{0mm}\includegraphics[angle=-90,scale=0.08]{Iratios-All-sn2.jpg}
%\vspace{0mm}
%\caption{%Iso-CO line intensity ratio-ratio diagram across all ThrUMMS DR6 data.  Each point is one \lbv\ data voxel where all 3 lines are above the colour-coded S/N ratio, from among 360,000 voxels with S/N$>$0.  Also shown is the grid of radiative transfer solutions from \S\ref{radxfer} for any ratio-ratio point.
%\label{iratios}}
%\vspace{0mm}
%\end{figure}

\vspace{1mm}Another well-known issue with the plane-parallel/LTE treatment is the assumption that both the \tco\ and \ttco\ lines' data arise from the same parcels of gas, even per voxel.  This is because the \tco\ line opacity is far higher than for \ttco\ ($\tau_{12}$ = 60\,$\tau_{13}$ assumed here; see \S\ref{intint}), causing the \tco\ emission to be heavily affected by radiative trapping and skewing the results proceeding from a na\"ive calculation based on an LTE line ratio, as discussed by \cite{hd15} and references therein.  Finally, in the inner Galaxy we know that the intrinsic abundance ratio $R_{13}$ drops from its value near the solar circle \citep[the 60 we use here;][]{gwb14} to a level nearer to 40 at half our distance from the Galactic Center.  Both of these objections mean that the column densities \nco\ we derive will likely be underestimated, but as we will show (\S\ref{convlaw}), the \nco\ values we do derive still imply a higher \ico\ to \nco\ conversion factor than is usually assumed in many other works.  Thus, the assumptions turn out to be not unreasonable, and our treatment is at least better (we would argue, substantially so) than the simple $X$ factor approach in common use.

%%%%%%%%%%%
%%   Section 3.2  %%
%%%%%%%%%%%
\subsection{Intensity Ratios}\label{intint}

\vspace{1mm}With the RGB rendering for the iso-CO species, Figures \ref{samplemos} and \ref{full121318-mom0} show a wide variety of line ratios (= colours) throughout the 4Q.  Similar renderings can be made for moments in $b$, i.e., the $lV$ diagrams ({\color{red}Fig.\,\ref{full-lv-combo}}), and these show the same kind of line ratio variations.  The variable ratios across all Sectors are summarised in {\color{red}Figure \ref{bigiratios}}.  At the most basic level, this shows that a constant $X$ factor cannot apply to all molecular clouds, or even to parts of clouds.

\vspace{1mm}For example, where the opacity in the \itco\ and \ittco\ lines is small enough, their large brightness ratio approaches the intrinsic gas-phase abundance ratio $R_{13}$ (typically $\sim$ 60$\pm$20 in the ISM, on the left edge of the diagram in Fig.\,\ref{bigiratios}).  In such locations, \itco\ is often relatively bright, and the common \tex\ for all 3 species will be even higher there.  Where \itco/\ittco\ is small, however (i.e., it approaches 1), the opacity in each line will be relatively large, and the \tex\ will be not much larger than the \tco\ brightness temperature.  Indeed, areas of relatively high \tex\ and low $\tau$ are commonly seen throughout the mosaics as ``red,'' while areas with high $\tau$ and low \tex\ are seen as ``blue.''  In such disparate conditions, it is not physically reasonable to assume that \itco\ will convert simply to \nco.

\vspace{1mm}While the voxel distribution in Figure \ref{bigiratios} has a range of values on each axis, allowing our RGB colour rendering to be informative, these ranges are nevertheless relatively small, and clearly peaked around ratios $I_{13}$/$I_{12}$ = 0.6$\pm$0.2 and $I_{18}$/$I_{13}$ = 0.35$\pm$0.10, where the 1$\sigma$ uncertainties are taken from the S/N$>$5 contours.  Moreover, this peak is more concentrated as the overall S/N goes up, suggesting that while the absolute \tex\ and $\tau$ values do vary by factors of $\sim$3--5 or so, it is not unreasonable to think of the 3 species as having a typical opacity ratio, and hence abundance ratio.  This is shown in the $R_{18}$ contours of Figure \ref{bigiratios}: the voxel distribution peaks around $R_{18}$ $\approx$ 4.0$\pm$1.5 when focusing on the highest S/N points.  For this abundance ratio, at least, it is probably more robust against our radiative transfer assumptions than any quantity involving \tco.  This is because the \ttco\ and \ceto\ opacities are (respectively) moderate and small, meaning that any differential radiative trapping, selective photodissociation, or other non-LTE effects may be small at this level of uncertainty.

\vspace{1mm}On the other hand, the relatively small range of observed $R_{18}$ values in Figure \ref{bigiratios} may be averaging over a number of effects known to affect the isotopologue abundance ratios \citep[see, e.g.,][]{f82,g85}.  This is a very complex subject that deserves its own treatment.

% Figure 3: sample XvsI
\begin{figure*}[t]
\vspace{0mm}
\centerline{\includegraphics[angle=0,scale=0.37]{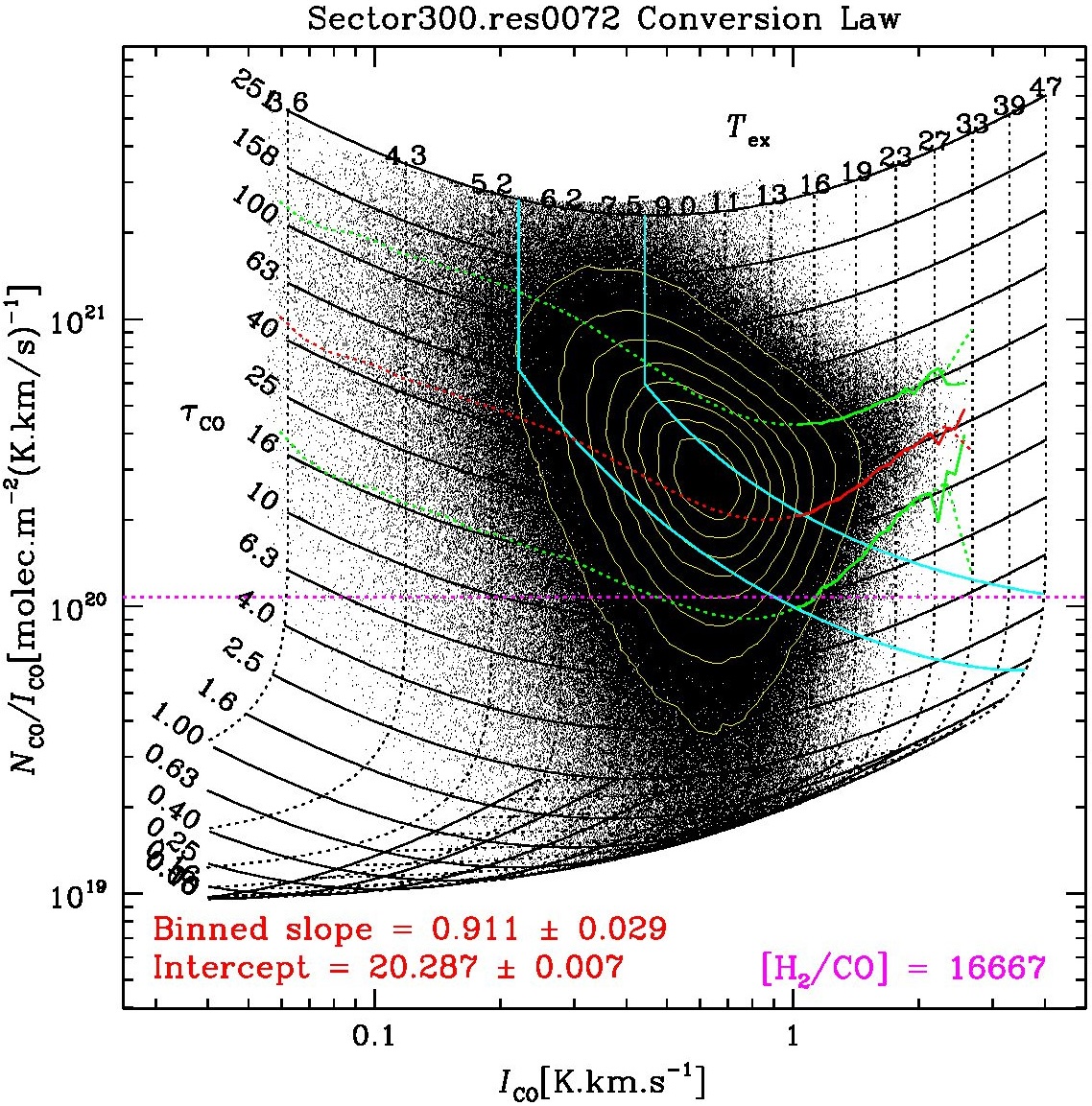}}
\vspace{-2mm}
\caption{\footnotesize Sample plot for Sector 300 of the distribution of radiative transfer solutions for \nco, derived from the 3 iso-CO species as a function of the observed \ico\ in each voxel (i.e., each pixel and velocity channel) of \lbv\ space at the native ThrUMMS angular resolution of 72$''$.  Similar plots for the other Sectors are shown in Appendix \ref{rta}.  The $>$10$^6$ data points here (black dots) are plotted in the $N$/$I$ vs $I$ plane, effectively an $X$ vs $I$ plot once a gas-phase $R_{12}$ = [\htwo]/[\tco] abundance ratio is chosen.  The magenta label and dotted line are then illustrative of where the standard $X$ factor would lie for the given $R_{12}$.  Thin yellow contours outline the distribution where the density of points is high, at levels of 10\% to 80\% of the peak density, in intervals of 10\%.  Underlying the plot is a grid of radiative transfer solutions, with solid curves at constant $\tau$ and dotted curves at constant \tex, as labelled.  The dotted red curve joins up the mean $X$ value in each of 30 equally-spaced bins of data in log$I$, while the dotted green curves join the $\pm$1$\sigma$ levels around the red means in each bin.  The cyan curves show the 1- and 2-$\sigma$ limits in the data: the vertical limits to the left are from the \tco\ noise limits; the curved limits dropping to the right come from the \ttco\ noise limits.  The red and green curves are shown as solid where a linear regression to the mean $\pm$ $\sigma$ values gives the results labelled in red, generally where the points have S/N $>$ 2--3.  Because the fit is made in $X$ vs $I$ space, one must add 1 to the labelled slope to obtain the corresponding value for $p$ in the conversion law, $N$ $\propto$ $I^p$.  In this case, $p$ = 1.91$\pm$0.03.  The intercept log$N_0$ applies at log$I$=0. $$ $$
\label{x300}}
\vspace{-6mm}
\end{figure*}

%%%%%%%%%%%
%%   Section 3.3  %%
%%%%%%%%%%%
\subsection{Optical Depth and Excitation Temperature Variations}\label{tauratios}

\vspace{1mm}We use the $N$/$I$ vs $I$ diagram, first introduced in Paper I, as our key diagnostic tool for the radiative transfer and conversion law analysis that follows; an example is shown in {\color{red}Figure \ref{x300}}.  With the addition of a given abundance ratio $R_{12}$ = [\htwo]/[\tco], the \nco/\ico\ ratio is also an effective $X$ factor, so we also use the term ``$X$vs$I$ diagram'' as a shorthand.  We see this by connecting \ntco\ to the total molecular mass surface density via
\begin{eqnarray}
	\Sigma_{\rm mol} & = & N_{\rm H_2}~\mu_{\rm mol}~m_{\rm H} \\			%% THREE
			%    1.8785936 x 10^{-24} Msun.pc-2/(mol.m-2)
	& = & 1.88\,{\rm M}_{\odot}\,{\rm pc}^{-2}~N_{\rm^{12}CO}~R_{12}/(10^{24}{\rm molecules\,m}^{-2})~, \nonumber %$\sim$ 59\,kg\,m$^{-2}$,
\end{eqnarray}
where $\mu_{\rm mol}$ = 2.35 for 9\% He by number and $m_{\rm H}$ is the mass of the H atom.  This is analogous to formulations common to extragalactic studies, namely the use of $\alpha_{CO}$ in $M_{\rm mol}$ = $\alpha_{\rm CO}L_{\rm CO}$ \citep{bwl13} instead of \nhtwo\ = \xco\ico\ as we prefer here.  Eq.\,(3) makes the scaling to $\Sigma_{\rm mol}$ explicit, namely
\begin{equation}
	\alpha = \mu_{\rm mol}~m_{\rm H}~R_{12}~X.						%% FOUR
\end{equation}
Thus, substituting $N$=$XI$ with a standard value of $X_{\rm CO}$ = 2$\times$10$^{24}$\,molecules\,m$^{-2}$\,(K\,\kms)$^{-1}$ into either Eq.\,(3) or (4), we recover an $\alpha_{\rm CO}$ = 3.8\,M\solar\,pc$^{-2}$\,(K\,\kms)$^{-1}$, also a standard value.  Either way, our radiative transfer treatment highlights the non-constancy of the $X$ or $\alpha$ factors.  In all figures involving the column density solutions \ntco, we use it interchangeably with $\Sigma_{\rm mol}$ for convenience, at a single $R_{12}$ = 10$^{4}$.  However, we discuss other values for $R_{12}$, and implications thereof, at various points in the text.

\vspace{1mm}The most convenient thing about the $X$vs$I$ diagram is that we can directly compare different areas on the sky, and see numerically how the line ratios translate into the \tnt\ solutions, portrayed as a grid in this diagram.  Thus, where \itco\ is bright but \ittco\ is not very, this translates into low $N$ and therefore also low $X$ values, and the solutions reside in the high-\tex, low-$\tau$ part of the grid.  In contrast, where \ittco\ approaches \itco\ and they are both at least somewhat bright, this gives large $N$, large $X$, and we fall into somewhat lower-\tex\ but higher-$\tau$ loci in the diagram.

\vspace{1mm}One can see this effect more intuitively in the \tnt\ parameter RGB-composite images, i.e., Figures \ref{samplemos} and {\color{red}\ref{fullTexZMtau-mom0}}.  The highest column densities \nco\ in such images render as the brightest green, also ranging from yellow (where \tex\ is also high, but $\tau$ less so) to cyan (where $\tau$ is also high, but \tex\ less so).  In contrast, there are areas where either \tex\ (red) or $\tau$ (blue) are bright, but \nco\ (green) is not, leaving the area rendered in shades of blue through purple to red.  In other words, high \nco\ requires somewhat high values of {\em both} \tex\ and $\tau$.  By intercomparing the iso-CO line composites with the \tnt\ composites, we observe that high \nco\ is generally seen to occur where both the \tco\ and \ttco\ emission are bright, but not usually where just one of these lines is prominent.

\vspace{1mm}In brief, the different colours can be intuitively understood by easily conceptualised physical conditions and radiative transfer effects: we refer to this mental lens as ``eyeball radiative transfer.''

% Figure 4, sample of N gradient
\begin{figure*}[ht]
\centerline{\includegraphics[angle=0,scale=0.052]{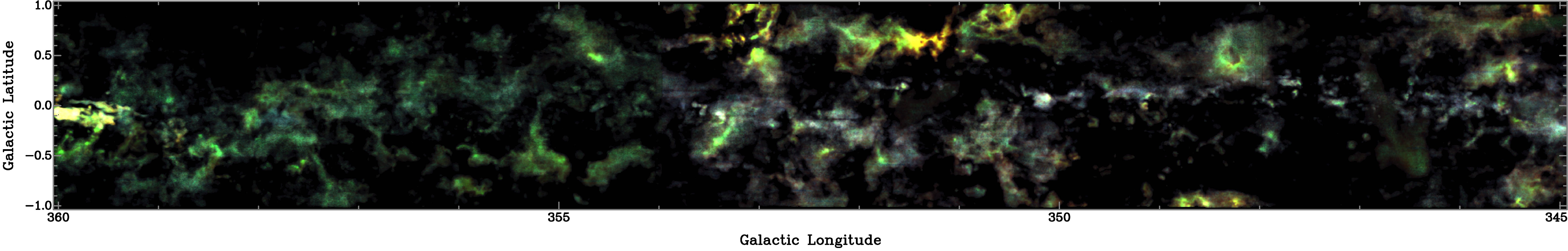}}
%\centerline{\includegraphics[angle=0,scale=0.052]{dr6-mosaicS678-TexZMtau.jpg}}
\vspace{-3.5mm}
%\centerline{\includegraphics[angle=0,scale=0.052]{dr6-mosaicS345-TexZMtau.jpg}}
\centerline{\includegraphics[angle=0,scale=0.052]{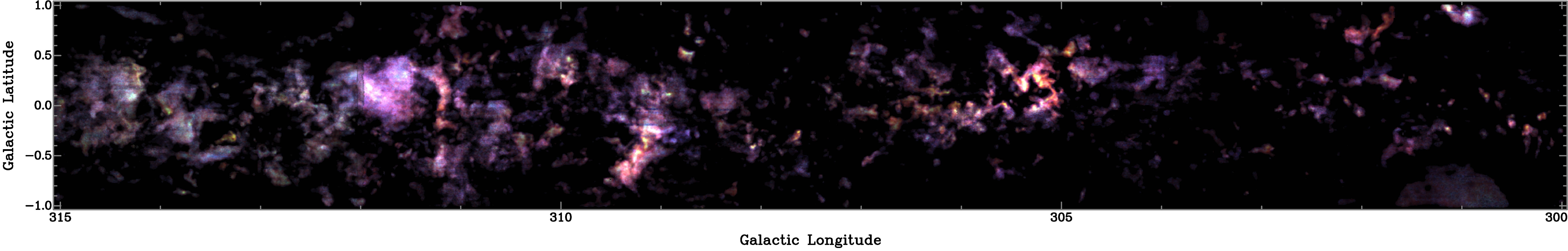}}
\vspace{-54mm}\hspace{12mm}{\bf {\color{red}\tex}\hspace{2mm}{\color{green}\nco}\hspace{2mm}{\color{cyan}$\tau_{\rm CO}$}}
\vspace{49mm}
\caption{\footnotesize More samples of the physical parameter composites (as labelled, similar to the lower panel of Fig.\,\ref{samplemos}), illustrating the overall gradient in \nco\ across the 4Q.  The colour scales here are identical in each panel, and the same as those in Fig.\,\ref{fullTexZMtau-mom0}. $$ $$
\label{Ngrad}}
\vspace{-9mm}
\end{figure*}

%%%%%%%%%%
%%   Section 4  %%
%%%%%%%%%%
\section{Atlas of Physical Properties}\label{atlas}

%%%%%%%%%%%
%%   Section 4.1  %%
%%%%%%%%%%%
\subsection{Column Density Maps}\label{coldens}

\vspace{1mm}The computation of solutions \tex,$\tau$ to Eq.\,(1) across a spatially-resolved map with 3$\times$10$^9$ independent ($l$,$b$,$V$) resolution elements, complete with a calculation of the partition function at each voxel and appropriately propagating uncertainties in each quantity, is a non-trivial exercise.  Moreover, in many locations the \ttco\ and especially \ceto\ can be quite weak, so there are S/N considerations in interpreting this analysis of our full-resolution data cubes.  (This is why we show cyan-coloured S/N limits in the $X$vs$I$ plots of Appendix \ref{rta}.)  Nevertheless, as described in \S\ref{method} we have performed this computation on the Sector cubes at {\bf {\em each}} 0.09\,\kms-wide voxel, as opposed to 1\,\kms\ velocity-binned versions of the data in Paper I and DR3.  After so computing the $\tau$ and \tex\ in each such voxel, we use Eq.\,(2) to derive cubes of column density per channel.

\vspace{1mm}The composite \tnt\ images in Figures \ref{samplemos} and \ref{fullTexZMtau-mom0} contain a lot of information.  While the mean \tex\ and $\tau$ values are around expected levels in most areas, the biggest surprise is just how large the \tco\ column density, or equivalently the molecular mass surface density, really is in many places.  The most extreme levels are around the CMZ: there, \ntco\ peaks at $\sim$ 2.5$\times$10$^{23}$\,{\rm m}$^{-2}$, which translates to $N_{H_2}\sim$ 2.5$\times$10$^{27}$\,{\rm m}$^{-2}$ and $\Sigma_{\rm mol}\sim$ 4900\,M\solar\,pc$^{-2}$.  Not far behind is the ministarburst G333 complex (see Paper II), with peak \ntco\ $\sim$ 2$\times$10$^{23}$\,{\rm m}$^{-2}$ and $\Sigma_{\rm mol}\sim$ 3700\,M\solar\,pc$^{-2}$.  Filling out the top 5 are NGC\,6334 (G351.4+0.6) at $\Sigma_{\rm mol}\sim$ 1700\,M\solar\,pc$^{-2}$, G345.5+0.3 at $\sim$1800\,M\solar\,pc$^{-2}$, and G327.3--0.6 at $\sim$ 2600\,M\solar\,pc$^{-2}$.  While impressive, these are not unique: there are many clouds ranging over 500--1500\,M\solar\,pc$^{-2}$ and even wider swaths down to 100\,M\solar\,pc$^{-2}$ and below.  The whole range is similar to the mass columns in parsec-scale dense clumps seen in previous surveys \cite[e.g.,][]{b18}, but is higher than what has traditionally been thought of as typical GMC mass columns of 100--200\,M\solar\,pc$^{-2}$ \citep{bwl13,hd15}, especially those derived from a single-valued $X$ factor.  Part of the reason for our higher $\Sigma_{\rm mol}$ values may also be due to ThrUMMS' much higher angular resolution than the classical wide-field studies.  %We can convert the above cool, opaque example to an equivalent $X_{\rm CO}$$\sim$ 8--40$\times$10$^{24}$\,m$^{-2}$/(K\kms), and so realise that the Milky Way's molecular gas mass may need to be revised upwards by a significant amount, at least in some locations; see the discussion in \S\ref{heavy}.

\vspace{1mm}Figure \ref{fullTexZMtau-mom0}, or as highlighted by {\color{red}Figure \ref{Ngrad}}, also illustrates very clearly that the overall column density gradually drops from $l$ = 360\degree\ to 300\degree, evident from the intensity gradient of the green shading across the 4Q.  This is not surprising, of course, since we have a longer path length through the Galactic disk towards the Center than in directions off to one side.  But this simple result is not at all evident in the iso-CO composite (Fig.\,\ref{full121318-mom0}), demonstrating in a very intuitive way that the radiative transfer treatment gives a much better physical representation of the overall molecular ISM than the emission lines do alone, radiative transfer assumptions notwithstanding.  % Quantify this in a figure??

%%%%%%%%%%%
%%   Section 4.2  %%
%%%%%%%%%%%
\subsection{Other Parameters and Higher Moments}\label{params}

\vspace{1mm}At the same time, we can also see that the excitation and opacity conditions are everywhere similar across the 4Q, since we find the same reddish-purple hues underlying any green cast at all longitudes.  From the data cubes underlying Figures \ref{Ngrad} or \ref{fullTexZMtau-mom0}, typical values for \tex\ range from 5 to 20\,K, with extreme values to 100\,K but a strong modal peak around 8\,K, and this distribution looks very similar everywhere.  For the \tco\ opacities, we find a broad range $\tau\sim$ 1--100+, where the extrema are not always well-determined, since they typically occur where the S/N in the \ttco\ data is low.  However, there is a broad peak in $\tau$ $\sim$ 20, but where values from 10 to 40 are quite common, and this pattern does not obviously vary with longitude.  This emphasises that the physical conditions {\em within} molecular clouds are likely to be very similar across much of the Galactic disk, but that the {\em amount} of molecular material in the disk is generally going to be larger towards the Center, as should be expected.

\vspace{1mm}The higher moments for both the emission lines and physical parameters are also instructive.  Although the \tco\ emission is more widely mappable than areas with radiative transfer solutions (due to the lower brightness and more limited emission area of the \ttco\ data), the \nco\ moments should nevertheless give a more representative sampling of the overall ISM kinematics than \tco, since moments in the latter could be distorted by opacity effects.  Comparing Figures \ref{full12co-mom1} and \ref{fullZM-mom1}, we find that the Galactic molecular velocity fields in each are broadly similar, yet do not agree in subtle ways.  For example, at some higher-latitude locations like NGC\,6334 and the Coalsack ($l$ = 300\degree--304\degree, $b$ $<$ 0\degree), the \tco\ emission extends to higher positive velocities than in \nco, presumably because it better traces the outermost gas in each cloud that least shares the bulk cloud kinematics.  In contrast, along the midplane in a number of locations, the \nco\ velocity field traces more negative values than \tco, presumably because these clouds lie at greater distances, but whose kinematic signal is masked in \tco\ by brighter (but lower column density) foreground emission.

\begin{figure}[t]
\hspace{-0.5mm}\includegraphics[angle=0,scale=0.11]{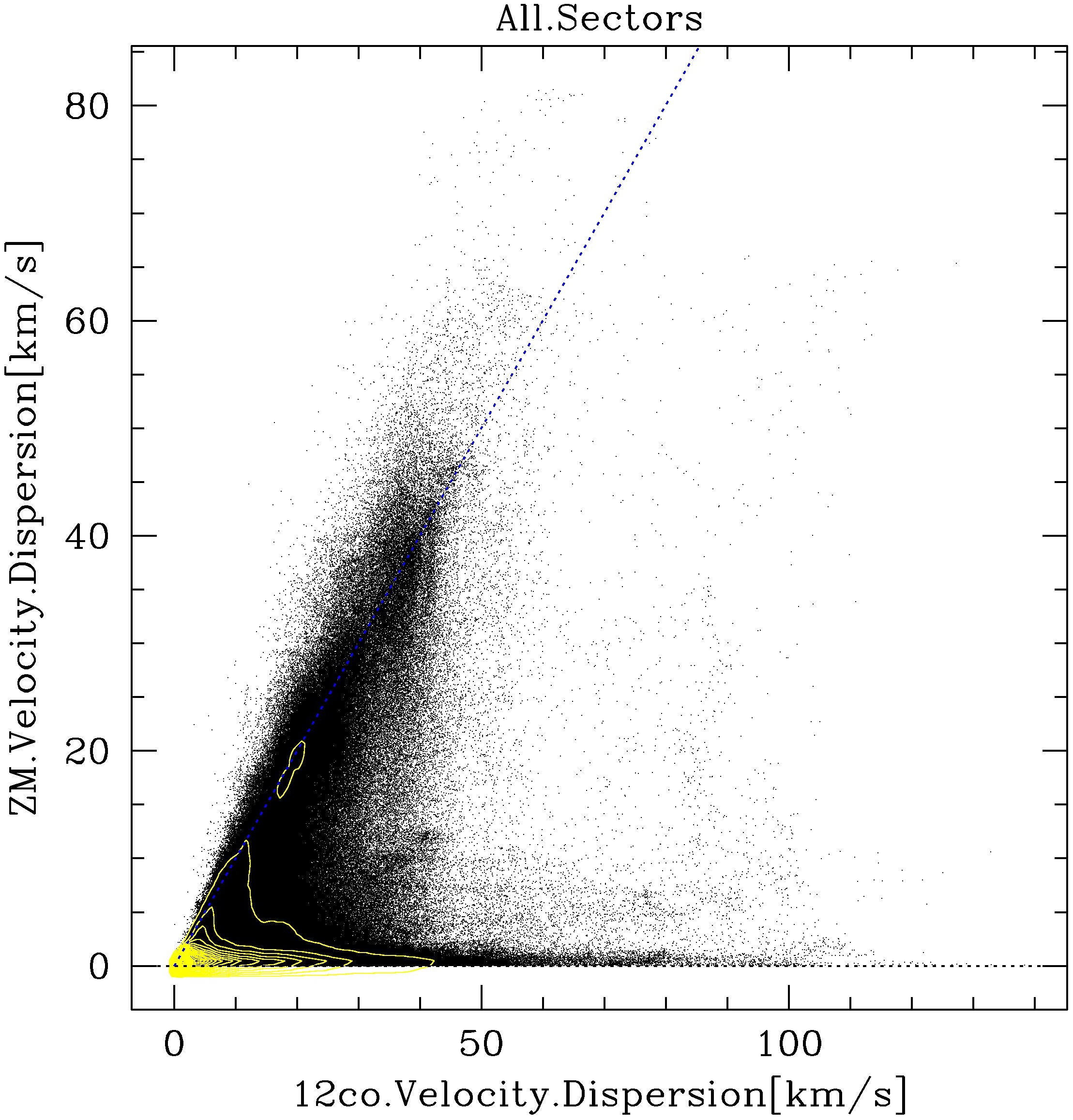}

\vspace{-1mm}
\caption{\footnotesize Comparison of \nco-weighted (labelled ``ZM'') and \tco-weighted velocity dispersions (2nd moments) across all 3$\times$10$^6$ \lb\ pixels.  Overlaid are yellow contours of pixel density at 2\%, 6\%, 10\%, ..., 98\% of the peak, showing that the data distribution in this space is not random, but rather, appears constrained to two separate, distinct domains.  The majority of pixels lie close to the dotted black line (along the x-axis); these pixels have $\sigma_V$(\nco) $\ll$ $\sigma_V$(\tco).  Nearly all other pixels lie close to the dotted blue line of equality, and so obey $\sigma_V$(\nco) $\approx$ $\sigma_V$(\tco).  $$ $$
\label{twodisps}}
\vspace{-13mm}
\end{figure}

\vspace{1mm}The differences become more striking when we consider the velocity dispersion of each tracer (Figs.\,\ref{full12co-mom2}, \ref{fullZM-mom2}).  Briefly, the \tco\ velocity dispersion distribution is {\em significantly} wider than that of \nco.  Indeed, in {\color{red}Figure \ref{twodisps}} we can see that there are two types of pixels: one where \sigv(\nco) $\approx$ \sigv(\tco) (the smaller fraction of all pixels), and another where \sigv(\nco) $\ll$ \sigv(\tco) (the much larger fraction).  Based on the distribution of points in this figure, we find a rough division between these two domains near \sigv(\nco) = 2\,\kms, although the value of this threshold is less important than its existence (see \S\ref{vdisp} for more details).  This dichotomy is likely a manifestation of the more diffuse (and more widely dispersed) molecular ISM being better traced by \tco, while the denser molecular ISM is better traced by \nco.  We show further in \S\ref{vdisp} that this is primarily an opacity effect, and underscores an important point: \tco\ can't be a good tracer of the overall conditions in molecular gas, but rather is biased towards tracing more diffuse molecular gas.  That is, since \sigv(\nco) does not track \sigv(\tco) in general, the majority small-\sigv\ domain seems to signpost those cloud envelopes where radiative trapping of \tco\ emission is relatively important.  This is likely also related to the nonlinear conversion laws we find with more formal calculations (see \S\ref{convlaw}).

% Figure 5: bimodal dispersions

\vspace{1mm}We also present $lV$ diagrams (integrals across all $b$ in the \lbv\ cubes) in \S\S\ref{LVmaps}--\ref{latmaps} for both the \tco\ and \nco\ cubes.  Such diagrams are a standard tool for analysis of such topics as spiral structure, the arm-interarm contrast, or kinematic distances.  Included are higher moments of these diagrams, i.e., integrals that are first and second moments in $b$, as well as the more usual zeroth moments.  Here again, while the \tco\ moments trace more features, the \nco\ maps highlight which of them contain the denser gas.

% Figure 6: slopes summary
\begin{figure*}[ht]
\centerline{\includegraphics[angle=0,scale=0.22]{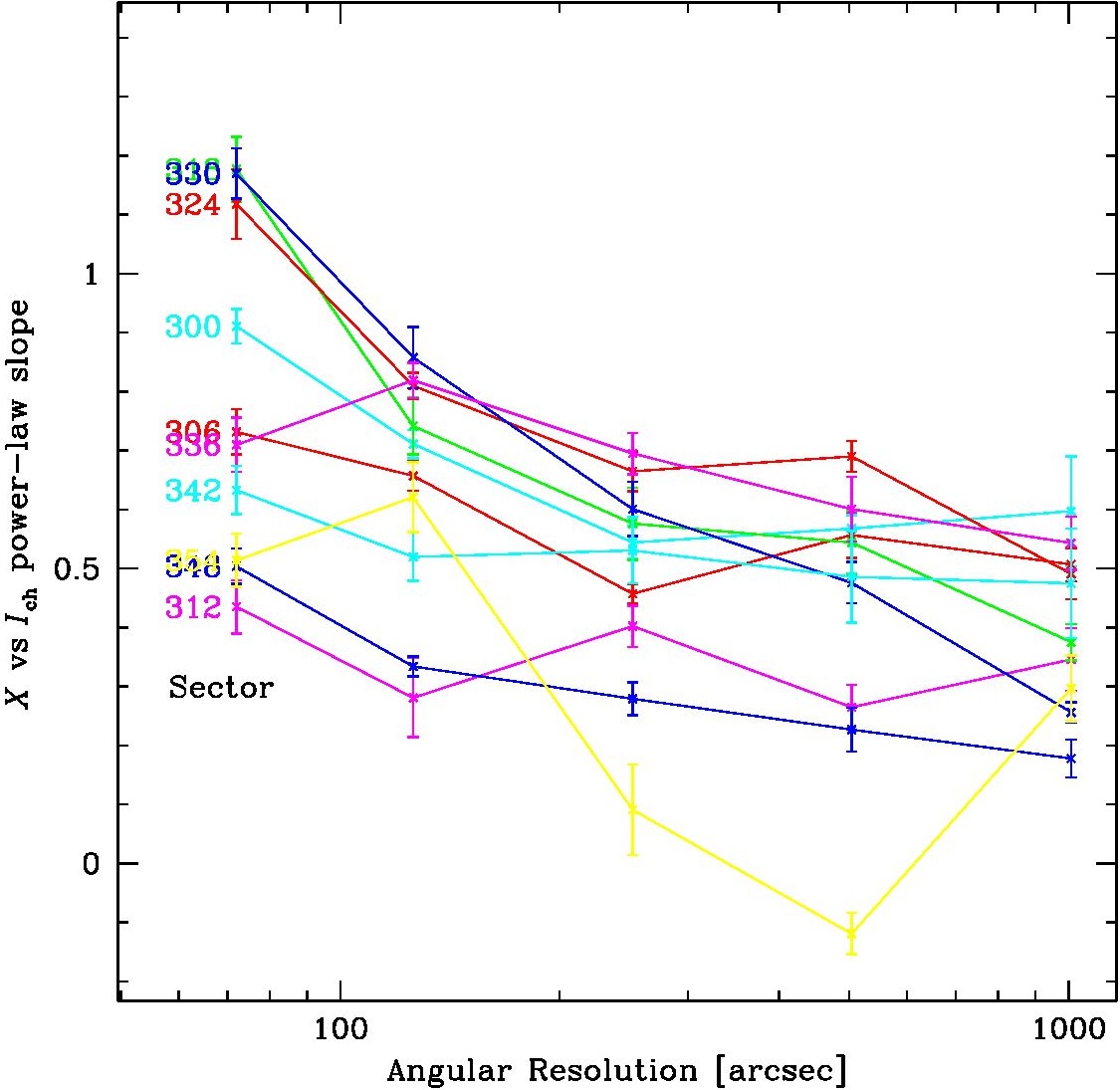}~~~~\includegraphics[angle=0,scale=0.22]{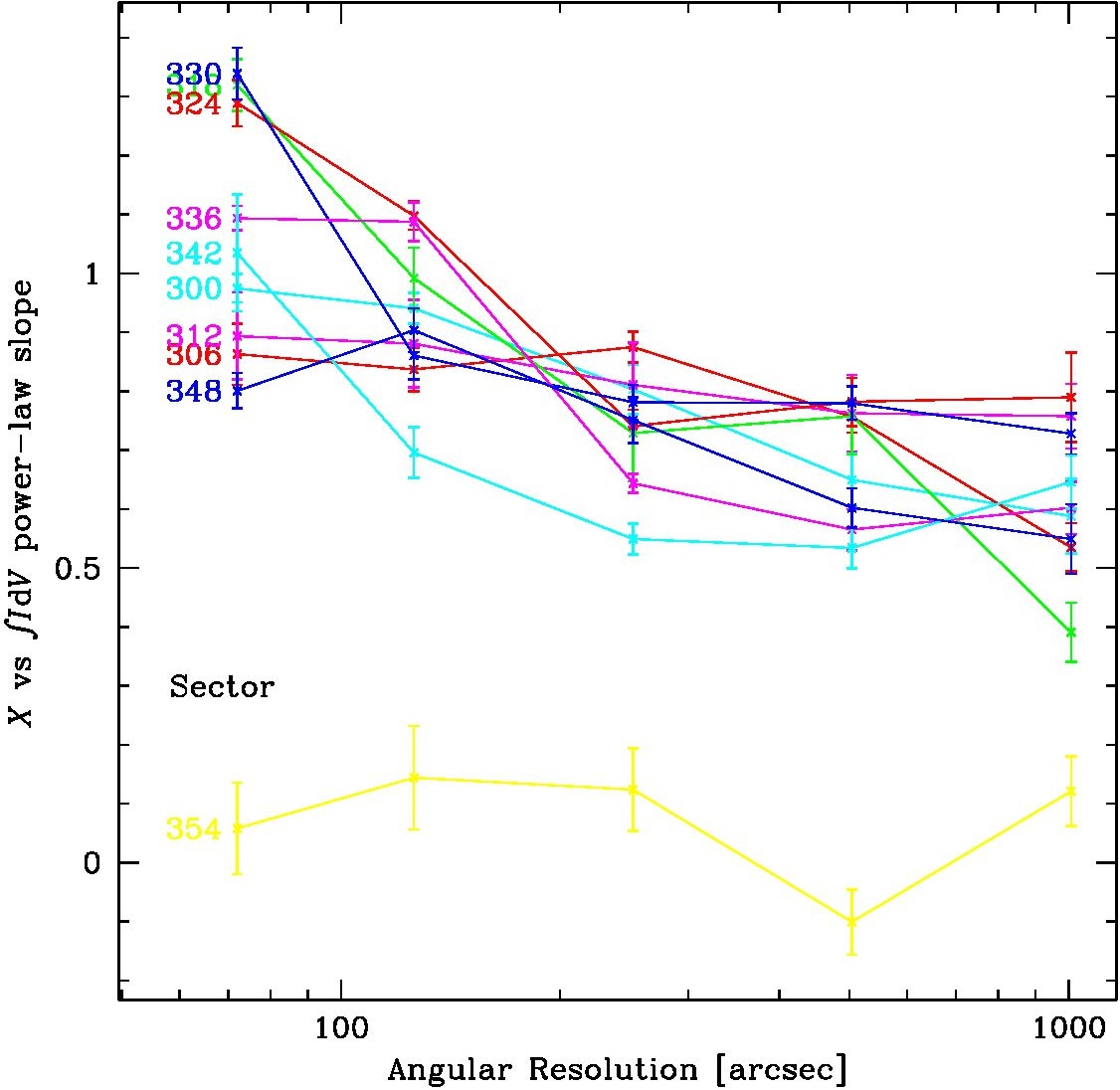}}
\vspace{-1.5mm}
\caption{\footnotesize Summary of conversion law analysis, where the index $p$ in Eq.\ (5) equals 1+the slope as measured in Fig.\,\ref{x300}, and its brethren described in Appx.\,\ref{rta}.  ({\em Left}) All measured slopes from the per-voxel analysis (panels of Figs.\,\ref{x300-12-multi}--\ref{x354-multi}) labelled by Sector.  ({\em Right}) All measured slopes from fits to the velocity-integrated data (panels of Figs.\,\ref{xcl300-12-multi}--\ref{xcl354-multi}). $$ $$
\label{slopes}}
\vspace{-9mm}
\end{figure*}

%%%%%%%%%%
%%   Section 5  %%
%%%%%%%%%%
\section{Global Conversion Laws}\label{convlaw}

%%%%%%%%%%%
%%   Section 5.1  %%
%%%%%%%%%%%
\subsection{The Formal Result}\label{heavy}

\vspace{1mm}Paper I presented the first radiative transfer analysis of the CO-isotopologue line ratios in the 3$\times$10$^9$ voxels of the DR3 PPV data cubes.  From this we found a distinctly non-linear relation was needed to convert \ico\ to \nco, $N$ $\propto$ $I^p$, where the average power-law index $p$ = 1.38$\pm$0.10 across the DR3 survey area.  This was done with 4-channel binning in the ThrUMMS data cubes in order to improve the S/N in the analysis.  In the CHaMP project, \cite{b18} used the same technique to analyse the higher-sensitivity iso-CO data at the full angular and velocity resolution of the Mopra telescope, and inferred an even steeper power-law index $p$ = 1.92$\pm$0.05, aggregated over all the CHaMP data.

\vspace{1mm}In Appendix \ref{rta}, we present the results of this same analysis but now applied to the full velocity resolution of the ThrUMMS data (broken down by Sector -- see Fig.\,\ref{x300} for an example).  We also do this for progressively convolved versions of the ThrUMMS data, in order to investigate the dependence of any conversion laws on angular, and hence physical, resolution.  In all cases, the analysis for S354 is affected by the bright, high-excitation \tco\ emission from the area around the Central Molecular Zone (CMZ), and so must be discounted in order to arrive at the disk-averaged cloud properties.  From the summary plot in {\color{red}Figure \ref{slopes}}, we see that the other 9 Sectors have a mean$\pm$SEM index $p$ = 1.82$\pm$0.10 at 72$''$ resolution (the leftmost 9 points in the left panel), lower than but consistent with the CHaMP result.

\vspace{1mm}Interestingly, the convolved ThrUMMS data yield progressively lower indices as the resolution is degraded.  At resolutions of 126$''$, 252$''$, 504$''$, and 1008$''$, we respectively obtain $p$ = 1.64$\pm$0.07, 1.53$\pm$0.04, 1.49$\pm$0.05, and 1.42$\pm$0.05, clearly an internally consistent trend in the ThrUMMS data, and also when combined with CHaMP.  We consider this significant because these angular resolutions correspond to relevant physical scales for many observational studies, both of Galactic molecular clouds and extragalactic cloud populations: that of the pc-scale clumps known to be the precursors of the star clusters that are the main products of star formation in these clouds.  Since a large fraction of the molecular material we map lies along the Scutum-Centaurus spiral arm of the Milky Way (see Appx.\,\ref{kinem}), we can take its average heliocentric distance of 3\,kpc as typical of the data.  Then these angular resolutions (including CHaMP's 37$''$, since most of the CHaMP clouds coincidentally also lie at $\sim$2--3\,kpc, but in the Carina Arm) correspond to physical resolutions of 0.54, 1.05, 1.83, 3.67, 7.33, and 14.7\,pc respectively.  %14.6607657

\vspace{1mm}In this context, we can understand this trend as a function of whether the intrinsic star formation properties of cluster-forming clumps are being adequately mapped.  At 1008$''$ \citep[the resolution of the Columbia-CfA survey;][]{dht01}, or 15\,pc resolution at 3\,kpc, one would have to say, probably not really, since there $p$ = 1.42 $\neq$ 1.  This must be even more true for many extragalactic studies where sub-kpc scale resolutions of nearby ($\sim$10\,Mpc) galaxies are at the cutting edge of modern observational capabilities \citep[e.g.,][]{L25}.  So at 100\,pc resolution (say), which obviously doesn't resolve the pc-scale star formation we observe in the Milky Way, nor even the high mass columns $\Sigma$ $>$ 1000\,M\solar\ that we find in some locations from the radiative transfer solutions, we might project the correct index to be $p$ $\approx$ 1.25 based on the above trend, but still not 1 (corresponding to an $X$vs$I$ slope of 0).

\vspace{1mm}At the same time, we can turn this argument around: even at 0.5\,pc, are we resolving all the important physics of star formation?  Certainly not!  For example, the low-mass, individual star formation that occurs in the $\sim$0.1--0.2\,pc-wide molecular filaments revealed by {\em Herschel} is also very widespread across the ISM.  So, our result of $p$ topping out around 2 doesn't mean that this power-law index can't rise to even higher values on even smaller scales.  But this underscores our main point: assuming $p$ = 1 at any scale is probably going to lead to various biases (in such things as mass scalings, etc.) that, where possible, should be avoided.  

\vspace{1mm}There are several other fundamental reasons to expect $p$$>$1, which we discuss in \S\ref{implics}.

%%%%%%%%%%%
%%   Section 5.2  %%
%%%%%%%%%%%
\subsection{A Practical Conversion Law, with Caveats}\label{practical}

\vspace{1mm}The radiative transfer analysis we have described, particularly the use of the $X$vs$I$ diagram (Fig.\,\ref{x300}, which we also call a ``V-plot,'' for velocity-resolved or per-channel), is a formally exact approach to obtaining the masses of molecular clouds.  However, this method includes the velocity resolution of the radio telescope's spectrometer as a factor in the calculations.  Obviously, different telescopes have different spectrometer hardware as part of their systems, and so, while we are confident that similar results could be obtained for $p$ at other facilities with similar angular resolutions, the normalisations $N_{0}$ will change with each instrument, limiting the practicality of this method to other investigators.

\vspace{1mm}Fortunately, there is a simple solution to this conundrum.  From the {\em velocity-integrated} analysis of \cite{b18}, ``I-plots'' not only provide a robust normalisation and avoids the channel-width issue, but are also more physically useful for characterising whole clouds (or portions thereof) on any scale.  In other words, the integral of the spectral line emission over all velocities, $\int$$I$d$V$, is the usual measurement made in any case.  One might object that integrals should be made only over the somewhat more narrowly-defined velocity ranges of individual clouds, rather than the much wider range of velocities representing the whole-disk emission at a given pixel.  However, as we have already described (\S\S\ref{params},\ref{vdisp}), the velocity dispersion (2nd moment) maps made from the \nco\ cubes have, for the vast majority of pixels, quite small dispersions.  For example, from Figure \ref{12vZMdisp} panel $d$, we see that 80\% of pixels have $\sigma_{V}$ $<$ 6\,\kms, 85\% are under 9\,\kms, and 90\% under 15\,\kms, all of which could easily be attributable to typical dispersions in single GMCs or cloud complexes.  Therefore, distortions to our statistics based on ``over-integration'' should be fairly minimal.  Then, while the velocity interval over which the integration is made cannot be formally defined, nor can an underlying grid of $\tau$,\tex\ solutions, we would argue that the radiative transfer approach and V-plot results strongly underpin the conclusions that follow.

\vspace{1mm}We have therefore also compiled the I-plot results for the Sector-by-Sector ThrUMMS data in \S\ref{vintlaws}, for all resolutions (native and convolved) as described above.  These are summarised in the right panel of Figure \ref{slopes}, which show an even tighter correlation with angular resolution among the 9 Sectors (not including the CMZ) than for the V-plots.  The mean$\pm$SEM result at 72$''$ is $p$ = 2.07$\pm$0.07, and at the progressively lower resolutions, 1.92$\pm$0.04, 1.84$\pm$0.03, 1.69$\pm$0.03, and 1.62$\pm$0.04; the last value is also $\neq$1.

\vspace{1mm}However for Sector 354, the I-plot $p$$\approx$1 (the slope is 0) at all resolutions.  This actually comports with several recent studies showing that molecular material in the brighter central regions of disk galaxies is dissimilar to clouds in their disks generally \citep[e.g.,][]{L25}, so our result for S354 is not surprising in that context.  But the difference between that and the other 9 Sectors is stark, and emphasises that interpretation of underresolved extragalactic data, where bright nuclear emission can overwhelm a fainter disk, is done at one's peril.

\vspace{1mm}More intriguingly, the CHaMP I-plots show an average $p$ = 1.68$\pm$0.06, lower than most of the ThrUMMS Sectors; see Figure \ref{chCLfits}.  The explanation seems to be that the more sensitive CHaMP maps include clouds which are of fairly low luminosity, where the nature of the radiative transfer solutions change.  This is visible in the $\tau$,\tex\ grid underlying the V-plots (Fig.\,\ref{x300}).  Even with extremely low-noise data, the $X$vs$I$ trend in such plots must eventually flatten when $I$ \lapp\ 0.3\,K\kms, because the \tnt\ solutions put a hard floor on the minimum $N$/$I$ ratio that is physically observable.  Apparently the CHaMP data have sufficient sensitivity in some areas for this effect to appear.  As well, we note that the ThrUMMS Sectors all cover much larger areas than the CHaMP maps, and as such, all of them have at least some features which are fairly bright, extending the span of high-$I$ data and ensuring that trend is well-sampled.  A number of the CHaMP maps of fainter clouds do not have such features.

\vspace{1mm}Thus, the CHaMP I-plots caution us that, while $p$ $\sim$ 2 is appropriate for large enough samples of molecular clouds as appear in the ThrUMMS maps, $p$ is expected to decline towards 1 as the faint-cloud population is better sampled.  As a practical matter, however, we can see that an index of 2 is appropriate for most large-scale (e.g., >1\,kpc) molecular cloud surveys.

\vspace{1mm}To conclude, the normalisation for the conversion law is best estimated by an average of the 9 ThrUMMS I-plots, described in \S\ref{vintlaws}.  Our recommended conversion law (in both Galactic and extragalactic formats) is then
\begin{eqnarray}   % 
	\Sigma_{\rm mol} & = & \mu_{\rm mol}~m_{\rm H}~R_{12}~N_{0}~I_{\rm^{12}CO}^{p} \\			%% FIVE
			%    1.8785936 x 10^{-24} Msun.pc-2/(mol.m-2); then x (9.1x10^{17}x10^{4}) = 1.709520176e-2
	& = & \alpha_{0}\left(\frac{R_{12}}{10^{4}}\right)\left[\frac{I_{\rm^{12}CO}}{\rm K\,km\,s^{-1}}\right]^{p}~, \nonumber %$\sim$ 59\,kg\,m$^{-2}$,
\end{eqnarray}
where the mean$\pm$SEM values of the coefficients are $N_{0}$ = (9.10$^{\times}_{\div}$1.41)$\times$10$^{17}$\,molecules\,m$^{-2}$, $\alpha_{0}$ = (1.71$^{\times}_{\div}$1.41)$\times$ 10$^{-2}$\,M\solar\,pc$^{-2}$, $p$ = 2.00$\pm$0.07 (the dispersions in these quantities are about 2$\times$ larger), and the other quantities are as previously given.  For example, at a pixel with \itco\ = 50\,K\kms, we obtain $\Sigma_{\rm mol}$ = 43\,M\solar\,pc$^{-2}$; at \itco\ = 200\,K\kms, the mass surface density rises to $\Sigma_{\rm mol}$ = 684\,M\solar\,pc$^{-2}$.

\vspace{1mm}This conversion law applies to \itco\ levels from 20--350\,K\kms\ in typical Galactic disk molecular clouds, and is of course calibrated to an angular/physical resolution of $\sim$0.6--1\farcm2/0.5--1\,pc.  At other resolutions, the normalisations $N_{0}$/$\alpha_{0}$ and index $p$ will vary somewhat from the above.  Furthermore, a local estimate of $R_{12}$ is also needed for an accurate $\Sigma_{\rm mol}$ to be obtained in a given situation, whether cloud or galaxy; see \cite{p21} for more background on this topic (they found an average value $R_{12}$ $\approx$ 16667 in the CHaMP clouds, but also that $R_{12}$ depends on \td).  Otherwise, one can use the information herein to estimate alternate formulations at other resolutions within the range discussed.

\vspace{1mm}Also, based on the radiative transfer solutions and more sensitive CHaMP data, we expect the conversion law index $p$ to drop to $\sim$1.5 or so for 10\,K\kms\ $<$ \itco\ $<$ 20\,K\kms\ and eventually to $\sim$1 for even lower \itco, with a concomitant change to higher values of $N_{0}$.  Among the nuclear molecular clouds, however, $p$ $\sim$ 1 at all resolutions, apparently due to the much higher excitation conditions there, which change the character of the radiative transfer conditions (i.e., distinctly higher \tex\ and lower $\tau$ than disk clouds).

% Figure 7: ZM-ld0 with BGT
\begin{figure}[t]
\vspace{0mm}
\centerline{\includegraphics[angle=0,scale=0.806]{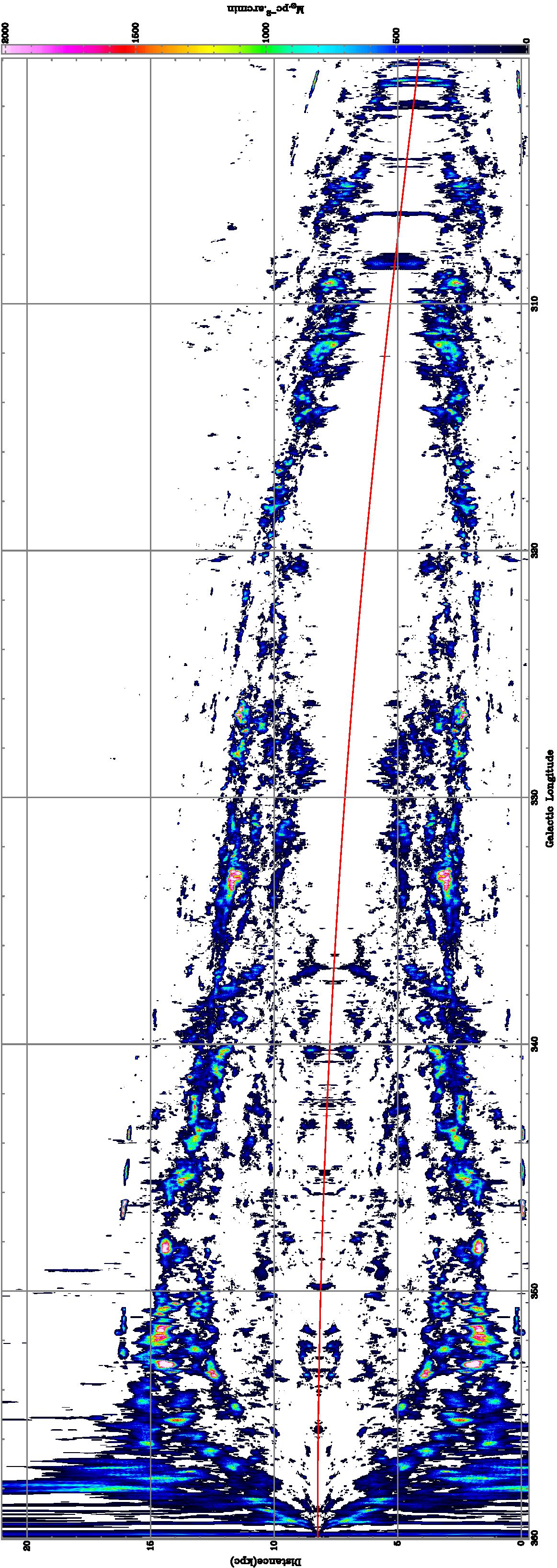}}
\vspace{-2mm}
\caption{\footnotesize Sample deprojection of \lv\ data (in this case, integrated \nco) into an \ld\ map using our favoured BGT distance scale (Table \ref{rotpars}).  The mirroring red tangent-point curve runs from (360\degree, 8.24\,kpc) to (300\degree, 4.12\,kpc) in the BGT model. $$ $$
\label{sample-ld0}}
\vspace{-9mm}
\end{figure}

%%%%%%%%%%
%%   Section 6  %%
%%%%%%%%%%
\section{Kinematic Analysis and Global 3D Architecture}\label{3d}\label{arch}

%%%%%%%%%%%
%%   Section 6.1  %%
%%%%%%%%%%%
\subsection{Refinements to Rotation Parameters for the Southern Milky Way}\label{newrotpars}

\vspace{1mm}With surveys of the Milky Way, especially ones at radio wavelengths which can potentially image the entire disk, there is a storied history of using the (presumed) orderly rotation of the Galaxy's constituent particles (i.e., stars, nebulae) to deproject our two-dimensional view within the Galactic Plane into some estimate of the Milky Way's 3D structure \citep[e.g.,][]{mb81}.  Naturally, we can explore this topic with the ThrUMMS data as well (see {\color{red}Appx.\,\ref{kinem}}).

\vspace{1mm}To do so, we need to examine the details of the orbital model.  One of the most widely-used set of rotation parameters comes from the BeSSeL project \citep{r19}, based on painstaking VLBI measurements of masers in many star-forming regions at various distances, but there are other efforts in this area as well, such as the VERA project \citep{o24}.  The {\em Gaia} mission has also expanded the reach of optical astrometry to heliocentric distances $>$1\,kpc, which has allowed more classical methods to better anchor Galactic rotation parameters \citep[e.g.,][]{bob23}.  But while {\em Gaia} studies are essentially all-sky, the maser VLBI work to date has been largely limited to the northern sky, due to the lack of sensitive VLBI antennas in the Earth's southern hemisphere.  As a result, for the 4Q we first critically review the various parameters before attempting any deprojection, as described in \S\ref{lsr}.

\vspace{1mm}Our reasoning was that, with the existing rotation parameters, the population of local molecular clouds (i.e., ones known to be at distances \lapp\ 200\,pc or so) should exhibit very small velocity offsets (if any) from a properly calibrated Local Standard of Rest (LSR), as adjusted by the VLBI results to date.  This turns out not to be true: there is a systematic sinusoidal residual in these clouds' \vlsr, visible in the \lv\ diagrams such as {\color{red}Figures \ref{full-lv0-12coZM-kHiau}--\ref{full-lv0-12coZM-kHbgt}}, and we solved for this additional adjustment from the ThrUMMS data.  We also experimented with the global rotation parameters $R_0$,$\Theta_0$ with different combinations of the BeSSeL, VERA, and {\em Gaia} results, to optimise the fit to all our 4Q data, primarily in the \nco\ \lv\ diagrams, but also with cross-checks from the \tco\ data.  Our favoured ``BGT'' combination of parameters, a hybrid of BeSSeL, {\em Gaia}, and ThrUMMS results, is tabulated with other combinations in {\color{red}Table \ref{rotpars}}.

% Figure 8: ZM-YX1 with BGT
\begin{figure}[t]
\vspace{0mm}
\centerline{\includegraphics[angle=0,scale=0.47]{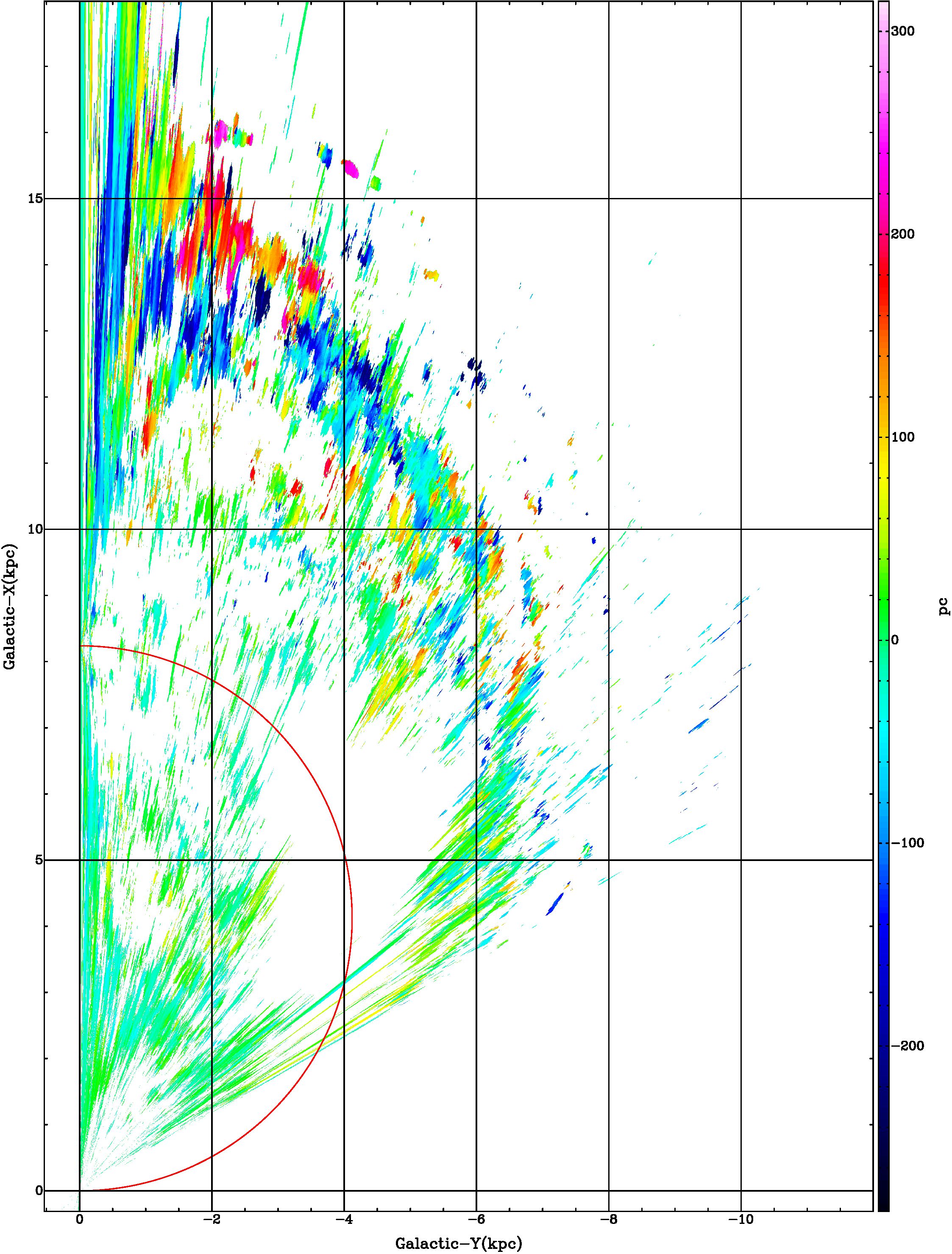}}
\vspace{-1mm}
\caption{\footnotesize Sample deprojection of \lv\ data (mean height of \nco) into an \xy\ map on the BGT distance scale.  The mirroring red tangent-point curve appears as a circle in this diagram. $$ $$
\label{sample-YX1}}
\vspace{-9mm}
\end{figure}

%%%%%%%%%%%
%%   Section 6.2  %%
%%%%%%%%%%%
\subsection{Kinematic Deprojections}\label{geom}

\vspace{1mm}Having settled on an optimal rotation model for the 4Q, we then constructed two different deprojections from \lbv\ space to true 3D physical space: an \lbd\ deprojection (i.e., heliocentric polar coordinates) and an \xyz\ deprojection (heliocentric Cartesian coordinates), as described in \S\ref{mwmaps}.  For purposes of this paper, this was actually done in 2D only, i.e., from \lv\ to either \ld\ or to \xy, and then $z$ recovered directly from $b$ in individual cases as discussed (the code developed to do this was designed to work in 2D or 3D, as needed).

% Figure 9: ZM-ld1h-zp with BGT
\begin{figure}[t]
\vspace{0mm}
\centerline{\includegraphics[angle=0,scale=0.826]{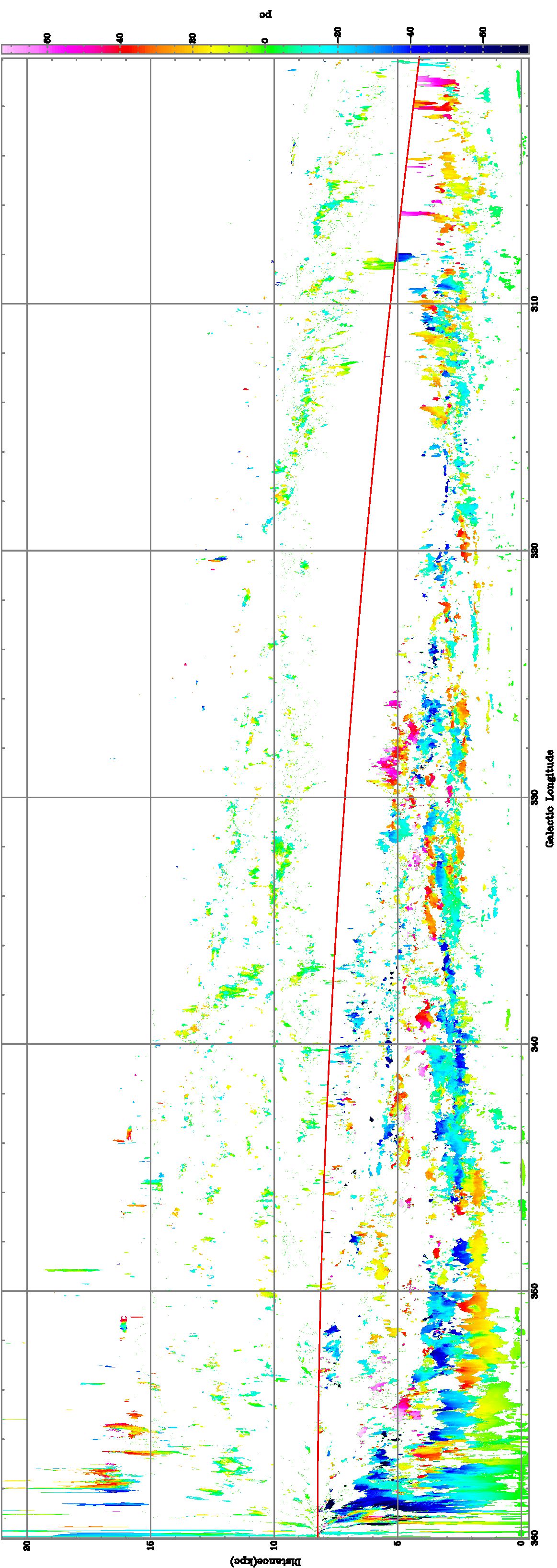}}
\vspace{-2mm}
\caption{\footnotesize Sample disambiguated $\bar{z}$ map from \nco\ data using the $\zeta^{+}$ function as described in \S\ref{nearfar}. $$ $$
\label{sample-ld1z}}
\vspace{-9mm}
\end{figure}

% Figure NOT REALLY: ZM-ld0-zp
%\begin{figure}[t]
%\vspace{0mm}
%\centerline{\includegraphics[angle=0,scale=0.806]{dr6-ZM-URCbgt-ld0-b1p3m.jpg}}
%\vspace{-1.5mm}
%\caption{The same \ld\ map of integrated \nco\ as in Fig.\,\ref{sample-ld0} but with the $\zeta^{+}$ filter applied.
%\label{sample-ld0z}}
%\end{figure}

\vspace{1mm}An example is shown in {\color{red}Figure \ref{sample-ld0}}, for the \nco\ \lv\ data integrated over all $b$ (0th moment) and deprojected onto an \ld\ frame.  Of course, for objects inside the solar circle, one obtains two solutions to the kinematic distance (``near'' and ``far''), so the original \lv\ data are partially duplicated in \ld\ space, and this is indicated in our \ld\ diagram by a red curve corresponding to the tangent-point distance of the kinematic solutions: in this projection, it is a kinematic mirror.  For the higher $b$-moments --- i.e., mean latitude $\bar{b}$, latitude dispersion $\sigma_{b}$ --- these can be further converted to $z$-moments --- mean height $\bar{z}$, thickness $\sigma_{z}$ --- from the computed distance scale, as described in \S\ref{height}.  {\color{red}Figure \ref{sample-YX1}} shows another example, but this time deprojected into \xy\ space.

%%%%%%%%%%%
%%   Section 6.3  %%
%%%%%%%%%%%
\subsection{Height Distributions and Disambiguation}\label{disamb}

\vspace{1mm}At this point in Galactic Plane surveys, one is faced with how to choose between the near and far distances, since only one can be physically acceptable for a given cloud.  Standard techniques include associating the object or region under study to matching data at other wavelengths (such as dust extinction measures, main sequence fitting of clusters, HI absorption or lack thereof, etc.)\ and using such data to give more definitive distances, or at least rule out one of the pair as less likely.  For example, \cite{dc20} used a combination of such techniques to compile a disambiguated list of clouds from the SEDIGISM survey in the 4Q.  Since ThrUMMS' latitude coverage is 2$\times$ that of SEDIGISM and our cloud definition algorithms would be based on different tracers, a similar effort here would be a useful comparison (Barnes et al., in prep.).  But such efforts depend crucially on how clouds or structures are defined, and according to the Krumholz principle, structure-agnostic procedures might be preferred.

\vspace{1mm}We have developed such a method and describe it in \S\ref{nearfar}.  The $\zeta^{+}$ likelihood function is a logically simple way to use both the scaled $b$-moments, i.e., $\bar{z}$ and $\sigma_{z}$, in either the \ld\ or \xy\ projections, in order to rapidly compute a near/far mask for any dataset on the same grid, such as the 0th-moment \nco\ maps.  The essential idea is to compare a cloud's (or actually, a pixel's) vertical size and mean height with the known scale height $z_{\rm sc}$ of the molecular layer \citep[19\,pc, according to][]{r19}, and design a numerical function of these 3 values that scores the likelihood of the combination being either near or far.  The discrimination is to then choose the more likely option at each pixel.  For example, the function will score a low likelihood if the pixel's height and size are too ``small'' or too ``large,'' but a higher likelihood if they are comparable to $z_{\rm sc}$.

\vspace{1mm}{\color{red}Figure \ref{sample-ld1z}} shows an example of what this looks like for the same $\bar{z}$ data as in Figure \ref{sample-YX1}, but in \ld\ space.  Where each pixel had its corresponding kinematic mirror pixel in Figures \ref{sample-ld0} \& \ref{sample-YX1}, this is no longer true in Figure \ref{sample-ld1z}: the $\zeta^{+}$ filtering has automatically masked out the lower-likelihood kinematic option among all mirror-pixel pairs.  As a gratifying bonus, this masked map has several desirable features that arise without any ``fine-tuning.''  (1)  Unreasonable ``far-large'' structures have been largely masked into the near domain, resulting in an appropriate $\bar{z}$ distribution on both sides of the mirror.  (2) Most pixels are placed in the near domain, as would be expected based on sensitivity arguments alone.  Nevertheless, there are also clearly some structures that seem to belong in the far domain, with small values for $\bar{z}$ and $\sigma_{z}$ even at the far distances --- they are probably too ``flat'' to be in the near domain.  (3) The discriminated structures are largely contiguous and mostly not random; the relatively small number of isolated pixels either below the mirror or above it turn out to be at the lower end of the \nco\ scale, possibly tracing the smallest clouds in our maps.

% Figure 10: mean heights vs distance
\begin{figure*}[ht]
\vspace{0mm}
\centerline{\hspace{-9mm}\includegraphics[angle=0,scale=0.16]{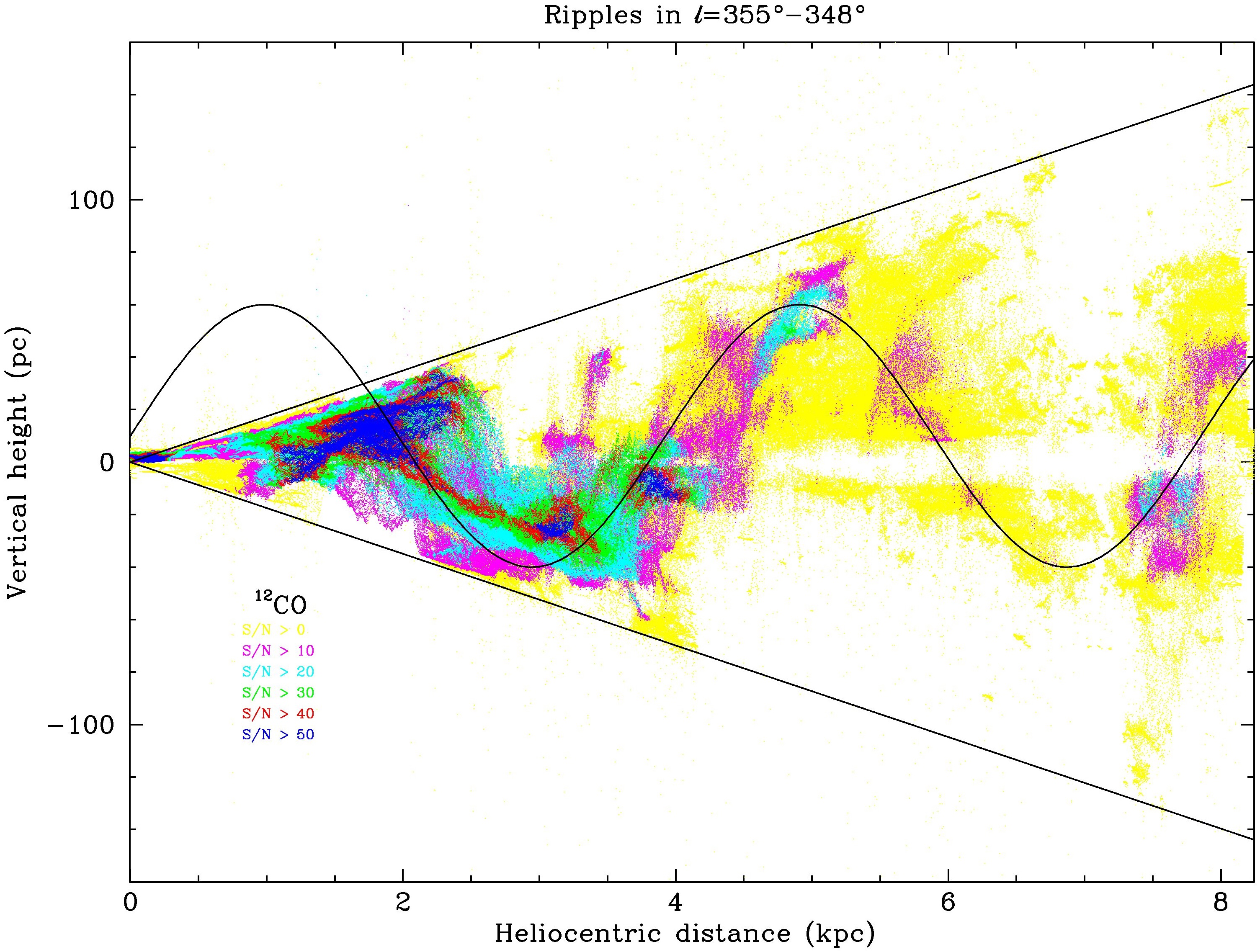}}
\vspace{-1mm}
\caption{\footnotesize Mean height vs distance for each pixel in 355\fdeg3 $>$ $l$ $>$ 348\fdeg0 from the \ld\ map of \tco.  The dots for each pixel are colour-coded by their S/N in the moment-0 map, as labelled, where the noise $\sigma_{\rm rms}$ $\approx$ 7\,K\,arcmin (the limiting S/N$>$0 for the yellow dots is actually more like S/N$>$4 given the thresholding inherent in the SAMed data).  Also overlaid is a rough visual fit of a sinusoid to the maximal \tco\ ridgeline.  This has wavelength 4\,kpc and amplitude 50\,pc, offset by $z$ = +10\,pc.  The sloping lines indicate ThrUMMS' latitude limits of $\pm$1\degree, with a vertical exaggeration of about 20:1. $$ $$
\label{awiggle}}
\vspace{-7mm}
\end{figure*}

\vspace{1mm}For the deprojected, $\zeta^{+}$-filtered, integrated \tco\ and \nco\ maps overall (e.g., Figs.\,\ref{12co-bgt-YX0zp} and \ref{ZM-bgt-YX0zp} in \S\ref{nearfar}, the $\zeta^{+}$-filtered versions of Figure \ref{sample-ld0} but in \xy\ space), we see something a little unexpected: as traced by the densest molecular material, there is {\bf \em only one} prominent spiral arm in the inner 4Q, approximately aligned with \cite{r19}'s fit for the Scutum-Centaurus Arm.  That is, despite our efforts to match the various rotation parameters to the 4Q data presented here, the spiral arm patterns fitted to the N hemisphere maser data and other prior works (in particular, the Sagittarius-Carina and Norma Arms) do not really do a good job of matching up with the overall location of the most massive molecular clouds in the inner 4Q.  This can be seen perhaps more clearly in Figures \ref{12co-bgt-YX0fl} (\tco) and \ref{ZM-bgt-YX0fl} (\nco), which also have the \cite{r19} spiral arms overlaid.  There, although the ridgeline of highest-$\Sigma$ material follows at least part of the Sct-Cen arm over about 40\degree\ of longitude, the rest at $l$ \lapp\ 320\degree\ meanders somewhat between the supposed Sct-Cen and Sgr-Car arm fits in a rather unsatisfying way.  \cite{r19}'s Norma arm seems not to correlate with anything significant in our deprojections.

\vspace{1mm}In contrast, the ridgeline of peak-$\Sigma$ or -\ico\ material does follow the pattern of optical/IR-derived dust features of \cite{z25} extremely well, and noticeably better than rotation models without the ThrUMMS -derived \uvw\ parameters for Solar motion (Table \ref{rotpars}).  Given that the BGT model was developed purely with the molecular data, this is very encouraging.  While we certainly realise that the ThrUMMS maps are not definitive, with the \cite{z25} results, they strongly suggest that the existing spiral arm models may need some significant revisions to match the 4Q data.

% Figure 11: Ara Finder
\begin{figure*}[ht]
\vspace{0mm}
\centerline{\includegraphics[angle=-90,scale=0.65]{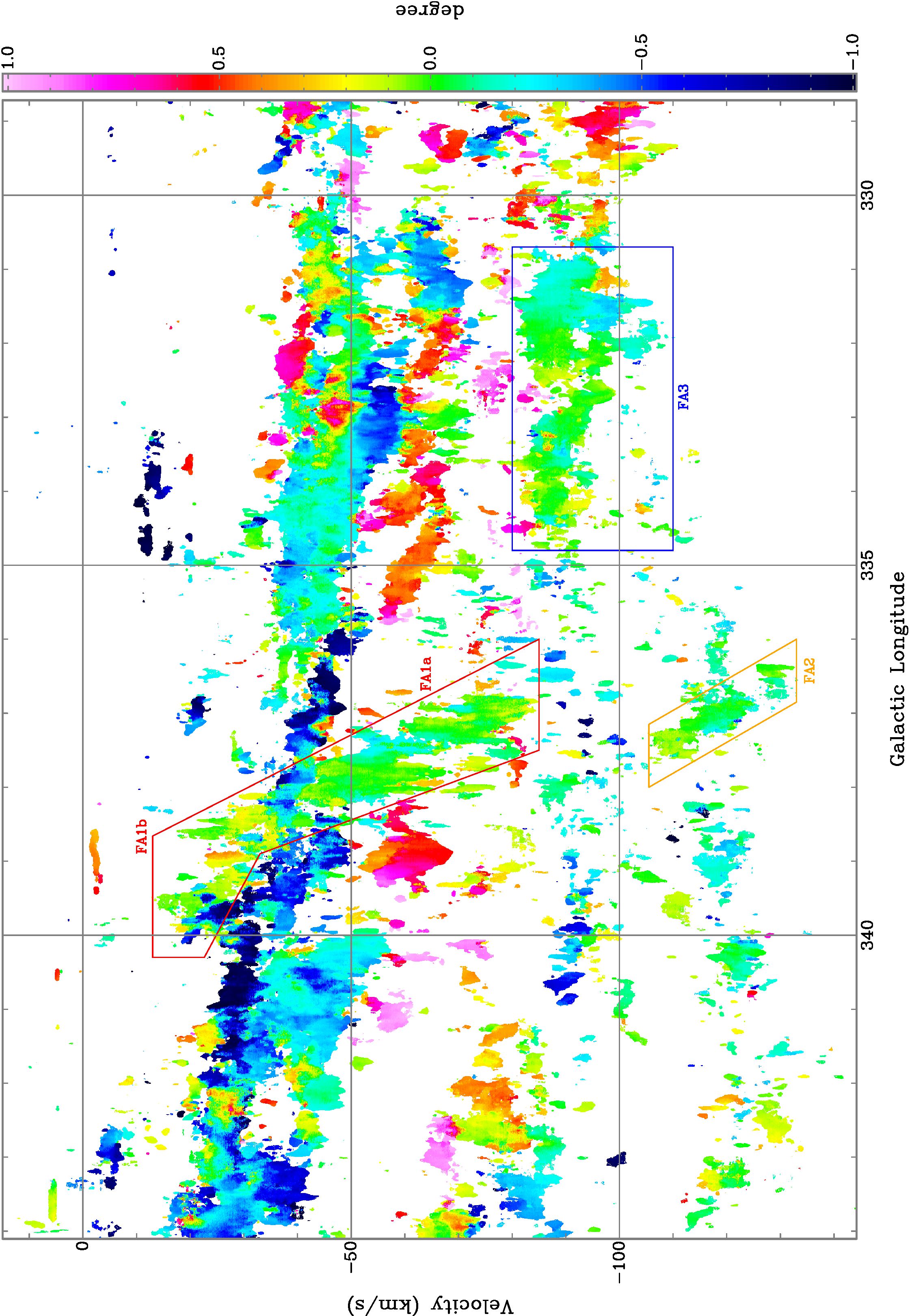}}
\vspace{-1.5mm}
\caption{\footnotesize Location of the Far Ara Clouds in \lv\ space, demarked by coloured polygons (FA1 = red, FA2 = orange, FA3 = blue) overlaid on the mean latitude $\bar{b}$ map for \nco, as in Fig.\,\ref{full-lv1-12coZM-rainbow}. $$ $$
\label{sample-ara}}
\vspace{-6mm}
\end{figure*}

%%%%%%%%%%%
%%   Section 6.4  %%
%%%%%%%%%%%
\subsection{Novel Structures Revealed}\label{wowie}

\vspace{1mm}In \S\ref{filtered} we discuss two of the most interesting features of the $\zeta^{+}$-filtered maps, apparently revealed here for the first time.  The first of these may already have been noticed in Figure \ref{sample-ld1z} and similar maps in the Appendices.  That is, there is a clear, widespread, and consistent pattern of ripples in the midplane of the Milky Way's molecular layer, at least within the solar circle --- see {\color{red}Figure \ref{awiggle}} for just one rendering.  These ripples are so prominent that we cannot conceive how the effects we see in our data could be some kind of observational artifact.  Furthermore, this conclusion is strongly reinforced by recent complementary observational and theoretical studies, which also suggest a possible origin: a gravitational disturbance from a recent passage of the Sgr dwarf galaxy \citep[e.g.,][]{pkd24,bin24,a25}.  See \S\ref{ripples} for a more detailed discussion of these ripples.

\vspace{1mm}The other intriguing aspect of our maps lies among the largest of the kinematically far clouds, in the direction of the constellation Ara: we call them the Far-Ara clouds, FA1--3 (see {\color{red}Fig.\,\ref{sample-ara}} and \S\ref{farside}).  While FA3 is probably a typical massive GMC complex on the inner edge of the Norma spiral arm \citep[see Paper II,][]{q15}, FA1 and 2 together present a rather unique aspect.  For a single molecular cloud structure, they have a {\em huge} velocity gradient, $\sim$110\,\kms\ over a $\sim$3\degree\ span, that is unprecedented outside the plunging orbits of clouds associated with the Galaxy's bar.  And this is probably the most feasible explanation: they seem to be arranged as if associated with the far end of the Near 3\,kpc Arm (N3A), just downstream from the far end of the bar.  They do not comport with any other spiral arm pattern at any equivalent position in the regular disk of the Milky Way: their projected size on the sky at a median kinematic distance of 12\,kpc is almost 1\,kpc, but the kinematic span in distance is much larger, $\sim$3\,kpc front-to-back.  And, for such a singular structure, they seem to have no molecular context, being virtually isolated in \lbv\ space.

\vspace{1mm}A more radical hypothesis is that FA1+2 represents a gas-rich, possibly infalling, dwarf satellite galaxy of the Milky Way.  It would be a rather large coincidence, but this dwarf would be located exactly in the Galactic Plane, and possibly rotating in the same Plane (explaining its velocity gradient), but at an indeterminate distance (20--300\,kpc) behind the Galactic disk.  We perform some elementary calculations in \S\ref{farside} and, while seeming a rather extraordinary idea, cannot rule it out on physical grounds.  Over the distance limits identified, its mass would range from about 1\% to 100\% of the Small Magellanic Cloud.  Although we favour the far-N3A location, this perplexing feature clearly needs further study.

%%%%%%%%%%
%%  Section 7  %%
%%%%%%%%%%
\vspace{2mm}
\section{Conclusions}\label{concl}

\vspace{0mm}We have presented the latest public updates (Data Release 6) to the ThrUMMS project, which include comprehensive \lbv\ = 60\degree$\times$2\degree$\times$$\sim$200\,\kms\ mosaics, and various 2D moments thereof, in the three species \tco, \ttco, \ceto\ across the Fourth Quadrant (4Q) of the Milky Way.  From these we have obtained the following results on the (1) local physical properties of Galactic molecular clouds, and (2) their global distribution as tracers of the Galaxy's spiral structure.

\vspace{1mm}$\bullet$ We have performed a complete LTE radiative transfer analysis of the iso-CO data and derived a number of additional parameters, including their \tex, \nco, and $\tau$ over this same data volume, which we also describe in detail.

\vspace{1mm}$\bullet$ We have used the iso-CO and \tnt\ data to derive updated \itco\ $\rightarrow$ \nco\  and $\rightarrow$ $\Sigma_{\rm mol}$ conversion laws for Milky Way molecular clouds, of the form $N$ or $\Sigma$ $\propto$ $I^{p}$.  For arcminute-resolution data (roughly corresponding to parsec scales) with integrated intensities 20\,K\,\kms\ $<$ \itco\ $<$ 350\,K\,\kms, the index $p$ = 2.00$\pm$0.07 in a consistent way over most of the inner 4Q, while the normalisations $N_{0}$ or $\alpha_{0}$ have respective mean$\pm$dispersion (9$^{\times}_{\div}$3)$\times$10$^{17}$\,molecules\,m$^{-2}$ or  (1.7$^{\times}_{\div}$3)$\times$10$^{-2}$\,M\solar\,pc$^{-2}$.  However, at lower resolutions/larger physical scales, $p$ drops to $\sim$1.5.  For nuclear clouds near the Galactic Center, $p$ $\sim$ 1 at all resolutions.

\vspace{1mm}$\bullet$ This conversion law gives higher values for \nco\ than the standard linear conversion with a single $X$ factor, typically by a factor of 2--3 in a given location.  Thus, a linear $X$ factor underestimates the column density.

\vspace{1mm}$\bullet$ The velocity dispersion maps in \lb\ are bimodal.  That is, \nco\ maps yield consistently lower velocity dispersions \sigv\ across large areas than do the \tco\ maps; elsewhere, the two dispersions are comparable.  An approximate division between the two domains can be drawn at a level \sigv\ = 2\,\kms\ for the \nco\ data.  Above this level, both maps seem to preferentially trace higher-opacity and -density star forming regions; below this level the \tco\ maps have a very wide range of \sigv\ which is typically $\gg$ that of the \nco, and seems to preferentially trace the more diffuse molecular gas.  This highlights an important difference between \tco\ as a molecular cloud tracer, and most other species: due to its much higher opacity, \tco\ picks up significant contributions from diffuse material around the star-forming portions of dense molecular gas.

\vspace{1mm}$\bullet$ We have also evaluated existing Galactic rotation models (such as from VLBI and {\em Gaia} studies) in the context of our new 4Q maps, and propose a slightly modified set of rotation parameters, the ``BGT'' model.  This model fits the \lv\ data in the 4Q much better than prior models by obtaining consistently small distances for known local ($d$$<$500\,pc) clouds, where non-rotational motions are surprisingly small, and better matching the Negative Velocity Envelope of molecular emission at most longitudes.

\vspace{1mm}$\bullet$ Using the BGT model, the \lv\ data products from the iso-CO and \tnt\ cubes have been further processed onto \ld\ and \xy\ grids by deprojecting their Galactic rotation via standard kinematic techniques.  For these deprojected maps, we have also developed a simple, fast discriminator, the $\zeta^{+}$ function, to automatically choose between near and far kinematic distances at each \lv\ $\rightarrow$ \ld\ or \xy\ pixel, based on the deprojected cloud height and size distributions.  These techniques have been combined to yield a number of new results on the global 3D geometry of the 4Q molecular cloud population, as follows.

\vspace{1mm}$\bullet$ For latitude-integrated quantities (0th moment, whether \ico\ or \nco), there is really only one prominent spiral arm in the 4Q, and most molecular gas doesn't clearly line up with the spiral arm patterns inferred from northern hemisphere data.  The highest column densities are somewhat aligned with the Scutum-Centaurus Arm as defined by \cite{r19} over longitudes $\sim$360\degree--325\degree, but at lower longitudes, the bulk of the molecular gas shifts towards the location of the Sagittarius-Carina Arm.  With a few exceptions, relatively little material is associated with either the Norma Arm or far-kinematic distances.

\vspace{1mm}$\bullet$ In latitude 1st moment maps, deprojecting onto a heliocentric distance scale reveals a striking, widespread, coherent series of ripples or undulations in the midplane of the Galactic molecular cloud distribution across much of the 4Q.  Roughly, these ripples have a wavelength 4\,kpc and amplitude 50\,pc, offset by $z$ $\approx$ +10\,pc, and apparently extend to the kinematic tangent-point distance at most longitudes.  They are consistent with other recent observational results, and potentially also with theoretical models of the most recent encounter of the Sgr dwarf with the Milky Way's disk.

\vspace{1mm}$\bullet$ The $\zeta^{+}$-filtered \ld\ height and size maps also reveal three very distinct, large \& flat structures which seem to be the most massive objects among the kinematically far cloud population, all lying in within a few degrees of each other in the direction of Ara.  Far-Ara cloud 3 (FA3) seems to be one of the most distant, ``normal'' for the disk, massive star-forming GMCs in our maps.  But the nature of FA1 and FA2, both also very flat, is much more unique and perplexing.  Together they exhibit a truly huge velocity gradient, $\sim$110\,kms\ across just 3\degree\ of longitude, nominally implying a front-to-back kinematic span of at least 3\,kpc ($d$ $\approx$ 10--13.5\,kpc) and a mass $\sim$10$^{7}$\,M\solar.  The properties may be consistent with a location near the far end of either the Milky Way's bar or Near 3\,kpc Arm, or conceivably, a gas-rich dwarf satellite galaxy of the Milky Way or a stripped remnant of the Sgr dwarf itself, of somewhat smaller mass to the SMC and located exactly behind the Galactic Plane at a poorly-constrained distance $\sim$20--300\,kpc.

%%%%%%%%%%
%%   Ackngmt   %%
%%%%%%%%%%
\begin{acknowledgments}
ThrUMMS owes its existence to the uncompensated contributions of many colleagues and all co-authors, past and present.  Without their heroic efforts, the impecunious PI would have never been able to realise this vision, which was also critically enabled by the superb engineering and welcome funds for the development of the MOPS digital filterbank.  At the same time, we lament the passing of Mopra from a peer-reviewed, publicly supported facility to a privately funded and operated enterprise, with very limited community access and concomitantly reduced productivity after 2016.  Before this change, Mopra was vigorously subscribed to by a diverse astronomical community, resulting in a wide scientific impact that was out of proportion to its funding needs.  This performance was enabled by the resourceful scientific and engineering staff at ATNF during 2002--15; we thank all these outstanding personnel for their support of the Mopra telescope and our observations.  PJB also gratefully acknowledges support from NSF grant AST-2206584, which allowed the revival of this project from its moribund state during much of 2016--22, and warmly thanks Robert Benjamin for illuminating discussions, including inspiring the latitude analysis which revealed the ripples.  Finally, we thank the anonymous referee for a very thoughtful and constructive report which enhanced and strengthened the discussion at several points.
\end{acknowledgments}

\vspace{1mm}Facilities: \facility{Mopra (MOPS)}

Software: {karma (Gooch 1997), Miriad (Sault et al.\ 1995), SuperMongo (Lupton \& Monger 2000)}

\clearpage



%%%%%%%%%%
%%  Appendix   %%
%%%%%%%%%%
\appendix
%\restartappendixnumbering %%not nec. in emulateapj
%%%%%%%%%
%   Section A1  %
%%%%%%%%%
\section{Full ThrUMMS DR6 Mosaics}\label{fullmos}

In this first Appendix we present an overview of the data products, and some general features revealed by considering their 4-dimensional nature.  That is, the data are functions of longitude, latitude, velocity, and the data quantity (line brightness, derived physical property) under consideration.

\subsection{Sky Intensity Composite Images in \lb\ and Image Ratios}\label{skymaps}

As described in \S\ref{lines}, the ThrUMMS data cubes are organised into 6\degree$\times$2\degree\ Sectors for processing convenience (e.g., file size, hardware limitations).  Both the 3D cubes and the spectral-line moments derived from them can nevertheless be fully mosaicked into 60\degree$\times$2\degree\ \lb\ moment maps or 60\degree$\times$2\degree$\times$342 \kms\ \lbv\ cubes.  However, in forming the mosaics, the Sector data are interpolated onto a much larger grid, making the mosaics a close yet inexact approximation to the primary Sector data.  This means that while the mosaics can be more convenient for presentation and display, analysis of the data therein is more properly performed on the Sector data.

\vspace{1mm}With that understood, here we present sample mosaic moment images and a discussion of some important features (i.e., based on the Sector data) in these various moments.  In many of the following figures, the display is divided into multiple panels to better fit on a printed page, whether 8.5$''\times$11$''$ or A4, oriented in portrait or landscape mode as convenient.  However, both the mosaics and the constituent Sector moments are available digitally in the relevant Data Releases.

\begin{sidewaysfigure*}[h]
\vspace{9mm}
\centerline{\includegraphics[angle=0,scale=0.068]{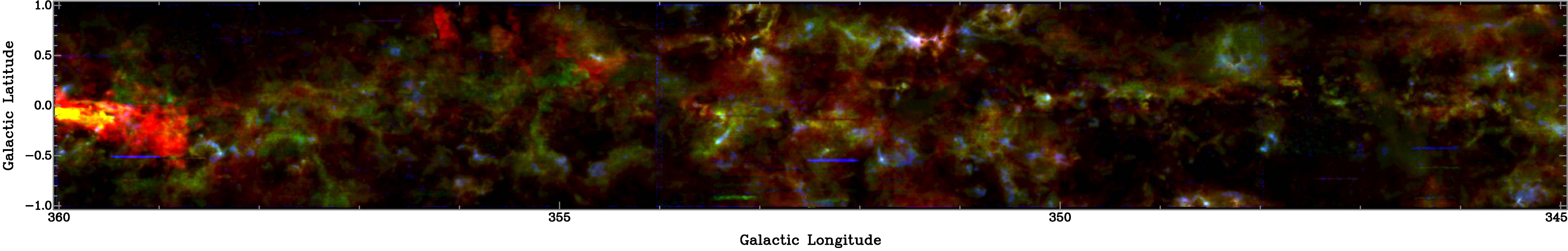}}
\centerline{\includegraphics[angle=0,scale=0.068]{dr6-mosaicS678-121318.jpg}}
\centerline{\includegraphics[angle=0,scale=0.068]{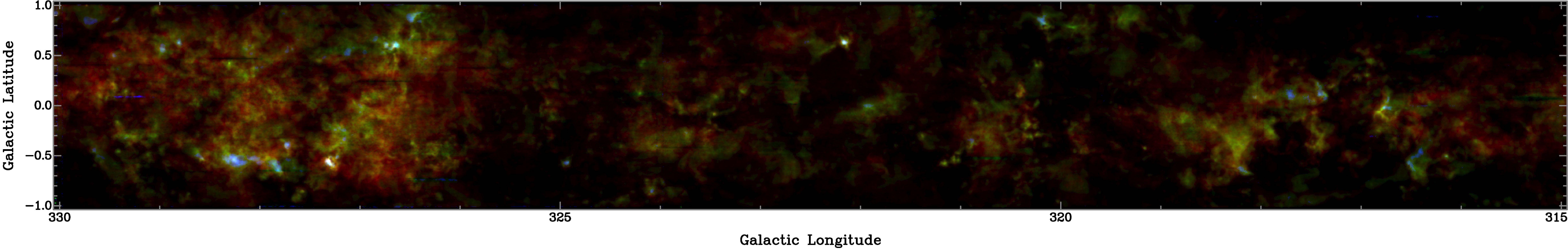}}
\centerline{\includegraphics[angle=0,scale=0.068]{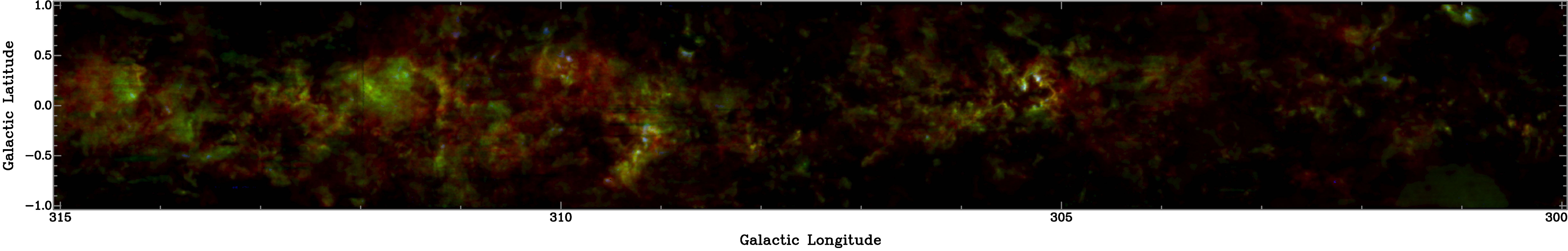}}
\vspace{-150mm}\hspace{12mm}{\bf {\color{red}\tco}\hspace{2mm}{\color{green}\ttco}\hspace{2mm}{\color{cyan}\ceto}}

\vspace{148mm}
\figcaption{\footnotesize Colour-composite images of the integrated intensity (zeroth moment in $V$) \lb\ mosaics in the three iso-CO lines, as labelled.  Each of the three mosaics is available digitally as part of DR6.  The data maximum, saturated colour, median error, and black levels in this image are respectively at 1137, 265, 1.91, --5\,K\kms\ (\tco); 229, 70, 0.75, --2.4\,K\kms\ (\ttco); and 33, 9.6, 0.37, 0.0\,K\kms\ (\ceto). $$ $$
\label{full121318-mom0}}
\end{sidewaysfigure*}

\vspace{1mm}{\color{red}Figure \ref{full121318-mom0}} shows the integrated intensities of all three iso-CO lines as a set of colour composite images.  In these \lb\ maps, the line emission has been integrated over all $V$ (the ``0th'' moment) with the smooth-and-mask method (SAM; Paper I).  Each species is then assigned to one colour channel (i.e., red, green, or blue), overlaid in data viewing software \citep[in our case, {\em kvis} from the {\sc Karma} package;][]{g97}, and rendered as an eps, pdf, or jpeg image.

\begin{sidewaysfigure*}[h]
\vspace{12mm}
\centerline{\includegraphics[angle=0,scale=0.068]{dr6-mosaicS8910-TexZMtau.jpg}}
\centerline{\includegraphics[angle=0,scale=0.068]{dr6-mosaicS678-TexZMtau.jpg}}
\centerline{\includegraphics[angle=0,scale=0.068]{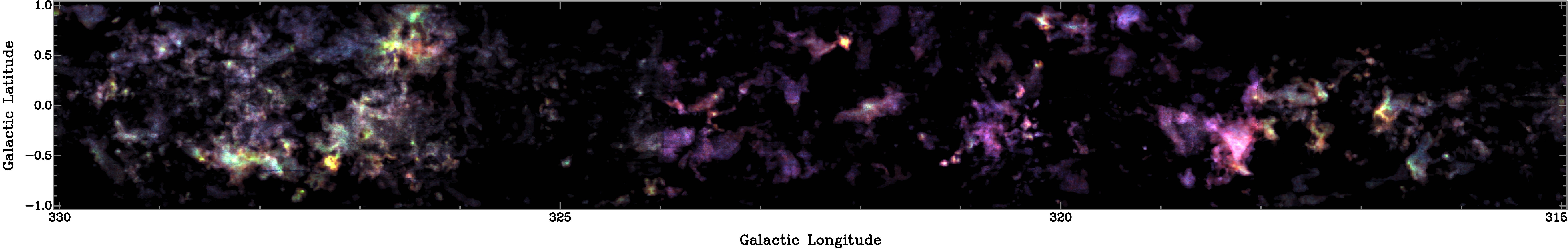}}
\centerline{\includegraphics[angle=0,scale=0.068]{dr6-mosaicS123-TexZMtau.jpg}}
\vspace{-150mm}\hspace{12mm}{\bf {\color{red}\tex}\hspace{2mm}{\color{green}\nco}\hspace{2mm}{\color{cyan}$\tau_{\rm CO}$}}

\vspace{148mm}
\figcaption{\footnotesize Colour-composite images of the physical parameter solutions in \lb\ mosaics (red = mean \tex, green = integrated \nco, blue = mean $\tau_{\rm ^{12}CO}$) from the analysis in \S\ref{radxfer}.  This makes manifest the radiative transfer effects implied in the iso-CO line ratios of Fig.\,\ref{full121318-mom0}.  The data maximum, saturated colour, median error, and black levels in these images are respectively at 3.16, 0.85, 0.054, --0.03\,K (mean \tex); 257, 44, 1.43, \& --0.4$\times$10$^{24}$\,molec\,m$^{-2}$ (\nco); and 5.53, 3.67, 0.0093, --0.11 (mean $\tau_{\rm ^{12}CO}$). $$ $$
\label{fullTexZMtau-mom0}} % THE Nco VALUES SHOULD BE 10x HIGHER??  Prolly not
\end{sidewaysfigure*}

\vspace{1mm}{\color{red}Figure \ref{fullTexZMtau-mom0}} contains a similar \lb\ mosaic to Fig.\,\ref{full121318-mom0}, but here we show a composite of the RGB = \tnt\ solutions to the radiative transfer analysis (\S\ref{radxfer}).  These have also been integrated across all $V$ to give one image per colour channel, but in this case the more physically relevant moments are the mean \tex, total $N$, and mean $\tau$ (respectively, the --1, 0, and --1 moments in {\sc Miriad} parlance).  This makes the \tex\ and $\tau$ scales numerically small, since the averaging is done over channels with no solution (taken as 0).  Nevertheless, the values are proportional to the contribution from non-zero \tex\ or $\tau$ channels.

\vspace{1mm}The use of colour in Figure \ref{full121318-mom0} makes the variable line ratios of Figure \ref{bigiratios} intuitive.  Numerically, the line ratios only vary by a factor of a few in each axis of Figure \ref{bigiratios}, yet this is enough to render strong colours with adjustable-contrast image displays.  In this way, the line ratio variations across the Galactic Plane clearly reveal distinct areas where opacity and excitation conditions change, and these changes correlate with the star-forming environment of the clouds.  We qualitatively review these relationships in this section; the quantitative radiative transfer results are discussed in more detail in Appendix \ref{rta}.

\vspace{1mm}Comparison of these two composite mosaics (Figs.\,\ref{full121318-mom0}, \ref{fullTexZMtau-mom0}) is very instructive.  As mentioned previously, the physical solutions (Fig.\,\ref{fullTexZMtau-mom0}) make manifest the radiative transfer implied in the variable iso-CO line ratios (Figs.\,\ref{bigiratios} and \ref{full121318-mom0}).  For example, intense \tco\ emission in well-known star-forming or HII regions, like NGC\,6334 ($l$$\sim$351\degree), the G333 complex, G305, or even in the Central Molecular Zone ($l$$>$358\degree), does not necessarily translate to high CO column density (green channel in physical composites).  Such areas also tend to have higher \tex\ (red channel in physical composites) and lower $\tau$ (blue channel), indicating warmer and more translucent gas.

\vspace{1mm}At the opposite extreme, where the \ttco\ is almost as bright as \tco, even where \tco\ is not particularly bright in an absolute sense, we obtain the highest column densities (green in Fig.\,\ref{fullTexZMtau-mom0}).  These clouds with lower \tex\ are widely distributed in areas well away from the warmer HII regions.  In many of these areas, \ceto\ also peaks up strongly near the $\tau$ peaks, emphasising their high opacity and column density, and indicating cooler and more opaque gas.

\vspace{1mm}The ratio-ratio diagram (Fig.\,\ref{bigiratios}) also conveys additional insights.  Points outside the grid of radiative transfer solutions, i.e., where either ratio is $>$1, could nominally indicate self-absorption in the more abundant species of each pair, neither of which are modelled in our radiative transfer code.  However, this mostly occurs for low-S/N voxels, and as a small percentage of all voxels.  For example, the fraction of voxels with $I_{13}$/$I_{12}$ $>$ 1 is 16.6\% for those voxels with S/N$>$2; 8.5\% for S/N$>$3; 5.2\% for S/N$>$4; and 3.0\% for S/N$>$5.  Absent noise, the true fraction of sightlines through the Milky Way with \tco\ self-absorption (over areas where \ttco\ is detectable) is then likely to be a few percent at ThrUMMS' resolution, probably indicating the average fractional sky coverage of the densest clumps among all molecular clouds, where such self-absorption is most likely to arise.

% Figure A1

% Figure A2

\vspace{1mm}Another feature of our radiative transfer calculations is the assumption of a fixed [\tco]/[\ttco] gas-phase abundance ratio, labelled as $R_{13}$ in Figure \ref{bigiratios}.  (In contrast, the [\ttco]/[\ceto] abundance ratio is explicitly solved for in our analysis, as labelled by the $R_{18}$ curves.)  The assumed $R_{13}$ = 60, while approximately true near the solar circle, is thought to vary systematically with Galactocentric radius, from $\sim$40 in the inner Galaxy to $\sim$80 or more in the outer, and so may affect our \nco\ results for the mostly inner-Galaxy clouds mapped by ThrUMMS.  However, changing the value of $R_{13}$ hardly affects the radiative transfer analysis at all, except for the conversion of final \tex\ and $\tau$ values to \ntco.  In that case, the curves displayed in Figure \ref{bigiratios} would change only at the top-leftmost corner of the diagram, where there are absolutely no data points.  So if $R_{13}$ = 40 were more accurate for a certain population of clouds, the change would be to merely shift the lowest-$\tau_{18}$ curves to a slightly more vertical aspect, and giving virtually the same \tex\ and $\tau$ solutions as before.  The final \ntco\ would be reduced, however, in proportion to the reduction in $R_{13}$.

\vspace{1mm}But in that case, the widely-assumed gas-phase abundance of \tco\ relative to \htwo\ of 2$\times$10$^{-4}$ would be even more incorrect than found by \cite{p21}.  For the large sample of mostly solar-circle clouds of CHaMP, they found that [\tco]/[\htwo] is a strongly-peaked function of dust temperature at 20\,K, peaking near 0.74$\times$10$^{-4}$ when assuming $R_{13}$ = 60.  Thus, the normal conversion of \nco\ to \nhtwo\ underestimates cloud masses by a factor of 3 or more.  If \nco\ should be even lower for inner-Galaxy clouds, this would make the anomaly even worse.

%\clearpage

%%%%%%%%%
%   Section A2  %
%%%%%%%%%
\subsection{The Global Velocity Field in ($l$,$b$)}\label{vfield}
\begin{sidewaysfigure*}[h]
\vspace{10mm}
\centerline{\includegraphics[angle=-90,scale=0.91]{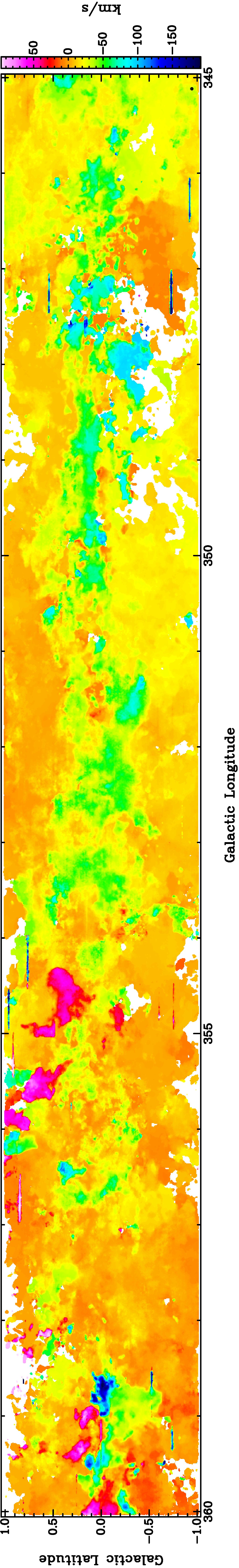}}
\centerline{\includegraphics[angle=-90,scale=0.91]{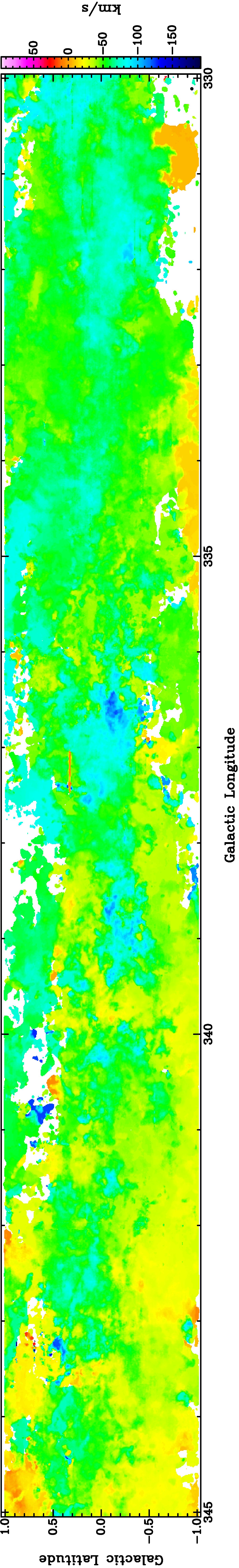}}
\centerline{\includegraphics[angle=-90,scale=0.91]{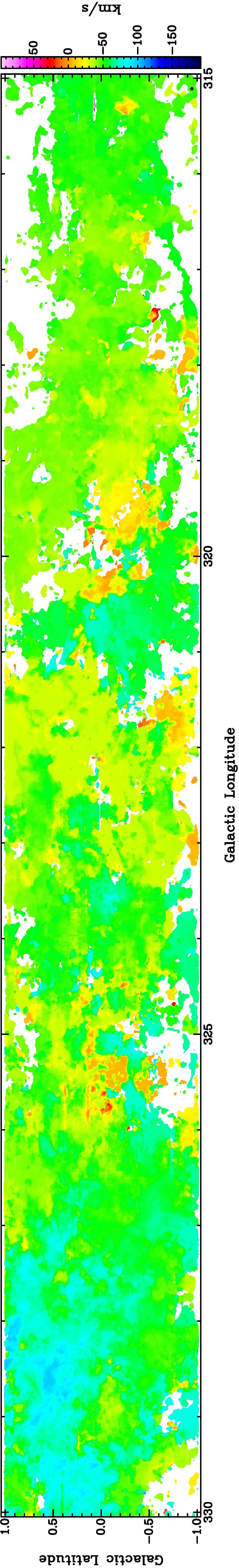}}
\centerline{\includegraphics[angle=-90,scale=0.91]{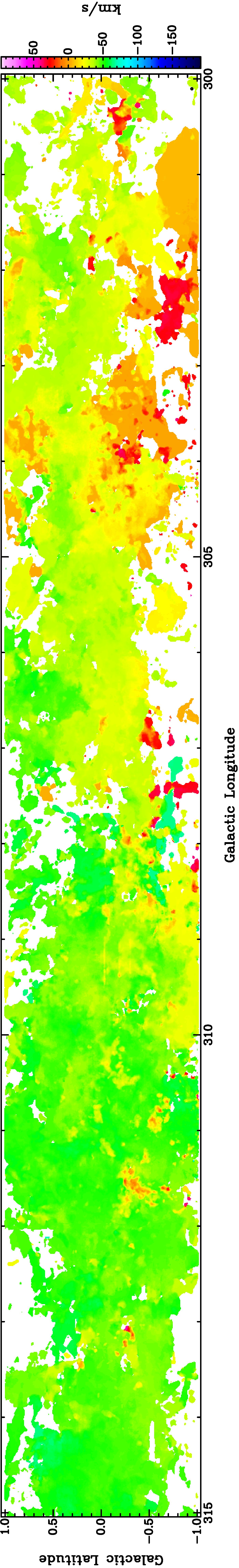}}
\vspace{0.3mm}
\figcaption{\footnotesize \lb\ mosaic of \tco\ 1st moment in $V$, the intensity-weighted mean \vlsr.  The displayed velocity scale (from --190 to +96 \kms) reflects all reliable \tco\ emission, i.e., ignoring data from artifacts such as ``weather stripes.'' $$ $$
\label{full12co-mom1}}
\end{sidewaysfigure*}

\begin{sidewaysfigure*}[h]
\vspace{13mm}
\centerline{\includegraphics[angle=-90,scale=0.91]{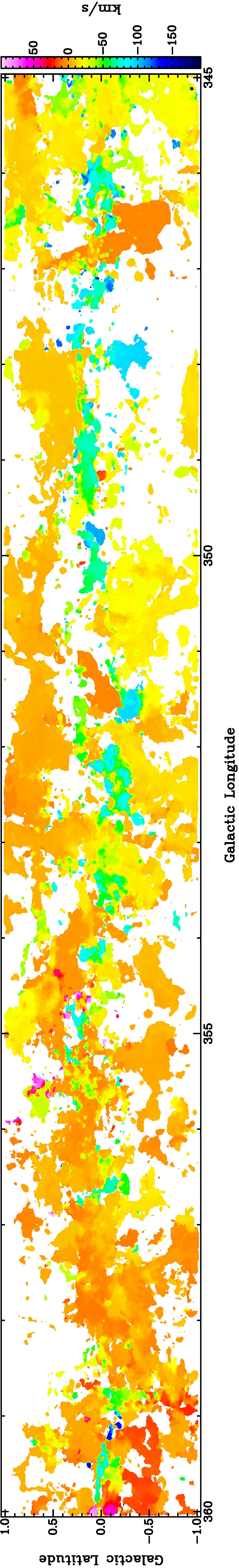}}
\centerline{\includegraphics[angle=-90,scale=0.91]{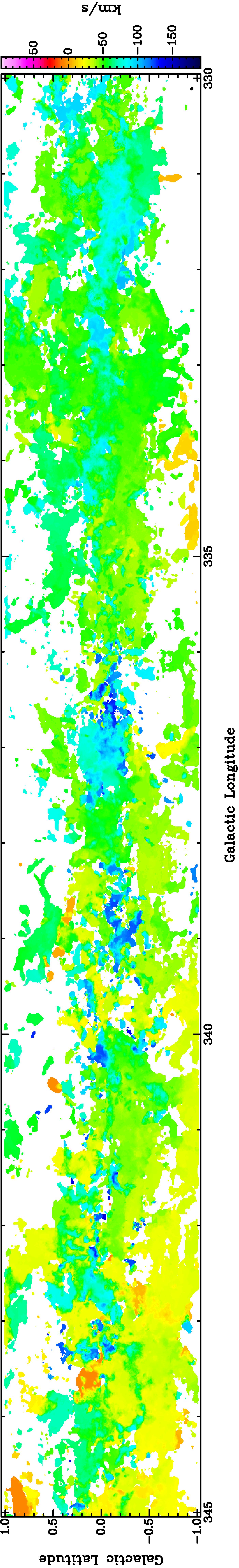}}
\centerline{\includegraphics[angle=-90,scale=0.91]{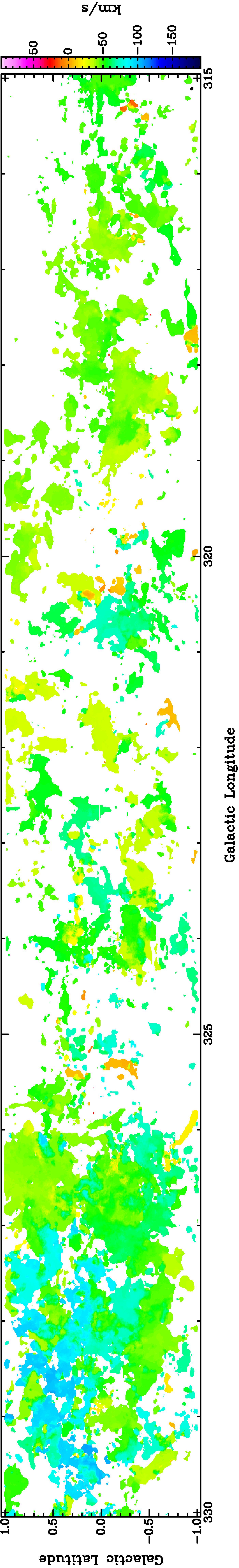}}
\centerline{\includegraphics[angle=-90,scale=0.91]{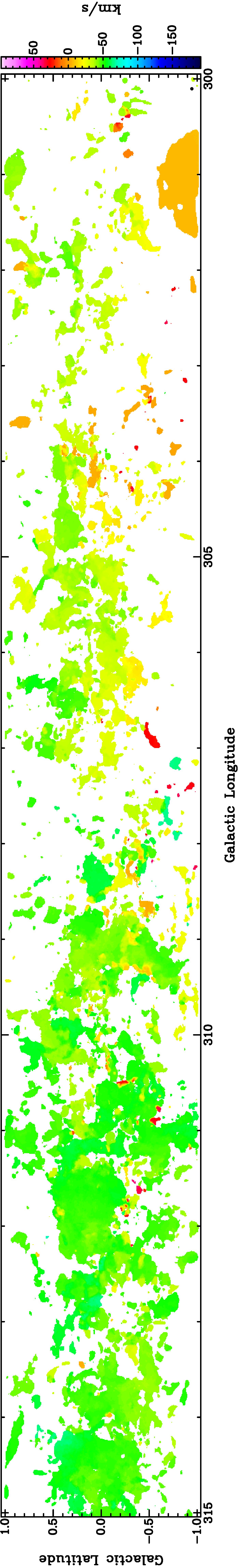}}
\vspace{0mm}
\figcaption{\footnotesize \lb\ mosaic of \nco\ 1st moment in $V$, the column-weighted mean \vlsr.  The displayed velocity scale is the same as in Fig.\,\ref{full12co-mom1} for ease of comparison.  The spatial coverage of the \nco\ data is, of course, less than that of the \tco\ data, since solving the radiative transfer equations requires reasonable S/N for both \tco\ and \ttco\ data. $$ $$
\label{fullZM-mom1}}
\end{sidewaysfigure*}

\begin{sidewaysfigure*}[h]
\vspace{11mm}
\centerline{\includegraphics[angle=-90,scale=0.91]{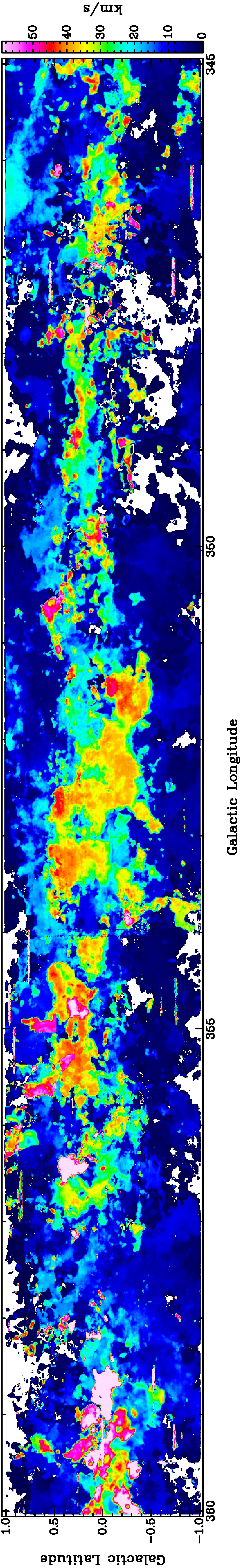}}
\centerline{\includegraphics[angle=-90,scale=0.91]{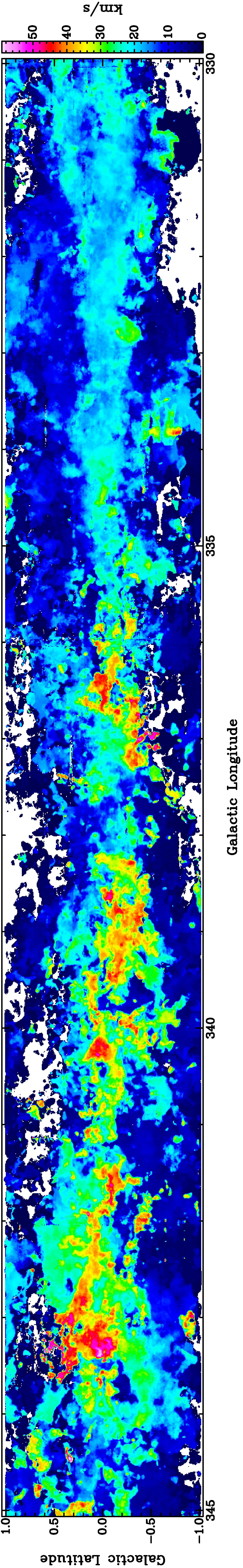}}
\centerline{\includegraphics[angle=-90,scale=0.91]{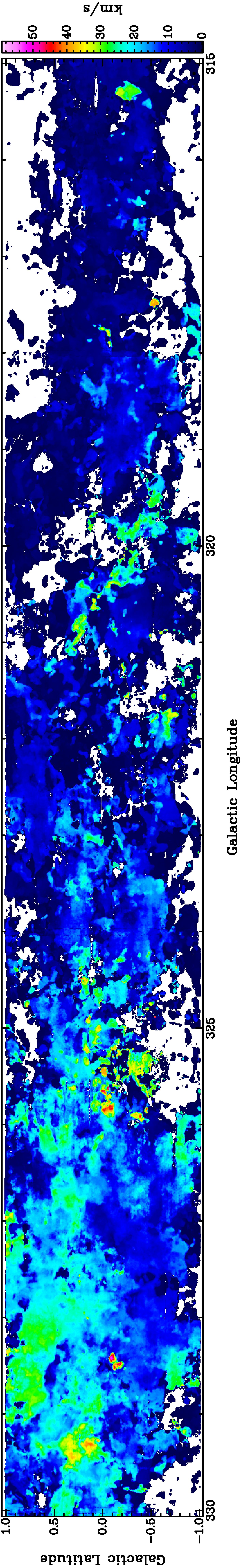}}
\centerline{\includegraphics[angle=-90,scale=0.91]{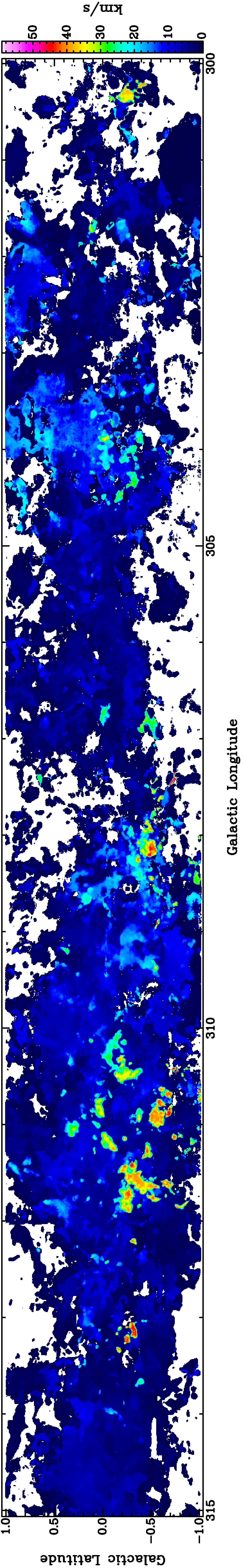}}
\vspace{0mm}
\figcaption{\footnotesize \lb\ mosaic of \tco\ 2nd moment in $V$, the intensity-weighted velocity dispersion \sigv.  The displayed velocity scale (0--59 \kms) reflects nearly all reliable \tco\ emission, i.e., ignoring data from artifacts such as ``weather stripes,'' and saturating only a few small areas (e.g., near the Galactic Centre) with \sigv\ up to 166.5\,\kms. $$ $$
\label{full12co-mom2}}
\end{sidewaysfigure*}

\begin{sidewaysfigure*}[h]
\vspace{11mm}
\centerline{\includegraphics[angle=-90,scale=0.91]{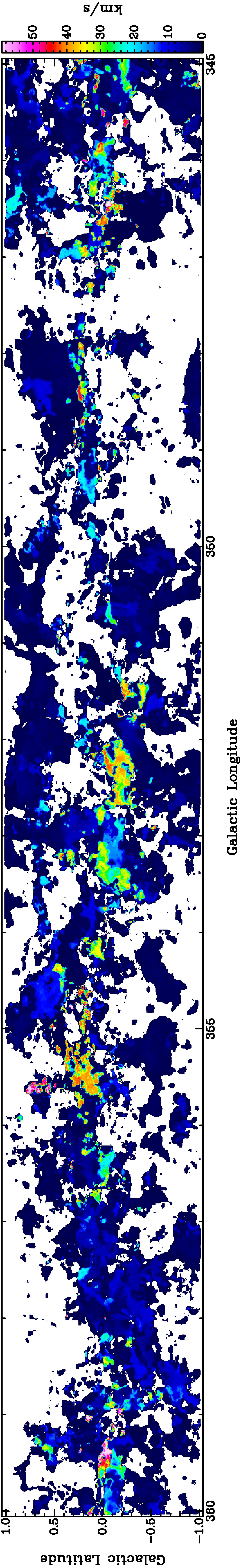}}
\centerline{\includegraphics[angle=-90,scale=0.91]{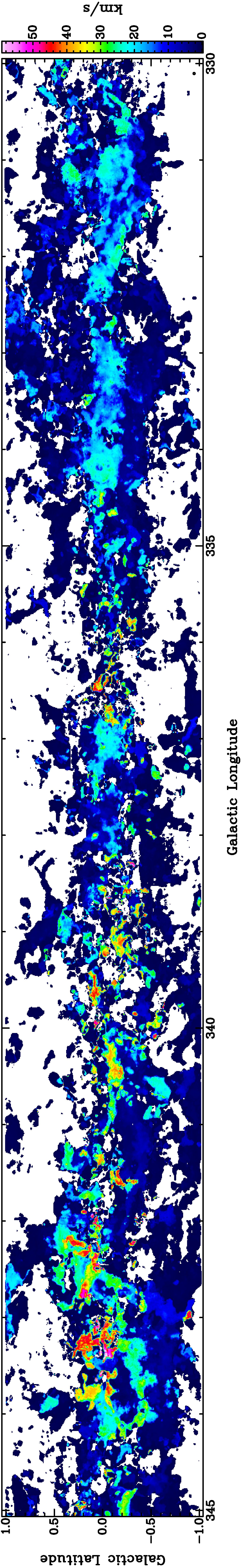}}
\centerline{\includegraphics[angle=-90,scale=0.91]{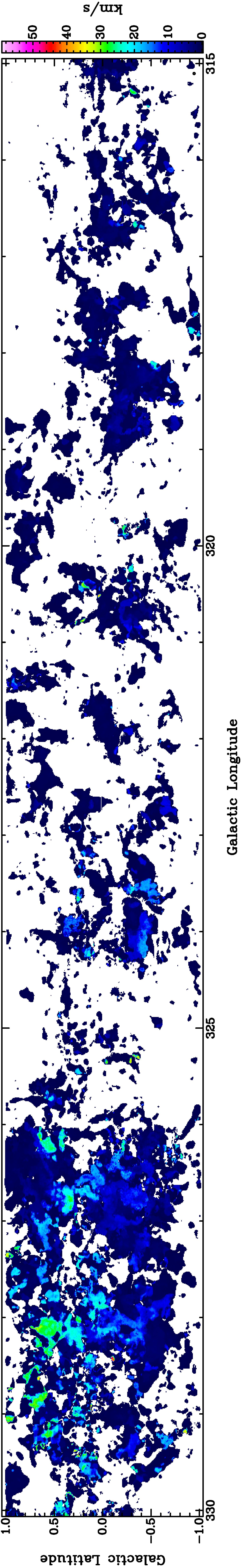}}
\centerline{\includegraphics[angle=-90,scale=0.91]{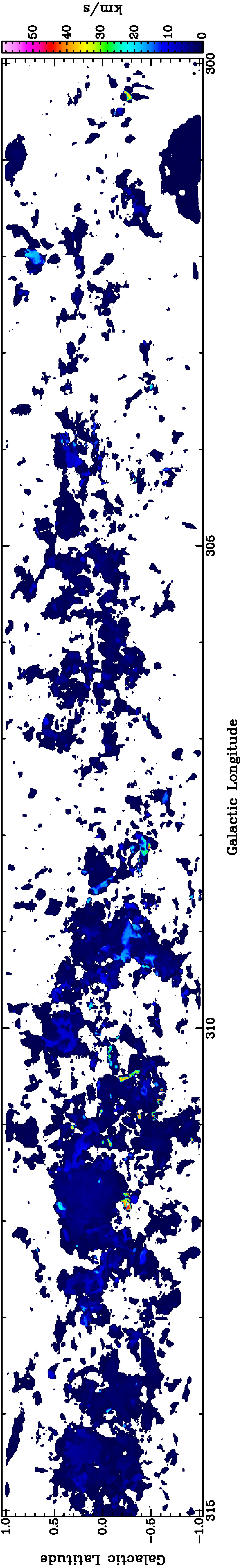}}
\vspace{0.4mm}
\figcaption{\footnotesize \lb\ mosaic of \nco\ 2nd moment in $V$, the column-weighted velocity dispersion \sigv.  The displayed velocity scale is the same as in Fig.\,\ref{full12co-mom1} for ease of comparison, and saturating only a few very small areas (e.g., near the Galactic Centre) with \sigv\ up to 86.7\,\kms.  The spatial coverage of the \nco\ data is less than that of the \tco\ data, as in Fig.\,\ref{fullZM-mom1}. $$ $$
\label{fullZM-mom2}}
\end{sidewaysfigure*}

In {\color{red}Figures \ref{full12co-mom1}--\ref{fullZM-mom2}}, we show some higher-moment \lb\ mosaics, namely the \tco-intensity- and total-column-weighted mean \vlsr\ (1st moment; Figs.\,\ref{full12co-mom1} and \ref{fullZM-mom1}, resp.) and the \tco\ and total-column velocity dispersion \sigv\ (2nd moment; Figs.\,\ref{full12co-mom2} and \ref{fullZM-mom2}, resp.).  In both pairs of moments, the \tco\ mosaic has wider areal coverage than \nco, since the latter requires good sensitivity also in the \ttco\ data.  Thus, there are large areas where \tco\ is clearly detected but \ttco\ is not.

\vspace{1mm}We can also calculate similar 1st and 2nd moments for the \ttco, \ceto, \tex, and $\tau_{\rm CO}$ cubes, and some of these are included in the DR6 files.  For our purposes, however, much of their information overlaps with that of the \tco\ and \nco\ moments, so we do not discuss them further here.

% Figure A3

% Figure A4

% Figure A5

% Figure A6

%\clearpage

\vspace{1mm}On the whole, the mean \vlsr\ maps reveal a number of common features related to the Milky Way's large-scale spiral arm structure, where the molecular cloud population is most concentrated, and shows the differential rotation of these arms in the Galactic disk.  For example, the clouds at the most negative velocities (cooler colours in the rendering of Figs.\,\ref{full12co-mom1} \& \ref{fullZM-mom1}) tend to cluster much more tightly around $b$ = 0\degree, since kinematically they also tend to be the furthest away from us, and so are projected at only small angles from the Galactic Plane (GP).  This can be seen in the centrelines of both Figures, most clearly at $l$ $\sim$ 328\degree--360\degree.  The lower-velocity material (warmer colours) is more widely-distributed in latitude, because the similar physical heights above \& below the GP translate into somewhat larger latitude spreads at the smaller distances of these clouds.  Conceptually, in this high depth-of-field \vlsr\ view, the nearer clouds can be visualised as a broad ($b$ $\sim$ $\pm$1\degree) screen through which one can peer at the more distant and narrowly-confined ($b$ $\sim$ $\pm$0.3\degree) cloud population.

\vspace{1mm}We discuss the velocity fields in more detail in Appendix \ref{kinem}.

%\clearpage

%%%%%%%%%
%   Section A3  %
%%%%%%%%%
\subsection{The Velocity Dispersion in \lb\ is Bimodal}\label{vdisp}
The velocity dispersion mosaics (Figs.\,\ref{full12co-mom2}, \ref{fullZM-mom2}) reveal further information about the molecular cloud population.  On the one hand, the \sigv\ values tend to peak along the same midplane along which we have the greatest depth of field as revealed by the \vlsr\ mosaics.  This is not so surprising since multiple prominent clouds along a given line of sight will add to the total \sigv\ measured, especially for the high-opacity \tco\ emission that can pick up more emission components throughout the GP.  While the \nco-\sigv\ map (Fig.\,\ref{fullZM-mom2}) shows the same effect, the highest \sigv\ values are more tightly clustered around the GP, since \nco\ is an opacity-corrected version of the \tco\ emission.  Even by itself, this suggests that the \tco\ opacity is an important factor that may affect our interpretations of such maps.

\vspace{1mm}Apart from the overall spatial distribution of \sigv, the statistics of \sigv\ values are very instructive.  That is, there are far more lines of sight with low \sigv\ in the GP than there are high-\sigv\ pixels.  A typical value for \sigv(\tco) in massive star-forming clumps is $\sim$2\,\kms\ \citep{b16}; in a large star formation complex, \sigv(\tco) can rise to 10\,\kms\ or more locally, and approach 20\,\kms\ globally \citep{b18}.  Therefore, over large areas of the Milky Way, the conventional wisdom is that a typical line of sight near the Galactic Plane will intersect many clouds at different \vlsr, contributing to widespread velocity confusion (i.e., large \sigv(\tco)) and often making identification of individual clouds difficult.  This can be seen, for example, in \cite{MML17}'s analysis of the CfA \tco\ survey, with an effective resolution of 8$'$.

\begin{figure*}[h]
\hspace{1.5mm}\includegraphics[angle=0,scale=0.106]{dr6-all-12covsZMv4-mom2.jpg}

\vspace{-84mm}
\hspace{2mm}(a)\hspace{85mm}(b)\includegraphics[angle=90,scale=0.103]{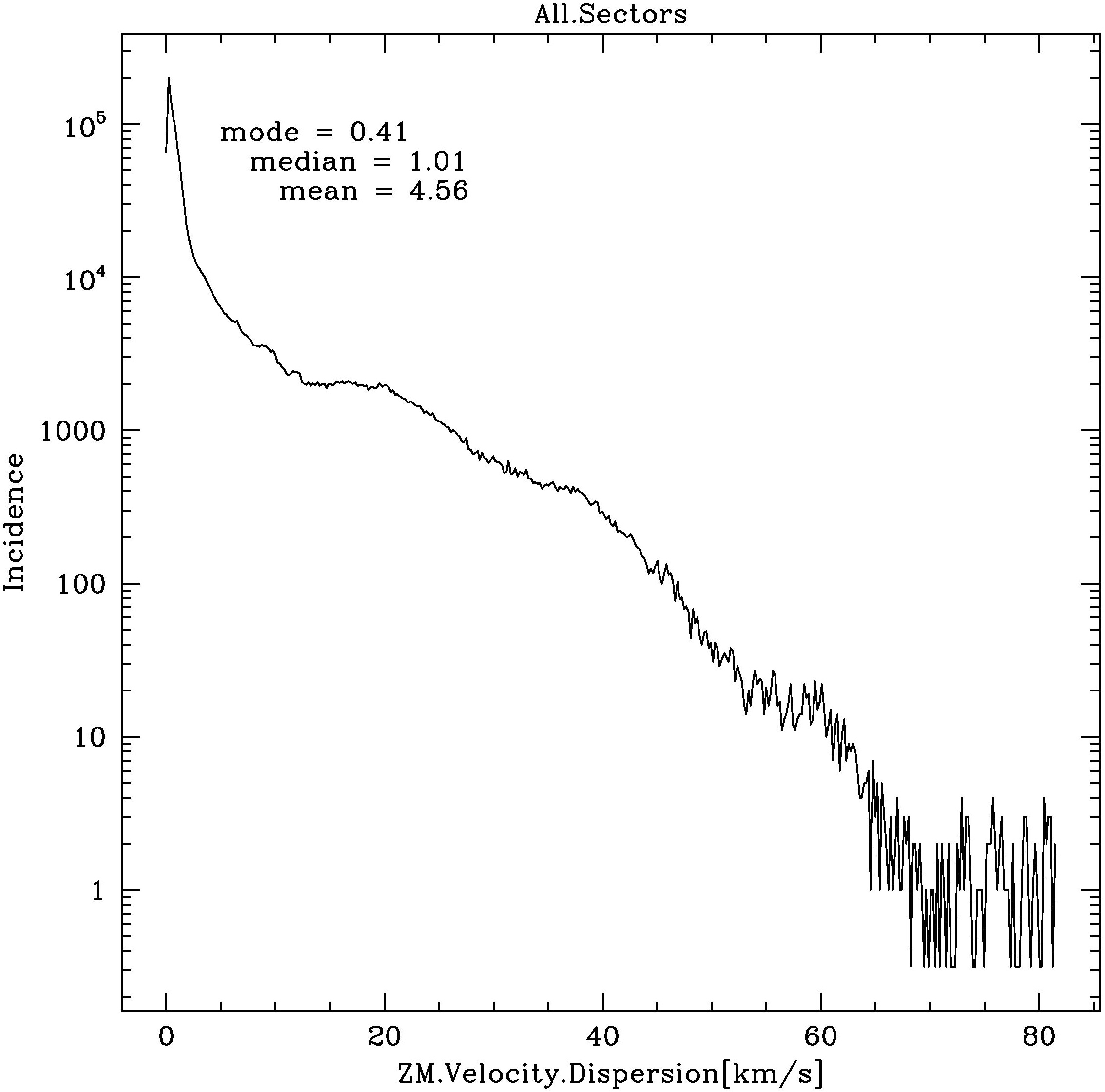}

\vspace{1mm}
\hspace{2mm}\includegraphics[angle=0,scale=0.103]{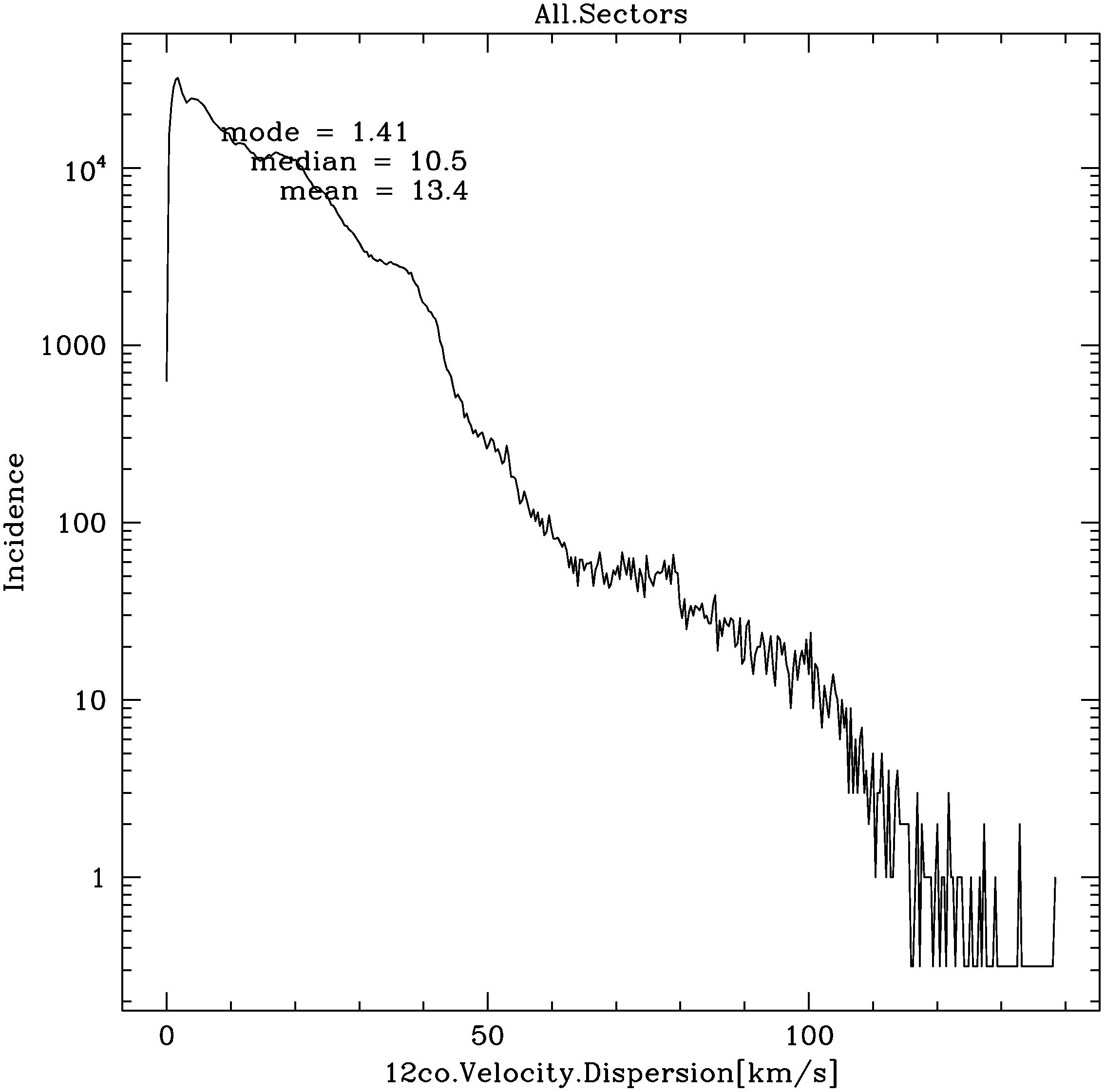}~~~~~\includegraphics[angle=0,scale=0.103]{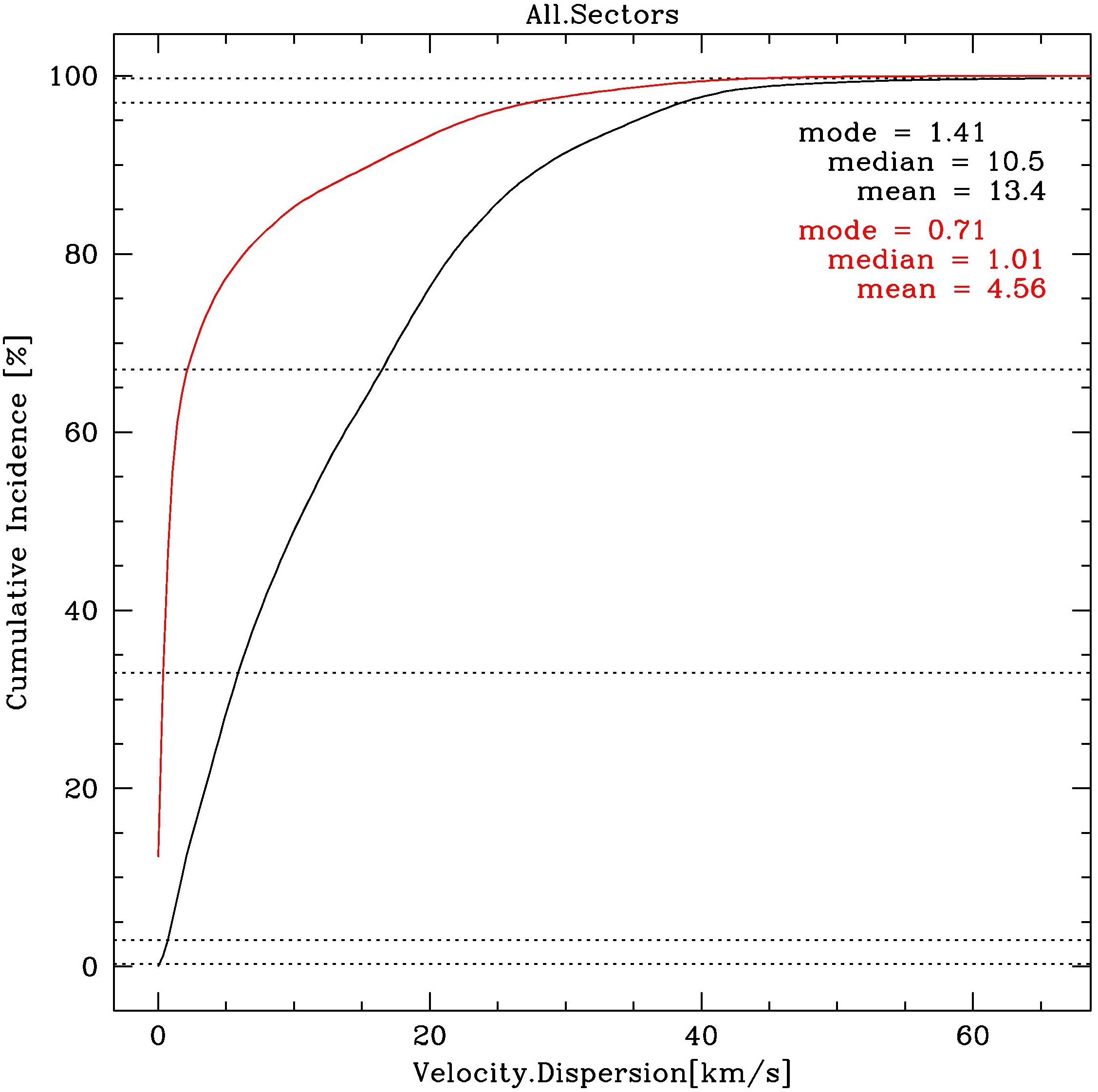}
\vspace{-4mm}

\hspace{2mm}(c)\hspace{84mm}(d)

\vspace{-1mm}
\caption{\footnotesize Comparison of \nco-weighted (labelled ``ZM'') and \tco-weighted velocity dispersions, i.e., the 2nd moment, across all \lb\ Sector data.  Panel $a$ is a copy of Fig.\,\ref{twodisps}.  Panels $b$ and $c$ show histograms of pixel incidence for each dispersion separately, labelled with statistics of their distributions.  Panel $d$ shows, overlaid for both dispersions with red=\nco\ and black=\tco, cumulative histograms of pixel incidence. $$ $$
\label{12vZMdisp}}
\vspace{-6mm}
\end{figure*}

\vspace{1mm}Instead, we see that already at 1\farcm2 resolution, the median \sigv(\tco) for ThrUMMS has dropped to 10.5\,\kms, while 80\% of all lines of sight have \sigv(\tco) $<$ 22\,\kms\ (panel $d$ of {\color{red}Fig.\,\ref{12vZMdisp}}).  This strongly suggests that, even for \tco, velocity confusion is largely a resolution effect, where multiple intrinsically ``small'' clouds contribute to the signal in a ``large'' beam.  However, this is not the entire story.  With our maps of \nco, velocity confusion is even further reduced, with a median \sigv(\nco) = 1.0\,\kms, 80\% of sightlines with \sigv(\nco) $<$ 6\,\kms, and 90\% with \sigv(\nco) $<$ 16\,\kms\ (panel $a$ of Fig.\,\ref{12vZMdisp}).

% Figure A7

\vspace{1mm}As mentioned in the main text, the \sigv\ distributions are also not random.  In particular, we can see the effect of the \tco\ opacity in a direct comparison, pixel for pixel, of the two dispersions (Fig.\,\ref{twodisps} = panel $a$ of Fig.\,\ref{12vZMdisp}).  There are clearly two relationships between the data in this dispersion-dispersion ($\sigma$-$\sigma$) space, as described in \S\ref{params}.  While the \tco\ dispersions still have a somewhat large range in the wide variety of locations that we sample, albeit limited to small values in most pixels, the corresponding \sigv(\nco) values clearly bifurcate into two domains.  One domain is where \sigv(\nco) \lapp\ \sigv(\tco) (the ``large-\sigv'' domain); the other domain is where \sigv(\nco) is quite small, either \lapp 2\,\kms\ in an absolute sense, or \sigv(\nco)/\sigv(\tco) \lapp\ 0.3 in a relative sense (the ``small-\sigv'' domain).  This dichotomy persists even where \sigv(\tco) ranges up to 50--100\,\kms, and there are wide areas in the GP where the ratio of dispersions \tco:\nco\ can exceed 10:1!

\begin{figure*}[h]
\centerline{\includegraphics[angle=0,scale=0.147]{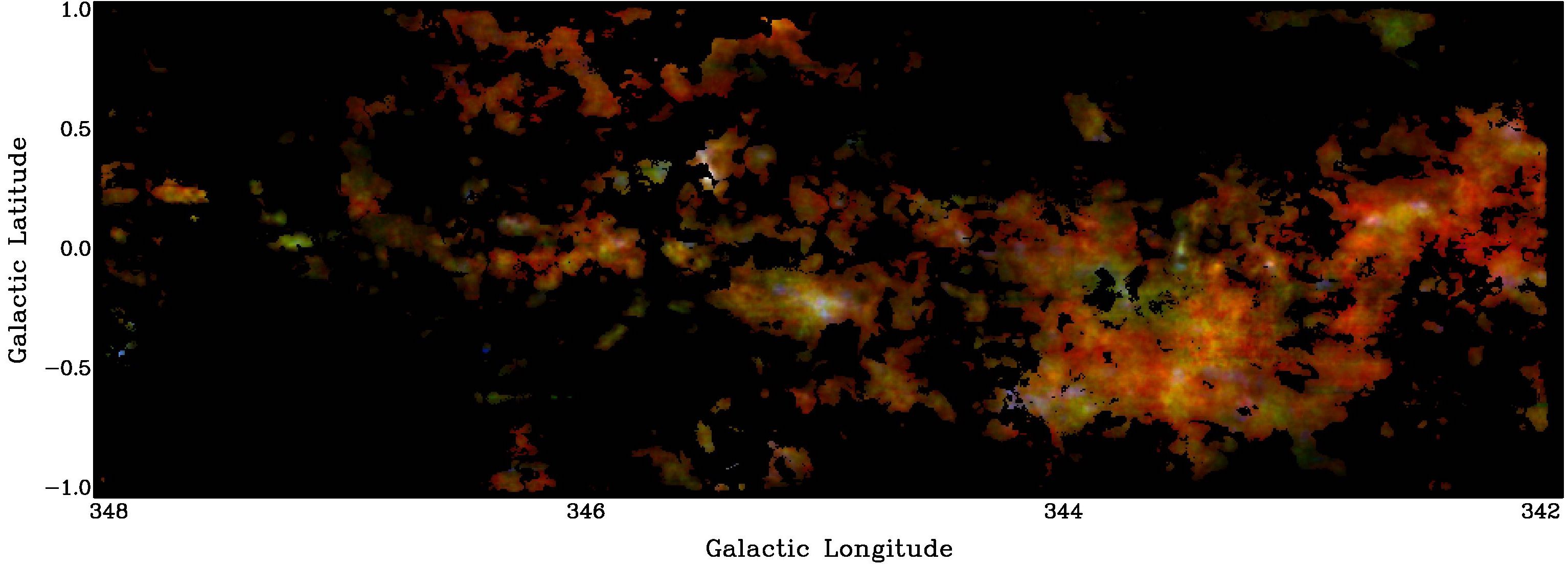}}
\vspace{-5mm}
\centerline{\includegraphics[angle=0,scale=0.147]{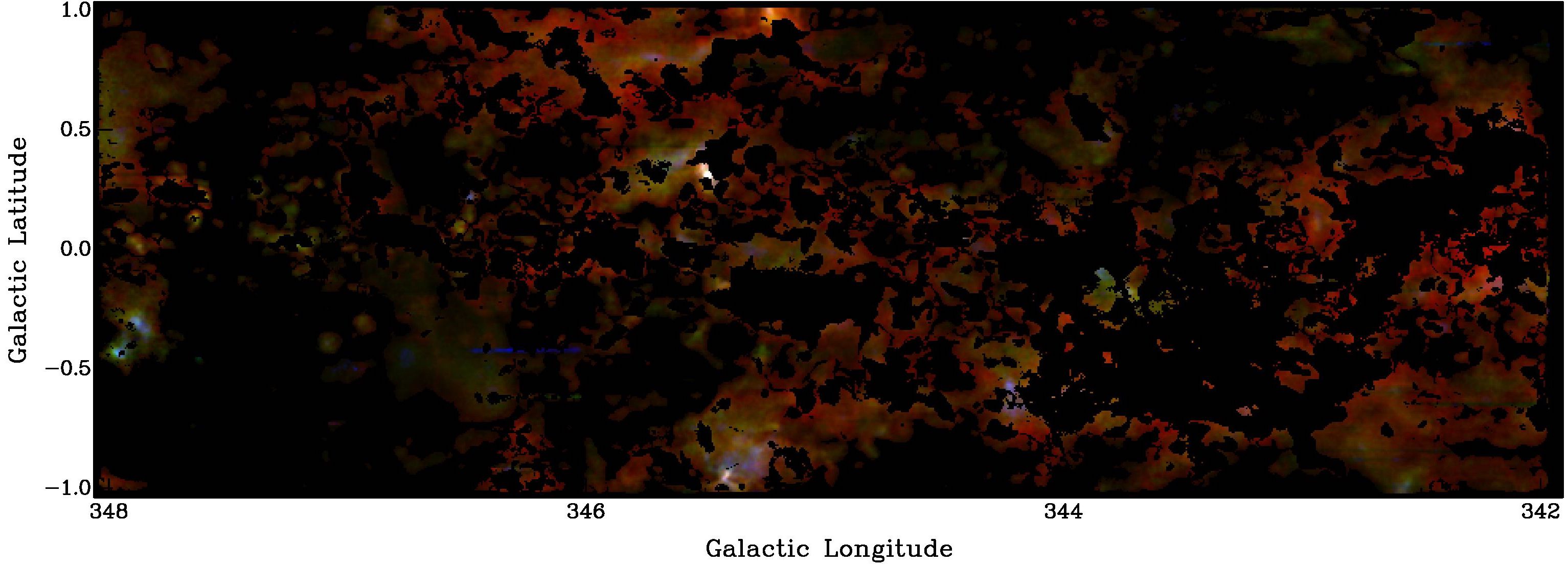}}
\vspace{-5mm}
\centerline{\includegraphics[angle=0,scale=0.147]{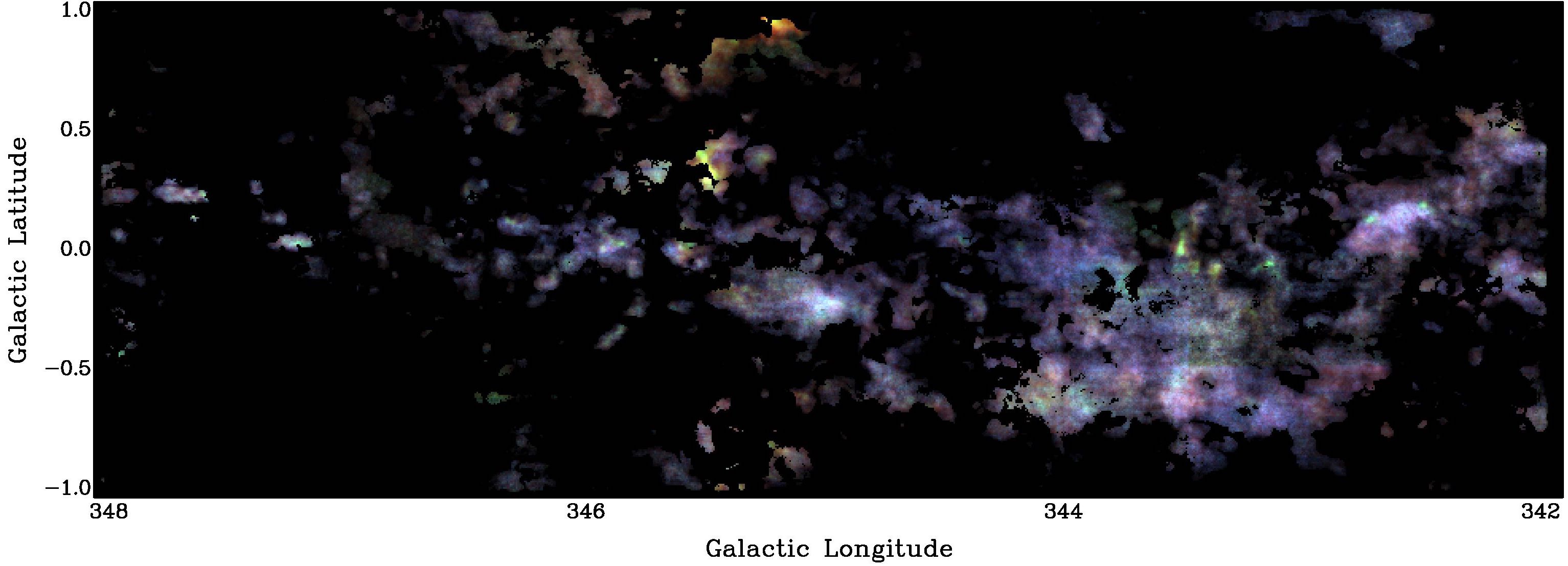}}
\vspace{-5mm}
\centerline{\includegraphics[angle=0,scale=0.147]{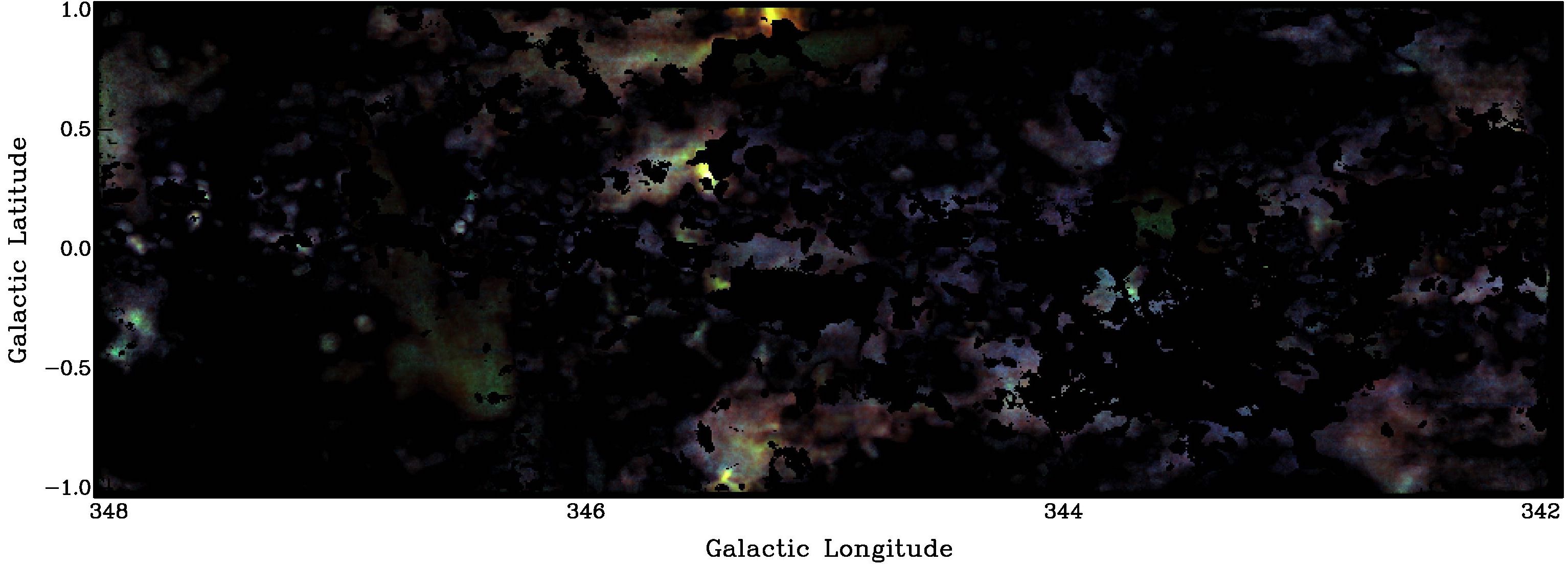}}

{\color{white}
\vspace{-214mm}\hspace{22mm}Large-\sigv(\nco)

\vspace{50mm}\hspace{22mm}Small-\sigv(\nco)

\vspace{50mm}\hspace{22mm}Large-\sigv(\nco)

\vspace{50mm}\hspace{22mm}Small-\sigv(\nco)} % end white

\vspace{51mm}
\caption{\footnotesize In Sector 342 as an example, illustration of the origin of the velocity dispersion distributions seen in Fig.\,\ref{12vZMdisp}.  Using RGB colour composites similar to Figs.\,\ref{full121318-mom0}, \ref{fullTexZMtau-mom0}, the panels show from top to bottom (with appropriate labels): the iso-CO composite masked to areas with \sigv(\nco) $>$ $\sigma_{\rm thr}$ = 2\,\kms; the same composite masked to areas with \sigv(\nco) $\leq$ $\sigma_{\rm thr}$; the (\tnt) composite masked to areas with \sigv(\nco) $>$ $\sigma_{\rm thr}$; and the same (\tnt) composite masked to areas with \sigv(\nco) $\leq$ $\sigma_{\rm thr}$.  For a given composite, the 3 colour channels are scaled to the same brightness and contrast levels in each mask, in order to normalise comparisons.  A similar relationship between \sigv(\nco) and the brightness distributions of each quantity exists in all the other Sectors as well. $$ $$
\label{sample-hilo}}
\end{figure*}

\vspace{1mm}That this is primarily an effect of opacity in the \tco\ emission can be seen as follows.  As an illustrative exercise, we can define $\sigma_{\rm thr}$ = 2\,\kms\ as one threshold between the large- and small-\sigv\ domains, regardless of the \sigv(\tco) values at the same pixels.\footnote{We could also set $\sigma_{\rm thr}$ to have slightly different values, which makes only a small difference to the appearance of the masked images, or instead set the threshold at a dispersion ratio $\sigma$/$\sigma_{\rm thr}$ = 0.3, which produces a slightly more fragmented appearance, but in any case, the results are similar to the description which follows.  Our point here is not so much the type or value of the threshold, but that a threshold, i.e., a boundary between two domains in $\sigma$-$\sigma$ space, exists at all.}  We can then mask any of the \lb\ moment maps, such as Figures \ref{full121318-mom0} and \ref{fullTexZMtau-mom0}, according to whether \sigv(\nco) is greater or less than $\sigma_{\rm thr}$ (or whether the ratio is greater or less than $\sigma$/$\sigma_{\rm thr}$).  The results for Sector 342 (used here as an example) are shown in {\color{red}Figure \ref{sample-hilo}}; the results for any other Sector are very similar.

% Figure A8

\vspace{1mm}In the top two panels of Figure \ref{sample-hilo} we compare the 3 iso-CO species' emission as a function of the $\sigma_{\rm thr}$ mask.  It is immediately clear that the large-\sigv\ domain is dominated by concentrations of bright \tco, coupled also with frequent bright \ttco.  In contrast, the more widespread (especially in other Sectors) small-\sigv\ domain tends to exhibit fainter \tco\ and (especially) \ttco\ emission, in larger, more diffuse structures than the large-\sigv\ clouds.

\vspace{1mm}In the bottom two panels of Figure \ref{sample-hilo} we understand better the cause of these effects.  While both the \tex\ and \nco\ distributions are generally somewhat warmer \& denser in the large-\sigv\ domain, it is the opacity distribution which is most strikingly different.  The large-\sigv\ domain clearly corresponds to a substantially more opaque cloud population, while the small-\sigv\ structures are much more translucent.  This makes sense of the scatter plot of \sigv\ values in panel $a$ of Figure \ref{12vZMdisp}, and in particular, the existence of a population of clouds where the \tco\ dispersion can be 10-20 or more times the \nco\ dispersion.  Such clouds are not very opaque and, while the \tco\ $\tau$ (and consequently, \sigv) values will still be large in an absolute sense, the \ttco\ lines are much narrower there, leading to overall much smaller \sigv(\nco).  In summary, the large- and small-\sigv\ domains are high- and low-opacity domains.

\vspace{1mm}There is a larger consequence for this understanding of how \tco\ emission should, or should not, be used to estimate \nco.  For a very large fraction of the lines of sight through a disk galaxy like the Milky Way, the \tco\ emission will strongly belong to one of the large-\sigv\ or small-\sigv\ domains, but hardly any such pixels will belong to some domain which looks like an average of these two.  Thus, using a single, average \tco\ $\rightarrow$ \nco\ conversion factor on a given \tco\ map will simultaneously give \nco\ values both too high and too low across this map.  Analysis based on such a conversion will therefore be suspect.  We have argued similarly in prior papers based on a more precise mathematical treatment of the radiative transfer solutions \citep[e.g.,][]{b18}, and present that analysis on the ThrUMMS DR6 data in Appendix \ref{rta}.

%\clearpage

% Figure A8.5
%\begin{figure*}[h]
%\vspace{0mm}
%\includegraphics[angle=0,scale=0.029]{dr6-all-12co-hist-mom2.jpg}~~~\includegraphics[angle=0,scale=0.029]{dr6-all-12covsZM-chist-mom2.jpg}
%
%\vspace{-3mm}(a)\hspace{88mm}(b)
%
%\vspace{1mm}
%\caption{\footnotesize . $$ $$
%\label{12vZMsigratios}}
%\end{figure*}

%%%%%%%%%
%   Section A4  %
%%%%%%%%%
\subsection{Longitude-Velocity Intensity Composites}\label{LVmaps}
A standard presentation and analysis tool for Galactic studies is the longitude-velocity or \lv\ diagram, where instead of integrating or averaging over all velocities to form moments, one does so over all latitudes.  This has the advantage of (partially) deprojecting any overlapping velocity components at one longitude from clouds at different heliocentric distances, which are Doppler-shifted to different \vlsr\ due to Galactic rotation.  (This assumes, of course, that a cloud's latitude contributes negligibly to the \vlsr.)

\begin{sidewaysfigure*}[h]
\vspace{15mm}
\centerline{\hspace{-1mm}\includegraphics[angle=-90,scale=0.90]{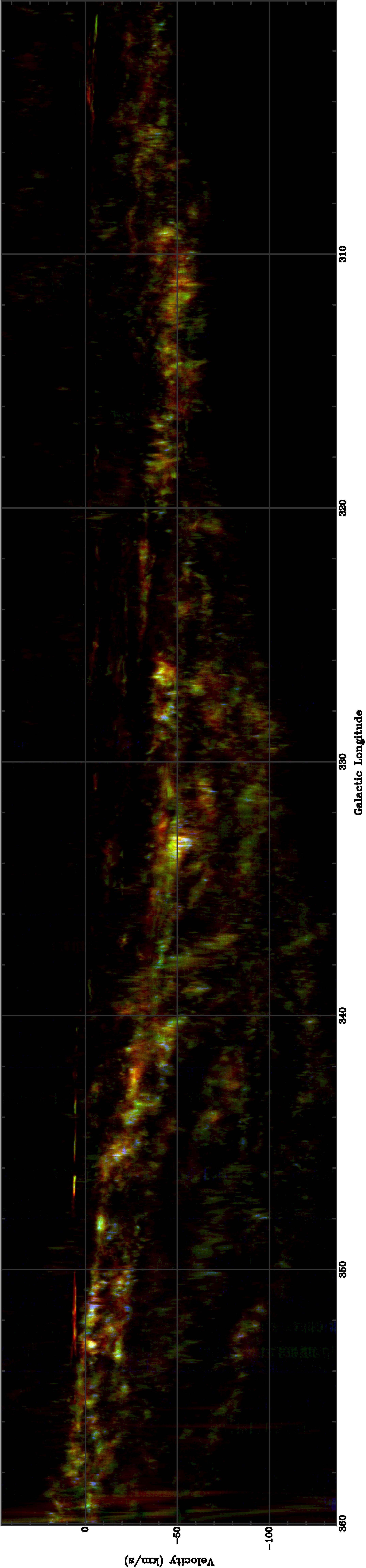}}
\centerline{\hspace{-1mm}\includegraphics[angle=-90,scale=0.90]{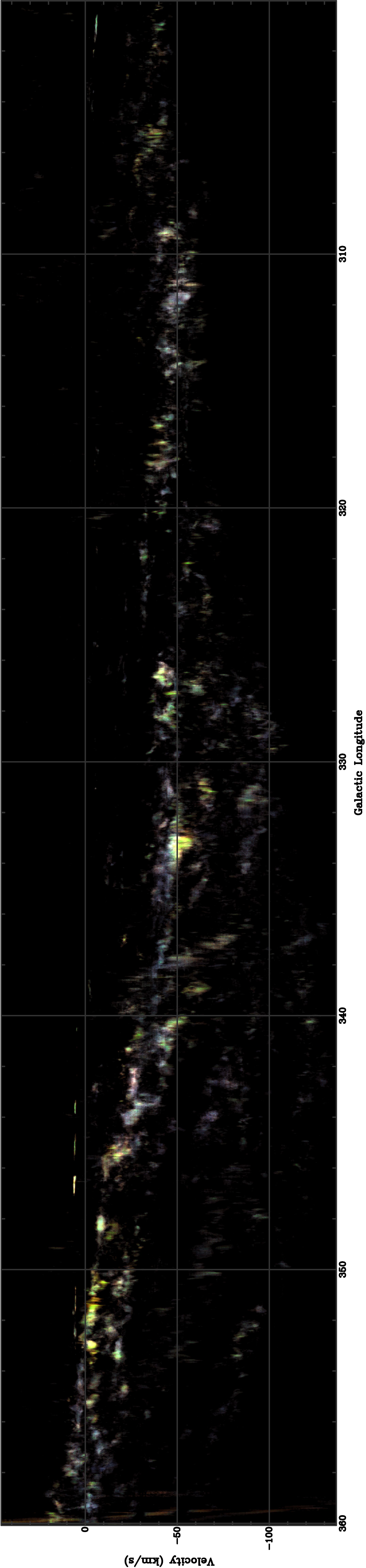}}
\vspace{-105mm}\hspace{108mm}{\bf {\color{red}\tco}\hspace{2mm}{\color{green}\ttco}\hspace{2mm}{\color{cyan}\ceto}}

\vspace{50mm}
\hspace{112mm}{\bf {\color{red}\tex}\hspace{2mm}{\color{green}\nco}\hspace{2mm}{\color{cyan}$\tau_{\rm CO}$}}

\vspace{50mm}
\caption{\footnotesize Two 60\degree$\times$181\,\kms\ ThrUMMS DR6 RGB-composite mosaics of ({\em top}) line intensity and ({\em bottom}) physical property, integrated or averaged across $b$ (zeroth or --first moment) as appropriate: i.e., $lV$ diagrams.  These are from the same data cubes as portrayed in Figs.\,\ref{full121318-mom0} and \ref{fullTexZMtau-mom0}, except restricted to a smaller \vlsr\ range in order to focus on the spiral arm emission.  Line ratio and physical property variations persist in these projections as well.  The data maximum and saturated colour levels %median error, and black levels , {\color{red}9.5}, --5; , {\color{red}?}, --2.4; , {\color{red}?}, 0.0
in these images are respectively at ({\em top}) 780.8 \& 623.7\,K\,arcmin (\tco), 233.4 \& 153.7\,K\,arcmin (\ttco); and 37.49 \& 15.85\,K\,arcmin (\ceto); ({\em bottom}) 7.095 \& 3.751\,K (mean \tex), %(ZM = 4058.62,1762.52 Msun/pc2 * arcmin) should be divided by 1.8785936e-24 to convert to Nco
21.6 \& 9.38$\times$10$^{26}$\,molec\,m$^{-2}$\,arcmin (\nco), and 21.07 \& 16.79 (mean $\tau_{\rm ^{12}CO}$). $$ $$ 
\label{full-lv-combo}}
\end{sidewaysfigure*}

\vspace{1mm}With our fully 3D iso-CO and \tnt\ data cubes, we can form RGB-composite \lv\ diagrams to reveal line ratio and physical property variations in this domain as well, similarly to the sky moments of Figures \ref{full121318-mom0} and \ref{fullTexZMtau-mom0}.  Two examples are presented in {\color{red}Figure \ref{full-lv-combo}}.

\vspace{1mm}The striking variations in line ratio and physical properties seen in Figures \ref{full121318-mom0} and \ref{fullTexZMtau-mom0} are clearly visible in these \lv\ diagrams too.  Additionally, we see these variations broken out into the various spiral arm features that have been so widely studied in the literature.  The most prominent of these, both in terms of the iso-CO intensity distribution and the most extreme values of the physical \tnt\ quantities, is the Scutum-Centaurus arm \citep[e.g.,][]{r19}.  The Sagittarius-Carina and Norma arms are also visible in some guise.

\vspace{1mm}Among other aspects, all the major spiral arm features have in common some relatively large \sigv, typically $\sim$10\,\kms, at least when viewed on this scale and integrated/averaged over all $b$ (the conclusions from \S\ref{vdisp} notwithstanding).  This renders as numerous ``vertical brushstroke'' features in these diagrams (given the aspect ratios of the \lv\ pixels) and is typical of distant, massive star-forming regions.  That is, they have relatively small longitude extents (\lapp1\degree) but large \vlsr\ extents.

\vspace{1mm}The biggest contrast with this property of the major spiral arms is seen in sharp ``horizontal'' features near \vlsr\ = $\pm$10\,\kms, extending over 1\degree--3\degree\ or more at a number of longitudes but with \sigv\ \lapp\,1\,\kms.  This is the known signature of low-mass molecular clouds that are relatively local to the Sun, typically 150--400\,pc; an example is the Coalsack at $l$=301\degree.  We see that these clouds also exhibit striking line ratio and physical property variations, even at the much smaller physical map scales afforded by their proximity.

% Figure A9

% Figure Aj
%\begin{figure*}[h]
%\centerline{
%\hspace{4mm}\includegraphics[angle=0,scale=0.05]{dr6-mosaicAll-TexZMtauLO-lv.jpg}}
%\vspace{-3mm}\centerline{
%\hspace{4mm}\includegraphics[angle=0,scale=0.05]{dr6-mosaicAll-TexZMtauHI-lv.jpg}}
%\vspace{0mm}
%\caption{{\bf LEAVE OUT?} A similar $lV$ diagram to Fig.\,\ref{full121318-lv} but for a \tex-\nco-$\tau$ composite like Fig.\,\ref{fullTexZMtau-mom0}.  Two panels are shown at different brightness to bring out bright and faint details in these high dynamic range data.  ({\em Top}) The data maximum, saturated colour, median error, and black levels in this image are respectively {\color{red}check all!} at 3.16, 0.85, {\color{red}9.5}, --0.03\,K (mean \tex); 257, 44, {\color{red}?}, \& --0.4$\times$10$^{24}$\,molec\,m$^{-2}$ (\nco); and 5.53, 3.67, {\color{red}?}, --0.11 (mean $\tau_{\rm ^{12}CO}$).  ({\em Bottom}) The saturation levels are ...
%\label{full121318-hilv}}
%\end{figure*}

%% Figure A10
\begin{sidewaysfigure*}[h]
\vspace{10mm}
\centerline{\hspace{-0.8mm}\includegraphics[angle=-90,scale=0.90]{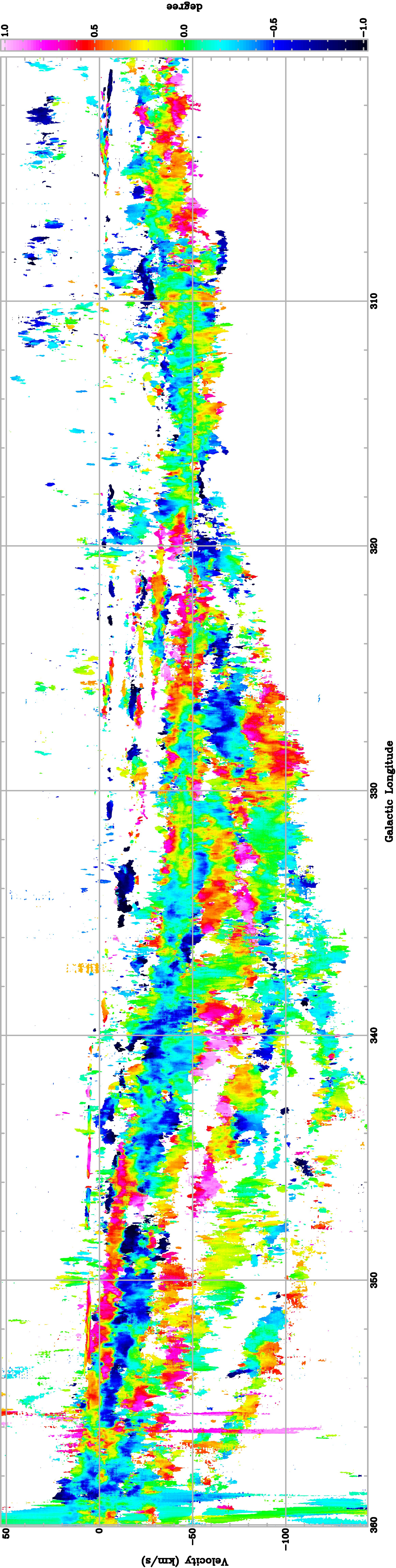}}
\vspace{2mm}
\centerline{\hspace{-0.8mm}\includegraphics[angle=-90,scale=0.90]{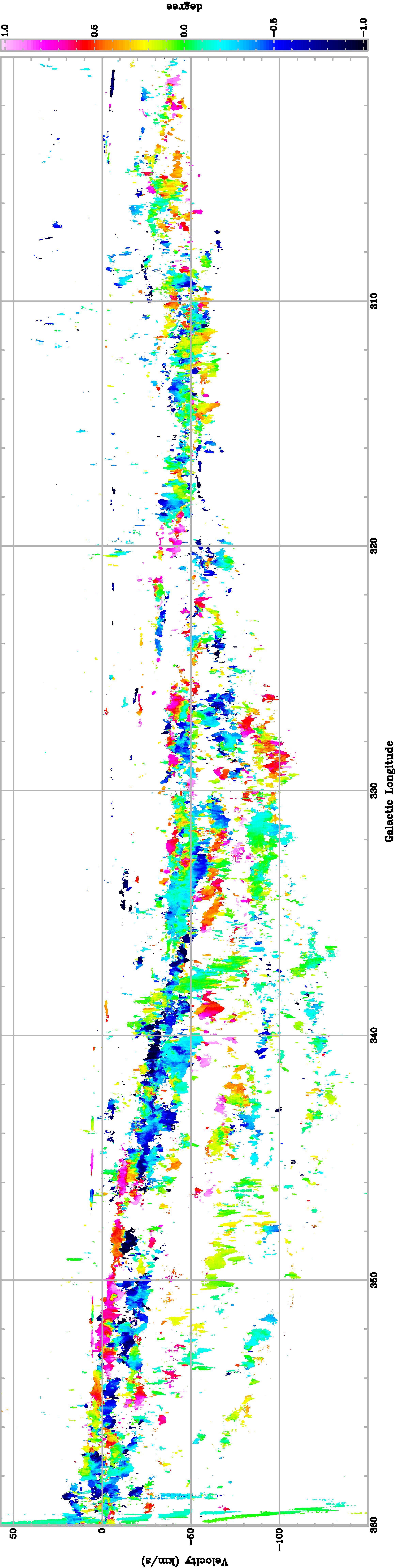}}
\vspace{2mm}
\caption{\footnotesize Two 60\degree$\times$200\,\kms\ ThrUMMS DR6 mosaics of ({\em top}) \tco-weighted and ({\em bottom}) \nco-weighted  mean latitude: i.e., first moment $lV$ diagrams integrated across $b$.  These are from the same data cubes as portrayed in Figure \ref{full-lv-combo}.  $$ $$
\label{full-lv1-12coZM-rainbow}}
\end{sidewaysfigure*}

% Figure A10-alt: choose this or prior
%\begin{sidewaysfigure*}[h]
%\centerline{\hspace{-1mm}\includegraphics[angle=-90,scale=0.92]{dr6-mosAll-12co-lv1-doppler.jpg}}
%\vspace{2mm}
%\centerline{\hspace{-1mm}\includegraphics[angle=-90,scale=0.92]{dr6-mosAll-ZM-lv1-doppler.jpg}}
%\vspace{2mm}
%\caption{Two 60\degree$\times$200\,\kms\ ThrUMMS DR6 mosaics of ({\em top}) \tco-weighted and ({\em bottom}) \nco-weighted  mean latitude: i.e., first moment $lV$ diagrams integrated across $b$.  These are from the same data cubes as portrayed in Figure \ref{full-lv-combo}.  
%\label{full-lv1-12coZM-doppler}}
%\end{sidewaysfigure*}

% Figure A11 for real
\begin{sidewaysfigure*}[h]
\vspace{10mm}
\centerline{\hspace{-1mm}\includegraphics[angle=-90,scale=0.90]{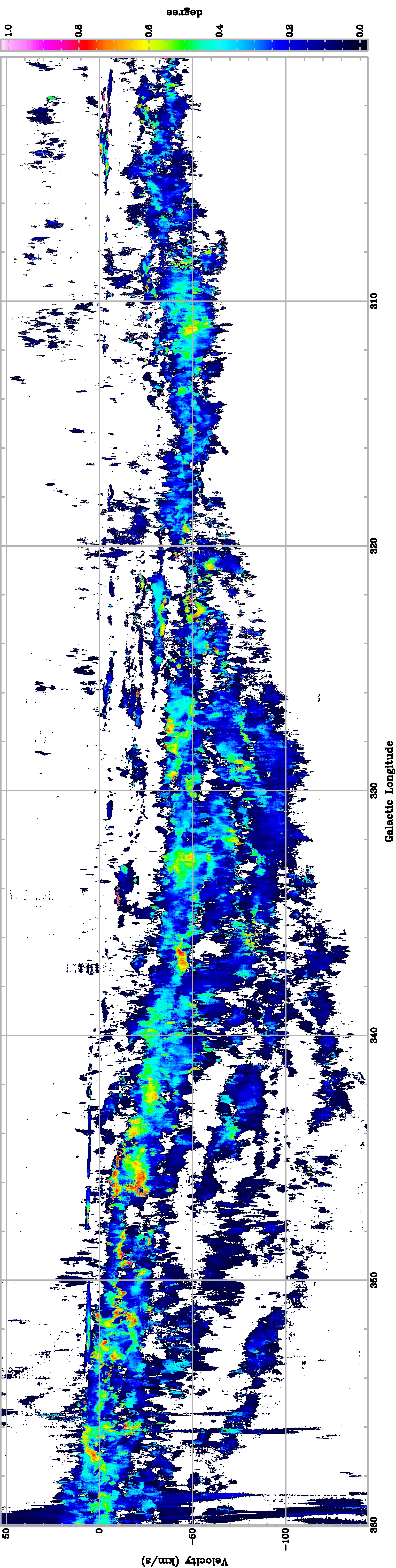}}
\vspace{2mm}
\centerline{\hspace{-1mm}\includegraphics[angle=-90,scale=0.90]{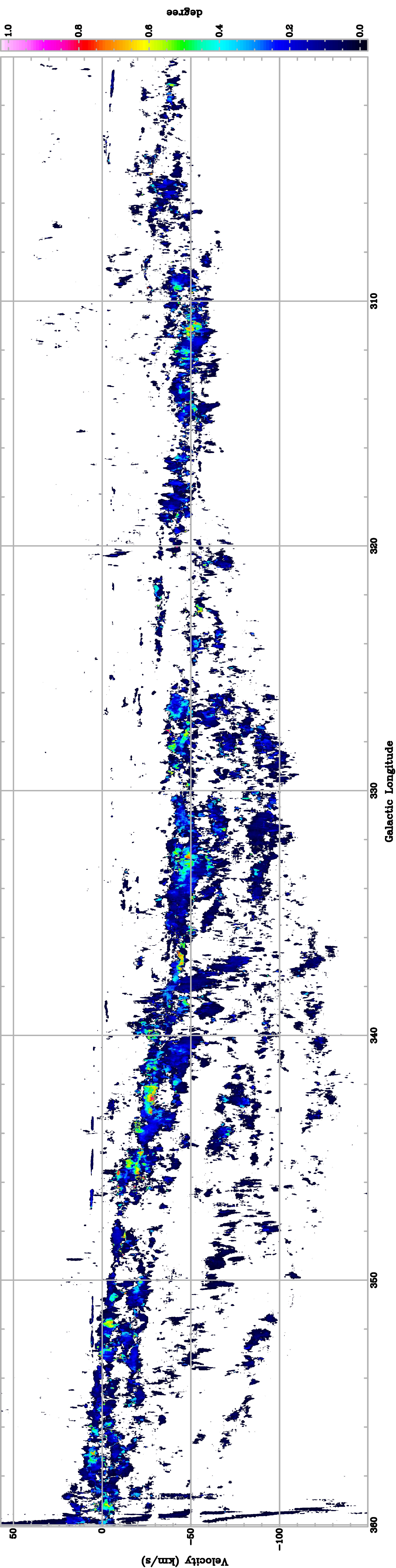}}
\vspace{2mm}
\caption{\footnotesize Two 60\degree$\times$200\,\kms\ ThrUMMS DR6 mosaics of ({\em top}) \tco-weighted and ({\em bottom}) \nco-weighted latitude dispersion: i.e., second moment $lV$ diagrams integrated across $b$.  These are from the same data cubes as portrayed in Figure \ref{full-lv-combo}.  $$ $$
\label{full-lv2-12coZM}}
\end{sidewaysfigure*}

%%%%%%%%%
%   Section A5  %
%%%%%%%%%
\subsection{Latitude Distributions in ($l$,$V$)}\label{latmaps}
As with sky maps (moments in $V$), analysis of information in standard \lv\ diagrams can also be extended to higher moments in $b$.  A first-moment \lv\ diagram gives the intensity-weighted mean latitude $\bar{b}$, while a second moment gives the latitude dispersion $\sigma_{b}$ of the emission or other quantity.  These moments can be calculated for any of the six cubes in hand, but we limit ourselves here to those for the \tco\ and \nco\ cubes, which we consider most instructive.  Other moments are available digitally.

\vspace{1mm}Similarly to Figures \ref{full12co-mom1}--\ref{fullZM-mom2}, the \tco\ mosaic has wider areal coverage than the \nco\ one, since the latter requires good sensitivity also in the \ttco\ data, resulting in more pixels where \tco\ is clearly detected but \ttco\ is not.  Nevertheless, the $\bar{b}$ ({\color{red}Fig.\,\ref{full-lv1-12coZM-rainbow}}) and $\sigma_{b}$ ({\color{red}Fig.\,\ref{full-lv2-12coZM}}) distributions generally follow each other reasonably well.

\vspace{1mm}In both cases, the mean latitude $\bar{b}$ is near 0\degr\ over a large fraction of \lv\ space, as expected for the ensemble of Galactic molecular clouds which are the most extreme Population I objects in the Milky Way: they essentially define the Galactic equator.  Where $\bar{b}$ deviates most noticeably from 0\degr\ coincides with clouds at the nearest distances, $d$ \lapp\ 3\,kpc (and smallest |\vlsr| \lapp\ 50\,\kms), since that is where their very narrow scale height \citep[$z_{\rm sc}$ = 19\,pc according to][]{r19} projects to the widest range of latitudes (1\degr\ = 50\,pc at 3\,kpc).  Further away, the clouds become progressively more confined to small |$b$|, as presaged in the discussion of \S\ref{vfield}.

\vspace{1mm}Likewise, the latitude dispersion $\sigma_{b}$ (which, for the smaller complexes and clouds, also corresponds approximately to the cloud sizes) is largest at small $d$, and smaller at large $d$.  Thus, a more or less typical cloud of physical size 10\,pc will subtend an angular extent of 0\fdeg6 at 1\,kpc, but only 0\fdeg1 at 6\,kpc.

\vspace{1mm}We return to the discussion of latitude structure after first exploring Galactic kinematics in Appendix \ref{kinem}.

\clearpage

%%%%%%%%%%%%%
%%      Appendix B     %%
%%%%%%%%%%%%%
\section{Iso-CO Radiative Transfer Analysis}\label{rta}

%%%%%%%%%
%   Section B1  %
%%%%%%%%%
\subsection{Sector-by-Sector Velocity-Resolved Results}\label{vreslaws}
In Paper I we first described our combined line ratio analysis, which enabled the discovery of a non-linear conversion law from \tco\ line brightness \ico\ to column density \nco, of the form \nco\ = $N_0$ \ico$^p$.  Our analysis pipeline was further refined using the higher-sensitivity CHaMP data \citep{b18} on a large collection of molecular clouds in the general direction of the Carina Arm.  These are mostly close to the Solar Circle at an approximate distance of $R_0$ (see Appx.\,\ref{kinem}) from the Galactic Centre, but ranging in heliocentric distance from a few relatively local clouds ($d$ $\sim$ 200\,pc), through many clouds around 2--3\,kpc (such as the 120\,pc-long $\eta$ Carinae GMC at 2.5\,kpc), and out as far as the massive, luminous SF region NGC\,3603, roughly $\sim$7\,kpc away.

\vspace{1mm}The value of the power-law index $p$ was found to be $>$1, but also depends somewhat on how it is measured.  At the full velocity and angular resolution of the CHaMP data, $p$ $\approx$ 1.9 and $N_0$ $\approx$ 10$^{20}$\,m$^{-2}$, while at lower velocity resolution (e.g., integrating all \tco\ emission from a cloud) $p$ is closer to 1.3 and $N_0$ drops to $\approx$ 4$\times$10$^{19}$\,m$^{-2}$.  The implication was that the vast literature using a single-valued $X$ factor to estimate the $N$/$I$ ratio for molecular clouds misses some essential radiative transfer physics, which tends to yield 2--3$\times$ larger cloud masses than the single $X$ factor does.

\vspace{1mm}We now apply those improved techniques to the iso-CO Sector data presented in \S\ref{fullmos}, with the aim of investigating whether (1) we can see any variation in the inferred conversion law coefficients with position in the Galaxy, such as Galactocentric distance, or (2) we can find any dependence of the conversion laws on the angular resolution of the data, within the very wide-area ThrUMMS maps.

\vspace{1mm}We consider the three iso-CO species' data in each Sector.  Following \cite{b18}, we solved, for each voxel in the \lbv\ space of the data, the three simultaneous radiative transfer equations for the species' common LTE excitation temperature \tex\ and 3 opacities $\tau$, assuming the intrinsic gas-phase abundance ratio $R_{13}$ = [\tco]/[\ttco] is everywhere equal to the local ISM value of 60.  This will of course not always be true, especially in the inner Galaxy where $R_{13}$ is thought to drop to a value around 40, but our approach gives us a common standard by which to judge variations in the conversion law with position.  When interpreting the results, we can then make post-facto allowances for a different $R_{13}$ in the inner Galaxy or elsewhere, while the analysis using $R_{13}$ = 60 gives limiting values.

\begin{figure*}[h]
\vspace{0mm}
\centerline{\includegraphics[angle=0,scale=0.12]{x300-72.jpg} \includegraphics[angle=0,scale=0.12]{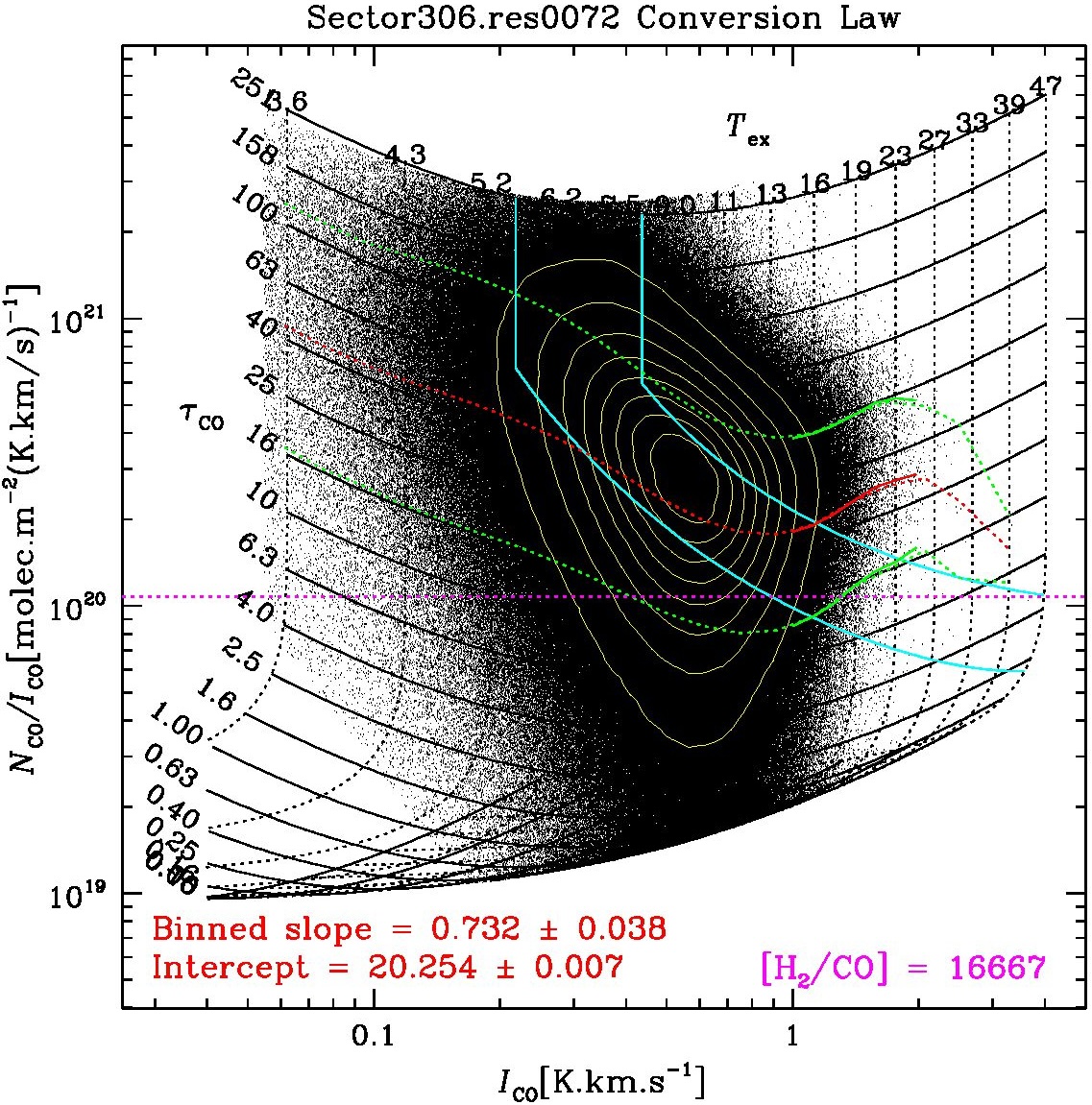} \includegraphics[angle=0,scale=0.12]{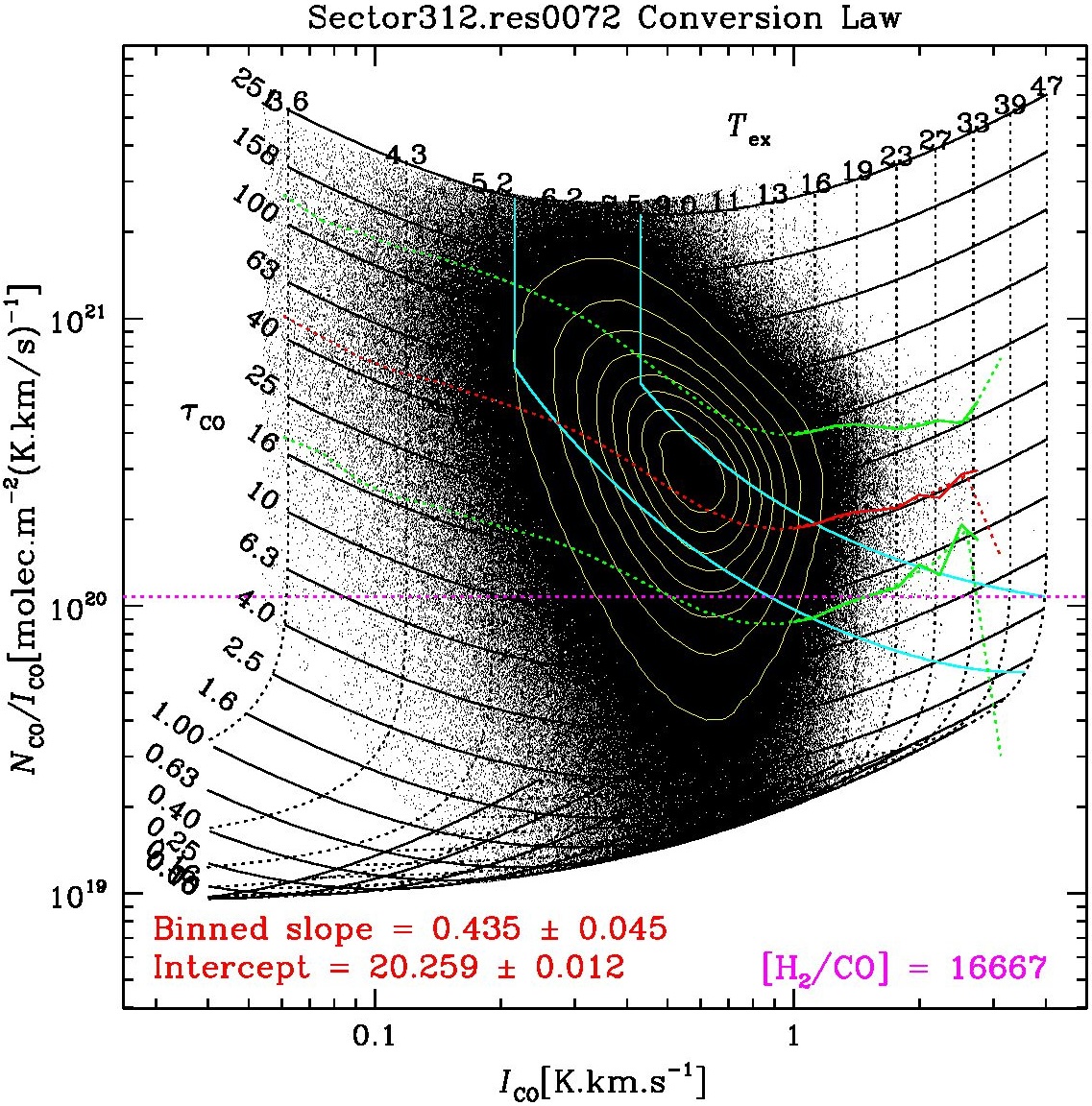}}
\vspace{-3mm}
\centerline{\includegraphics[angle=0,scale=0.12]{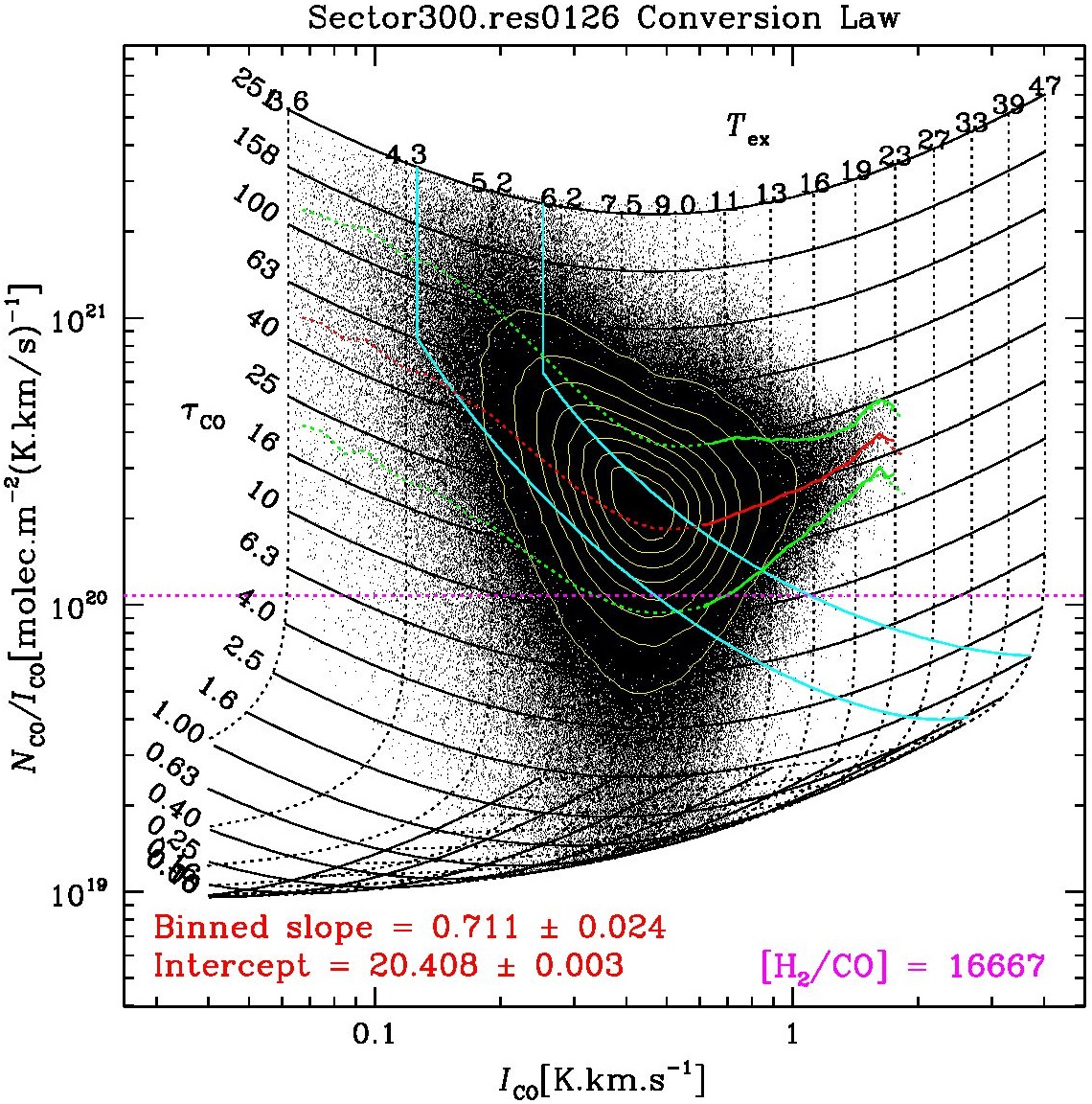} \includegraphics[angle=0,scale=0.12]{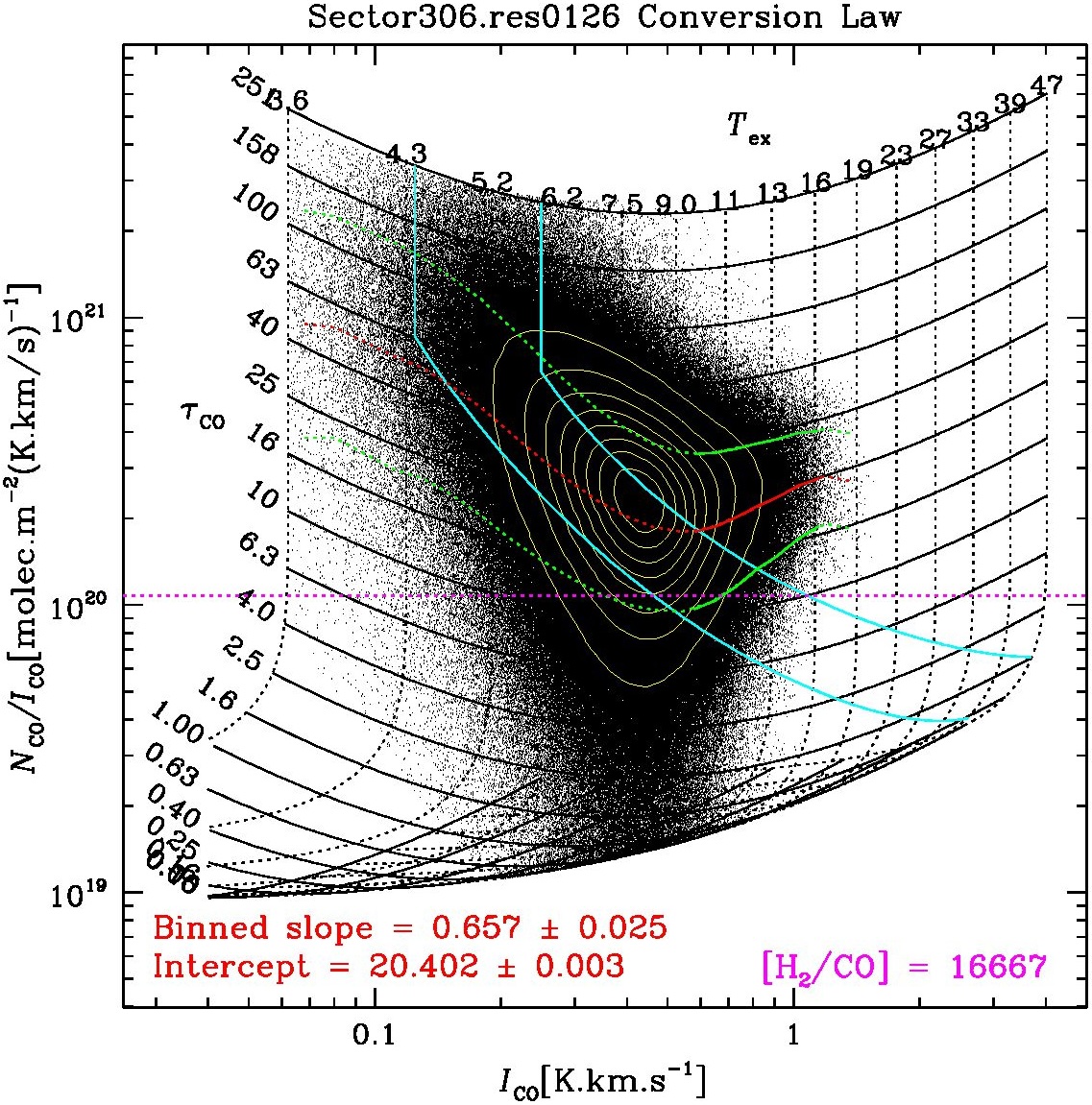} \includegraphics[angle=0,scale=0.12]{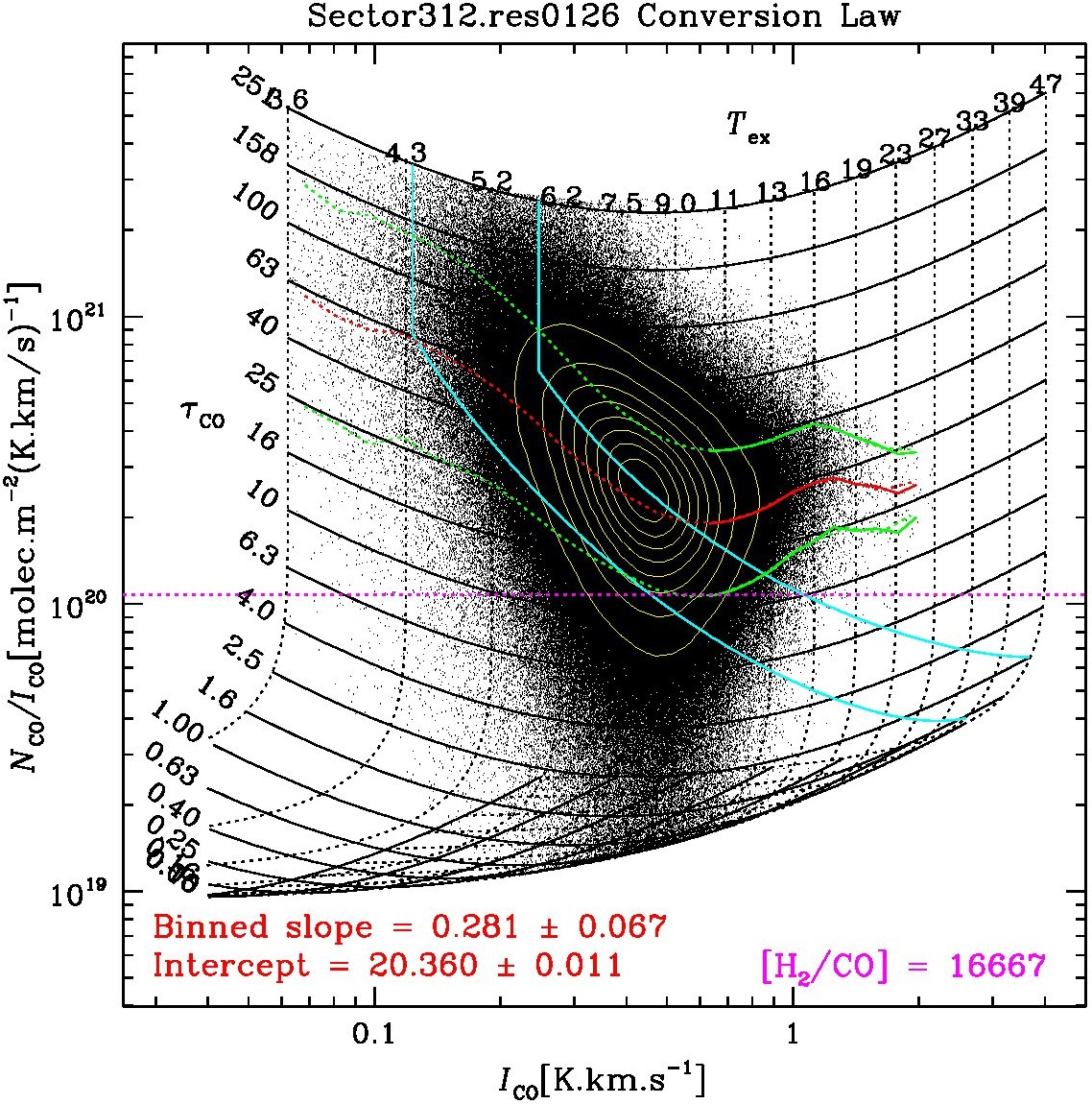}}
\vspace{-3mm}
\centerline{\includegraphics[angle=0,scale=0.12]{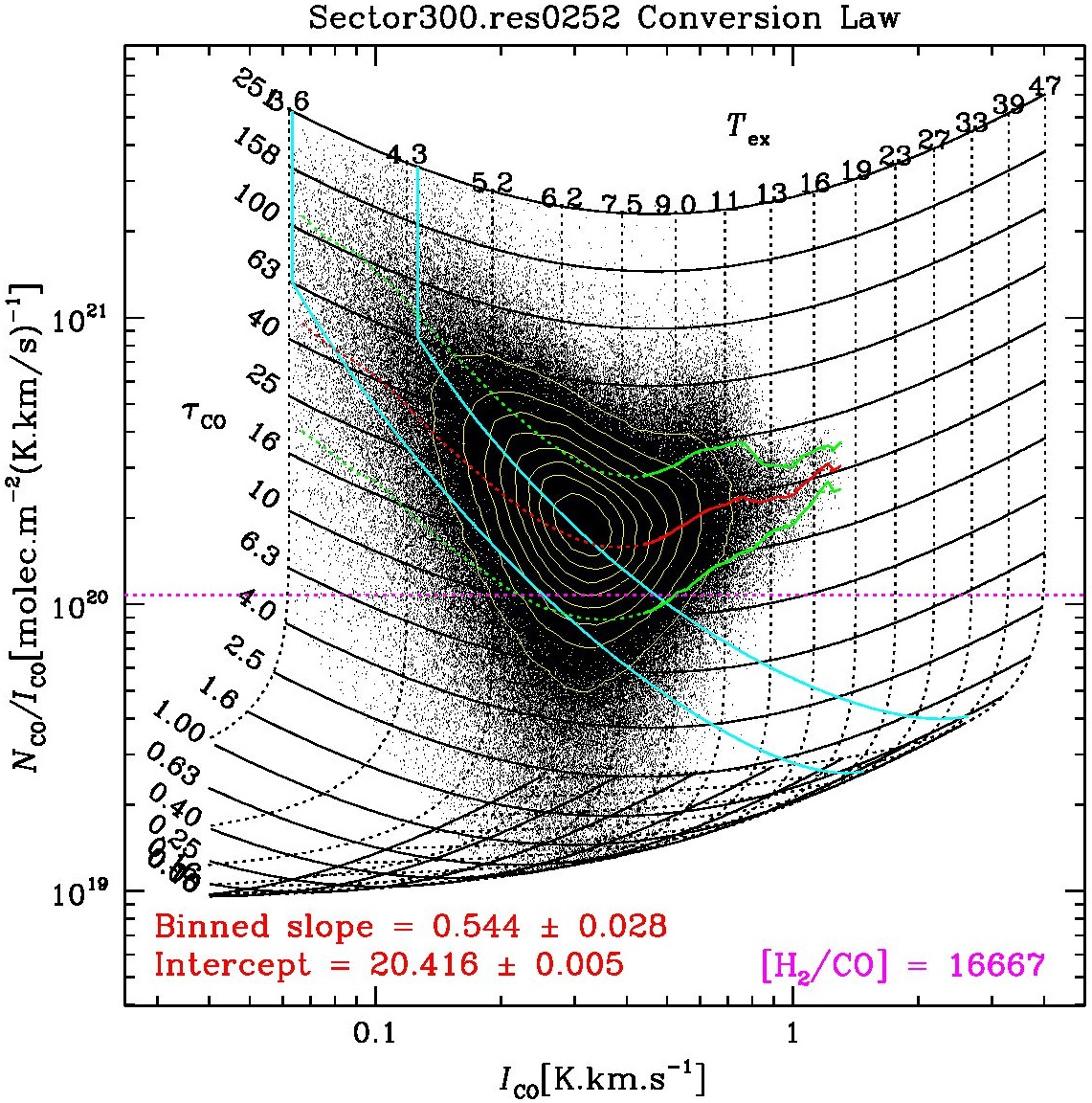} \includegraphics[angle=0,scale=0.12]{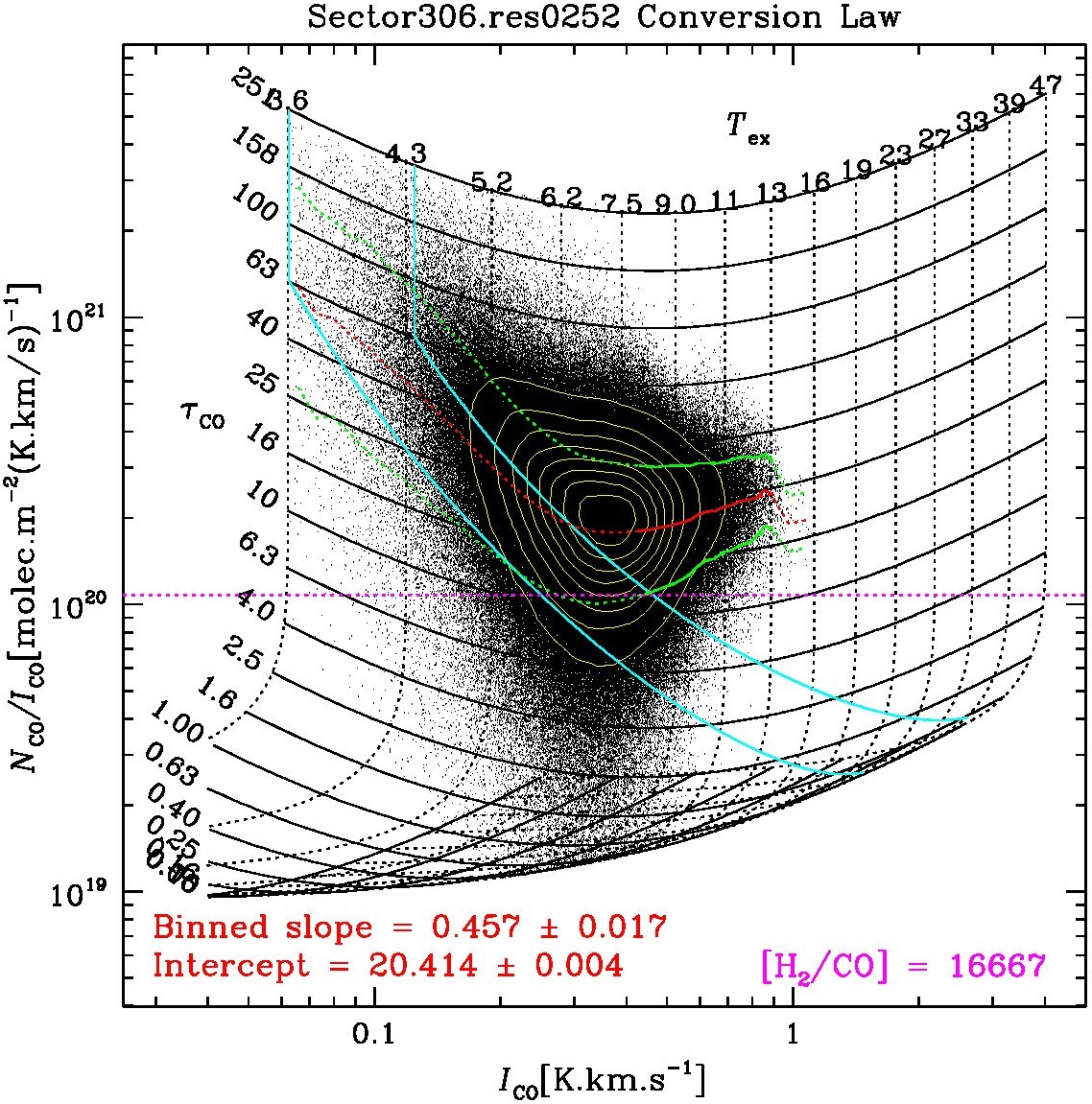} \includegraphics[angle=0,scale=0.12]{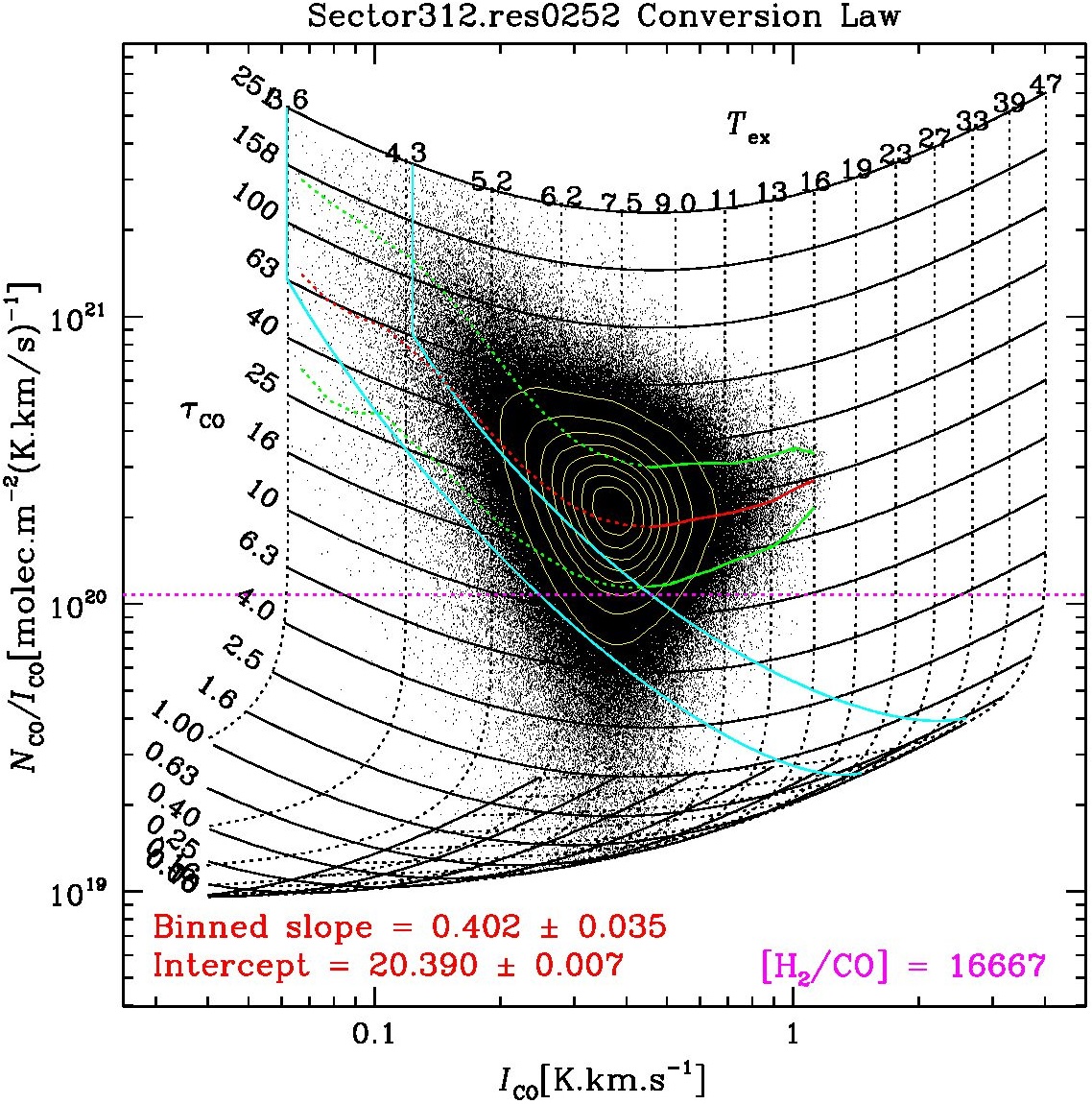}}
\vspace{-3mm}
\centerline{\includegraphics[angle=0,scale=0.12]{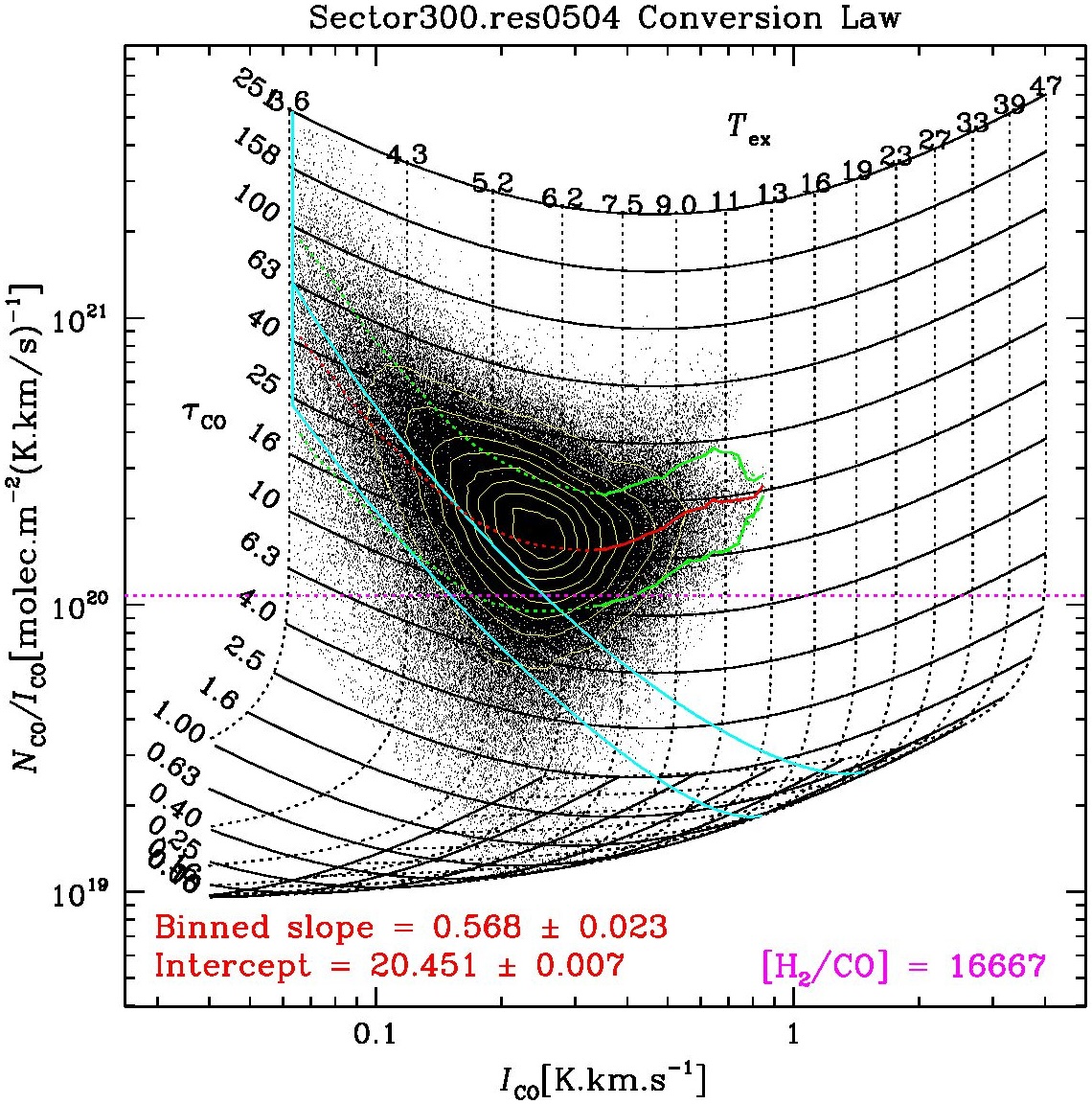} \includegraphics[angle=0,scale=0.12]{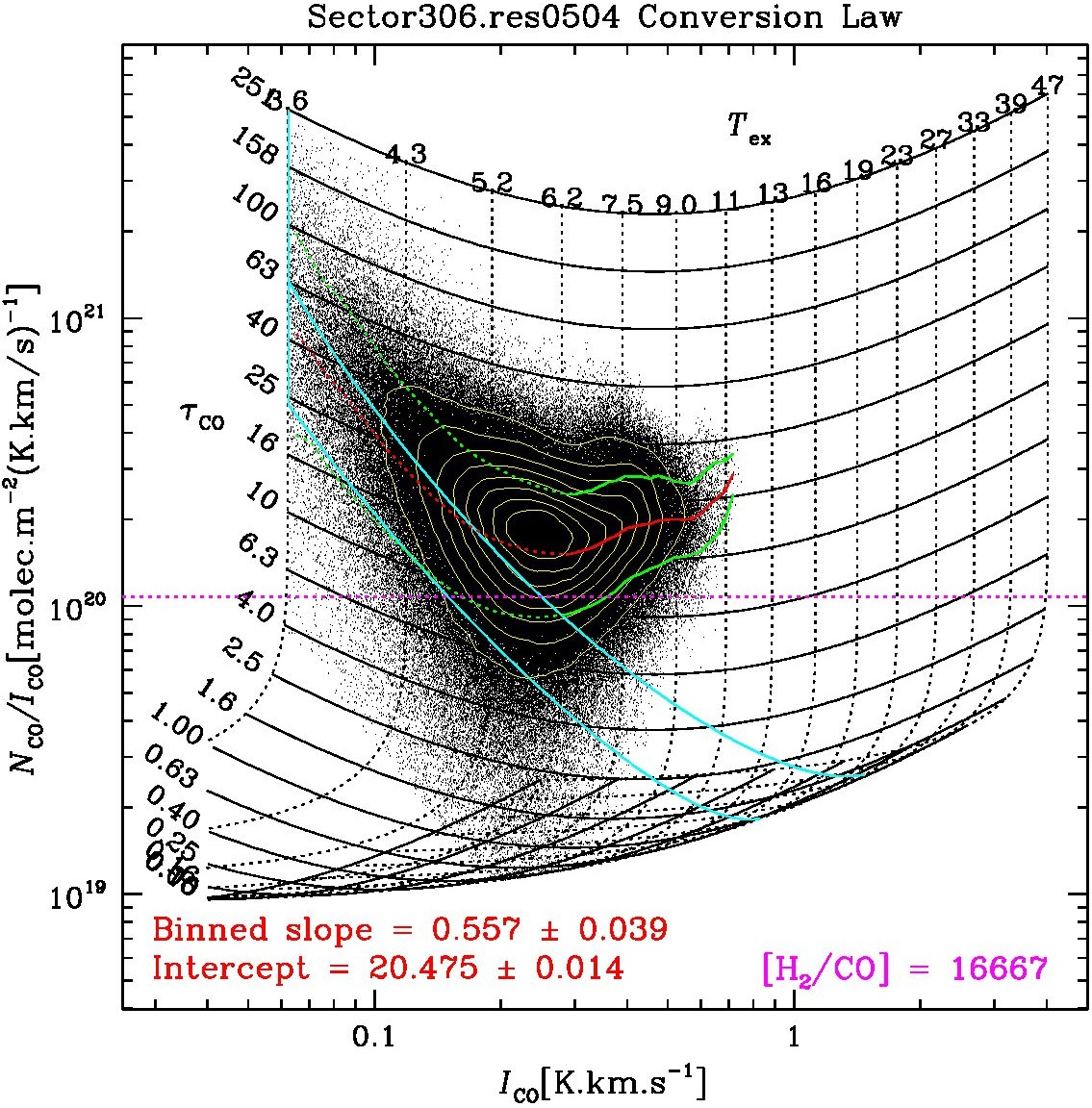} \includegraphics[angle=0,scale=0.12]{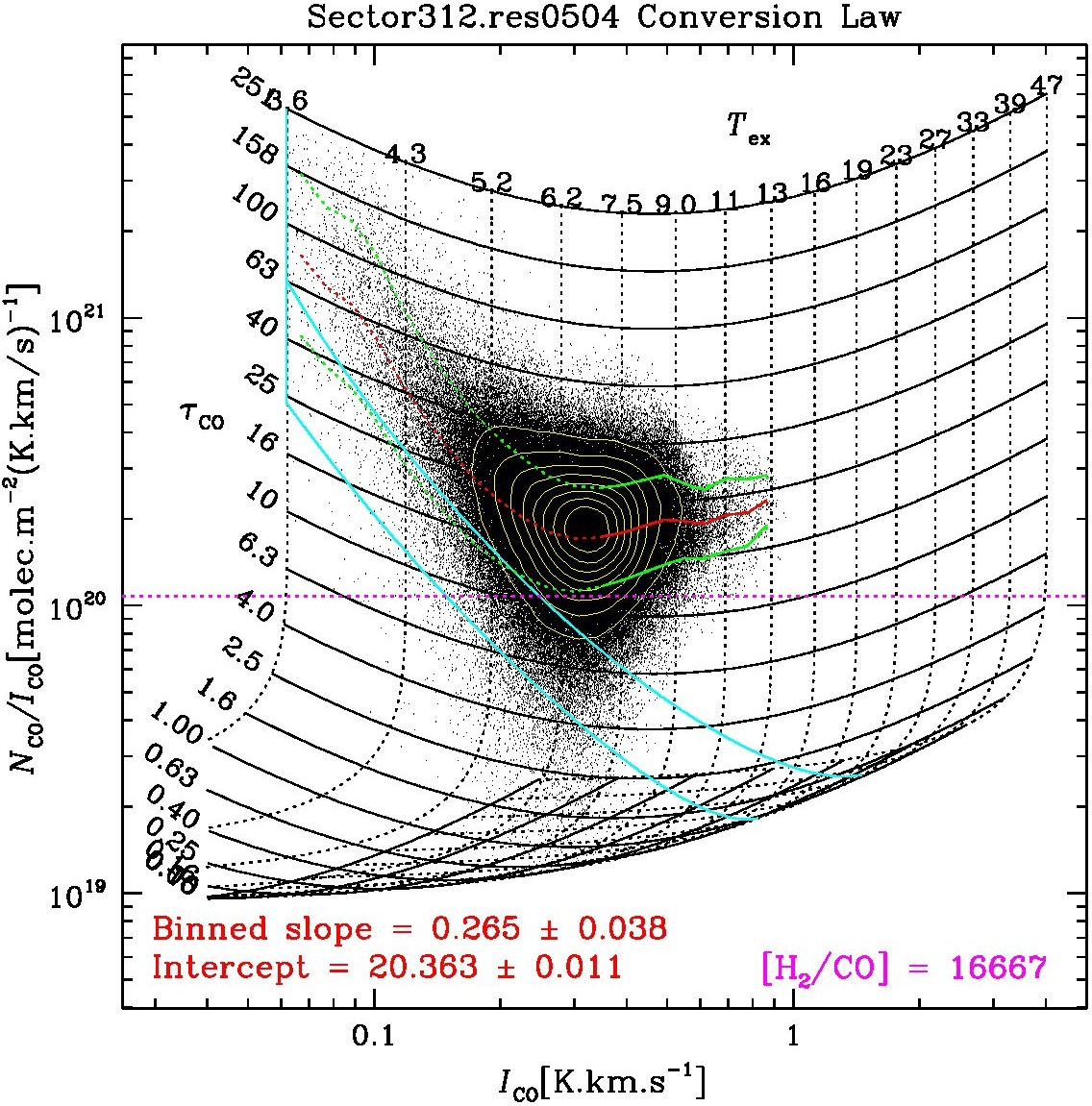}}
\vspace{-3mm}
\centerline{\includegraphics[angle=0,scale=0.12]{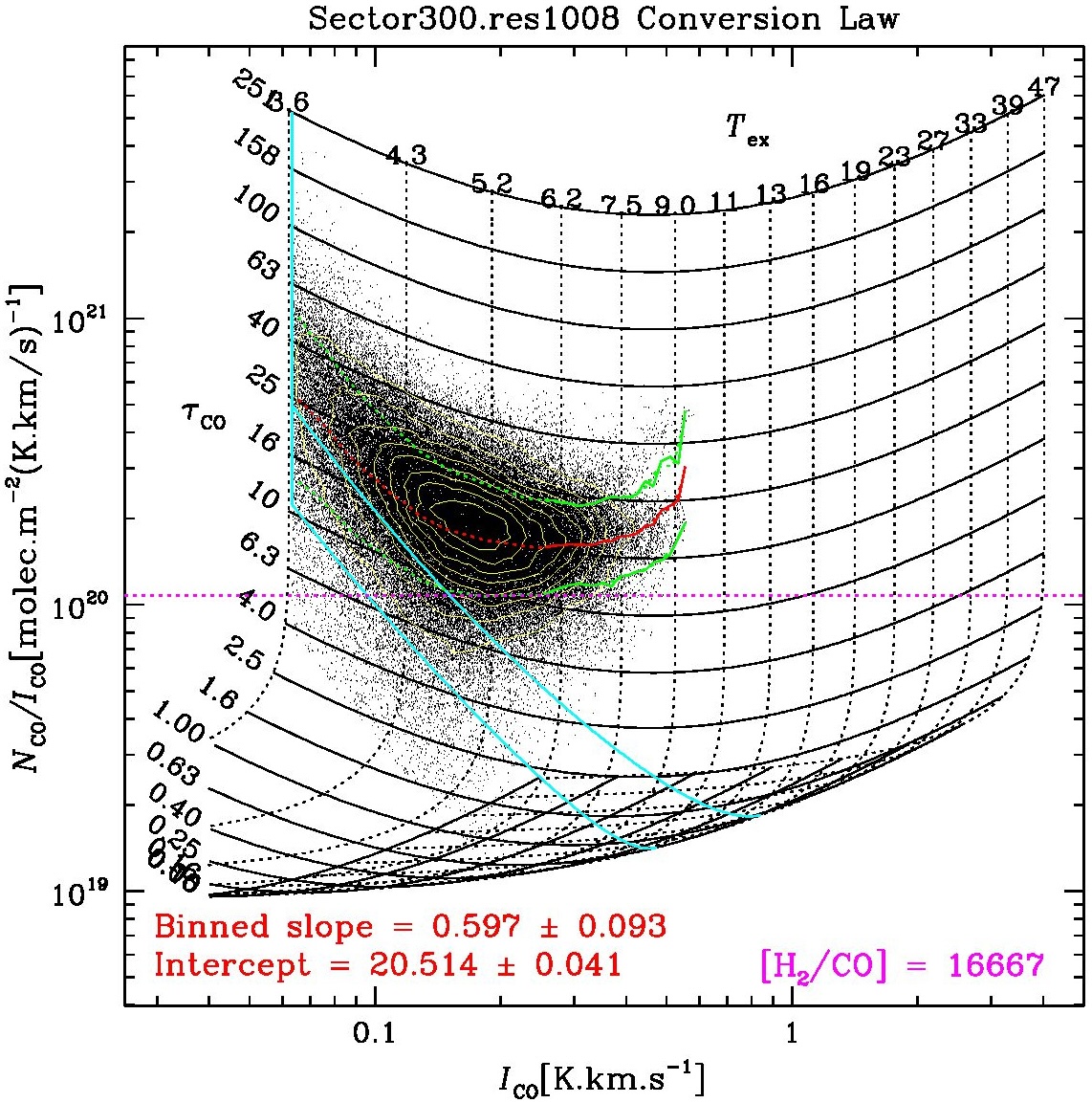} \includegraphics[angle=0,scale=0.12]{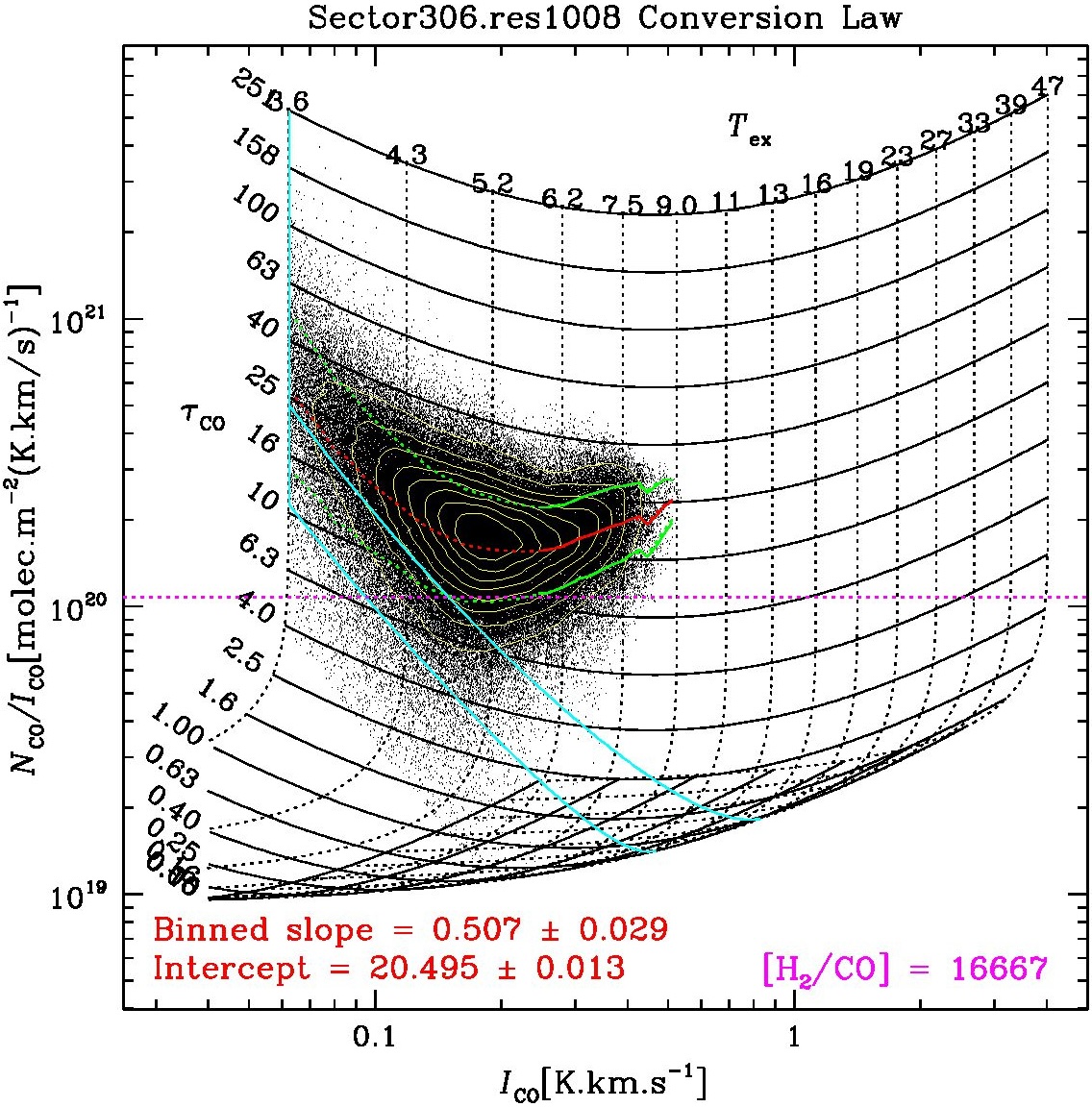} \includegraphics[angle=0,scale=0.12]{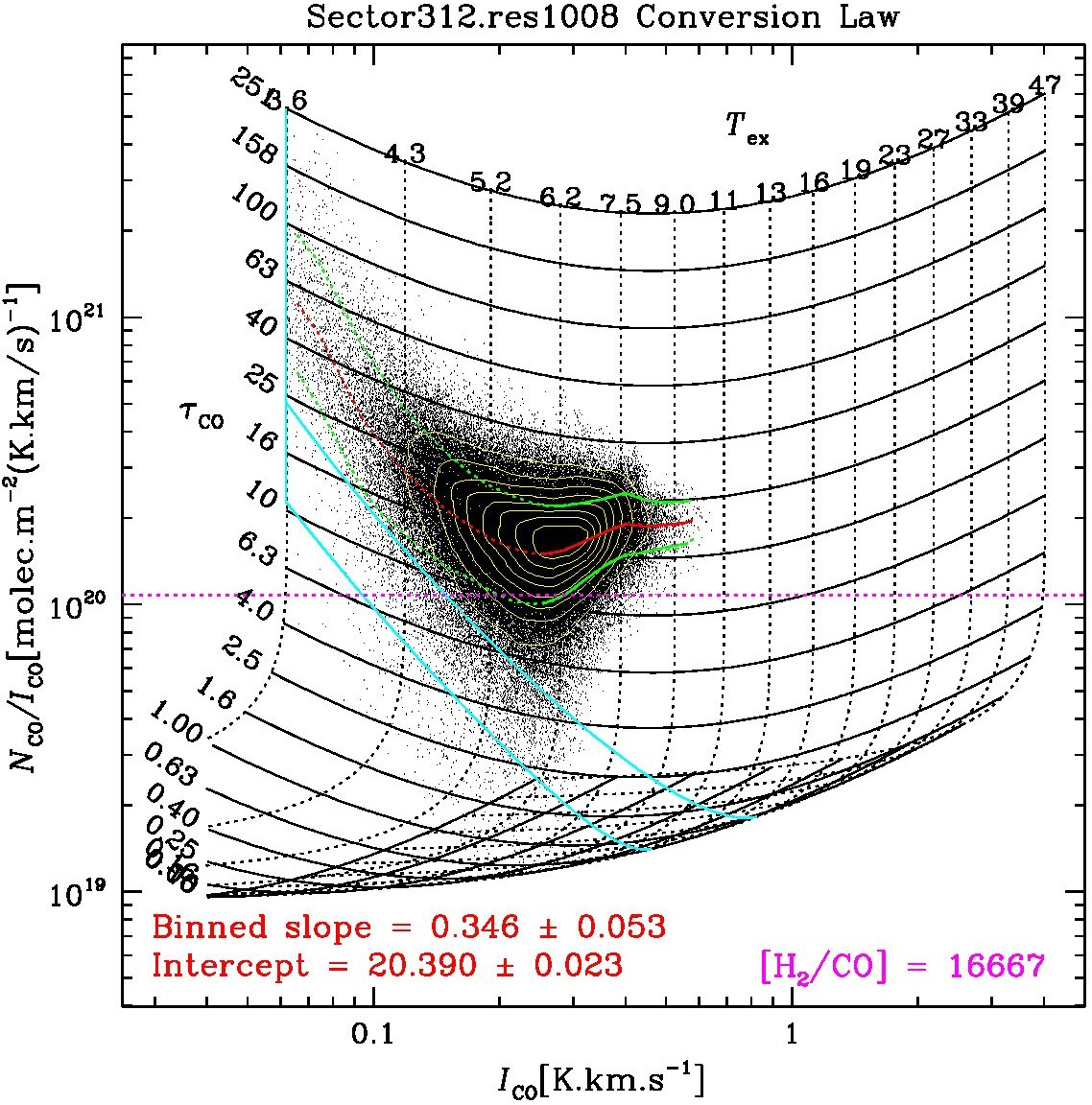}}
\vspace{-1mm}
\caption{\footnotesize Similar plots to Fig.\,\ref{x300}, but for Sectors 300, 306, and 312 (left, middle, right  columns respectively).  The top panel in each column is at ThrUMMS' native resolution of 72$''$.  The second  and subsequent rows show the analysis results for the cubes progressively convolved to a resolution of 126$''$, 252$''$, 504$''$, and 1008$''$, the last being the same resolution as that of the CfA survey \citep{dht01}. $$ $$
\label{x300-12-multi}}
\vspace{0mm}
\end{figure*}

\begin{figure*}[h]
\vspace{0mm}
\centerline{\includegraphics[angle=0,scale=0.12]{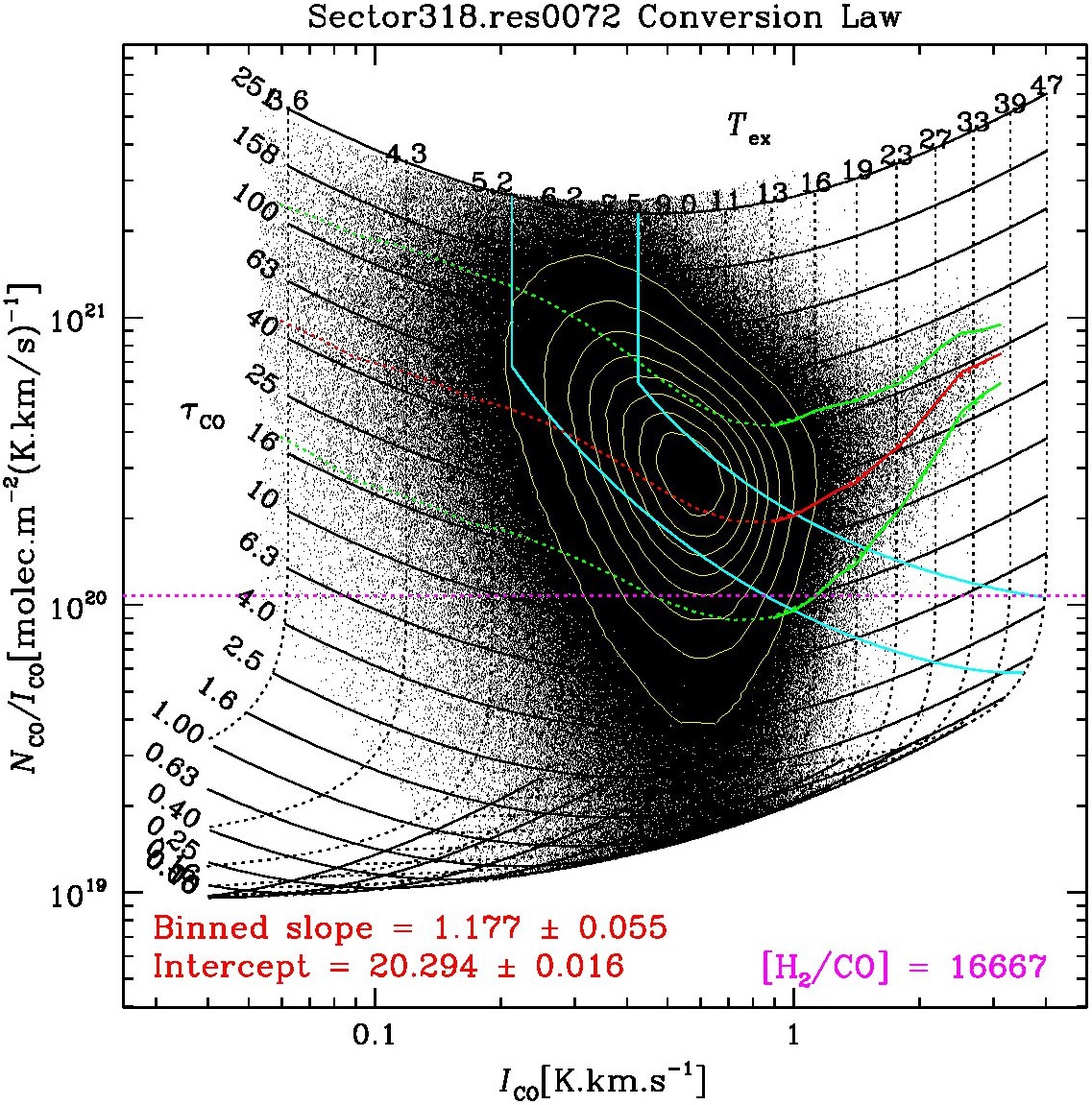} \includegraphics[angle=0,scale=0.12]{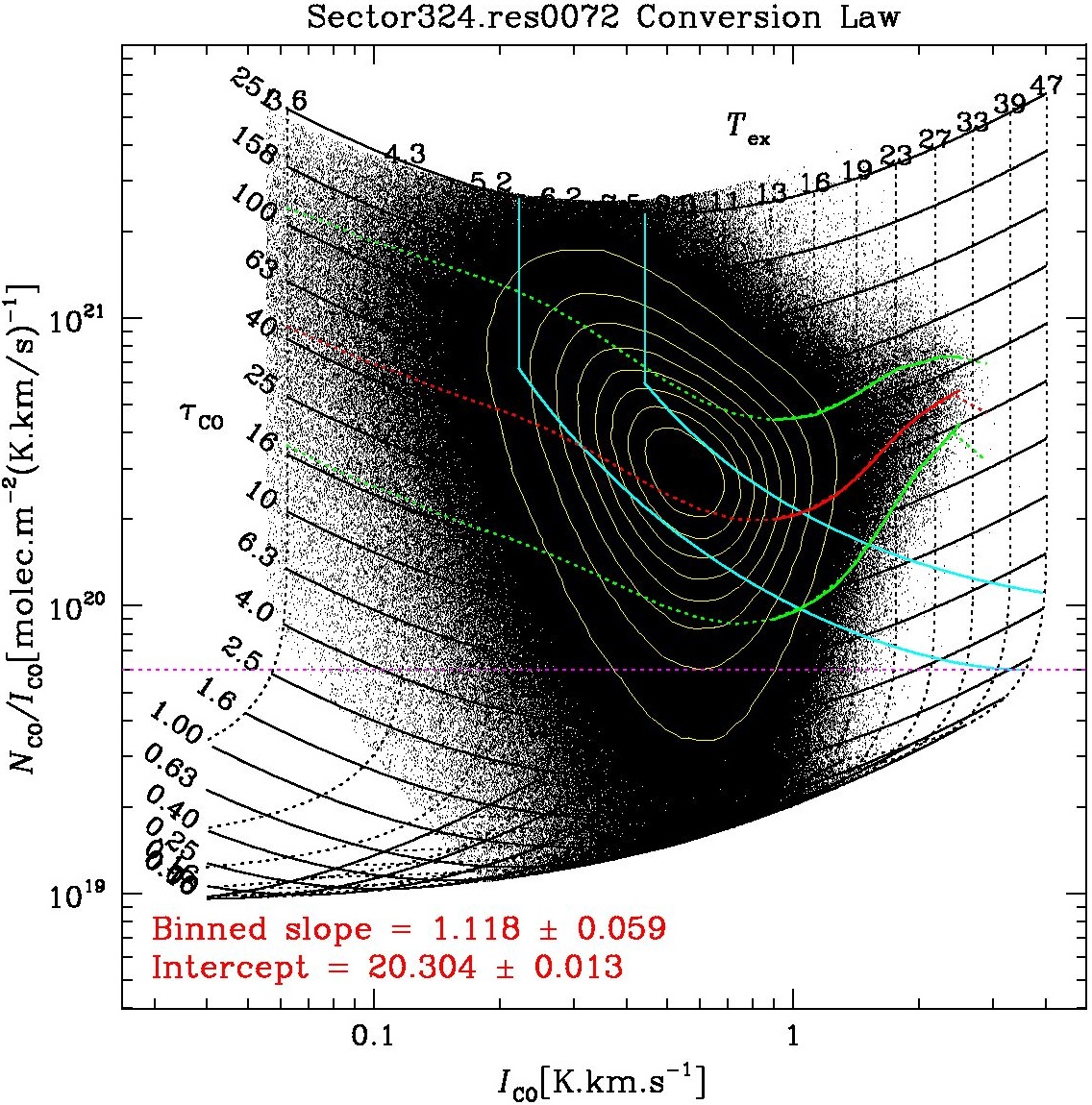} \includegraphics[angle=0,scale=0.12]{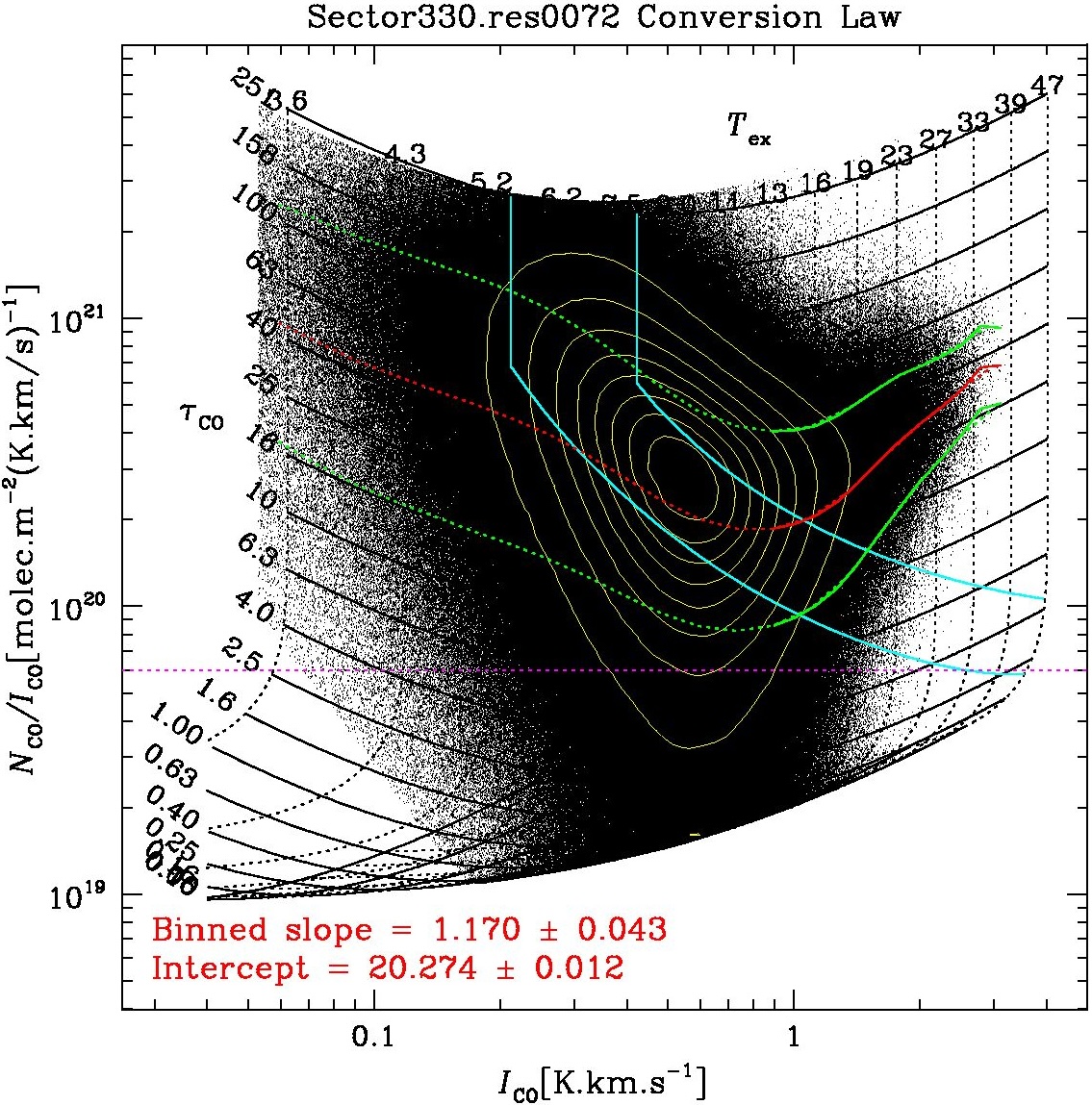}}
\vspace{-3mm}
\centerline{\includegraphics[angle=0,scale=0.12]{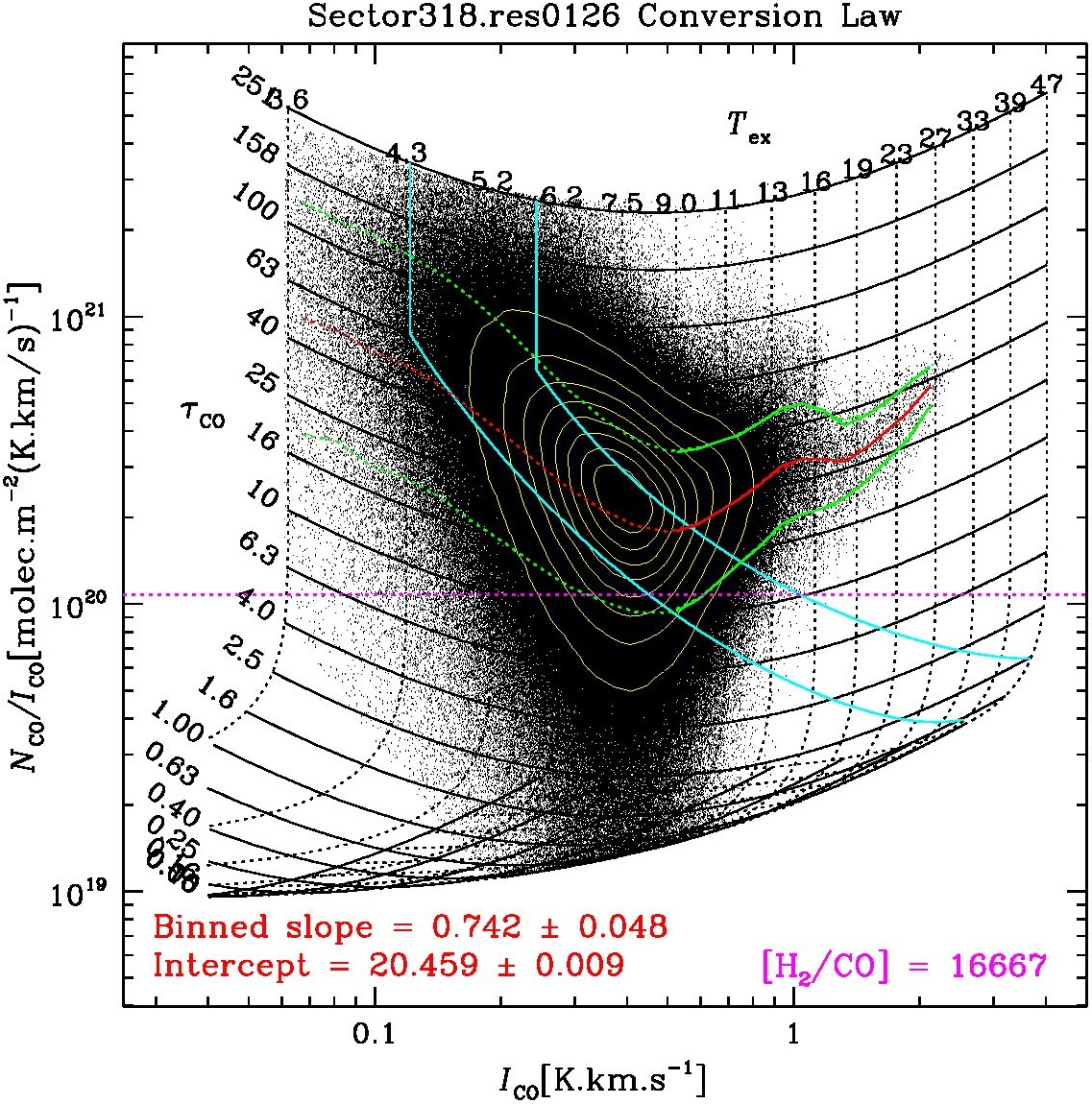} \includegraphics[angle=0,scale=0.12]{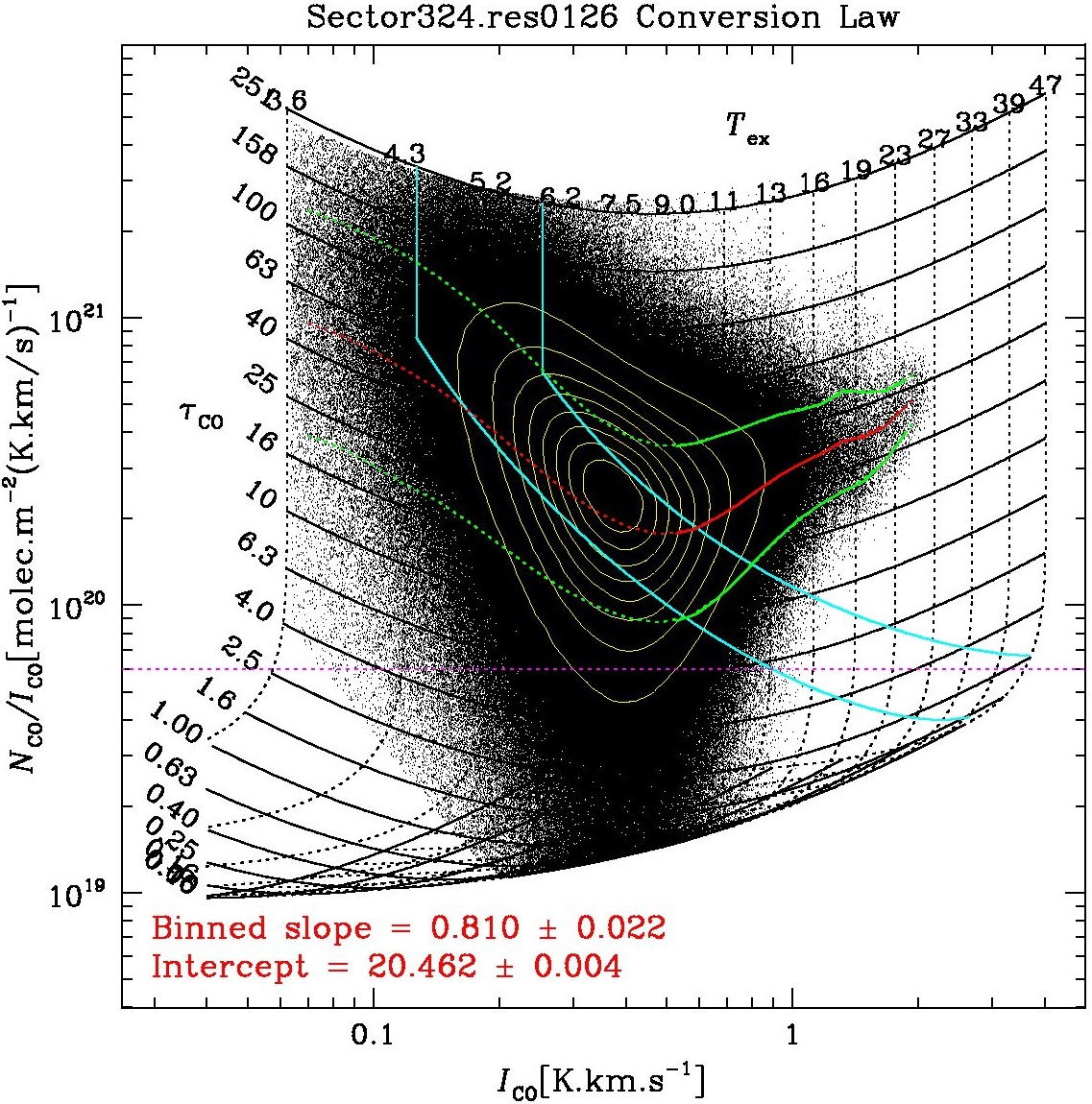} \includegraphics[angle=0,scale=0.12]{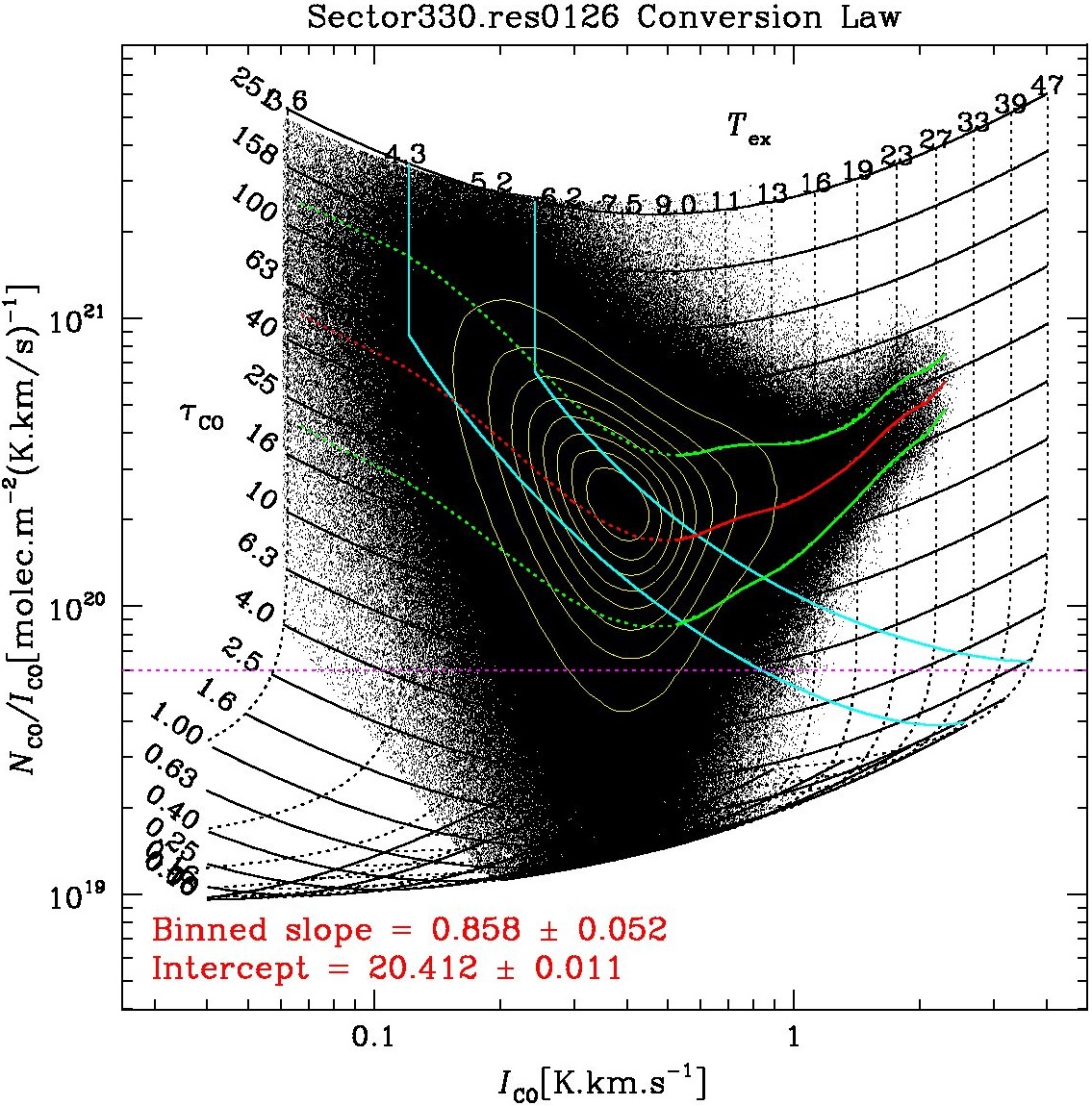}}
\vspace{-3mm}
\centerline{\includegraphics[angle=0,scale=0.12]{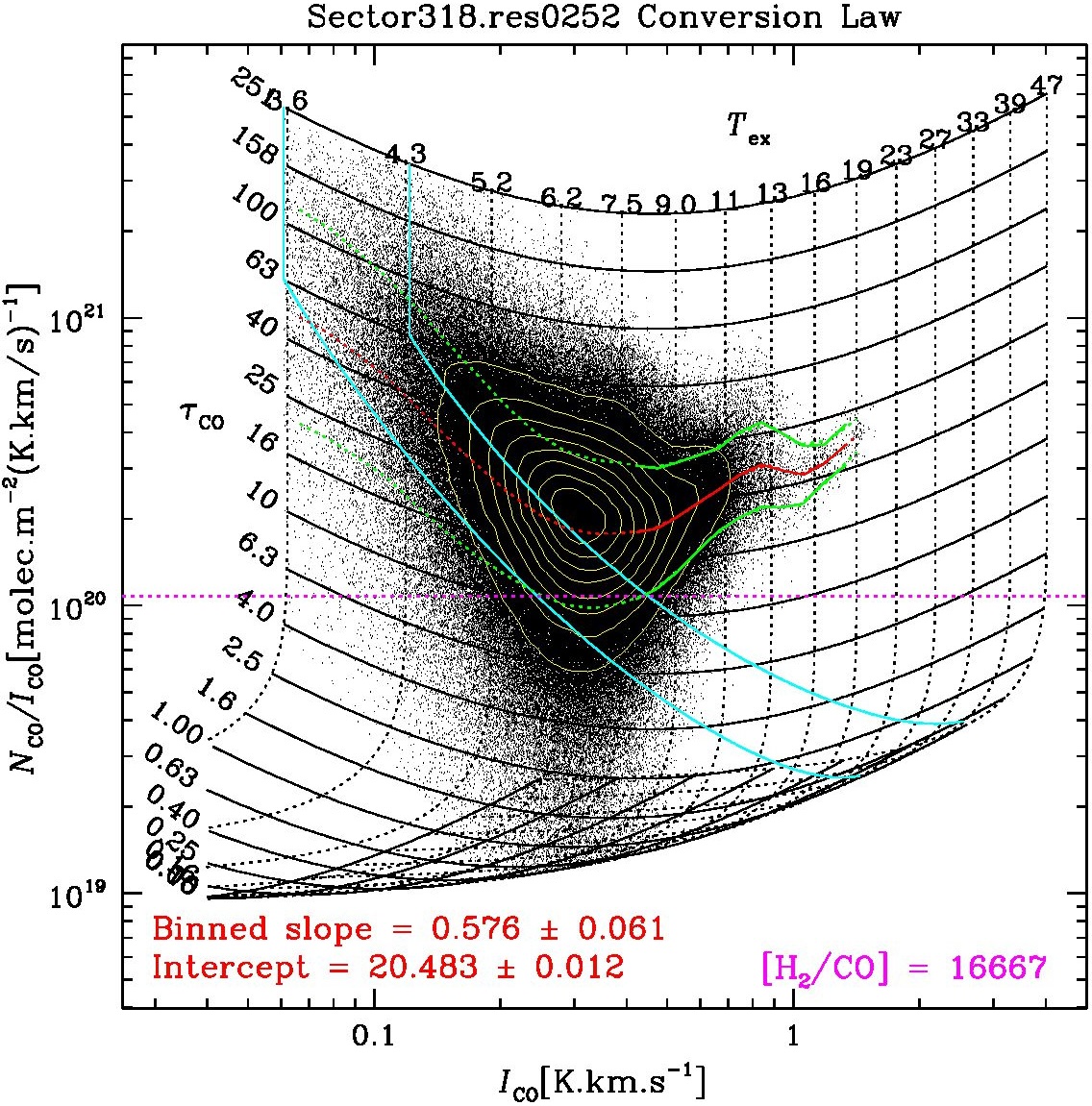} \includegraphics[angle=0,scale=0.12]{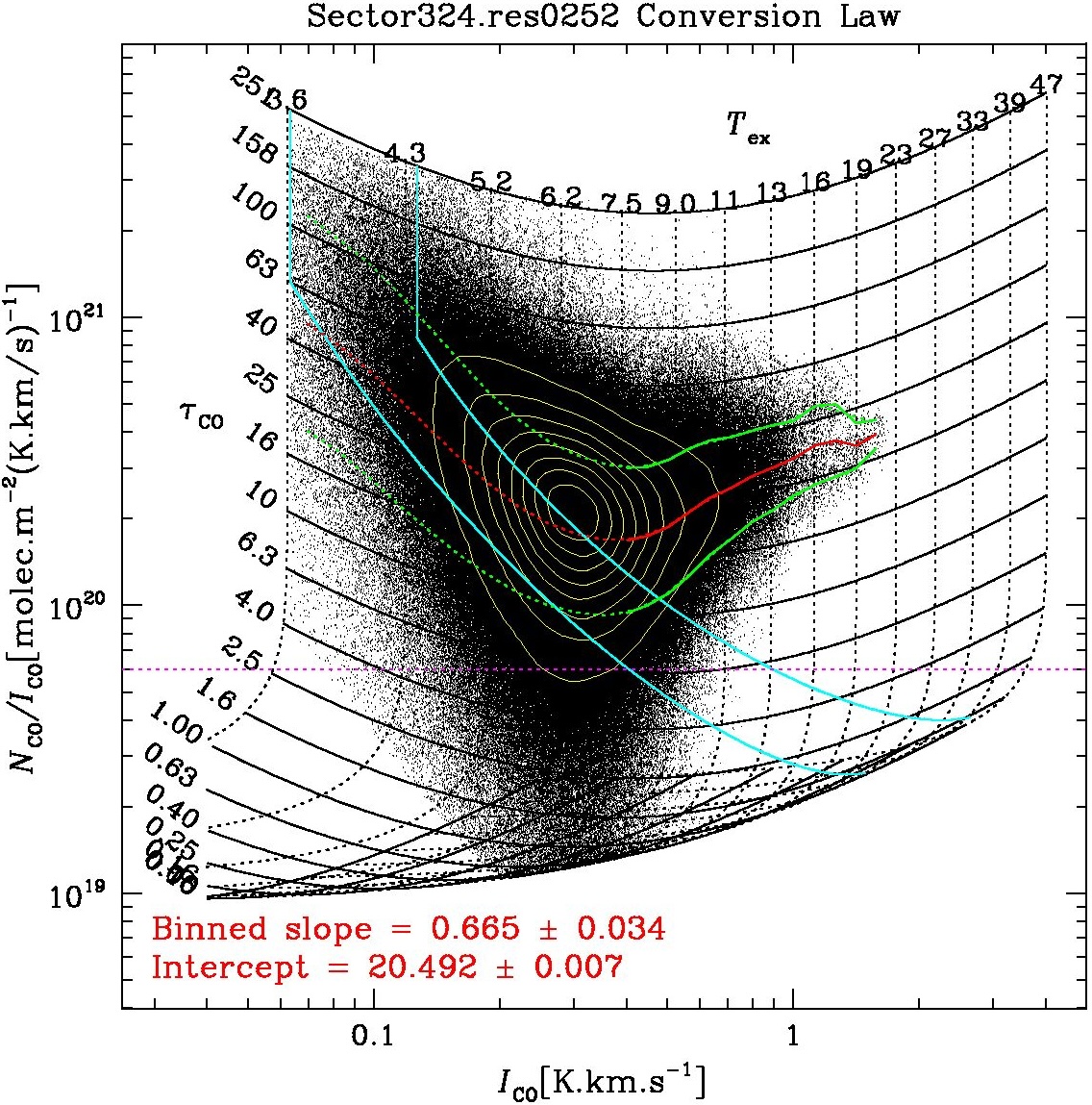} \includegraphics[angle=0,scale=0.12]{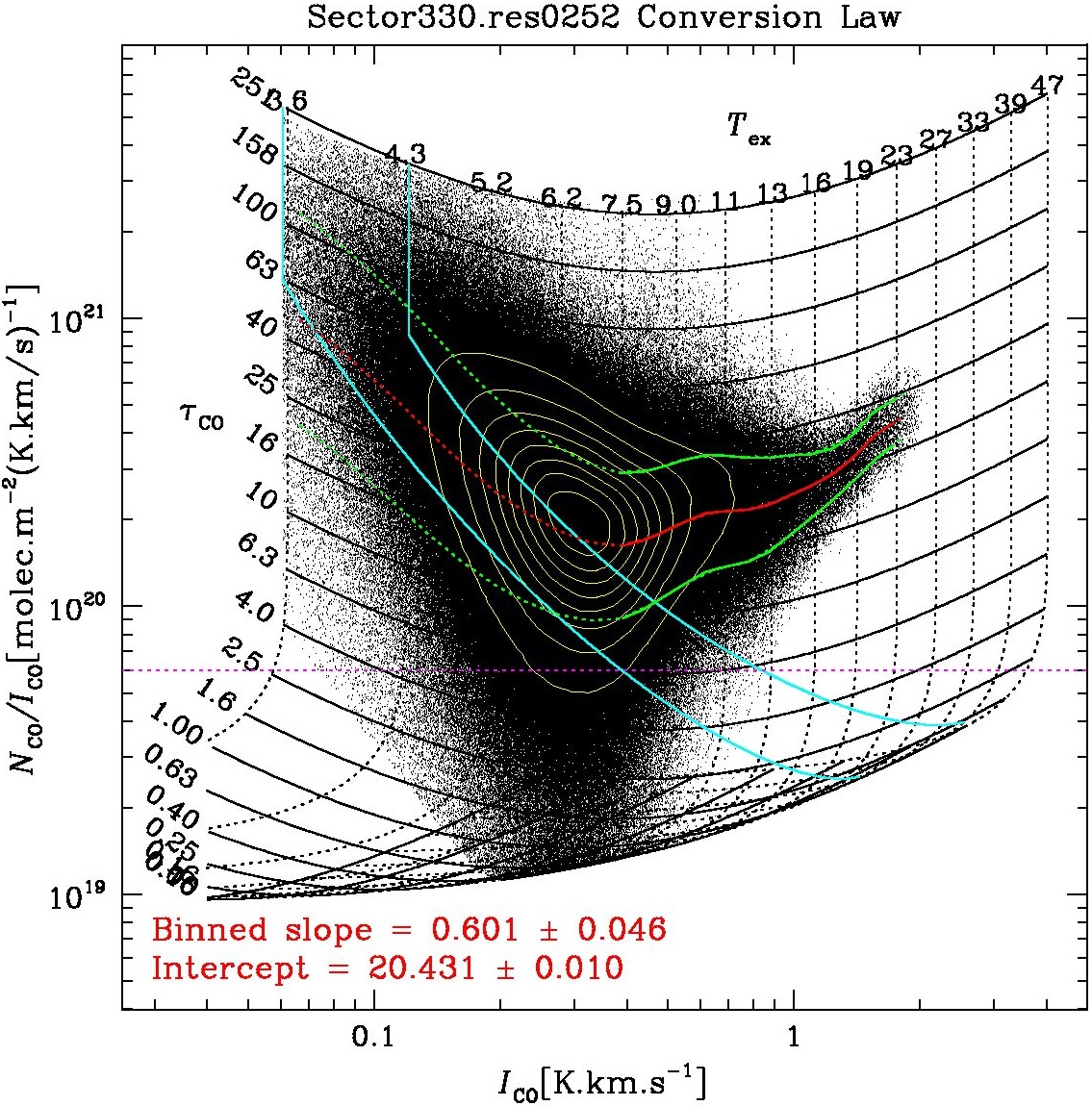}}
\vspace{-3mm}
\centerline{\includegraphics[angle=0,scale=0.12]{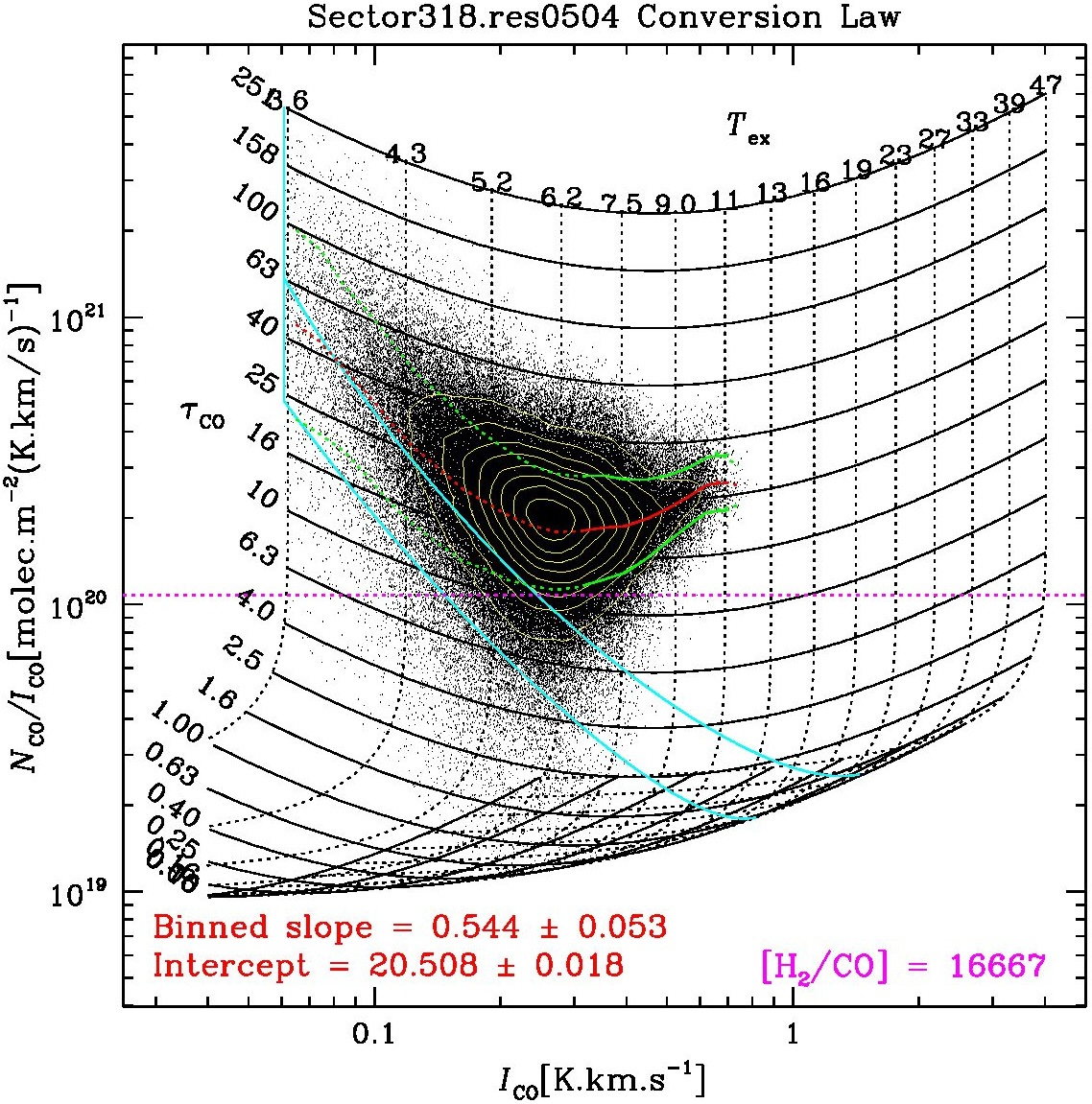} \includegraphics[angle=0,scale=0.12]{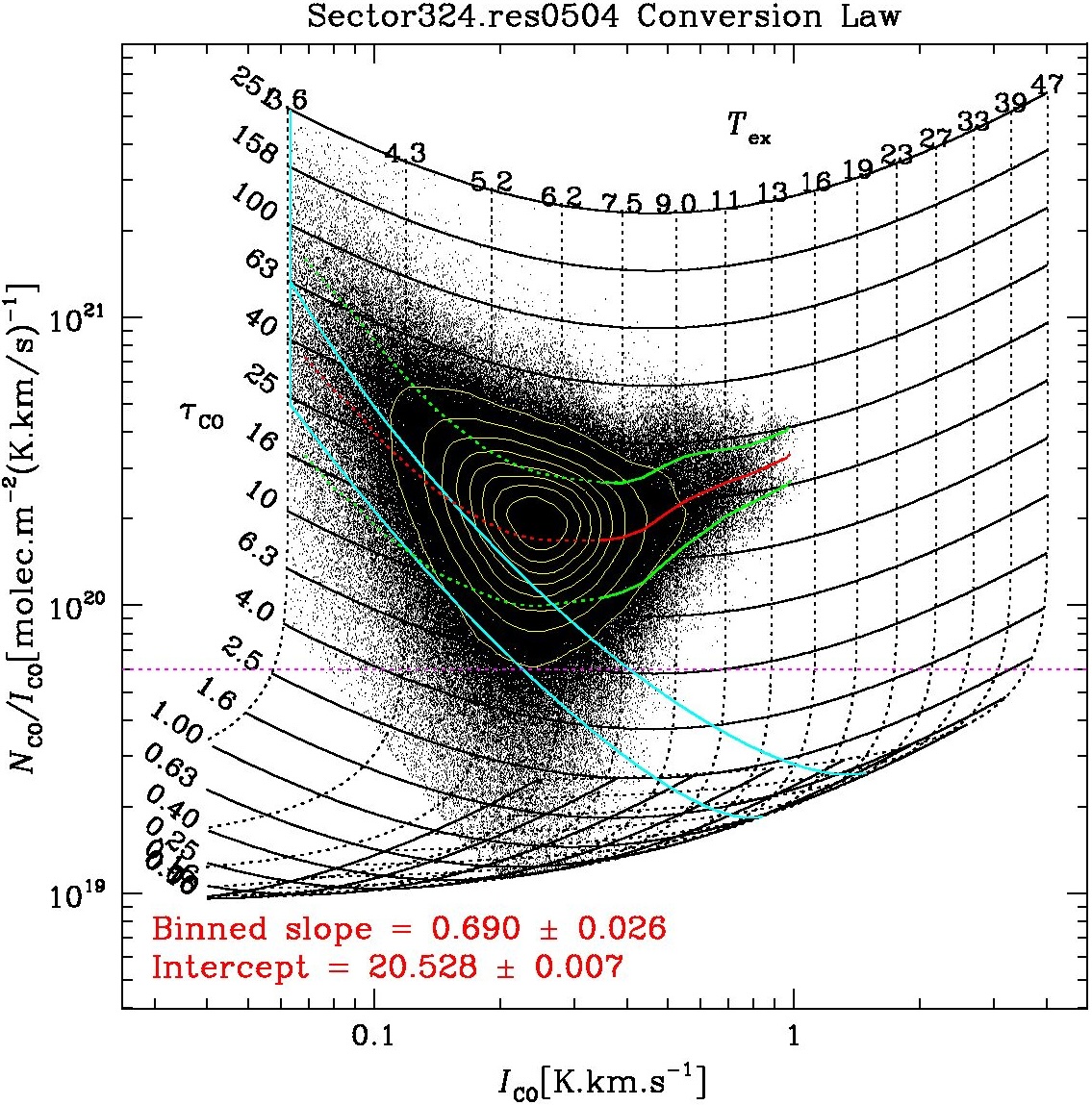} \includegraphics[angle=0,scale=0.12]{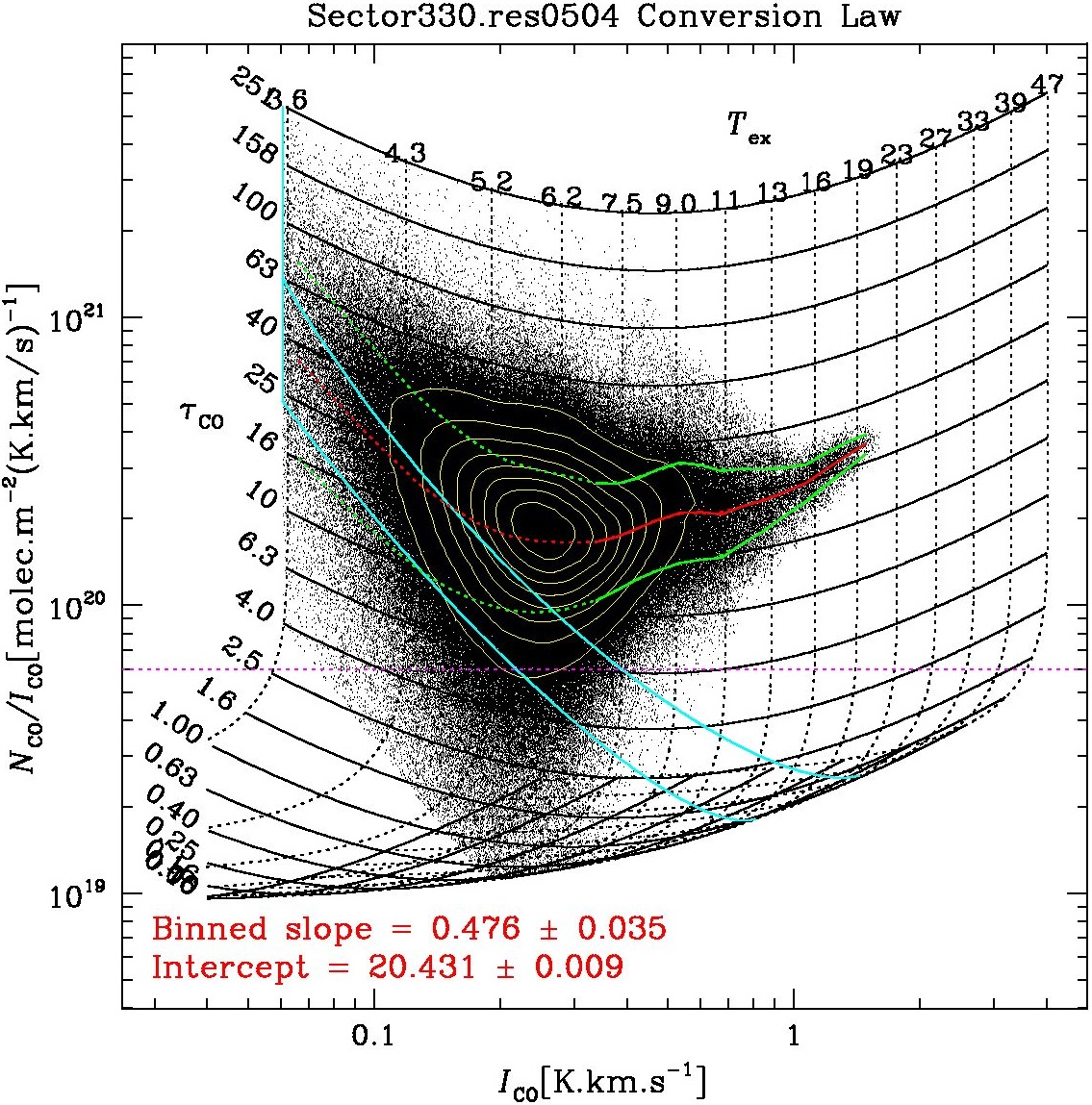}}
\vspace{-3mm}
\centerline{\includegraphics[angle=0,scale=0.12]{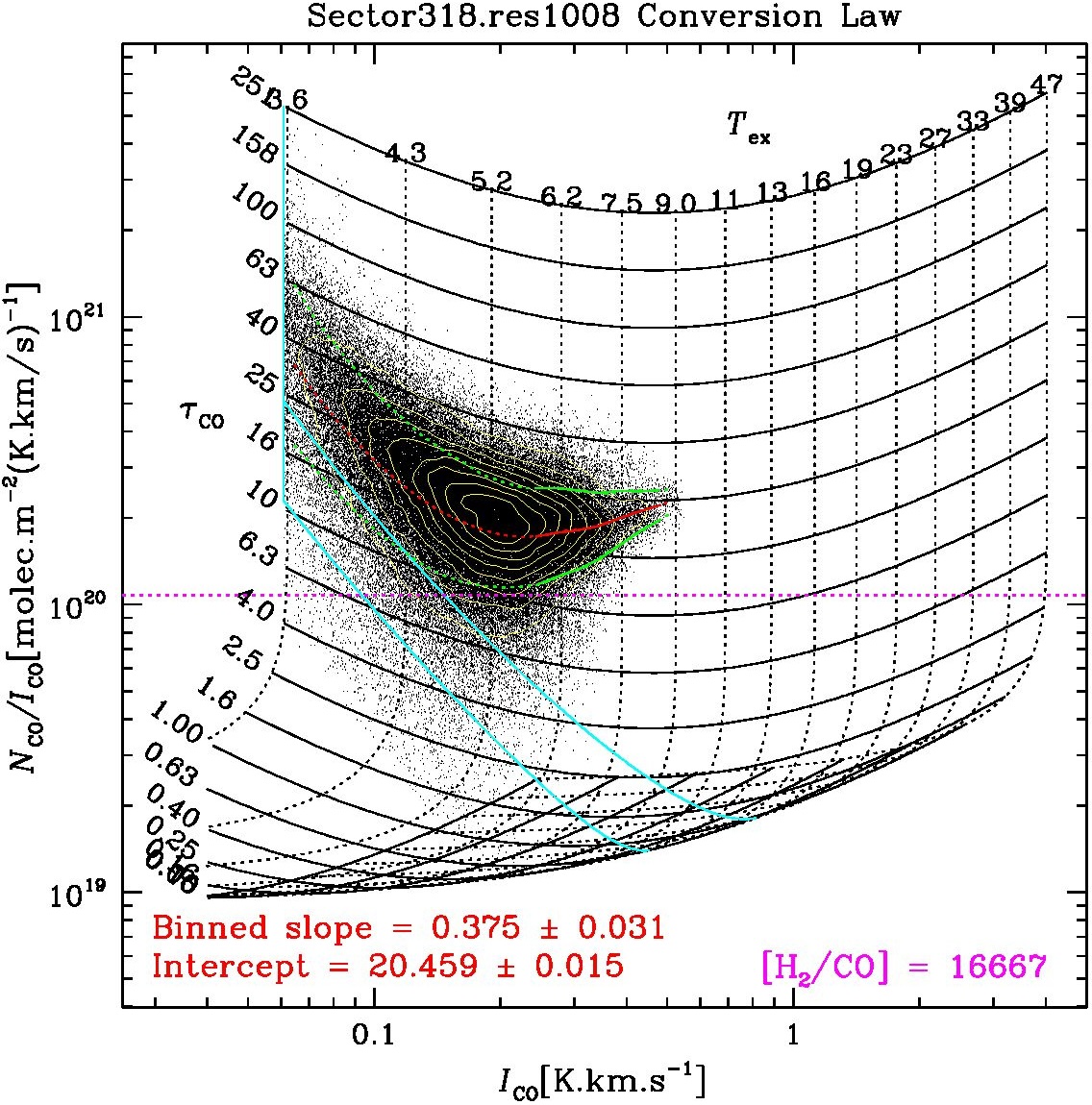} \includegraphics[angle=0,scale=0.12]{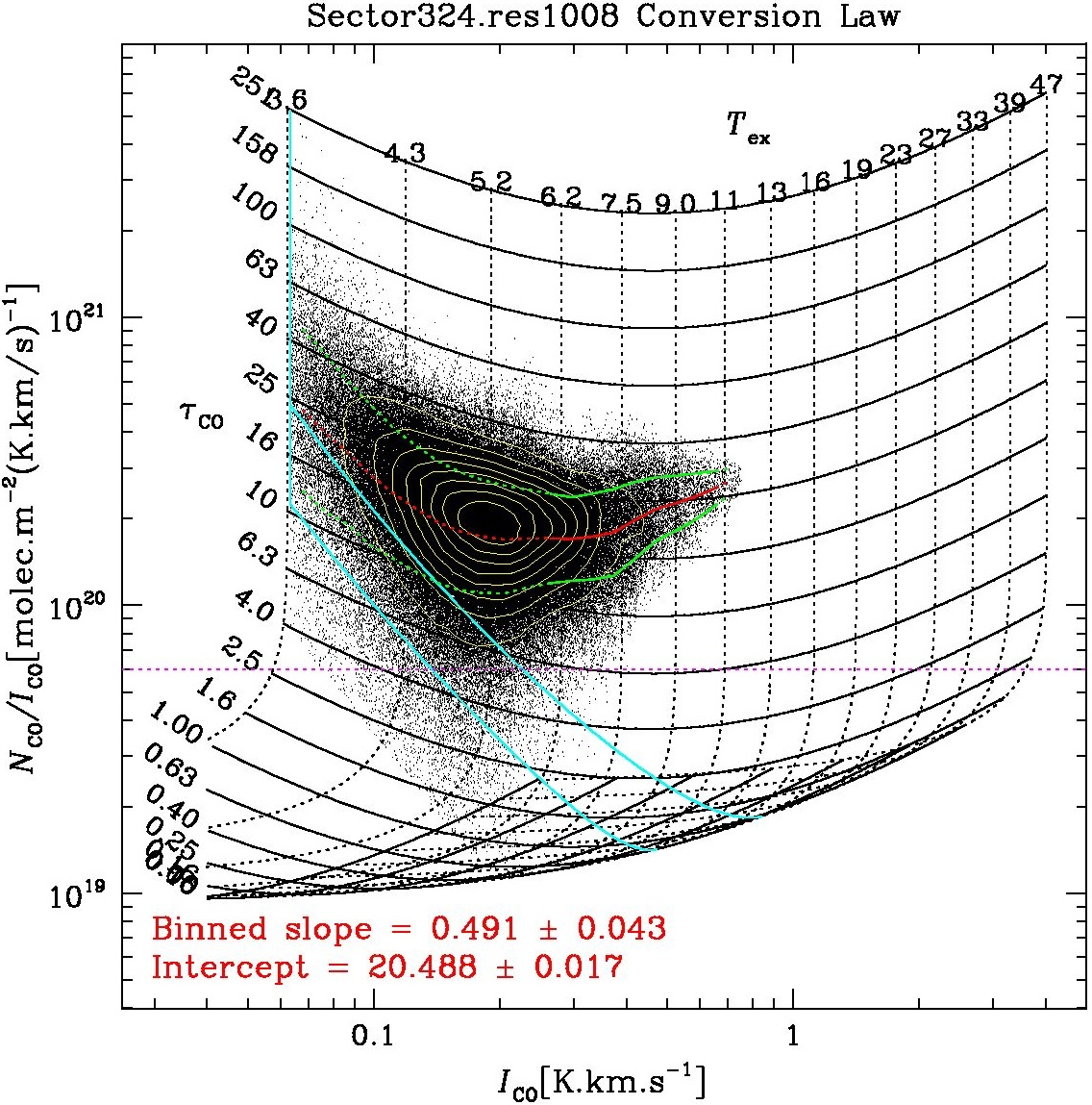} \includegraphics[angle=0,scale=0.12]{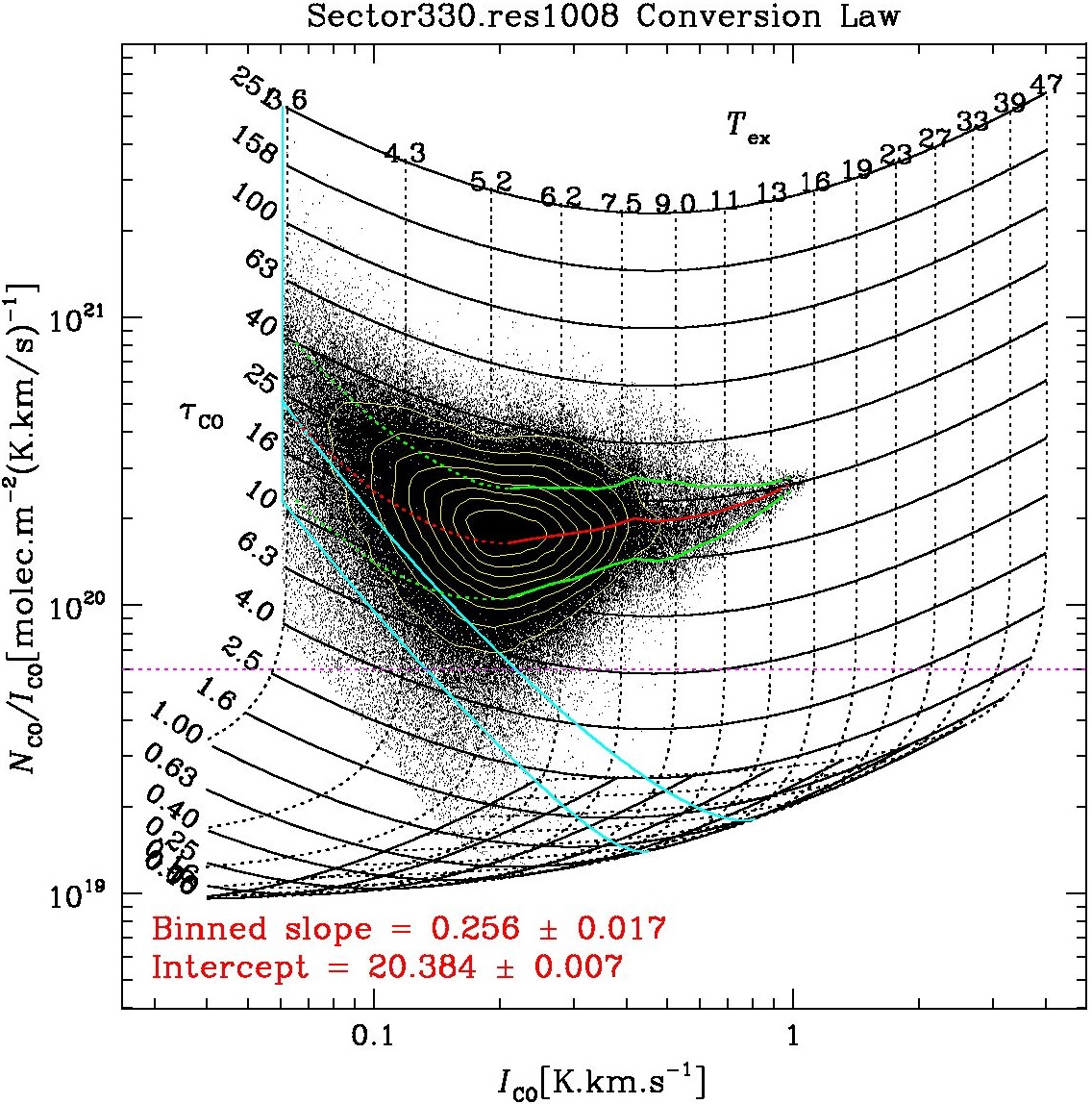}}
\vspace{-1mm}
\caption{\footnotesize Similar plots to Fig.\,\ref{x300-12-multi}, but for Sectors 318, 324, and 330 (left, middle, right  columns respectively).  For S324--S354, the gas-phase [\htwo]/[\tco] abundance level was drawn at 3$\times$10$^4$. $$ $$
\label{x318-30-multi}}
\vspace{0mm}
\end{figure*}

\begin{figure*}[h]
\vspace{0mm}
\centerline{\includegraphics[angle=0,scale=0.12]{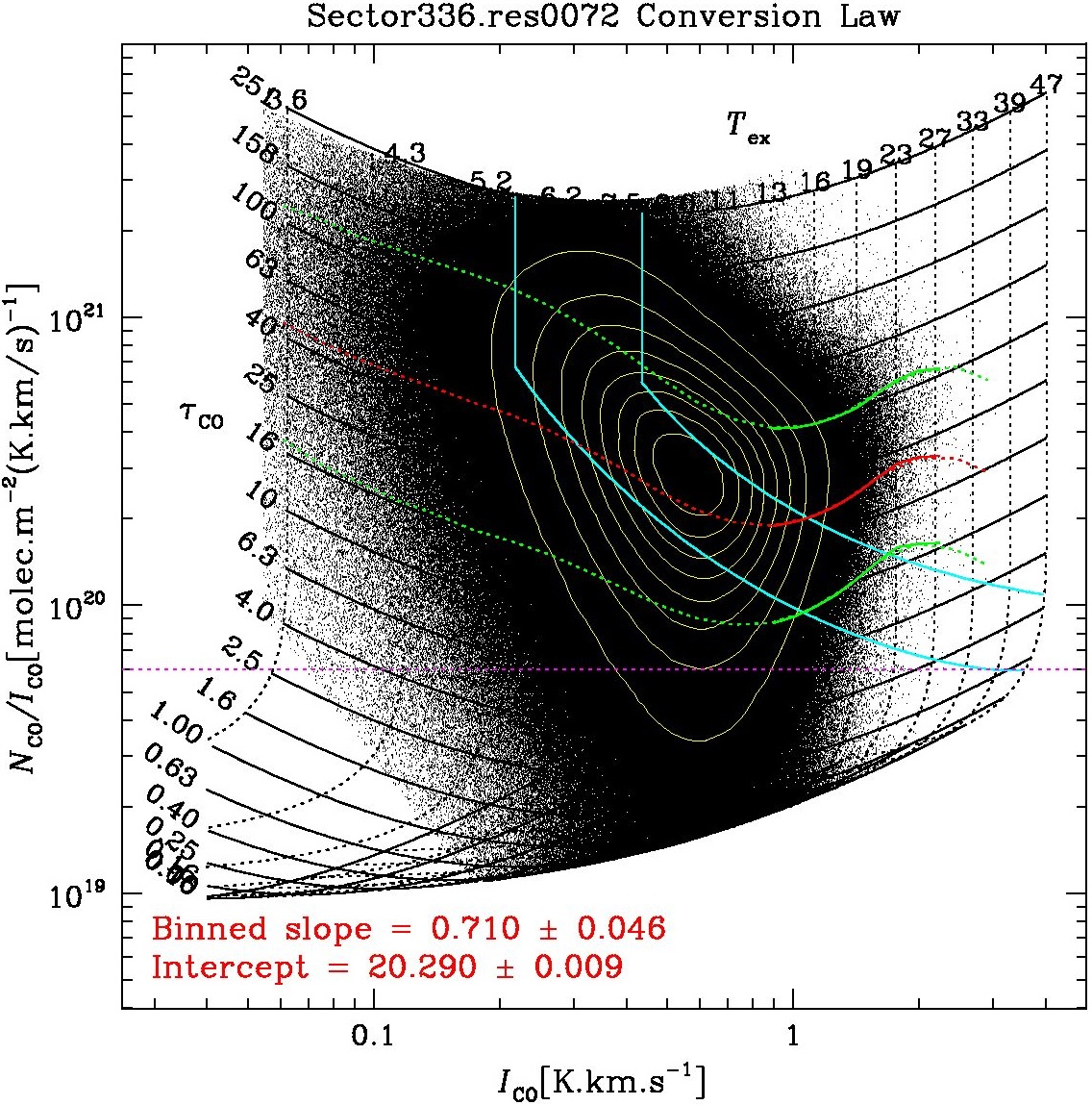} \includegraphics[angle=0,scale=0.12]{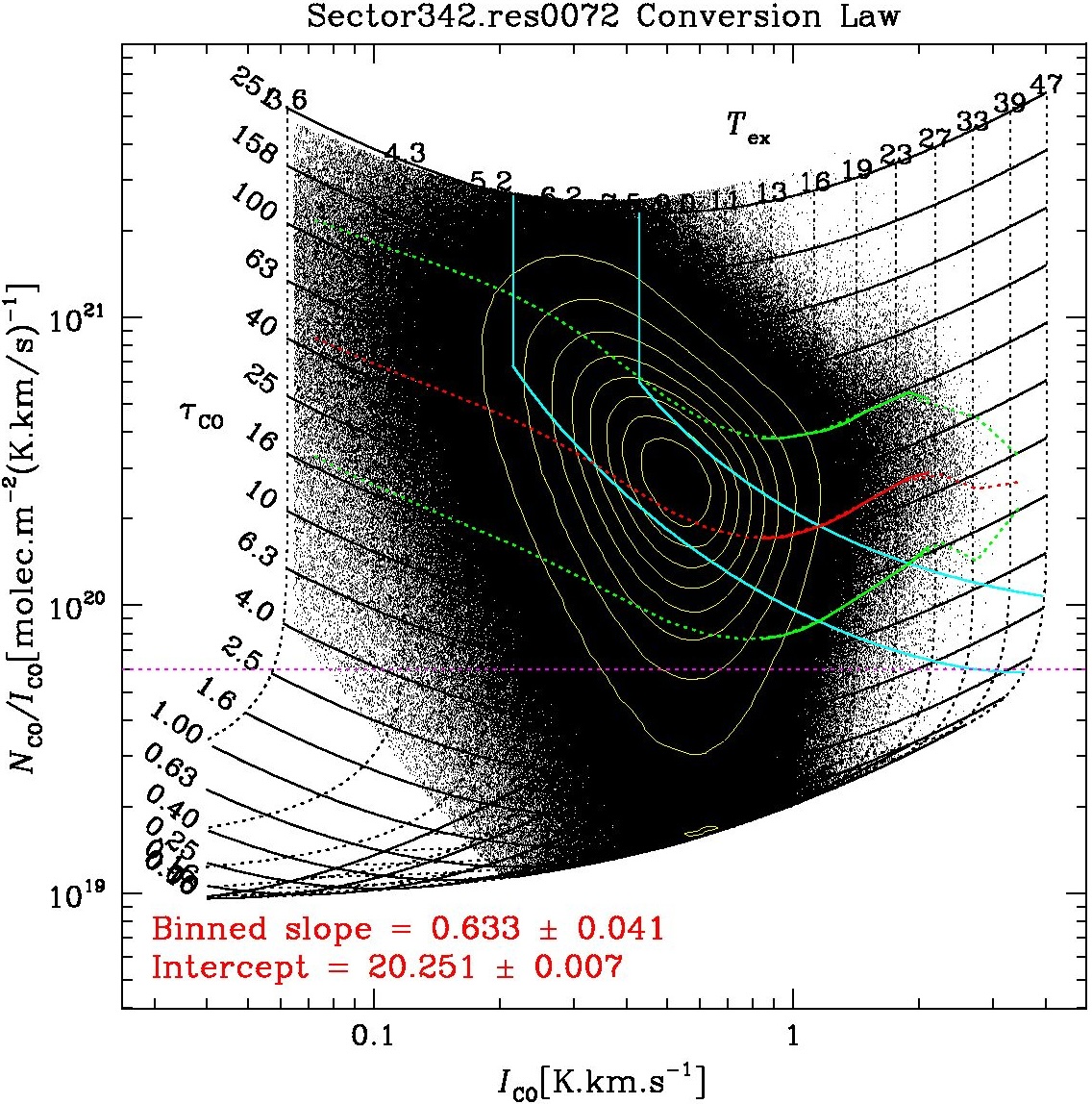} \includegraphics[angle=0,scale=0.12]{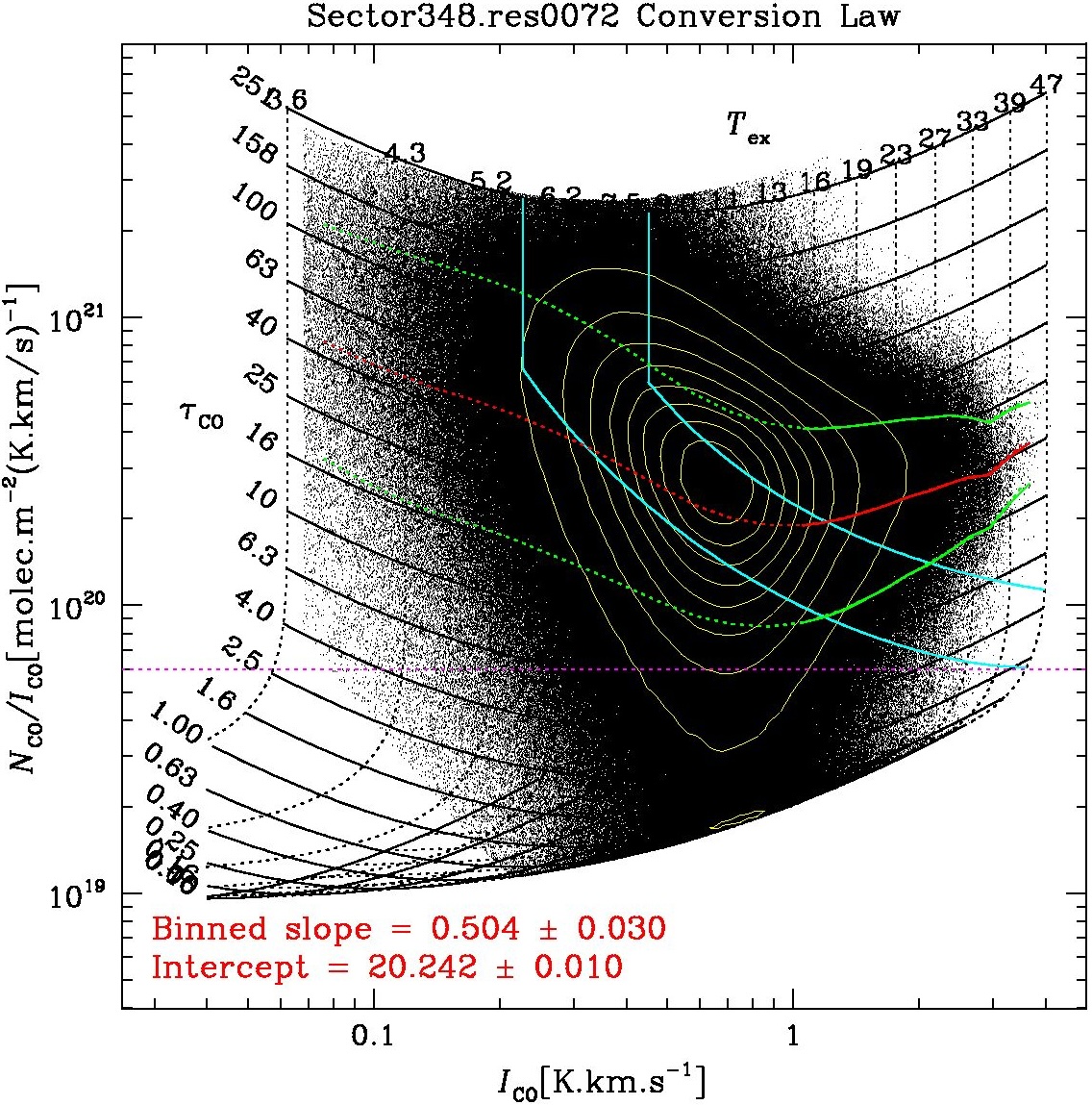}}
\vspace{-3mm}
\centerline{\includegraphics[angle=0,scale=0.12]{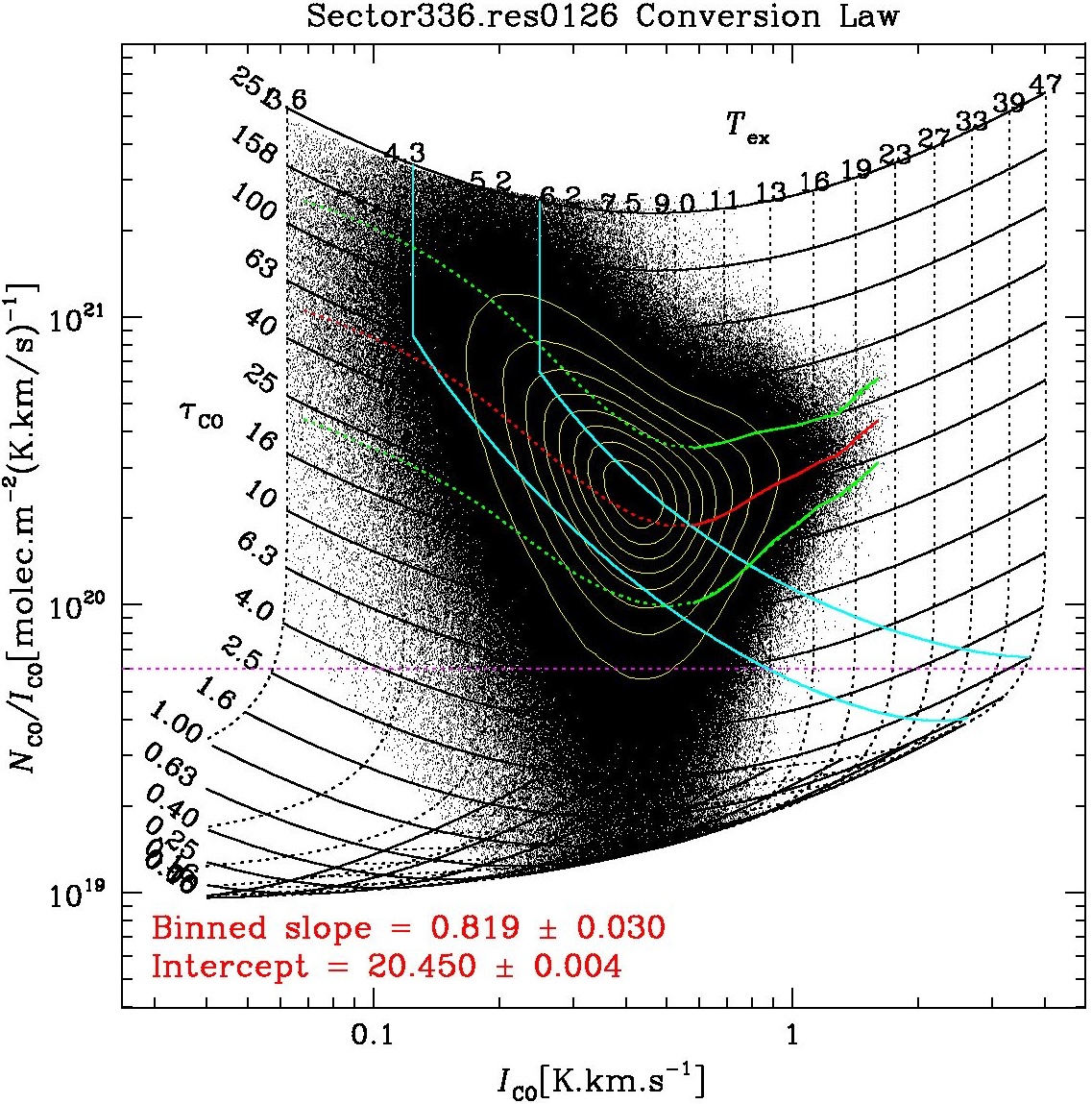} \includegraphics[angle=0,scale=0.12]{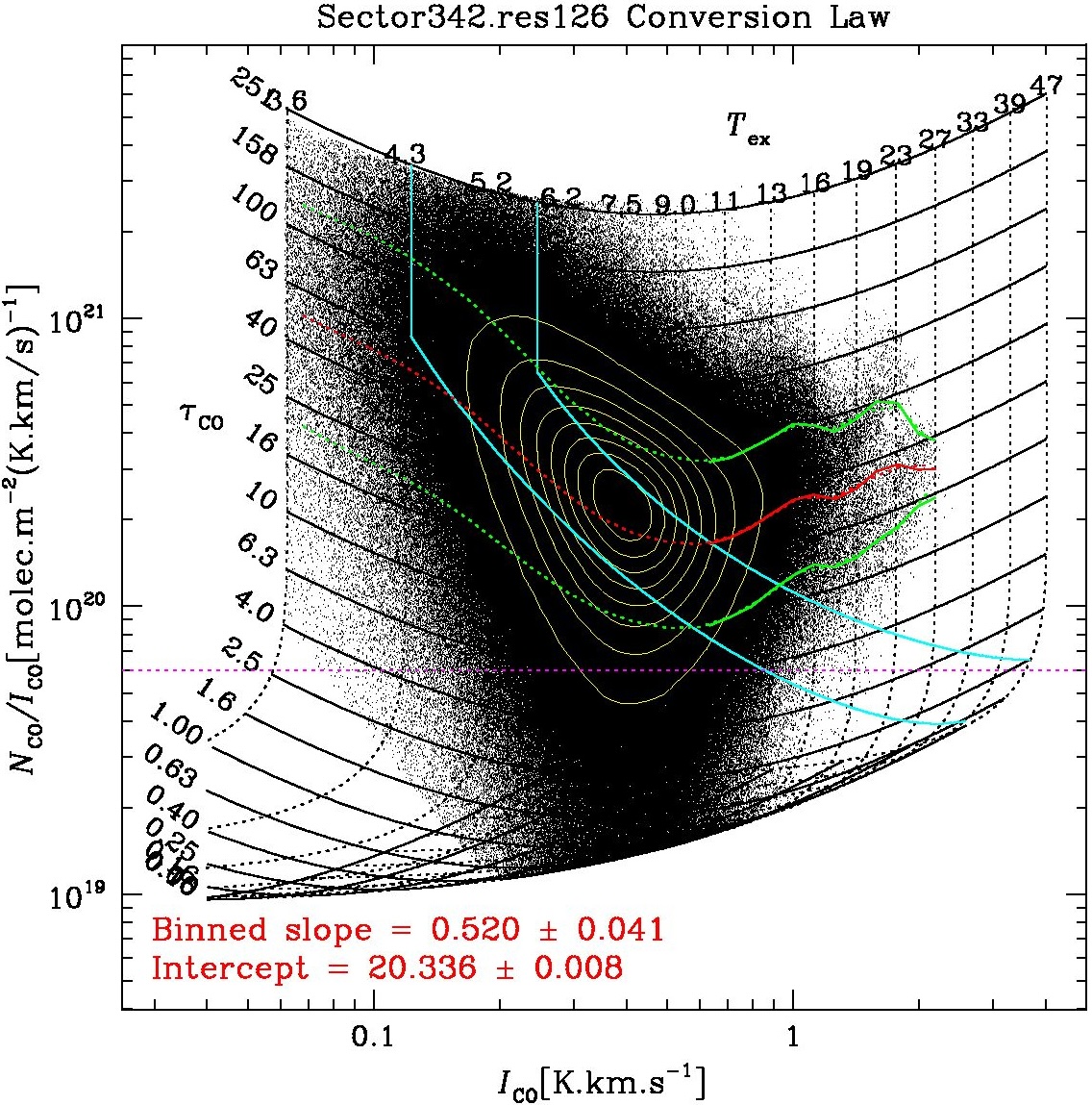} \includegraphics[angle=0,scale=0.12]{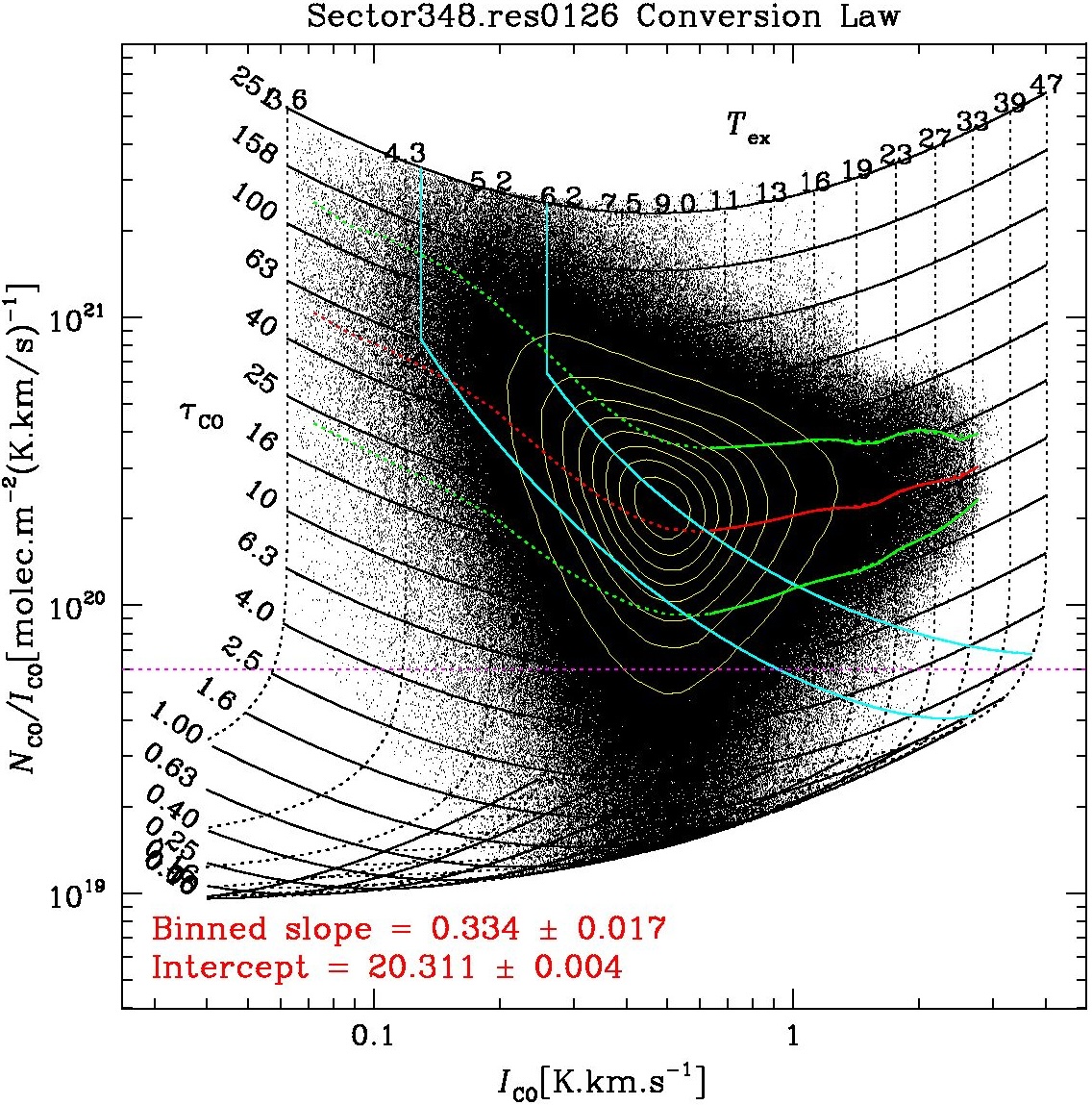}}
\vspace{-3mm}
\centerline{\includegraphics[angle=0,scale=0.12]{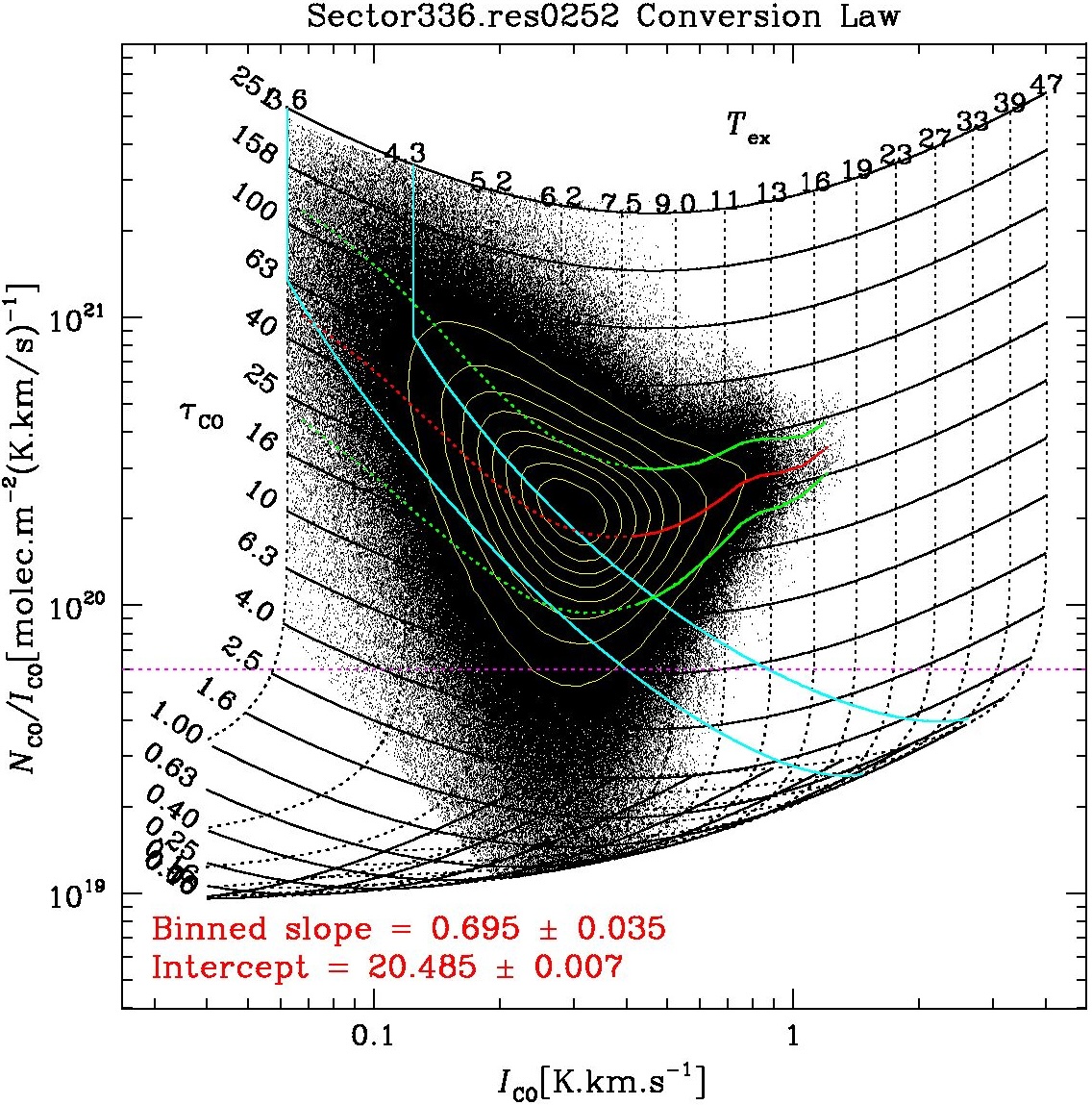} \includegraphics[angle=0,scale=0.12]{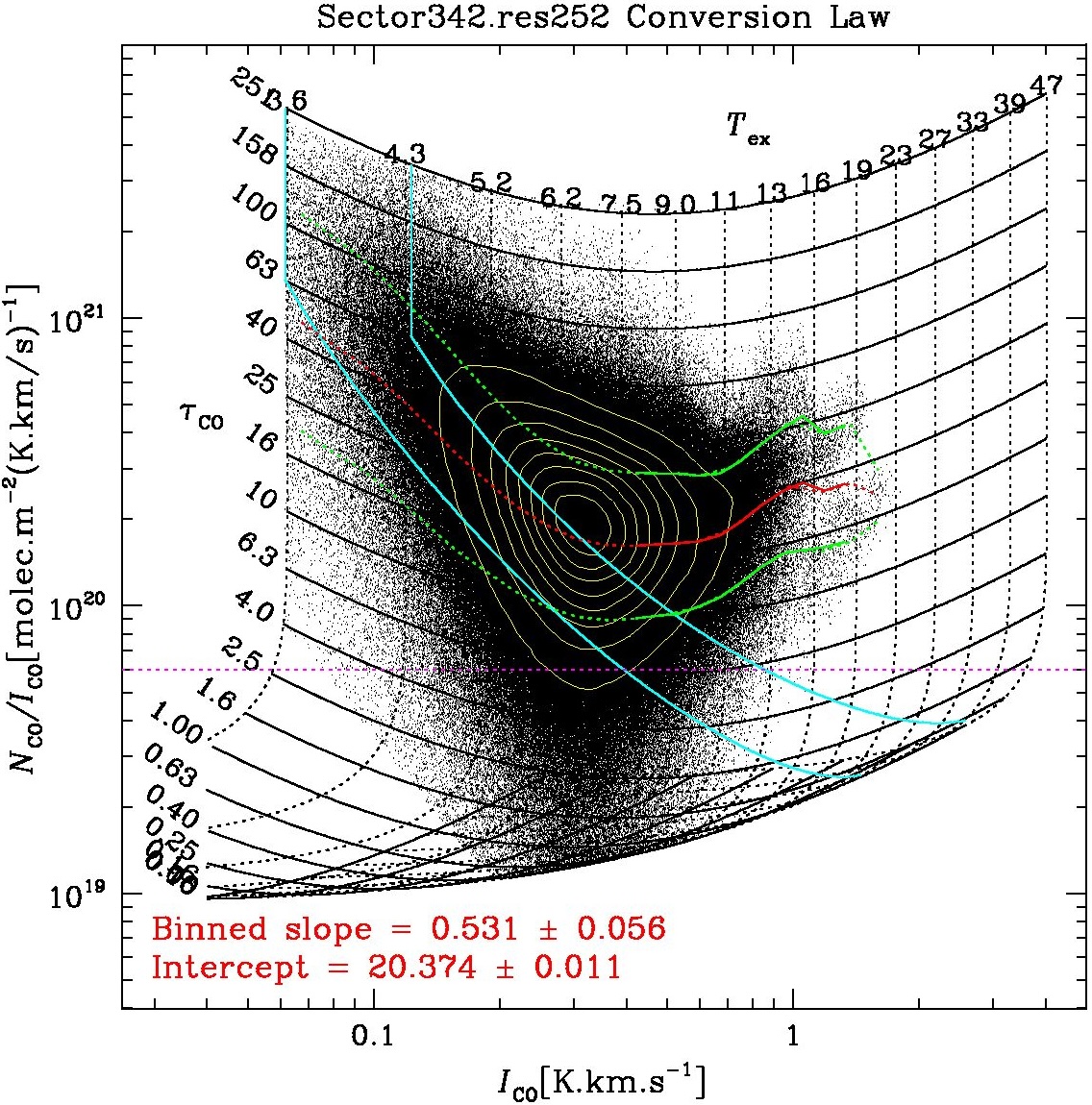} \includegraphics[angle=0,scale=0.12]{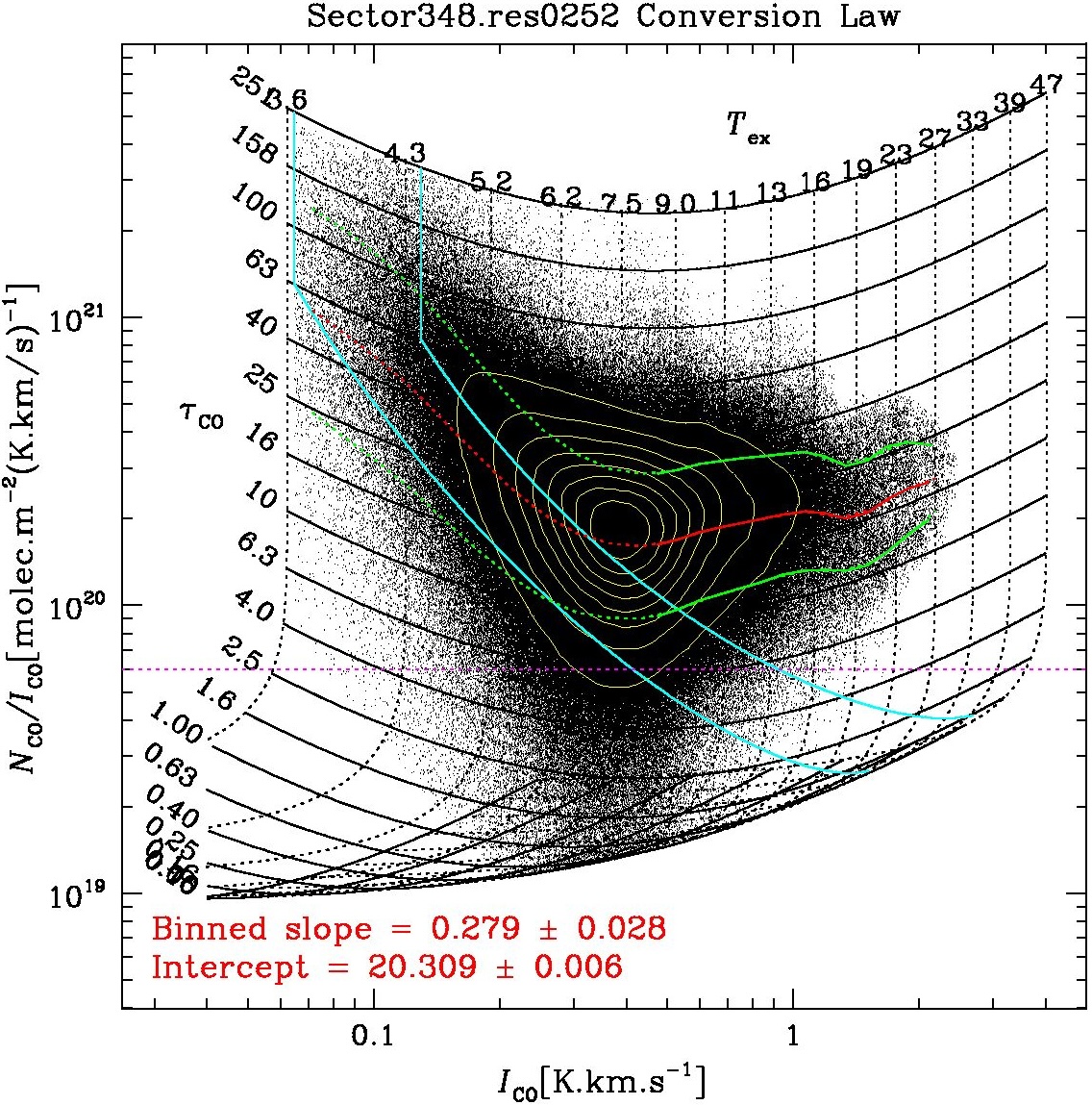}}
\vspace{-3mm}
\centerline{\includegraphics[angle=0,scale=0.12]{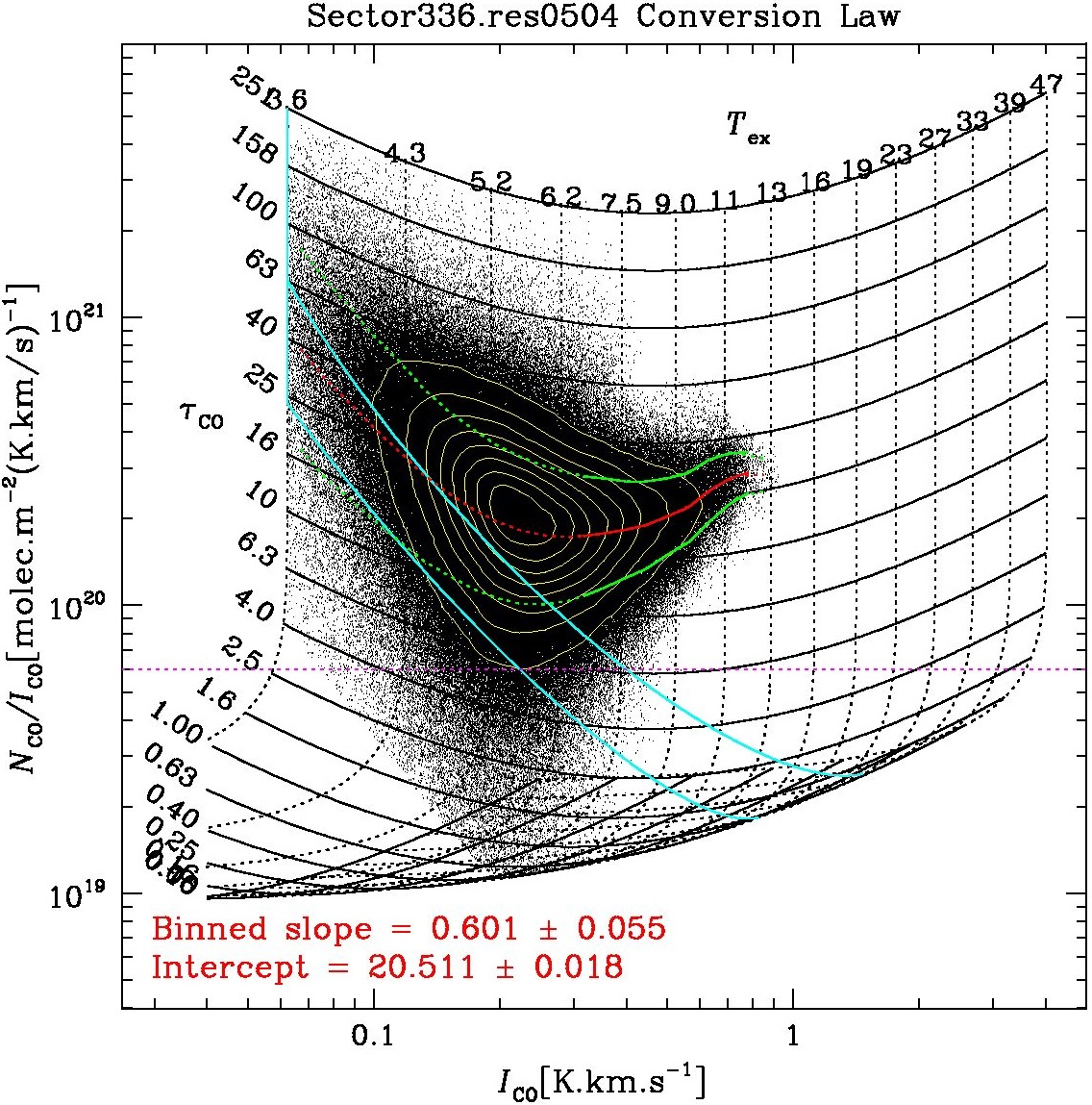} \includegraphics[angle=0,scale=0.12]{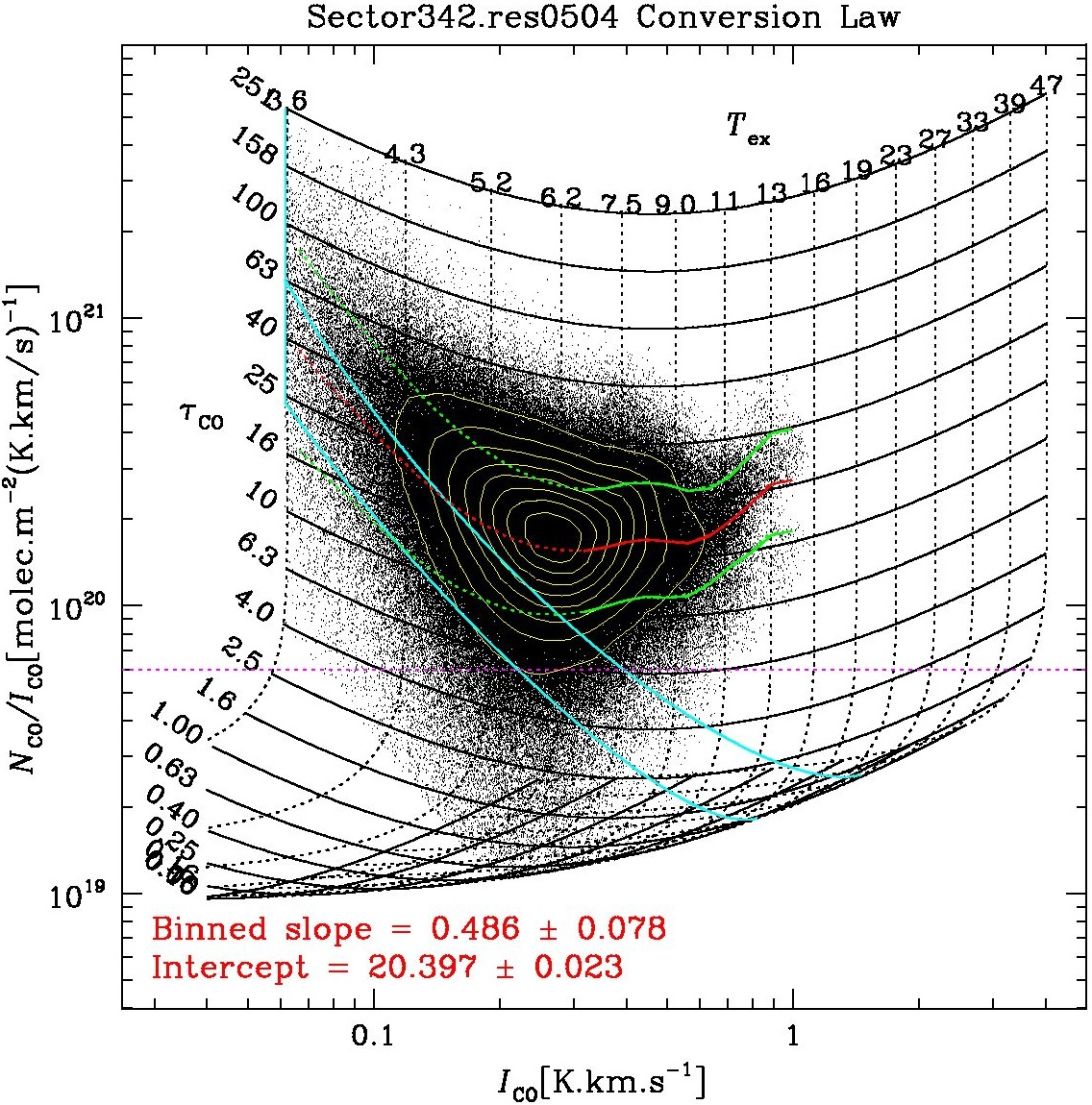} \includegraphics[angle=0,scale=0.12]{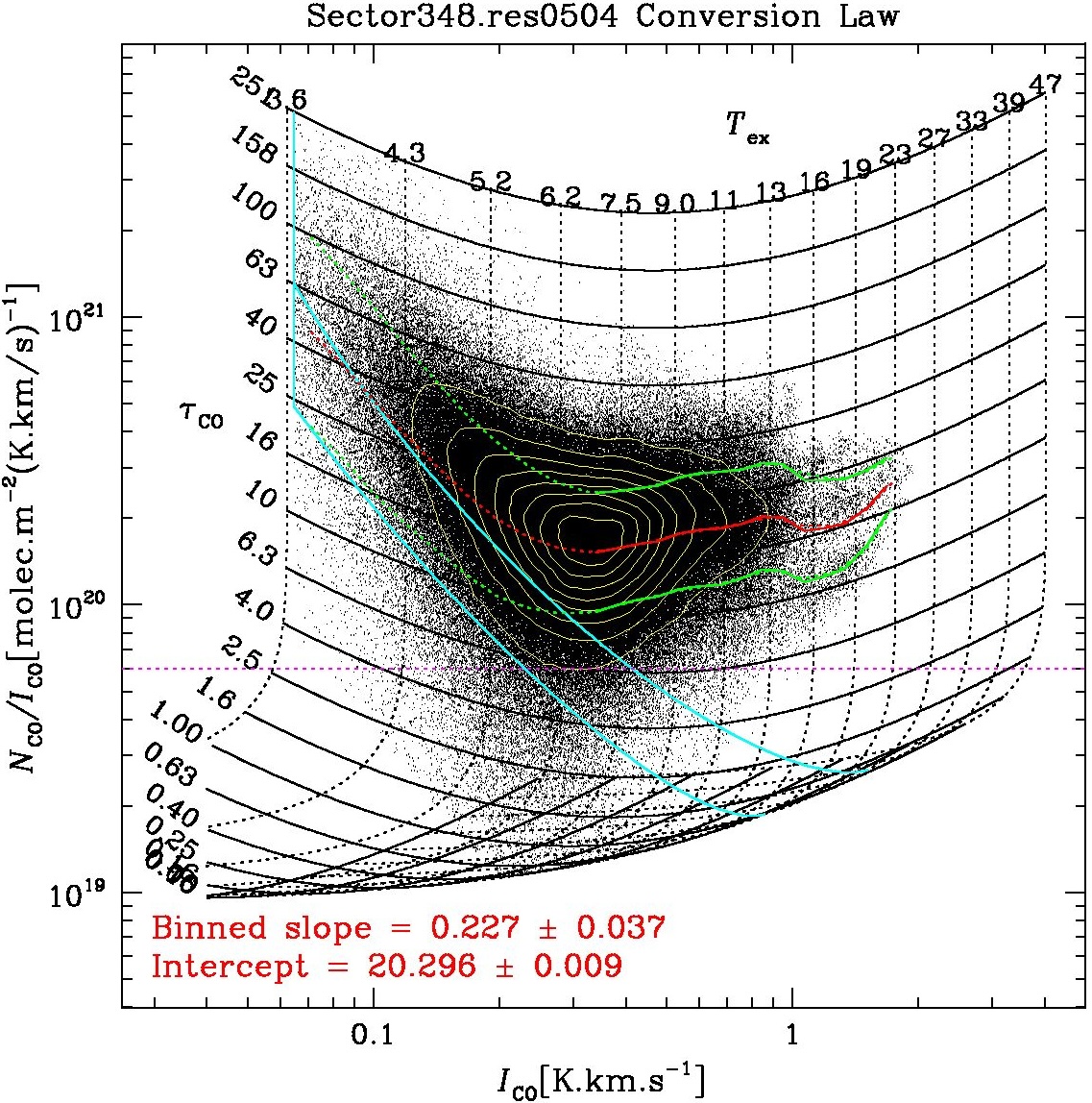}}
\vspace{-3mm}
\centerline{\includegraphics[angle=0,scale=0.12]{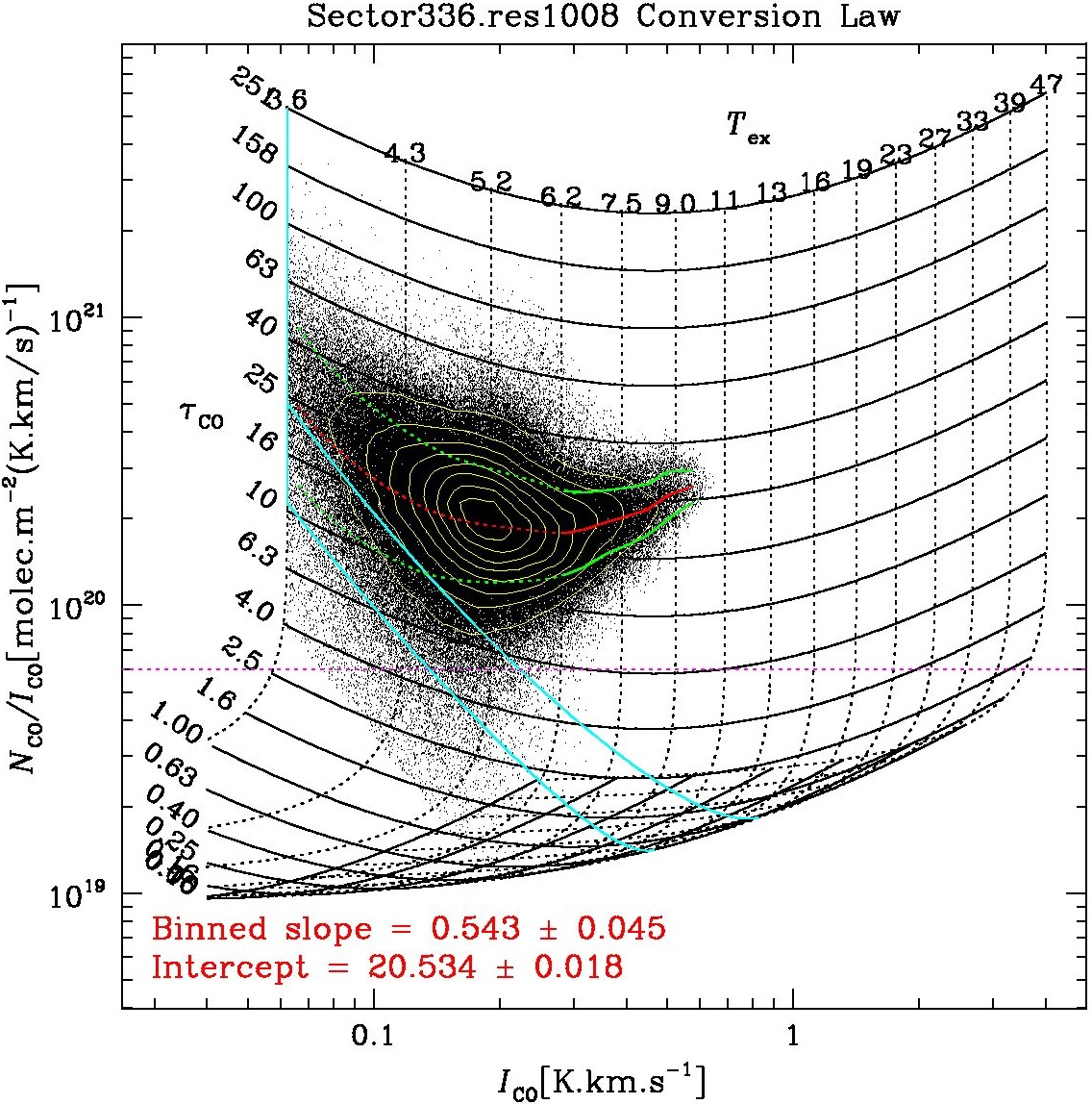} \includegraphics[angle=0,scale=0.12]{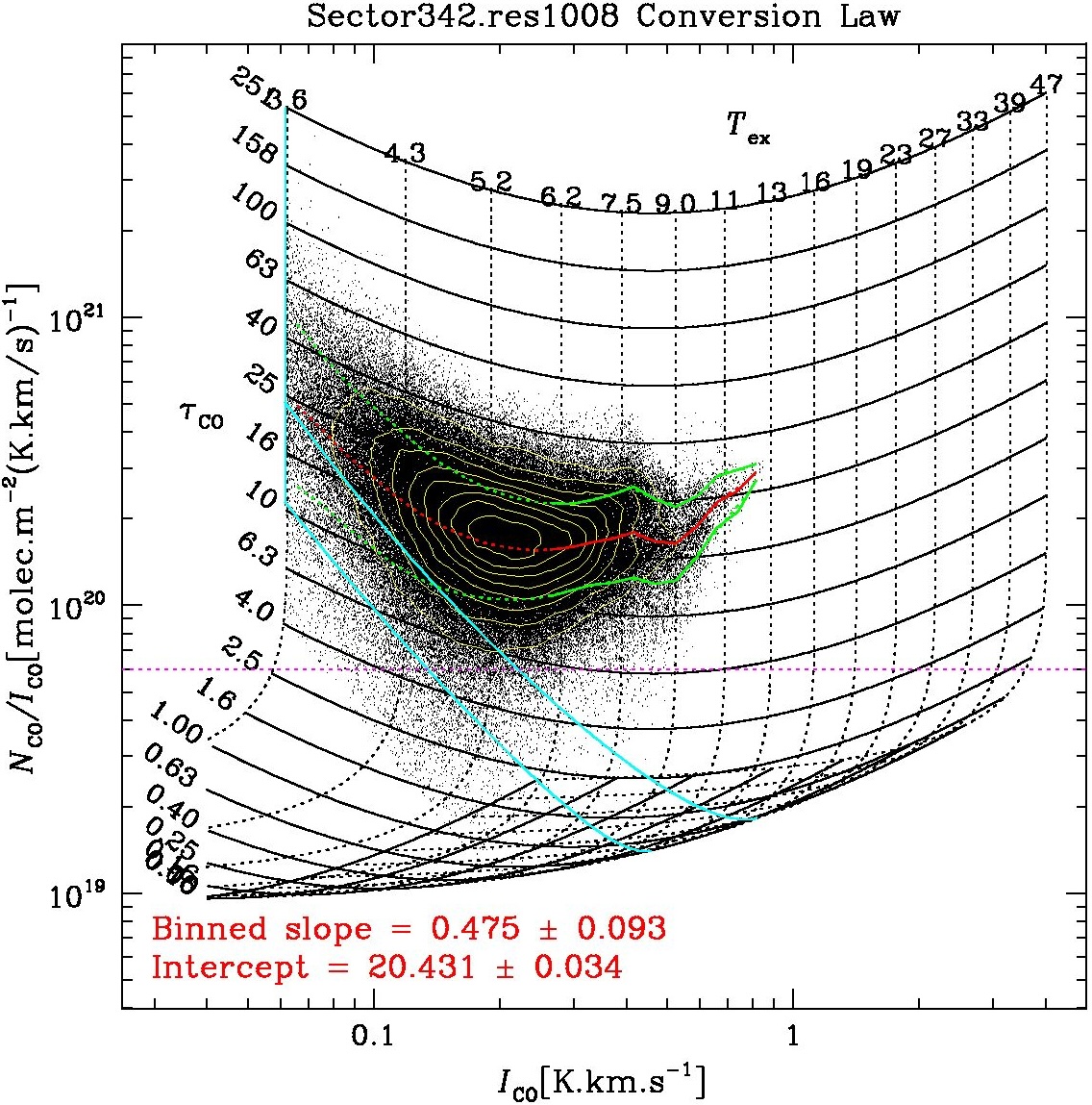} \includegraphics[angle=0,scale=0.12]{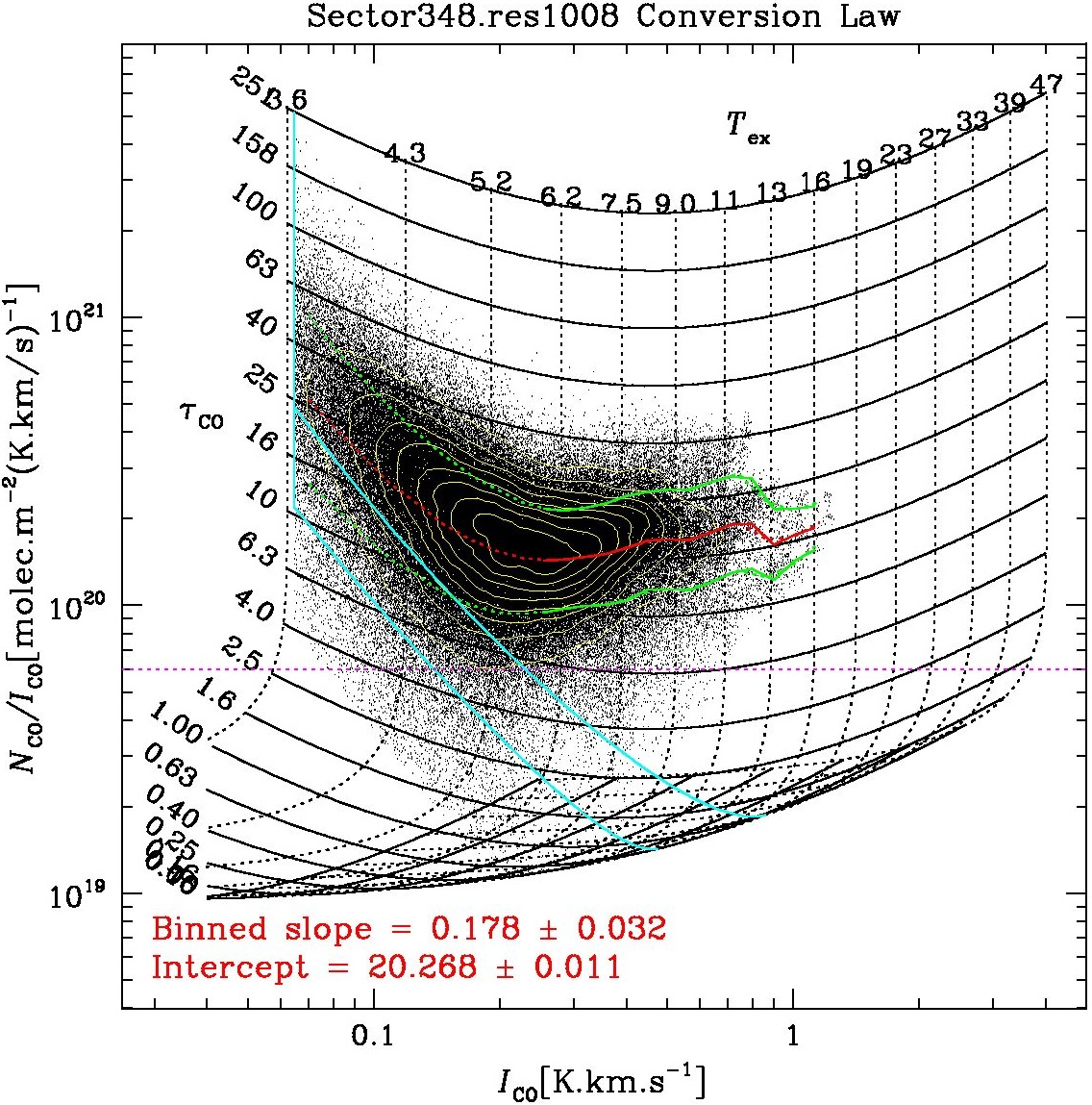}}
\vspace{-1mm}
\caption{\footnotesize Similar plots to Fig.\,\ref{x318-30-multi}, but for Sectors 336, 342, and 348 (left, middle, right  columns respectively). $$ $$
\label{x336-48-multi}}
\vspace{0mm}
\end{figure*}

\begin{figure*}[h]
\vspace{0mm}
\centerline{\includegraphics[angle=0,scale=0.12]{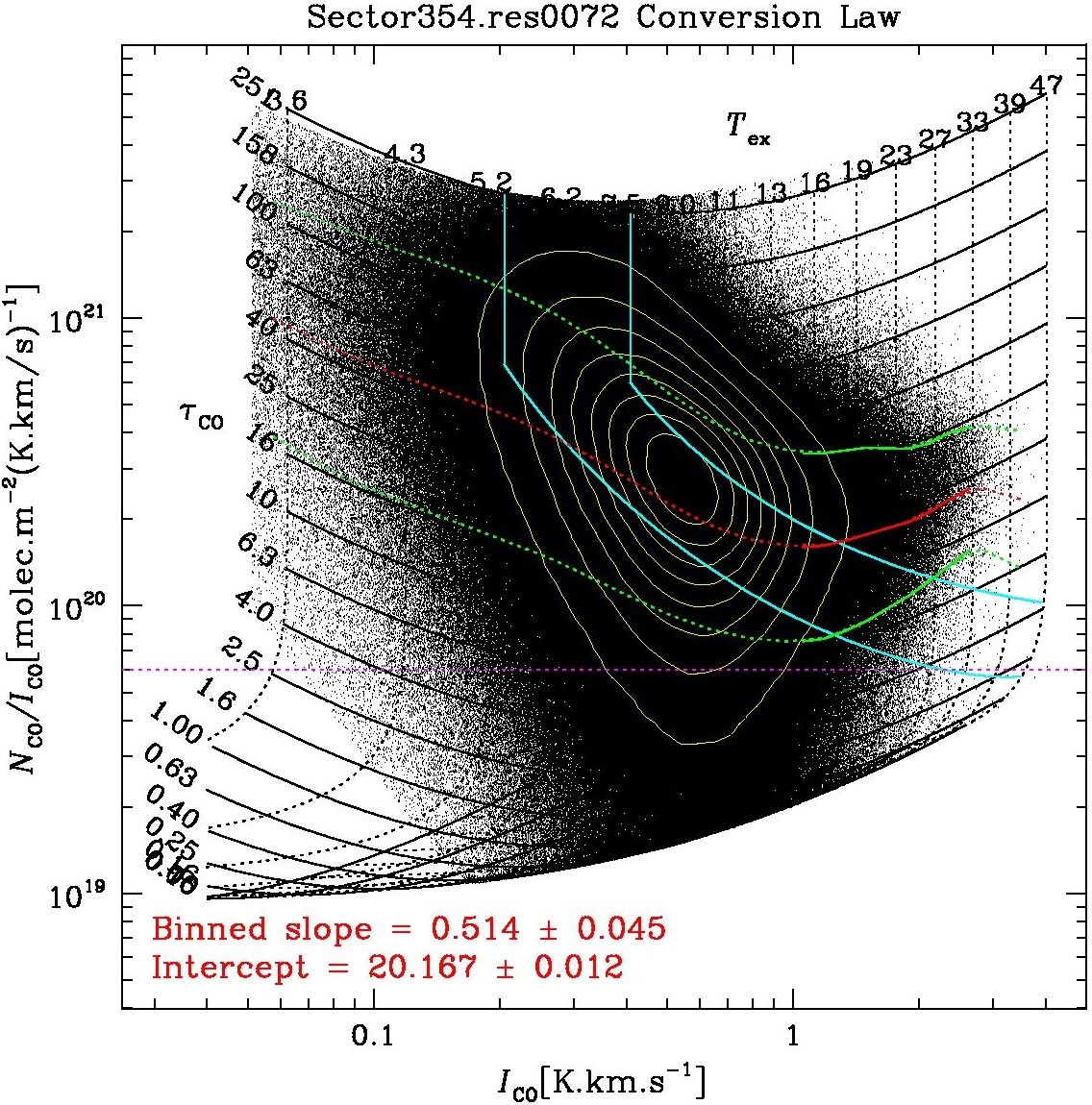} \includegraphics[angle=0,scale=0.12]{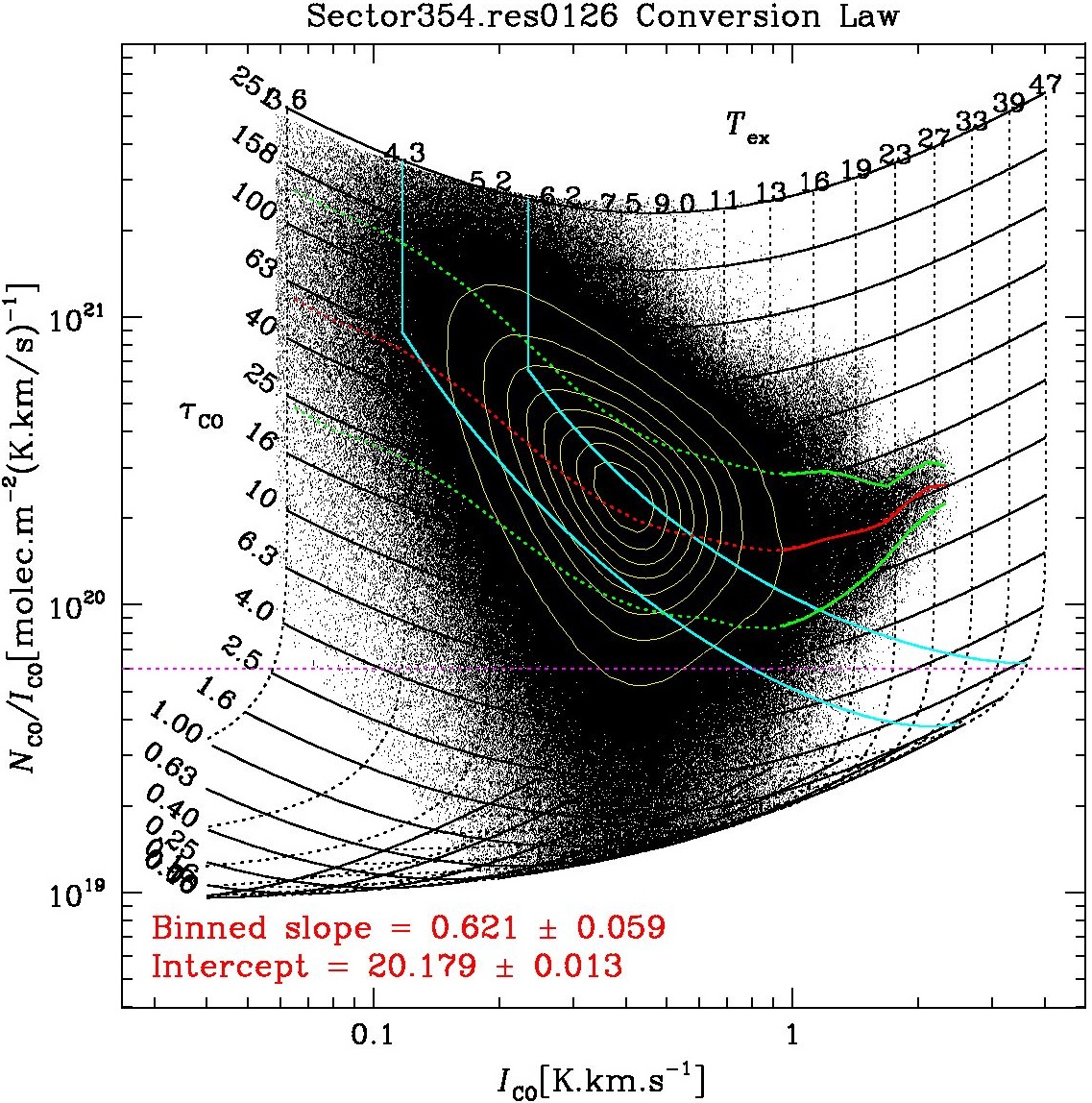} \includegraphics[angle=0,scale=0.12]{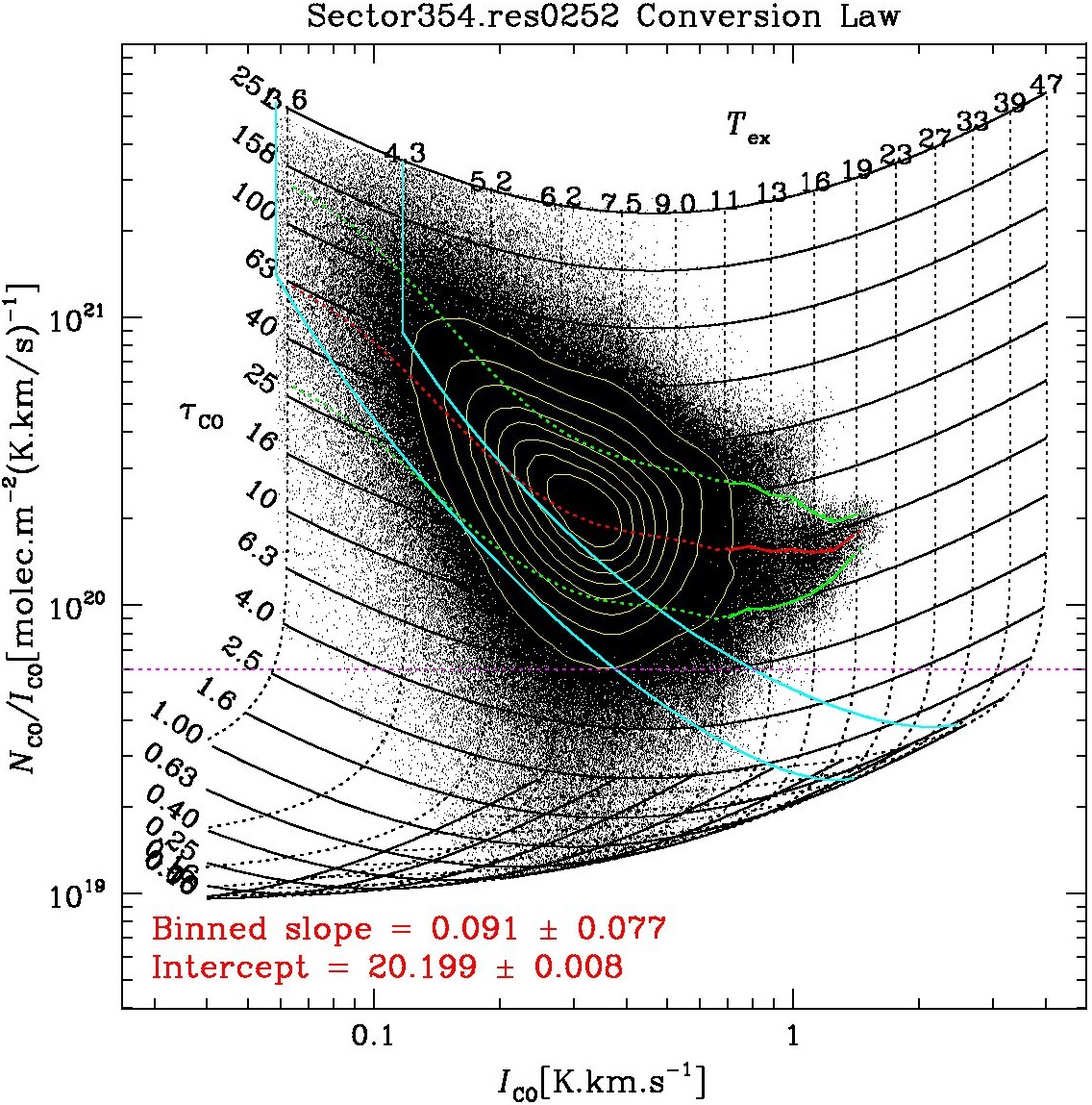}}
\vspace{0mm}
\centerline{\includegraphics[angle=0,scale=0.12]{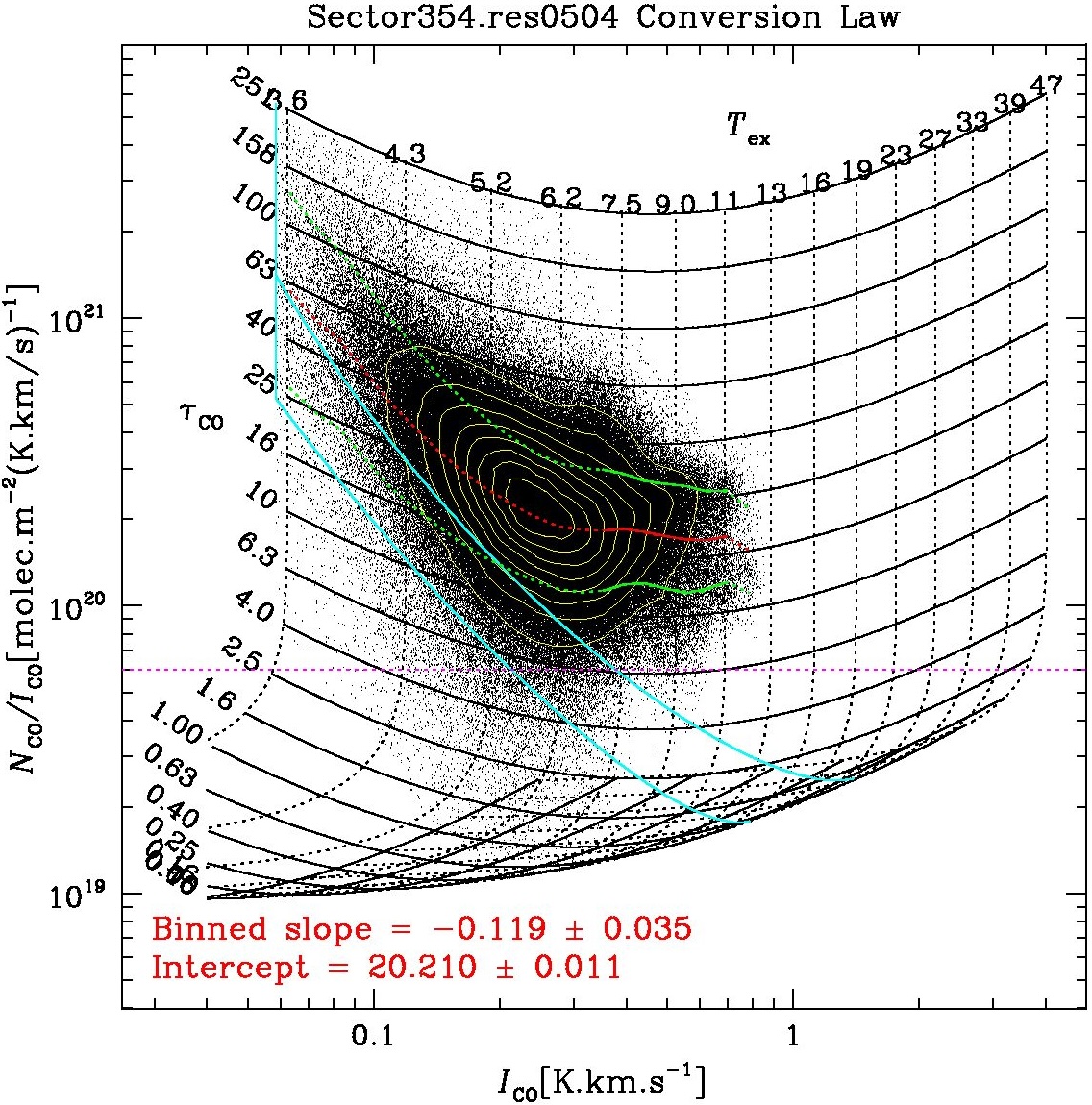} \includegraphics[angle=0,scale=0.12]{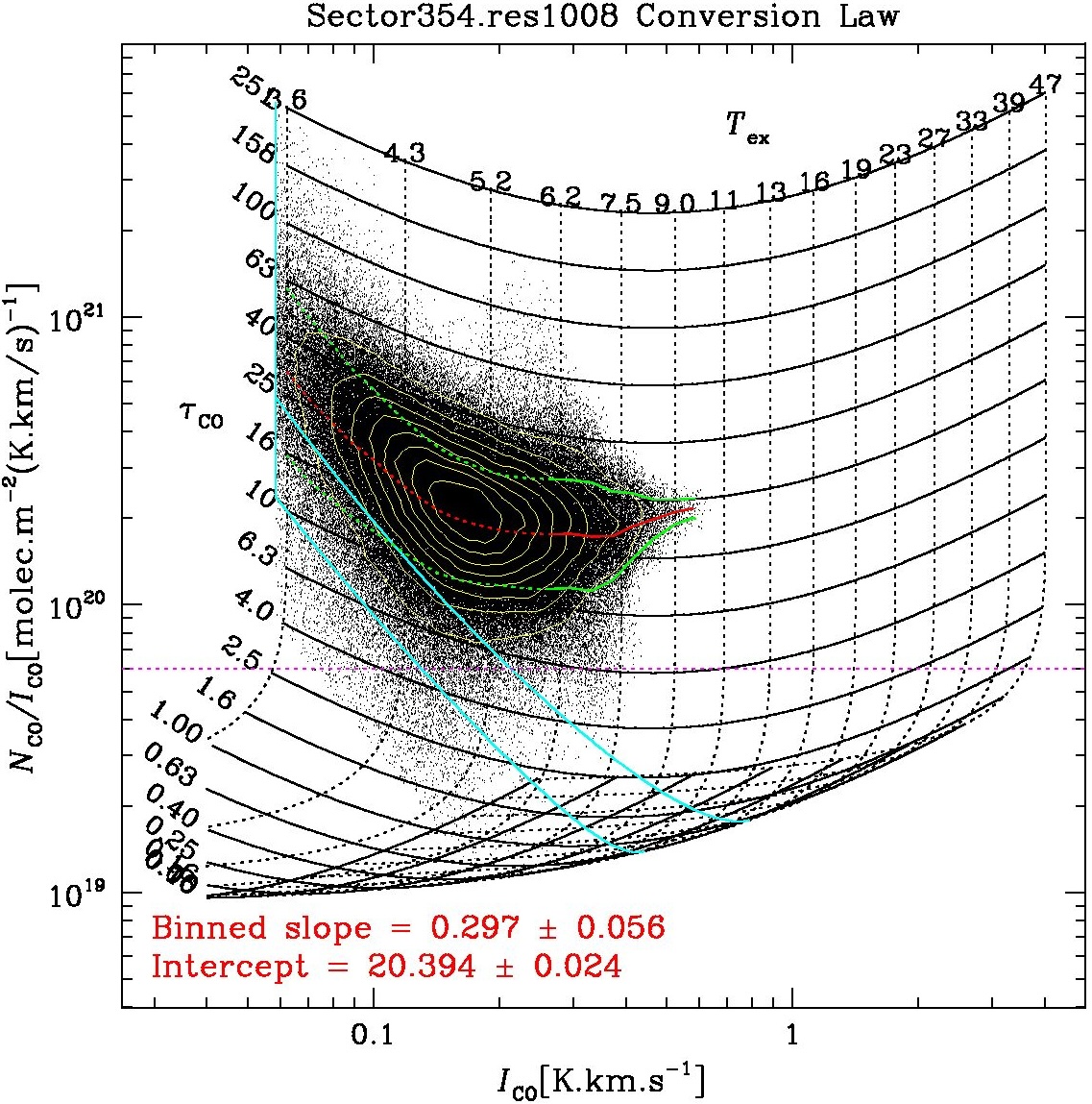}}
\vspace{-1mm}
\caption{\footnotesize Similar plots to Fig.\,\ref{x300-12-multi}, but for Sector 354.  The panels are at the same 5 progressive resolutions as in the previous 3 Figures. $$ $$
\label{x354-multi}}
\vspace{-7.5mm}
\end{figure*}

\vspace{1mm}In {\color{red}Figure \ref{x300}} we show a sample plot of how we measure the conversion law.  Similar plots are obtained for the other 9 sectors, which are shown at the top of each column of panels in {\color{red}Figures \ref{x300-12-multi}--\ref{x354-multi}}.  In those same Figures, the results in the rows below the top one are for progressively greater angular convolutions of the starting iso-CO data, as labelled; the full radiative transfer analysis was then redone in each case, giving the results for a slope and intercept as labelled in red of each panel.  

% Figure B1 (+11 for Appx A in aastex): S300--S312 XvsI

% Figure B2: S318-S330 XvsI

% Figure B3: S336-S348 XvsI

% Figure B4: S354 XvsI

%%%%%%%%%
%   Section B2  %
%%%%%%%%%
\subsection{Comparison with Velocity-Integrated Analysis}\label{vintlaws}
\begin{figure*}[h]
\vspace{0mm}
\centerline{\includegraphics[angle=0,scale=0.12]{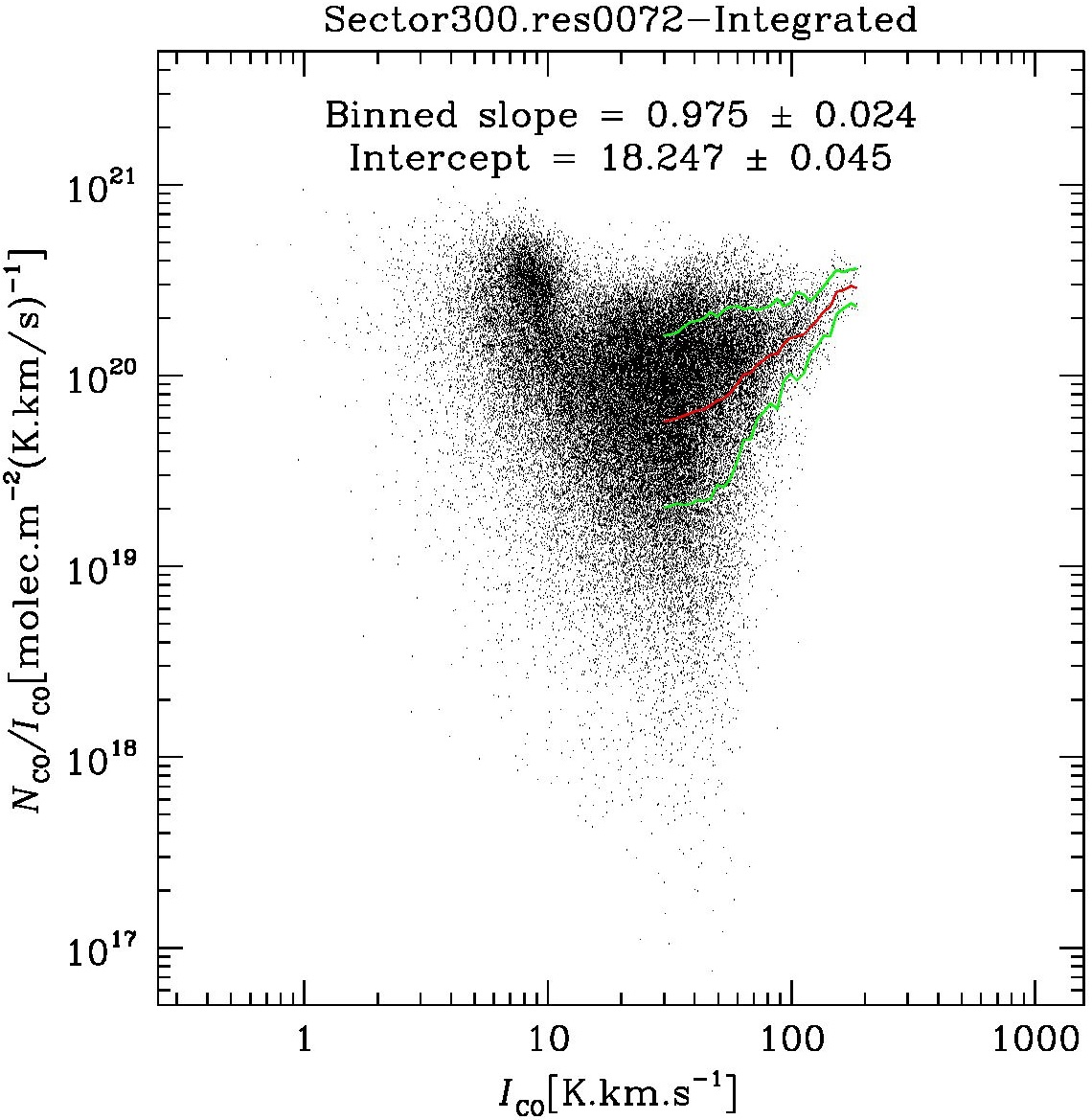} \includegraphics[angle=0,scale=0.12]{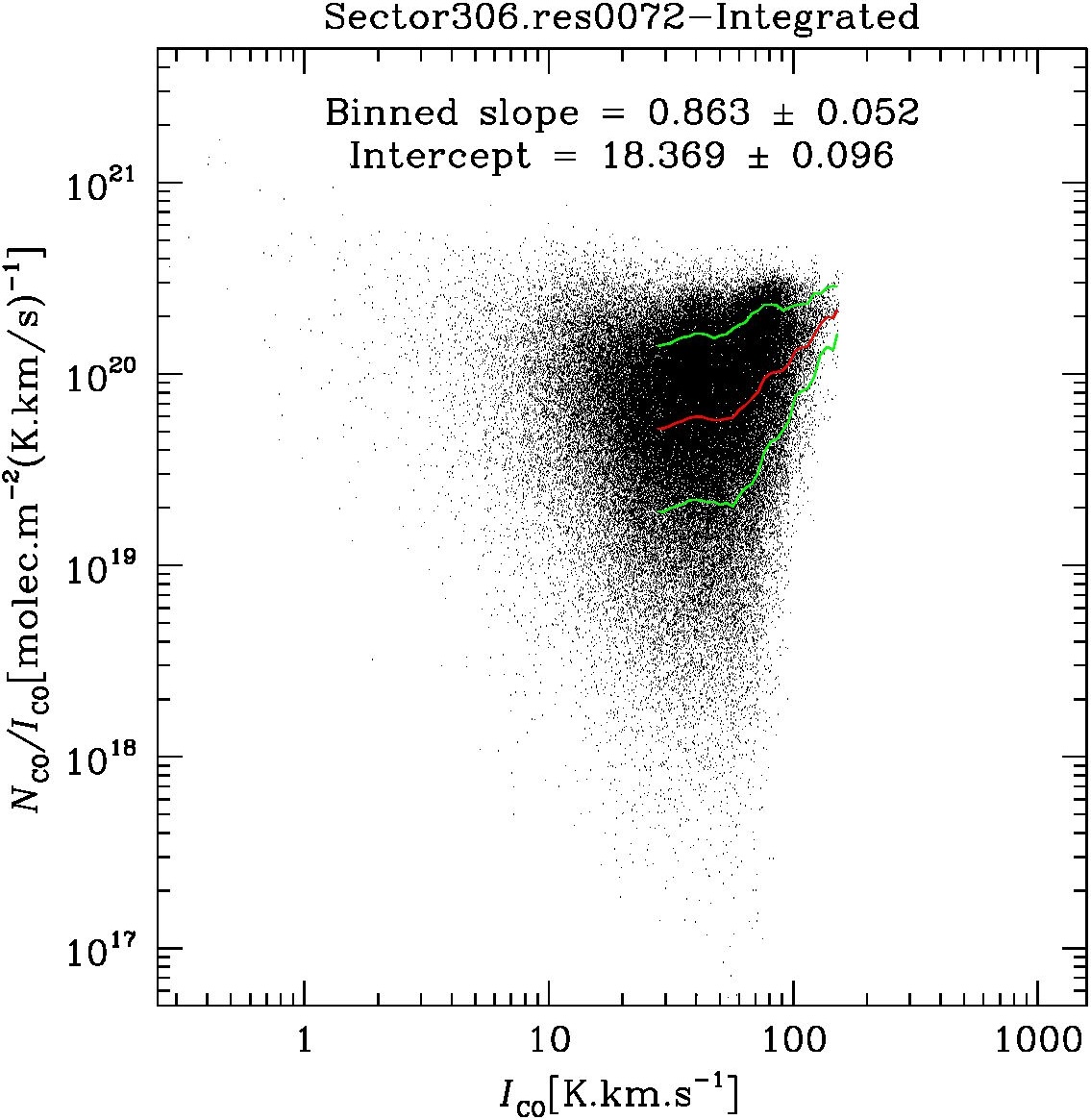} \includegraphics[angle=0,scale=0.12]{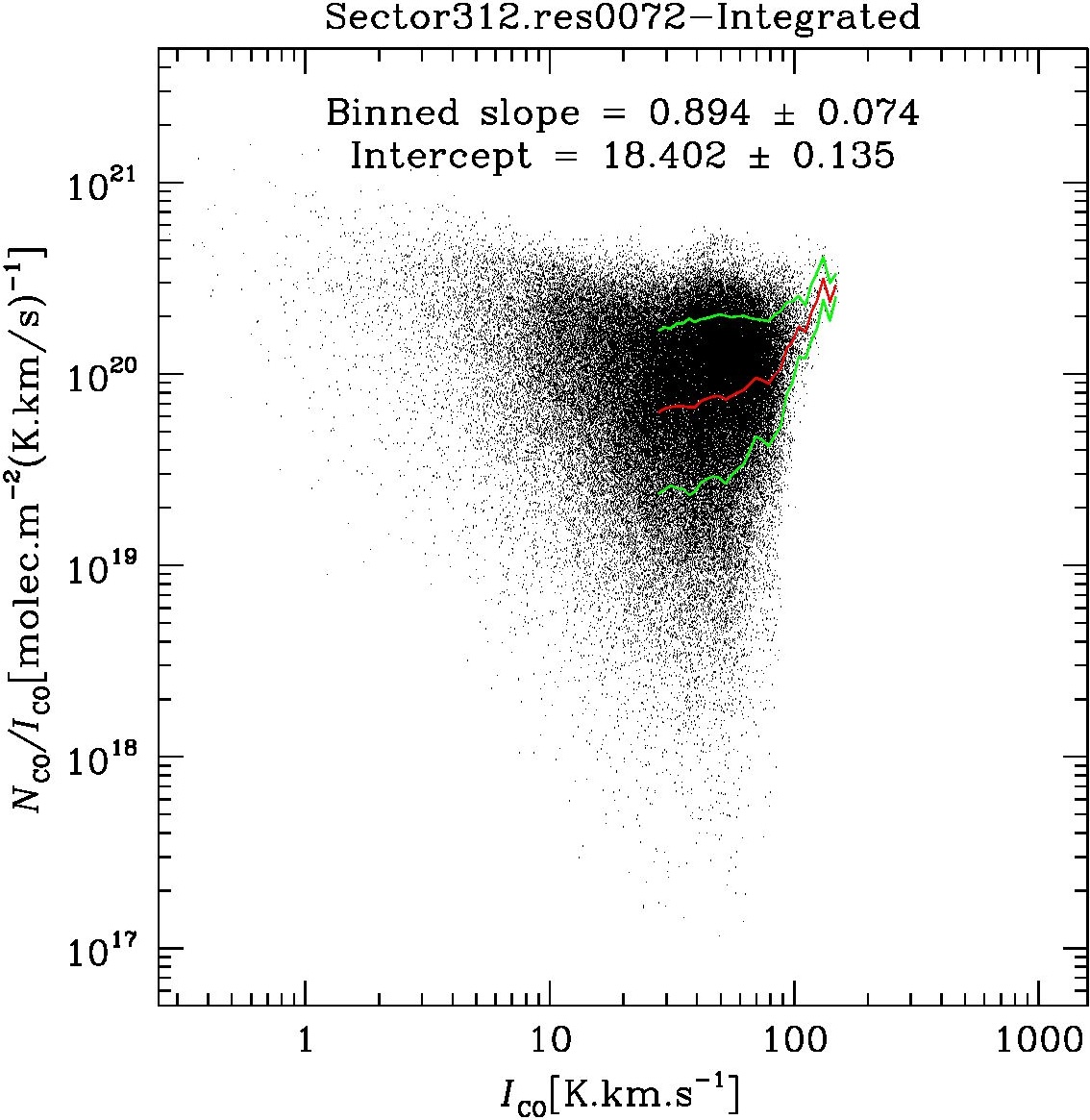}}
\vspace{-3mm}
\centerline{\includegraphics[angle=0,scale=0.12]{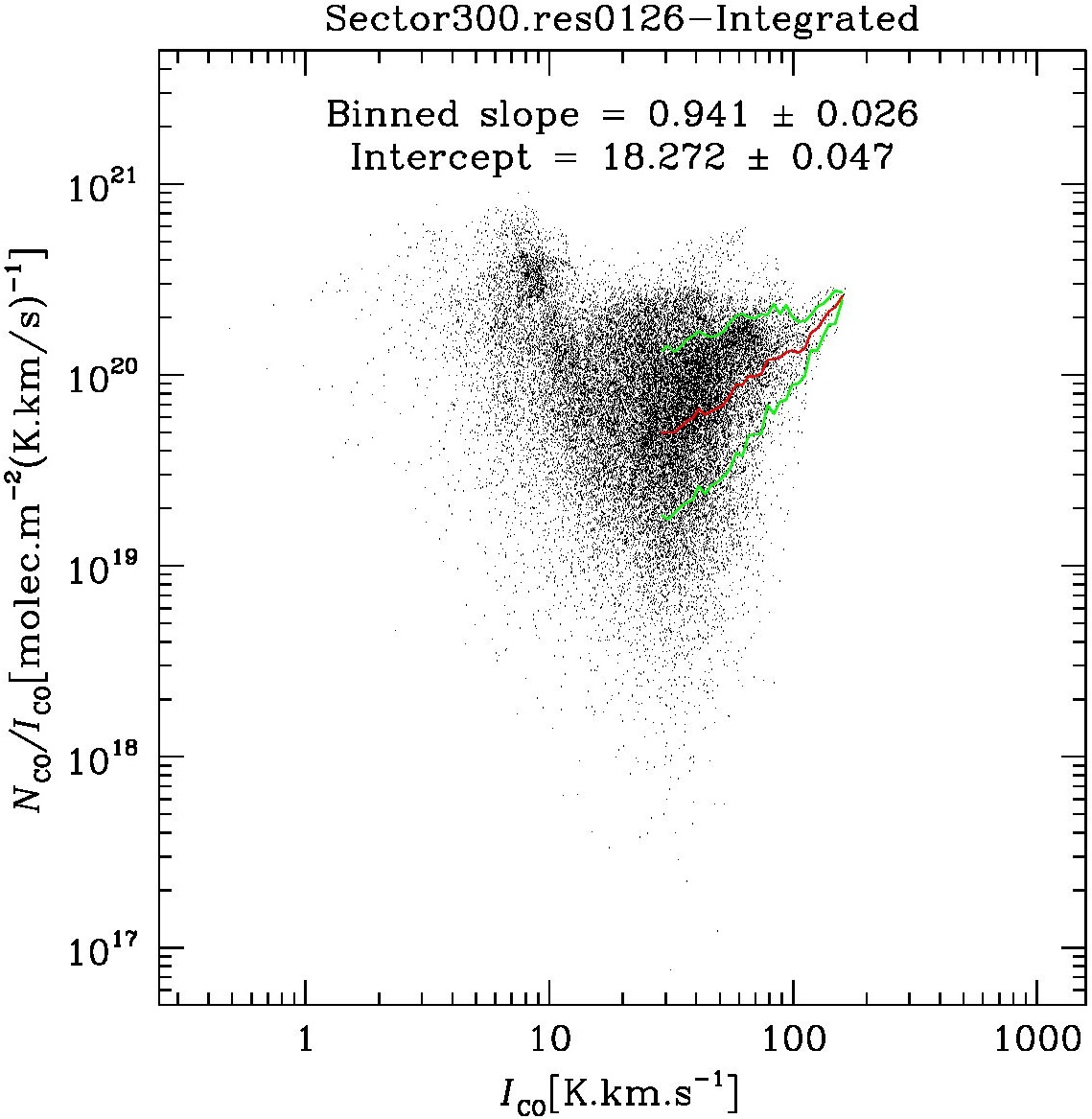} \includegraphics[angle=0,scale=0.12]{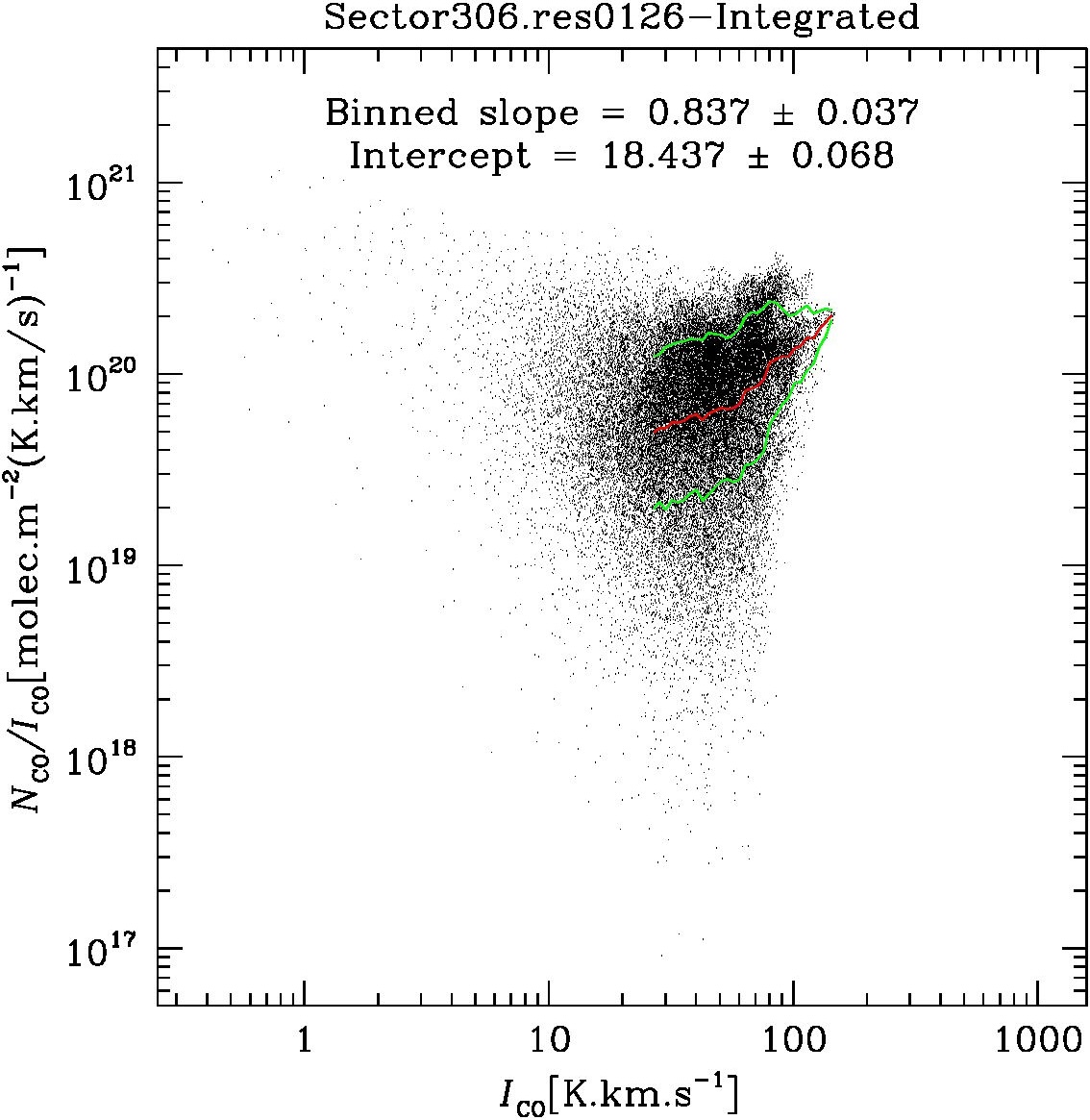} \includegraphics[angle=0,scale=0.12]{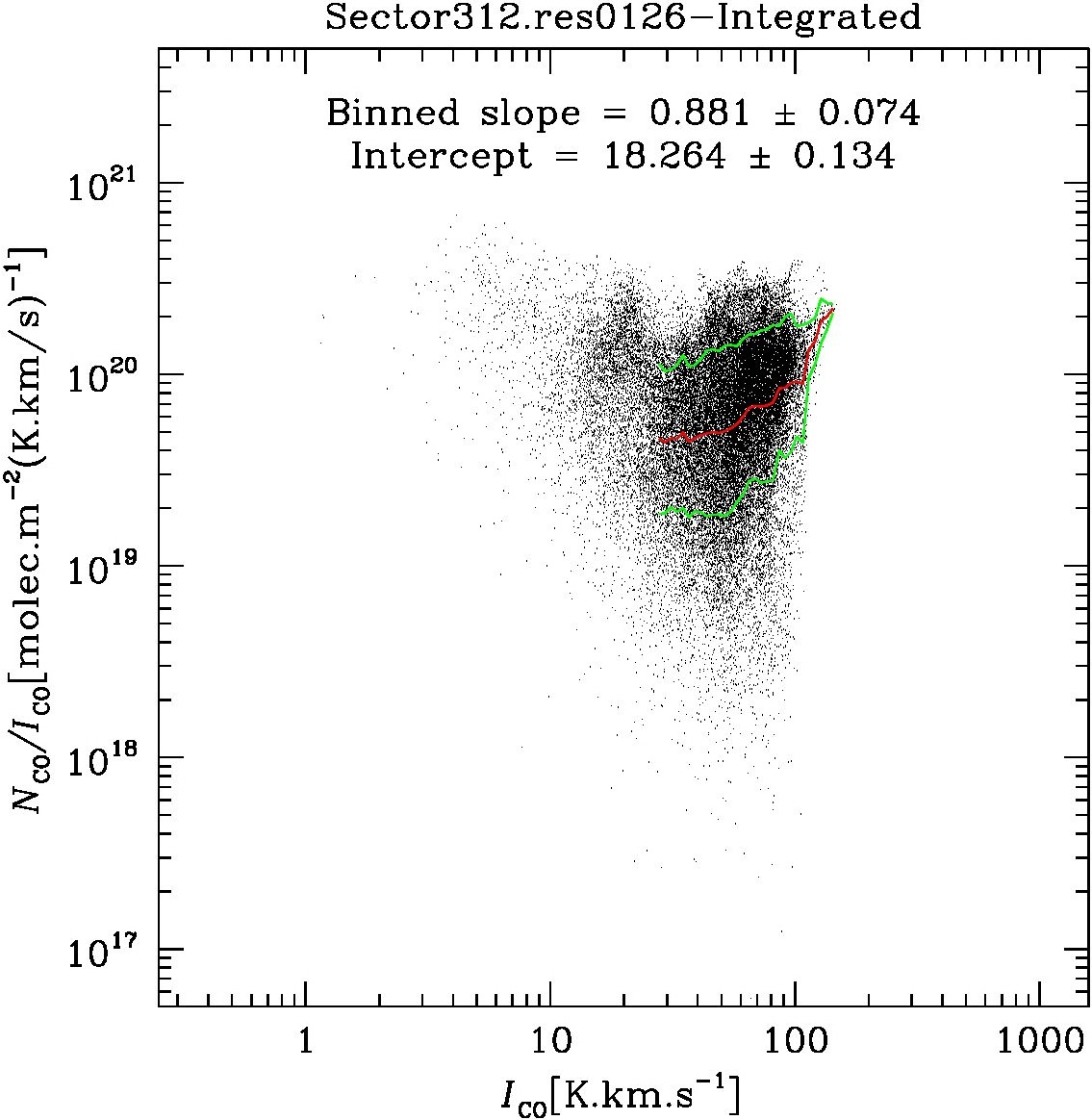}}
\vspace{-3mm}
\centerline{\includegraphics[angle=0,scale=0.12]{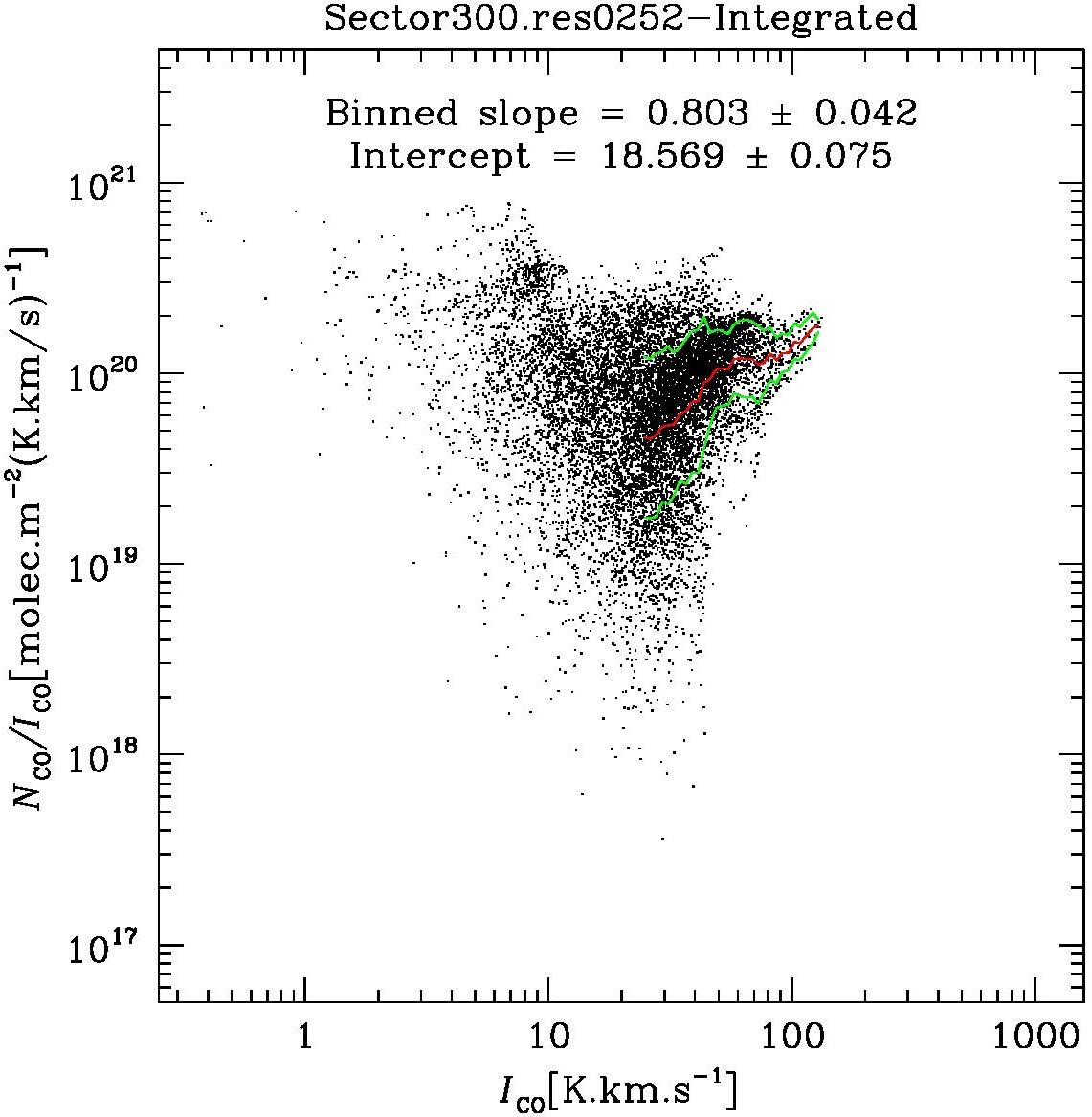} \includegraphics[angle=0,scale=0.12]{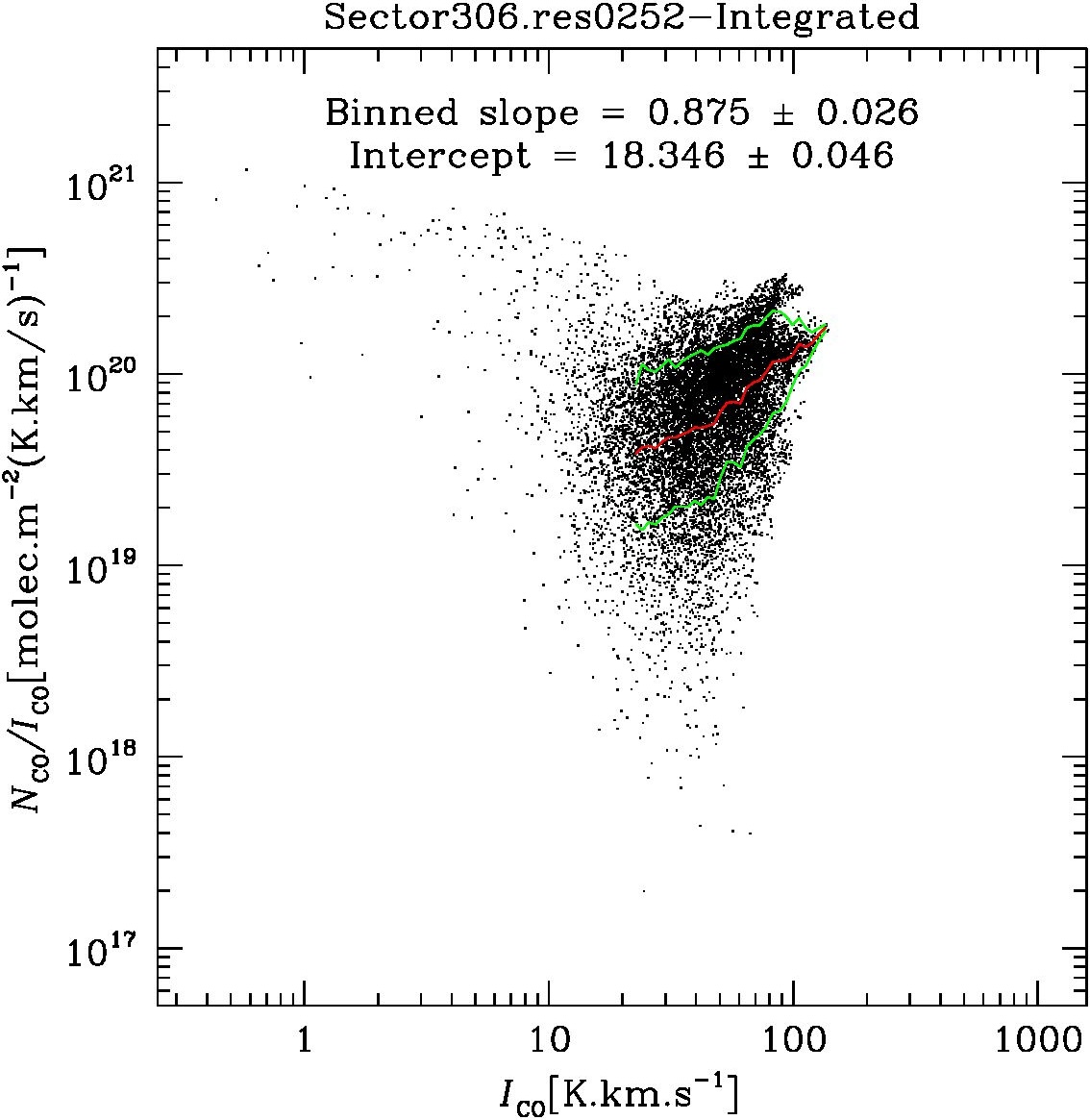} \includegraphics[angle=0,scale=0.12]{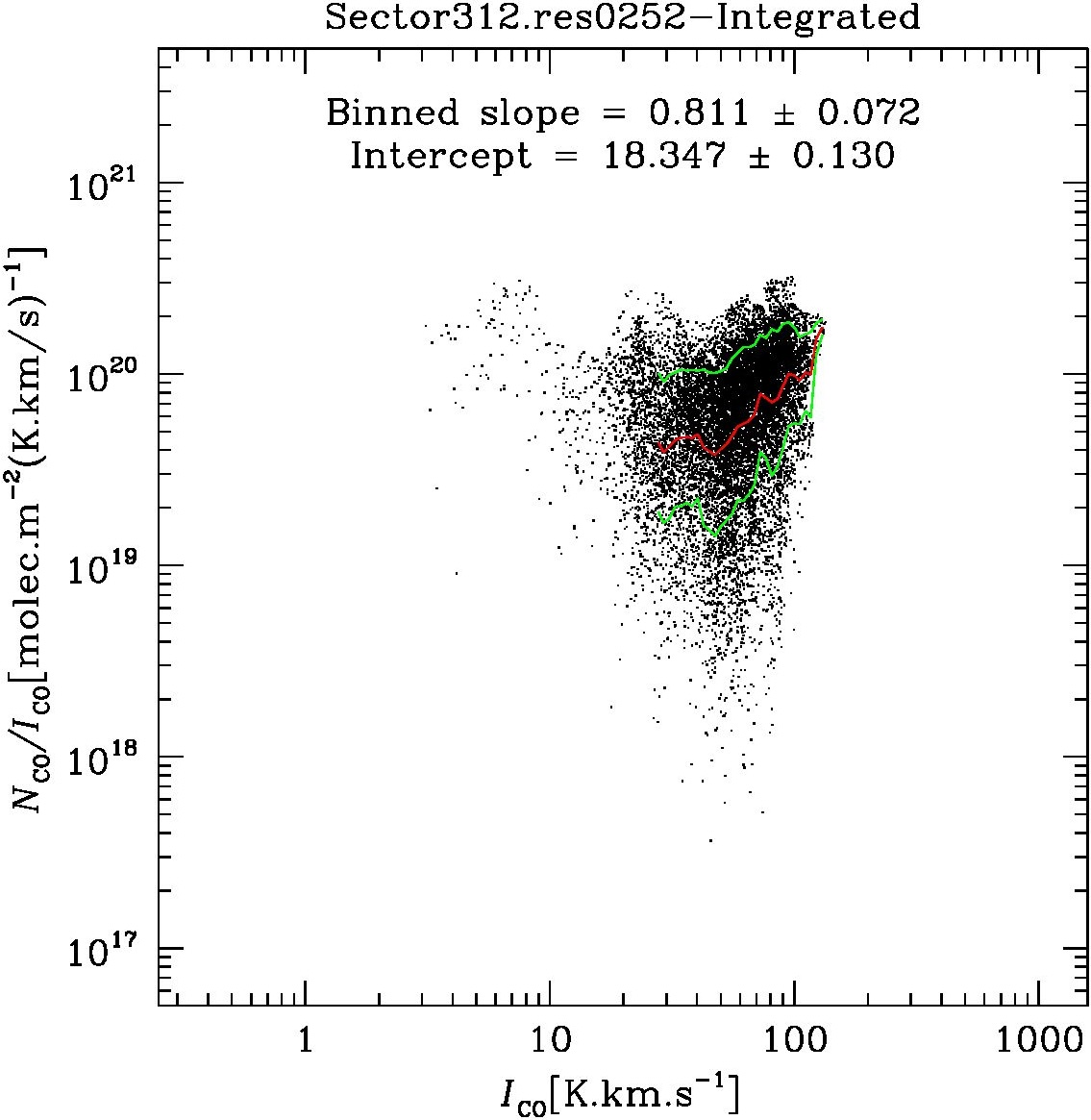}}
\vspace{-3mm}
\centerline{\includegraphics[angle=0,scale=0.12]{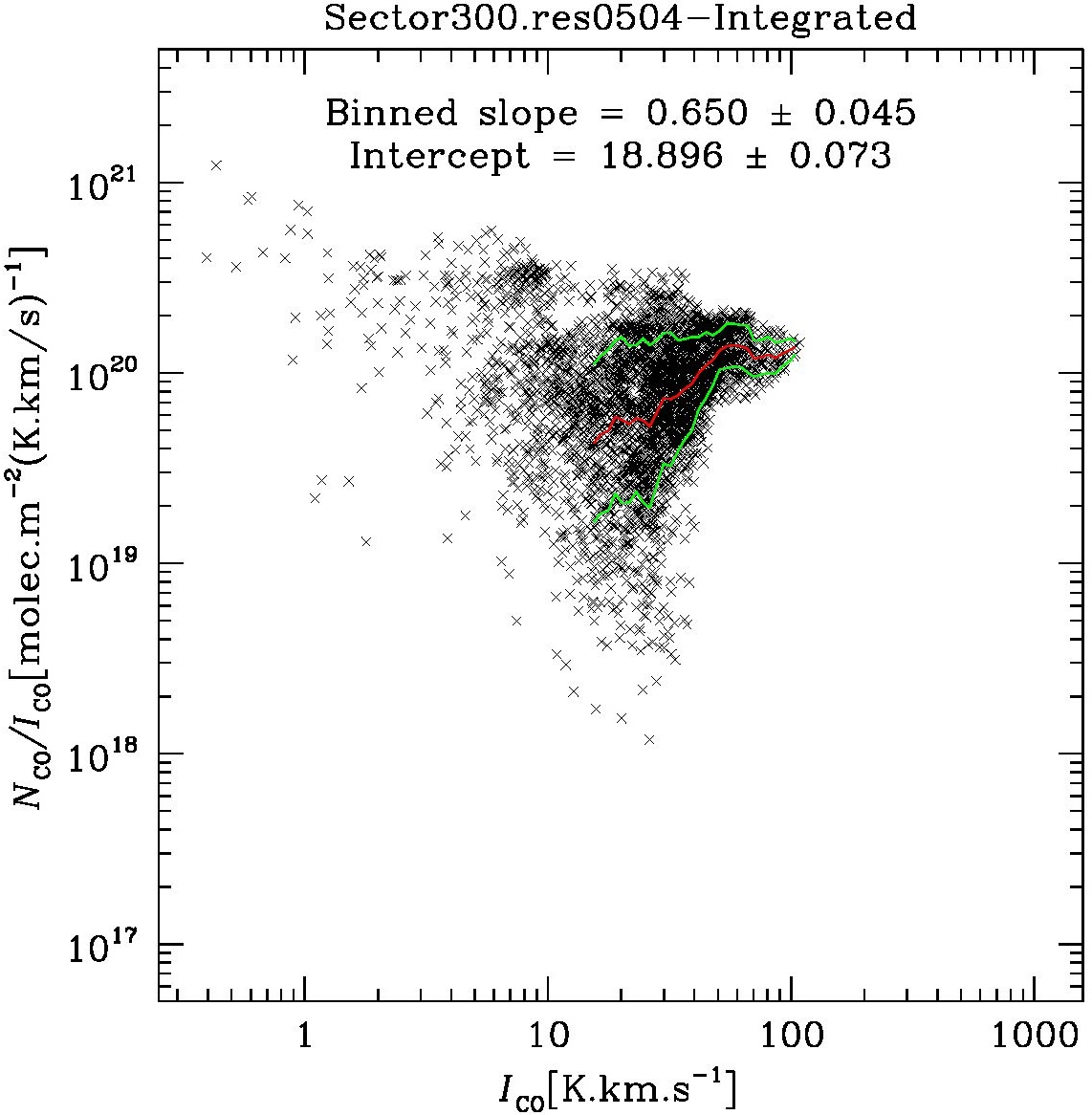} \includegraphics[angle=0,scale=0.12]{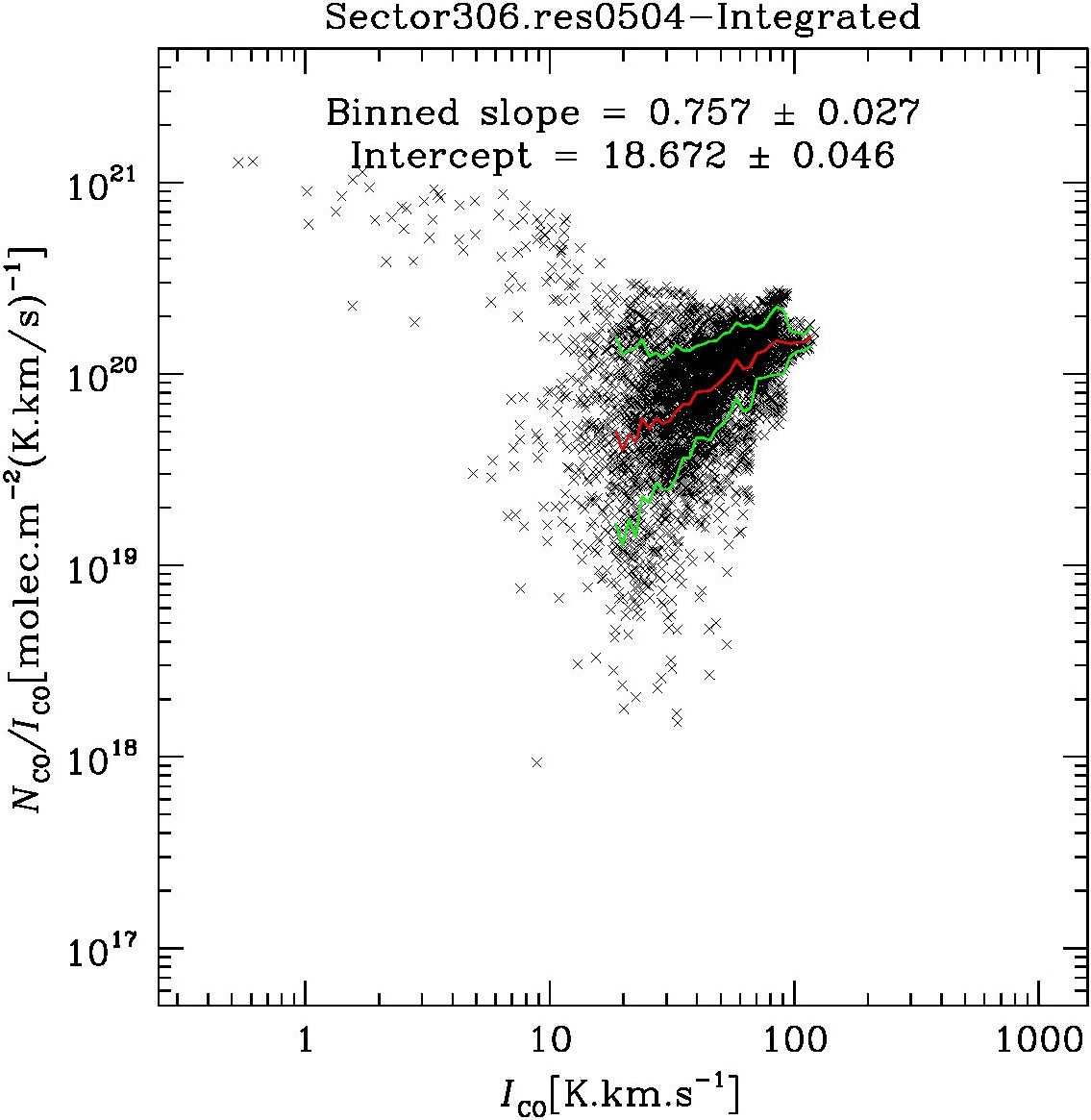} \includegraphics[angle=0,scale=0.12]{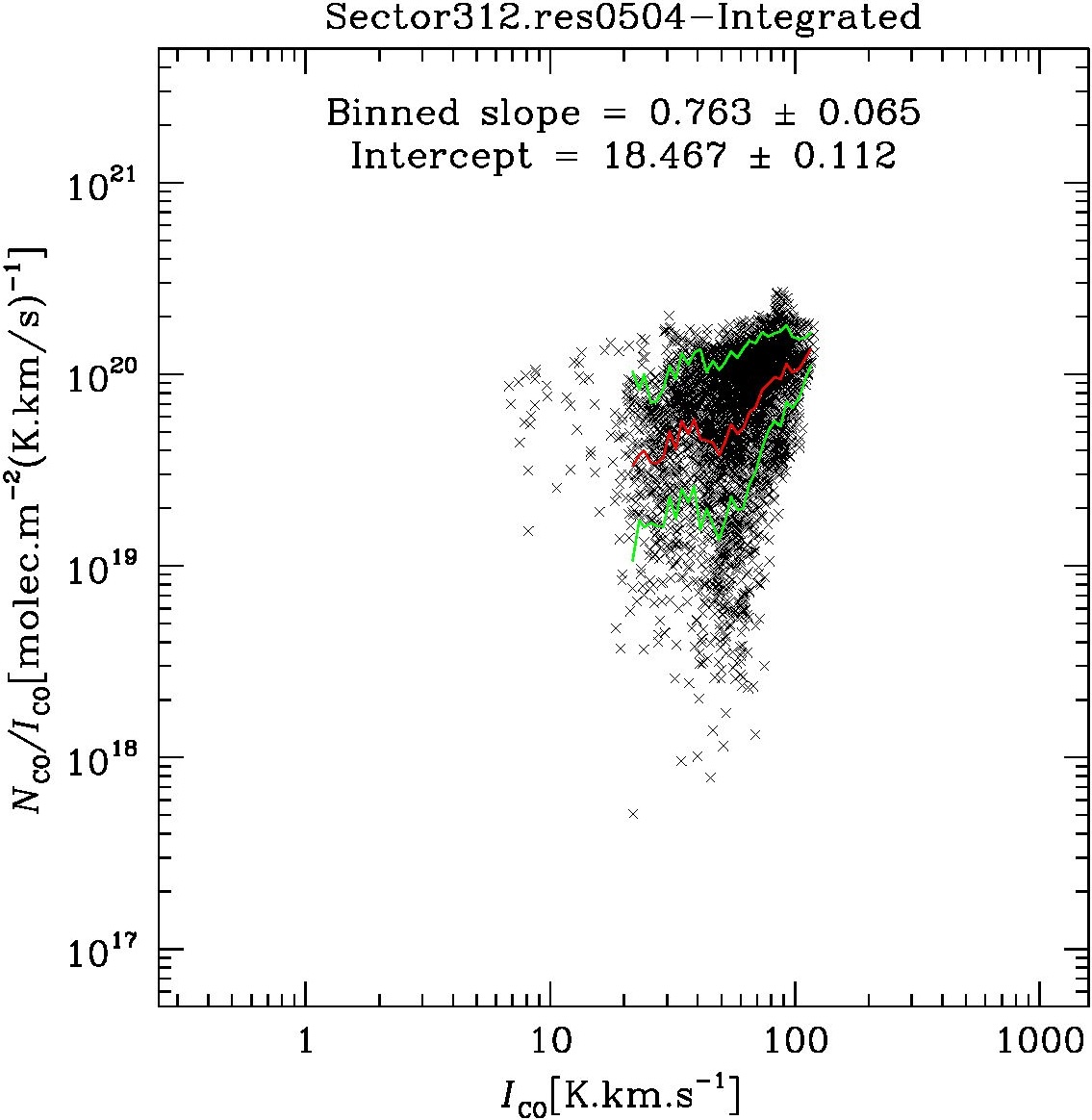}}
\vspace{-3mm}
\centerline{\includegraphics[angle=0,scale=0.12]{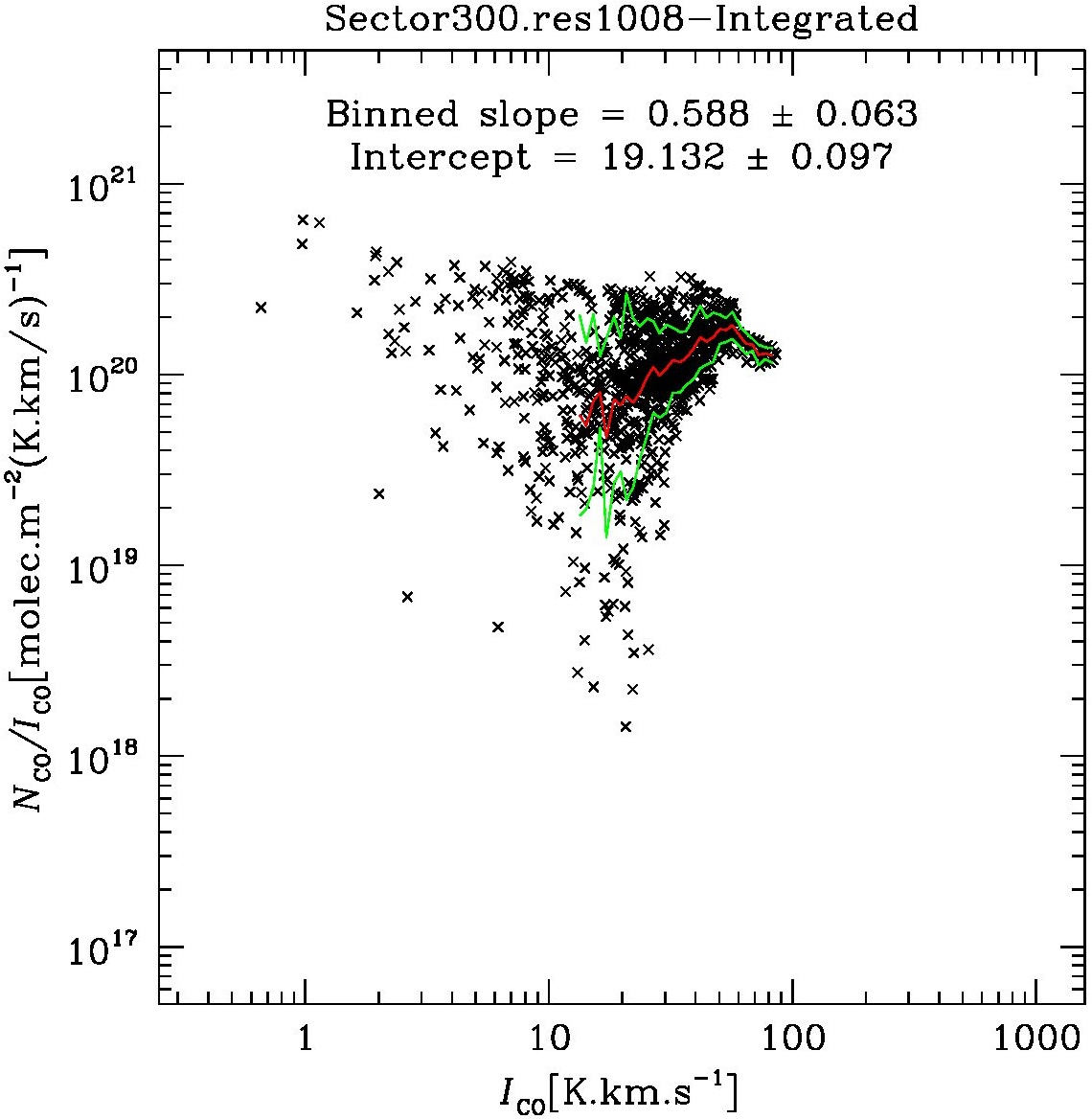} \includegraphics[angle=0,scale=0.12]{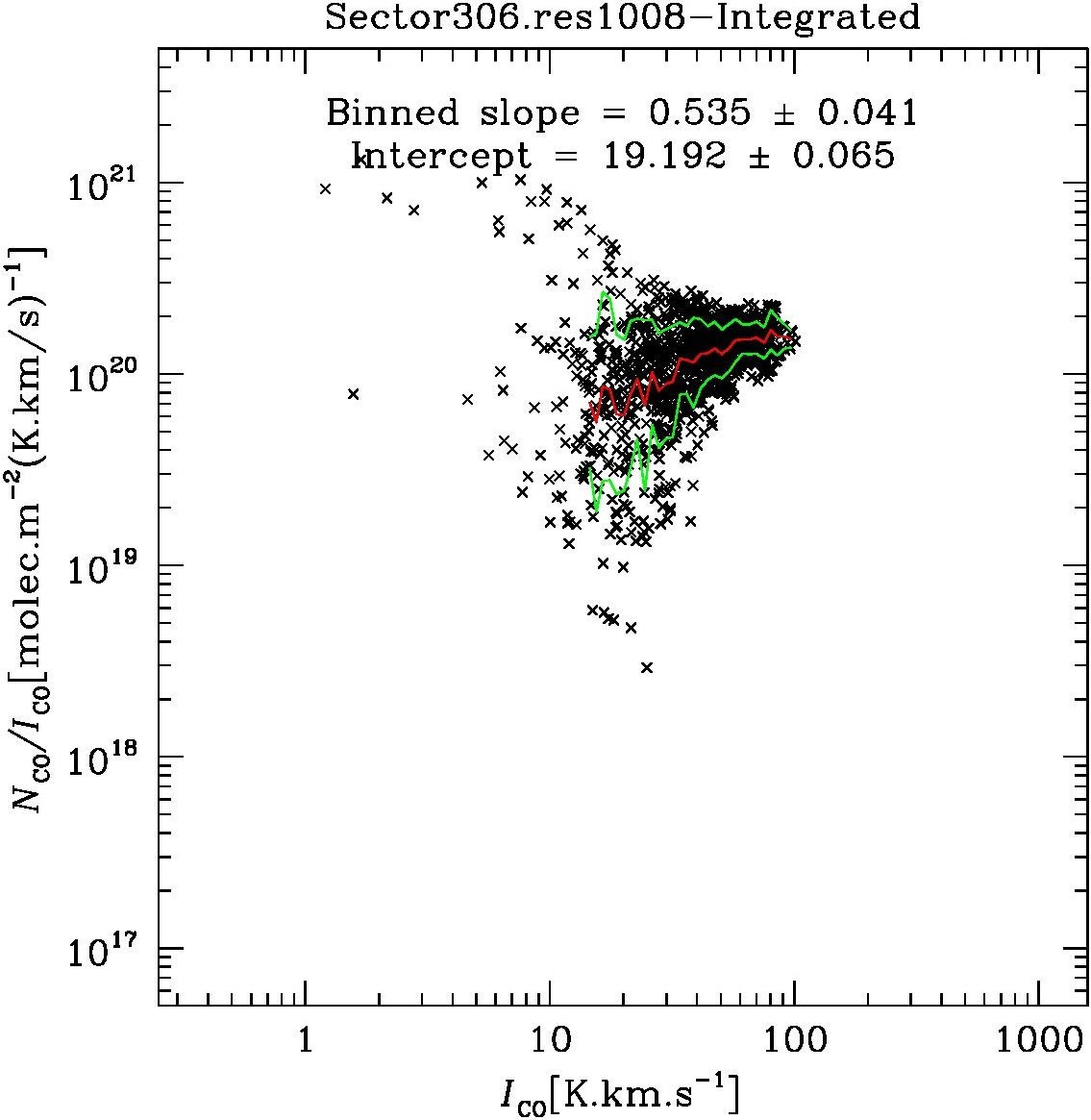} \includegraphics[angle=0,scale=0.12]{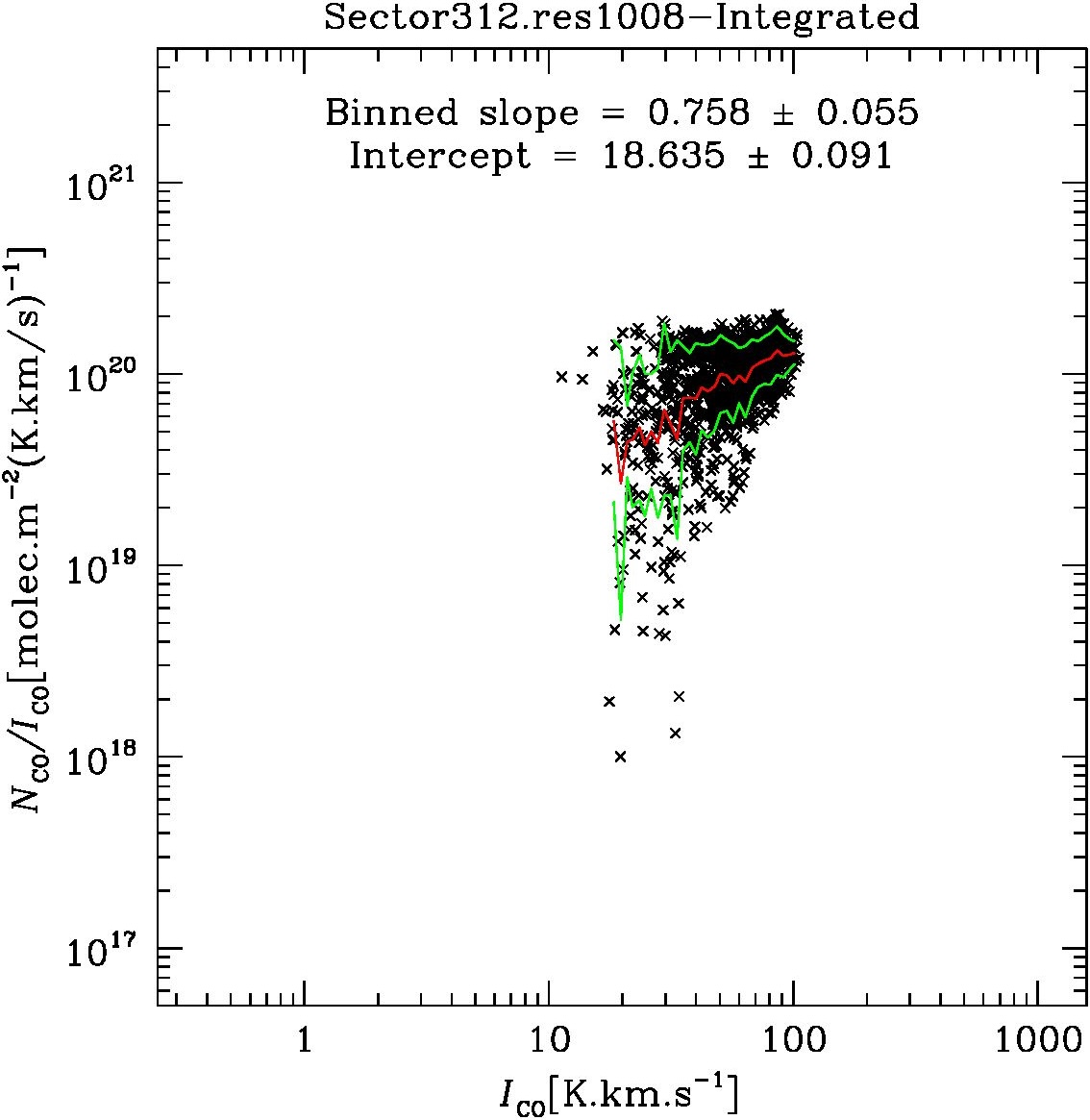}}
\vspace{-1mm}
\caption{\footnotesize \nco/\ico\ ratio vs. \ico\ as in Figure \ref{x300-12-multi}, i.e., arranged in columns by Sector and in rows by resolution, but where each quantity is integrated over all velocity channels, regardless of whether \nco\ is defined in all the channels where \ico\ is detected.  We don't include a ($\tau$, \tex) grid here, since the velocity integration blurs the grid due to the variable number of channels used for each point. $$ $$
\label{xcl300-12-multi}}
\vspace{0mm}
\end{figure*}

\begin{figure*}[h]
\vspace{0mm}
\centerline{\includegraphics[angle=0,scale=0.12]{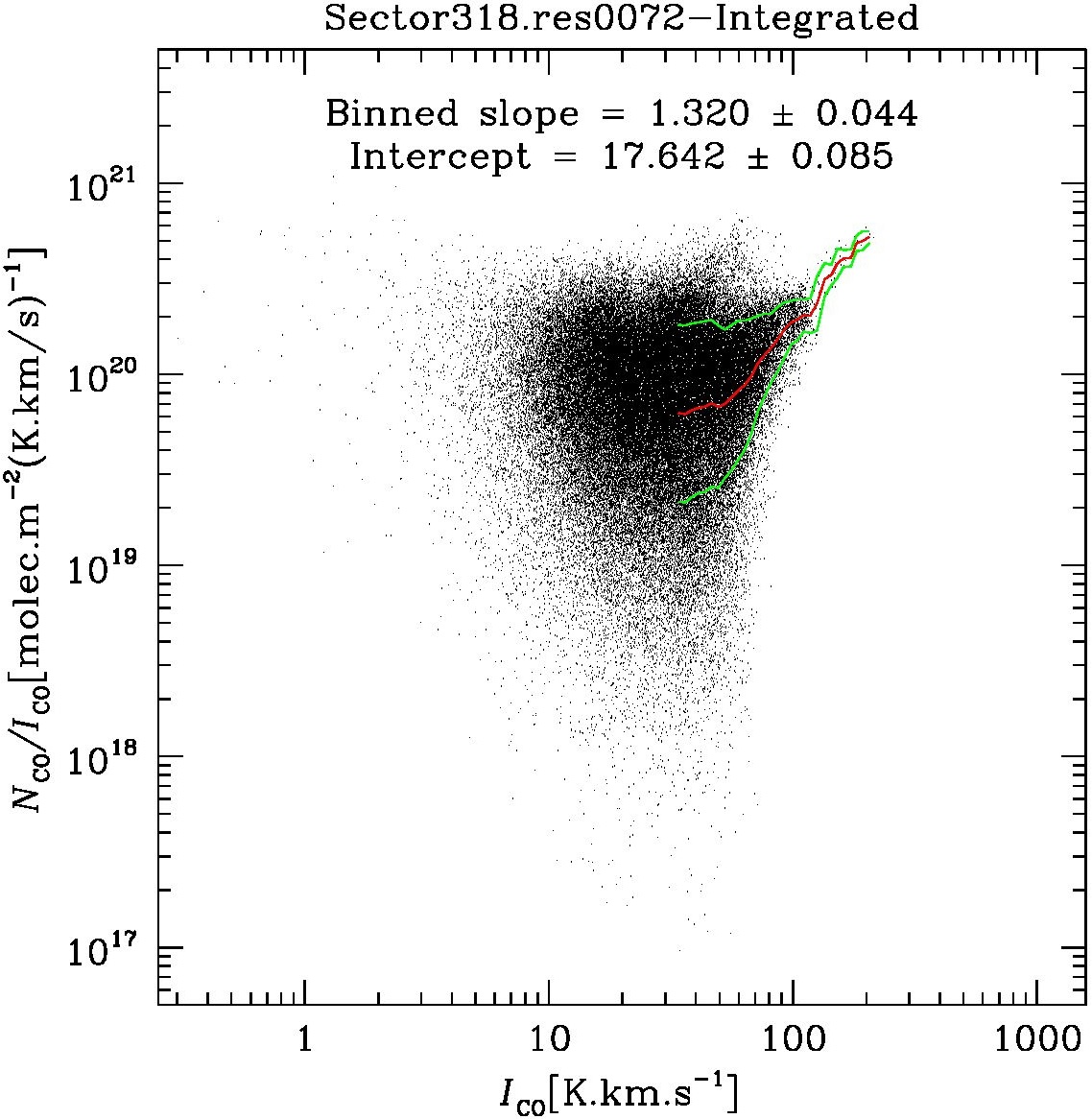} \includegraphics[angle=0,scale=0.12]{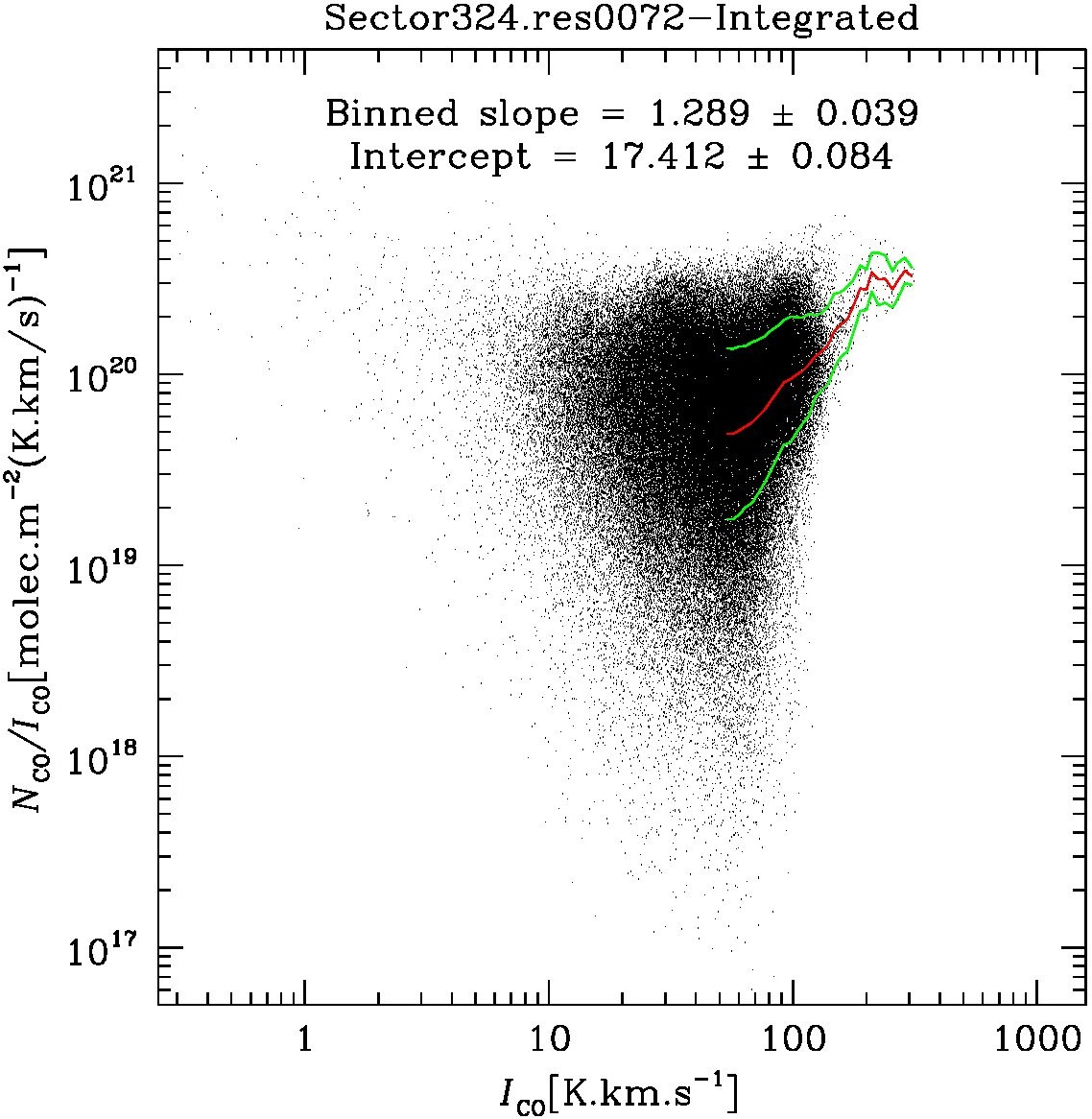} \includegraphics[angle=0,scale=0.12]{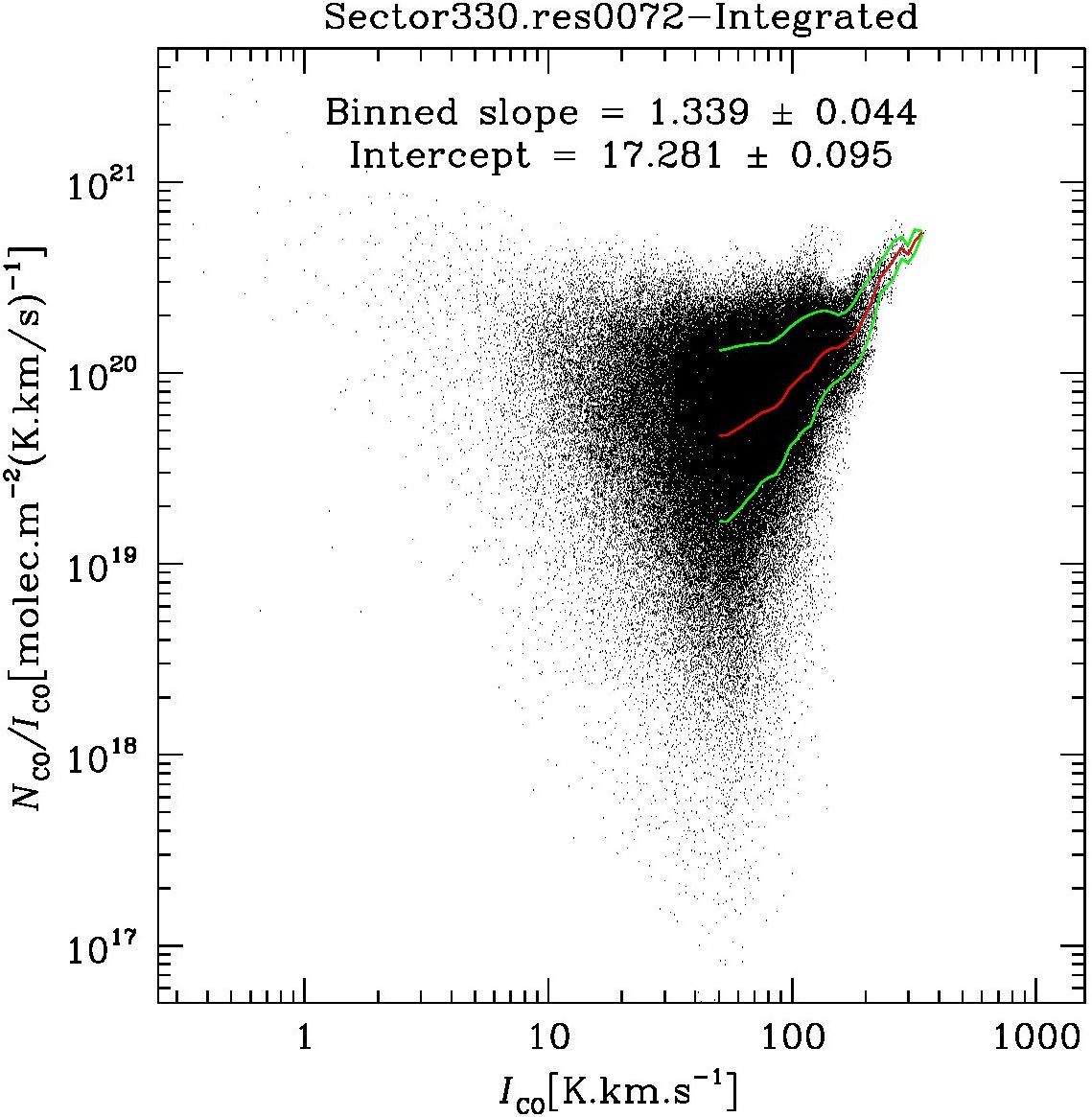}}
\vspace{-3mm}
\centerline{\includegraphics[angle=0,scale=0.12]{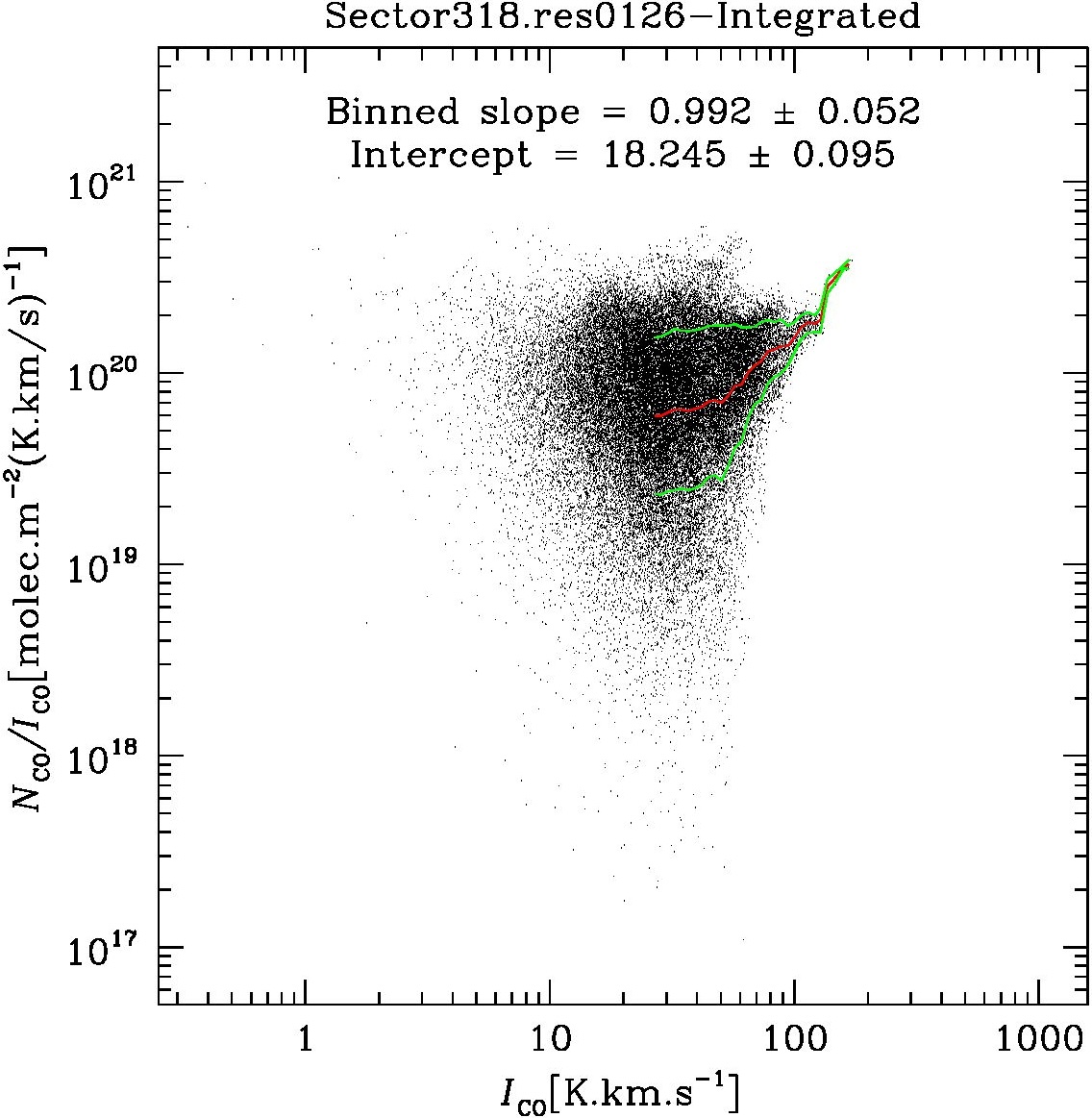} \includegraphics[angle=0,scale=0.12]{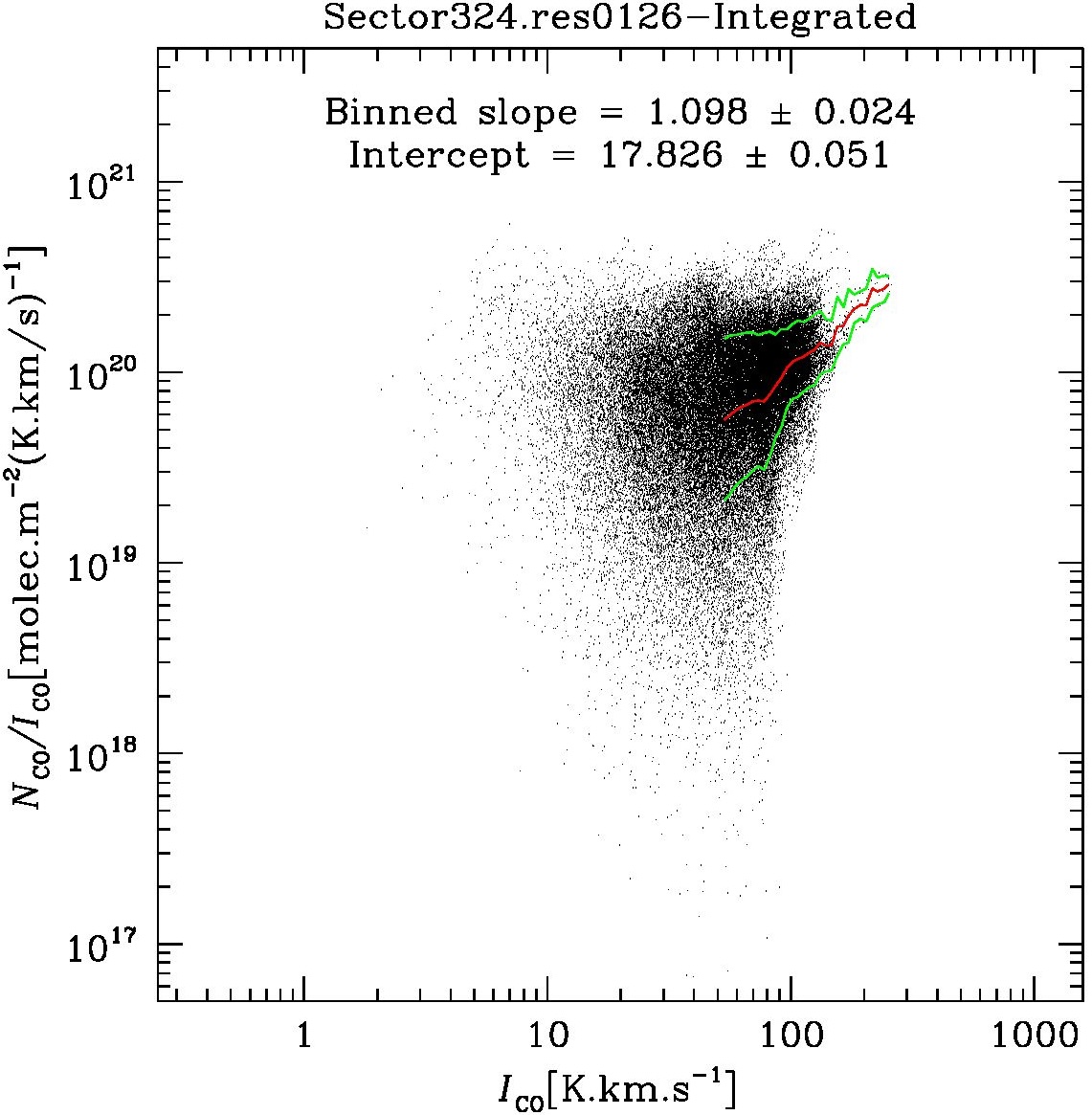} \includegraphics[angle=0,scale=0.12]{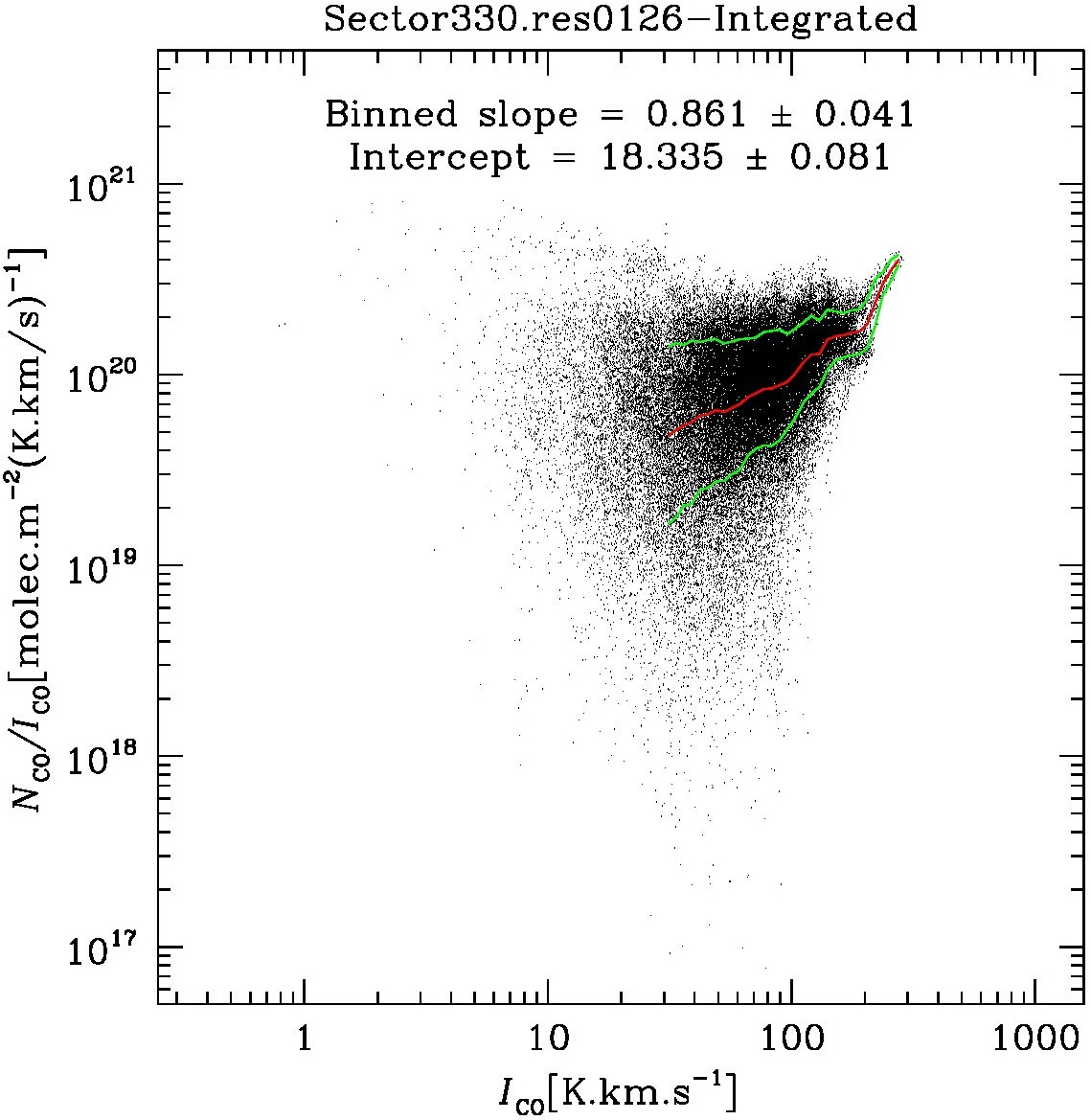}}
\vspace{-3mm}
\centerline{\includegraphics[angle=0,scale=0.12]{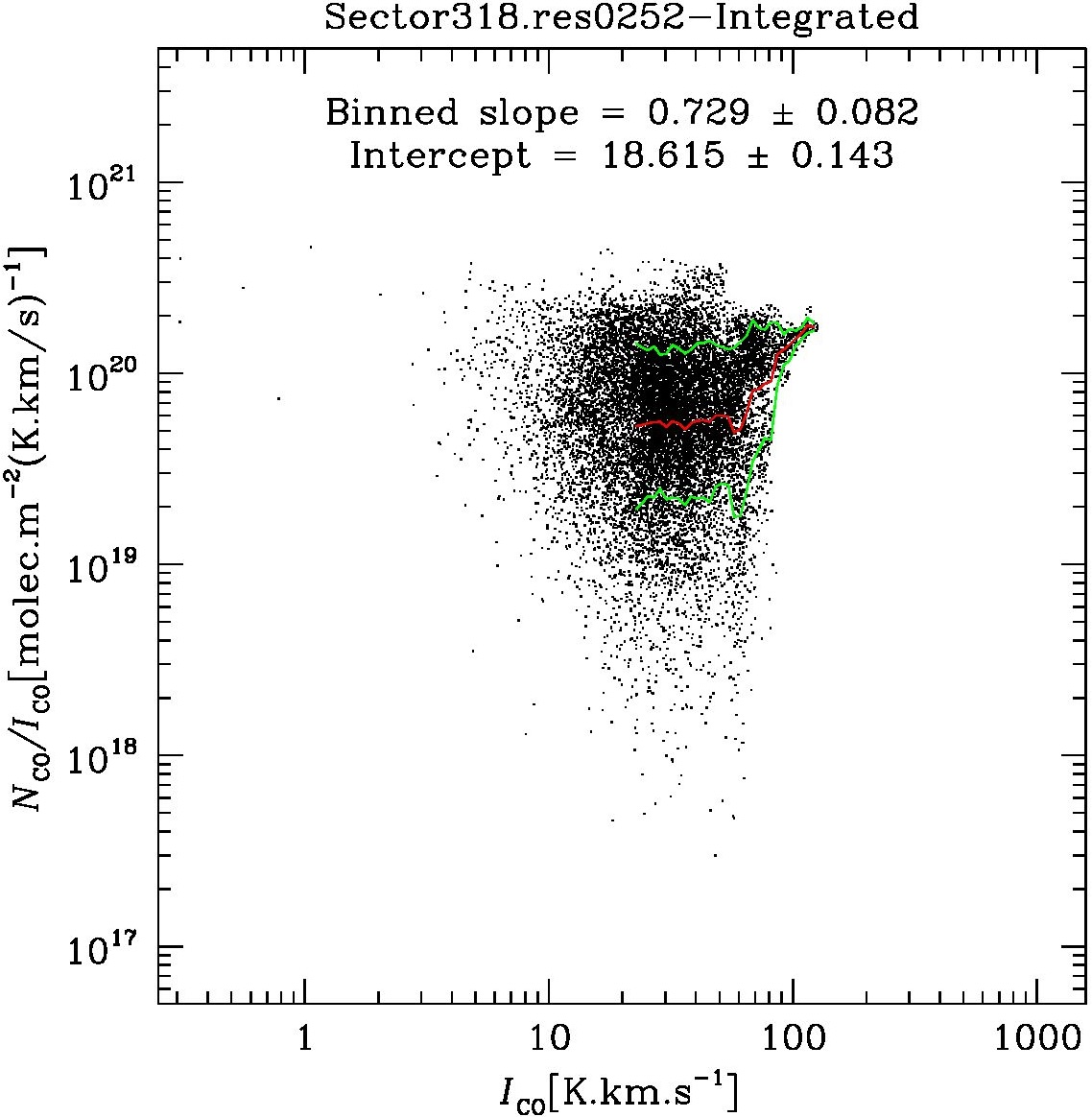} \includegraphics[angle=0,scale=0.12]{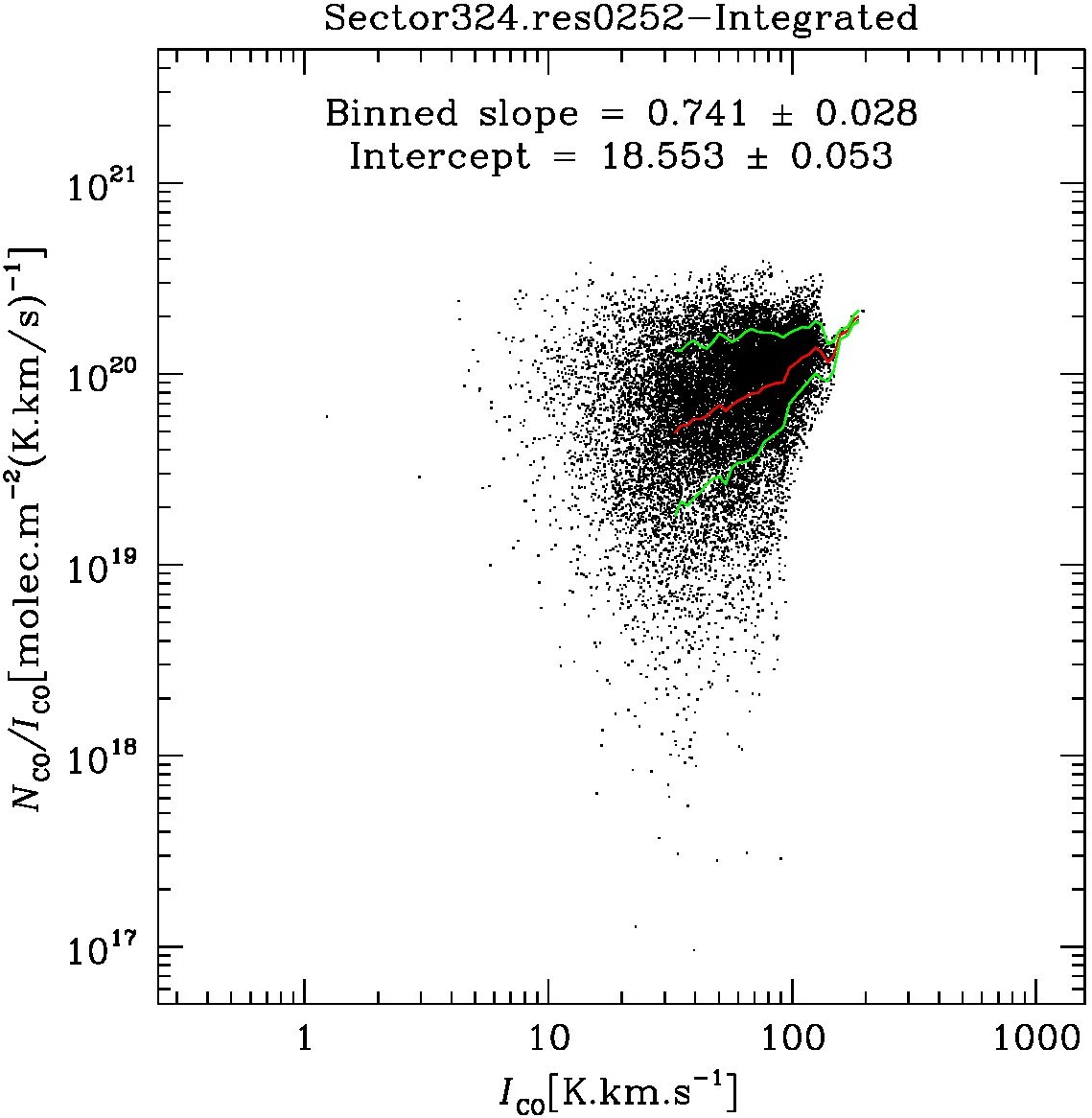} \includegraphics[angle=0,scale=0.12]{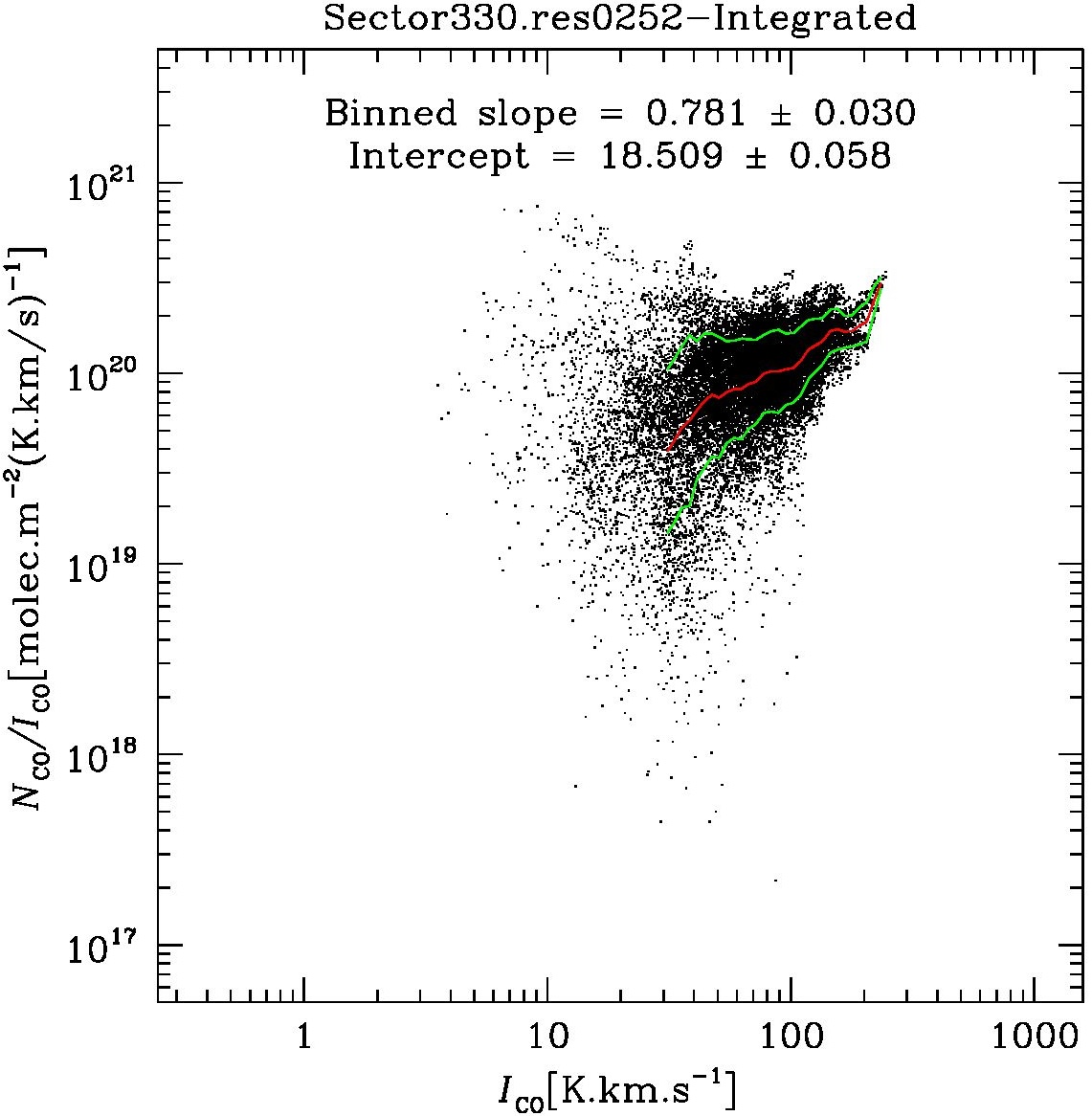}}
\vspace{-3mm}
\centerline{\includegraphics[angle=0,scale=0.12]{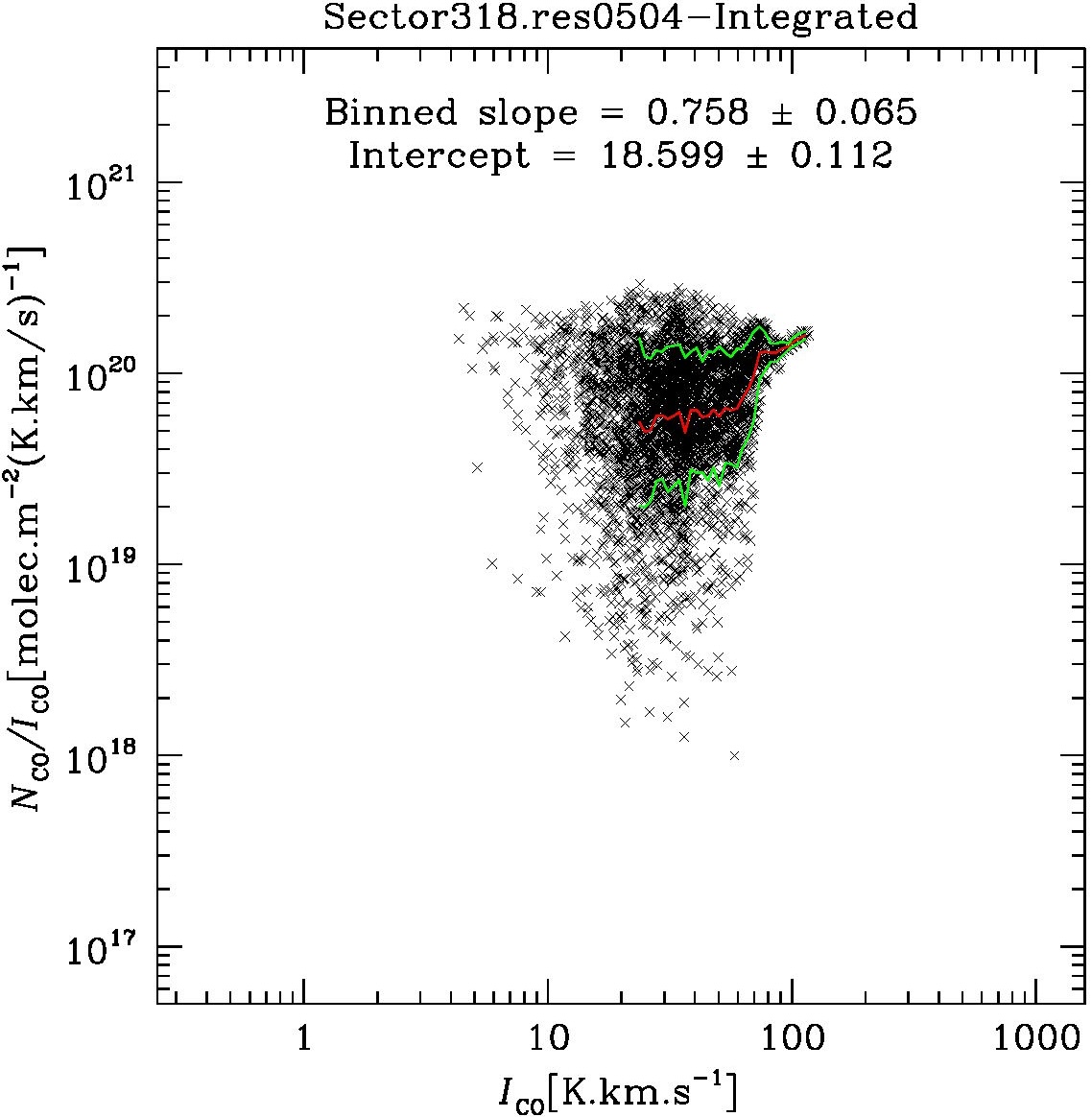} \includegraphics[angle=0,scale=0.12]{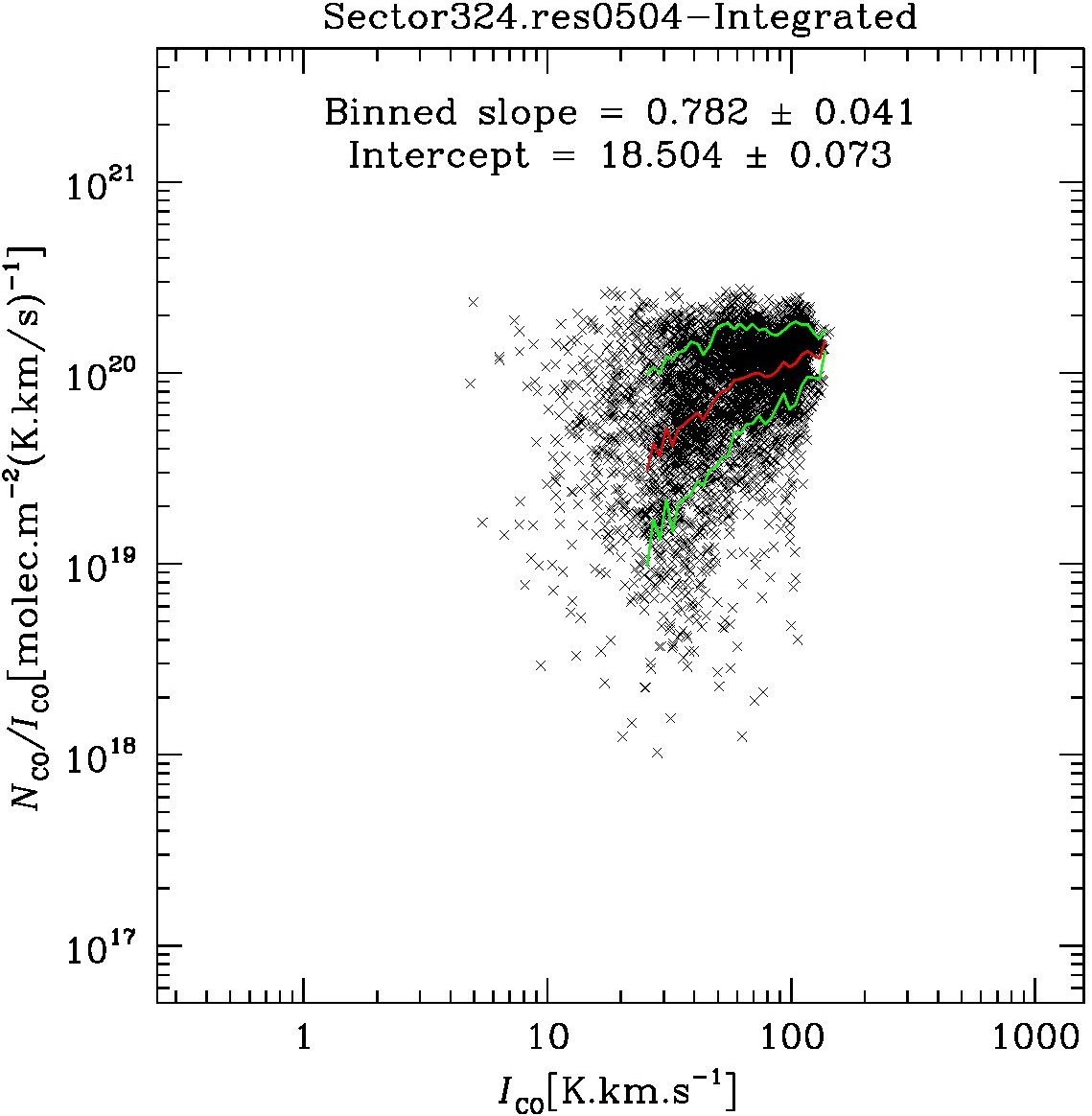} \includegraphics[angle=0,scale=0.12]{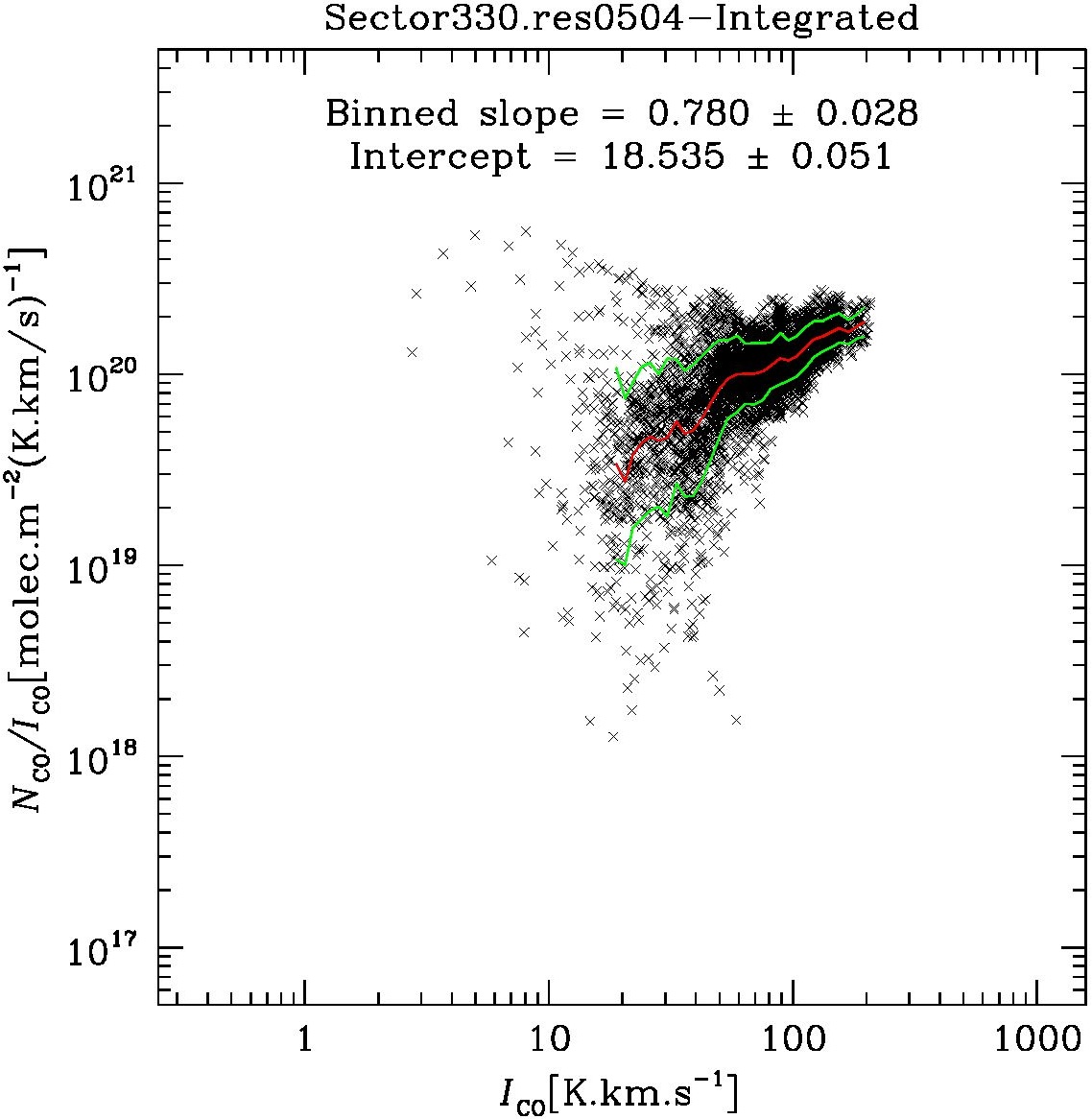}}
\vspace{-3mm}
\centerline{\includegraphics[angle=0,scale=0.12]{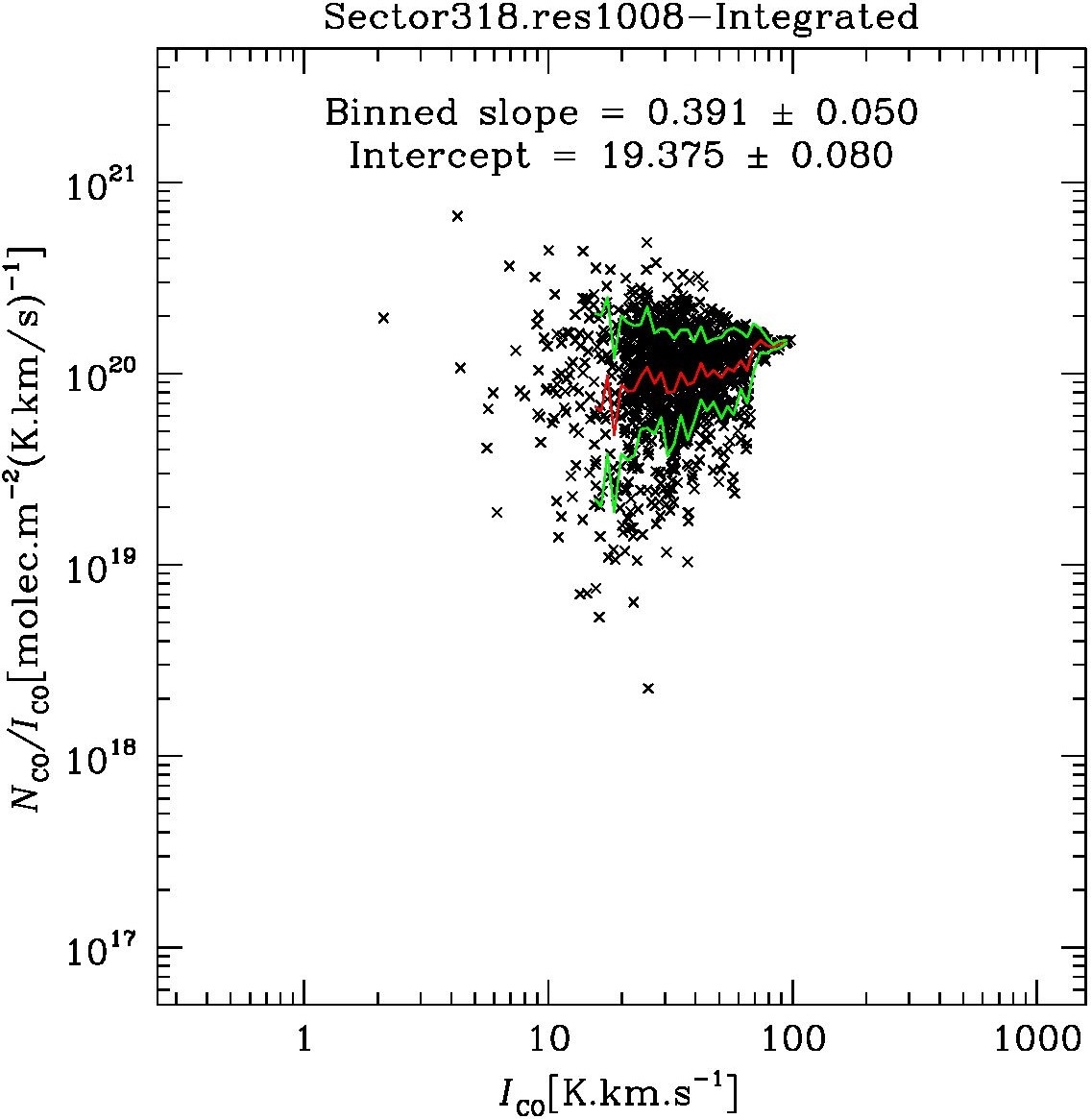} \includegraphics[angle=0,scale=0.12]{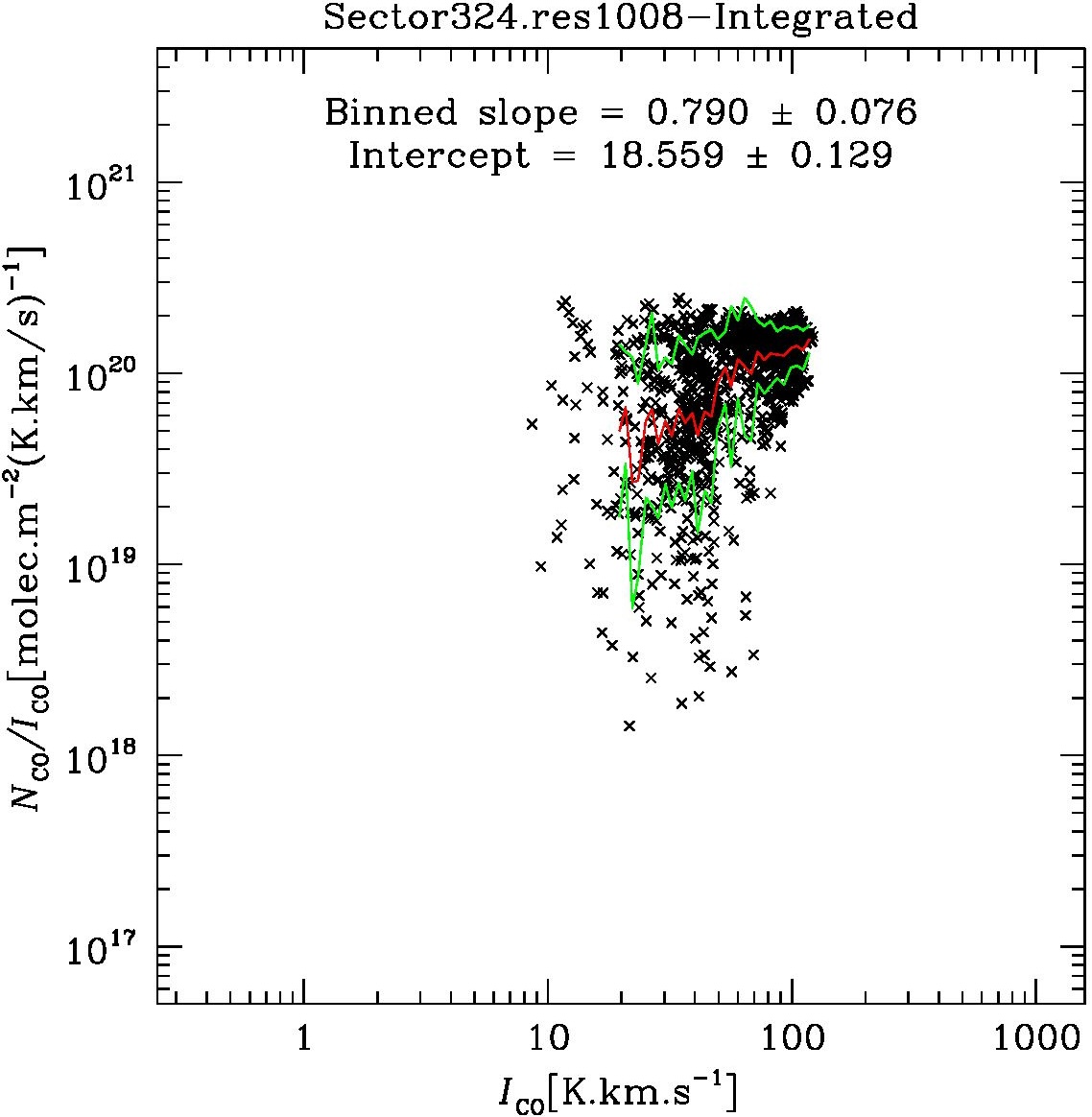} \includegraphics[angle=0,scale=0.12]{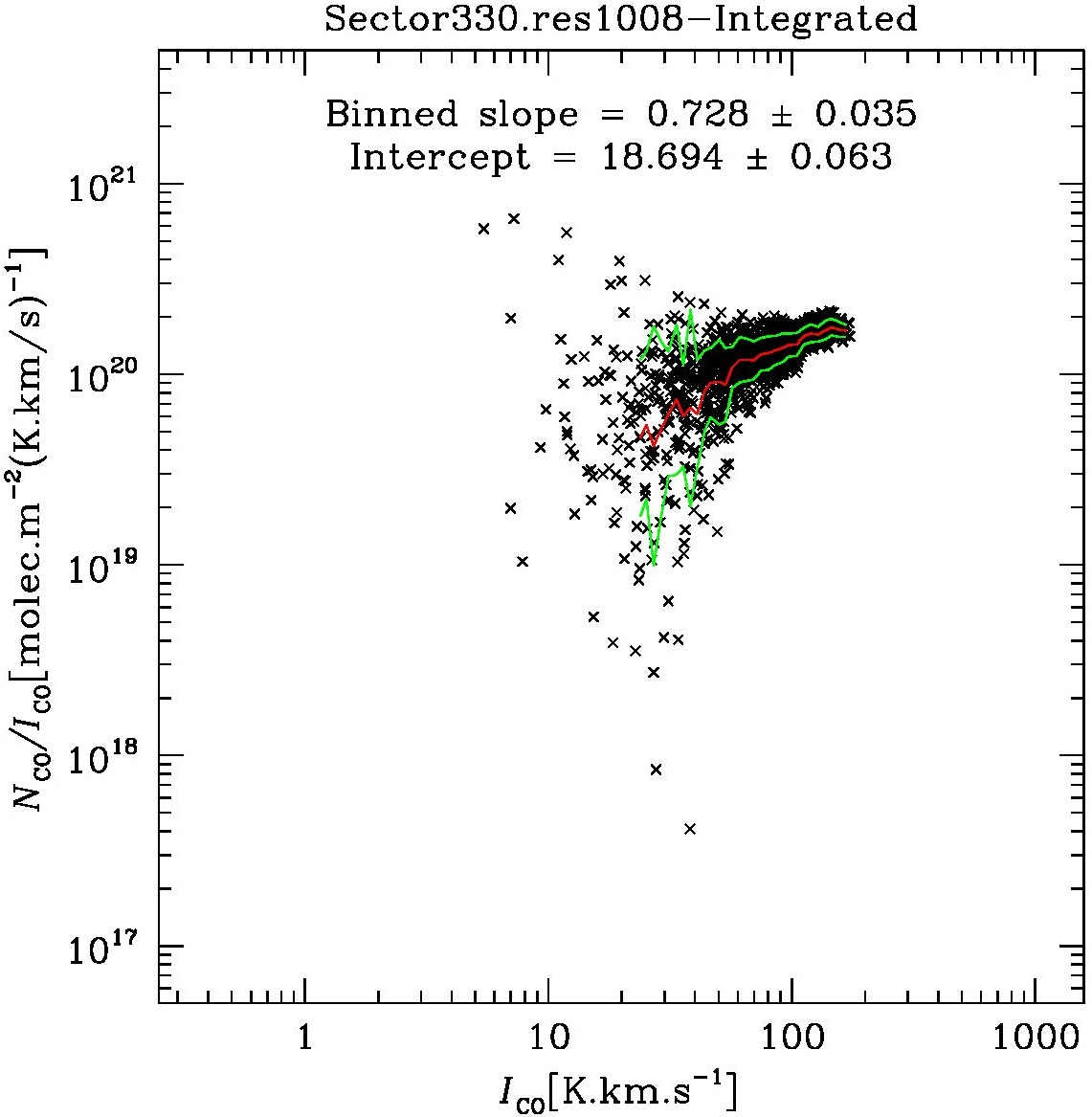}}
\vspace{-1mm}
\caption{\footnotesize Similar plots to Fig.\,\ref{xcl300-12-multi}, but for Sectors 318, 324, and 330 (left, middle, right  columns respectively). $$ $$
\label{xcl318-30-multi}}
\vspace{0mm}
\end{figure*}

\begin{figure*}[h]
\vspace{0mm}
\centerline{\includegraphics[angle=0,scale=0.12]{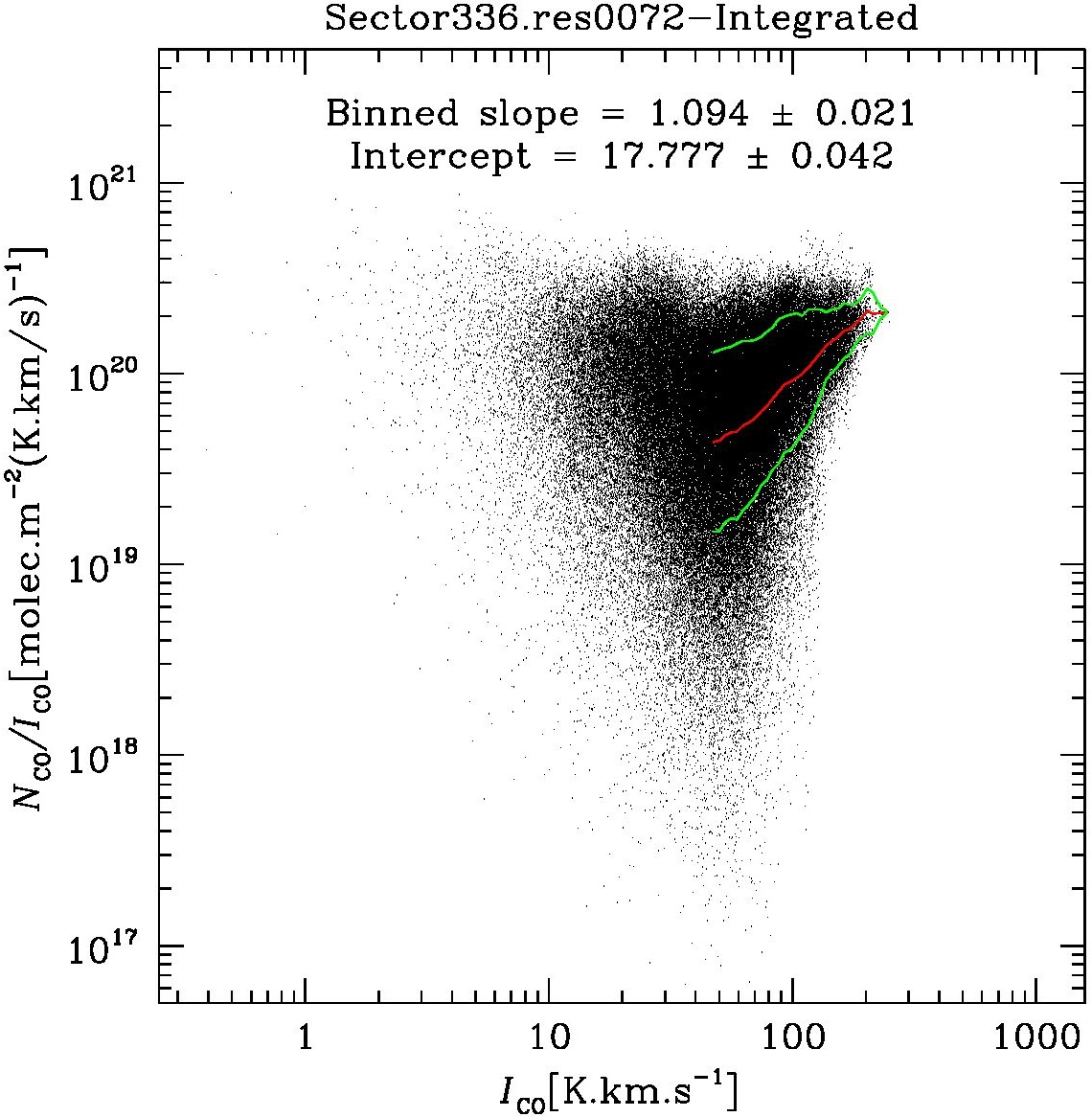} \includegraphics[angle=0,scale=0.12]{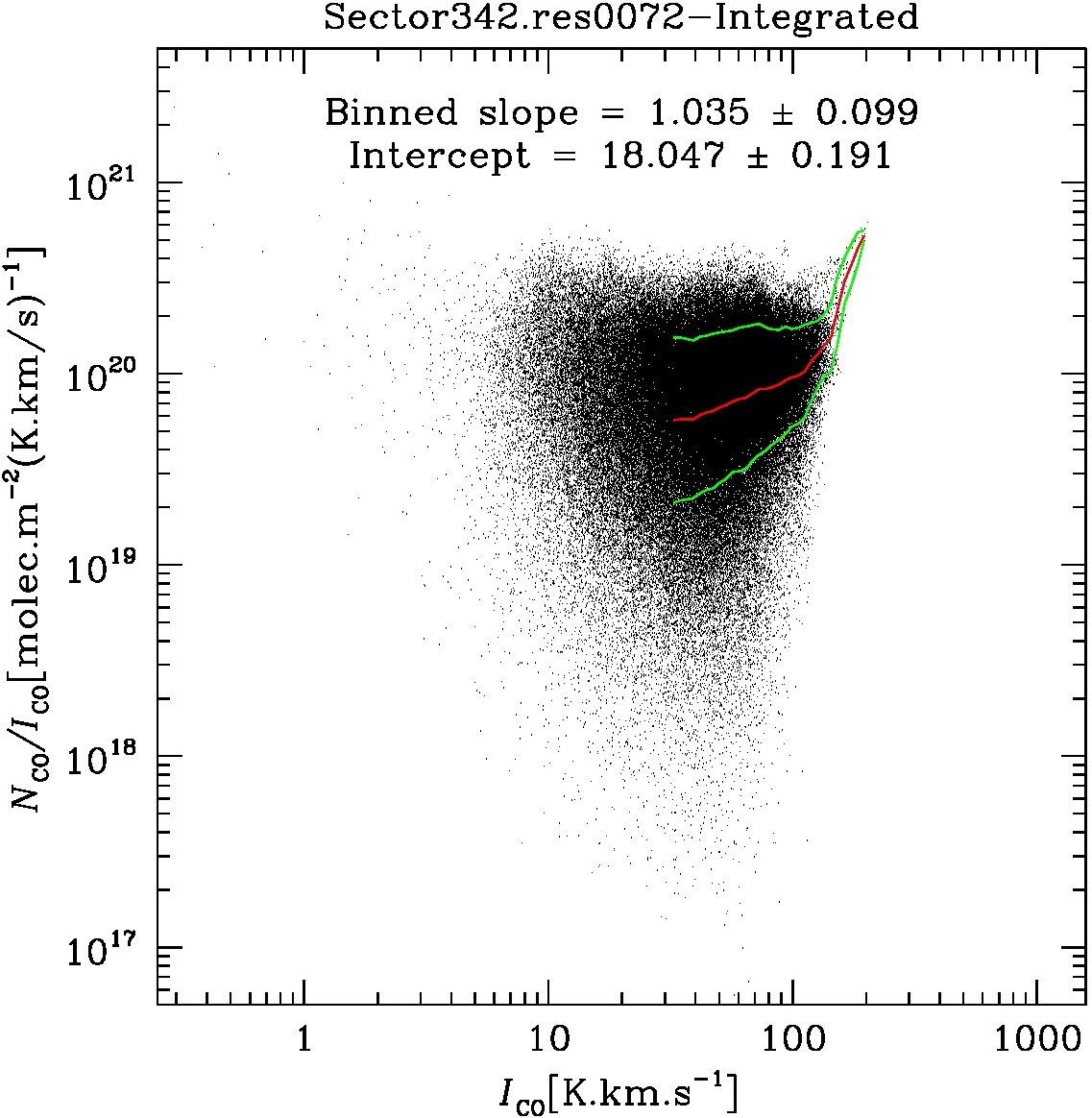} \includegraphics[angle=0,scale=0.12]{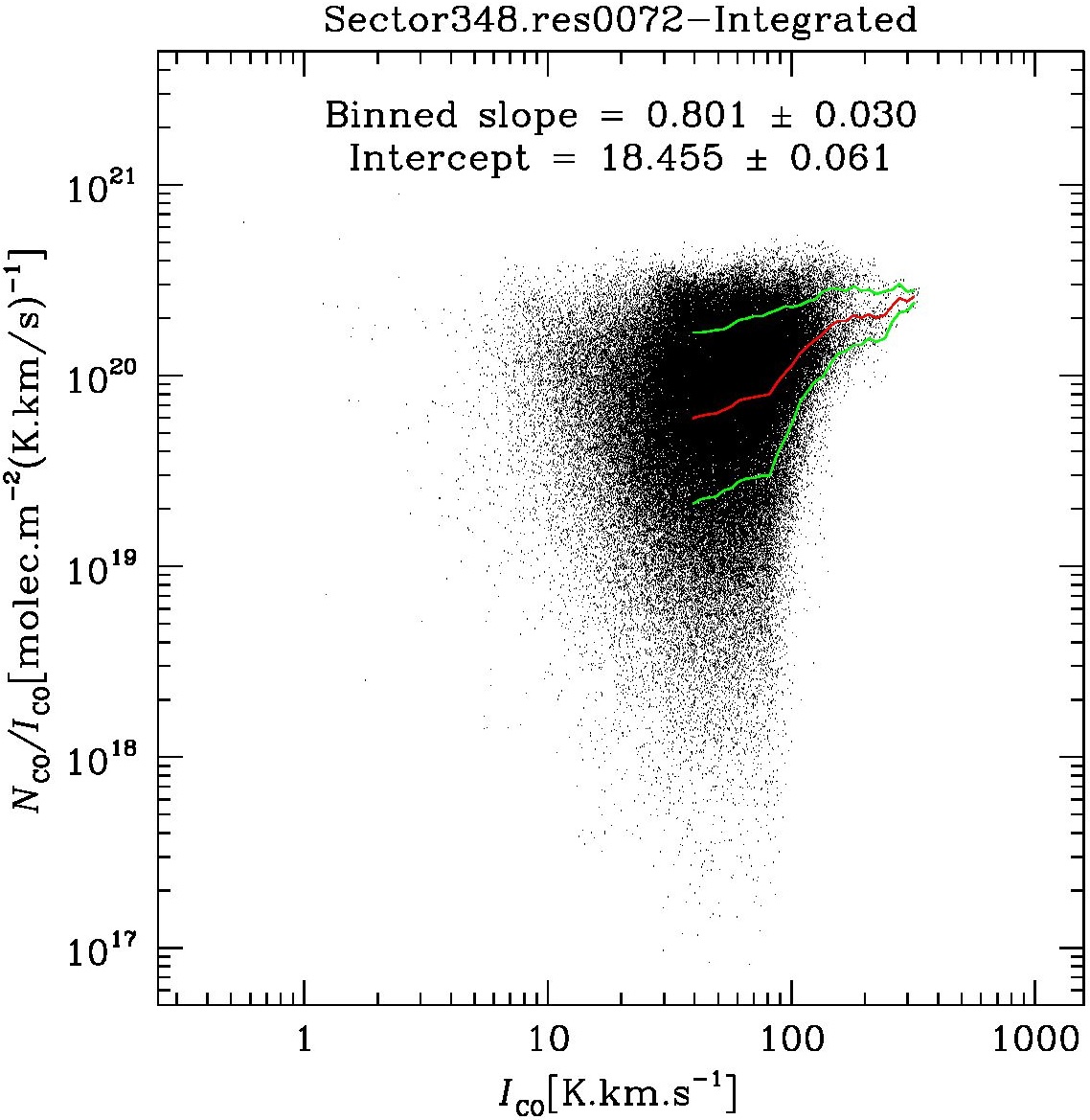}}
\vspace{-3mm}
\centerline{\includegraphics[angle=0,scale=0.12]{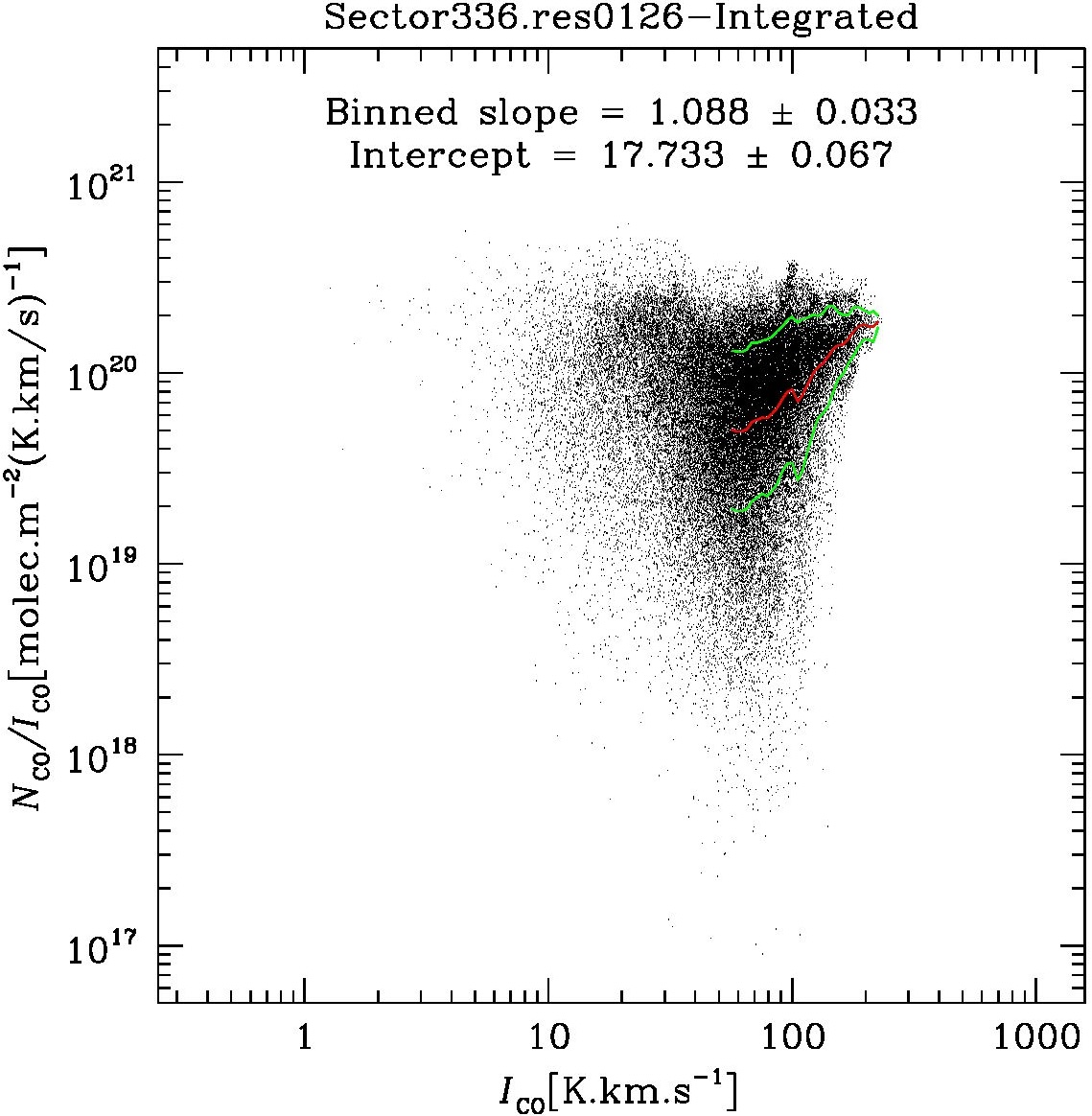} \includegraphics[angle=0,scale=0.12]{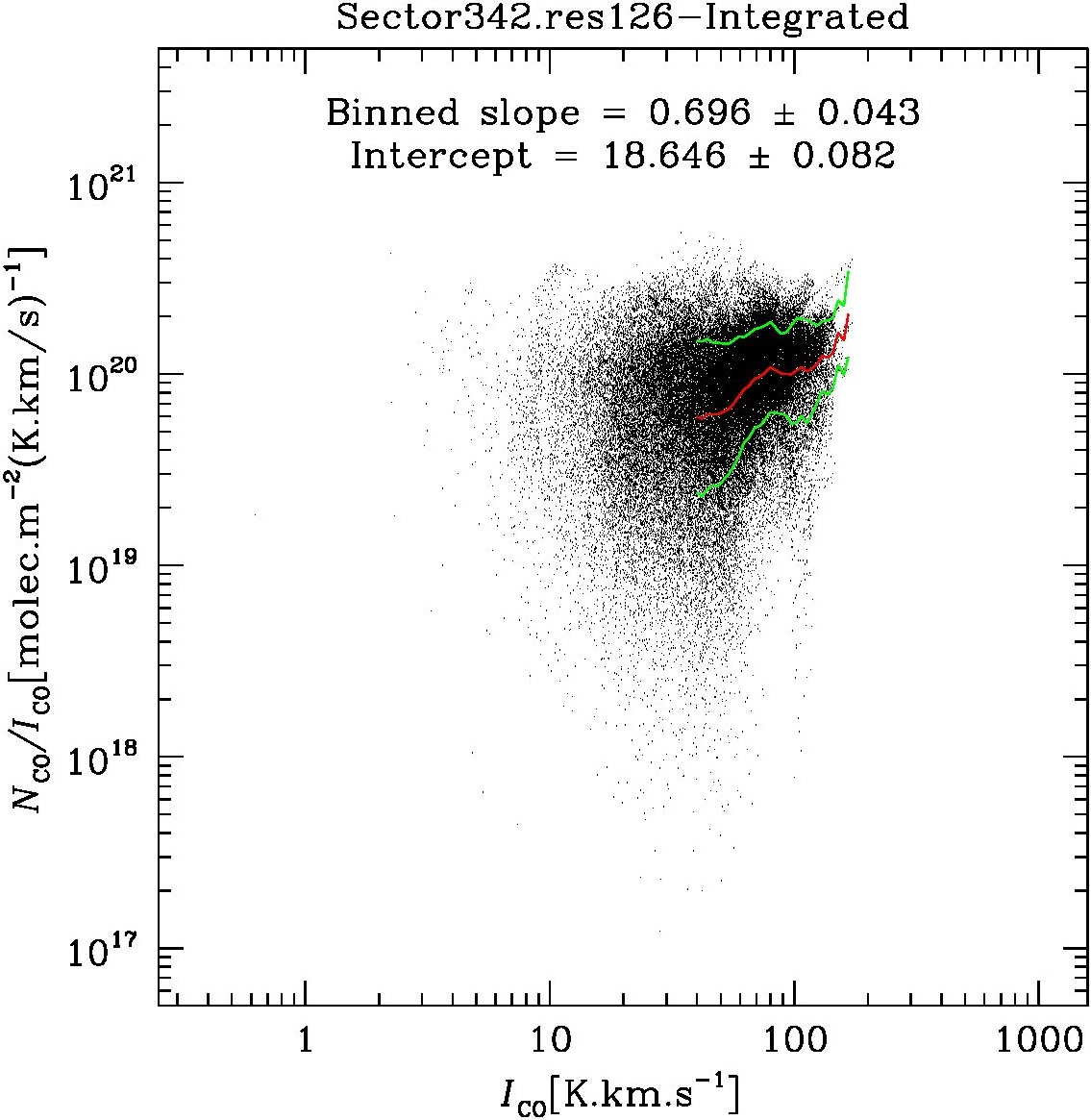} \includegraphics[angle=0,scale=0.12]{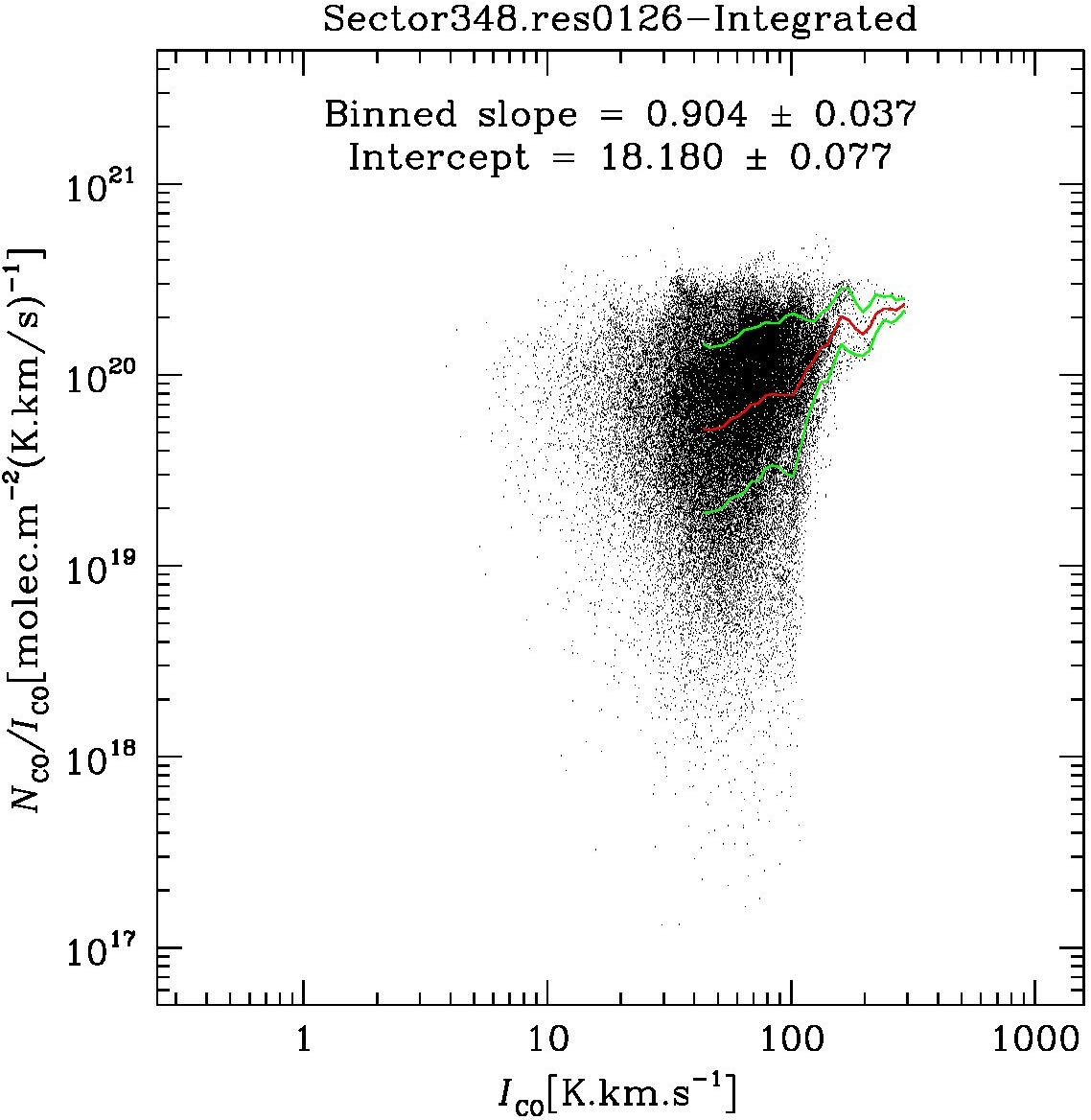}}
\vspace{-3mm}
\centerline{\includegraphics[angle=0,scale=0.12]{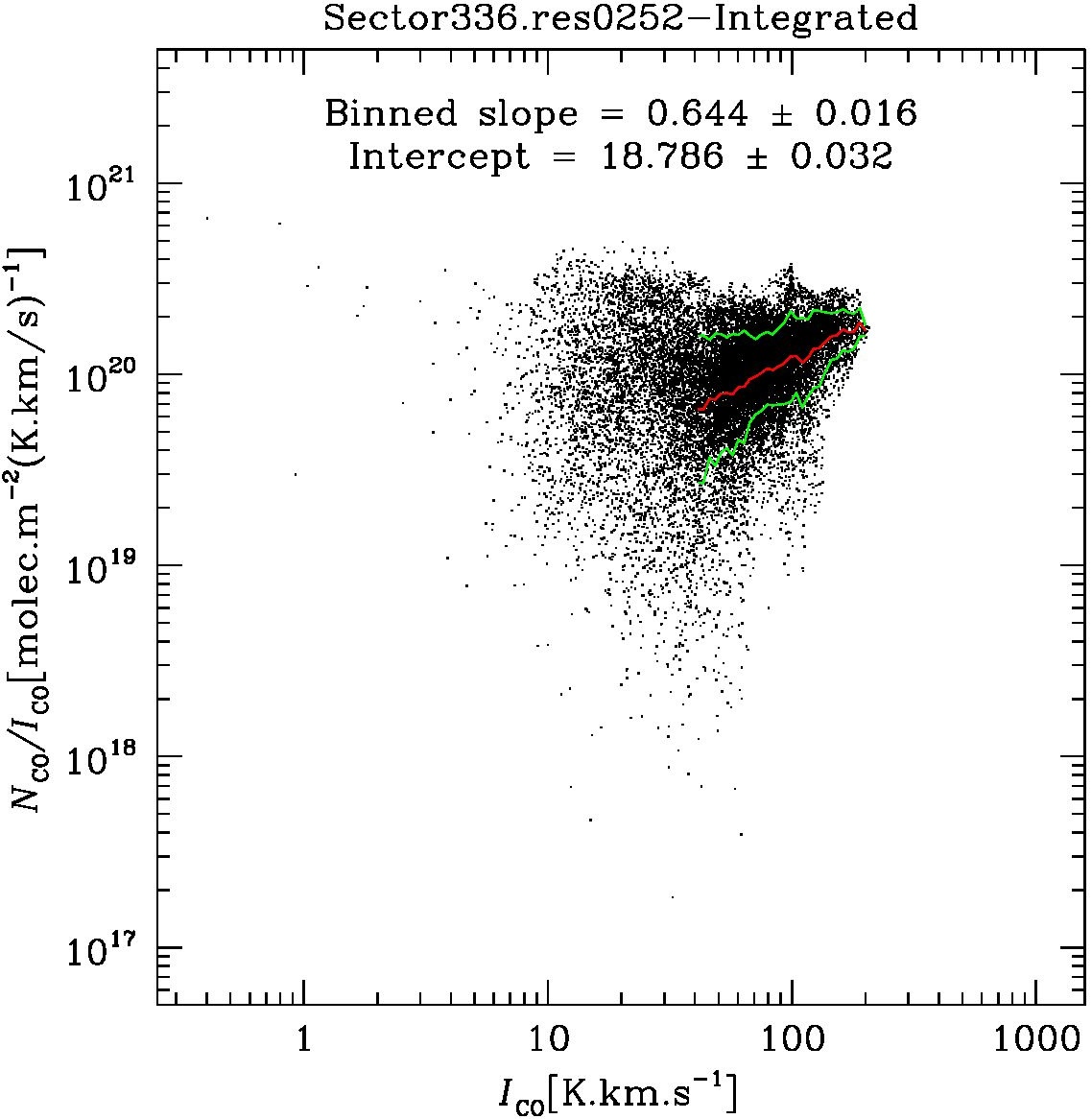} \includegraphics[angle=0,scale=0.12]{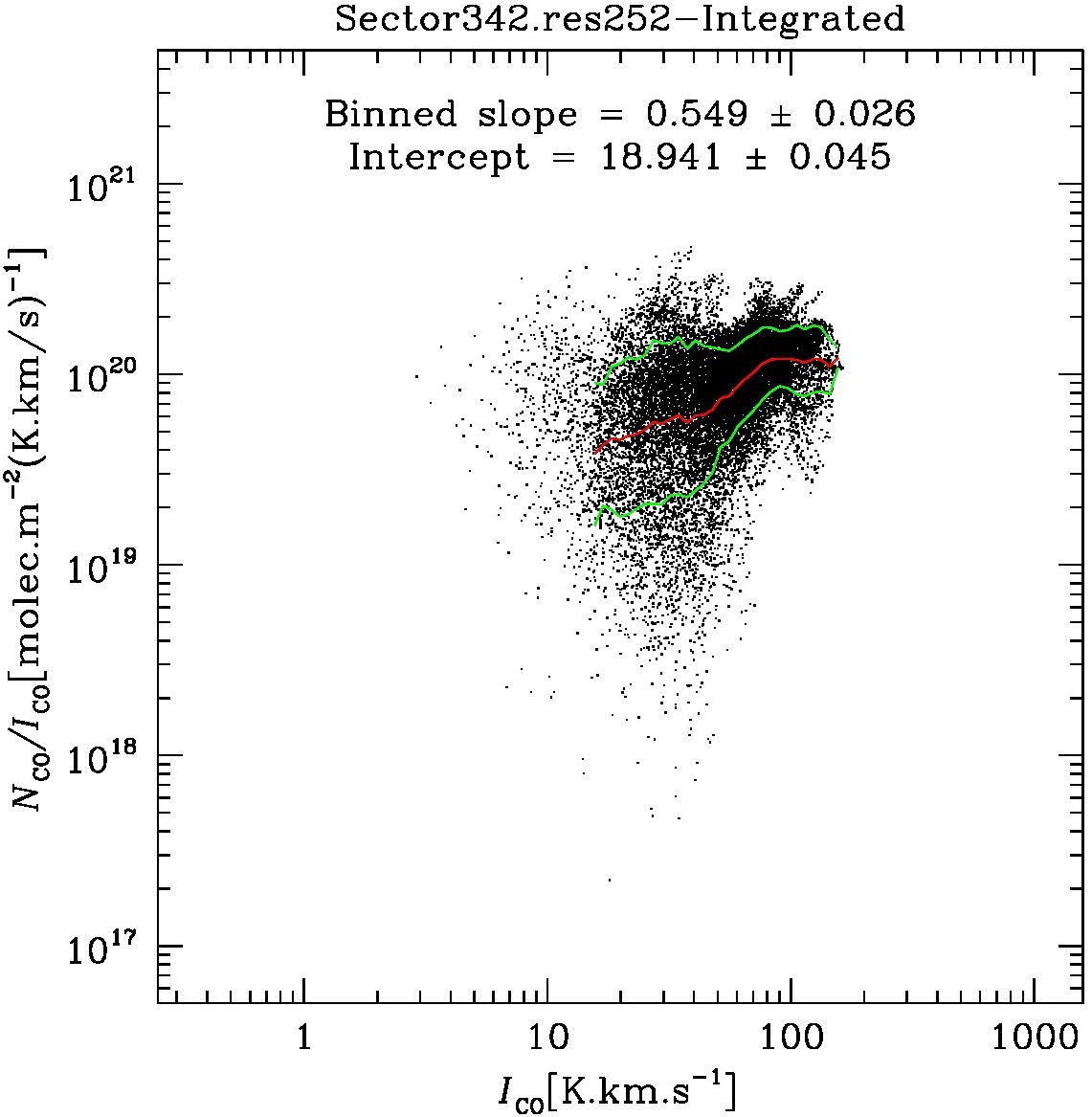} \includegraphics[angle=0,scale=0.12]{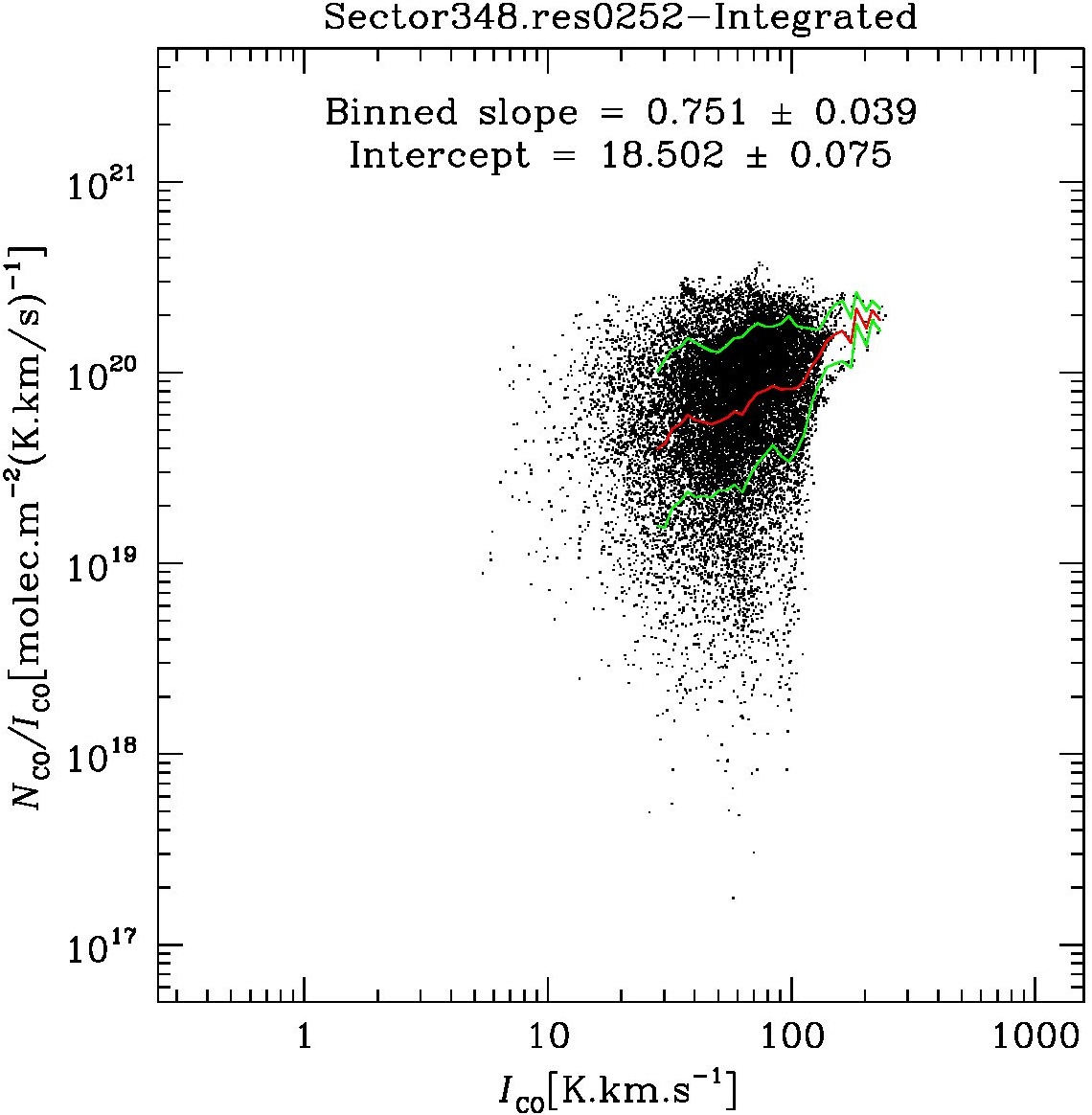}}
\vspace{-3mm}
\centerline{\includegraphics[angle=0,scale=0.12]{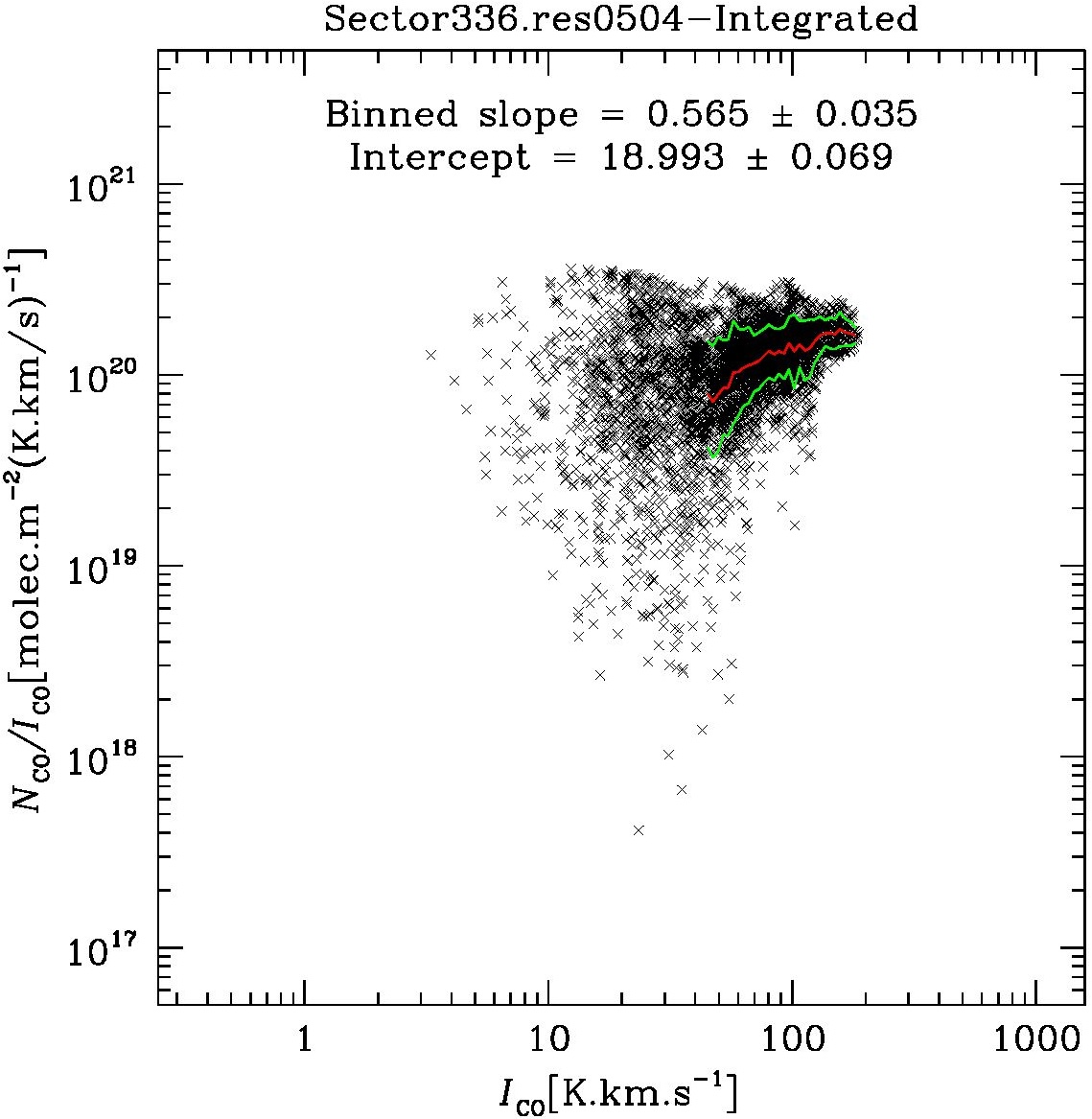} \includegraphics[angle=0,scale=0.12]{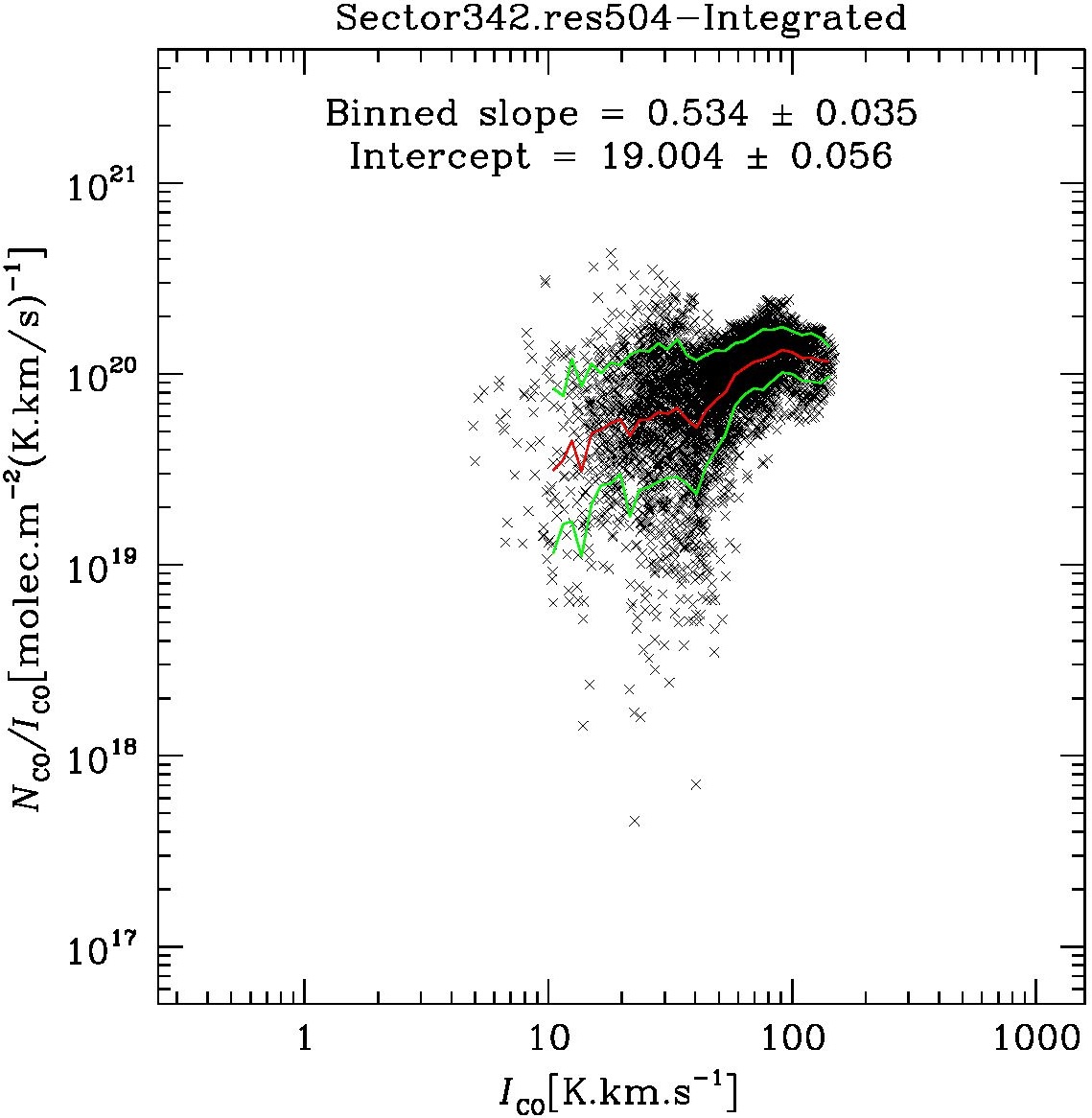} \includegraphics[angle=0,scale=0.12]{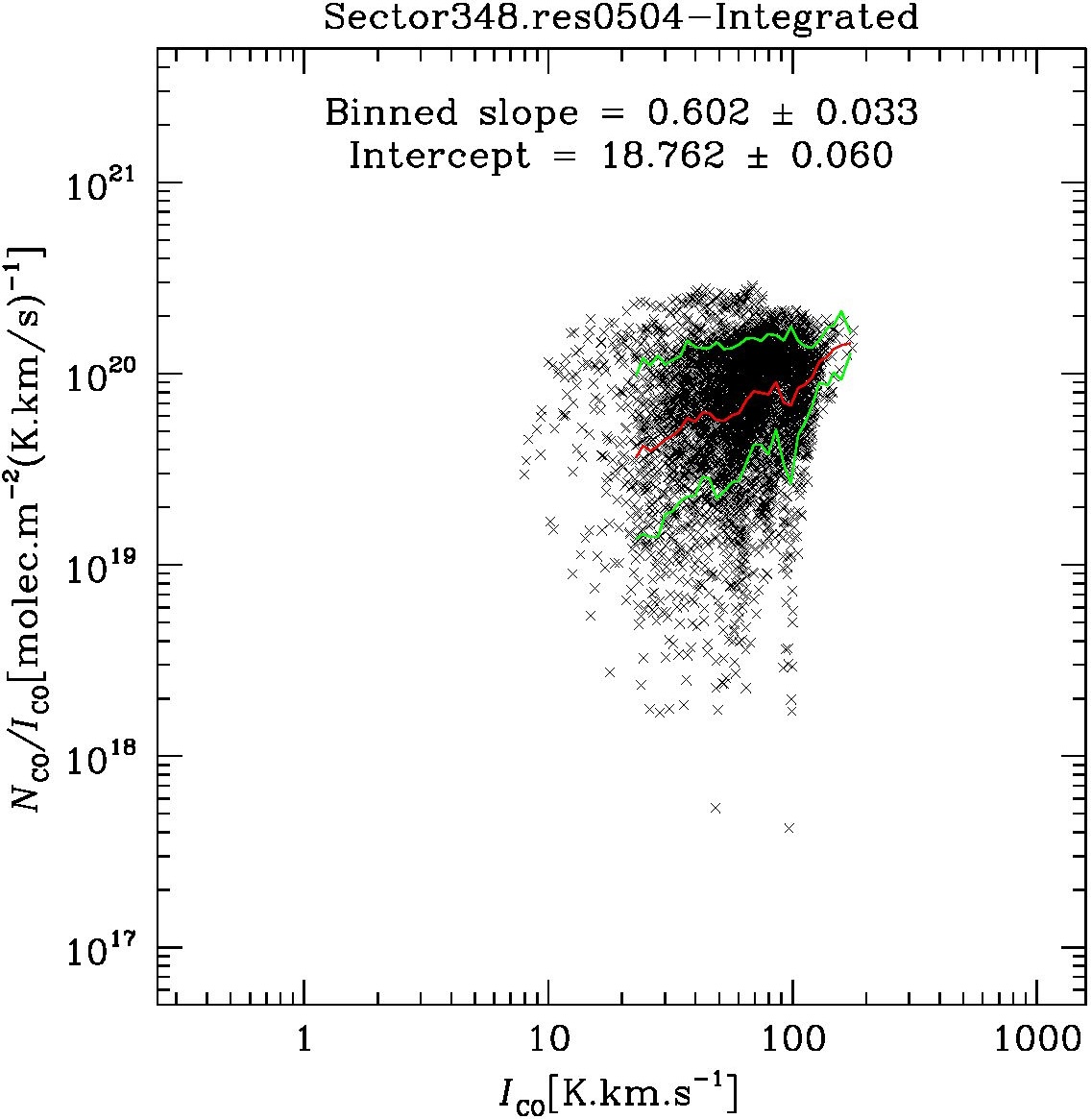}}
\vspace{-3mm}
\centerline{\includegraphics[angle=0,scale=0.12]{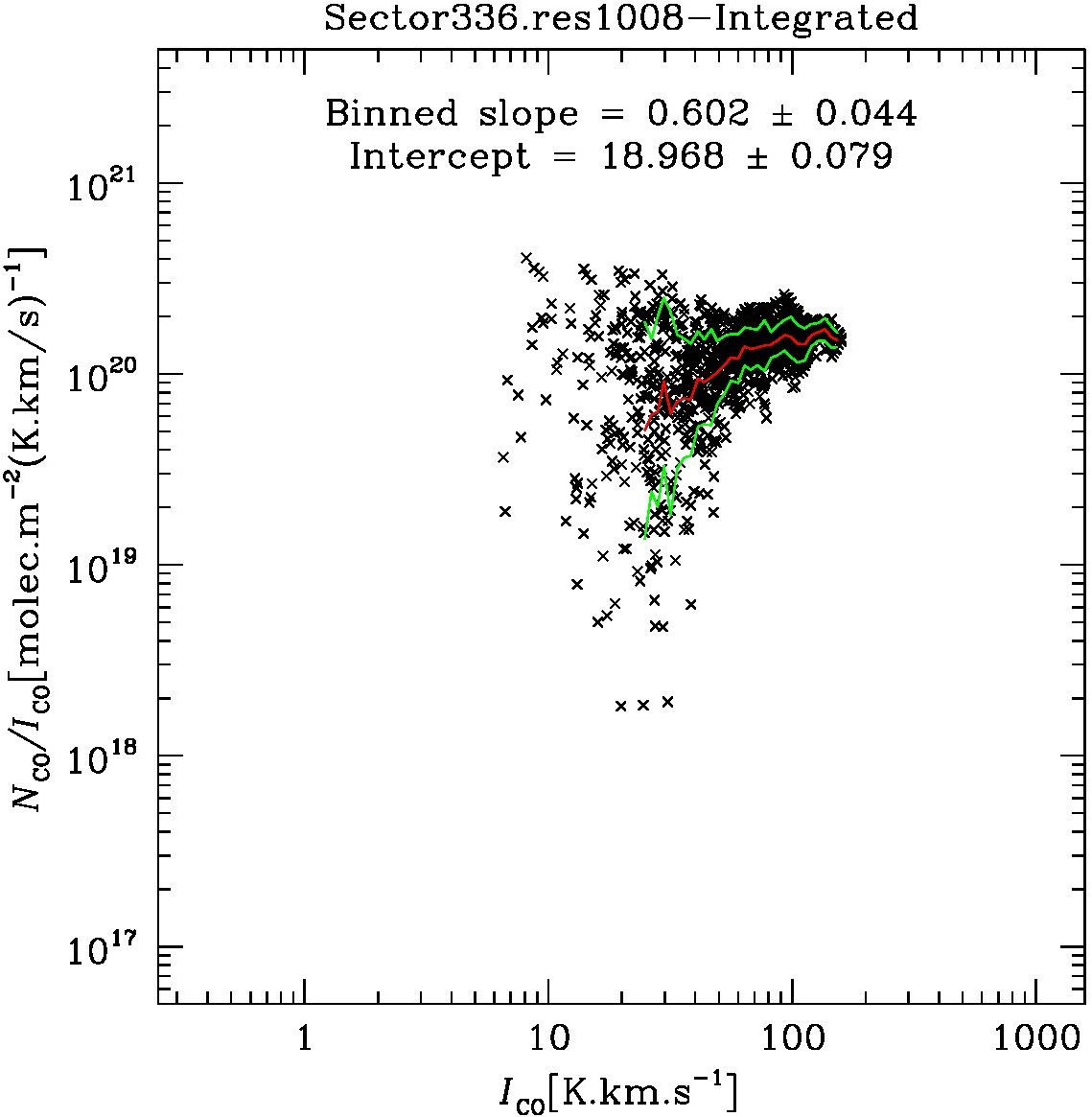} \includegraphics[angle=0,scale=0.12]{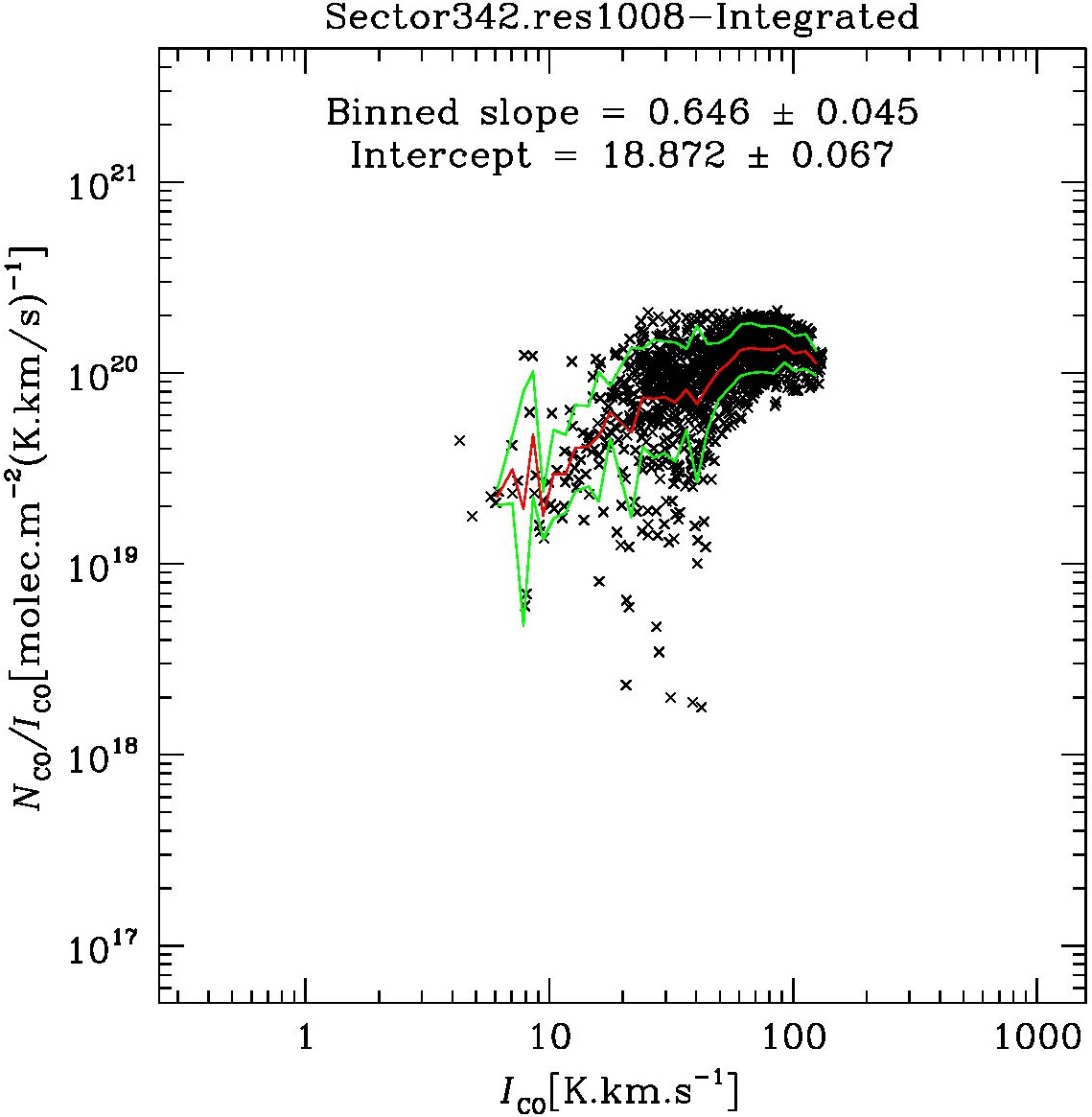} \includegraphics[angle=0,scale=0.12]{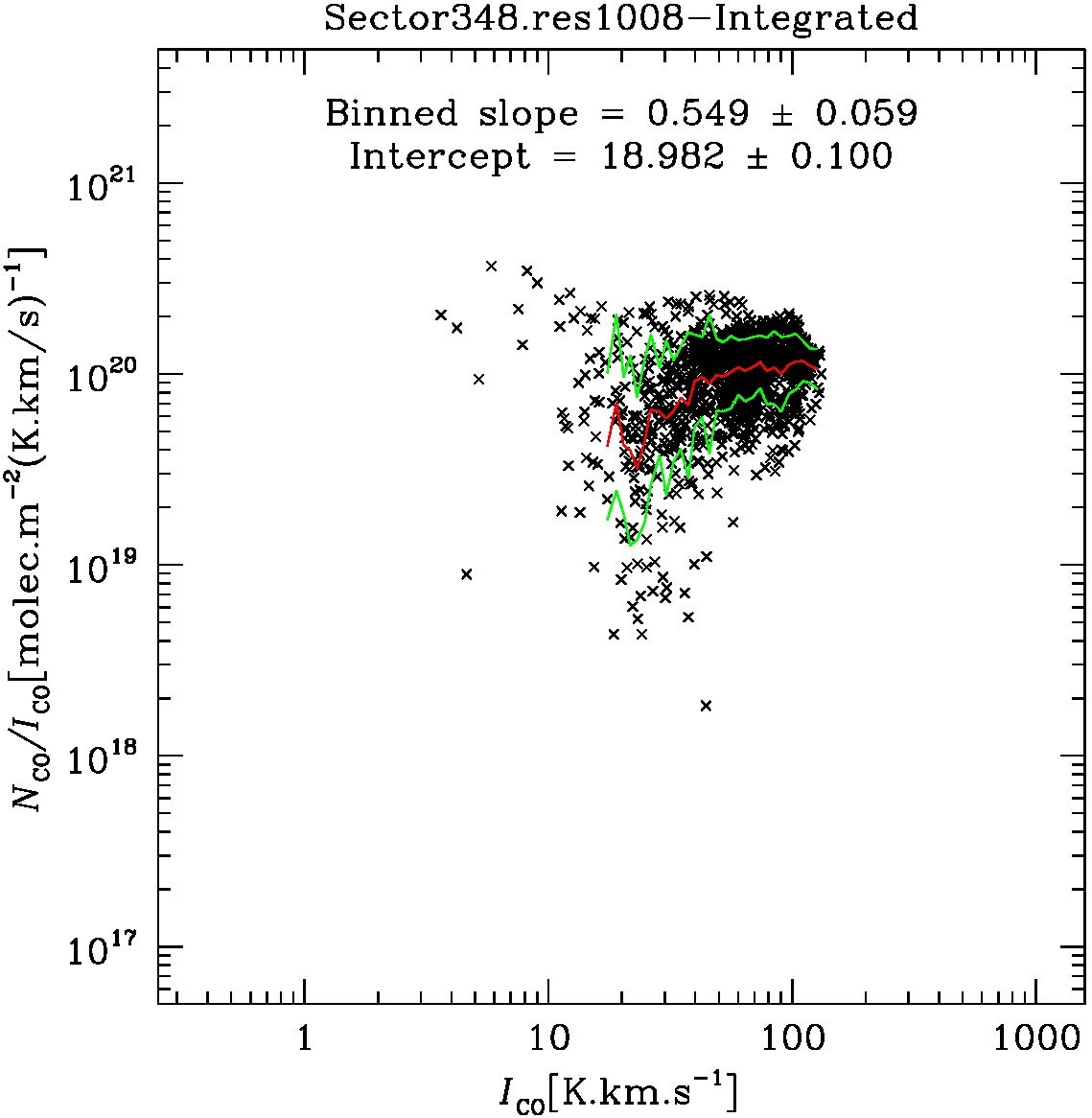}}
\vspace{-1mm}
\caption{\footnotesize Similar plots to Fig.\,\ref{xcl300-12-multi}, but for Sectors 336, 342, and 348 (left, middle, right  columns respectively). $$ $$
\label{xcl336-48-multi}}
\vspace{0mm}
\end{figure*}

\begin{figure*}[h]
\vspace{0mm}
\centerline{\includegraphics[angle=0,scale=0.12]{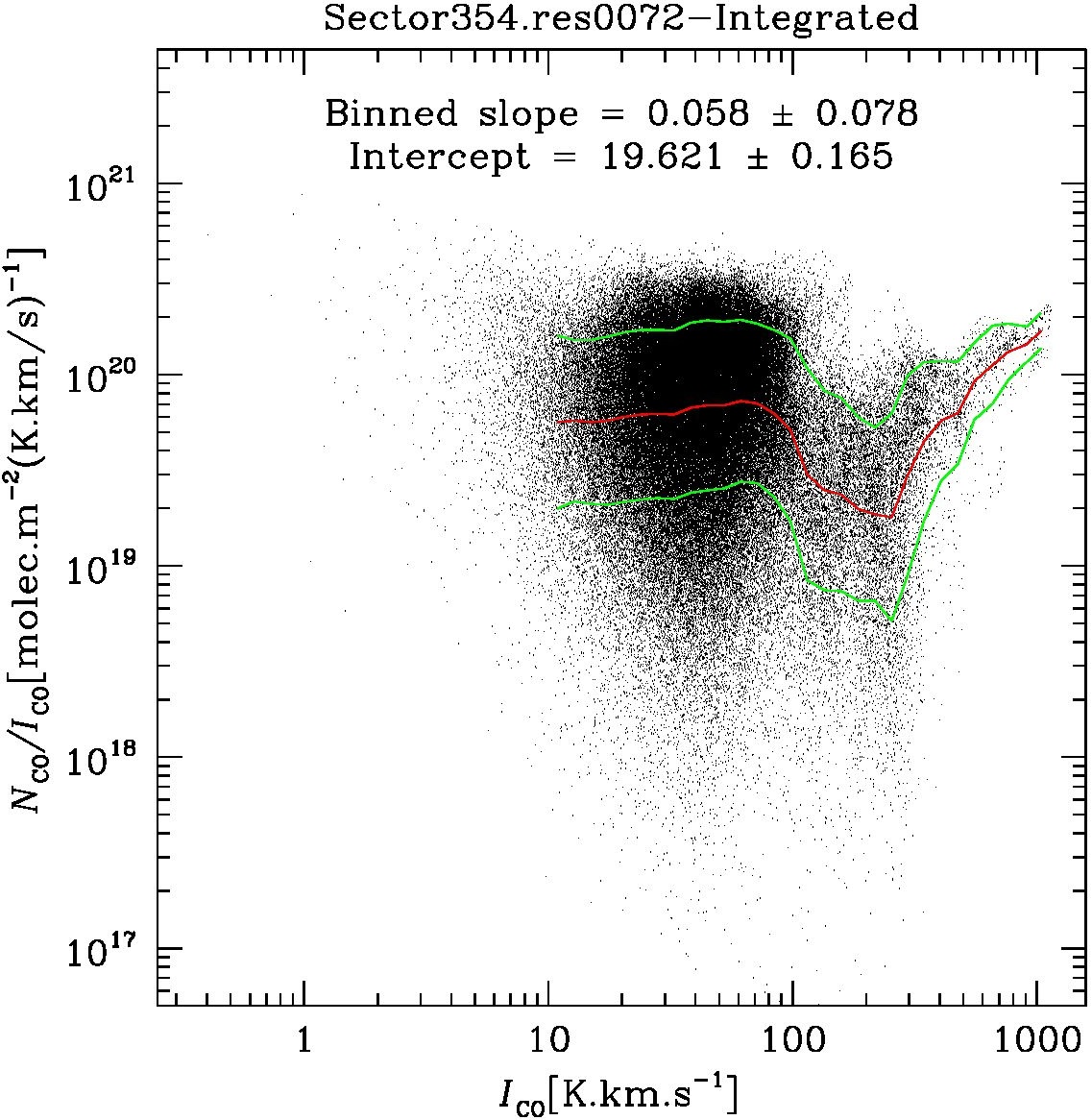} \includegraphics[angle=0,scale=0.12]{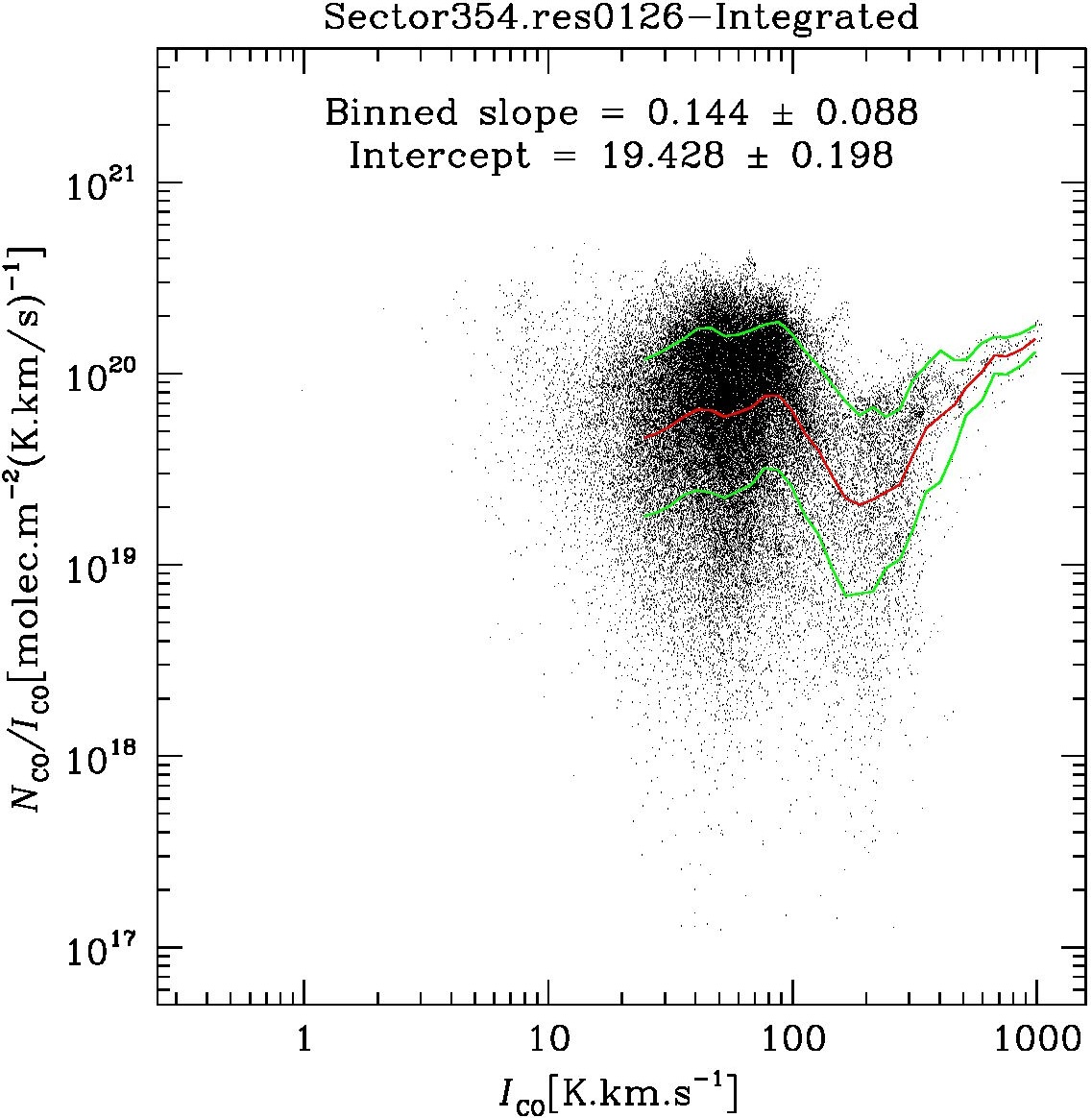} \includegraphics[angle=0,scale=0.12]{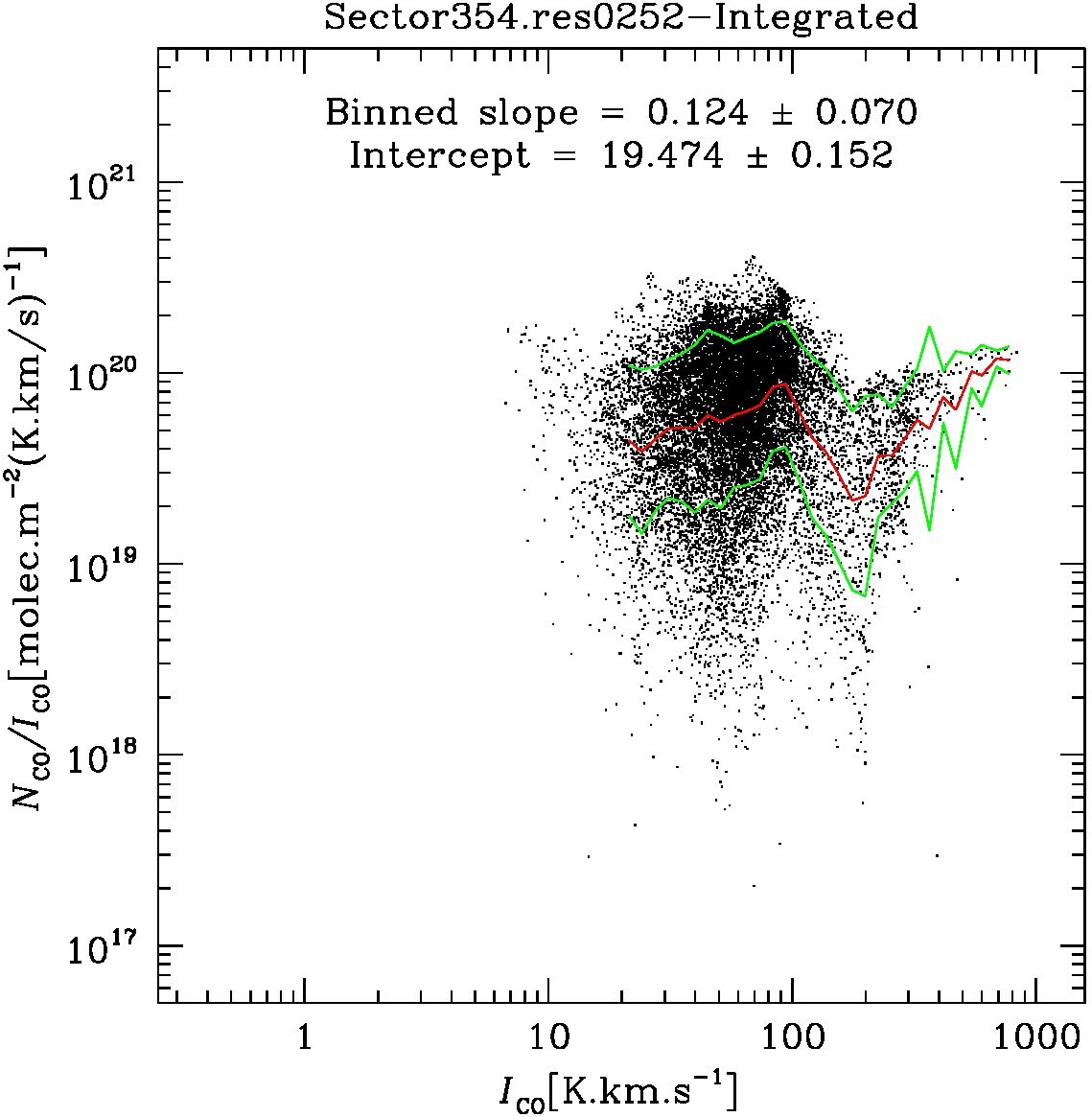}}
\vspace{0mm}
\centerline{\includegraphics[angle=0,scale=0.12]{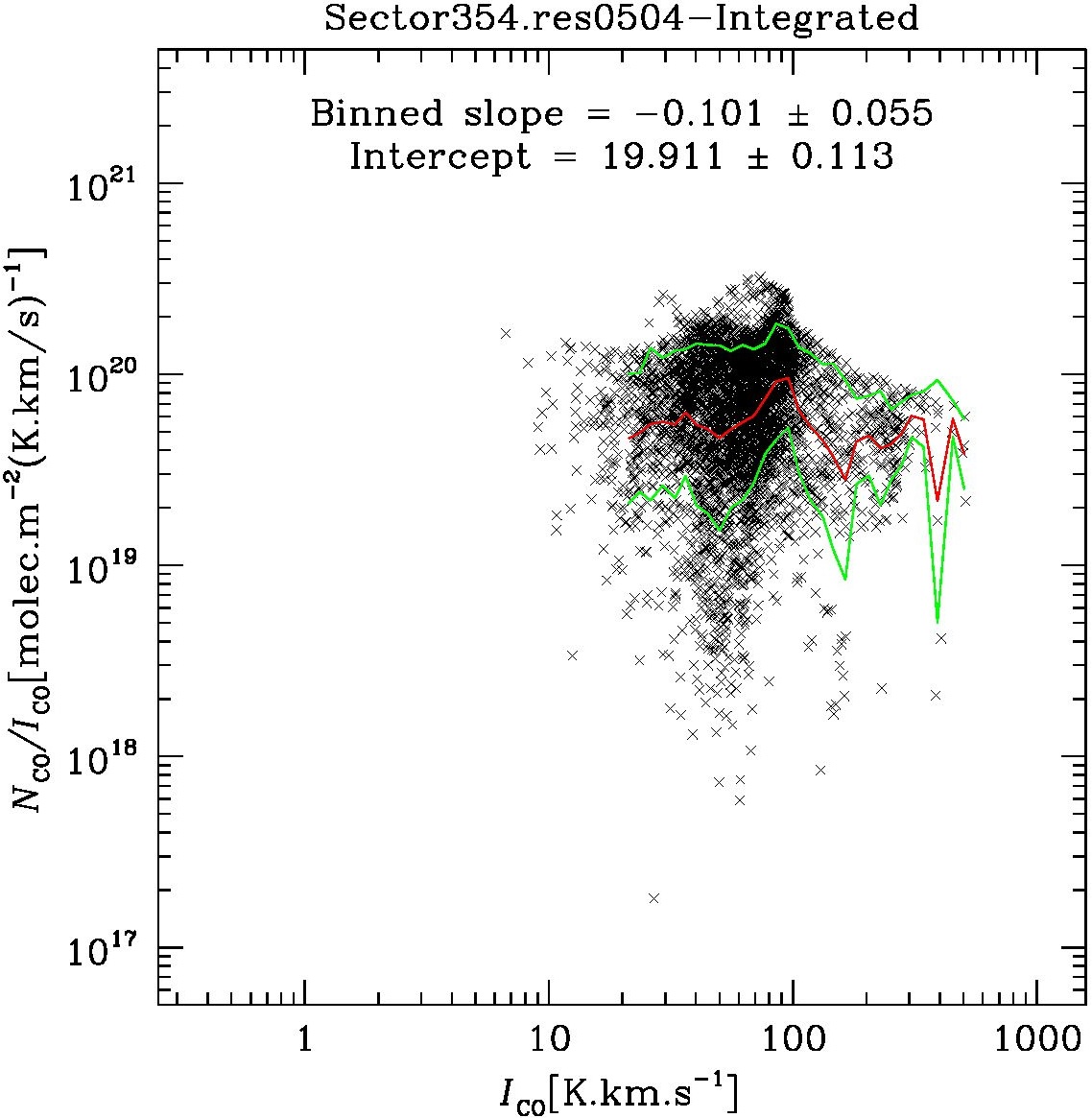} \includegraphics[angle=0,scale=0.12]{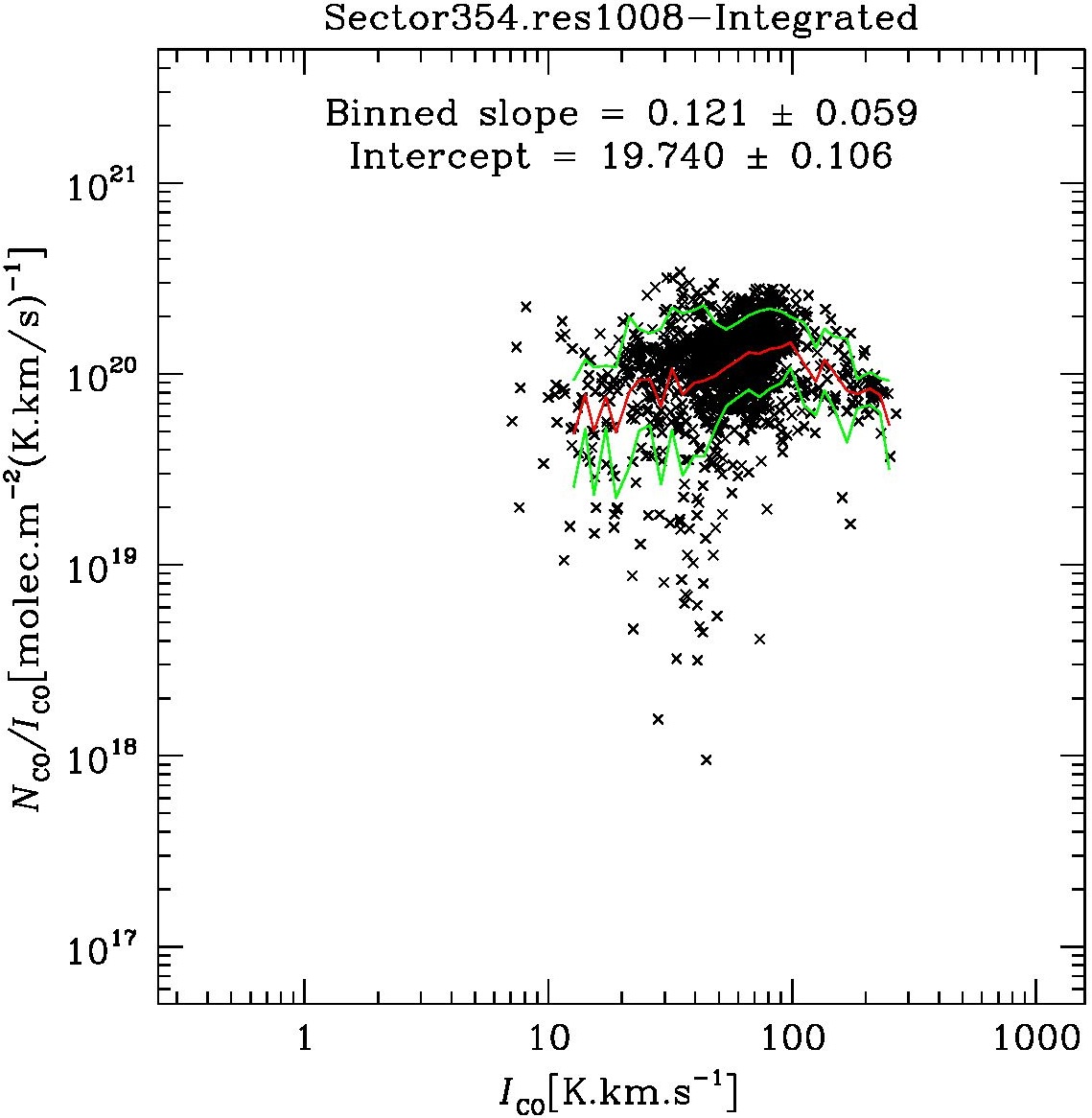}}
\vspace{-1mm}
\caption{\footnotesize Similar plots to Fig.\,\ref{xcl300-12-multi}, but for Sector 354.  The panels are at the same 5 progressive resolutions as in the previous 3 Figures. $$ $$
\label{xcl354-multi}}
\vspace{-1mm}
\end{figure*}

The \nco\ vs.\ \ico\ analysis above is a strict treatment, within the limitations of the assumptions of plane-parallel LTE and $R_{13}$ = 60, since each \lbv\ voxel in each observed species is compared directly to solve for the physical parameter triad (\tnt).  Put another way, each (\itco,\ittco) data pair maps precisely to a point on a 3D (\tnt) surface, represented by the (\tex,$\tau$) grid in Figures \ref{x300}--\ref{x354-multi}.  As a practical matter, however, it is the observed \tco\ integrated line intensity which is most often used in the literature to estimate molecular cloud masses and related quantities.  This is most especially true in the extragalactic domain, where the typical angular resolution and sensitivity limits preclude a per-channel approach, and is common in many large-scale Galactic Plane surveys as well.  In other cases, species like \ttco, \hcop, or HCN are sometimes used as proxies for surface density, star formation rate, or other quantities, but again usually as velocity-integrated quantities.  Thus, to broaden the utility of our treatment, we also computed the equivalent integrated relations as prescribed by \cite{b18}, for each Sector and resolution as before, and these are shown in {\color{red}Figures \ref{xcl300-12-multi}--\ref{xcl354-multi}}.  The velocity-integrated approach gives a less formal, semi-empirical relation because the velocity integration occurs over a different number of channels for each pixel.

% Figure B5: S300--S312 XclvsI

% Figure B6: S318--S330 XclvsI

% Figure B7: S336--S348 XclvsI

% Figure B8: S354 XclvsI

% Figure B9: CHaMP fits summary
\begin{figure*}[ht]
\vspace{0mm}
\centerline{\includegraphics[angle=0,scale=0.27]{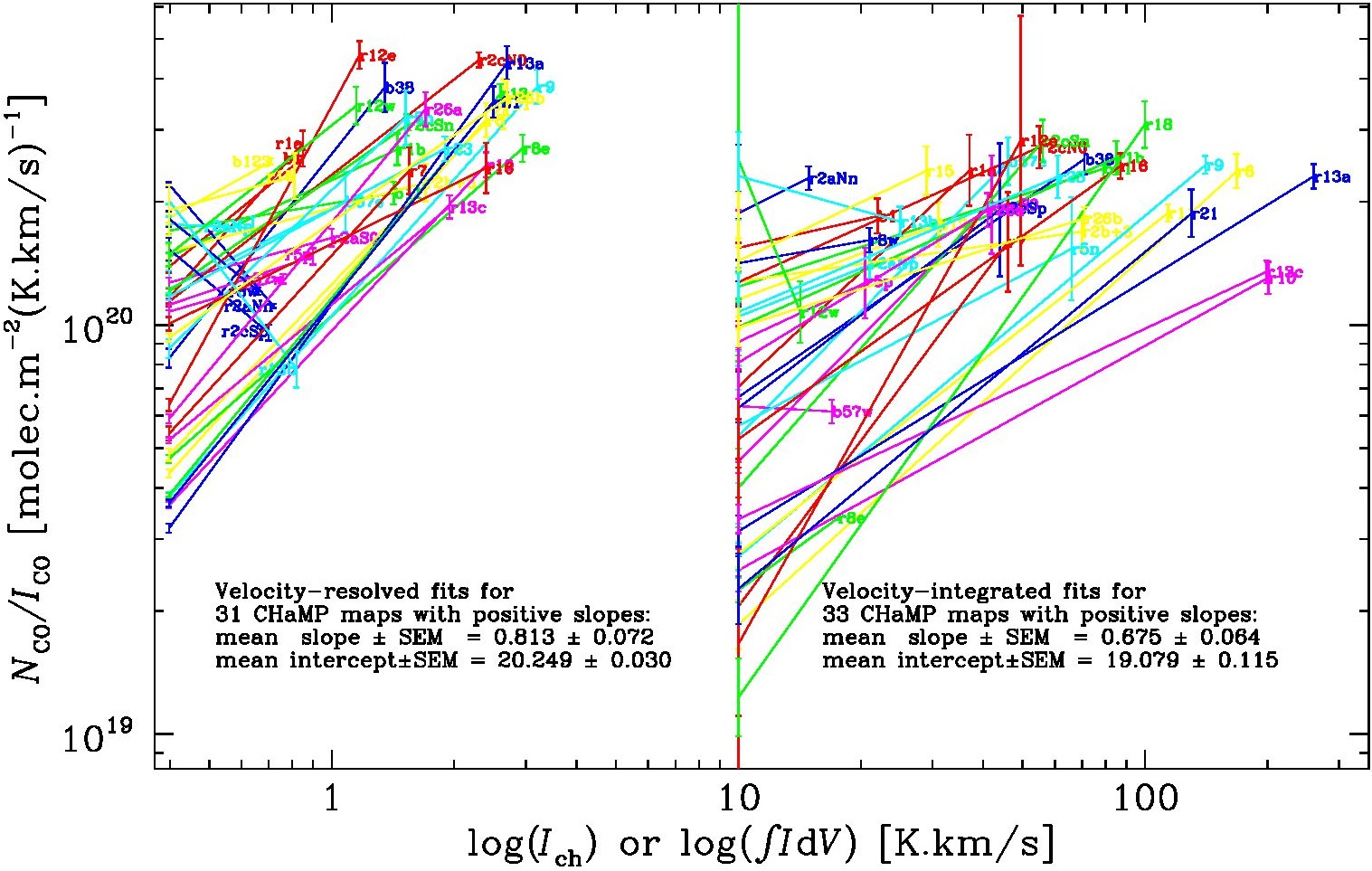}}
\vspace{0mm}
\caption{\footnotesize Direct overlay of all conversion law fits from CHaMP maps on the same scale \citep[from][]{b18}, shown as representative power-laws in log-log space, and labelled by Region or BYF designator.  Because of the velocity integration, the I-fits are naturally located to the right of the V-fits in the $X$=$N$/$I$ vs $I$ diagram.  The lower limits on the $I$ scale among the V-plots are set by the average noise levels in the respective data, while the upper-$I$ limits come from the brightest pixels; the lower $I$ limits among the I-plots are typically where the data runs out, rather than where they become noise-affected.  The error bars in $X$ for both groups are from the fit uncertainties in $N_{0}$ at the lower-$I$ end, and from uncertainties in the slope (= $p$--1) at the upper-$I$ end.  The axis scales are chosen to be square, so that slopes of 1 make a 45\degree\ angle to the axes. $$ $$
\label{chCLfits}}
\vspace{-1mm}
\end{figure*}

\vspace{1mm}For ease of comparison, summary plots of the slopes as a function of Sector and resolution are shown in Figure \ref{slopes}, for both the velocity-resolved and -integrated approaches (hereafter referred to as V-plots and I-plots, respectively).  These make manifest several interesting trends from the individual $X$ vs $I$ panels.  First, there is a clear resolution effect on the fitted power laws in all Sectors except S354, for both the V- and I-plots, in the sense that the fitted $X$ vs $I$ slope decreases as the resolution is degraded.\footnote{Recall from Fig.\,\ref{x300} that the fitted $X$ vs $I$ slope is by definition equal to $p$--1, where $p$ is the index in $N$ = $N_{0}$$I^{p}$ and the normalisation $N_{0}$ is where the fit crosses log$I$ = 0.}  In contrast, Sector 354's individual $X$ vs $I$ plots show a very different distribution of voxels or pixels compared to the other Sectors.  This is entirely due to the presence of the Central Molecular Zone (CMZ) within $\sim$1\degree\ of the Galactic Centre (see Fig.\,\ref{full121318-mom0}), where extremely bright \tco\ emission fundamentally changes the (\tnt) calculation.  There, the much brighter \tco\ than elsewhere pushes the distribution of voxels in the V-plots into a part of the (\tex,$\tau$) grid with lower $\tau$ and higher \tex\ compared to the other Sectors, or even compared to other voxels within S354 but away from the CMZ.  Thus, the $X$ vs $I$ slope {\em and} normalisation are much lower in S354 because the fit is now distorted away from more typical cloud conditions, and this distortion persists through the various convolutions and also in the I-plots.

\vspace{1mm}For the other Sectors, we see that the decrease in the fitted slope is driven by the convolution smearing out the brightest emission in each Sector, so that such areas blend with surroundings that have an intrinsically lower slope and/or a variety of normalisations $N_{0}$.  However, under typical (\tex,$\tau$) conditions, the slope remains distinctly positive (i.e., $p$ is distinctly $>$1) whether one is sampling the brighter or less bright portions of such clouds (the fitting, shown by the solid red lines in the V-plots, is restricted to points above the S/N limits, shown by the cyan curves).  Thus, in both panels of Figure \ref{slopes}, $p$ trends down from $\sim$2 to $\sim$1.5 as the resolution goes from 72$''$ to 1008$''$.

\vspace{0.5mm}The second noticeable result in Figure \ref{slopes} is that the slopes' trend in the I-plots (right panel) is above the trend in the V-plots (left panel), i.e., $p_{I}$ $\approx$ $p_{V}$+0.25 for Sectors 300--348.  This seems to be a bigger difference than the scatter among slope values {\em within} either the V- or I-plots: for example, for the lower 9 Sectors, the V-plot slopes at 72$''$ have mean$\pm$SEM = 1.07$\pm$0.07 while those for the I-plot slopes are 0.82$\pm$0.10, roughly a 3$\sigma$ difference.  At the other resolutions, differences between the I and V slopes are even clearer, with each group $\sim$5--6$\sigma$ apart; collectively, the differences are about 10$\sigma$, so this seems a real effect.

\vspace{0.5mm}This means that there is both an angular-resolution and velocity-resolution effect on the derived conversion laws.  Presumably, at even higher angular resolution, the power-law index $p$ is even larger.  In a related analysis of the CHaMP clouds \citep{b18}, the better-than-Nyquist-sampled data have an angular resolution of 37$''$ (the Mopra telescope's native resolution at 110\,GHz, as opposed to the beam-sampled ThrUMMS data) and used the same spectrometer, so a comparison should be instructive.  While their maps were too small to conduct angular resolution experiments, \cite{b18} did find a consistent velocity-resolution effect on the fitted conversion laws, ranging from the same per-channel analysis as shown in Figures \ref{x300-12-multi}--\ref{x354-multi} through a series of broader binnings in velocity, up to a full $V$-integrated analysis like that in Figures \ref{xcl300-12-multi}--\ref{xcl354-multi}, effectively filling in the gap in velocity resolution between our V- and I-plots.  The extrema of the CHaMP results are representatively shown in {\color{red}Figure \ref{chCLfits}}.  

\vspace{0.5mm}Surprisingly however, when averaged over the 303 pc-scale CHaMP clumps observed at higher sensitivity than in ThrUMMS, \cite{b18} found the opposite relationship, $p_{V}$ $>$ $p_{I}$.  Note that Figure \ref{chCLfits} cites post-facto averaging of their results across 36 maps: their data-aggregated values for a single conversion law index were $p_{V}$ = 1.924$\pm$0.052 and $p_{I}$ = 1.273$\pm$0.019.  We discount the latter value, since as they noted, the individual normalisations $N_{0}$ varied quite widely between clouds, making their aggregated value for $p_{I}$ artificially low.  The average $p_{I}$ in Figure \ref{chCLfits} across their maps is consistent with their results, $p_{I}$ = 1.67$\pm$0.06.  Similarly, our average $p_{V}$ in Figure \ref{chCLfits} across their maps is consistent with their results, $p_{V}$ = 1.81$\pm$0.07, and also with their aggregated index as above.  Then, for CHaMP, we roughly have a relationship $p_{I}$ $\approx$ $p_{V}$--0.2, and are left with the question: Does the value of $p$ go up when integrating the line emission over all velocity channels, like we see in the ThrUMMS data, or go down, as in CHaMP?

\begin{figure*}[ht]
\vspace{-1mm}
\centerline{\includegraphics[angle=0,scale=0.30]{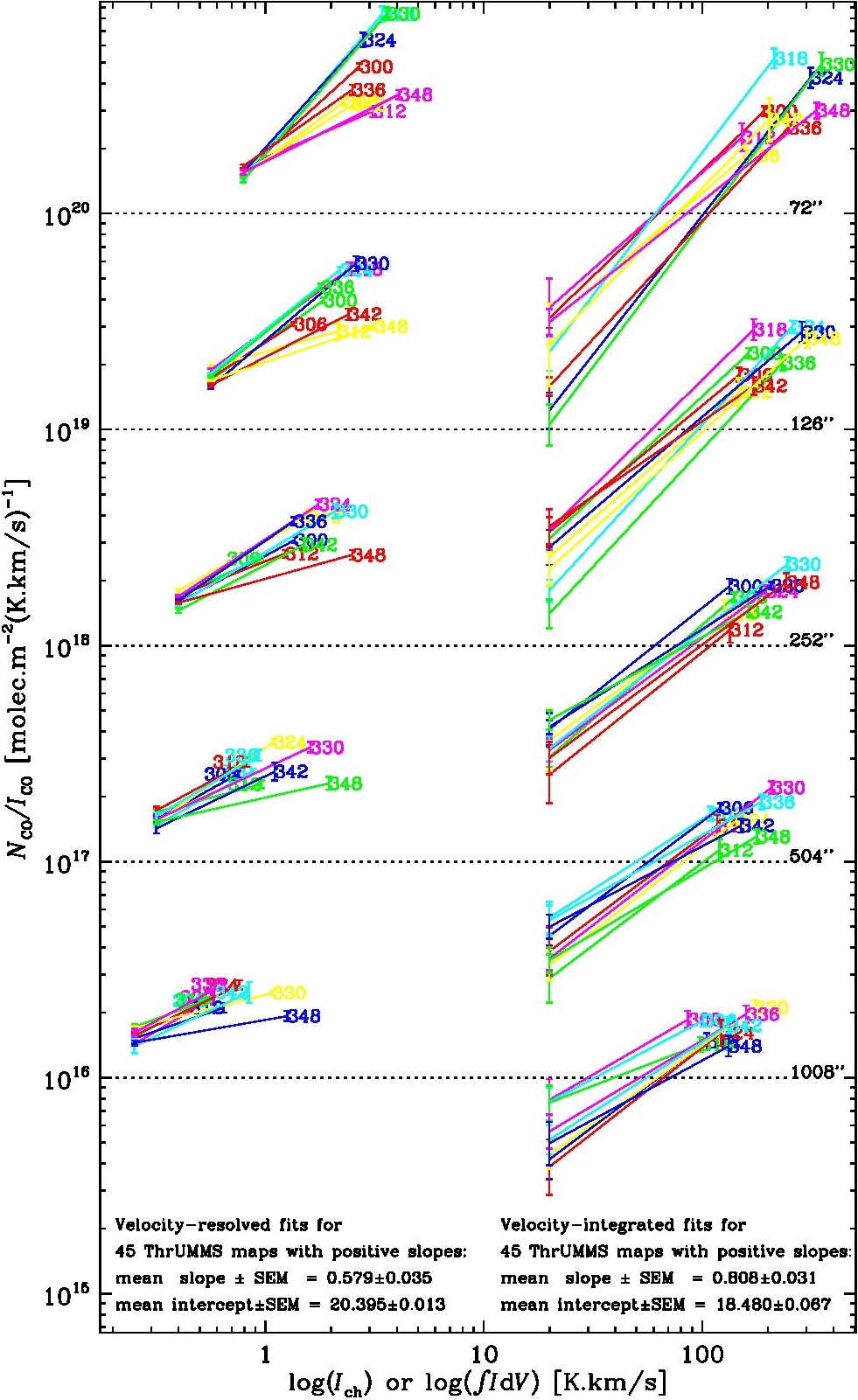}}
\vspace{0mm}
\caption{\footnotesize Direct overlay of all conversion law fits from ThrUMMS Sectors (Figs.\,\ref{x300-12-multi}--\ref{x354-multi} and \ref{xcl300-12-multi}--\ref{xcl354-multi}), but presented similarly to Fig.\,\ref{chCLfits}.  The scale is set for the results from the finest-resolution data (dotted line labelled 72$''$).  For the convolved data, each group (labelled 126$''$, ..., 1008$''$) is successively offset one order of magnitude lower for clarity of display, but they should be understood to lie on the same scale.  The noise limits at the lower-$I$ end for each group at 72$''$ are higher than for the CHaMP data (Fig.\,\ref{chCLfits}), but in the case of the V-fits, these limits go to lower levels with each successive convolution.  The I-fit lower-limits do not decrease because these are more affected by the lack of data, rather than just noise. $$ $$
\label{thrCLfits}}
\vspace{0mm}
\end{figure*}

\vspace{0.5mm}Given that the telescope and the analysis procedures are all identical, the data and results should be equivalent except for the differences already noted.  Therefore, the explanation for this otherwise small discrepancy should lie in these known differences, such as the higher sensitivity of the CHaMP data, or the larger size of the ThrUMMS maps which encompass some of the brightest molecular clouds in the Galaxy.  The different angular resolutions, 37$''$ vs.\ 72$''$, may also make a difference.\footnote{At a distance of 3\,kpc, where the highest mass concentrations lie according to the kinematic analysis in Appendix \ref{kinem}, these beamsizes correspond to scales of 0.54 and 1.05\,pc.}  To investigate this, we overlay the ThrUMMS conversion law fits in {\color{red}Figure \ref{thrCLfits}} to mimic the plots in Figure \ref{chCLfits}.  % 37''x3000pc = 111,000 au =

\vspace{1mm}Considering first the V-plots, we can readily see how the more sensitive CHaMP data (Fig.\,\ref{chCLfits}) allow us to discern an intrinsically steeper conversion law down to fainter levels than are accessible in the ThrUMMS maps.  The higher noise levels in the latter constrain the portion of the (\tex,$\tau$) grid that can provide reliable solutions, producing a kind of ``pinching'' effect at the low-$I$ end of each fit, and flattening the fits overall in Figure \ref{thrCLfits}.

\vspace{1mm}At the same time, the individual CHaMP I-plots \citep[see][their Appendix A]{b18} suggest that, especially in the fainter clumps, there is an inherent flattening of the conversion law to smaller slopes at lower $I$.  Indeed, the (\tex,$\tau$) grid itself becomes distorted and compressed at low \tex\ or $\tau$ (see Fig.\,\ref{x300}), effectively {\em requiring} the conversion law to flatten at low $I$, with sufficient sensitivity.  This means that we should expect the I-plots in sensitive maps to have smaller slopes and indices $p$ than in the respective V-plots, especially of fainter clouds whose gas conditions lie predominantly in the distorted part of the (\tex,$\tau$) grid.

\vspace{1mm}In contrast, for the ThrUMMS maps we see that the brightest emission in each Sector will preserve the higher-slope behaviour at the highest $I$ levels, and this would be a more dominant effect in the higher-dynamic-range $I$ maps.  But also, because each Sector contains some bright emission and there are no Sectors with only-faint clouds, we lack a sampling of lower-slope data in the I-plots, giving a larger average slope/$p$ than for the ThrUMMS V-plots

\vspace{1mm}In summary, we argue that the more sensitive CHaMP maps provide the physically more accurate trend, where the fitted conversion law slope/index is more a accurate characterisation of the radiative transfer in a V-plot of sensitive data, but does not provide practical general $I$$\rightarrow$$N$ conversions.  In the same maps, the value of the slope/index necessarily drops in an I-plot of sensitive data, but such plots do provide a more practical way to apply the conversion law fits to general molecular line map data.  The ThrUMMS fits, however, give somewhat depressed values for the true slope/index of the inherent conversion law in V-plots, due primarily to their somewhat lower sensitivity, and these V-plot fits are made even less accurate as the resolution becomes poorer.  The ThrUMMS I-plots, however, are somewhat immune to the sensitivity limitations especially at higher resolutions, since they have high dynamic range and preserve the intrinsic radiative transfer behaviour over practical line brightness ranges.

% Figure B10: ThrUMMS fit summaries

\vspace{1mm}While the above arguments may seem a bit squishy, the index values in the preferred CHaMP V-plots and ThrUMMS I-plots at 72$''$ are actually quite close: 1.81$\pm$0.07 or 1.92$\pm$0.05 depending on how the CHaMP data are averaged, and 2.07$\pm$0.07 for ThrUMMS.  A simple average of the CHaMP-aggregate \citep{b18} and ThrUMMS values is used below.  The normalisation must be taken from the I-plots, since the V-plots are inherently connected to the velocity resolution in the data, a function of the spectrometer only and not intrinsic to clouds many kpc away.  The average ThrUMMS I-plot 72$''$ value among the lower-9 Sectors (i.e., not including S354) is then log$N_{0}$$\pm$SEM = 17.96$\pm$0.15.  %CHaMP V-plot value inappropriate  
Our overall recommendation for a general conversion law is then %% 17.959112 +- 0.446312 ==> 9.10148e17 */ 2.79455, cube-root=1.40854
\begin{eqnarray}
	N_{\rm ^{12}CO} & = & N_{0}I_{\rm ^{12}CO}^{p}~~,~~~~~{\rm where}					\nonumber	\\
	N_{0} & = & (9.10^{\times}_{\div}1.41)\times10^{17}\,{\rm molecules\,m}^{-2}~~~~~{\rm and}				\\			% EQ.B1
	p & = & 2.00\pm0.07 														\nonumber
\end{eqnarray}
for parsec-resolution molecular line data (specifically, for \tco), valid over a range of \itco\ = 20--350\,K\kms.  To this must be added a factor for the gas-phase abundance of \tco\ relative to \htwo, widely taken to be  $R_{12}^{-1}$ = 1$\times$10$^{-4}$, but probably somewhat lower at 0.6$\times$10$^{-4}$ when averaged over a wide range of gas conditions \citep[see, e.g.,][]{p21}.  The main environmental dependency of $R_{12}$ seems to be on the dust temperature $T_{d}$, with a peak $R_{12}^{-1}$ at $T_{d}$ = 20\,K.  This dependency in turn likely reflects the local physics of the gas: e.g., at lower $T_{d}$, there is probably extensive freeze-out of CO onto dust grains; at higher $T_{d}$, higher irradiation likely is contributing to CO dissociation.

%One must also note that, in any of these cases, the trend is often dominated by the brightest emission in a given map, whether a 5$'$-square, high-sensitivity CHaMP map of an individual clump, all the way to a multi-degree expanse from ThrUMMS covering a large GMC.  And it is in the brightest emission that the power-law index seems to be steepest, consistently near $p_{\rm bright}$=2.  In the fainter clouds, the index does seem to drop somewhat closer to linearity, but remains consistently above it with a typical $p_{\rm fainter}$=1.6.

%%%%%%%%%
%   Section B3  %
%%%%%%%%%
\subsection{Broader Implications of New Conversion Laws}\label{implics}
We see strong support for these new conversion laws from several other issues generally related to molecular cloud masses.  First, it is well-known that the high \tco:\ttco\ abundance ratio, and hence high opacity ratio of their lines, will likely distort an LTE calculation.  Specifically, radiative trapping of the \tco\ photons means that, in general, it is unlikely to be sampling exactly the same gas volumes as the \ttco\ emission, even per voxel.  In the presence of temperature stratification of clouds from (e.g.) external irradiation, this could produce $X$vs$I$ solutions that are biased to higher-\tex\ and lower-$\tau$ loci, effectively flattening our derived $p$ solutions.  On this basis alone, our $p$ values are very likely to be lower limits to the true radiative transfer solutions.

\vspace{1mm}Second, in some early work, it was argued that a single $X$ factor was consistent with the apparently virial balance (i.e., between gravity and internal turbulence/magnetic support) of molecular clouds \citep[e.g., see][and references therein]{bwl13}.  Later however, it became apparent that on clump scales, molecular clouds are not generally in virial balance \citep{bm92,b11,b16}.  Instead, it was found that, for ensembles of clouds or clumps, the virial-$\alpha$ (the ratio between a mass required for virial balance and the actual mass measured by other means) is a broadly decreasing function of cloud mass, with $\alpha$$\sim$10 or more for lower-mass clumps (unbound by gravity) and only dropping to $\sim$1 for masses \gapp\ 10$^{3}$\,M\solar.  While the \cite{bm92} interpretation was that this was evidence for pressure confinement of clouds (e.g., by CO-dark gas, or an HI envelope), essentially including surface terms in the virial theorem, an alternative is for super-linear conversion laws to give larger measured cloud masses, and lower $\alpha$, consistent with the present results.  Either way, our LTE analysis makes {\bf\em no} assumptions about the virial state of the gas: it is all on a per-voxel, i.e., per pixel and per channel, basis.  Thus, one could use our \nco\ maps to directly calculate $\alpha$ over any given area in a less biased way than has been done traditionally.

\vspace{1mm}Third, non-LTE calculations \citep[e.g.,][]{g11} reveal that population inversions, or sub-thermal excitation, can occur in low-$J$ levels for CO.  The sense of the correction for this effect is to give slightly higher CO column densities compared to LTE, which suggests once again that our derived $p$ values are, if anything, lower limits.  However, \cite{g11} show that under most cloud conditions that we observe, the corrections are relatively small, $\sim$0.2 in the log or smaller.

\clearpage

%%%%%%%%%%%%%
%%      Appendix C     %%
%%%%%%%%%%%%%
\section{Galactic Kinematics}\label{kinem}

%%%%%%%%%
%   Section C1  %
%%%%%%%%%
\subsection{Solar Motion and Rotation Parameter Fitting}\label{lsr}
In this Appendix we examine in detail what kinematic clues we can discern from the ThrUMMS data, in order to compare and refine models of Galactic rotation and the Solar motion with respect to the Local Standard of Rest (LSR).  We use this information to obtain better distances to all clouds and global spiral features, so improving our knowledge of the Milky Way's architecture as a whole.  However, we do not discuss in this paper the high-velocity features near the Galactic Centre, $l$ $>$ 355\degree, which are known features of non-circular orbits associated with the Galactic Bar in the innermost parts of the Milky Way's disk.  Such work is deferred to a future study; here we focus on circular rotation in the majority of the disk.

\vspace{1mm}We start with the IAU definition of the \vlsr\ scale from over 50 years ago.  In this convention, the Sun was estimated to be moving in the general direction of Vega ($\alpha$ = 18$^h$, $\delta$ = +30\degree, B1900) at around 20\,\kms, compared to a collection of nearby stars (the Local Standard of Rest).  Then, in a cartesian frame centred on the Sun (where +$x$ points towards the Galactic Centre, +$y$ points in the direction of orbital motion $l$=90\degree, and +$z$ points to the north Galactic pole $b$=+90\degree), the Sun's peculiar motion relative to LSR converts to components (\uvw) along the same axes, as given in {\color{red}Table \ref{rotpars}}, line 1.  The problem of Galactic kinematics then centres on determining accurate values for (1) these three components, which in principle can be determined locally, plus (2) the Sun's Galactocentric distance $R_0$ and orbital speed (also called the Galactic rotation) $\Theta_0$, which need to be determined globally.  In general, however, the rotation is actually a function of distance, $\Theta$($R$), since the mass distribution in the disk is non-uniform.  $\Theta$($R$) also must be measured globally, but is sometimes assumed to be constant where convenient, i.e., $\Theta$($R$) = $\Theta_0$, since it is found to be relatively flat across a wide range of $R$, including near the Sun.  For example, the BeSSeL project found $\Theta$ to vary by only $\sim$10\,\kms\ across $R$ $\approx$ 6--14\,kpc \citep{r19}.  However, a flat $\Theta$ gives very poor kinematic distance solutions interior to $R$ $\sim$ 4\,kpc, so a more physically reasonable rotation curve is highly desirable.  Fortunately, there exist good parametrisations for $\Theta$($R$); hereafter, we use the Universal Rotation Curve formulation of \cite{p96} as implemented by \cite{r19}.  The underlying caveat is that these global and local parameters are not completely separable, since the angular velocity of the Sun's Galactic rotation $\Omega$\solar\ = $(\Theta_{0}+v_{0})/R_0$ connects the local ($v_0$) and global ($\Theta_0$,$R_0$) parameters.  Thus, the BeSSeL results listed in Table \ref{rotpars}, line 2, were obtained with a global $\chi^2$ minimisation, and so are probably the most robust solutions to date for all parameters simultaneously, at least for the northern Milky Way where their data are concentrated.  Nevertheless, separate results for (1) or (2) can add useful information to this discussion.

\vspace{1mm}Using the standard treatment of Galactic rotation \citep[e.g.,][]{mb81}, we project IAU-based heliocentric distances onto our \lv\ diagrams.  Here we primarily benchmark our considerations with the \nco\ \lv\ map, with the rationale that, as a less biased tracer of mass than the individual iso-CO spectral lines, \nco\ is the preferred tracer for this exercise compared to \tco, for the reasons described in (e.g.) the \sigv\ discussion of \S\ref{vdisp}.  In other words, the \nco\ structures that we map should be better indicators of fundamental Galactic mass and potential distributions, which should be the ultimate drivers of any other tracers that derive from them (such as star formation indicators or non-rotational motions, which we wanted to avoid overfitting).  Once we have optimised our rotation parameters to \nco, supplemented in some cases by the \tco\ features where (for example) no \nco\ data exist, we will be in a position to examine how the other tracers line up with the rotation model inferred from this approach.

\begin{sidewaysfigure*}[h]
\centerline{\hspace{-1mm}\includegraphics[angle=-90,scale=0.84]{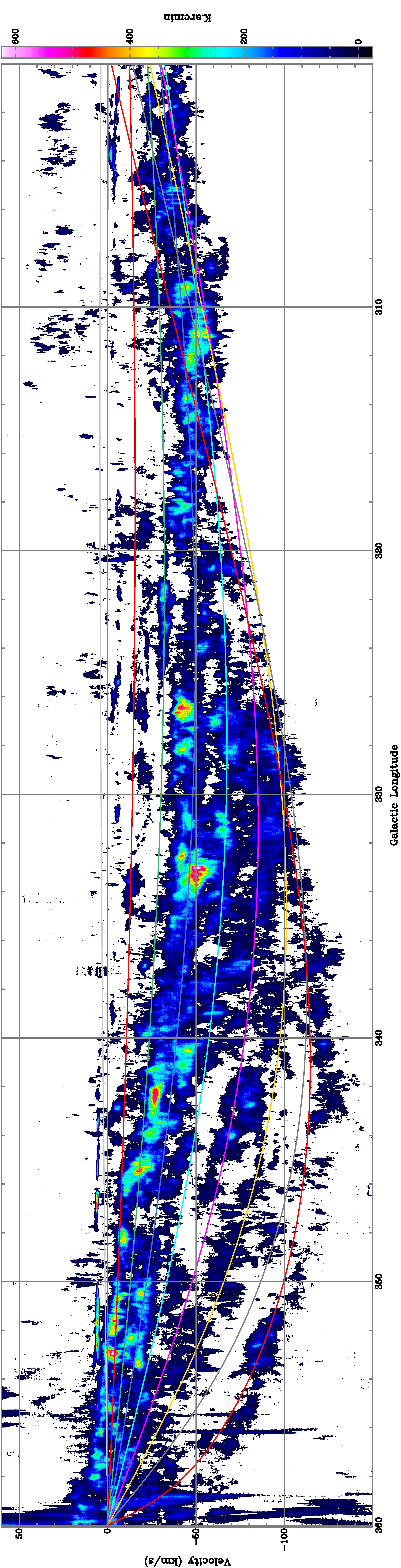}}
\vspace{2mm}
\centerline{\hspace{-1mm}\includegraphics[angle=-90,scale=0.84]{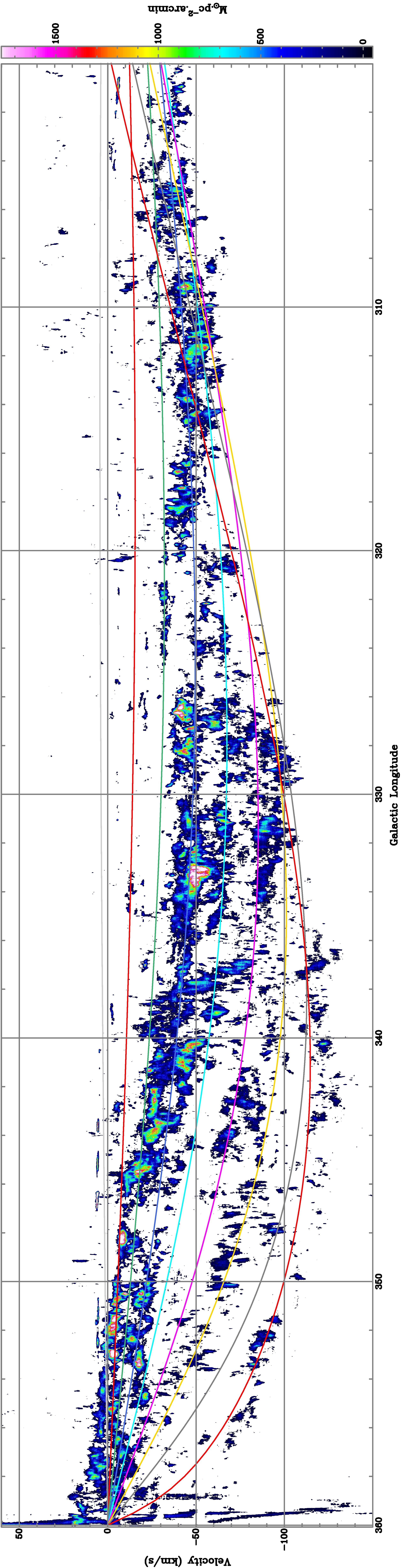}}
\vspace{2mm}
\caption{\footnotesize Two 60\degree$\times$210\,\kms\ ThrUMMS DR6 mosaics of latitude-integrated ({\em top}) \tco\ emission and ({\em bottom}) \nco: i.e., standard zeroth-moment $lV$ diagrams.  [For computational convenience, the \nco\ scale is labelled as the equivalent mass surface density $\Sigma$ = $f$\nco, where $f$ = 1.88$\times$10$^{-24}$\,M\solar\,pc$^{-2}$/(molec m$^{-2}$), although we already know that this underestimates the gas-phase \tco\ to \htwo\ conversion \citep{b18}.]  The \tco\ and \nco\ moments are from the same data cubes as portrayed in Figure \ref{full-lv-combo}, but are here overlaid by curves at a series of fixed heliocentric distances.  These start at --0.3\,kpc (light grey; see text for an explanation), then run from 0\,kpc (dark grey, very close to the \vlsr\ = 0\,\kms\ gridline) in steps of 1\,kpc (red, green, ...) to 8\,kpc (red again).  They reflect the IAU-standard definition of LSR with the original conversions to (\uvw), but using the \cite{r19} values for $R_0$, $\Theta_0$, and their Universal Rotation Curve (URC) (line 1 of Table \ref{rotpars}). $$ $$
\label{full-lv0-12coZM-kHiau}}\vspace{-16.5mm}
\end{sidewaysfigure*}

\vspace{1mm}Thus, in {\color{red}Figure \ref{full-lv0-12coZM-kHiau}}'s top and bottom panels, we show the same \tco\ and \nco\ \lv\ maps as respectively displayed in Figure \ref{full-lv-combo}'s top (red channel) and bottom (green channel) panels.  These are overlaid by contours of IAU-LSR-based heliocentric distances, i.e., with the IAU definition of (\uvw) but using \cite{r19}'s solutions for $R_0$ and $\Theta_0$, a reasonable hybrid approach as noted above.  We point out two immediate problems with these distance contours: (1) the local clouds around $\pm$10\,\kms\ deviate systematically from the \vlsr=0\,\kms\ coordinate that they should follow, e.g., near $l$ $\approx$ 300\degree--306\degree and 340\degree--353\degree, and (2) the negative \vlsr\ envelope (hereafter NVE) of the data distribution, representing the tangential velocity at each longitude, systematically exceeds the tangent-point distance for the contours over $l$ $\approx$ 300\degree--313\degree\ (by about 20\,\kms), while the NVE fails to reach the distance contour extrema over $l$ $\approx$ 330\degree--356\degree\ (by as much as 30\,\kms).

\vspace{1mm}Such discrepancies have been understood for some time to require improvements to our understanding of both the scale of the Milky Way ($R_0$ and $\Theta_0$) and of the Sun's peculiar motion with respect to LSR.  In fact, this has been a major industry among many groups and individuals for several decades, which we cannot properly review here.  Instead, we cite only a few representative works to illustrate the issues germane to the present discussion.

\vspace{1mm}The BeSSeL project has computed, using VLBI data of masers in massive star-forming regions, a sequence of progressively better-characterised solutions to these parameters \citep[the latest being][]{r19}.  The VERA project has similarly applied VLBI techniques to directly measure the proper motion of Sgr A*, and hence solutions for $R_0$ and ($\Theta_0+v_0$) \citep{o24}.  Both of these projects, however, are largely based on northern-hemisphere VLBI

%%%%%%%%%%%
%%      Table 1     %%
%%%%%%%%%%%
\begin{deluxetable}{lcccccc}
\tabletypesize{\footnotesize}
\tablecaption{Kinematic parameters for Solar motion \& Galactic rotation}
%\tablewidth{0pt}
\tablehead{
\colhead{Kinematic} & \colhead{Galactocentric} & \colhead{Orbital Speed} & & \multicolumn{3}{c}{Solar Motion\tablenotemark{a}} \\
\cline{5-7}
\colhead{Model} & \colhead{Radius $R_0$} & \colhead{$\Theta_0$} & & \colhead{$u_0$} & \colhead{$v_0$} & \colhead{$w_0$} \\
\colhead{(Reference)} & \colhead{(kpc)} & \colhead{(\kms)} & & & \colhead{(\kms)} & 
}
\startdata
IAU-LSR & 8.15\tablenotemark{b} & 236.3\tablenotemark{b} & & 10.27 & 15.32 & 7.74 \\
BeSSeL \citep{r19} & 8.15 & 236.3\tablenotemark{c} & & 10.6 & 10.7\tablenotemark{c} & 7.6 \\
BeSSeL+ThrUMMS & 8.15\tablenotemark{d} & 241.3\tablenotemark{e} & & 2.1\tablenotemark{e}  & 5.7\tablenotemark{e} & 7.6\tablenotemark{d} \\
BeSSeL+Gaia \citep{bob23} & 8.24\tablenotemark{f} & 257.3\tablenotemark{f} & & 10.6\tablenotemark{d} & 10.7\tablenotemark{d,f} & 7.6\tablenotemark{d} \\
BeSSeL+Gaia+ThrUMMS & 8.24\tablenotemark{f} & 262.3\tablenotemark{g} & & 2.1\tablenotemark{g} & 5.7\tablenotemark{g} & 7.6\tablenotemark{d} \\
BeSSeL+VERA \citep{o24} & 8.55\tablenotemark{h} & 248.3\tablenotemark{h} & & 10.6\tablenotemark{d} & 10.7\tablenotemark{d,h} & 7.6\tablenotemark{d} \\
BeSSeL+VERA+ThrUMMS & 8.55\tablenotemark{h} & 253.3\tablenotemark{i} & & 2.1\tablenotemark{i} & 5.7\tablenotemark{i} & 7.6\tablenotemark{d} 
\enddata
\tablenotetext{a}{Additional motion relative to LSR, parallel to cartesian Galactic coordinates $x$,$y$,$z$.} 
\tablenotetext{b}{Assumed values from BeSSeL; the IAU definition of LSR converts to the (\uvw) shown on this line.} 
\tablenotetext{c}{For BeSSeL, ($\Theta_0+v_0$) = 247\,\kms\ is more strongly constrained than either $\Theta_{0}$ or $v_{0}$ separately.} 
\tablenotetext{d}{Assumed values from BeSSeL.} 
\tablenotetext{e}{Assuming ($\Theta_0+v_0$) from BeSSeL and optimising $v_0$ for local clouds; $u_0$ is optimised separately.} 
\tablenotetext{f}{Assumed values from the Gaia-Cepheids analysis, combining their solution for ($\Theta_0+v_0$) = 268\,\kms\ with the BeSSeL value for $v_0$.} 
\tablenotetext{g}{As with Note f, but assuming the B+T values for $u_0$ \& $v_0$.} 
\tablenotetext{h}{For VERA, $\Omega$\solar\ = $(\Theta_{0}+v_{0})/R_0$ = 30.30\,\kms/kpc is {\bf far} more strongly constrained than the individual parameters.  Their separate solution for $R_0$ (as listed, $\pm$15\%) then constrains ($\Theta_0+v_0$) = 259.0\,\kms, where we solve for $\Theta_0$ assuming the BeSSeL value for $v_0$.} 
\tablenotetext{i}{As with Note h, but assuming the B+T values for $u_0$ \& $v_0$.}
\label{rotpars}
\end{deluxetable}

\noindent networks, and so have necessarily focused on the Galaxy's 1st \& 2nd Quadrants (0$<$$l$$<$180\degree), although this will eventually be rectified in the future.  \cite{bob23} uses a different approach: recent, high-precision Gaia astrometry and kinematic data on a large sample of Cepheids on the Solar circle.  While this example is truly global, the Gaia data do not reach the intrinsic kinematic precision of the VLBI data; indeed, uncertainties from all efforts continue to forestall a definitive set of results.  Therefore, and even without VLBI or Gaia, the global ThrUMMS data can provide useful insights to Galactic kinematics.

\begin{sidewaysfigure*}[h]
\centerline{\hspace{-1mm}\includegraphics[angle=-90,scale=0.84]{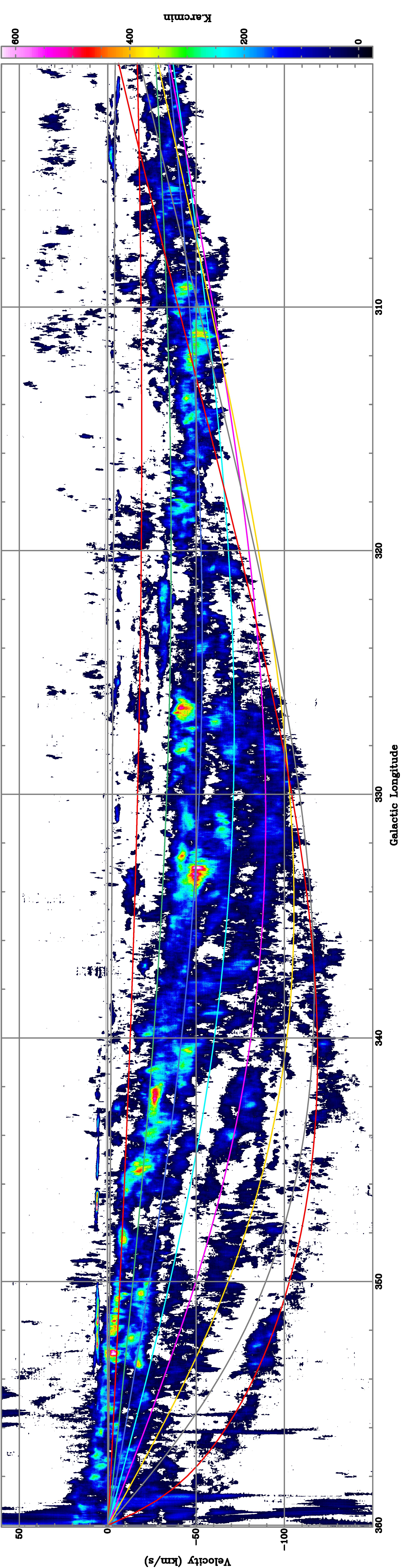}}
\vspace{2mm}
\centerline{\hspace{-1mm}\includegraphics[angle=-90,scale=0.84]{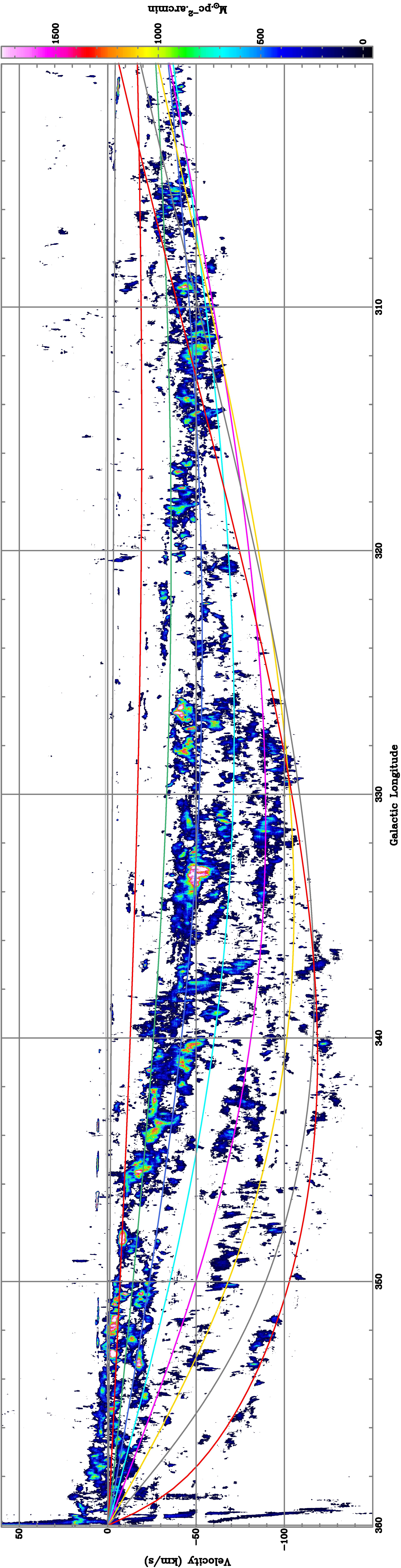}}
\vspace{2mm}
\caption{\footnotesize Same as Fig.\,\ref{full-lv0-12coZM-kHiau}, except that the heliocentric curves were obtained using the full BeSSeL solution from \cite{r19} (line 2 of Table \ref{rotpars}). $$ $$
\label{full-lv0-12coZM-kHbes}}\vspace{0mm}
\end{sidewaysfigure*}

\vspace{1mm}In {\color{red}Figure \ref{full-lv0-12coZM-kHbes}} we therefore display the same data as in Figure \ref{full-lv0-12coZM-kHiau} but with distance contours using \cite{r19}'s full 5-parameter solution (see Table \ref{rotpars}).  The net effect of this change is to noticeably improve the fit of the distance curves to the data, especially over $l$ $\approx$ 300\degree--313\degree.  At these low longitudes, both issues identified above are improved: (1) the $d$=0\,kpc curve passes nicely through the middle \vlsr\ range of these local clouds, while (2) the degree to which the tangential \vlsr\ of the data exceeds the model's tangent-point \vlsr\ has been reduced, by about a third, to $\sim$10--15\,\kms.

\vspace{1mm}Looking at Table \ref{rotpars}, we can see that this improvement was largely wrought by the reduction in $v_0$, from 15.3 to 10.7\,\kms, since neither $u$ nor $w$ changed appreciably from the IAU definitions.  This is understandable since a smaller $v_0$ means that the Sun is moving towards $l$=90\degree\ more slowly than the original LSR definition, so any objects near $l$=90\degree\ would have larger (more positive) \vlsr\ on this new scale.  Conversely, objects near $l$=270\degree\ should have more negative \vlsr\ on this new scale, just as we see.

\vspace{1mm}However, the fits over the higher longitude range, $l$ $\approx$ 330\degree--356\degree, actually get worse with the BeSSeL parameters.  While portions of the gap between the NVE and the further distance curves could be attributed to a simple lack of clouds at these locations, or perhaps due to sensitivity limitations at large distances, we find such arguments somewhat unsatisfactory.  More importantly, the local clouds at $l$ $\approx$ 340\degree--353\degree\ lie further from the $d$=0\,kpc curve for the \cite{r19} model than for the IAU model.  Therefore, we view these mismatches largely as a modelling problem.

\vspace{1mm}Despite the considerable efforts of the BeSSeL group to obtain robust solutions for the Milky Way's rotation, it should not be too surprising that the \cite{r19} parameters are only a moderately good fit to the southern Milky Way, which has not been as accessible to VLBI techniques.   But starting with these parameters, we reasoned that a less sophisticated exploration of parameter space could yield useful improvements.  In particular, the local clouds could be argued as perhaps the best possible samplers of the LSR in the disk, since they necessarily define the most extreme Population I objects that are available for study, and as a consequence, should be the least disturbed from a putative average circular motion around the Galaxy.  Some workers in this field might disagree with the last statement, but as an exercise, we wanted to see what this presumption might teach us.  Our motivation was primarily to improve the $d$=0\,kpc fitting to the local clouds.

% Figure C1 (+21 for Appcs A,B in aastex) -- IAU

% Figure C2 -- BeSSeL

% Figure C3 -- B+ThrUMMS
\begin{sidewaysfigure*}[h]
\centerline{\hspace{-1mm}\includegraphics[angle=-90,scale=0.84]{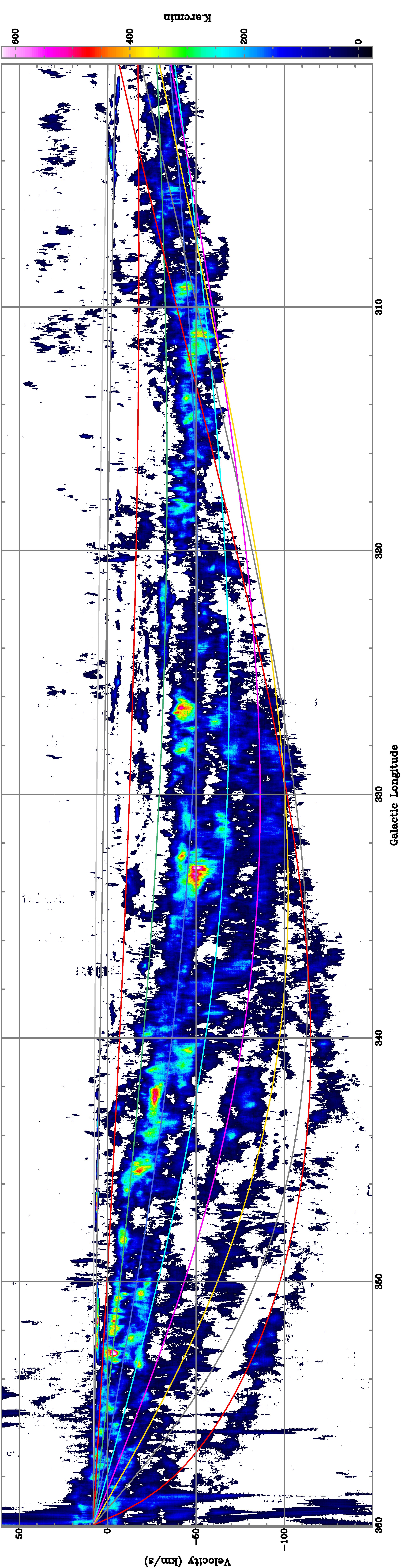}}
\vspace{2mm}
\centerline{\hspace{-1mm}\includegraphics[angle=-90,scale=0.84]{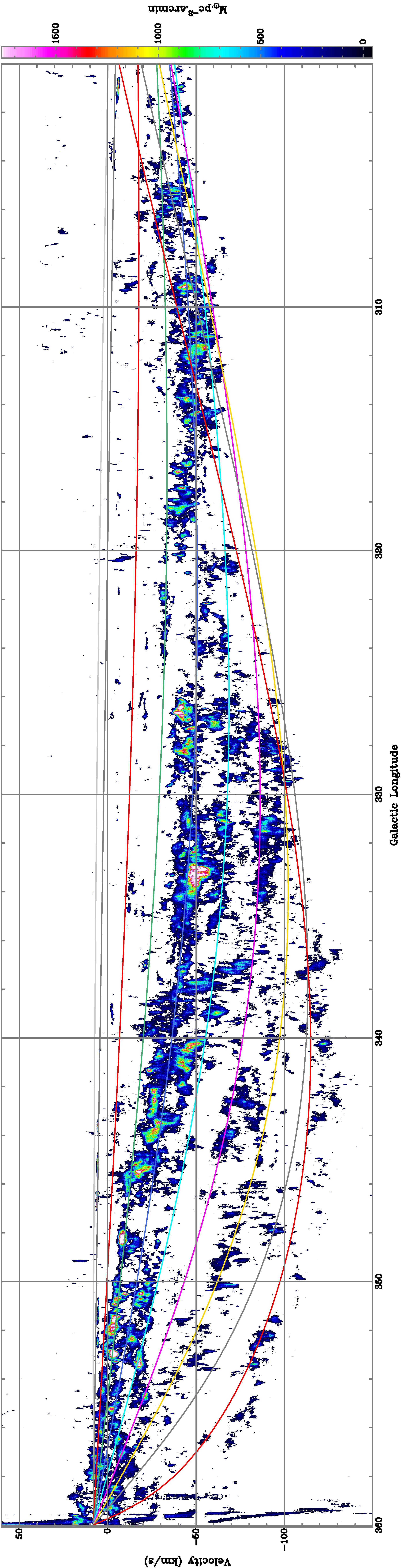}}
\vspace{2mm}
\caption{\footnotesize Same as Fig.\,\ref{full-lv0-12coZM-kHbes}, except that the heliocentric curves are now from a modified BeSSeL solution, optimised to fit the 0\,kpc curve (dark grey) to the local clouds near $l$ $\approx$ 300\degree--306\degree and 340\degree--353\degree\ (line 3 of Table \ref{rotpars}). $$ $$
\label{full-lv0-12coZM-kHthr}}\vspace{0.2mm}
\end{sidewaysfigure*}

%\clearpage

\vspace{1mm}Therefore, choosing to leave the \cite{r19} $R_0$ and $\Theta_0$ values as given, we iteratively varied the (\uvw) values until we obtained a ``best'' fit of the $d$=0\,kpc curve to the local clouds, as judged by eye.  Actually, we only varied $u$ and $v$, since $w$ (motion in the $z$ direction) contributes so little to this problem.  The result, listed in Table \ref{rotpars} as the BeSSeL+ThrUMMS model (line 3), is illustrated in {\color{red}Figure \ref{full-lv0-12coZM-kHthr}}.

\vspace{1mm}This produced a gratifying improvement to the fits of the distance curves to the data.  We see in Figure \ref{full-lv0-12coZM-kHthr} that the local clouds' \lv\ locus is very well fit by the $d$=0\,kpc curve, with deviations of only $\pm$2\,\kms.  To emphasise this, we have also added a ``--0.3\,\kms\ curve'' in light grey (this light grey curve is also shown in Fig.\,\ref{full-lv0-12coZM-kHbes} for reference, but is a little hard to make out there since it lies close to the \vlsr=0\,\kms\ gridline).  This curve is obviously unphysical as a distance, but is numerically allowed by the kinematic equations.  We include it here to represent a maximal distance uncertainty, and corresponding \vlsr-fit uncertainty, to our LSR fitting.

\vspace{1mm}In order to optimise the 0\,kpc curve fit, we had to reduce the values of {\bf both} $u_0$ and $v_0$ from the \cite{r19} values.  Reducing $u_0$ is the only way to improve the LSR fit to the local clouds at high longitudes, $l$ $\approx$ 340\degree--353\degree.  This is understandable since a smaller $u_0$  in the LSR means that the Sun is moving towards $l$=0\degree\ more slowly than the IAU-LSR value; thus, objects in that direction should have a higher (more positive) \vlsr\ on this new scale, as we need.  However, reducing $u_0$ also affects the fitting at the lower longitudes, so we had to reduce $v_0$ as well in order to recover good local cloud fits near $l$=300\degree, which in turn required a small adjustment to $u_0$, and so on.  The (\uvw) results in line 3 of Table \ref{rotpars} are the most optimised for the local cloud fitting.

\vspace{1mm}Note that we did not try to account for any individual or averaged cloud peculiar motions in the \uv\ fitting, such as from local dynamics, turbulence, streaming motions, etc., which are on the order of 7--10\,\kms.  The optimisation was thus a pure fit to the local clouds' \vlsr\ envelope.  Because the fitting residuals quoted above are so small, much smaller than the typical cloud-to-cloud velocity dispersions, we must conclude that non-rotational motions are surprisingly small for the local cloud population.  As an unexpected bonus, this result is actually more robust than fitting to ThrUMMS data alone might have been expected.  A cursory inspection of CfA \citep{dht01} and/or FUGIN \citep{um17} data reveals a sinusoidal \vlsr\ pattern in the local clouds at other longitudes as well, which conforms remarkably well to our \uv\ fit.

\vspace{1mm}The final result for the local clouds is excellent across all longitudes; however, we now have a somewhat significant tension with the \cite{r19} result for $u_0$.  While there is some intrinsic slop in the $v_0$ fitting due to the primary VLBI constraint being on either the $\Omega$\solar\ or ($\Theta_0+v_0$) values, it is more difficult to explain away the change to $u_0$.  On the other hand, the smaller $u_0$ conveys an additional improvement to fitting the NVE at high longitudes as a free bonus, albeit reducing the data gap by only 20\% or so.

% Figure C4.0 -- B+VERA
%\begin{sidewaysfigure*}[h]
%\centerline{\hspace{-1mm}\includegraphics[angle=-90,scale=0.92]{dr6-mosAll-12co-lv0-kHbv.jpg}}
%\vspace{2mm}
%\centerline{\hspace{-1mm}\includegraphics[angle=-90,scale=0.92]{dr6-mosAll-ZM-lv0-kHbv.jpg}}
%\vspace{12.3mm}
%\caption{Same as Fig.\,\ref{full-lv0-12coZM-kHbes}, except that the heliocentric curves are now from a hybrid BeSSeL+VERA solution as described in the text.
%\label{full-lv0-12coZM-kHbv}}
%\end{sidewaysfigure*}

% Figure C4.1 -- B+V+ThrUMMS
\begin{sidewaysfigure*}[h]
\centerline{\hspace{-1mm}\includegraphics[angle=-90,scale=0.84]{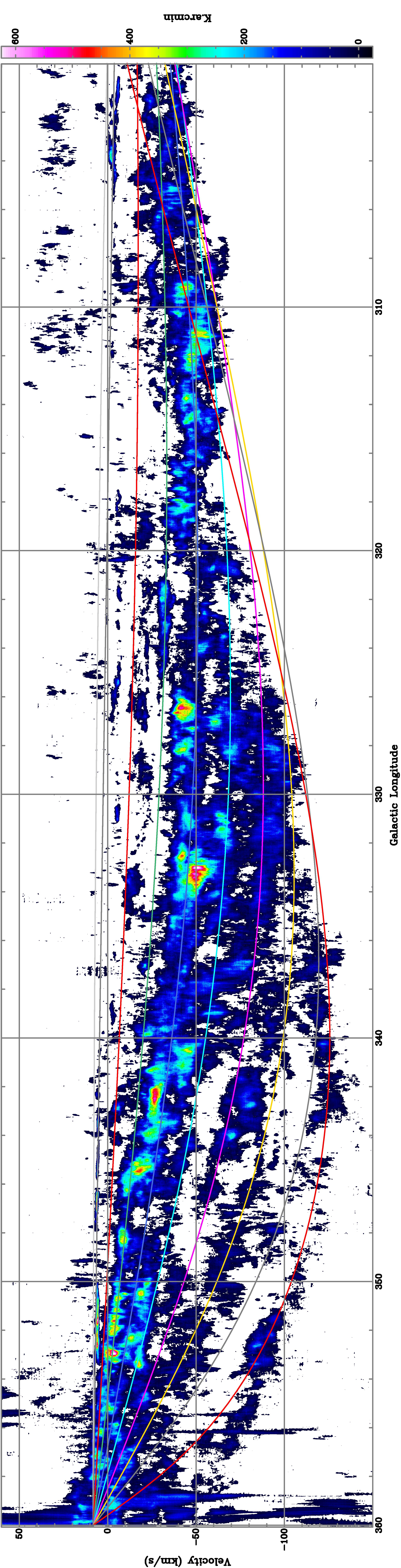}}
\vspace{2mm}
\centerline{\hspace{-1mm}\includegraphics[angle=-90,scale=0.84]{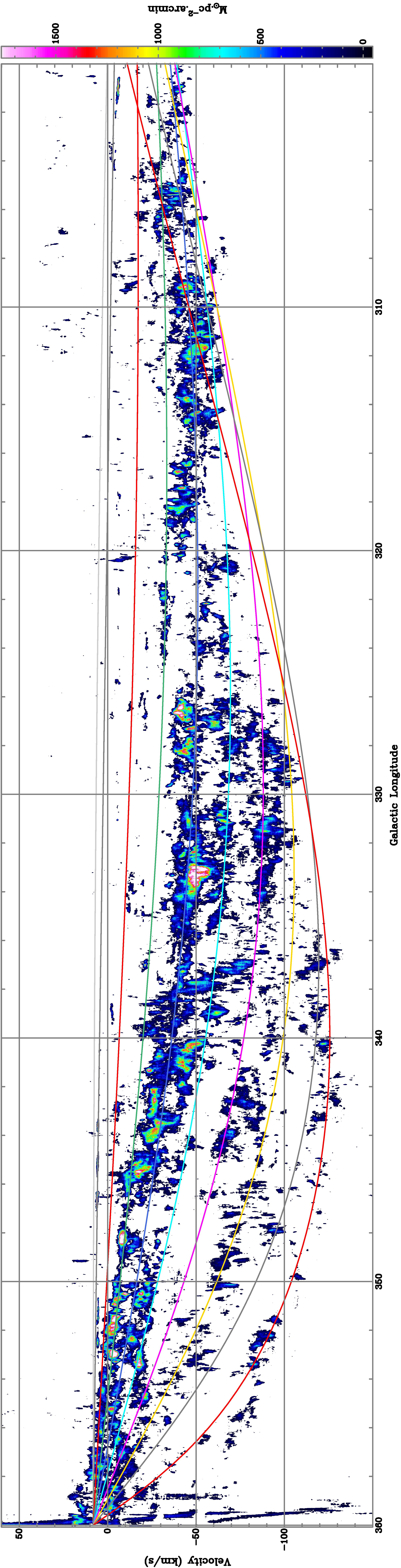}}
\vspace{2mm}
\caption{\footnotesize Same as Fig.\,\ref{full-lv0-12coZM-kHthr}, except that the heliocentric curves are now from a hybrid BeSSeL+VERA solution as described in the text, plus optimised fits to the local clouds near $l$ $\approx$ 300\degree--306\degree and 340\degree--353\degree\ (line 7 of Table \ref{rotpars}). $$ $$
\label{full-lv0-12coZM-kHbvt}}\vspace{0mm}
\end{sidewaysfigure*}

% Figure C5.0 -- B+Gaia
%\begin{sidewaysfigure*}[h]
%\centerline{\hspace{-1mm}\includegraphics[angle=-90,scale=0.92]{dr6-mosAll-12co-lv0-kHbg.jpg}}
%\vspace{2mm}
%\centerline{\hspace{-1mm}\includegraphics[angle=-90,scale=0.92]{dr6-mosAll-ZM-lv0-kHbg.jpg}}
%\vspace{12.3mm}
%\caption{Same as Fig.\,\ref{full-lv0-12coZM-kHbes}, except that the heliocentric curves are now from a hybrid BeSSeL+Gaia-Cepeids solution as described in the text.
%\label{full-lv0-12coZM-kHbg}}
%\end{sidewaysfigure*}

% Figure C5.1 -- B+G+ThrUMMS
\begin{sidewaysfigure*}[h]
\centerline{\hspace{-1mm}\includegraphics[angle=-90,scale=0.84]{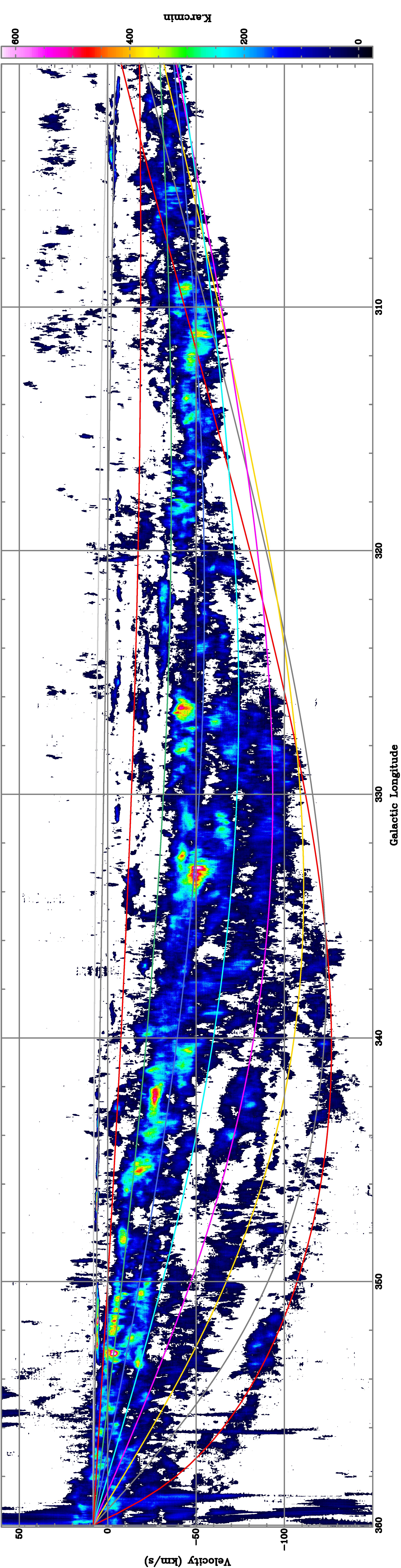}}
\vspace{2mm}
\centerline{\hspace{-1mm}\includegraphics[angle=-90,scale=0.84]{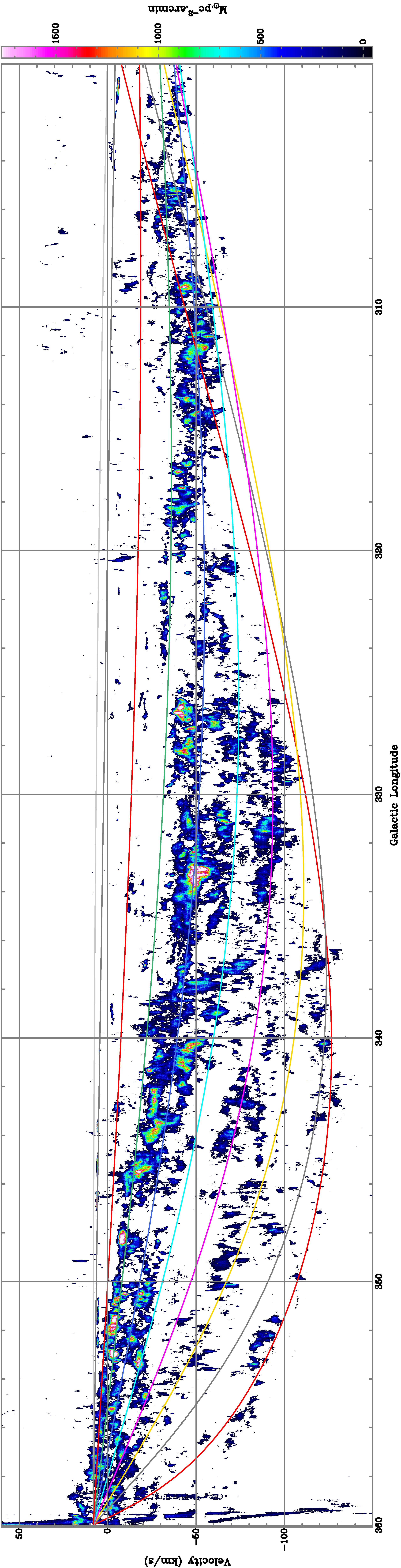}}
\vspace{2mm}
\caption{\footnotesize Same as Fig.\,\ref{full-lv0-12coZM-kHthr}, except that the heliocentric curves are now from a hybrid BeSSeL+Gaia-Cepheids solution as described in the text, plus optimised fits to the local clouds near $l$ $\approx$ 300\degree--306\degree and 340\degree--353\degree\ (line 5 of Table \ref{rotpars}). $$ $$
\label{full-lv0-12coZM-kHbgt}}\vspace{0mm}
\end{sidewaysfigure*}

\vspace{1mm}We also examine results from \cite{o24}, who provide the most highly constrained rotation parameter to date, the Sun's angular velocity around Sgr A*, $\Omega$\solar\ = 30.30$\pm$0.02\,\kms/kpc = ($\Theta_0+v_0$)/$R_0$.  Their separate parameters are less well-constrained, $R_0$ = 8.55\,kpc and ($\Theta_0+v_0$) = 259.0\,\kms ($\pm$15\% for both), but their larger $R_0$ and $\Theta_0$ values than \cite{r19}'s harken back to earlier versions of the BeSSeL project, and should at least be instructive.  This reasoning is also supported by \cite{bob23}'s Gaia results, which give $R_0$ = 8.24$\pm$0.20\,kpc and ($\Theta_0+v_0$) = 268$\pm$8\,\kms.

\vspace{1mm}Interestingly, during our local-cloud \vlsr-optimisation experiments, we noticed that changing the (\uvw) values really only had a meaningful effect on the nearby (\lapp3\,kpc) distance curves.  While the further distance curves also moved around slightly with changes to $u$ and $v$, these changes were relatively small, certainly compared to the effect on the local cloud fitting.  In contrast, changing the values of $R_0$ and $\Theta_0$ had very little effect on the nearby distance curves, but a more profound one on the further curves.  In other words, the fitting of each subset of parameters (i.e., the local \uvw\ vs.\ the global $R_0$,$\Theta_0$) is indeed mostly orthogonal to fitting the other subset, as originally argued.  Thus, the (\uvw) parameters optimised for the BeSSeL solutions were equally the best values for (\uvw) for the other projects' $R_0$,$\Theta_0$.

\vspace{1mm}In summary, we compared all 3 projects' (BeSSeL, Gaia-Cepheids, VERA) results for $R_0$,$\Theta_0$ with both the unmodified BeSSeL (\uvw) results (lines 4 \& 6 in Table \ref{rotpars}) and the ThrUMMS-modified results (lines 5 \& 7).  We found that the ThrUMMS-modified version of each kinematic model created a much-preferred fit to the local clouds and LSR, and show the hybrid BeSSeL+VERA+ThrUMMS distance contour overlays (the ``BVT'' model) on our \lv\ data in {\color{red}Figure \ref{full-lv0-12coZM-kHbvt}}, and the BeSSeL+Gaia+ThrUMMS version (the ``BGT'' model) in {\color{red}Figure \ref{full-lv0-12coZM-kHbgt}}.  While we were primarily focused on the LSR optimisations, the larger $R_0$ \& $\Theta_0$ values of the VERA \& Gaia studies compared to the BeSSeL (and especially IAU) values also significantly improved the distance curves' fit to the NVE of the data distribution, most noticeably at the low longitudes. %(compare Figs.\,\ref{full-lv0-12coZM-kHbes} and \ref{full-lv0-12coZM-kHbv}).  

\vspace{1mm}Of all the model combinations, we prefer the BGT model (line 5 of Table \ref{rotpars}) since it has the best effects in $\tfrac{3}{4}$ of our \lv\ data.  The net effect is to (1) significantly improve the LSR model to kinematically fit the local cloud population at all longitudes, and (2) make strong improvements to matching the distance contours to the data distribution of all clouds at low longitudes.  The fit in (1) was improved from $\pm$5--10\,\kms\ with the standard IAU parameters or (in some locations) with the BeSSeL parameters, to \lapp$\pm$2\,\kms\ over all $l$ mapped here.  At low longitudes, the mismatch in (2) was reduced from $\sim$20\,\kms\  with the IAU parameters, or from $\sim$15\,\kms\ with the unmodified BeSSeL parameters, to $\sim$5\,\kms\ with our BGT hybrid parameters.  At high longitudes, the mismatch in (2) remained $\sim$30\,\kms\ for all models.  The main issue with our BGT model is the large tension between the new fitted value for $u_0$ and the BeSSeL value; ours is 8.5\,\kms\ smaller.  It is to be hoped that future VLBI work in the southern hemisphere might resolve both the $u_0$ tension and the NVE mismatch at high longitudes between data and models.

%\vspace{1mm}{\em 1 table and 5 figures for this subsection}

%\clearpage

%%%%%%%%%
%   Section C2  %
%%%%%%%%%
\subsection{Deprojected Milky Way Maps}\label{mwmaps}
\vspace{1mm}The effort to finesse Galactic rotation models was made in order to maximise the scientific utility of the next three steps in our analysis.  The first is to deproject the \lv\ maps discussed so far, and their progenitor \lbv\ cubes, into longitude-distance space (either  ($l$,$d$) or  ($l$,$b$,$d$) as appropriate), effectively a polar coordinate mapping of the 4th Quadrant, or even into the more intuitive cartesian space.  Traditionally, this deprojection is performed on a cloud-by-cloud basis.  In other words, for many Milky Way surveys, cloud catalogues are generated from the \lbv\ data as lines in a spreadsheet or text list, based on some cloud-identifying algorithm.  The Galactic rotation equations are then applied to the clouds' \lv\ coordinates in order to tabulate distances.  Often, these routines will generate both of the kinematically allowed near and far distances, and then other algorithms will be applied to make a determination of which distance to prefer --- see \cite{dc20} for an example.  They used the SCIMES algorithm with SEDIGISM \ttco\ \jto\ data to make the cloud identifications in \lbv\ space, and a number of astrophysical filters to disambiguate the near/far solutions.

\vspace{1mm}The same approach can also be used on any of the ThrUMMS data cubes: not just the \tco\ or \ttco\ observed data, but also the \nco\ cubes, and we plan future works to explore those results.  However, we also wanted to experiment with a structure-agnostic procedure.  That is, to do the deprojection directly onto an \lbd\ cube, voxel by voxel, without prejudging what constitutes a molecular ``cloud'' and only then solving for a distance.  The intent is to apply statistical methods to the \lbd\ cubes and then look at the overall structure of the Milky Way so discerned, e.g., to more intuitively define spiral-arm and interarm regions and compare cloud properties between.

\begin{sidewaysfigure*}[h]
\vspace{6.4mm}
\centerline{\includegraphics[angle=-90,scale=0.835]{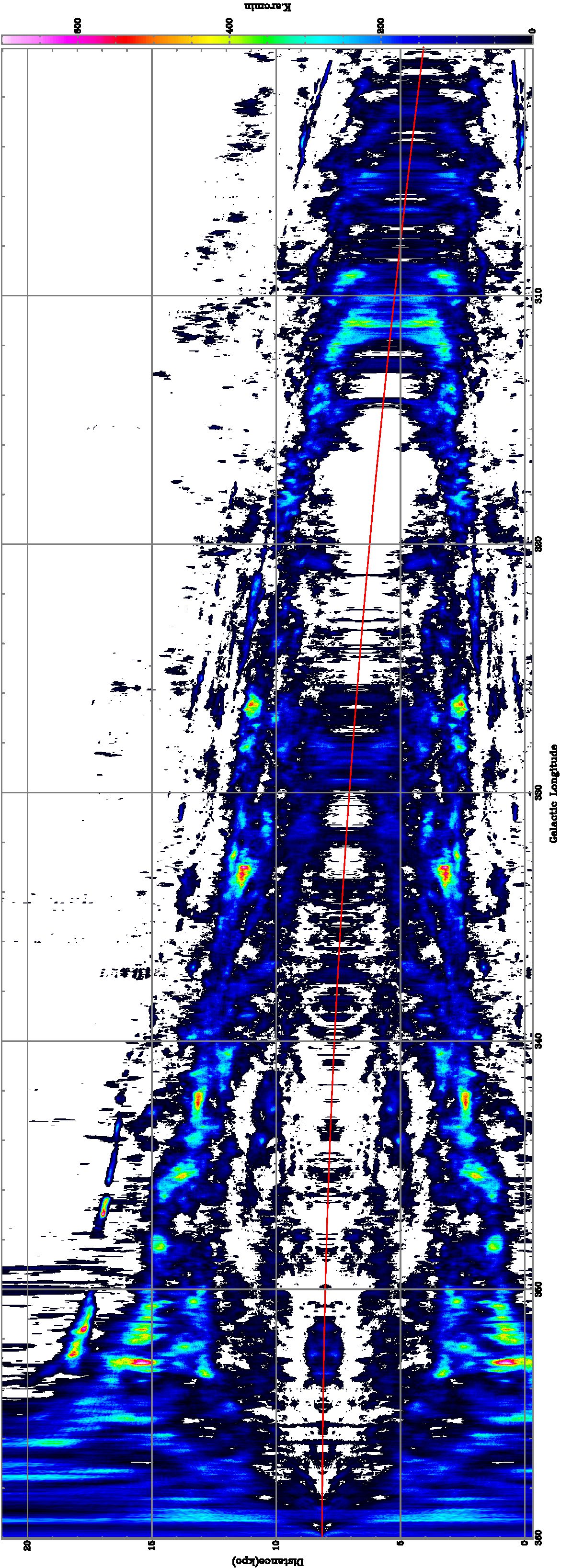}\hspace{1.6mm}}
\vspace{-3mm}
\centerline{\includegraphics[angle=-90,scale=0.84]{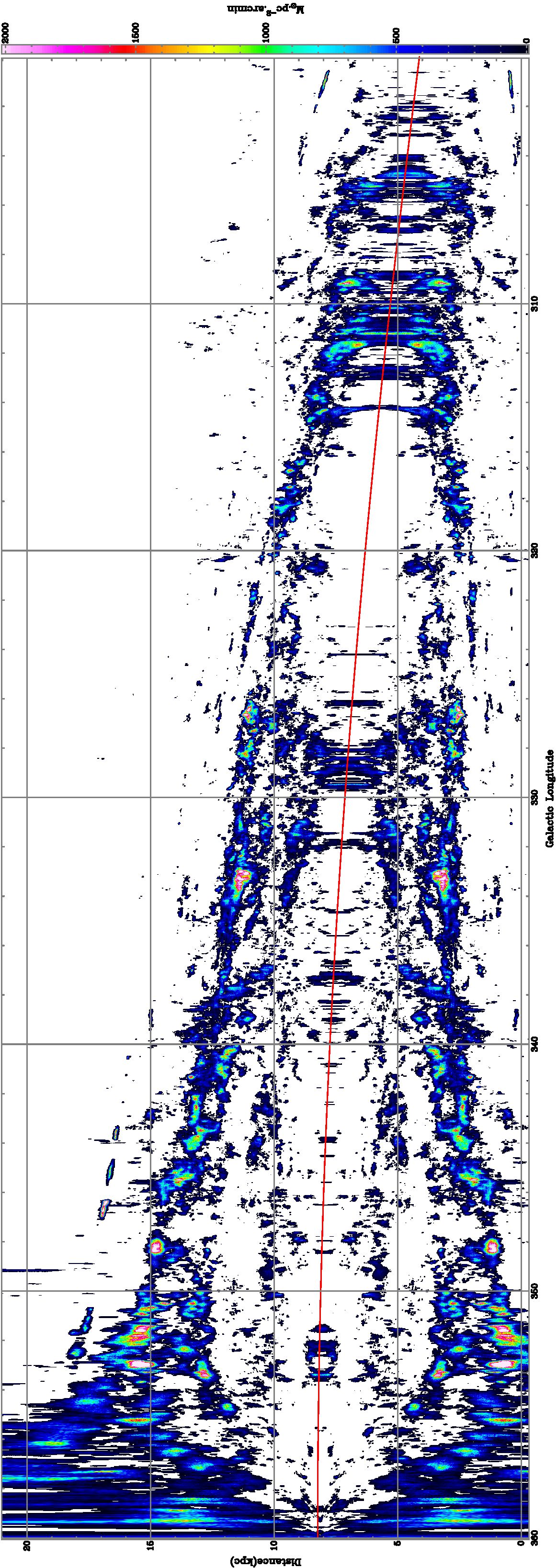}}
\caption{\footnotesize ({\em Top}) \tco\ \lv\ data as in Figs.\,\ref{full-lv-combo} and \ref{full-lv0-12coZM-kHiau}--\ref{full-lv0-12coZM-kHbgt}, but deprojected onto the IAU distance scale (Fig.\,\ref{full-lv0-12coZM-kHiau} contours) as an \ld\ map.  This includes both the near and far kinematic distance solutions, which are mirrored along a sinusoidal curve (in red) from (360\degree,8.15\,kpc) to (300\degree,4.075\,kpc), at the tangent-point distance for each $l$.
({\em Bottom}) Same as the top panel but with the \nco\ \lv\ data. $$ $$
\label{iau-ld0}}\vspace{0mm}
\end{sidewaysfigure*}

\begin{sidewaysfigure*}[h]
\vspace{6mm}
\centerline{\includegraphics[angle=-90,scale=0.835]{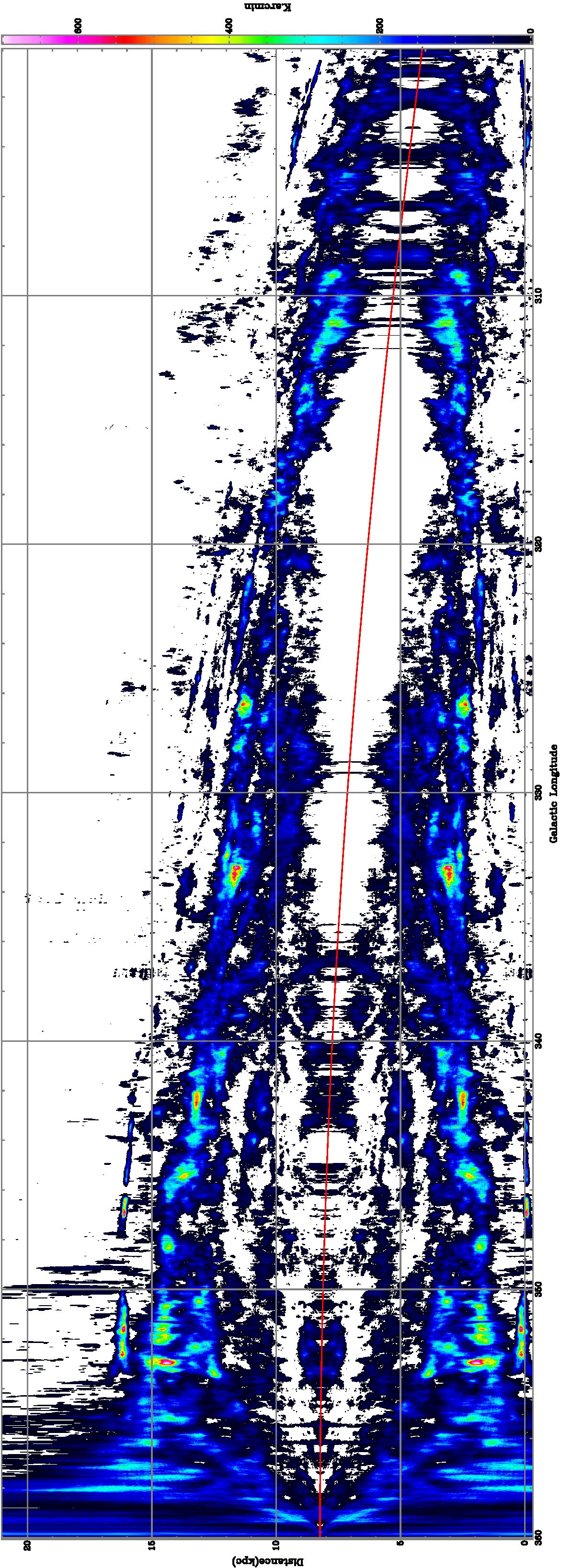}\hspace{1.6mm}}
\vspace{-3mm}
\centerline{\includegraphics[angle=-90,scale=0.84]{dr6-ZM-URCbgt-ld0.jpg}}
\caption{\footnotesize Similar \ld\ maps to Fig.\,\ref{iau-ld0} but deprojected onto the BGT distance scale (Fig.\,\ref{full-lv0-12coZM-kHbgt} contours) as an \ld\ map.  The mirroring red tangent-point curve runs from (360\degree,8.24\,kpc) to (300\degree,4.12\,kpc) in the BGT model.  ({\em Top}) \tco\ data.  ({\em Bottom}) \nco\ data. $$ $$
\label{bgt-ld0}}\vspace{-0.9mm}
\end{sidewaysfigure*}

\vspace{1mm}We show in {\color{red}Figure \ref{iau-ld0}} the result of deprojecting the \tco\ and \nco\ \lv\ maps (integrated across all $b$ at each pixel) onto the IAU distance scale displayed in Figure \ref{full-lv0-12coZM-kHiau}, and compare this with a similar deprojection of both \lv\ maps from Figure \ref{full-lv0-12coZM-kHbgt} (the BGT model) in {\color{red}Figure \ref{bgt-ld0}}.  Note that either deprojection doubles the data volume inside the solar circle, since we are displaying all valid near and far distance solutions for the \lv\ data in the ($l$,$d$) map.  These near and far solutions are graphical mirrors of each other across a curve representing the tangent-point distance as a function of $l$.

\vspace{1mm}This comparison of the two distance models makes clear the improvements to distance solutions for the local clouds ($d$ $\approx$ 0\,kpc) as well as better-modelling the Galactic rotation for near-tangent clouds across $l$ $\sim$ 300\degree--313\degree.  That is, for the local clouds, the IAU model loses many of them beyond the --0.3\,kpc computational limit, while the BGT model recovers them and places them close to the 0\,kpc coordinate.  Similarly, for the clouds near the NVE at low longitudes, the IAU model smears many of them across the tangent-curve mirror over large distance ranges, while the BGT model dramatically reduces such artifacts.  Although we focused mainly on using the \nco\ features to optimise the BGT model parameter fits, slight adjustments based on the more widespread \tco\ emission helped refine the fitting even further, and made the improvements to the \tco\ \ld\ map even more convincing.  In particular, the match to the local clouds near \vlsr\ = 0\,\kms\ is even more striking in \tco\ than in \nco, especially at the higher longitudes, while the tangent curve mirror shows a cleaner separation between the near and far clouds in the BGT model than in the IAU one, for either data set.

% Figure C6: 12co- and ZM-ld0 with IAU

% Figure C7: 12co- and ZM-ld0 with BGT

\begin{figure*}[h]
\centerline{\includegraphics[angle=0,scale=0.923]{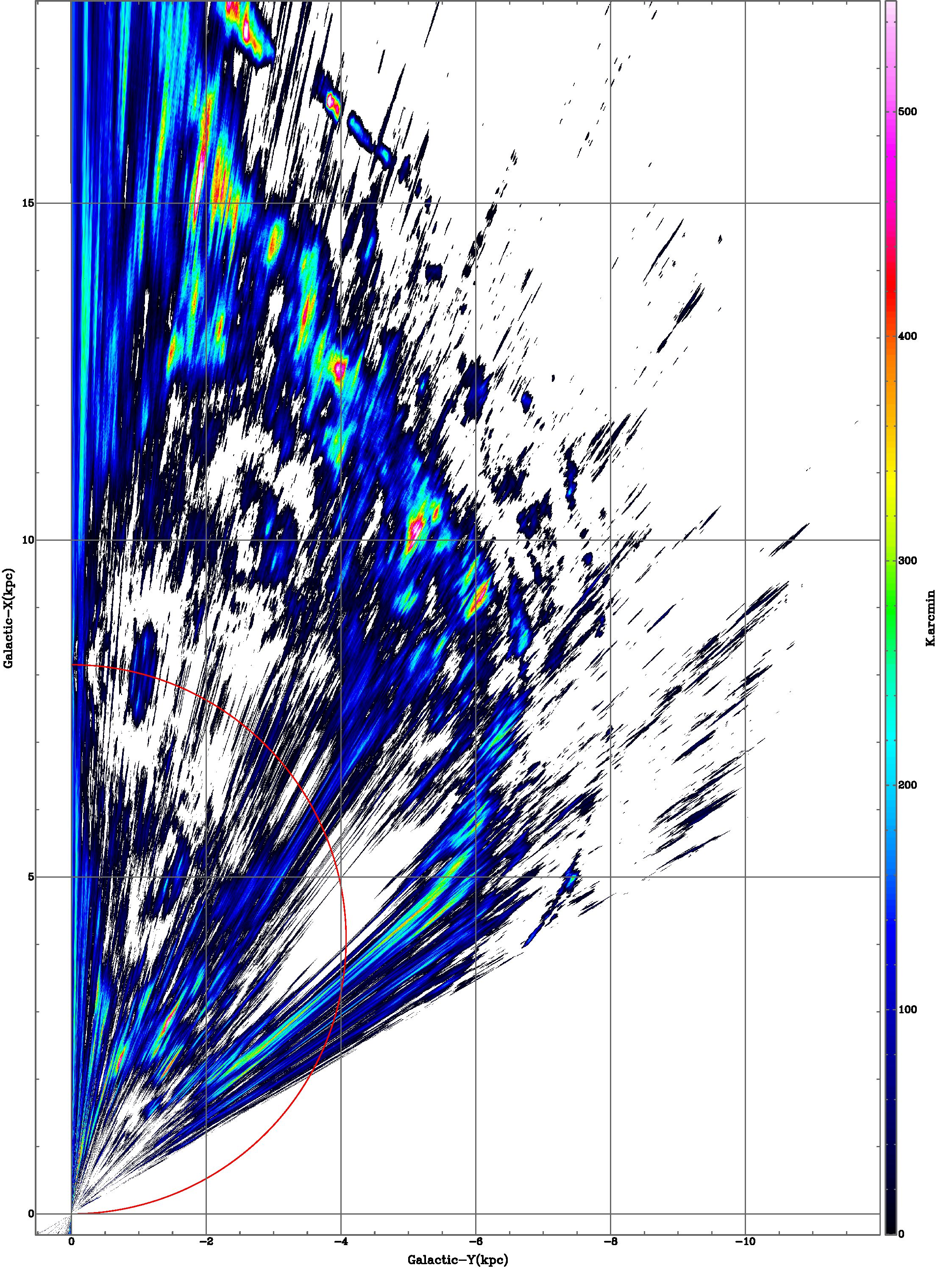}}
\vspace{-2mm}
\caption{\footnotesize Cartesian deprojection, based on the IAU model for Galactic rotation (Fig.\,\ref{full-lv0-12coZM-kHiau} contours), of the \tco\ \lv\ data into Galactic \xy\ space, where the +$x$ direction points from the Sun up to the Galactic Centre (GC) and the +$y$ direction from the Sun left towards $l$=90\degr.  The Sun is at coordinates (0,0), the GC at (8.15,0), and the red circle denotes the locus of tangent-point distances in this model. $$ $$
\label{12co-iau-YX0}}
\end{figure*}

\vspace{1mm}There is another advantage to this approach: with a distance model, we can (conceptually, at least) just as easily project cartesian $x$,$y$ coordinates onto the \lv\ data as we can the $d$ coordinate, and so deproject the data into \xyz\ space.  Numerically, the cartesian deprojection is a little more tricky since it is a 2D problem, namely from \lv\ to \xy, whereas the polar deprojection is only 1D, from $V$ to $d$.  Nevertheless, both transformations are tractable, and we show what the corresponding \xy\ deprojections look like in {\color{red}Figures \ref{12co-iau-YX0}} to {\color{red}\ref{ZM-bgt-YX0}}. %{\em (YX0: 12co/ZM-iau, 12co/ZM-bgt)}

\vspace{1mm}Here again we can clearly see how adoption of different rotation models affects the resulting distance solutions and changes the deprojected appearance of large-scale structures in the Milky Way.  Note again, however, that we have applied {\bf no} filters to these maps to discriminate between near and far distance solutions for any given \lv\ pixel.  Effectively, each of the original \lv\ pixels inside the Solar circle is represented in two different \ld\ or \xy\ pixels in each \ld\ or \xy\ map, one at the near kinematic distance and one at the far distance.  Outside the Solar circle, of course, there is only a far kinematic distance solution.

\begin{sidewaysfigure*}[h]
\vspace{10mm}
\centerline{\includegraphics[angle=-90,scale=0.84]{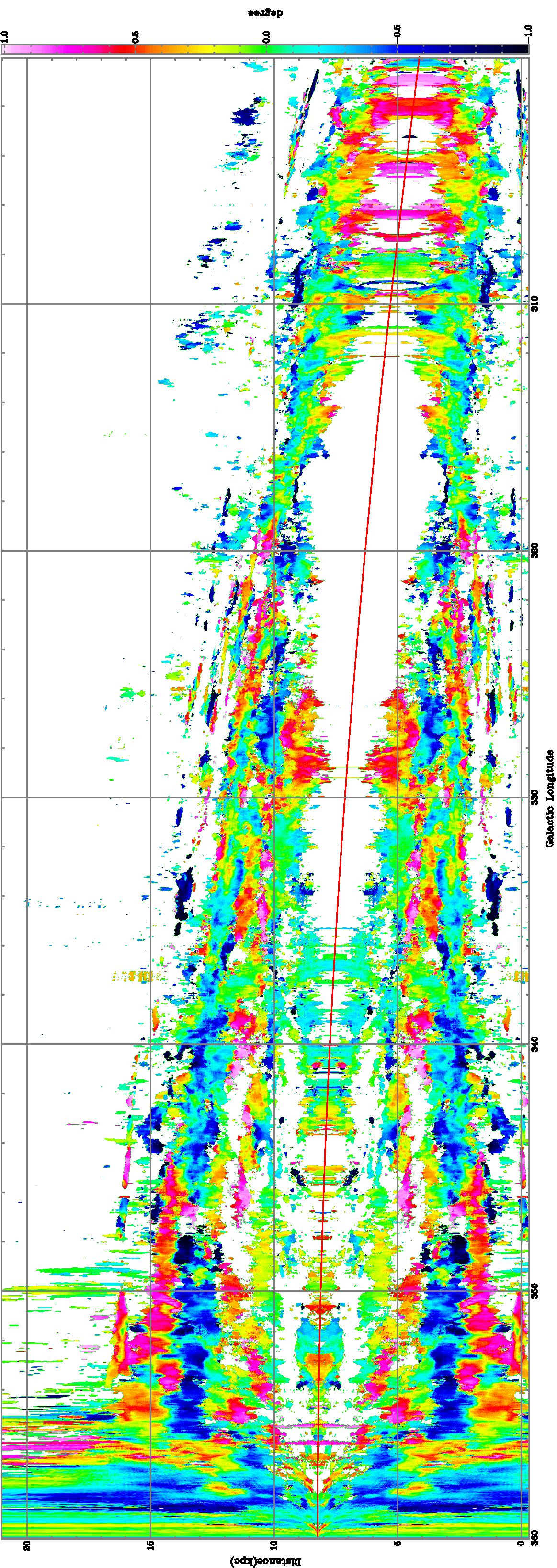}\hspace{0mm}}
\vspace{-3mm}
\centerline{\includegraphics[angle=-90,scale=0.84]{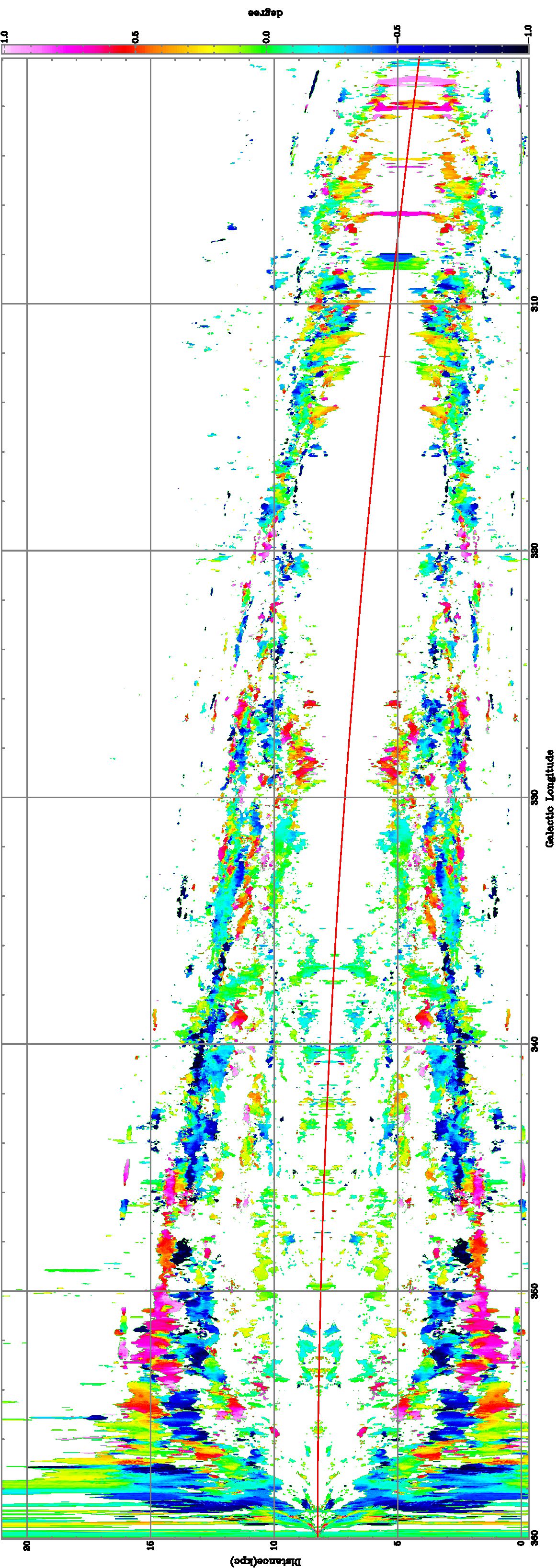}}
\caption{\footnotesize Similar \ld\ maps to Fig.\,\ref{bgt-ld0} (i.e., using the BGT model) but for the 1st latitude moment (mean latitude $\bar{b}$) in the \tco\ ({\em top}) and \nco\ ({\em bottom}) data. $$ $$
\label{both-ld1}}
\end{sidewaysfigure*}

\begin{sidewaysfigure*}[h]
\vspace{10mm}
\centerline{\includegraphics[angle=-90,scale=0.84]{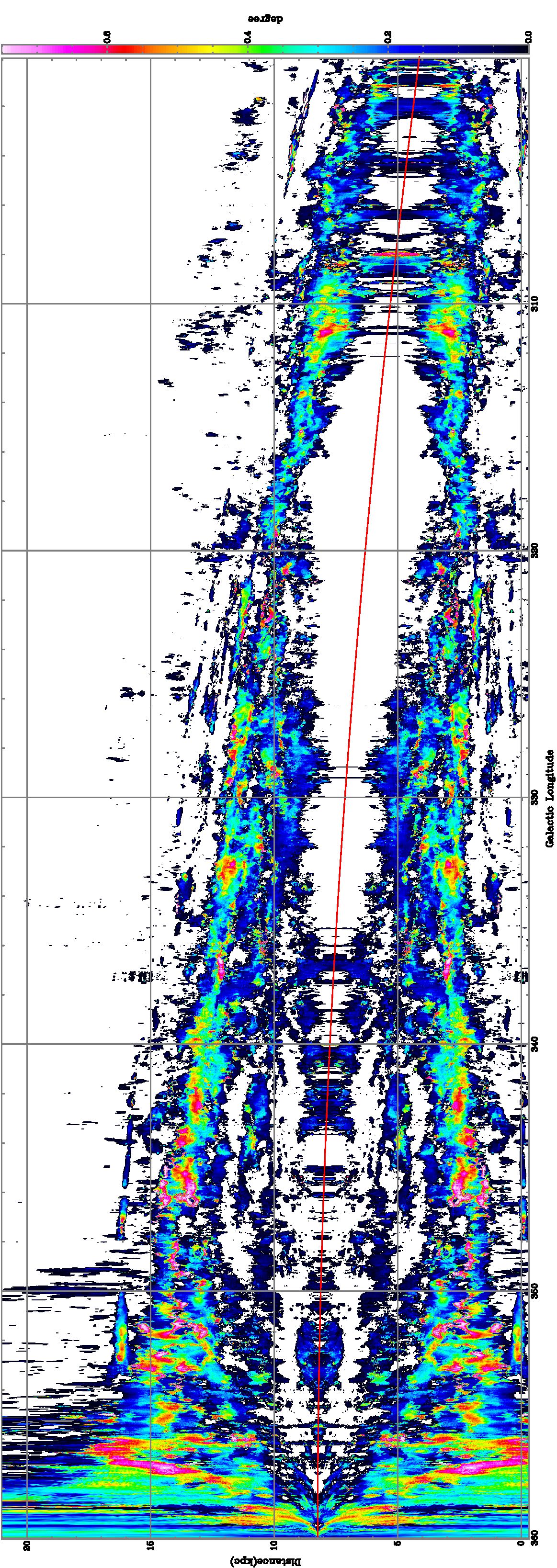}\hspace{0mm}}
\vspace{-3mm}
\centerline{\includegraphics[angle=-90,scale=0.84]{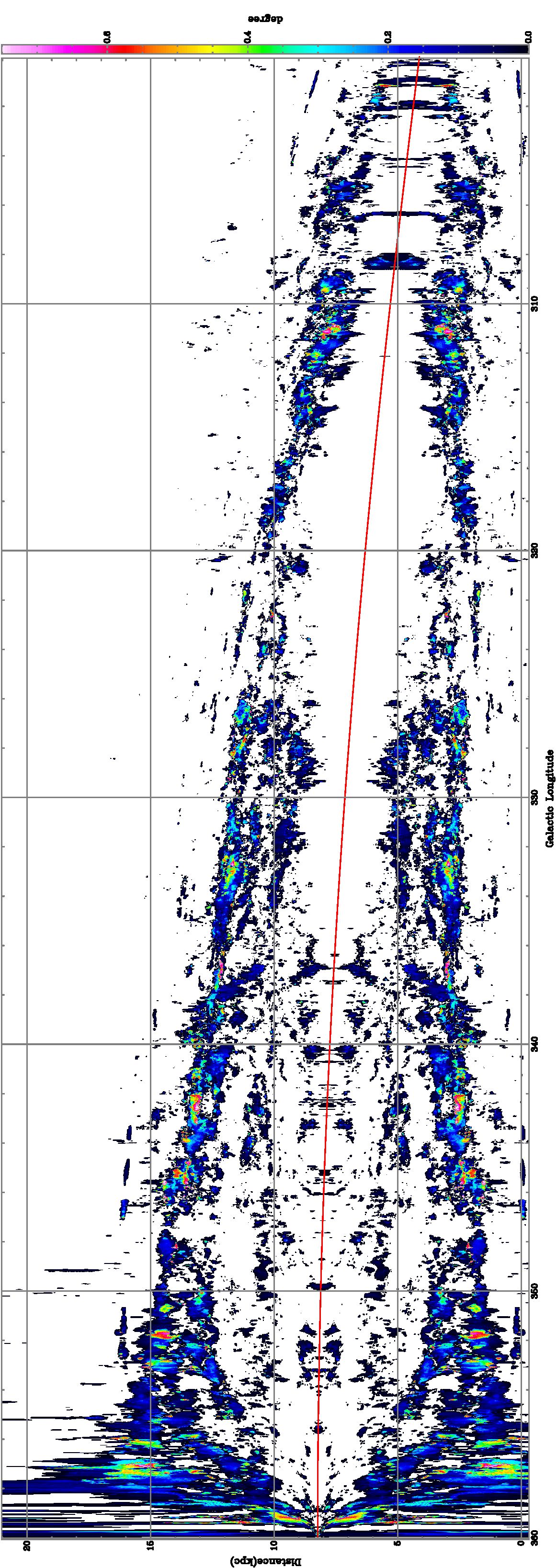}\hspace{0mm}}
\caption{\footnotesize Similar \ld\ maps to Fig.\,\ref{both-ld1} but for the 2nd latitude moment (latitude dispersion $\sigma_b$) in the \tco\ ({\em top}) and \nco\ ({\em bottom}) data. $$ $$
\label{both-ld2}}
\end{sidewaysfigure*}

\vspace{1mm}At this point, we omit any further discussion of other distance models, and continue in \S\ref{height} with the significance of higher-moment \ld\ maps shown in {\color{red}Figures \ref{both-ld1}} (mean latitude $\bar{b}$) and %{\color{red}\ref{12co-YX1}} %{\color{red}\ref{ZM-YX1}} 
{\color{red}\ref{both-ld2}} (latitude dispersion $\sigma_b$) %{\color{red}\ref{12co-YX2}} %{\color{red}\ref{ZM-YX2}}
which were constructed using only the preferred BGT model.

% Figure C8: 12co-YX0 with IAU

% Figure C9: ZM-YX0 with IAU
\begin{figure*}[h]
\centerline{\includegraphics[angle=0,scale=0.923]{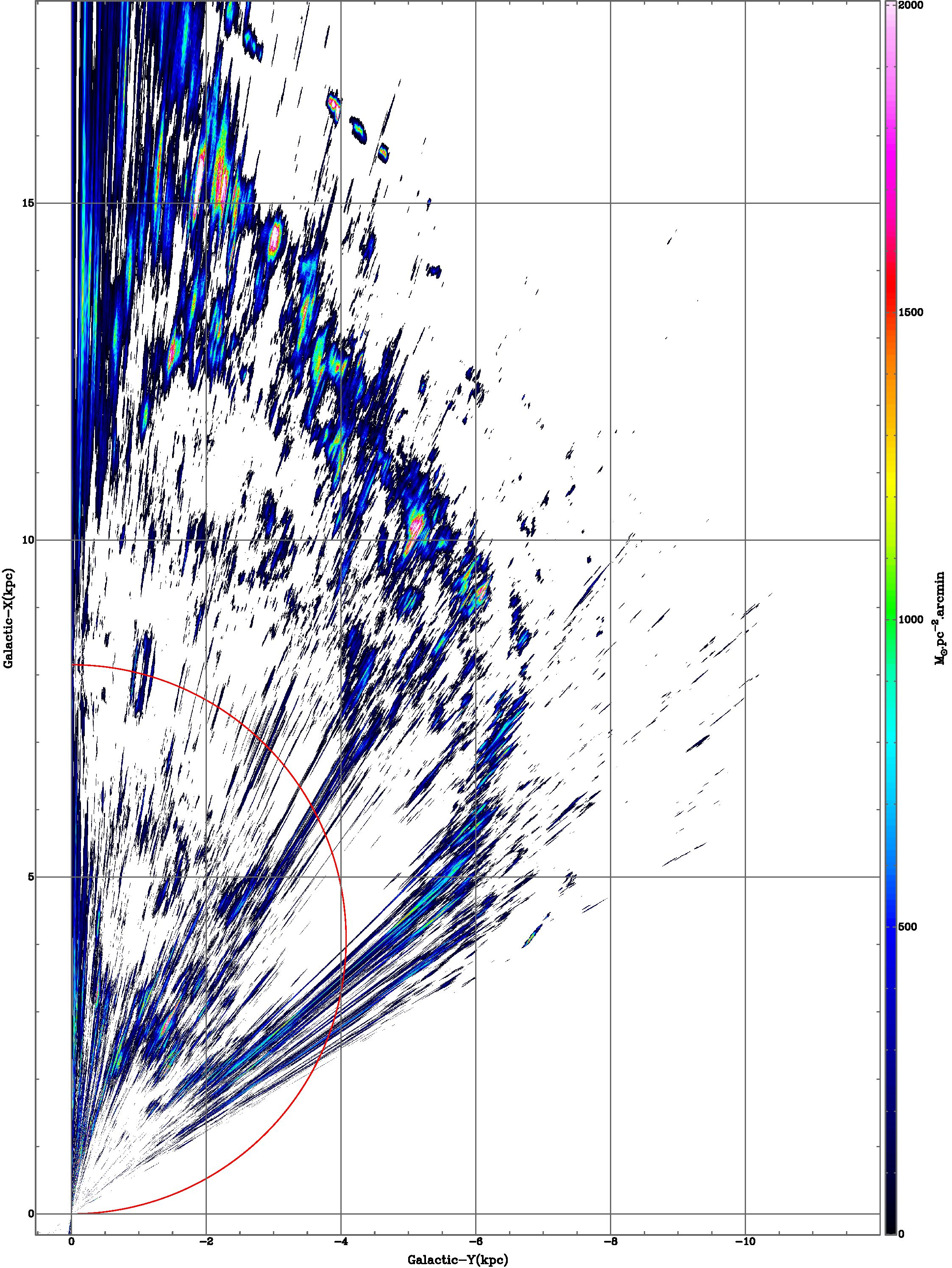}}
\vspace{-2mm}
\caption{\footnotesize Same as Fig.\,\ref{12co-iau-YX0} but for the \nco\ data. $$ $$
\label{ZM-iau-YX0}}
\end{figure*}

% Figure C10: 12co-YX0 with BGT
\begin{figure*}[h]
\centerline{\includegraphics[angle=0,scale=0.92]{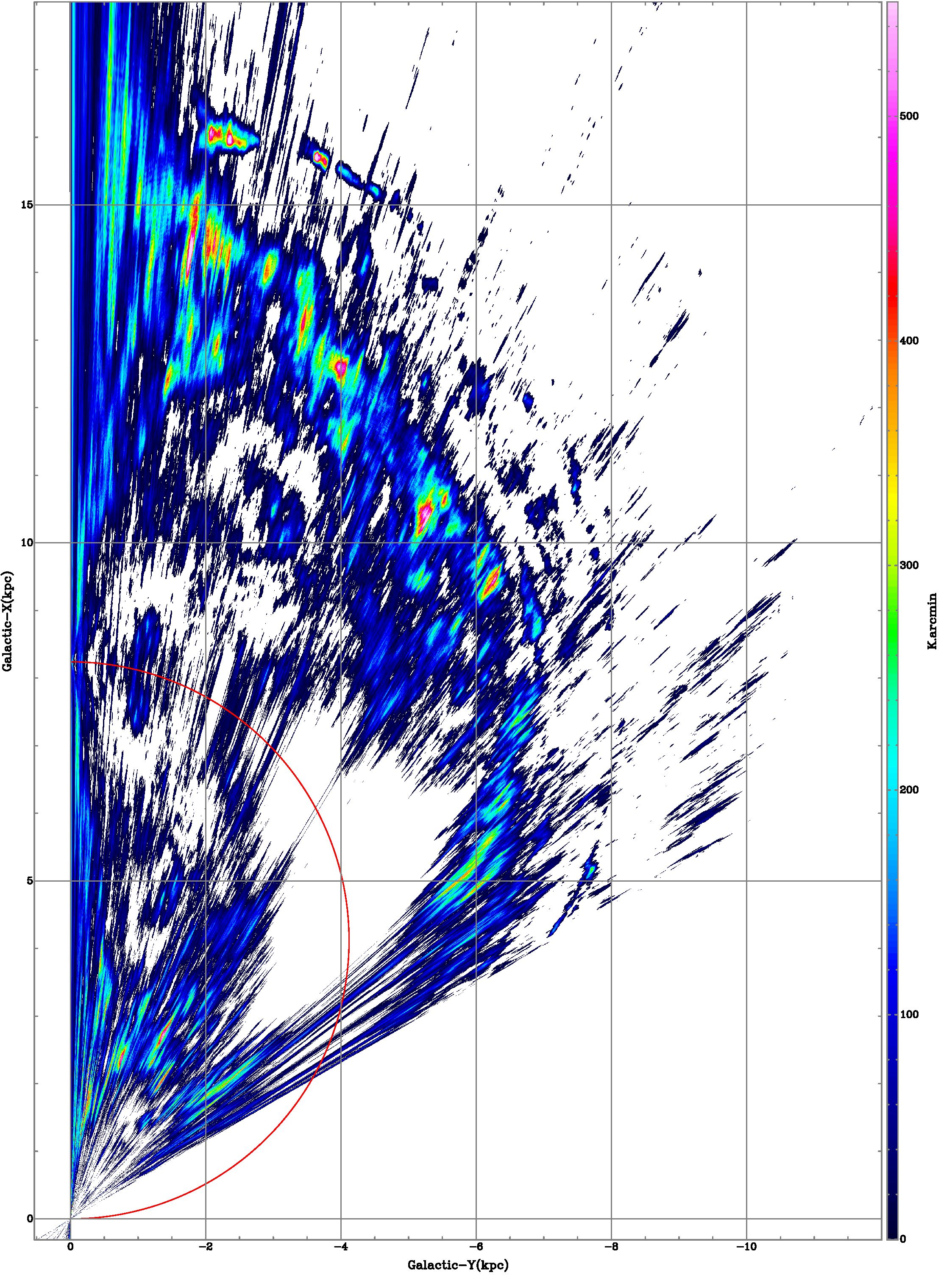}}
\vspace{-2mm}
\caption{\footnotesize Same as Fig.\,\ref{12co-iau-YX0} but using the BGT rotation model (Fig.\,\ref{full-lv0-12coZM-kHbgt} contours), where the Galactic Centre is at coordinates (8.24,0). $$ $$
\label{12co-bgt-YX0}}
\end{figure*}

% Figure C11: ZM-YX0 with BGT
\begin{figure*}[h]
\centerline{\includegraphics[angle=0,scale=0.92]{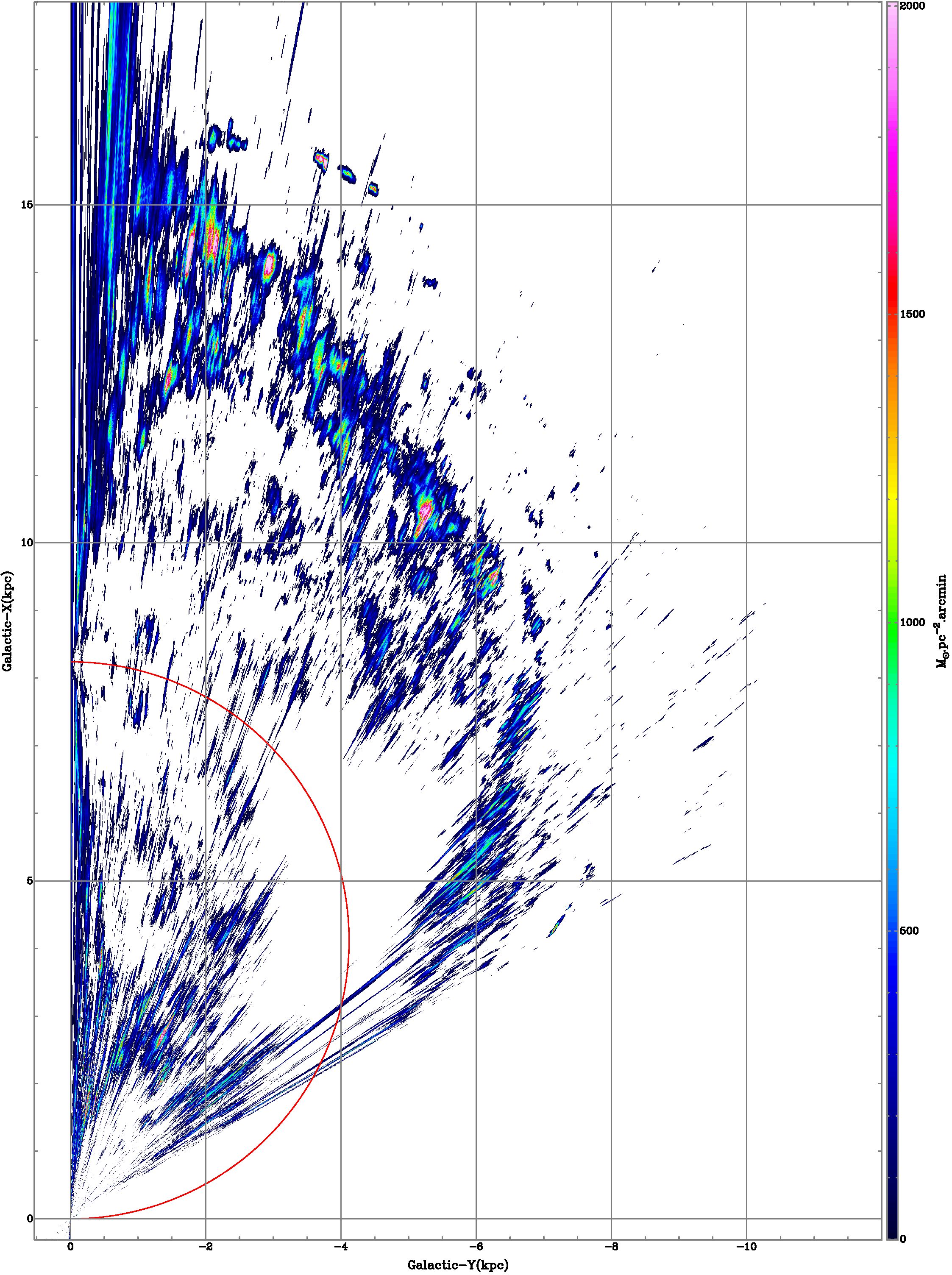}}
\vspace{-2mm}
\caption{\footnotesize Same as Fig.\,\ref{12co-bgt-YX0} but for the \nco\ data. $$ $$
\label{ZM-bgt-YX0}}
\end{figure*}

% Figure C12 -- 12co+ZM ld1 with BGT

% Figure C13 -- 12co+ZM ld2 with BGT

%{\em (4 nore figures 12co-YX1, ZM-YX1, 12co-YX2, ZM-YX2 would be 14--17, but prolly not)}
%{\em So prolly 8 figures total for this subsection}

%\clearpage

%%%%%%%%%
%   Section C3  %
%%%%%%%%%
\subsection{Height Distributions}\label{height}

\vspace{1mm}Having derived a proper distance scale by deprojecting the full \lbv\ cubes into \lbd\ or \xyz\ space allows the second step in this global analysis, which is to convert $b$ in the cubes to $z$, the physical height above or below the Galactic Plane.  This rescaling can be applied equally to the various 2D latitude moments as well.

\vspace{1mm}We consider first the height conversions for latitude-integrated maps (0th moments), where we must be a little careful with units, and how to interpret them.  For the spectral line cubes like \tco, the native voxel units are in brightness [K].  While the common integrated intensity (0th moment over $V$ as a function of $l$,$b$) gives the usual [K\,\kms] units, the latitude integral (0th moment over $b$ as a function of $l$,$V$) will integrate to [K\,arcmin] or other angular unit.  For the physical parameter data, excitation temperature \tex\ or opacity $\tau$ cubes will give similar \lv\ moments, with units [K\,arcmin] or [ratio\,arcmin].  For \nco\ or equivalently $\Sigma$, however, since the voxel unit is [molecules\,m$^{-2}$] or [M\solar\,pc$^{-2}$], the native integral over $b$ will have units [M\solar\,pc$^{-2}$\,arcmin] or similar.  These units can be seen in the 0th moment Figure labels so far.  The point is that these 2D moments are still proper pixel-by-pixel functions of the \lv, \ld, or \xy\ coordinates, but summed over all corresponding $b$ voxels from the cubes with non-zero values.

\vspace{1mm}When we change these from latitude integrals to height integrals, however, we must first convert the per-voxel d$b$ from whatever angular unit to radians, then multiply by distance $d$ so that we get a properly-scaled height integral/sum: $\int$$Q$d$z$ = $\int$$Qd$\,d$b$, for quantity $Q$.  The units for the $T_b$, \tex, or $\tau$ \ld\ maps will then be [K\,pc] or [ratio\,pc] as expected, but the column/surface density \ld\ maps will have units [M\solar\,pc$^{-2}$\,pc] = [M\solar\,pc$^{-1}$].  The latter is now a linear mass or column density $\Lambda$ (e.g., M\solar\,pc$^{-1}$ in the $l$ direction), summed through the Galactic Plane at that \ld\ or \xy\ coordinate, i.e., $\Sigma$ integrated over the height $z$ in the data cube.

\vspace{1mm}This does some unexpected things to the $\Lambda$ data values in the \ld\ or \xy\ maps.  To see this, note that column or surface densities, as observed, are distance-independent quantities: a well-resolved cloud of uniform $\Sigma$, size $R$, and distance $d$ will be observed to have the same $\Sigma$ at 2$d$, but size $R$/2 (assuming it is still well-resolved).  That is, the $\Sigma$ value per pixel will be unchanged, as will the pixels' angular size.  But the pixels' physical scale at 2$d$ is twice that at $d$, or 4$\times$ the area, while there will only be 1/4 the number of pixels spanning the cloud.  Thus, the cloud will integrate to exactly the same observed mass at $d$ as at 2$d$, as it must.

\vspace{1mm}For \nco\ or $\Sigma$ data in \lbd\ space, the pixel values are individually derived from the per-pixel \lbv\ radiative transfer solutions \citep{b15}.  Projecting a $\Sigma$\lbv\ cube to \lbd\ space does not change the pixel units, nor do they change when converting a $\Lambda$\lv\ map to \ld.  But they do change when projecting the $b$ axis to $z$, since the physical scale subtended by the fixed angular pixel size must scale with the kinematic distance.  Thus, in an \ld\ map previously integrated over all $b$, at each $d$ coordinate the pixel's fixed $b$ size will scale to a different $z$ size, and so the unit [M\solar\,pc$^{-2}$\,arcmin] must change to [M\solar\,pc$^{-2}$\,arcmin $\times$ (pixel's pc/arcmin scale at $d$)].  Obviously, this pc/arcmin conversion increases with $d$, so the \ld\ map in $z$ units will, beyond a certain distance $d_1$ where the conversion factor = 1, have higher pixel values than the original \ld\ map in $b$ units, and lower pixel values than the original within that same distance.  The conversion factor is 1 where $d_1$ (in pc) equals the angular conversion factor to radians; e.g., when the angular unit is arcmin, $d_1$ = (60 arcmin/deg)*(180/$\pi$) = 3.44\,kpc.

\vspace{1mm}In particular, this means that {\bf all} far-distance $\Lambda$ values in an \ld\ map are numerically larger than their near-distance counterparts, when converted from a $b$ integral to a $z$ integral.  This is {\em consistent} with $\Sigma$ per pixel being conserved with distance, because we have turned an angular integration into a linear one.  For example, consider a cloud structure placed at its near-$d$ with angular size $b$, corresponding to a physical size $z$, and integrating to a mass $M$.  If its far-mirror cloud is at 2$d$, it is still observed to have size $b$, but this is now 2$z$, integrating to a mass 4$M$, or to 2$M$ per $l$-angle.

\vspace{1mm}For higher-moment latitude maps such as the intensity-, column-, or mass-weighted $\bar{b}$ or $\sigma_b$, the units are in the native latitude units and have their much more intuitive meaning of $Q$-weighted mean latitude or latitude dispersion.  Visually, however, the higher-moment \ld\ maps in $z$ units will have the same numerical conversion as described above, compared to the same maps in $b$ units.

\begin{sidewaysfigure*}[h]
\centerline{\includegraphics[angle=-90,scale=0.84]{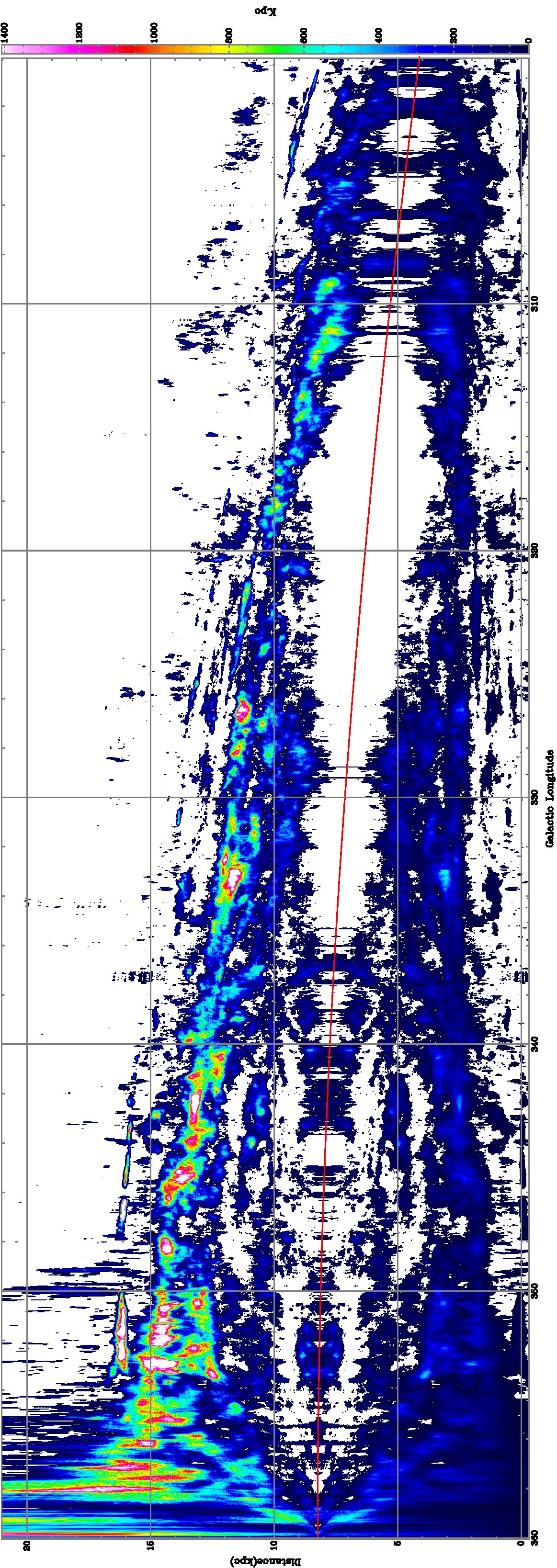}\hspace{0mm}}
\vspace{-3mm}
\centerline{\includegraphics[angle=-90,scale=0.84]{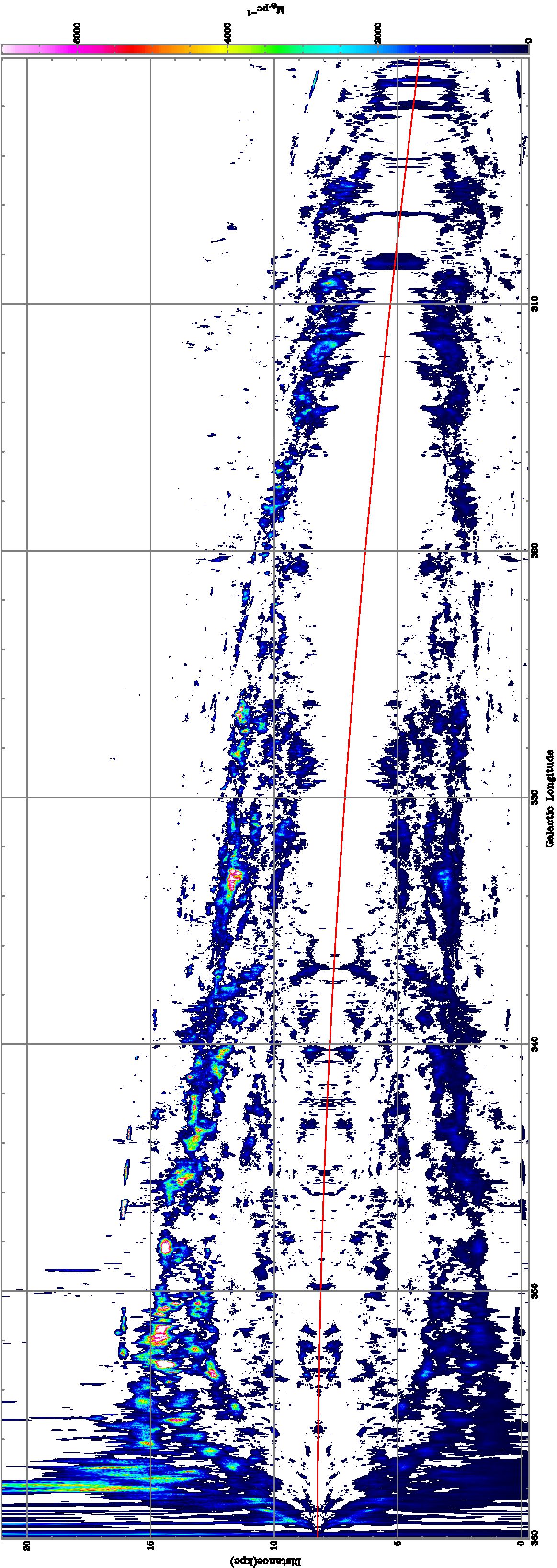}\hspace{0mm}}
\caption{\footnotesize Similar \ld\ maps to Fig.\,\ref{bgt-ld0} (integrated over all $b$) but where the latitude integration has been rescaled to height via the distance ($y$ axis).  This gives the average line brightness or linear density through the Galactic Plane.  ({\em Top}) \tco\ and ({\em bottom}) \nco\ data. $$ $$
\label{both-ld0h}}
\vspace{-10.5mm}
\end{sidewaysfigure*}

\begin{sidewaysfigure*}[h]
\centerline{\includegraphics[angle=-90,scale=0.84]{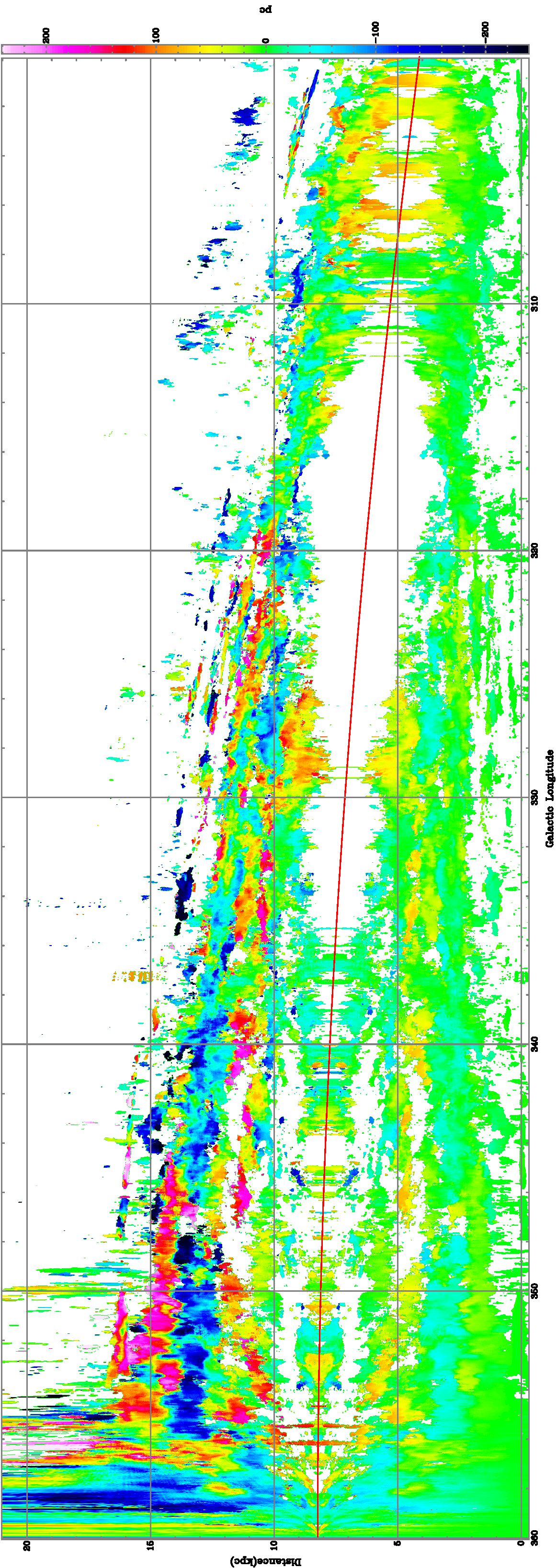}\hspace{0mm}}
\vspace{-3mm}
\centerline{\includegraphics[angle=-90,scale=0.84]{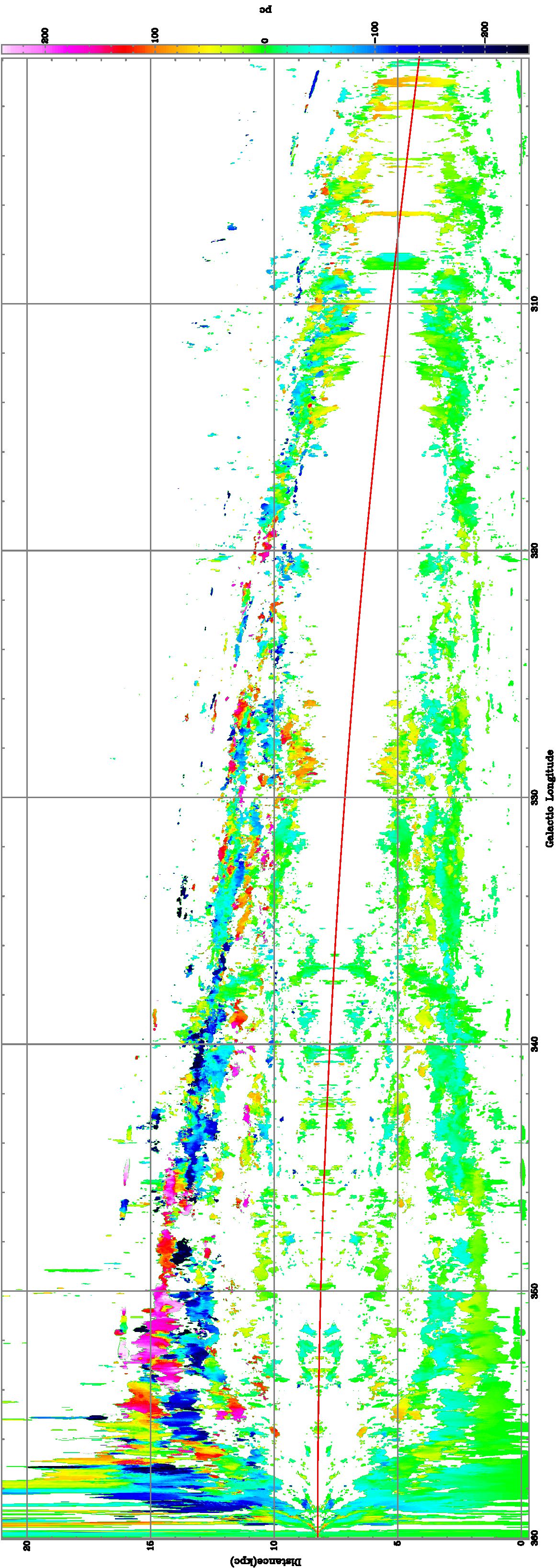}\hspace{0mm}}
\caption{\footnotesize Similar \ld\ maps to Fig.\,\ref{both-ld1} (mean latitude $\bar{b}$) but where the latitude has been rescaled to height via the distance ($y$ axis).  This gives a height distribution which becomes progressively wider as $d$ increases, $\pm$300\,pc or more at far-kinematic distances, even while it remains small ($\pm$50\,pc) where $d$ \lapp\ 3\,kpc.  ({\em Top}) \tco\ and ({\em bottom}) \nco\ data. $$ $$
\label{both-ld1h}}
\vspace{-11mm}
\end{sidewaysfigure*}

\begin{sidewaysfigure*}[h]
\centerline{\includegraphics[angle=-90,scale=0.84]{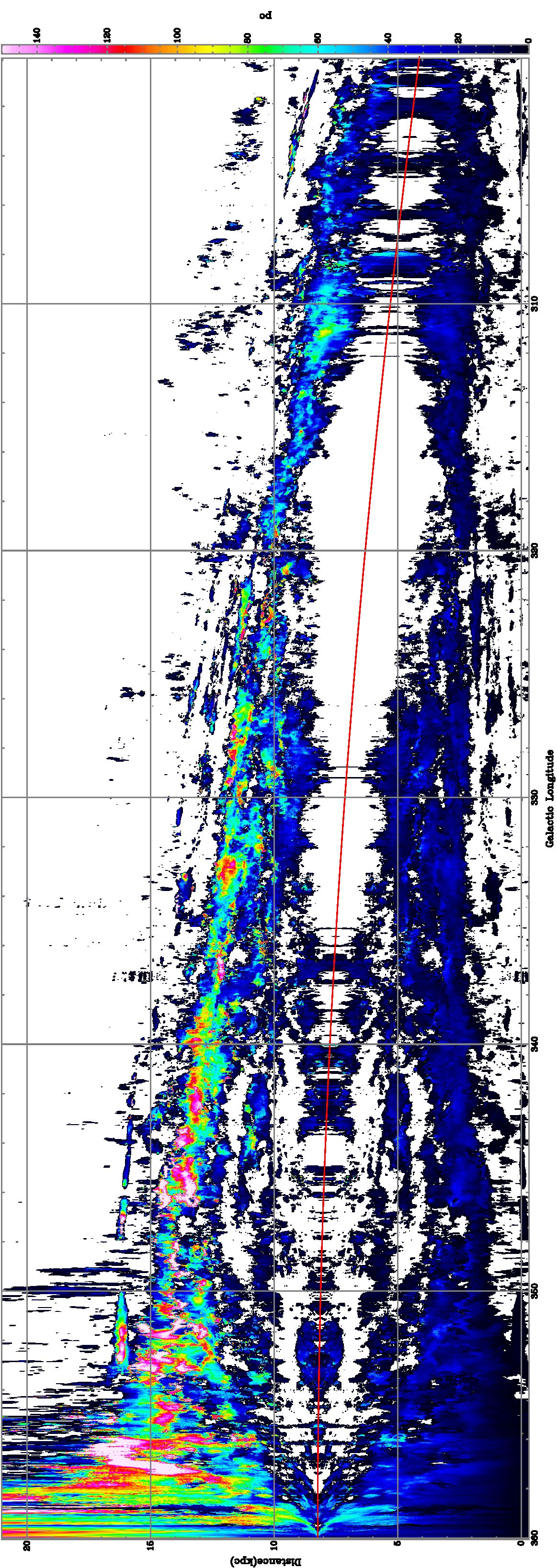}}
\vspace{-3mm}
\centerline{\includegraphics[angle=-90,scale=0.84]{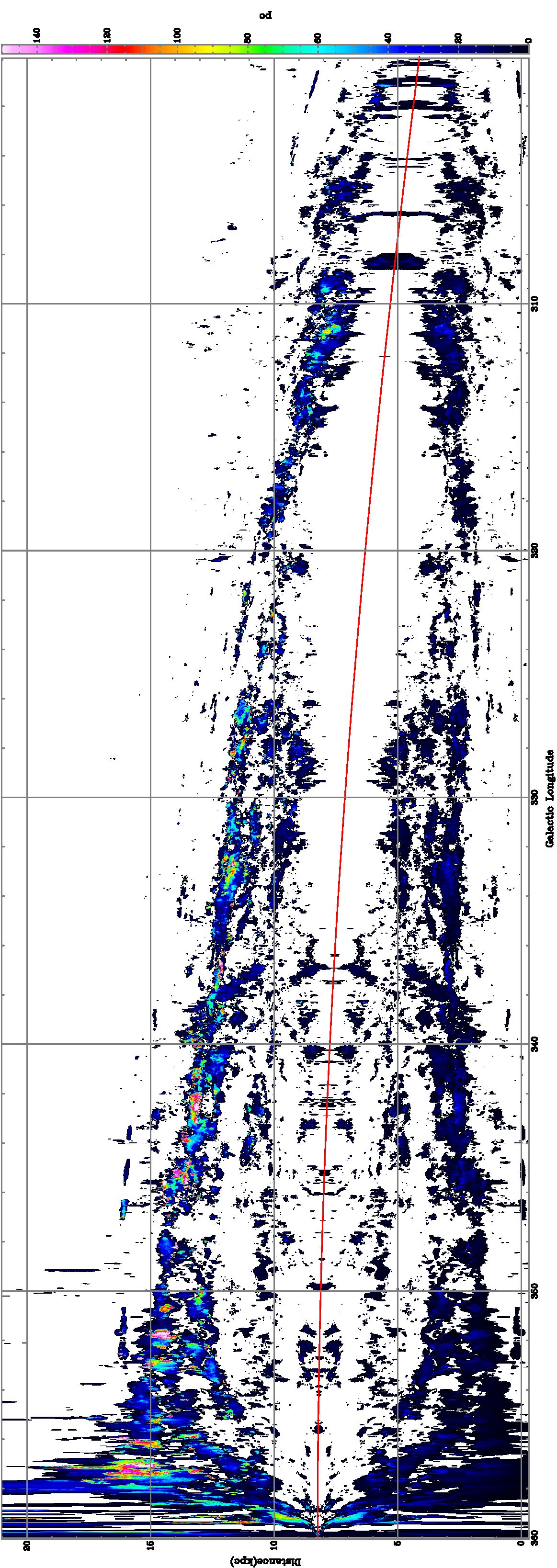}}
\caption{\footnotesize Similar \ld\ maps to Fig.\,\ref{both-ld2} (latitude dispersion $\sigma_b$) but where the angular dispersion has been rescaled to a physical size via the distance ($y$ axis), as in Fig.\,\ref{both-ld1h}.  This gives a height dispersion which becomes progressively wider as $d$ increases, up to $\sim$230\,pc at far-kinematic distances, even while it remains small (\lapp30\,pc) where $d$ \lapp\ 3\,kpc.  ({\em Top}) \tco\ and ({\em bottom}) \nco\ data. $$ $$
\label{both-ld2h}}
\vspace{-11mm}
\end{sidewaysfigure*}

\vspace{1mm}With this understanding, we show in {\color{red}Figure \ref{both-ld0h}} %to {\color{red}\ref{ZM-YX0h}} 
the moment-0 height-integrated \ld\ %and \xy\ 
maps (we omit the corresponding \xy\ maps until \S\ref{filtered} to focus more clearly on the near/far masking here and in \S\ref{nearfar}).  As might be expected, these look quite similar to the latitude-integrated maps in Figure \ref{bgt-ld0}, %\ref{12co-bgt-YX0}, and \ref{ZM-bgt-YX0}, 
except that the numerical values increase linearly with distance from the Sun due to the increased vertical scaling in the integrals.  This is perhaps more intuitive in the higher moment maps {\color{red}Figures \ref{both-ld1h}} \& {\color{red}\ref{both-ld2h}}, %{\color{red}\ref{ZM-YX2h}}, 
where the heights simply scale with distance for a fixed $\pm$1\degr\ range in latitude.  %{\em (will be 18--20?  the other 6 figures YX0h, YX0h, YX1h, YX1h, YX2h, YX2h would be 21--26, but prolly not)}

% Figure C14: 12co- and ZM-ld0h with BGT

% Figure C15: 12co- and ZM-ld1h with BGT

% Figure C16: 12co- and ZM-ld2h with BGT

\vspace{1mm}One should immediately note that for most pixels at near distances, the typical $\bar{z}$ (Fig.\,\ref{both-ld1h}) and $\sigma_z$ (Fig.\,\ref{both-ld2h}) values are both generally comparable with those expected for a thin molecular layer, \lapp50\,pc or so.  In contrast, many of the far distances give values for each typically much larger than would be expected from such a thin layer.  Unsurprisingly, this means that our data probably sample the near-side clouds much more extensively than the far-side population, so that for most \lv\ pixels, we should more likely prefer them located below the tangent curve mirror in an \ld\ diagram than above it.  We quantify such a procedure next.

%\clearpage

%%%%%%%%%
%   Section C4  %
%%%%%%%%%
\subsection{Near/Far Masking}\label{nearfar}

\vspace{1mm}With the new height scalings apparent in Figures \ref{both-ld0h}--\ref{both-ld2h}, %\ref{ZM-YX2h}, 
we are led to the third step in this analysis, that of using the height distributions themselves as a statistical basis for near-far distance discrimination.  This was inspired by the concept of a distance probability density function (DPDF) pioneered by \cite{eb13}.  In their case, they used structure-finding algorithms to associate spectroscopic kinematic distances with their continuum survey sources, and a sophisticated Bayesian analysis of multiple distance priors to obtain ``best'' distance estimates for each identified source.  Here, we do something much simpler: we start with the spectroscopy and have all the kinematic information we need (and structural information too, if we wanted to use it), which is evaluated directly (\S\S\ref{lsr}--\ref{mwmaps}) and separately from the distance filtering.  We then develop a binary near/far distance likelihood estimator based on only one prior, the scale height of the molecular layer.  This estimator is evaluated at each \ld\ or \xy\ pixel from the inherent height ($\bar{z}$) and size ($\sigma_z$) information in the data, meaning that any pixel-to-pixel structures that already exist in the data are treated completely agnostically.

\vspace{1mm}The molecular layer of the Galactic Plane is known to be physically thin, and at least in the inner Galaxy (interior to the Solar circle), it doesn't have significant warps of the $\sim$kpc scale seen in the outer Galaxy: it is globally flat.  Thus, \cite{r19} found the exponential scale height of the massive young SFR masers they studied to be $z_{\rm sc}$ = 19\,pc.  Young Population I samples give only slightly larger scale heights, e.g., $z_{\rm sc}$ $\sim$ 45\,pc for OB stars in the solar neighbourhood \citep{r00}, while older Pop I samples can have $z_{\rm sc}$ \gapp\ 100\,pc or more, depending on the sample \citep{mb81}.  Our maps of $\sigma_{z}$ (e.g., see Fig.\,\ref{both-ld2h}) give values commensurate with an extreme Pop I value, especially after the near/far filtering described here, so we formulate our approach to scale with \cite{r19}'s $z_{\rm sc}$.\footnote{Perhaps puzzlingly, the value inferred by \cite{dht01} is ``large,'' $z_{\rm sc}$ $\sim$ 90\,pc.  We believe such values overstate the true molecular scale height for at least two reasons.  First, the angular resolution of the CfA survey is 7--14$\times$ larger than ours, and as we can see in our maps, there is a huge amount of structure below that scale.  Second, even in our maps we can see that \tco\ gives larger height dispersions $\sigma_{z}$ than does the true \nco\ distribution.  A possible third reason for this difference is discussed in \S\ref{filtered}.}  Regardless of the actual number, we posit that the Galaxy's molecular clouds hew to the kinematically coldest Pop I sample, and so should have $z_{\rm sc}$ in the lower end of the Pop I range.  Although we use \cite{r19}'s value for our numerical near/far filtering, to some extent this is fungible through how we weight the inputs (i.e., both the first and second height moments) to the masking.

\begin{figure}[t]
\hspace{68mm}{\includegraphics[angle=0,scale=0.21]{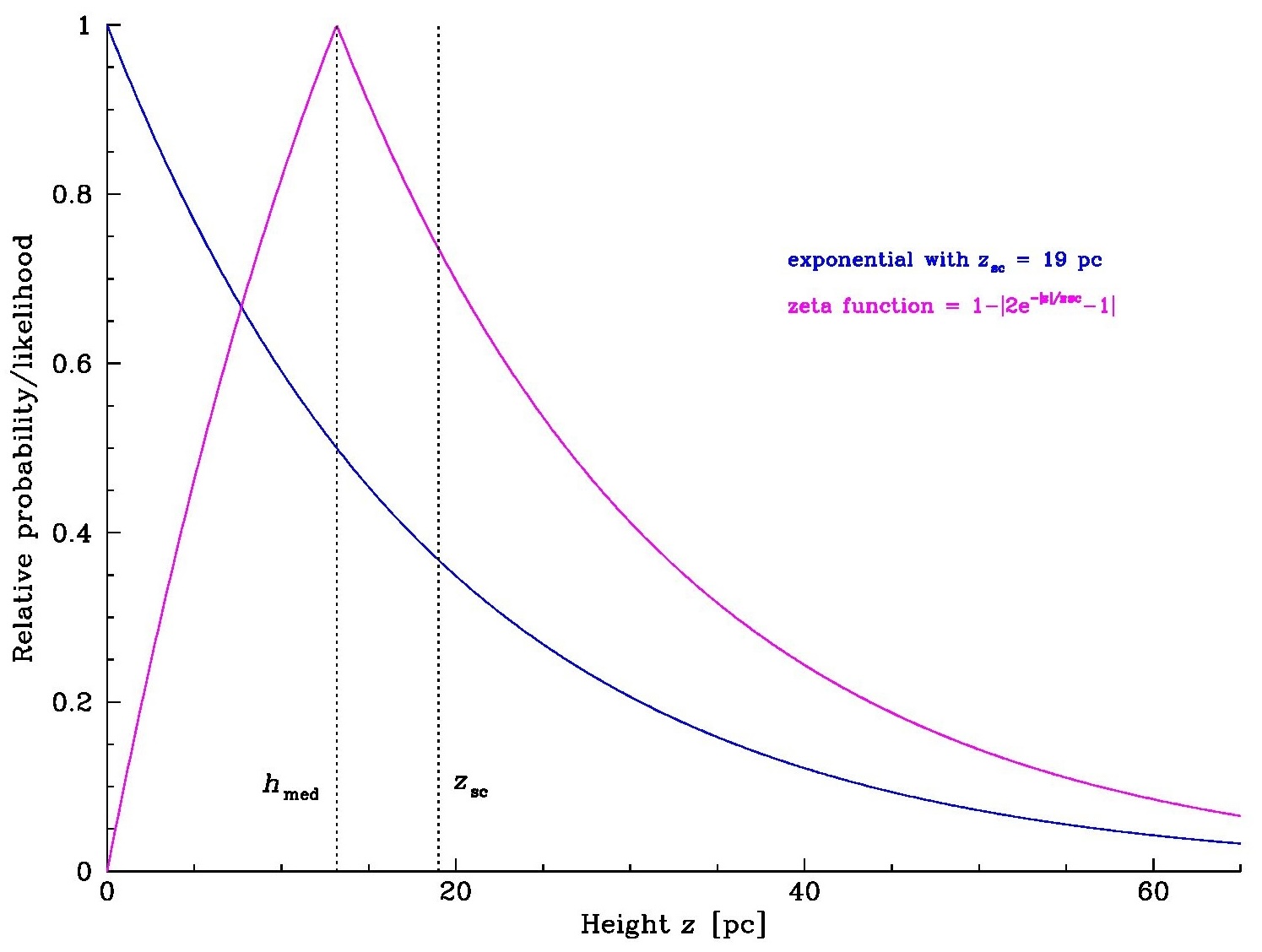}}

\vspace{-55mm}
\parbox[top]{60mm}{\caption{ \label{scalehts}}
\footnotesize Exponential height distribution of molecular clouds (blue) and corresponding $\zeta$ function (magenta) of the relative likelihood of observing a given cloud at height $z$.}

\vspace{33mm}
\end{figure}

\vspace{1mm}To filter by height, we want a function $\zeta$ for the relative likelihood of a cloud having a mean height $\bar{z}$ above or below the GP, given its assumed intrinsic Pop I height distribution.  We assume the distribution is a simple exponential
\begin{equation}
	P(z) = \frac{1}{z_{\rm sc}}e^{-|z|/z_{\rm sc}}~~,				% EQ.C1
\end{equation}
where the probability of a cloud lying at $z$$\pm$d$z$\,pc above or below the GP is $P$($z$)d$z$, and $\int_{0}^{\infty}$$P$d$z$ = 1 as written.  For such a distribution, the median height (where half of all clouds lie closer to the GP than this height, and the other half lie further from the GP) is $h_{\rm med}$ = ln(2)\,$z_{\rm sc}$.  Thus if $z_{\rm sc}$ = 19.0\,pc, $h_{\rm med}$ = 13.2\,pc.  An unnormalised version of this function is shown as a blue curve in {\color{red}Figure \ref{scalehts}}.

\vspace{1mm}To discriminate between near and far kinematic distances based on cloud or pixel height, we want to use the given $P$ to estimate whether a near distance is more or less likely than a far distance.  We therefore suppose that the more likely height for a pixel is the one that, near or far, is projected to be closer to the median height than the other.  Thus, distances which give heights that are either $\ll$$h_{\rm med}$ or $\gg$$h_{\rm med}$ are considered less likely than distances with heights nearer to $h_{\rm med}$.  We quantify this as the $\zeta$ function, formed by integrating Eq.\,C1 both from 0 to $h_{\rm med}$ and from $\infty$ to $h_{\rm med}$, and joining those segments at $h_{\rm med}$:
\begin{equation}
	\zeta(z) = 1-|2e^{-|z|/z_{\rm sc}}-1|~~.				% EQ.C2
\end{equation}
This is shown as a magenta curve in Figure \ref{scalehts}, where the choice of $h_{\rm med}$ as the most likely height is encoded by the factor 2 in the formula.  This is evaluated for a given \lv\ pixel at each implied near/far distance and height, and the results compared.  Numerically, whichever height (near or far) gives a larger $\zeta$ indicates the more likely distance.

% Figure C17: scaleheights  (C21 or C27 with the prior YX figures)

\vspace{1mm}Note that we do not use Eq.\,C1 directly to estimate the likelihood of a cloud being at height $z$, because this would universally favour near distances for all $b$ and disfavour all far distances, since far clouds would all have higher $z$, and thus, lower likelihoods.  This is true also of {\em any} monotonically decreasing function of $z$: in order to engineer any near/far discrimination, we need a function like $\zeta$ that is anchored to 0 at both $z$ = 0 and $\infty$, and has a peak value somewhere in between.

% Figure C18: 12co- and ZM-ld1h with zeta masking  (C28 with the prior YX figures)
\begin{sidewaysfigure*}[h]
\centerline{\includegraphics[angle=-90,scale=0.84]{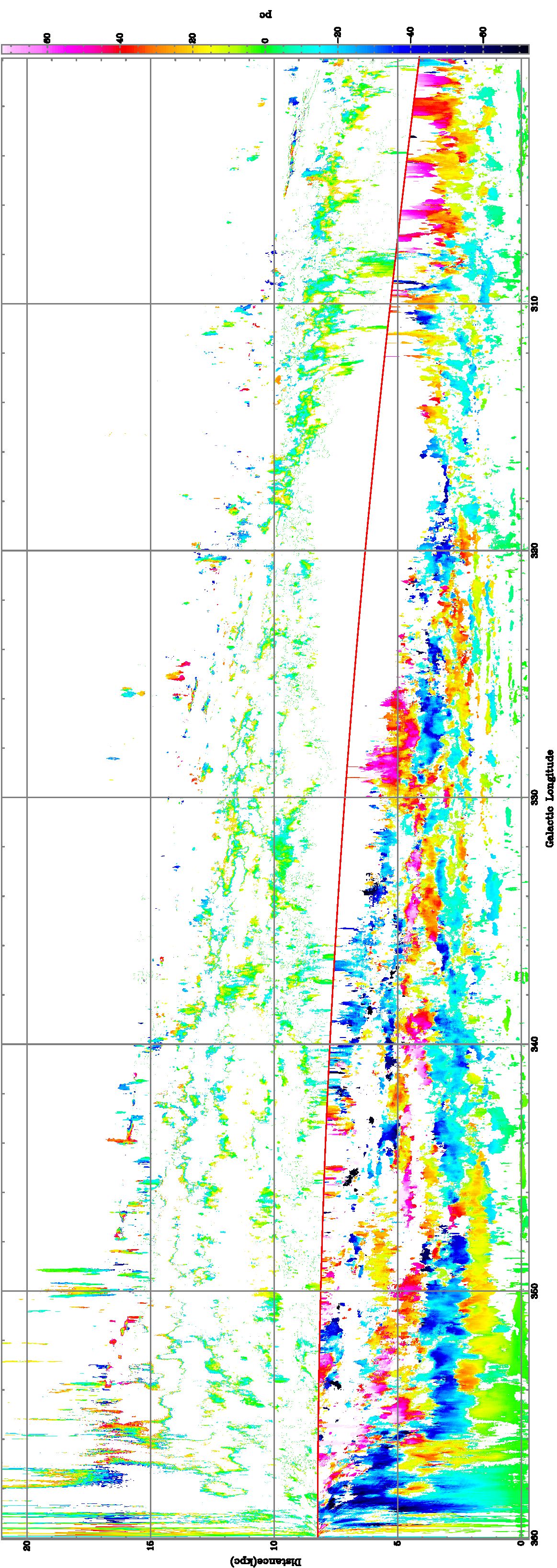}\hspace{0mm}}
\vspace{-3mm}
\centerline{\includegraphics[angle=-90,scale=0.84]{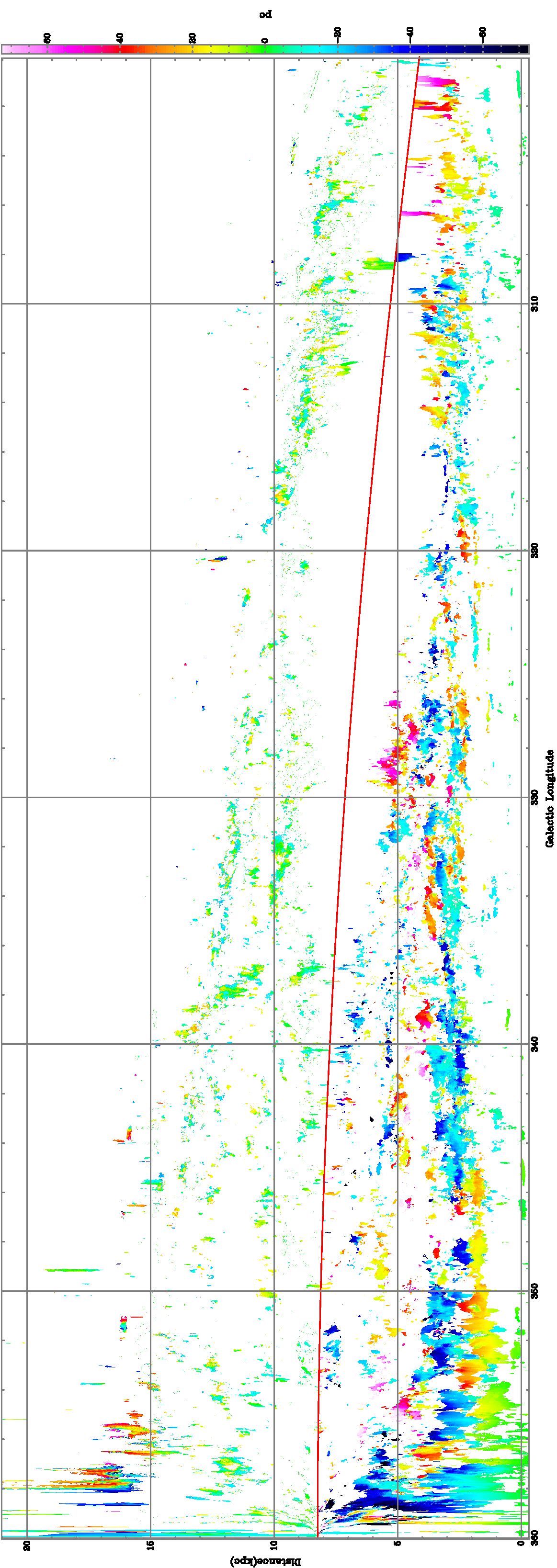}\hspace{0mm}}
\caption{\footnotesize Similar \ld\ maps to Fig.\,\ref{both-ld1h} (mean height $\bar{z}$) but where the heights have been masked via the $\zeta$ function.  This gives a mean height distribution which is much more comparable to the known thickness of the Galaxy's molecular layer, a few 10s of pc, at both near (below the tangent-mirror curve) and far (above the curve) distances.  ({\em Top}) \tco\ and ({\em bottom}) \nco\ data. $$ $$
\label{both-ld1h-b1m}}
\vspace{-11mm}
\end{sidewaysfigure*}

% Figure C19: 12co- and ZM-ld2h with zeta masking  (C29 with the prior YX figures)
\begin{sidewaysfigure*}[h]
\centerline{\includegraphics[angle=-90,scale=0.84]{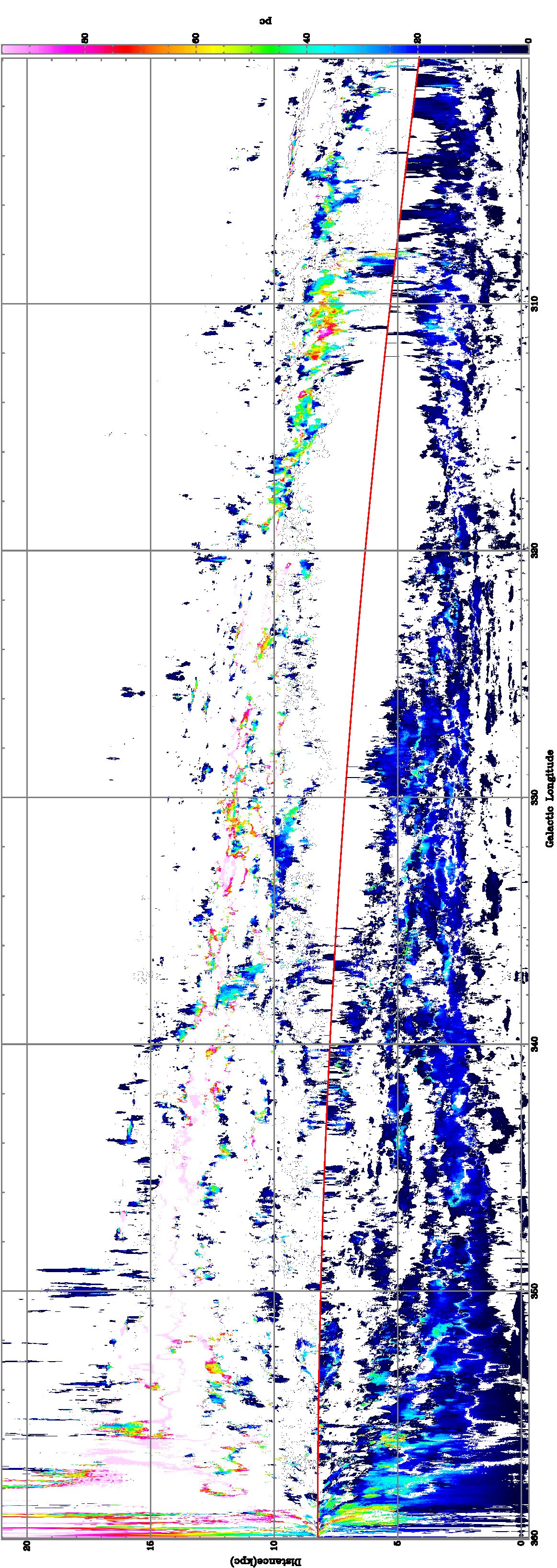}}
\vspace{-3mm}
\centerline{\includegraphics[angle=-90,scale=0.84]{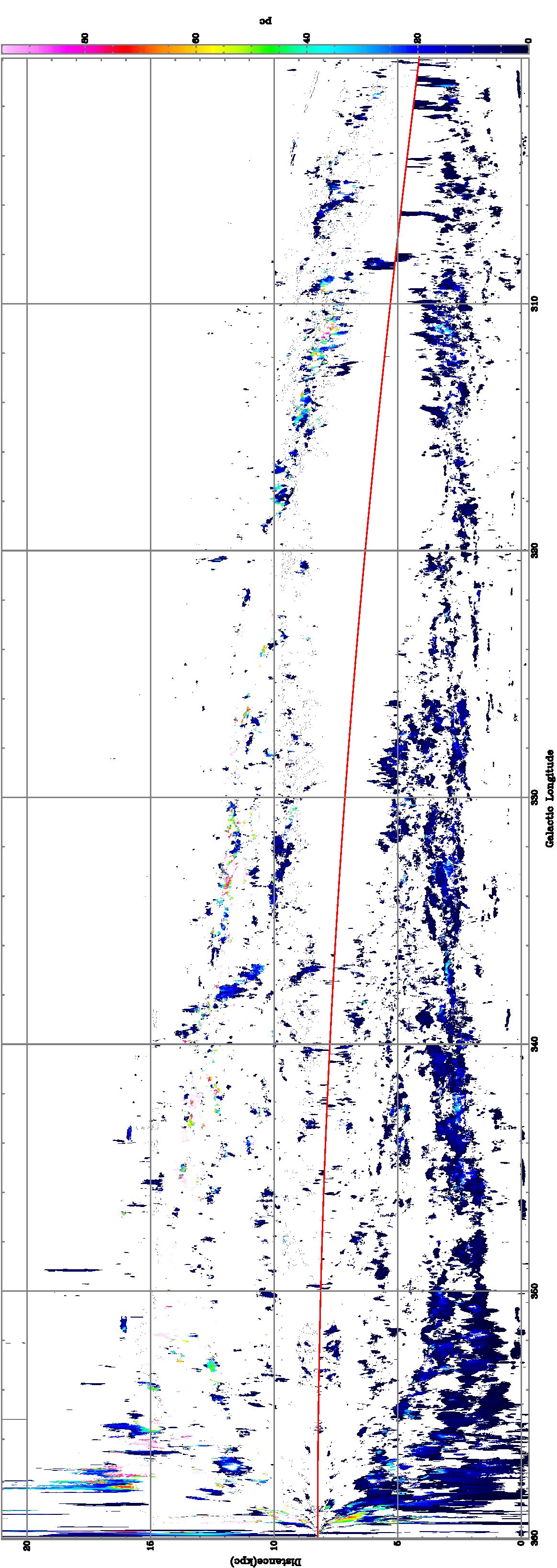}}
\caption{\footnotesize Similar \ld\ maps to Fig.\,\ref{both-ld2h} (height dispersion $\sigma_z$) but where the sizes have been masked via the $\zeta$ function based on the mean heights $\bar{z}$.  While this masks out most unrealistically large clouds at far distances, some still remain.  ({\em Top}) \tco\ and ({\em bottom}) \nco\ data. $$ $$
\label{both-ld2h-b1m}}
\vspace{-11.2mm}
\end{sidewaysfigure*}

\vspace{1mm}For example, suppose that at a given longitude, the tangent distance is 6\,kpc.  At this longitude, suppose further than a pixel in an \lv\ diagram (such as the mean latitude $\bar{b}$ diagram in Fig.\,\ref{both-ld1}) has a \vlsr\ which indicates either a near distance of 4\,kpc or a far distance of 8\,kpc (recall that near/far distance pairs are symmetric about the tangent distance).  Let us also suppose that this pixel has $\bar{b}$ = 0\fdeg65, in which case the near distance implies a mean height above the GP of $\bar{z}$ = 45\,pc, whereas a far distance implies $\bar{z}$ = 90\,pc.  Being only 3.4$\times$ larger than $h_{\rm med}$, we intuit that the near height makes the near distance more likely on this basis alone, whereas at 6.8$h_{\rm med}$, the far height makes the far distance less likely.  Inserting these heights into the $\zeta$ function, we find that $\zeta$(45) = 0.19 while $\zeta$(90) = 0.018, a roughly 10$\times$ smaller likelihood and comporting well with our intuition.

\vspace{1mm}A more marginal example is also instructive.  Suppose another pixel at the same longitude as the above has \vlsr\ indicating $d$ = 3 or 9\,kpc, and $\bar{b}$ = 0\fdeg13 indicating  $\bar{z}$ = 7 or 21\,pc.  The choice in this case is not intuitively obvious, but we easily compute $\zeta$(7) = 0.62 and $\zeta$(21) = 0.66, showing the far distance is slightly more likely here.

\vspace{1mm}An obvious advantage of this approach is how easily it can be applied to the whole dataset at a stroke, compared to the much more laborious tabulation of ``clouds'' and applying non-trivial astrophysical filters thereto: we show examples in {\color{red}Figures \ref{both-ld1h-b1m}} and {\color{red}\ref{both-ld2h-b1m}}.  This is not to devalue more traditional methods, merely to point our where our alternative performs well.

\vspace{1mm}In contrast, we can identify two ways in which this initial approach is suboptimal.  One is that the method is ``pixelly,'' that is, sometimes subject to local numerical fluctuations between nearby pixels that give unphysically sharp near/far solutions.  The other is that it tends to move a large fraction of clouds at low $\bar{b}$ to far distances.  Both of these effects are noticeable in Figure \ref{both-ld1h-b1m}.  There is also the related issue that the $\zeta$ function uses only the mean height $\bar{z}$ information and not the height dispersion $\sigma_z$ information, with the result that the $\sigma_z$ map (Fig.\,\ref{both-ld2h-b1m}) is less optimally masked than the $\bar{z}$ map (Fig.\,\ref{both-ld1h-b1m}).  That is, more unphysically large structures remain unmasked at far distances.  However, the latter issue can be addressed satisfactorily as described next, albeit by introducing additional parameters.

\vspace{1mm}We need an additional filtering function based on the vertical dispersion $\sigma_z$, which to some extent should also be related to vertical cloud size or thickness.  For example, a cloud with a large $\sigma_z$ but small $\bar{z}$ is more likely to be nearby, not far.  However, we do not have a simple observational size distribution to fall back on.  A principal reason for this is that most large molecular clouds have structure on all scales, from $\sim$100\,pc or more (GMCs) down to the sub-pc clumps, cores, and filaments that are seen to form individual or small groups of stars.  In addition, for many large-scale molecular cloud surveys, the size distribution of clouds they catalogue is strongly dependent on the angular resolution of the telescope used, from several arcminutes to a few arcseconds.  For our purposes, however, we can focus mainly on the middle scales, since our main objectives are (1) to favour placing the middling-large clouds at their near distances, in order to avoid physically unreasonable ``large-far'' clouds, and (2) avoid treating truly large GMCs as structures perpendicular to the Galactic Plane, since they almost always lie along it.  For the same reason as not using $P$ (Eq.\,C1) for $\zeta$ (Eq.\,C2), we therefore do not want a monotonic function which continues to rise to higher likelihoods as the size/dispersion parameter drops to 0, since this would again disfavour all far distances.  Finally, we want a formulation which can be combined with the $\zeta$ function for computational and logical simplicity.

\vspace{1mm}We therefore suppose that, in the ThrUMMS data, clouds will have an overall size distribution that is of the same order as their height distribution.  Obviously, clouds many times taller than $h_{\rm med}$ (i.e., in the $z$ dimension) are not physically likely, while clouds' physical sizes will have a minimum corresponding to ThrUMMS' angular resolution (1\farcm2, or about 1\,pc at 3\,kpc).  Thus, the $\sigma_z$ distribution itself should be a suitable proxy for cloud sizes.  We define 
\begin{equation}
	\zeta^{+}(z) = 1-|2e^{-(|\bar{z}|+w\sigma_{z})/z_{\rm sc}}-1|~~					%% EQ.C3
\end{equation}
similarly to $\zeta$, but which combines the two scaled $b$-moments $\bar{z}, \sigma_{z}$ to evaluate the $\zeta^+$ likelihood.  This will skew the results for $\zeta$ away from far distances where $\bar{z} \ll h_{\rm med}$ but $\sigma_{z} \sim h_{\rm med}$, exactly as desired.  However, the simple sum $|\bar{z}|+\sigma_{z}$ in the exponent is somewhat too aggressive towards eliminating far distances, so we also include a weighting factor $w$$<$1 for the combination.  After a little experimentation (described below), we found that $w$=0.3 was most satisfactory, in the sense that most of the believably flat and narrow features in the $\bar{z}$ and $\sigma_{z}$ maps stayed at their far distances, and most of the believably local clouds stayed local.  Thus, we evaluate the $\zeta$ function at each pixel in any $b$-moment map, including the 0th moments $\Sigma$ or $\Lambda$, with
\begin{equation}
	z \equiv |\bar{z}|+0.3\sigma_{z}~~					%% EQ.C4
\end{equation}
at each corresponding pixel in the $\bar{z}$ and $\sigma_{z}$ maps.

\vspace{1mm}The experimentation with $\zeta^+$ consisted mainly of varying the weighting factor $w$ in Equation C3, but also considering changes in $z_{\rm sc}$.  We considered $w$=0 (which recovers Eq.\,C2), 0.2, 0.3, 0.4, 0.5, and 1.  The problem with $w$=0 is the strong tendency to move all features at low $|b|$ to far kinematic distances, as described earlier.  At the opposite extreme, $w$=1 strongly prefers all features be placed at near distances, also mentioned above.  Nevertheless, the actual masking that results in either case is somewhat patchy, in the sense that structures appearing to be contiguous in the umasked \ld\ maps do not consistently and contiguously resolve to be on one side or the other of the tangent mirror.  While the overall balance from far to near shifted consistently as $w$ was raised from 0 to 1, the degree of contiguousness in the near/far masking at first rose, then fell again, as $w$ increased.  Depending on the feature one is most interested in, the optimal $w$ (where such a feature is most contiguously near or far) is slightly different, but generally $w$ $\approx$ 0.2--0.4 gives reasonable and believable results.  (We tended to focus on the Far Ara clouds --- see \S\ref{farside} --- in evaluating $w$.)  The patchiness, however, is always present at some level, and was somewhat irreducible (but see remarks below).  Overall though, we found that $w$=0.3 gives the most optimal near/far discrimination.

\vspace{1mm}In contrast, changes in $z_{\rm sc}$ produced more dramatic results.  As can be inferred from Figure \ref{scalehts}, as $z_{\rm sc}$ increases, the $\zeta$ function peak at $h_{\rm med}$ moves to higher $z$.  This means that the likelihood for any $z$ already less than $h_{\rm med}$ decreases, while the likelihood at any $z$ remaining above $h_{\rm med}$ increases: the discrimination shifts to favouring far distances.  Conversely, if $z_{\rm sc}$ decreases, the discrimination will favour near distances more.  Thus, changes in $z_{\rm sc}$ work oppositely to changes in $w$, and adopting a different $z_{\rm sc}$ only means that the optimal $w$ value will be forced to change oppositely, if the same contiguousness criterion is adhered to.  This criterion means that the individual values for $z_{\rm sc}$ and $w$ are less important than the combination.

% Figure C20: 12co- and ZM-ld1h with zeta+ masking  (C24 or C30 with the prior YX figures)
\begin{sidewaysfigure*}[h]
\centerline{\includegraphics[angle=-90,scale=0.84]{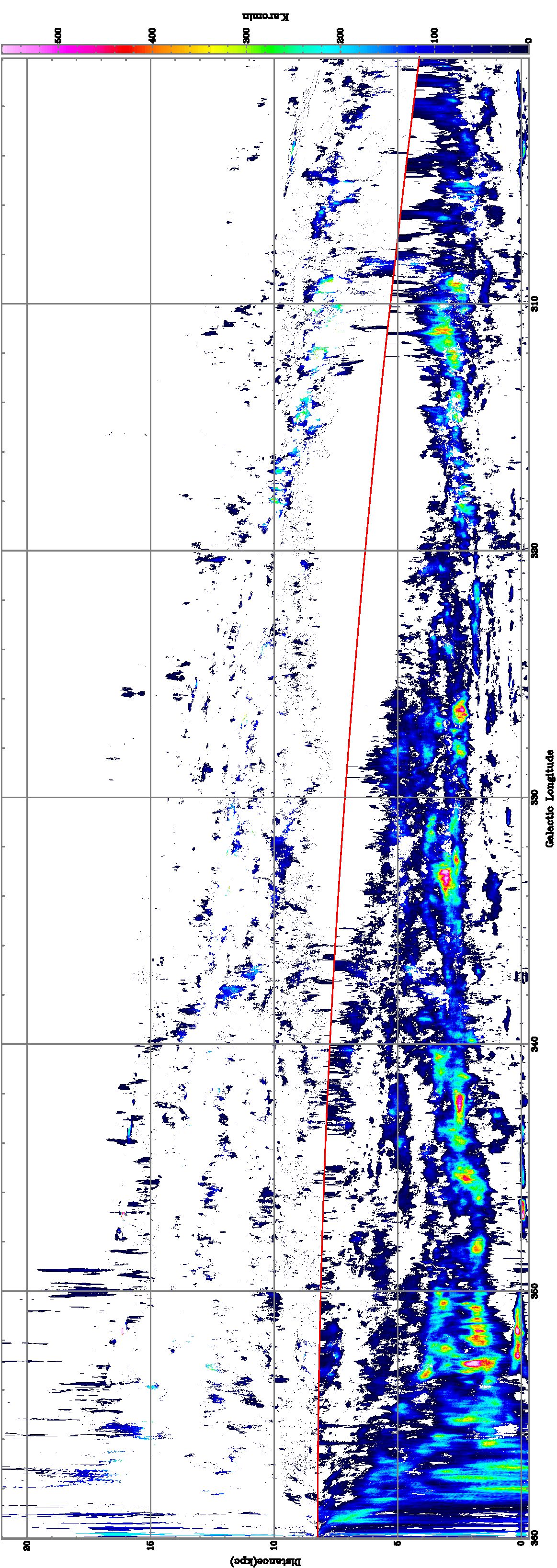}\hspace{0mm}}
\vspace{-3mm}
\centerline{\includegraphics[angle=-90,scale=0.84]{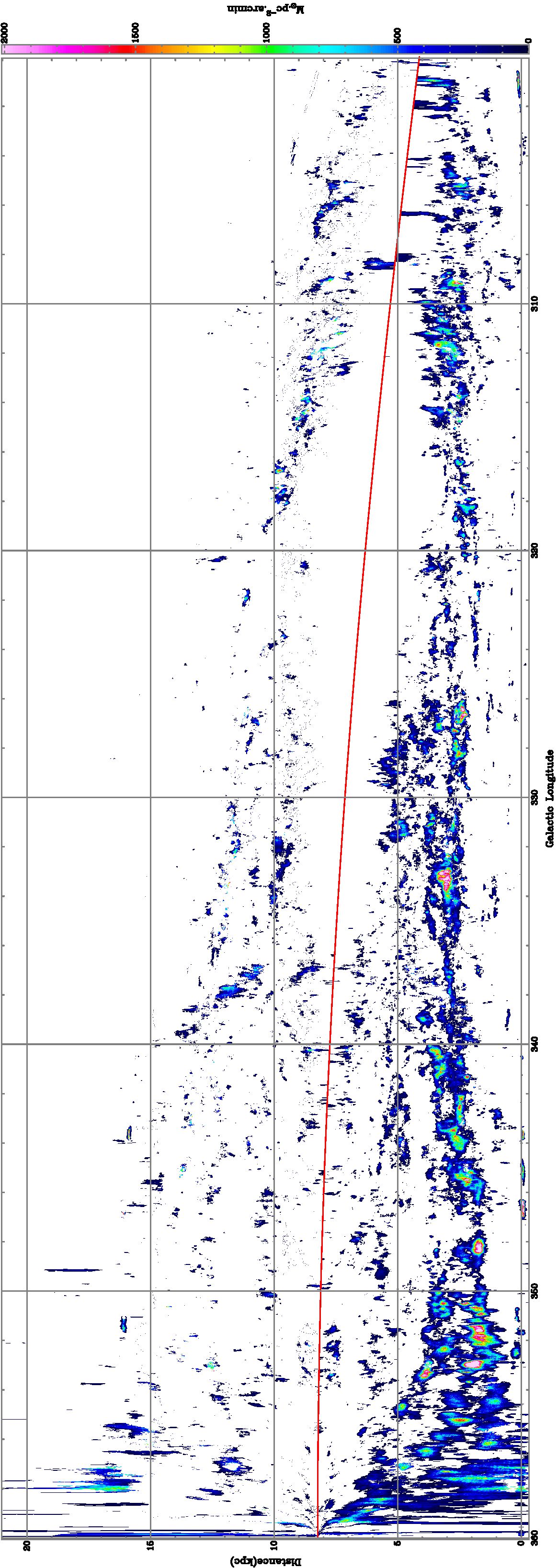}\hspace{0mm}}
\caption{\footnotesize Similar \ld\ maps to Fig.\,\ref{bgt-ld0} (latitude-integrated \tco\ emission and \nco\ \ld\ diagrams) but where the integrals have been masked via the $\zeta^+$ function.  This gives a first-order near/far deprojection of molecular clouds across the 4th Quadrant, based on the known thickness of the Galaxy's molecular layer, a few 10s of pc, at both near (below the tangent-mirror curve) and far (above the curve) distances.  ({\em Top}) \tco\ and ({\em bottom}) \nco\ data. $$ $$
\label{both-ld0-b1p3m}}\vspace{-12mm}
\end{sidewaysfigure*}

% Figure C21: 12co- and ZM-ld1h with zeta+ masking  (C25 or C31 with the prior YX figures)
\begin{sidewaysfigure*}[h]
\centerline{\includegraphics[angle=-90,scale=0.84]{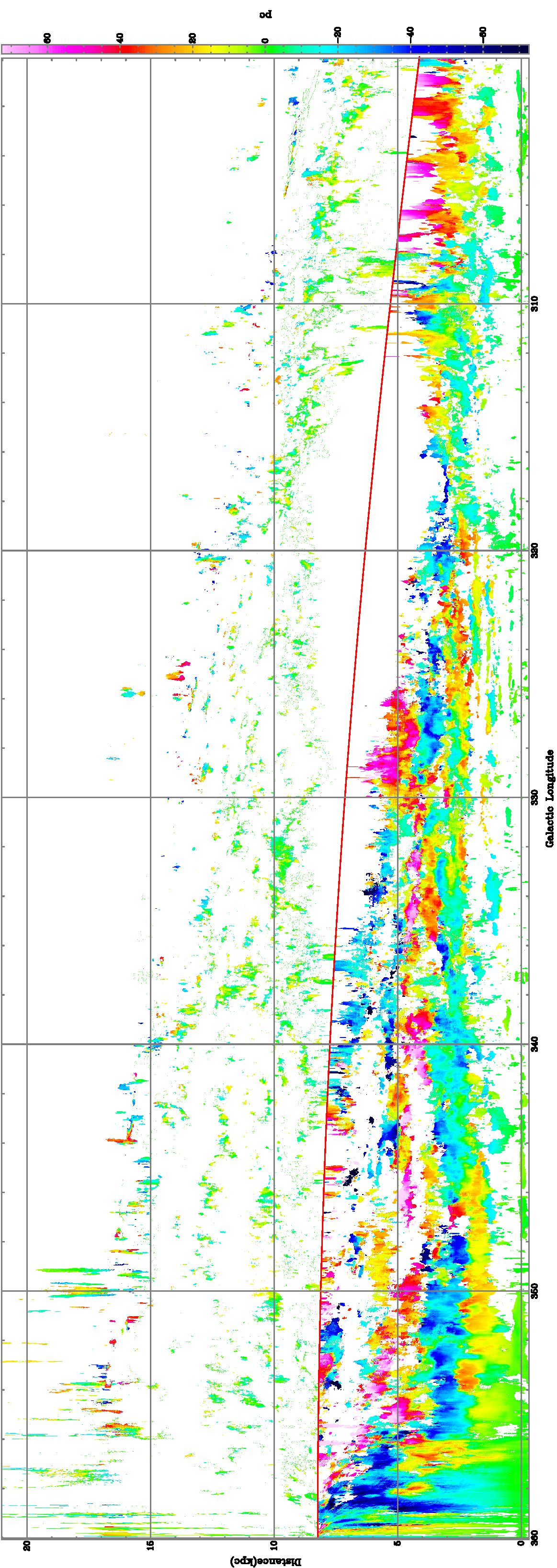}\hspace{0mm}}
\vspace{-3mm}
\centerline{\includegraphics[angle=-90,scale=0.84]{dr6-ZM-URCbgt-ld1h-b1p3m.jpg}\hspace{0mm}}
\caption{\footnotesize Similar \ld\ maps to Fig.\,\ref{both-ld1h-b1m} (mean height $\bar{z}$) but where the heights have been masked via the $\zeta^+$ function, our best near/far deprojection of molecular clouds across the 4th Quadrant.  ({\em Top}) \tco\ and ({\em bottom}) \nco\ data. $$ $$
\label{both-ld1h-b1p3m}}\vspace{-6mm}
\end{sidewaysfigure*}

% Figure C22: 12co- and ZM-ld2h with zeta+ masking  (C26 or C32 with the prior YX figures)
\begin{sidewaysfigure*}[h]
\centerline{\includegraphics[angle=-90,scale=0.84]{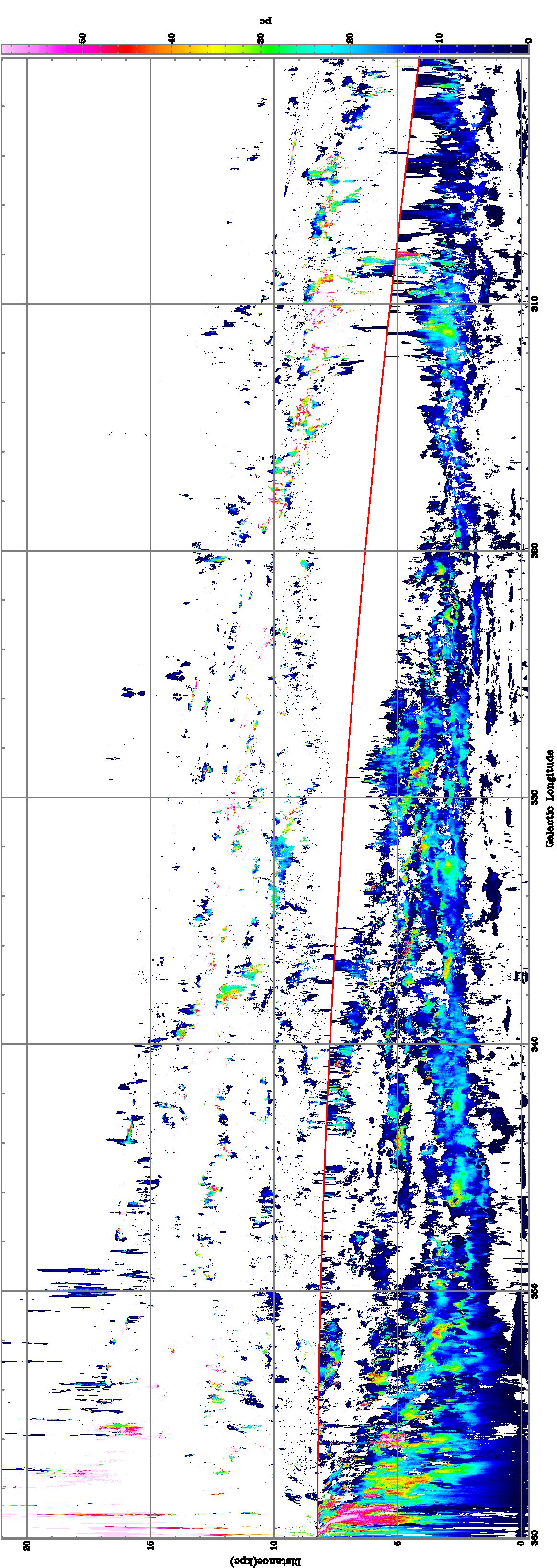}}
\vspace{-3mm}
\centerline{\includegraphics[angle=-90,scale=0.84]{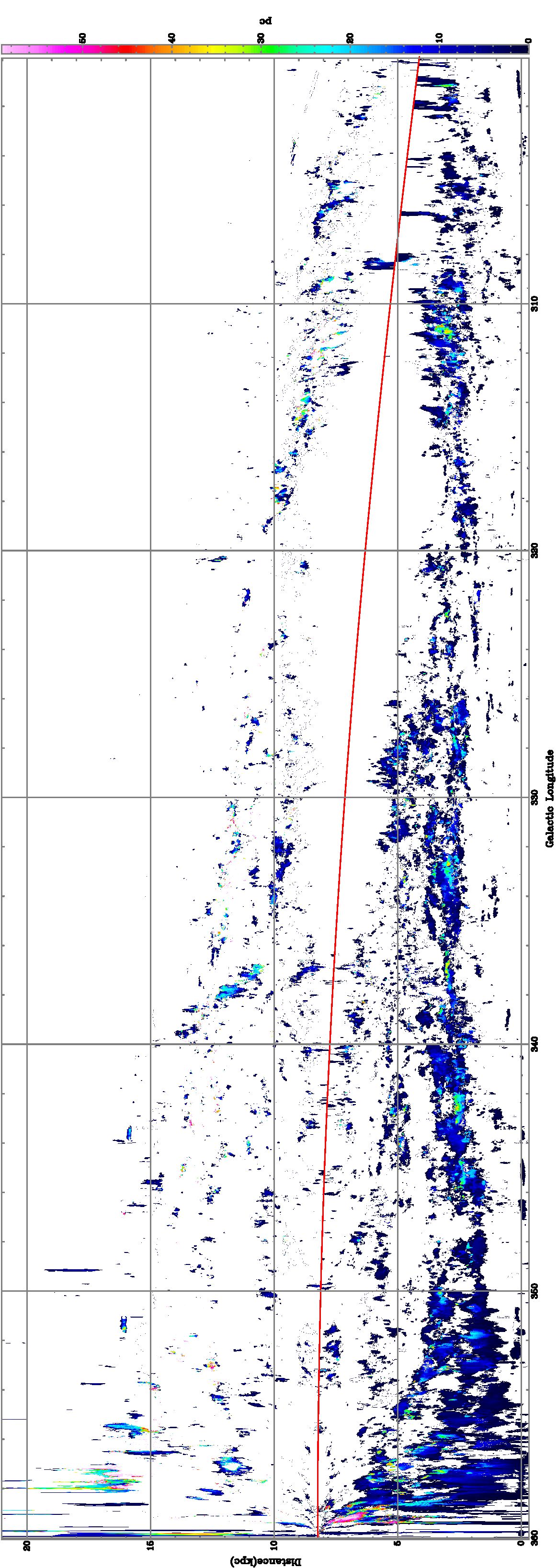}}
\caption{\footnotesize Similar \ld\ maps to Fig.\,\ref{both-ld2h-b1m} (height dispersion $\sigma_z$) but where the widths have been masked via the $\zeta^+$ function, our best near/far deprojection of molecular clouds across the 4th Quadrant.  ({\em Top}) \tco\ and ({\em bottom}) \nco\ data. $$ $$
\label{both-ld2h-b1p3m}}\vspace{-6mm}
\end{sidewaysfigure*}

% Figure C23: Final zeta+ filtered 12co-YX0
\begin{figure*}[h]
\centerline{\includegraphics[angle=0,scale=0.92]{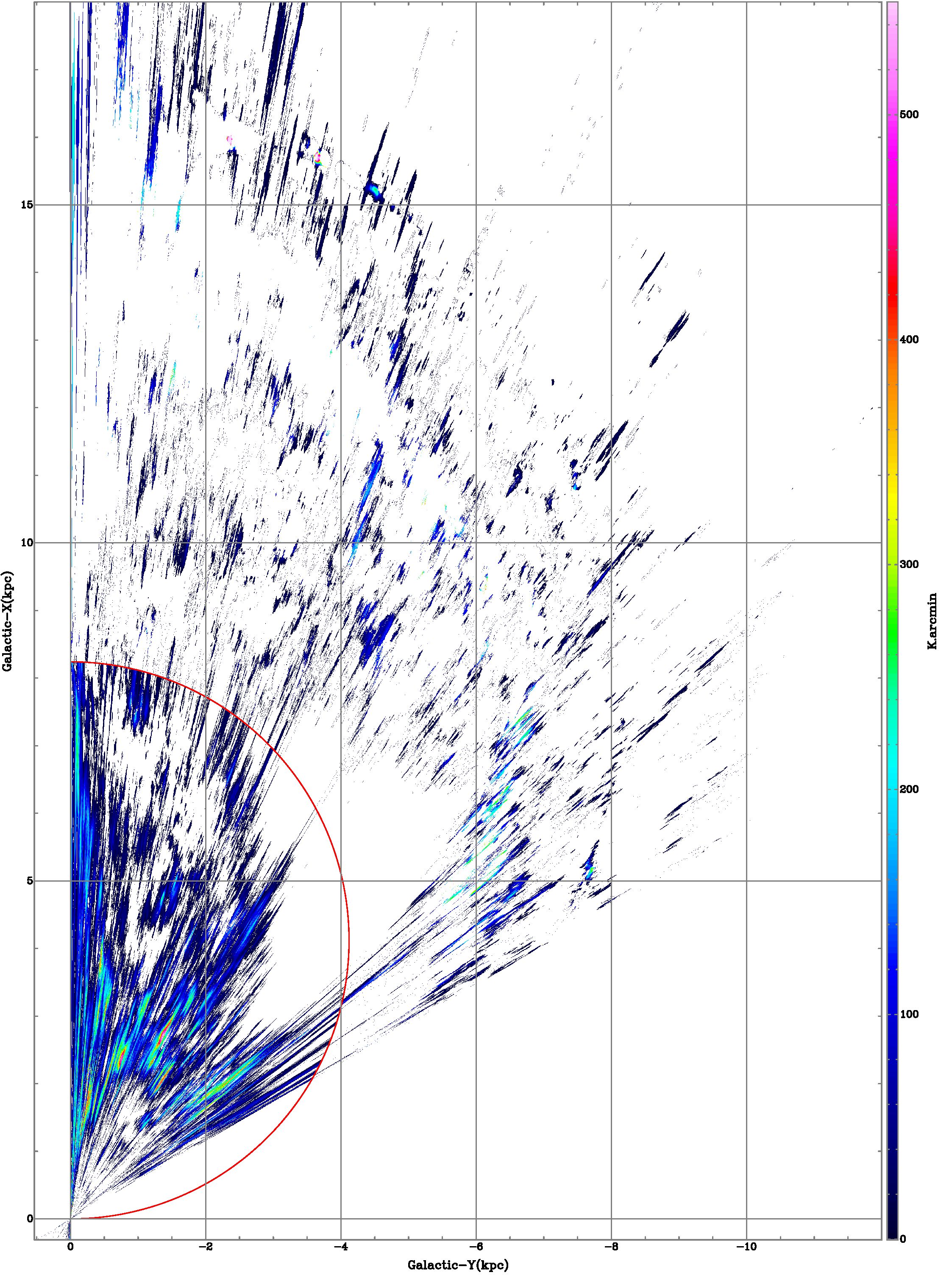}}
\caption{\footnotesize Same as Fig.\,\ref{12co-bgt-YX0} (a Cartesian deprojection of the \tco\ \lv\ total intensity map into Galactic \xy\ space) but now filtered by the near/far distance discriminator $\zeta^+$ (Eq.\,C3). $$ $$
\label{12co-bgt-YX0zp}}
\end{figure*}

% Figure C24: Final zeta+ filtered ZM-YX0
\begin{figure*}[h]
\centerline{\includegraphics[angle=0,scale=0.92]{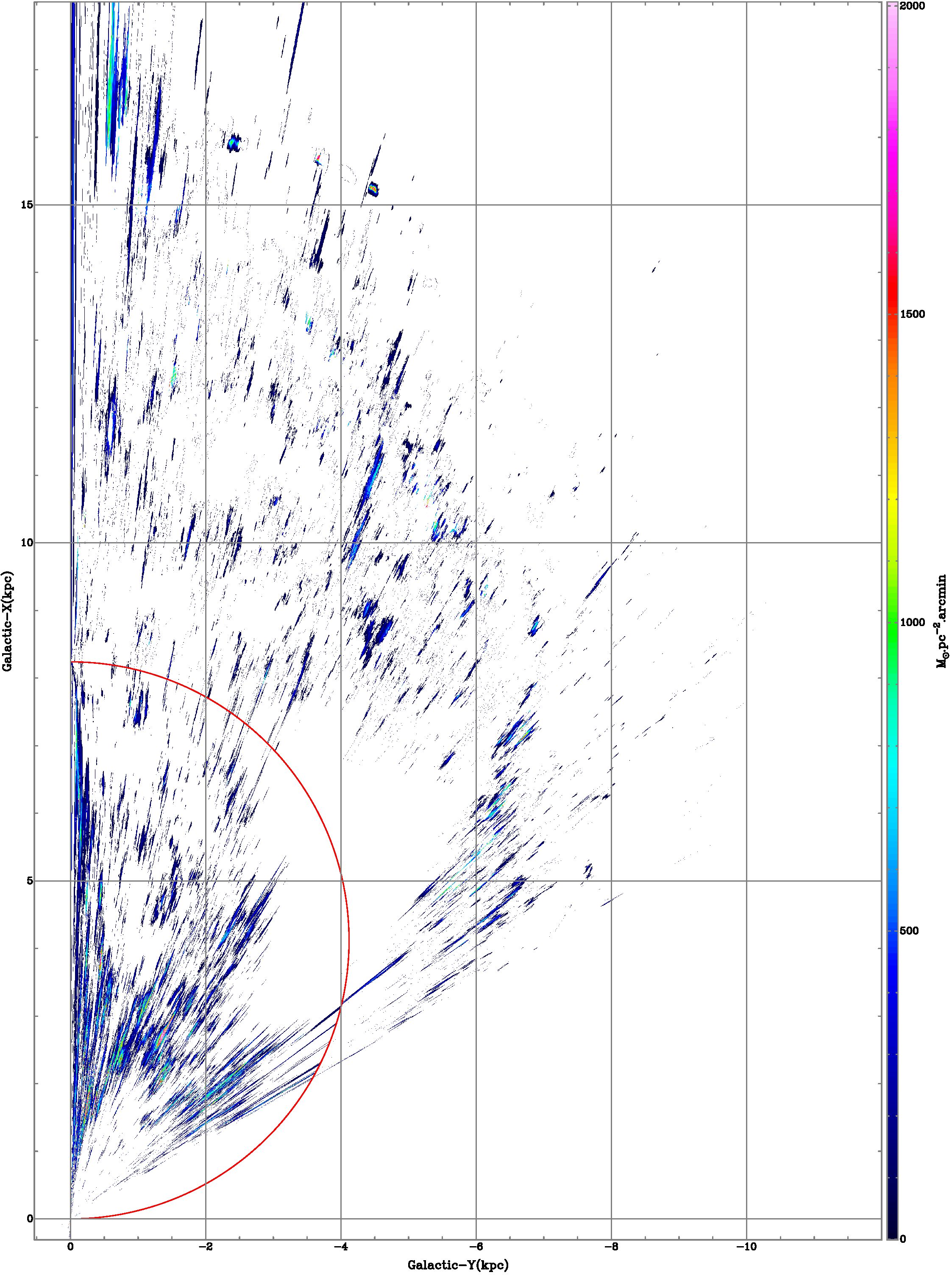}}
\caption{\footnotesize Same as Fig.\,\ref{12co-bgt-YX0zp} but for the latitude-integrated \nco\ map. $$ $$
\label{ZM-bgt-YX0zp}}
\end{figure*}

% Figure C25: Final zeta+ filtered 12co-YX1
\begin{figure*}[h]
\centerline{\includegraphics[angle=0,scale=0.92]{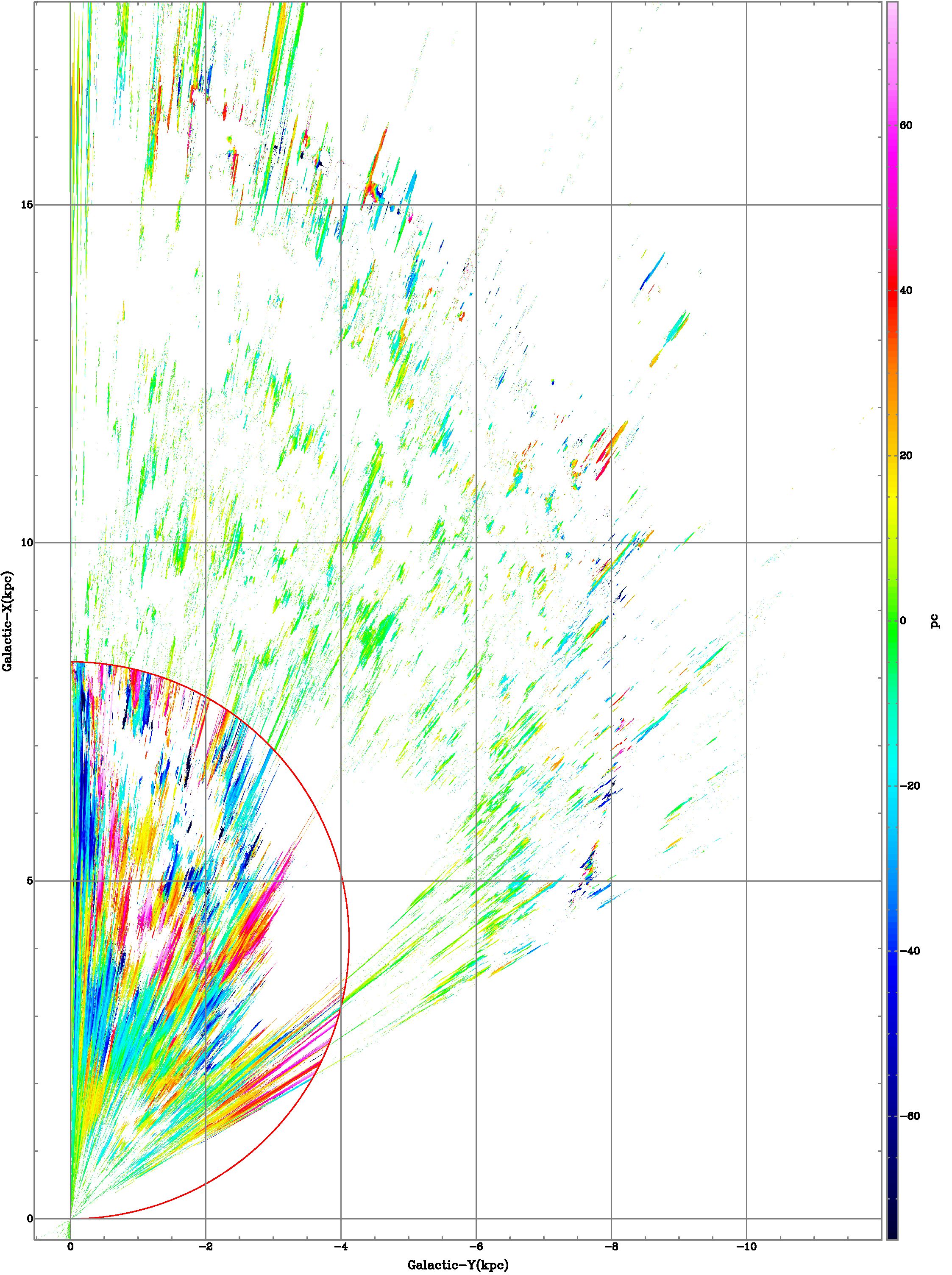}}
\vspace{-69mm}
\hspace{3mm}\includegraphics[angle=0,scale=0.105]{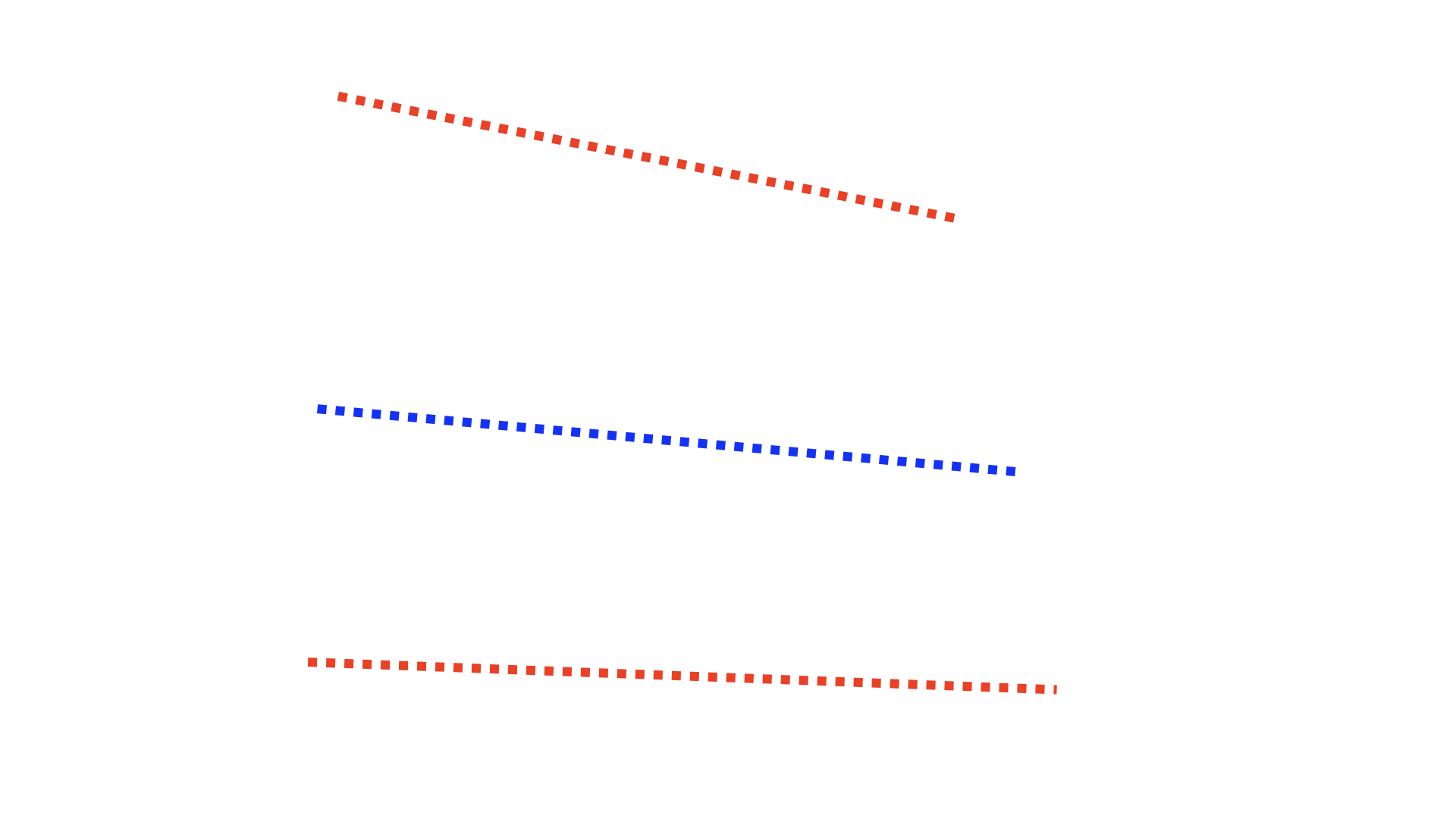}

\vspace{28.6mm}
\caption{\footnotesize As for Fig.\,\ref{12co-bgt-YX0zp} but now the \tco-weighted mean height $\bar{z}$, filtered by the near/far distance discriminator $\zeta^+$ (Eq.\,C3).  Also overlaid here is an approximate pattern linking the successive extrema in the height excursions, as discussed in \S\ref{ripples}. $$ $$
\label{12co-bgt-YX1zp}}
\end{figure*}

% Figure C26: Final zeta+ filtered ZM-YX1
\begin{figure*}[h]
\centerline{\includegraphics[angle=0,scale=0.92]{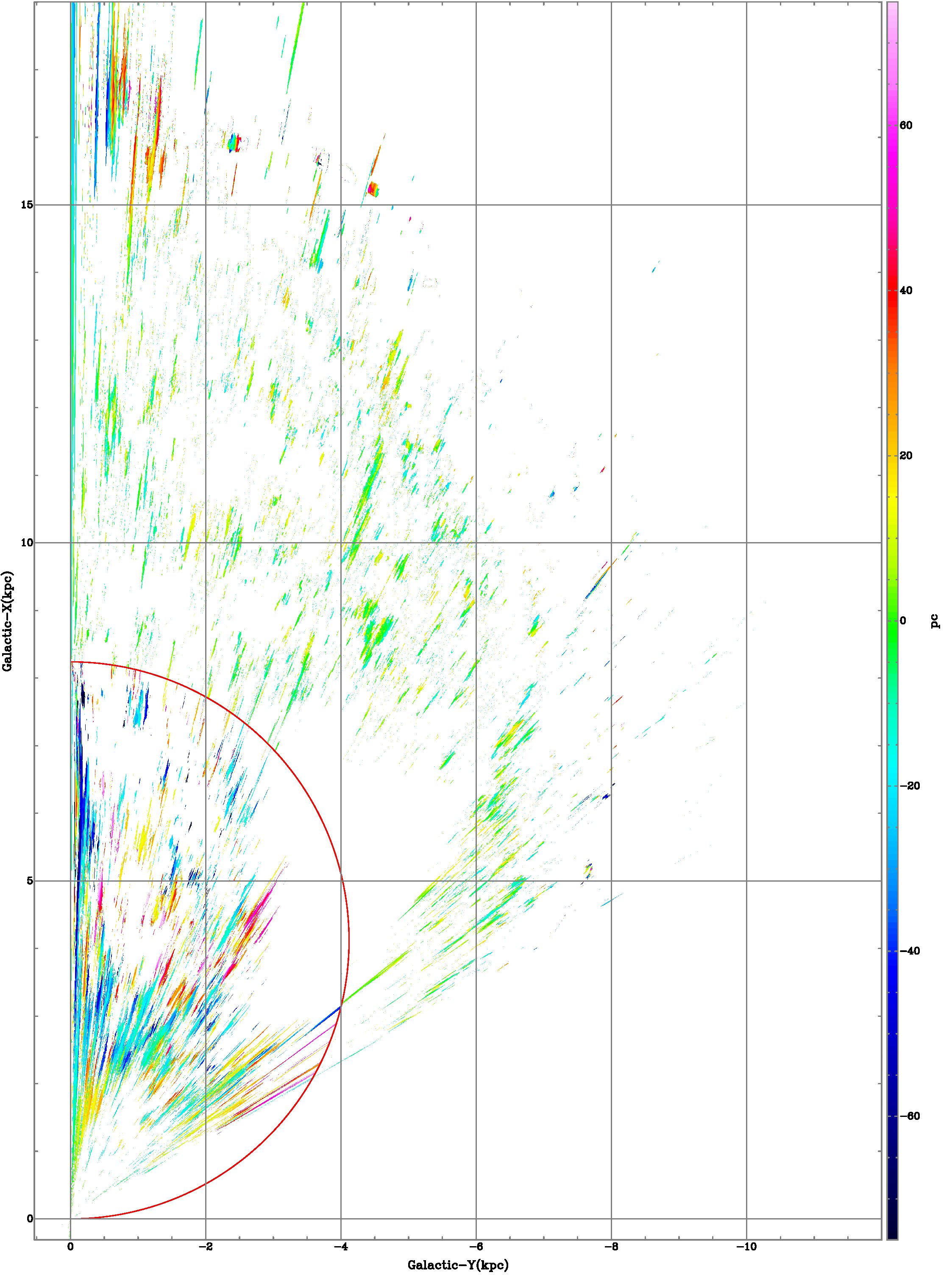}}
\vspace{0.1mm}
\caption{\footnotesize Same as Fig.\,\ref{12co-bgt-YX1zp} but for the latitude-integrated \nco\ map. $$ $$
\label{ZM-bgt-YX1zp}}
\end{figure*}

% Figure C27: Final zeta+ filtered 12co-YX2
\begin{figure*}[h]
\centerline{\includegraphics[angle=0,scale=0.92]{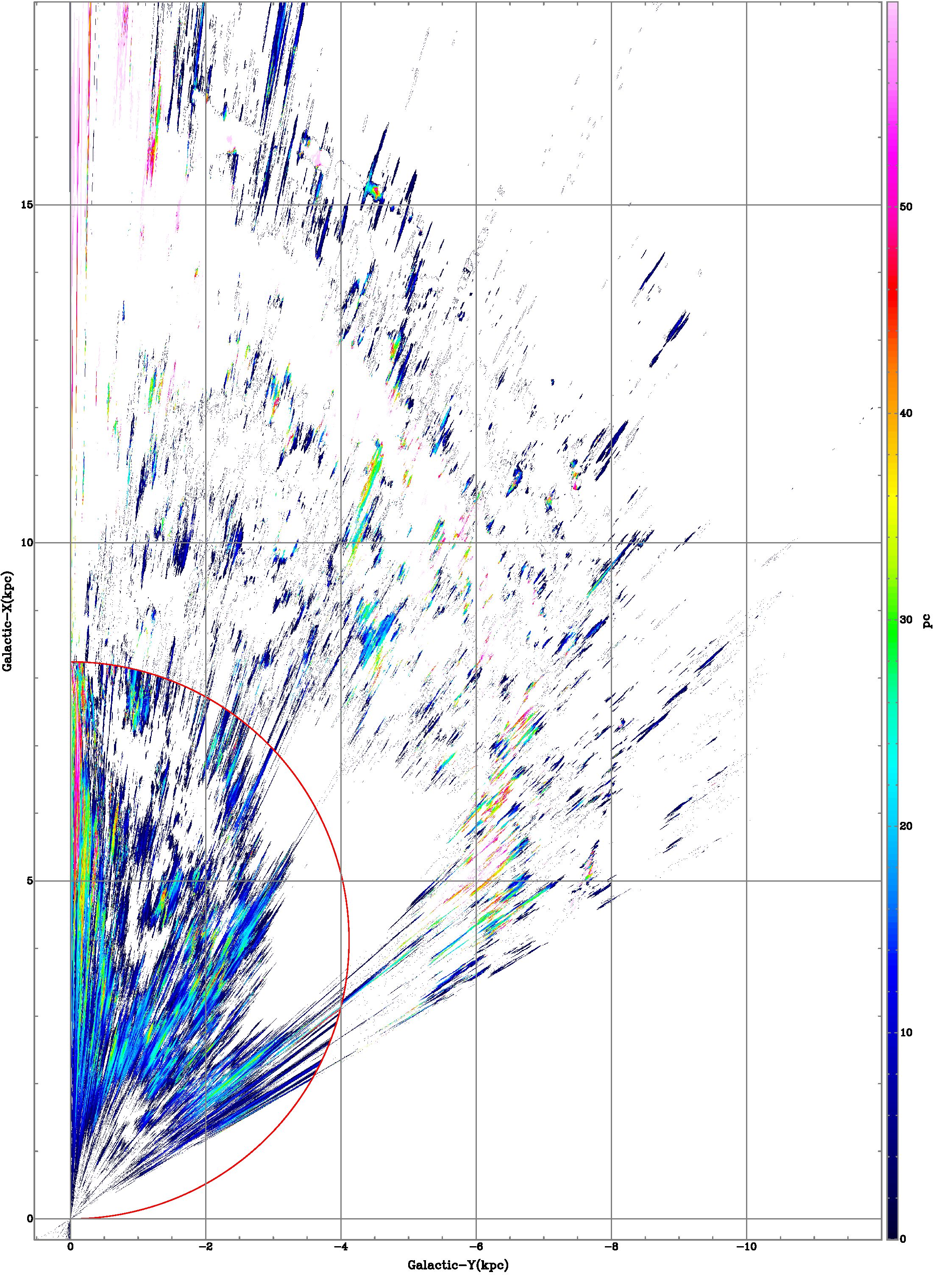}}
\vspace{-0.1mm}
\caption{\footnotesize As for Fig.\,\ref{12co-bgt-YX1zp} but now the \tco-weighted thickness $\sigma_z$, filtered by the near/far distance discriminator $\zeta^+$ (Eq.\,C3). $$ $$
\label{12co-bgt-YX2zp}}
\end{figure*}

% Figure C28: Final zeta+ filtered ZM-YX2
\begin{figure*}[h]
\centerline{\includegraphics[angle=0,scale=0.92]{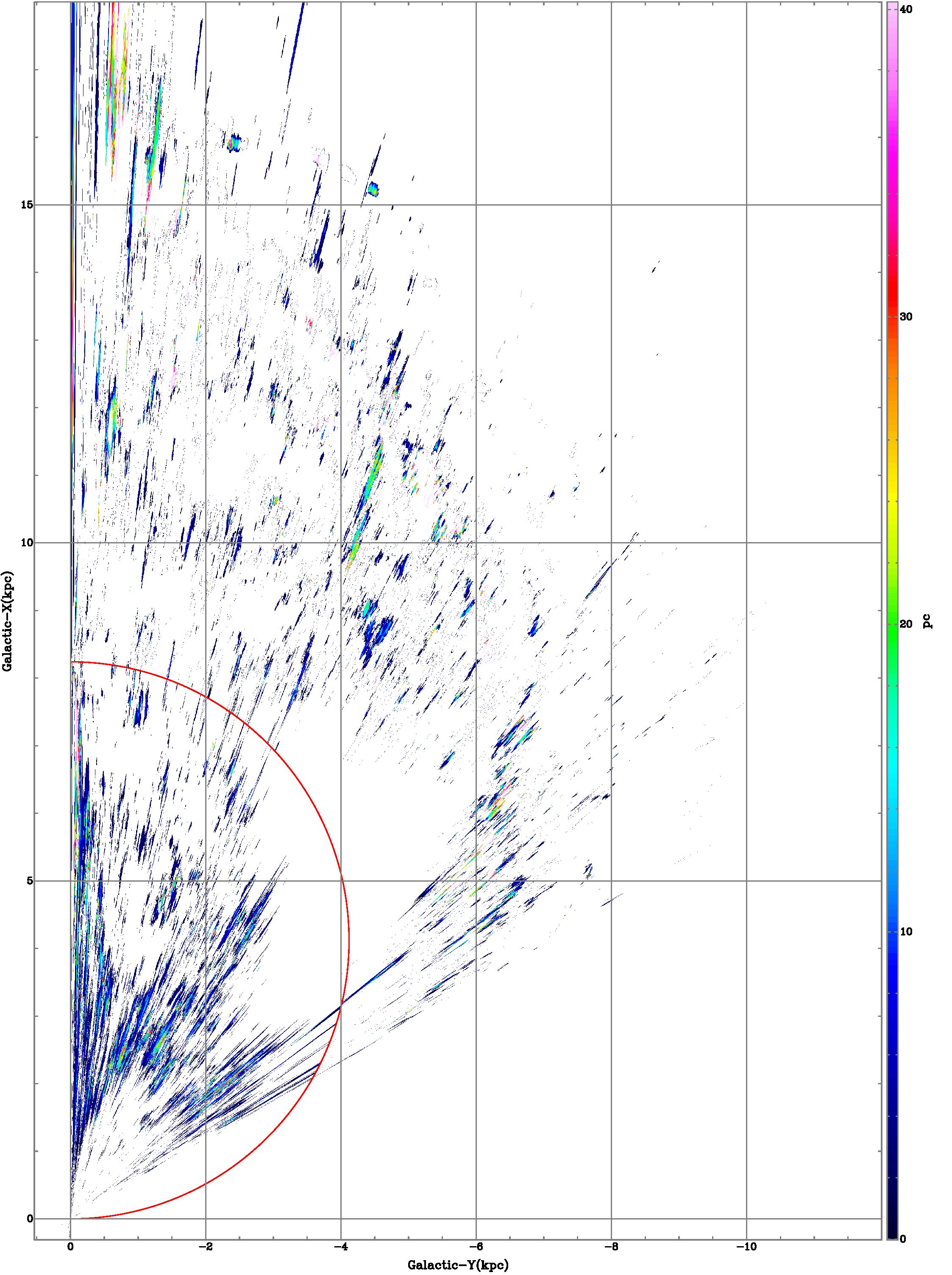}}
\vspace{0mm}
\caption{\footnotesize Same as Fig.\,\ref{12co-bgt-YX2zp} but for the latitude-integrated \nco\ map. $$ $$
\label{ZM-bgt-YX2zp}}
\end{figure*}

% Figure C29: Floor-masked 12co-YX0
\begin{figure*}[h]
\centerline{\includegraphics[angle=0,scale=0.921]{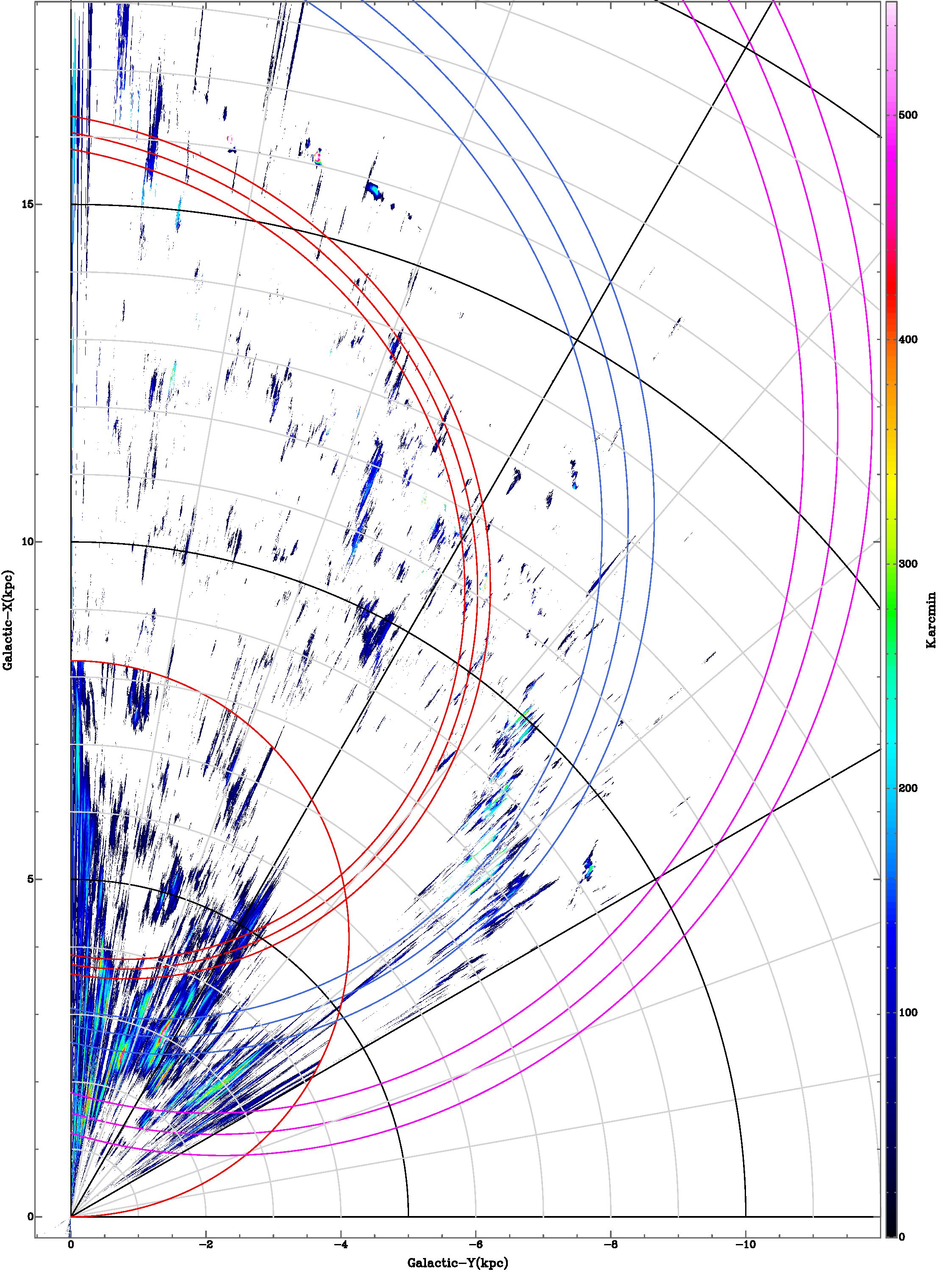}}
\vspace{-2mm}
\caption{\footnotesize Same as Fig.\,\ref{12co-bgt-YX0zp} but with pixels below a cutoff of 20\,K\,arcmin, about 2.5\% of the peak intensity and 3 times the noise level of $\sigma_{\rm rms}$ $\approx$ 7 K\,arcmin, being masked out.  Also overlaid are the Norma (red), Scutum-Centaurus (blue), and Sagittarius-Carina (magenta) spiral arm patterns of \cite{r19} and a heliocentric-polar coordinate grid, despite the data being on the same \xy\ grid as the last few figures. $$ $$
\label{12co-bgt-YX0fl}}
\end{figure*}

% Figure C30: Floor-masked ZM-YX0
\begin{figure*}[h]
\centerline{\includegraphics[angle=0,scale=0.921]{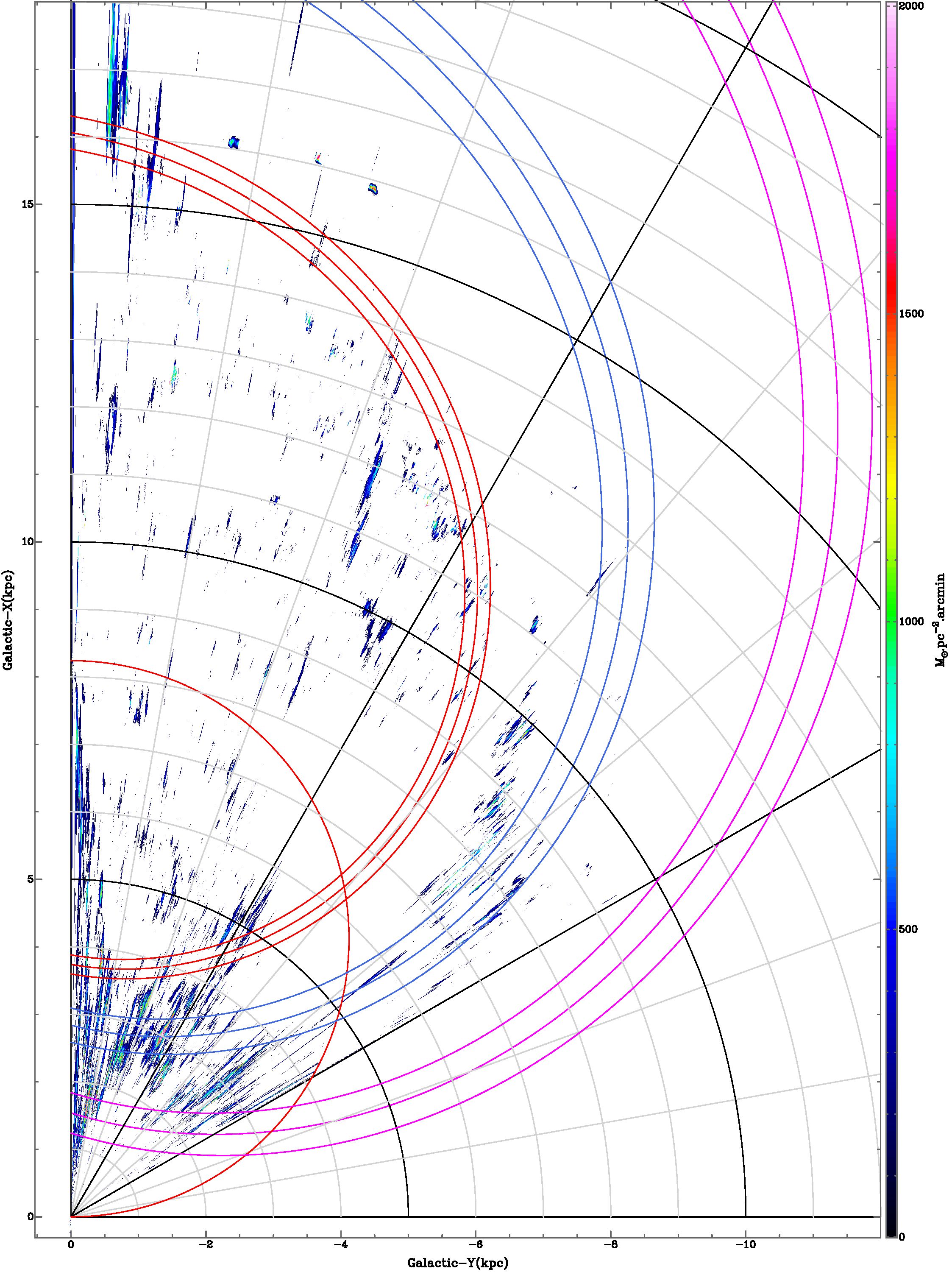}}
\vspace{-2mm}
\caption{\footnotesize Same as Fig.\,\ref{ZM-bgt-YX0zp} but with pixels below a cutoff of 100\,M\solar\,pc$^{-2}$\,arcmin, about 2.5\% of the peak intensity and 30 times the noise level of $\sigma_{\rm rms}$ $\approx$ 3\,M\solar\,pc$^{-2}$\,arcmin, being masked out.  We also show the same \cite{r19} spiral arms and polar grid as in Fig.\,\ref{12co-bgt-YX0fl}. $$ $$
\label{ZM-bgt-YX0fl}}
\end{figure*}

\vspace{1mm}All this being said, the $\zeta^+$ function appears to give a conceptually simple and computationally fast way to achieve a first-order deprojection of larger-scale Galactic structure from \lv\ data.  The final $\zeta^+$-filtered results for all the $b$-moments (i.e., integrated, mean, and dispersion maps for each of \tco\ and \nco) are shown in {\color{red}Figures \ref{both-ld0-b1p3m}--\ref{both-ld2h-b1p3m}} for the \ld\ deprojection, to be compared with the other \ld\ maps already presented.  For completeness, we also show the same 6 results in {\color{red}Figures \ref{12co-bgt-YX0zp}--\ref{ZM-bgt-YX2zp}} for the \xy\ deprojection.

\vspace{1mm}As a final note, we can see that the pixellation/patchiness issue is more pronounced in the \xy\ maps than in \ld, apparently a side-effect of the projection.  There seem to be many small features (i.e., $\sim$beamsized in longitude) which prefer to be assigned near or far distances regardless of what their neighbouring features do.  We experimented with small median filters (boxes of size 5$\times$5 to 11$\times$11 pixels) to mitigate the masking fungibility.  That is, we median-filtered the $\zeta^+$ function itself, prior to it being applied to the unmasked moment data, which did not have their values altered.  This rearranged some fraction of small features from near to far or vice versa, but had no systematic impact on the large-scale near/far patterns.  To some extent, this occurs where Equation C4 gives similar $\zeta^+$ values at both the near and far values for $|\bar{z}|$ and $\sigma_{z}$, and the filtering method becomes somewhat indeterminate (as in the second example of computing $\zeta$ described above).  Only with larger median filters (e.g., a 21$\times$21 pixel box) does the overall near/far discrimination break more cleanly.

\vspace{1mm}However, with this filter size we are averaging over 7 beamwidths in longitude (and over an indeterminate scale in latitude for \lv\ maps), so potentially lumping in small features at the wrong kinematic distance if they happen to lie close on the sky to large clouds at the opposite distance.  Thus, the ``finely-sprinkled dust'' effect may be more of an aesthetic problem than a numerical or physical one.  Few of them are actually sub-beam features; many of them seem to be just smaller, less massive clouds.  Apparently our data are sufficiently sensitive to find a widely-distributed small-cloud population in the disk of the Milky Way.

\vspace{1mm}And perhaps this is the simplest explanation: much of this effect is confined to low integrated intensity/column density.  That is, when a modest floor based on the moment-0 maps is used to mask the data in any of Figures \ref{both-ld0-b1p3m}--\ref{ZM-bgt-YX2zp}, namely at cutoffs of 20\,K\,arcmin or 100\,M\solar\,pc$^{-2}$\,arcmin, these faint and somewhat incoherent features all but disappear.  We show the modified moment-0 maps as examples in {\color{red}Figures \ref{12co-bgt-YX0fl}} and {\color{red}\ref{ZM-bgt-YX0fl}}, but the reader can imagine the same masks being applied to the higher moments (Figs.\,\ref{12co-bgt-YX1zp}--\ref{ZM-bgt-YX2zp}) as well.  In the remainder of this paper, however, we do not explicitly discuss this small-cloud population any further, since by itself, it is not so informative about the more prominent features of the overall cloud population and the general architecture of the Milky Way.

%%%%%%%%%
%   Section C5  %
%%%%%%%%%
\subsection{Notable Features of the Filtered Maps}\label{filtered}

\vspace{1mm}Having focused on optimising our kinematic distance determinations and near/far discrimination based on the information in the \lv\ and \ld\ diagrams, we now explore some of the features revealed in these maps.  While there are several features worthy of further study, such as the local low-mass cloud population used to anchor our revised LSR definition, or the fitting of spiral arm patterns to the most massive clouds, we particularly want to highlight two novel and striking large-scale features which, to our knowledge, are revealed here for the first time.

\subsubsection{Ripples Everywhere!}\label{ripples}

\begin{figure}[b]
\vspace{0mm}\hspace{69mm}\includegraphics[angle=0,scale=0.105]{I12.jpg}

\vspace{-4mm}\hspace{69mm}\includegraphics[angle=0,scale=0.105]{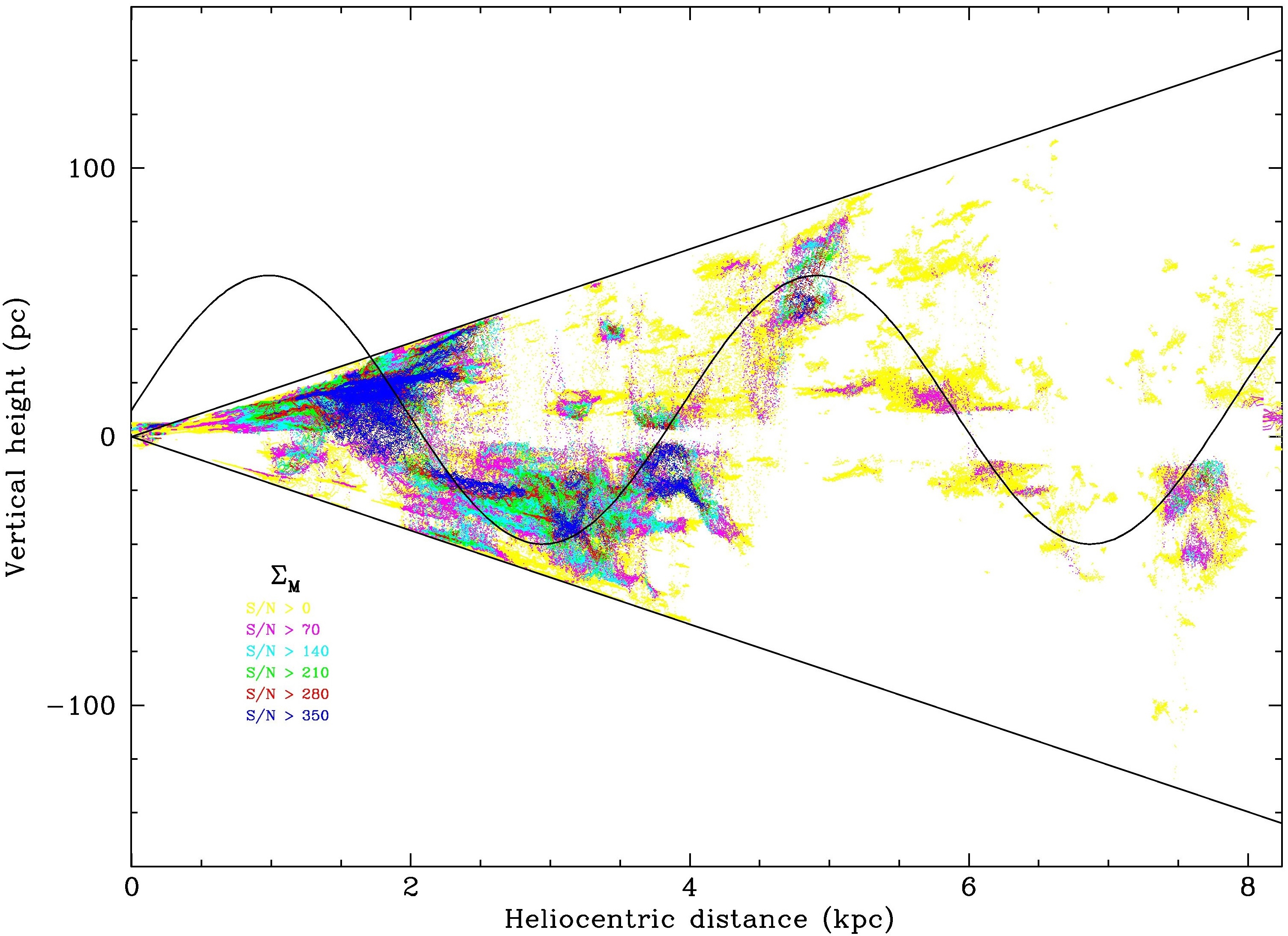}

\vspace{-115mm}
\parbox[top]{56mm}{\caption{ \label{wiggles}}
\footnotesize Mean height vs distance for each pixel in 355\fdeg3 $>$ $l$ $>$ 348\fdeg0 from the \ld\ maps in Fig.\,\ref{both-ld1h-b1p3m}: ({\em top}) \tco, and ({\em bottom}) \nco.  The dots for each pixel are colour-coded by their S/N in the respective moment-0 maps (Fig.\,\ref{both-ld0-b1p3m}), as labelled, where for the \tco\ data, the noise $\sigma_{\rm rms}$ $\approx$ 7\,K\,arcmin, while for \nco\ we have $\sigma_{\rm rms}$ $\approx$ 3\,M\solar\,pc$^{-2}$\,arcmin (the limiting S/N$>$0 for the yellow dots is actually more like S/N$>$4 given the thresholding inherent in the SAMed data).  Also overlaid in both panels is the same, rough visual fit of a sinusoid to the maximal \tco\ or \nco\ ridgeline.  This has wavelength 4\,kpc and amplitude 50\,pc, offset by $z$ = +10\,pc.  The sloping lines indicate ThrUMMS' latitude limits of $\pm$1\degree, with a vertical exaggeration of about 20:1.}

\vspace{50.8mm}
\end{figure}

\vspace{1mm}The first of these is easy to see --- a very prominent undulation in the mean height of the molecular cloud population, most clearly between heliocentric distances 2--7\,kpc and longitudes $\sim$347\degree--355\degree\ in the \ld\ map of Figure \ref{both-ld1h-b1p3m}, and equivalent locations in the \xy\ maps (Fig.\,\ref{12co-bgt-YX1zp} as suggested by the dotted lines, and Fig.\,\ref{ZM-bgt-YX1zp}).  The direction and amplitude of this undulation/wave/ripple is first to positive heights, perhaps $\sim$+20\,pc at around 2\,kpc; then down to --30\,pc at $\sim$3.5\,kpc; then to +40\,pc at 5\,kpc; and beyond that, the pattern becomes somewhat less distinct, but the trend is back down to about --40\,pc.  The same pattern appears in both the \tco\ data and the \nco.  If this is some sort of wave, we are nominally looking at a wavelength of a few kpc and an amplitude of a few 10s of pc, although the apparent amplitude may be underestimated with our $\pm$1\degree latitude limit.  To see this, we show in {\color{red}Figure \ref{wiggles}} plots of the mean height vs distance, at any longitude from this sector, plus a fit by eye with an approximate sine wave.

% Figure C31: mean heights vs distance

\vspace{1mm}The wave pattern in Figure \ref{wiggles} is very striking, even without the sine overlay.  Indeed, it suggests that we could have missed some nearby clouds ($d$ $<$ 2\,kpc) at more positive latitudes (up to $b$ $\sim$ +4\degree) that would enhance and extend this pattern.  Figure \ref{wiggles} also suggests that the pattern might persist in the anticenter direction, where data from other surveys covering perhaps $l$ $\approx$ 160\degree$\pm$20\degree\ in the second and third quadrants, extending down to $b$ $\sim$ --4\degree, would be very instructive.  And correspondingly, it further suggests that any similar wavelike pattern could possibly be seen in profile (i.e., directly on the sky) towards longitudes $l$ $\approx$ 70\degree$\pm$20\degree\ and/or $l$ $\approx$ 250\degree$\pm$20\degree.  This last idea is remarkably close to the (somewhat scattered) wave pattern that has been widely noticed in, for example, the Nanten and CHaMP surveys of molecular clouds along the Carina Arm in $l$ = 290\degree$\pm$10\degree\ \citep{b11}.  With an apparent amplitude $\sim$1\degree\ and many of the clouds in those surveys at $d$ = 2--3\,kpc, this translates to a projected amplitude of $\sim$40\,pc, encouragingly close to the value in Figure \ref{wiggles}.

%\clearpage

\vspace{1mm}Actually, the ripples are even more widespread in the ThrUMMS data --- this pattern also extends fairly clearly to lower longitudes, albeit somewhat more patchily.  Looking at Figure \ref{both-ld1h-b1p3m}, in order of prominence the corrugation patterns run roughly from 325\degree--331\degree; 300\degree--310\degree; and 332\degree--347\degree; and typically from distances of $\sim$2\,kpc to where the tracer runs out, or up to the tangent-distance, whichever is reached first.  Indeed, in Figure \ref{wiggles} the wave is arguably traced all the way to the Galactic Centre.  Thus, apart from the moderate patchiness, these ripples display large-scale coherence across the 4Q.  Another example of this is that the ripples seem to be roughly parallel to lines of constant heliocentric distance.  One might suspect this is some kind of observational artifact, but in the native \lv\ data (Fig.\,\ref{full-lv1-12coZM-rainbow}) there is no particular alignment of the ripples with the $l$ or $V$ coordinates.  Instead of a curious cosmic coincidence, there may be a more mundane physical explanation for this pattern.  The alignment of the ripples' phase is also roughly along the inferred pattern of the Scutum-Centaurus (especially) and Sagittarius-Carina (somewhat) spiral arms as defined by \cite{r19}, which for parts of their length nearest the Sun, do indeed run at somewhat constant heliocentric distance.

\vspace{1mm}It is important to note that this ripple pattern is so pervasive across the 4Q that it cannot possibly be an artifact of the distance-filtering.  Partly this is because our $\zeta^+$ function tends to place most of the \lv\ data at near distances, and so we can already clearly see the same undulating pattern in the native $\bar{b}$ map of Figure \ref{full-lv1-12coZM-rainbow}.  Furthermore, if this pattern were affected (or even created) by the filtering, that process would tend to place some of the undulations at their far kinematic distances, making the nearside ripples less well-defined (we see something like this in maps computed with $w$=0; see \S\ref{nearfar}) and the farside pattern potentially hard to discern at all.  However, the filtering is based on height, and with an amplitude of $\sim$40\,pc, the ripple pattern statistically {\em must} lie at mostly nearside distances, otherwise the equivalent farside pattern would have an amplitude \gapp2$\times$ greater, inconsistent with the discussion (in \S\ref{nearfar}, and next) of a relatively small $z_{\rm sc}$.

\vspace{1mm}There is another intriguing facet of this wave, presaged by the discussion in \S\ref{nearfar} about the value of $z_{\rm sc}$.  That is, the wave itself could partially explain the apparent discrepancy between the \cite{r19} and \cite{dht01} values.  Although the ripples seem to have an amplitude of 50\,pc or so, approaching \cite{dht01}'s value, the thickness/scale height of the molecular layer {\em at any given distance} is much smaller (about 10--20\,pc, as apparent in Figs.\,\ref{12co-bgt-YX2zp} and \ref{ZM-bgt-YX2zp}), commensurate with \cite{r19}'s value.  One can easily imagine how the CfA survey data could be averaged over the whole Galactic Plane to yield a larger $z_{\rm sc}$, even if the intrinsic thickness at a given location is much less, if the midplane itself has large-scale corrugations on the order of 50\,pc.

% Figure C32: thickness distributions
\begin{figure*}[t]
\centerline{\includegraphics[angle=0,scale=0.101]{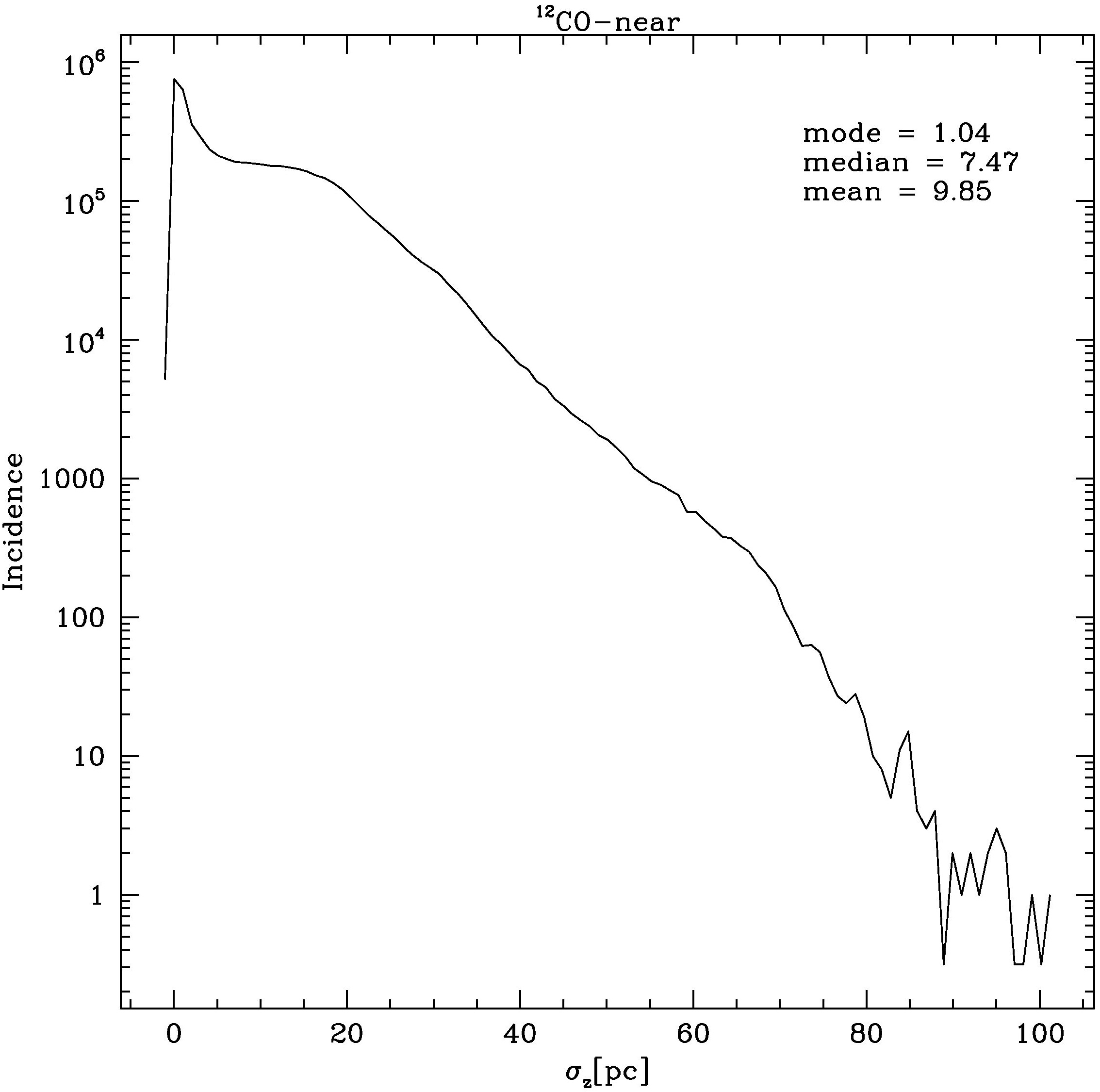}\,\includegraphics[angle=0,scale=0.101]{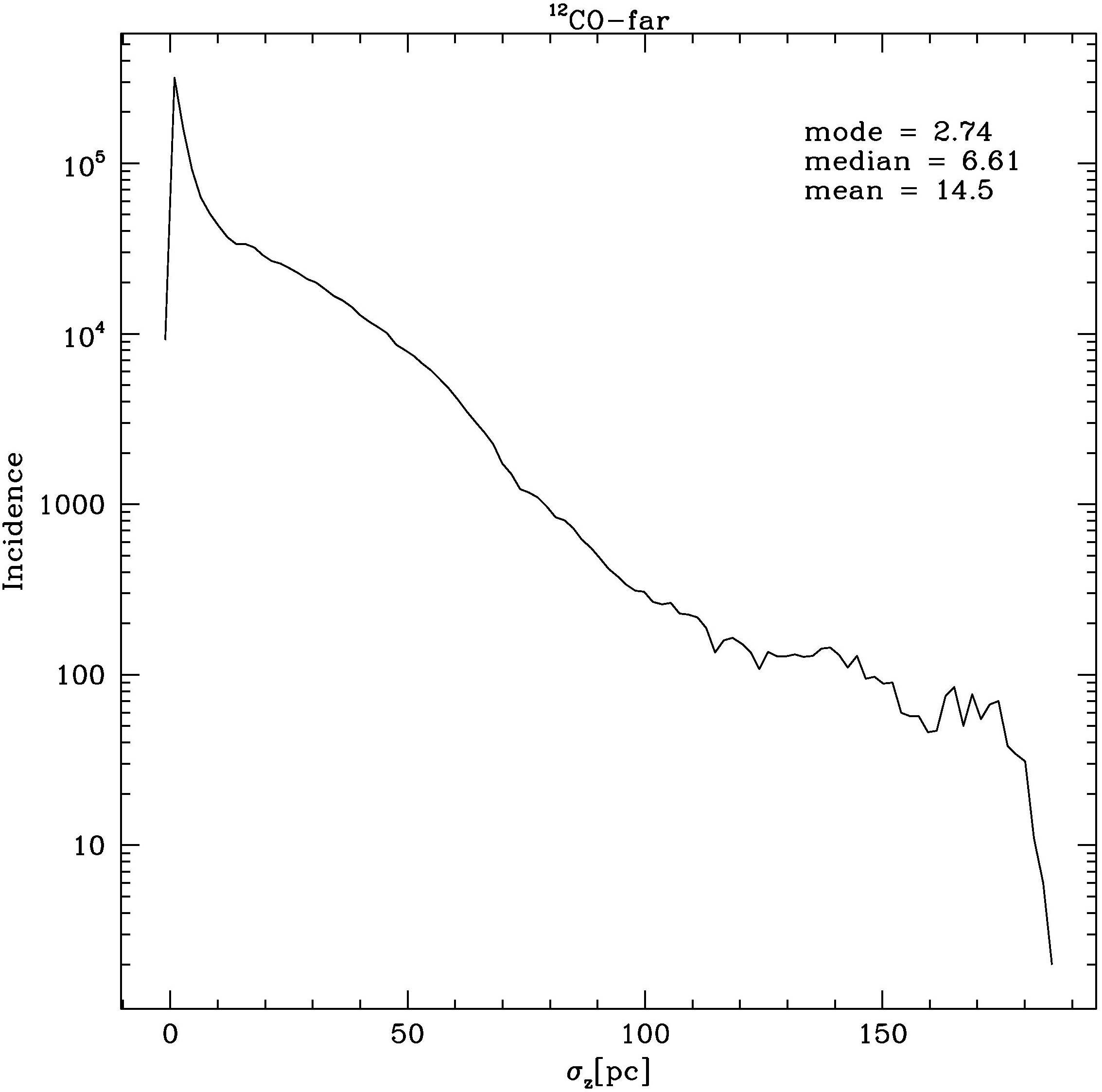}}
\centerline{\includegraphics[angle=0,scale=0.101]{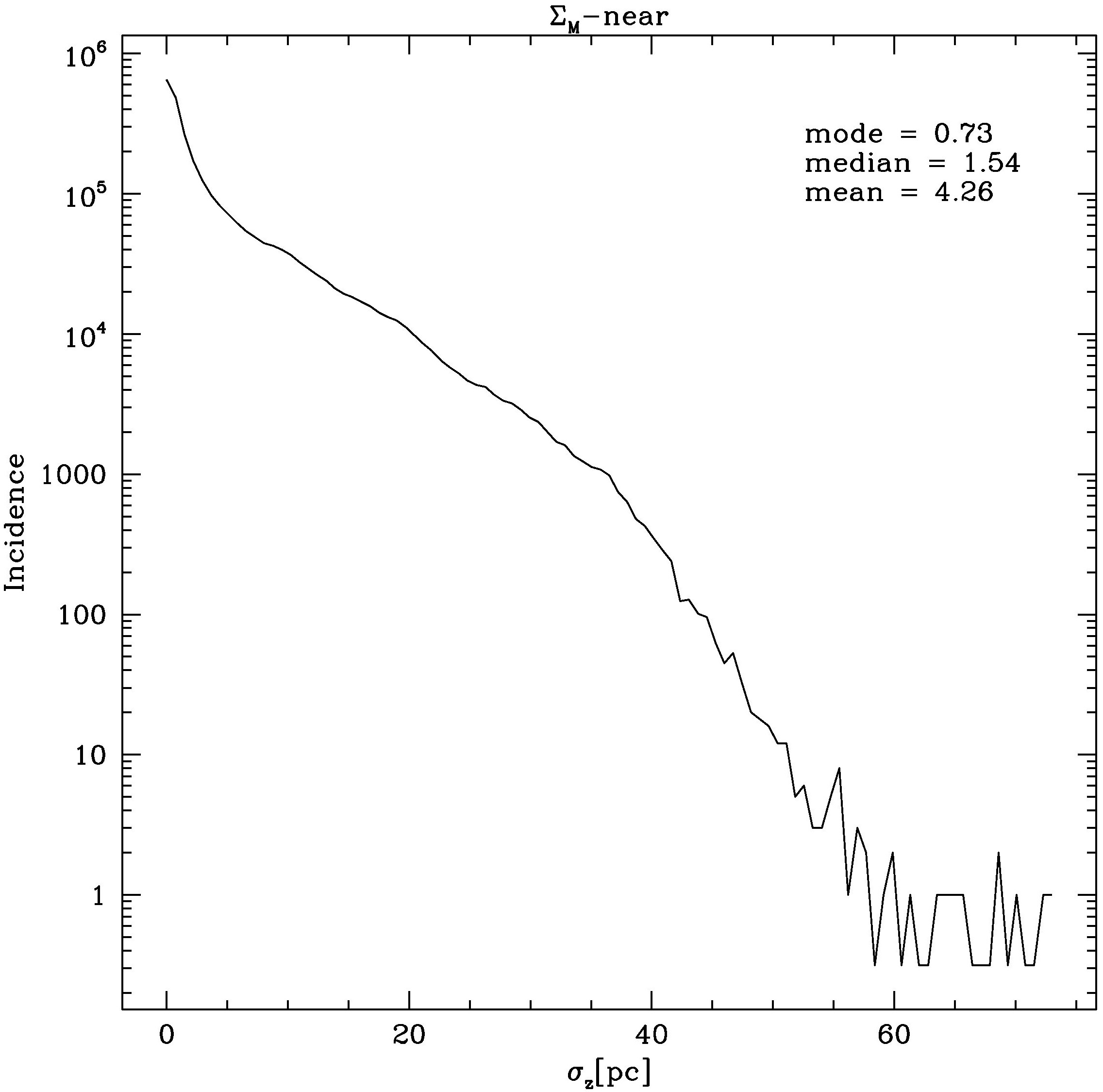}\,\includegraphics[angle=0,scale=0.101]{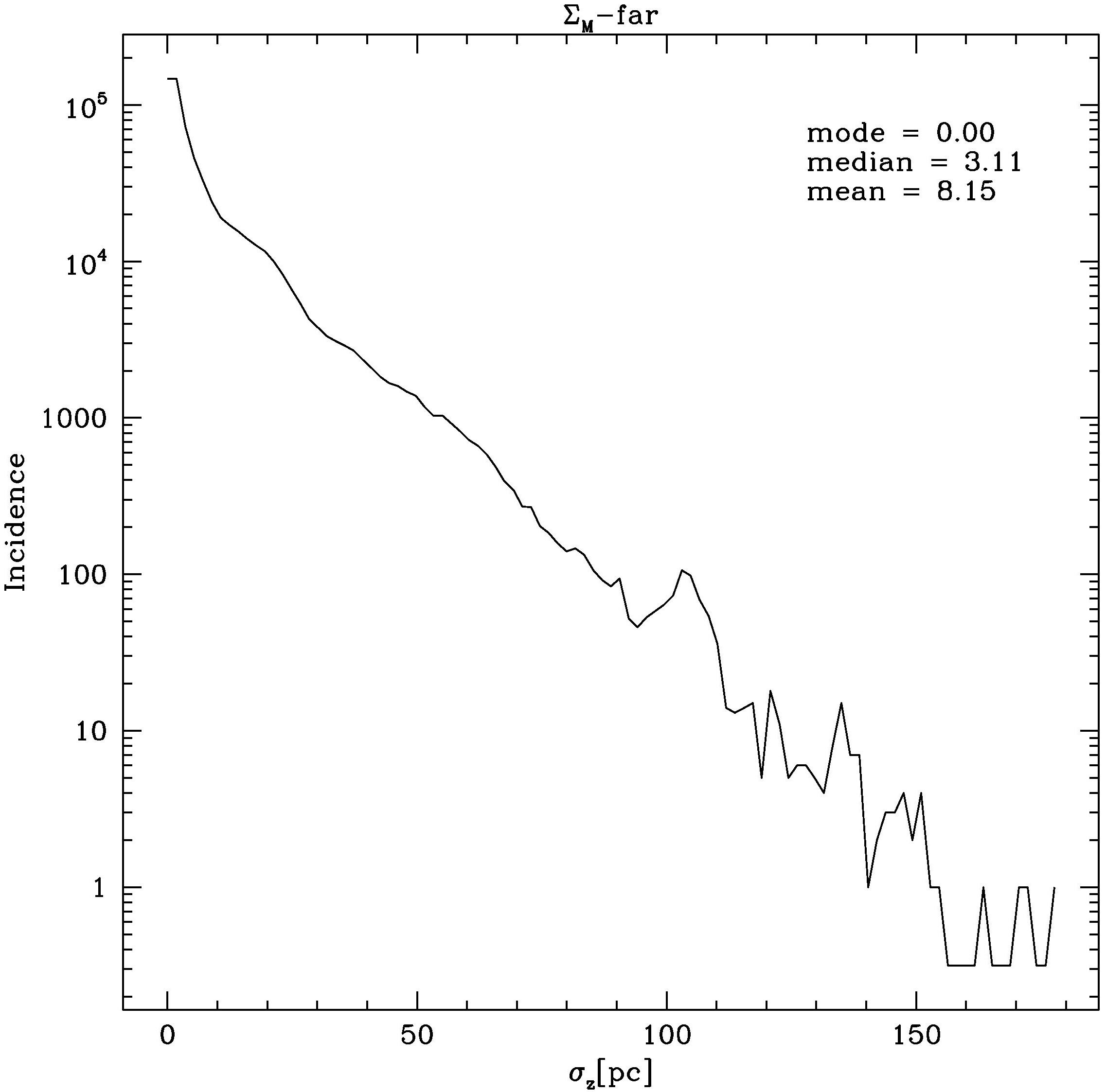}}
\vspace{-1mm}
\caption{\footnotesize Sample distributions of the $\sigma_{z}$ per pixel (2nd latitude moments after application of the $\zeta^{+}$ filter) from Figs.\,\ref{12co-bgt-YX2zp} and \ref{ZM-bgt-YX2zp}.  Each panel shows one subset of the $\sigma_{z}$ data, as labelled, and each histogram contains a well-defined exponential segment in its central portions.  The inverse slopes in these segments correspond to a cloud thickness scale length, respectively $\sim$7.9, 16.5, 4.8, and 12.6\,pc. $$ $$
\label{thickness}}
\vspace{-12mm}
\end{figure*}

\vspace{1mm}To make this clear, we show in {\color{red}Figure \ref{thickness}} examples of the $\sigma_{z}$ distribution in subsets of all pixels in Figs.\,\ref{12co-bgt-YX2zp} and \ref{ZM-bgt-YX2zp}.  These represent the projected vertical thickness of molecular clouds within the midplane, which one would expect to have typical values somewhat less than the scale height of the molecular layer itself.  In all cases, whether traced by \tco\ or \nco, or measured within the tangent distance or beyond it, the mean or median values are a few to 15\,pc.  Interestingly, the histograms all approximate exponential functions, where the inverse slopes give a scale length for each distribution, and these also are on the order of 10\,pc.  These examples lend support to smaller {\em local} scale heights, i.e., $z_{\rm sc}$ $\sim$ 20\,pc as per \cite{r19}, for the Galaxy's molecular layer, even while a sine wave of amplitude $A$ has a {\em global} dispersion $A/\sqrt{2}$, or $\sim$35\,pc for our ripples with $A$ = 50\,pc.  The latter is similar to the scale height of the nearby OB star population as mentioned in \S\ref{nearfar}, suggesting mutual consistency.

\vspace{1mm}Returning to the varying height of the midplane, we are not aware of any previous study defining such widespread and large-scale ripples in the molecular layer of the Milky Way.  However, a number of recent studies using other tracers have shown evidence of similar waves in some portions of the Solar neighbourhood.  One of the largest-scale examples is based on {\em Gaia} and other data for two stellar populations (young giant stars and classical Cepheids) extending several kpc from the Sun's position \citep{pkd24}.  Remarkably, their results show strong evidence for a similar wave {\em exterior} to the Sun's Galactocentric orbit to what we see {\em interior} to the Sun's orbit, specifically, one which is oriented roughly along either Galactocentric circles or the spiral arm pattern.  Moreover, from their kinematic data they make a strong case that this is indeed an outwardly-propagating wave with oscillatory motions, rather than a merely static corrugation.  \cite{pkd24} further speculate that, since their stellar populations are young, and so cannot be kinematically relaxed, this wave motion might have been inherited from these stars' birth clouds.  In response, we note that their wave is approximately in-phase with the molecular ripple, so this supposition seems reasonable.  However, the \cite{pkd24} wave is larger than our ripples, with a half-wave length of $\sim$3\,kpc (i.e., it is one-sided with an equivalent wavelength 6\,kpc) and has a one-sided amplitude 150--200\,pc.  Although larger than the molecular wave in Figure \ref{wiggles}, these values may not be inconsistent due to the higher mass surface densities in the inner Galaxy compared to the outer domain.  That is, a wave propagating through a medium with a decreasing density should be expected to develop longer wavelengths and larger amplitudes, if kinetic energy is being conserved.  For example, this is exactly the manifestation seen in the simulation of \cite{bin24}.  At the very least, this potential complementarity between the two apparent wave signatures is intriguing and mutually reinforcing.

\vspace{1mm}What could provide a physical origin for such an oscillating Galactic midplane?  Many recent studies have found that a disturbance, such as a recent minor merger with the Milky Way, might be very capable of producing this effect in a Galactic disk.  For example, both \cite{bin24} and \cite{a25} modelled the repeated pericentre passages of the Sgr dwarf galaxy, the most recent only $\sim$35\,Myr ago according to \cite{bin24}.  It would be nice if we could use our rough sinusoid to constrain such dynamical models of the Galactic disk, but as both studies show, fully accounting for all the mass distributions (including, e.g., dark matter) in the disk is very challenging.  \cite{pkd24} and references therein also discuss merger events, and several other possible origins as well.  But the Milky Way is a complex system, with many interacting dynamical parts, and further analysis along these lines is well beyond the scope of this paper.

\vspace{1mm}A much cleaner example of a disk system is that of Saturn's rings.  The same physics (although on a {\em far} smaller scale!) also seems to operate in its C Ring, which has apparently been ``ringing like a bell'' from a heretofore unknown impactor since it collided with Saturn's ring plane in 1983 Sept 19 \citep{f25}.  That is, the impactor has induced an $m$ = 1 disturbance in the ring plane, which winds up over time to produce the spiral wave pattern we now see.  We suppose that neither the Saturn nor Milky Way cases are unusual in the Universe, and that disk oscillations are likely to be a common phenomenon, once techniques to detect them are improved.

\subsubsection{The Far Ara Clouds --- Large Distant Structures, Or A New Neighbour?}\label{farside}

\vspace{1mm}The second novel feature in the ThrUMMS data is more subtle than the widespread ripples, but upon noticing its existence, cannot be unseen.  Recall from \S\S\ref{latmaps} and \ref{nearfar} that the general pattern for the latitude distribution is to shrink from $\pm$1\degree\ at the nearer distances to a much smaller range (typically $\pm$0\fdeg3) as one runs towards the tangent distances or the Galactic centre.  However, there are a set of clouds in $l$ = 331\degree--340\degree\ where both the mean latitude $\bar{b}$ and latitude dispersion $\sigma_b$ remain close to zero over a very wide range of \vlsr.  The longitudes place the clouds close to the common corner of the constellations Norma, Scorpius, and Ara, but since the first two constellations already have large-scale Galactic features named after them, we dub these the ``Far Ara Clouds.''  Because of their size, they are all likely to be GMCs.  These features are most evident in the \nco\ panel of Figure \ref{full-lv1-12coZM-rainbow} as three separate clouds in \lv\ space.  More detailed zoom-ins appear in {\color{red}Figure \ref{ara}}, from which we define:

% Figure C33: FA finder
\begin{figure*}[t]
\centerline{\includegraphics[angle=-90,scale=0.34]{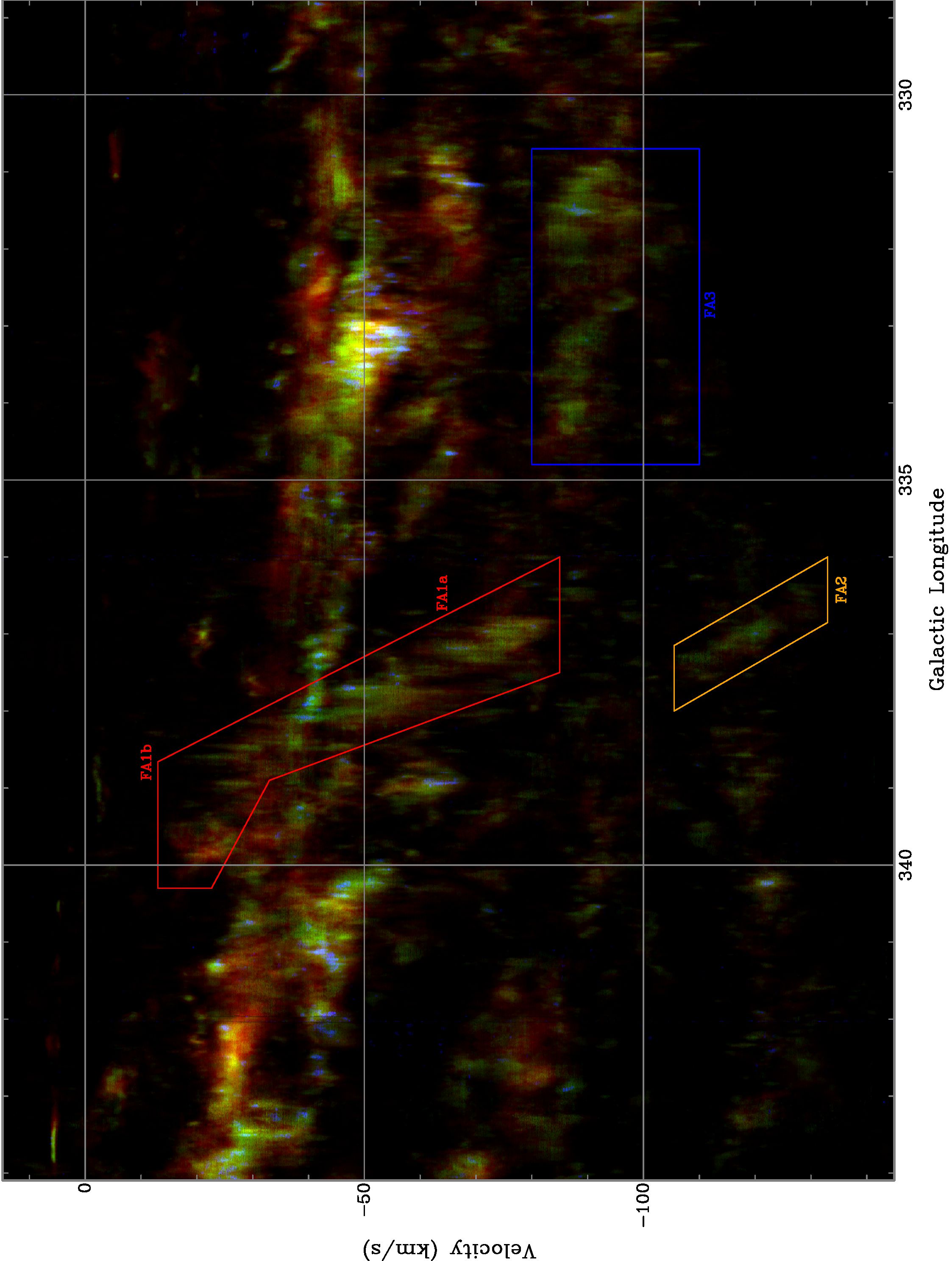}\,\includegraphics[angle=-90,scale=0.34]{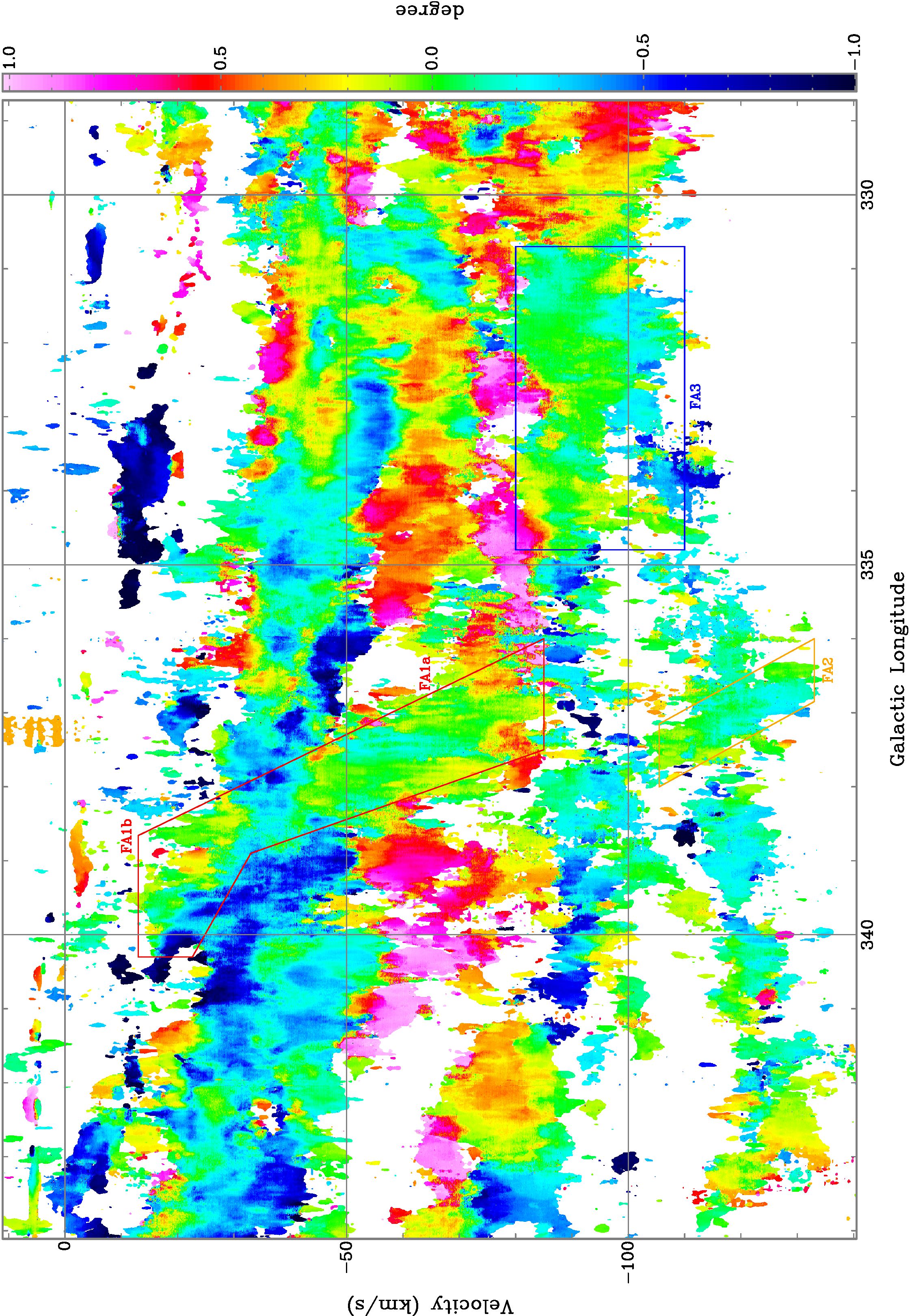}}
\vspace{-2.5mm}
\centerline{\includegraphics[angle=-90,scale=0.34]{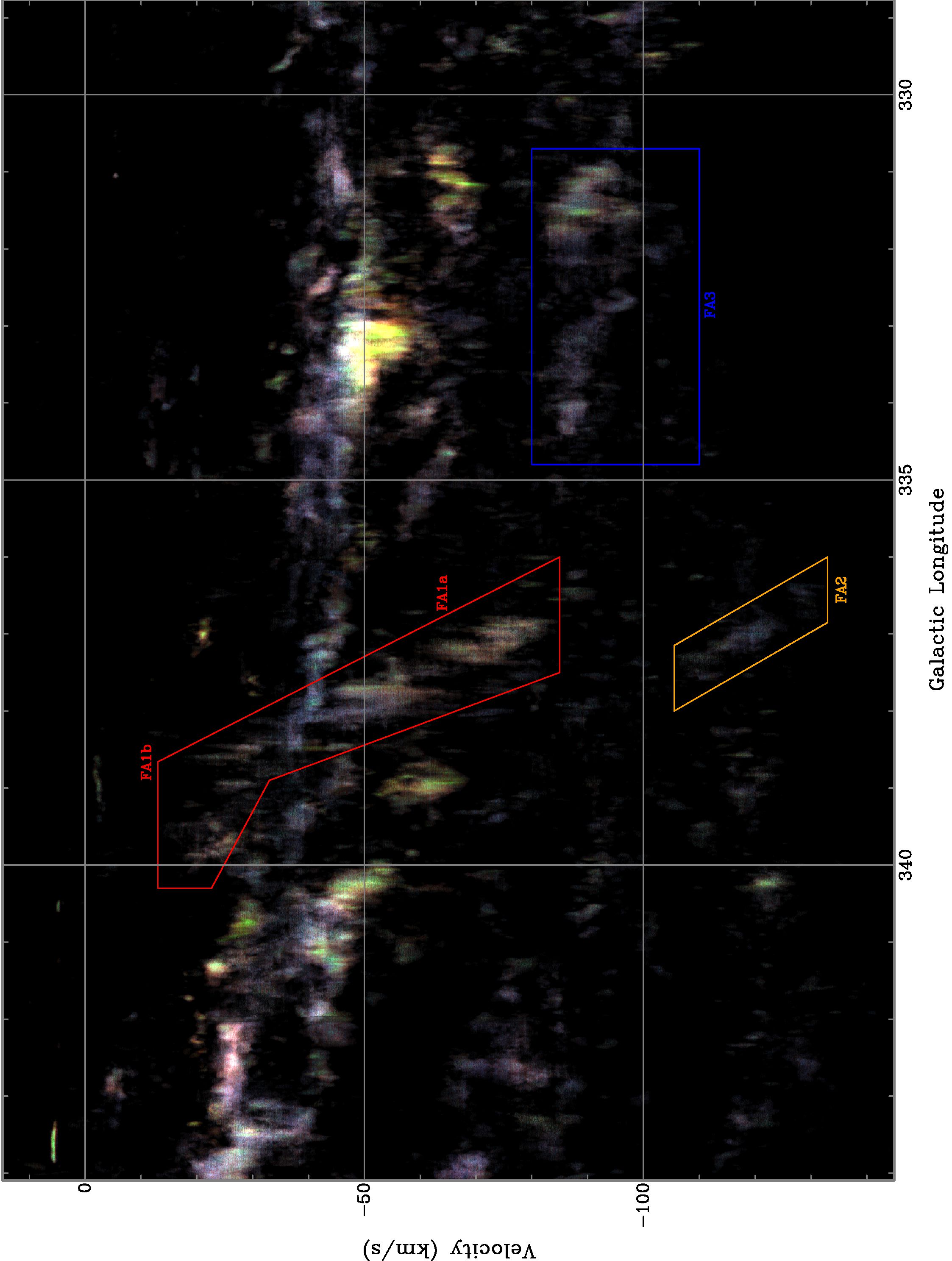}\,\includegraphics[angle=-90,scale=0.34]{FAzoom-ZM.jpg}}
\vspace{-1mm}
\caption{\footnotesize Finder charts for the Far Ara Clouds in \lv, demarked by coloured polygons in each panel (FA1 = red, FA2 = orange, FA3 = blue) which are used as cutouts for further analysis.  ({\em Left column}) RGB colour composites of the \tco, \ttco, and \ceto\ integrated intensities ({\em top}) and the mean-\tex, total \nco, and mean-$\tau$ ({\em bottom}), as in Fig.\,\ref{full-lv-combo}.  {(\em Right column}) Mean latitude $\bar{b}$ maps for ({\em top}) \tco\ and ({\em bottom}) \nco, as in Fig.\,\ref{full-lv1-12coZM-rainbow}. $$ $$
\label{ara}}
\vspace{-7mm}
\end{figure*}

\vspace{1mm}{Cloud 1 (hereafter FA1) spans $l$ = 336\fdeg7--340\fdeg3 and \vlsr\ = --85 to --13\,\kms, wholly within Ara's borders.  The equivalent deprojected coordinates are \xy\ $\approx$ (+10--13\,kpc, --4 to --5\,kpc).  FA1 is the most massive of the three clouds based on the moment-0 latitude integrals, but may be separable into two unequal parts, FA1a and FA1b; %(Figs,\ref.{both-ld0-b1p3m}. \ref{ZM-bgt-YX0zp}, \ref{ZM-bgt-YX0zp}).

\vspace{1mm}Cloud 2 (FA2) at $l$ = 335\fdeg7--337\fdeg7, \vlsr\ = --135 to --105\,\kms\ (also inside Ara), \xy\ $\approx$ (+8.3\,kpc, --3.5\,kpc); and

\vspace{1mm}Cloud 3 (FA3) at $l$ = 330\fdeg7--334\fdeg8 and \vlsr\ = --110 to --80\,\kms\ (just outside Ara, across the border with Norma) and \xy\ $\approx$ (+8.5\,kpc, --4.5\,kpc).}

\vspace{1mm}Using the \nco\ panel as a key, the clouds' signatures are also easily seen in \tco\ (the top panels of Figs.\,\ref{full-lv1-12coZM-rainbow} and \ref{ara}) and both \lv\ maps of $\sigma_b$ (Fig.\,\ref{full-lv2-12coZM}).  They are likely to be intrinsically far away, kinematically speaking, since they are very flat compared to most other clouds at the more easily-detected near distances.  With our height-based near/far discrimination, this makes them rather unlikely to be at a near distance (typical likelihoods $\zeta^+$ $\sim$ 0.2$\pm$0.2, compared to far values $\zeta^+$ $\sim$ 0.8$\pm$0.2), since then their sizes and heights would be constrained to less-credible small values.  If this placement can be confirmed, they would then be the largest and most massive clouds that we detect beyond the tangent distance in the 4Q, their \vlsr\ placing them 9--14\,kpc from the Sun, beyond any other believable location for a far-kinematic cloud.\footnote{We discount the features visible in Figs.\,\ref{12co-bgt-YX0fl} and \ref{ZM-bgt-YX0fl} around coordinates \xy\ = (+4--8\,kpc, $\sim$--6.5\,kpc), or longitudes $\sim$300\degree--320\degree, since their $\zeta^+$ values are more marginal (near-$\zeta^+$ $\sim$ 0.35$\pm$0.3 vs far-$\zeta^+$ $\sim$ 0.65$\pm$0.3), meaning their placement at far distances, away from otherwise similar nearside features around \xy\ $\approx$ (+2\,kpc, --2.5\,kpc), is not as clear-cut.}  Note also that their flatness and relative contiguousness persist with any values of $w$ or $z_{\rm sc}$, and that they are clearly visible among, and clearly different types of clouds compared to, their neighbours even in \lv\ space, once one knows what to look for.

\vspace{1mm}As such, and because each of the three clouds/GMCs is large and massive, possibly forming the only large-scale features of the farside molecular disk, they merit a closer inspection.  For example, how do they compare in physical properties to more typical nearside clouds?  We use the above \lv\ limits to make cutouts of each cloud and separate them from their brighter and more prominent foreground siblings.

\begin{figure*}[t]
\vspace{0mm}
{\includegraphics[angle=-90,scale=0.325]{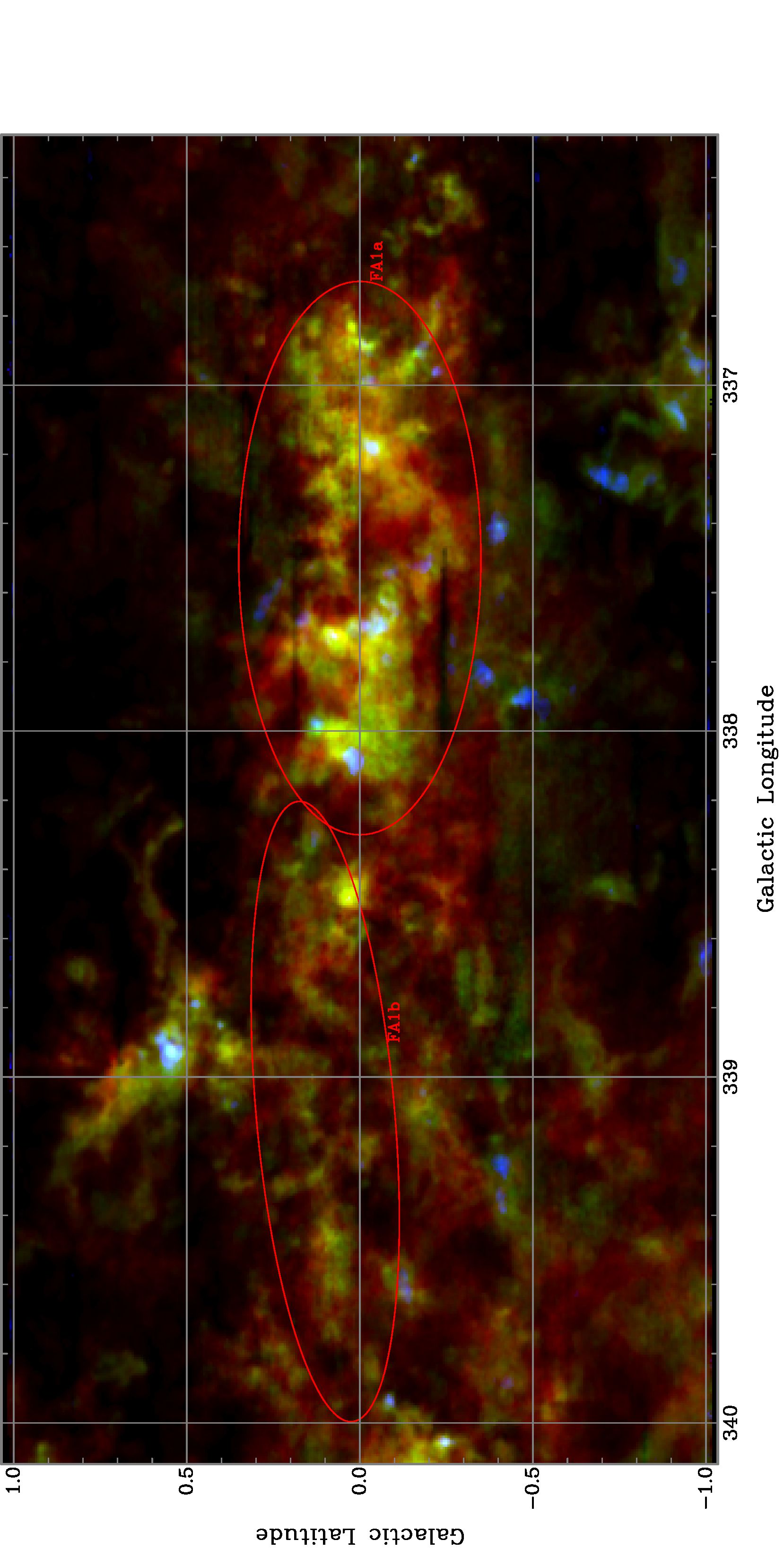}\includegraphics[angle=-90,scale=0.325]{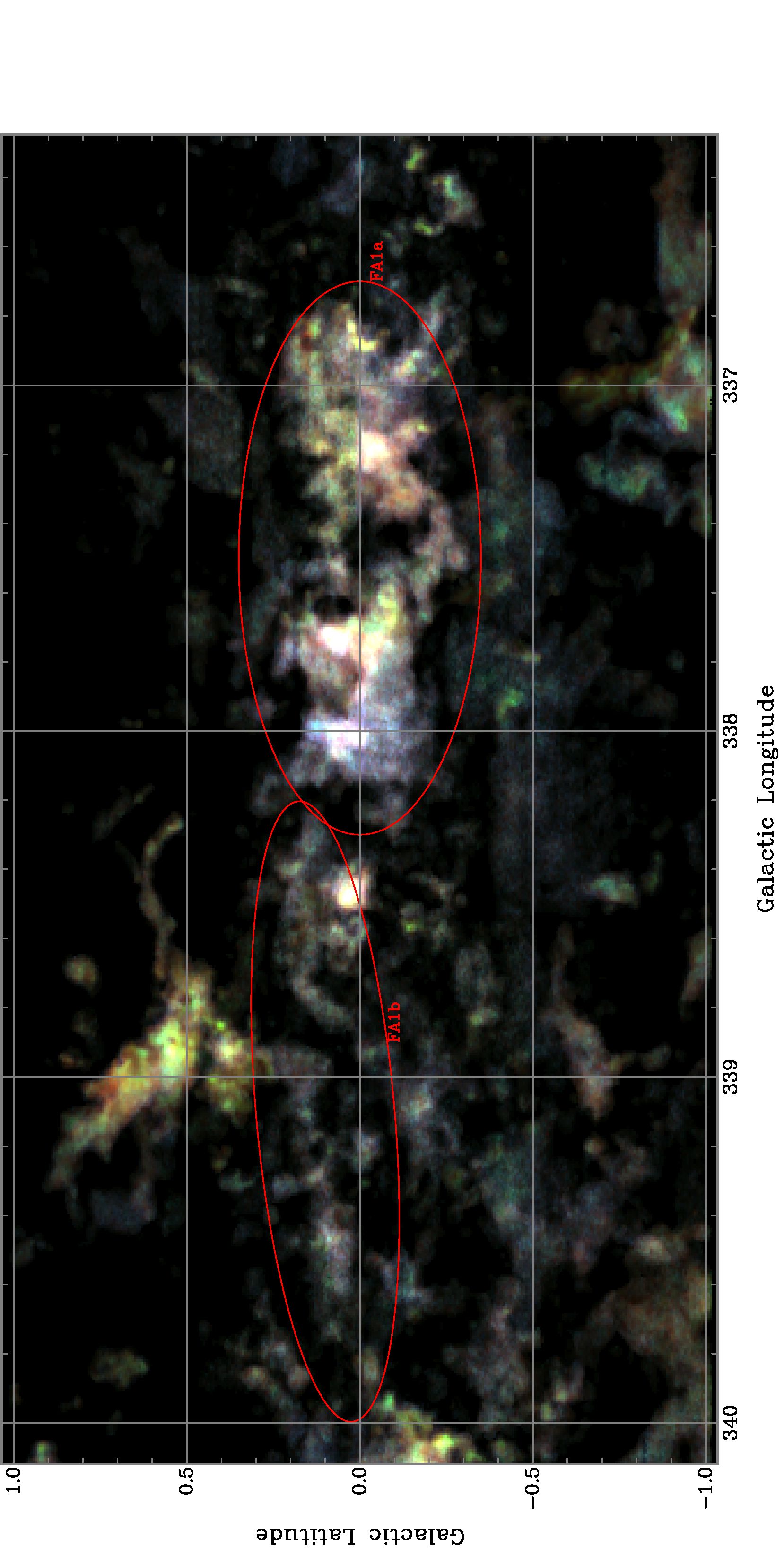}}

{\vspace{-2.7mm}\includegraphics[angle=-90,scale=0.325]{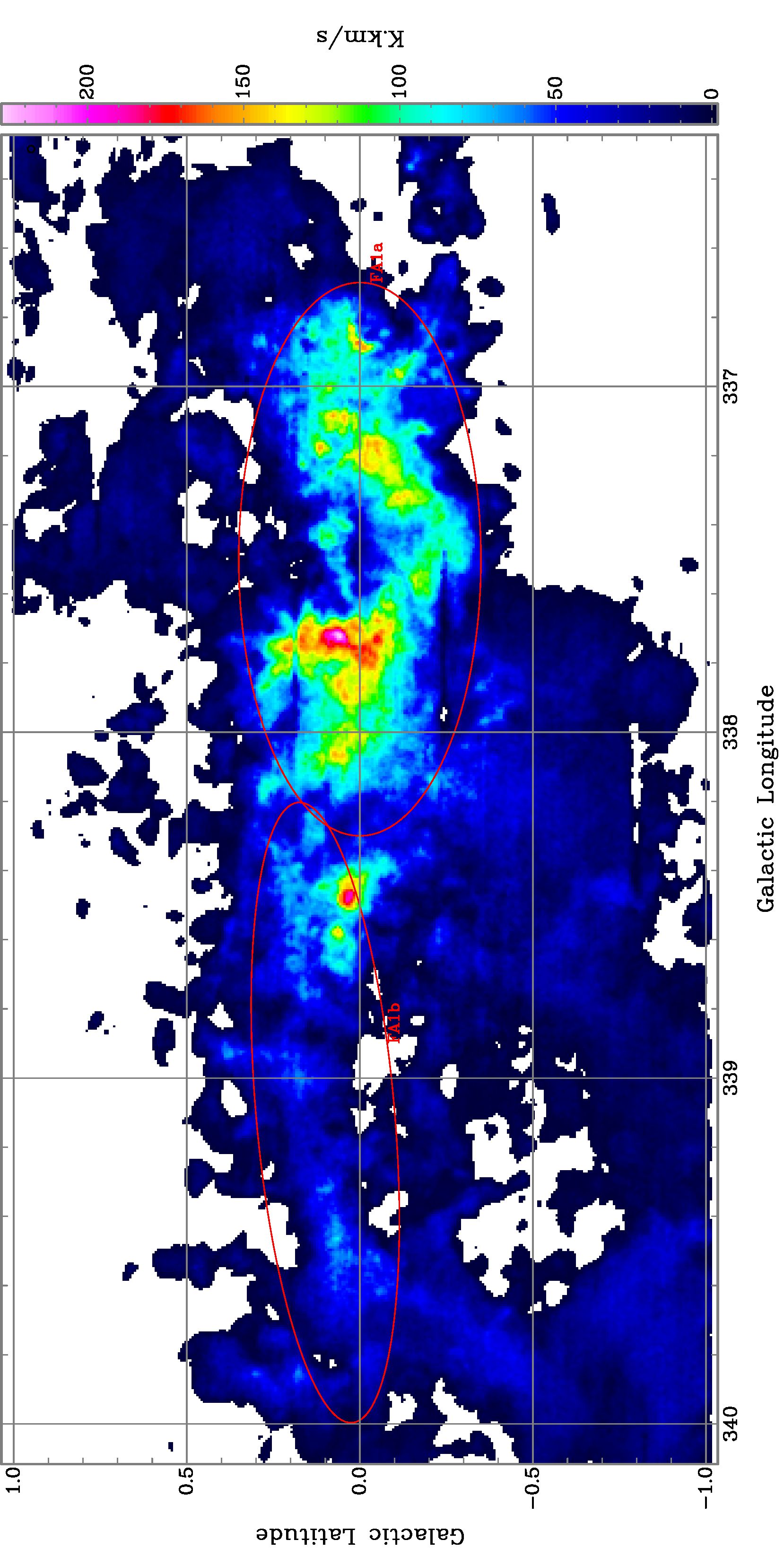}\includegraphics[angle=-90,scale=0.325]{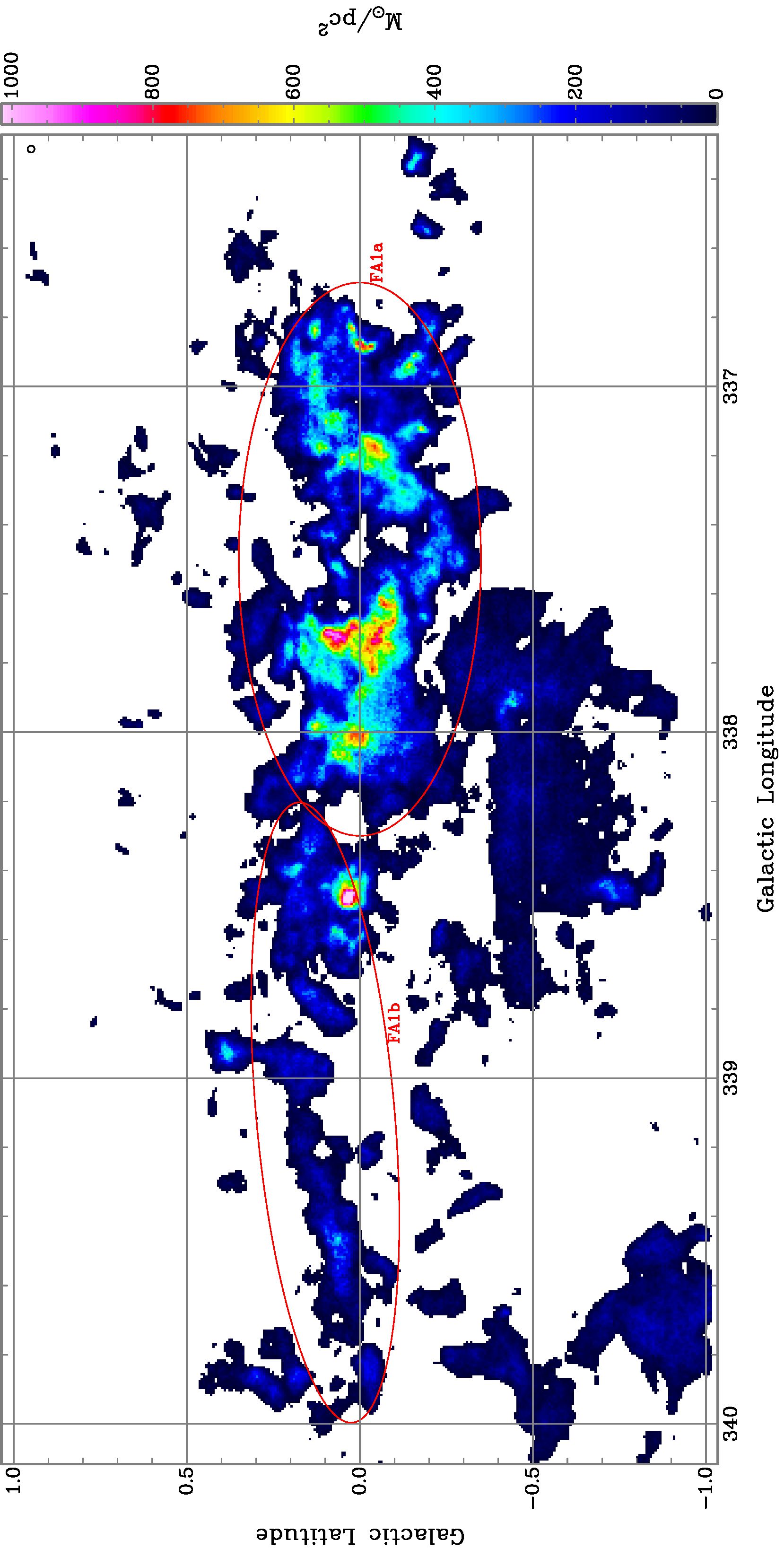}}

{\vspace{-2.7mm}\includegraphics[angle=-90,scale=0.325]{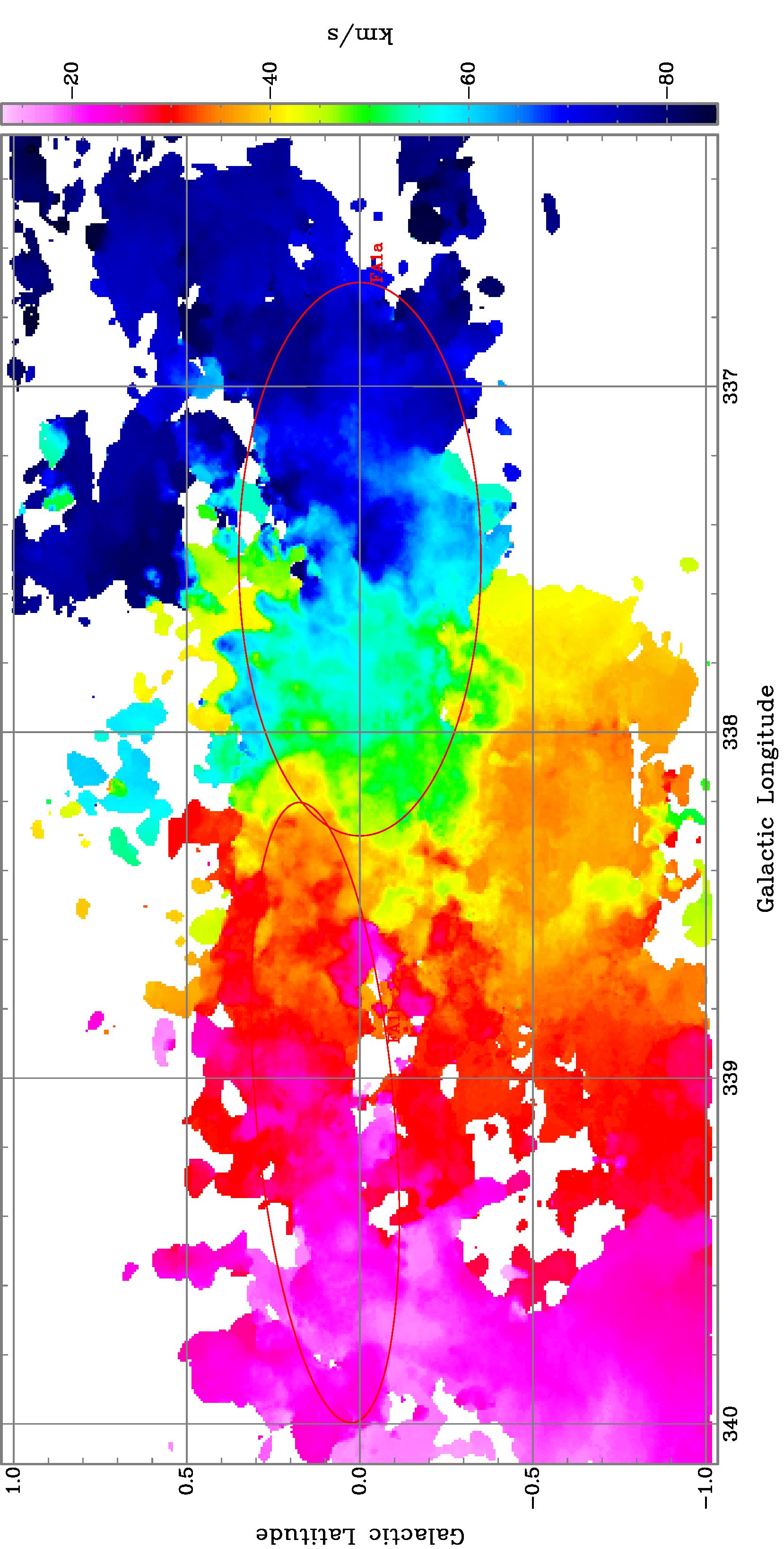}\includegraphics[angle=-90,scale=0.325]{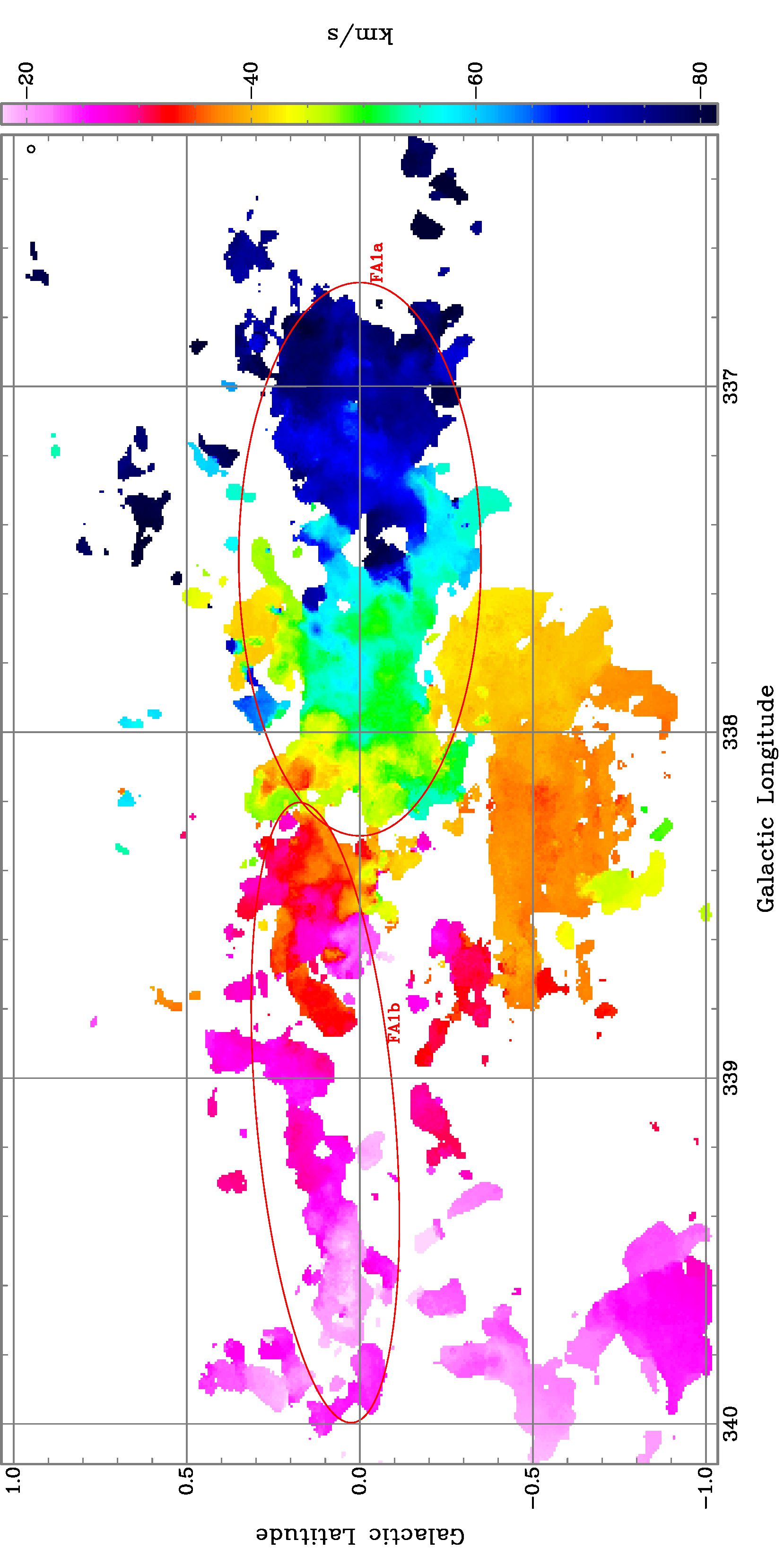}}

{\vspace{-2.7mm}\includegraphics[angle=-90,scale=0.325]{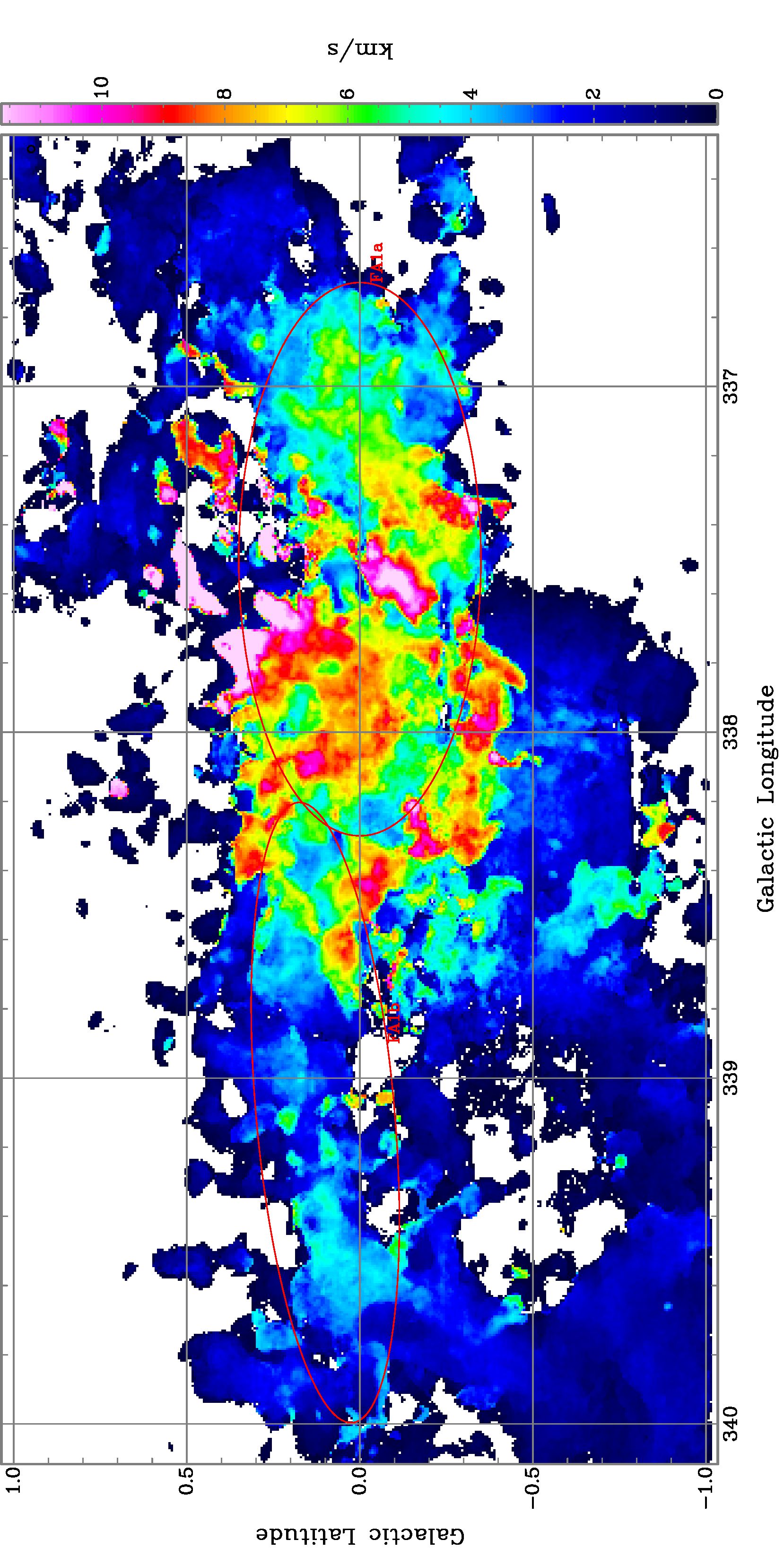}\includegraphics[angle=-90,scale=0.325]{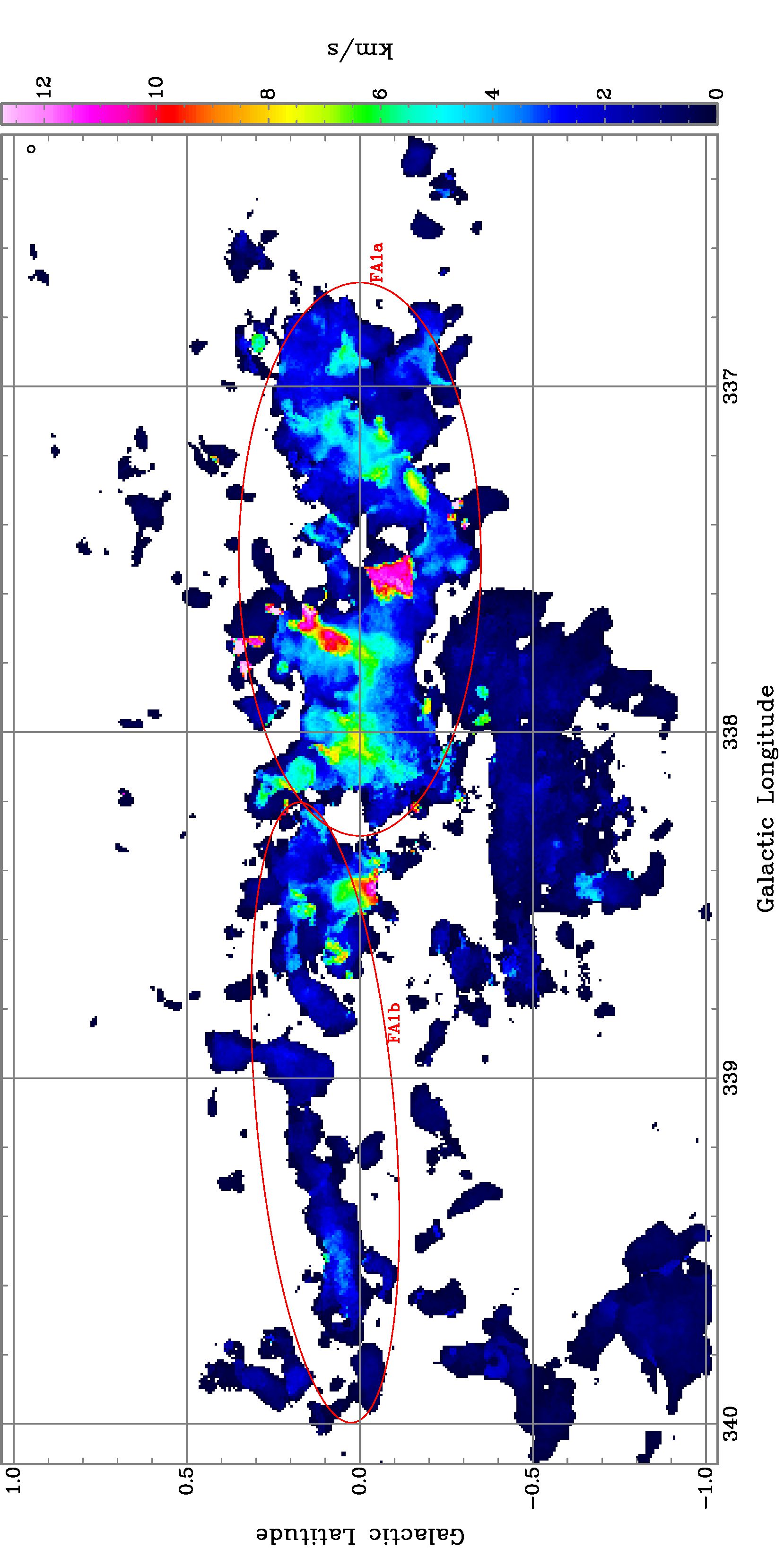}}

\vspace{-1mm}
\caption{\footnotesize On-sky \lb\ moments of the area around FA1.  In the top row we show RGB composites integrated over all \vlsr\ of ({\em left}) the \tco, \ttco, and \ceto\ intensities as in Fig.\,\ref{full121318-mom0}, and ({\em right}) the mean-\tex, total \nco, and mean-$\tau$ as in Fig.\,\ref{fullTexZMtau-mom0}.   The bright cloud at (338\fdeg9,+0\fdeg5) is an apparently foreground feature ($d$ $\sim$ 4\,kpc) compared to FA1.  In rows 2--4, we show separate moments of FA1, masked to the \lv\ polygon cutout shown in Fig.\,\ref{ara}, of ({\em left column}) the \tco\ and ({\em right column}) \nco\ cubes, respectively the 0th, 1st, and 2nd moments in each row.  To guide the eye, the nominal components FA1a and FA1b are shown approximately by labelled red ellipses in each panel.  Much of the non-FA1 structure seen in rows 2--4, particularly the \tco\ data, comes from a distinct signature of part of the Scutum-Centaurus Arm at a consistent kinematic distance of 3\,kpc and $\bar{b}$ $\approx$ --0\fdeg5. $$ $$
\label{FA1moms}}
\vspace{-9mm}
\end{figure*}

\vspace{1mm}We start with FA1, and show in {\color{red}Figure \ref{FA1moms}} some sample moment maps from its cutout (the red polygon in Fig.\,\ref{ara}).  The moments are extracted from the {\em unfiltered} \lbv\ data, since the $\zeta^+$ function uses a statistical approach to placing individual pixels, which should only be taken as a guide towards identifying contiguous clouds, which we are now attempting.  In these panels, we see that FA1 looks quite ordinary in most respects, except for its very large velocity gradient, 72\,\kms\ over about 3\degree\ in $l$, and its large distance.  At its median far-kinematic value of 12\,kpc, its overall properties are also correspondingly large.  Integrating the \nco\ map of FA1 (right column, second row in Fig.\,\ref{FA1moms}) within --0\fdeg3 $<$ $b$ $<$ +0\fdeg5 yields a total mass of about 7.7$\times$10$^6$\,M\solar, somewhat extreme even for a GMC.  Similarly, the longitude extent projects to a length on the sky of over 600\,pc.  Even more perplexing is the implied front-to-back size of 3\,kpc from the range of kinematic distances, which is longer than any other GMC or Giant Molecular Filament catalogued to date.

% Figure C34: FA1 moments

\vspace{1mm}The situation is not much improved if one prefers (despite FA1's extreme flatness) a near kinematic distance.  Although the mean $d$ is then only 3\,kpc, its kinematically implied front-to-back depth is still $\sim$2.5\,kpc, and the projected $l$ extent is still a considerable 150\,pc.  Although its total mass then integrates to only 4.8$\times$10$^5$\,M\solar, more typical for a large GMC, the size problem remains, and its $\sigma_z$ $\sim$ 10\,pc is then quite small if its depth is to be believed.

\vspace{1mm}More importantly, the velocity gradient is difficult to fit with most other spiral arm patterns seen elsewhere in the Galaxy.  At the disfavoured near distance, FA1's pitch angle would be negative (about --45\degree), making that location even more unlikely.  At the far distance, its PA is very high ($\sim$60\degree, best seen in Figs.\,\ref{12co-bgt-YX0fl} \& \ref{ZM-bgt-YX0fl}).  As such, it might be a rather extreme example of a ``feather,'' seen to come off spiral arms in some grand-design spirals like M51.  Those feathers don't usually make such large angles with tangentiality, but at least there is a possible analogue in a similar but smaller ($\sim$1\,kpc long) feature in the Sagittarius Arm \citep{k21}.  At $\sim$3.7\,kpc, the combined length of FA1+2 is nevertheless a challenge to understand, especially given its rather uniform thickness $\sigma_z$ $\sim$ 20\,pc.

% Figure C35: FA2 moments
\begin{figure*}[t]
\centerline{\includegraphics[angle=-90,scale=0.30]{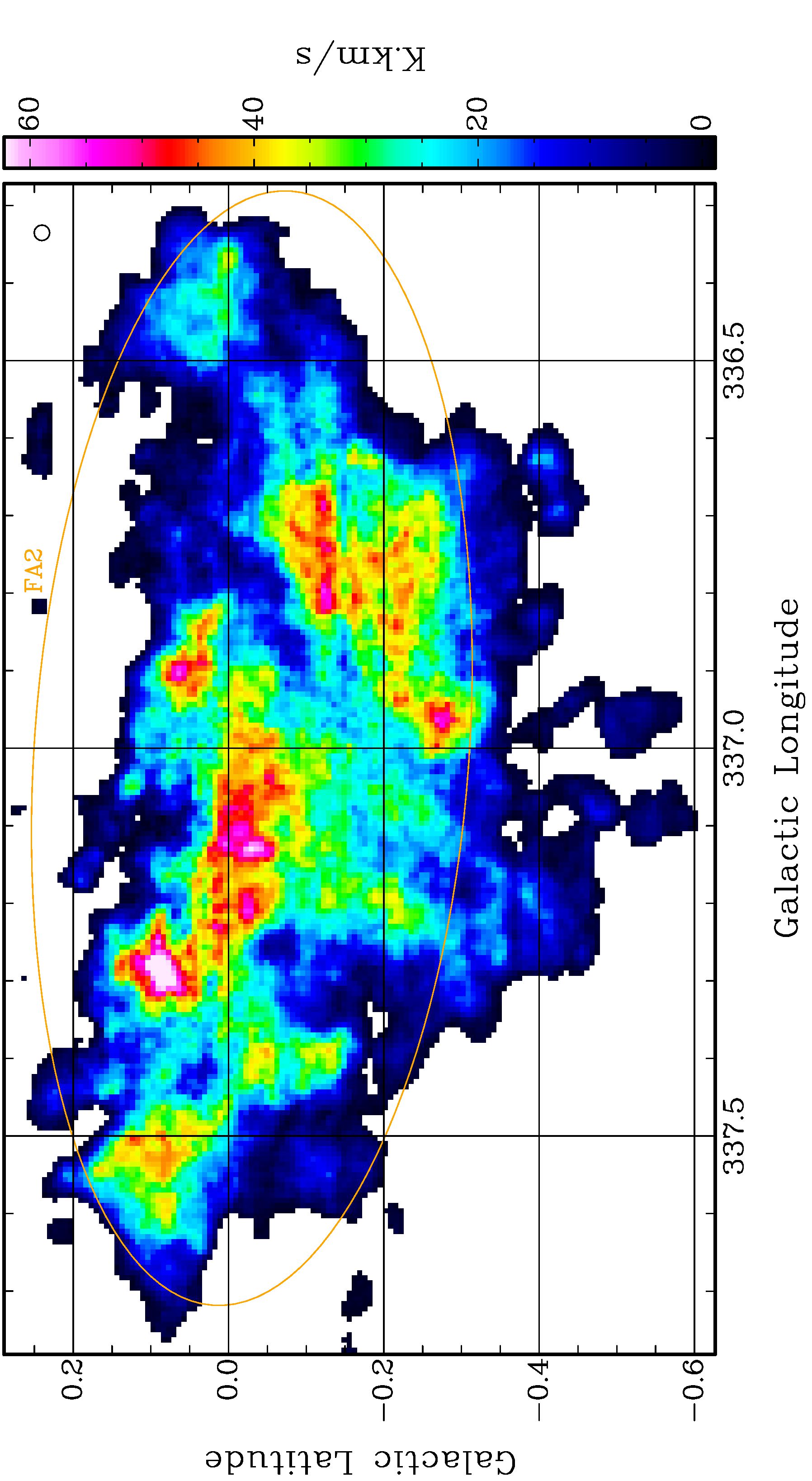}~~\includegraphics[angle=-90,scale=0.30]{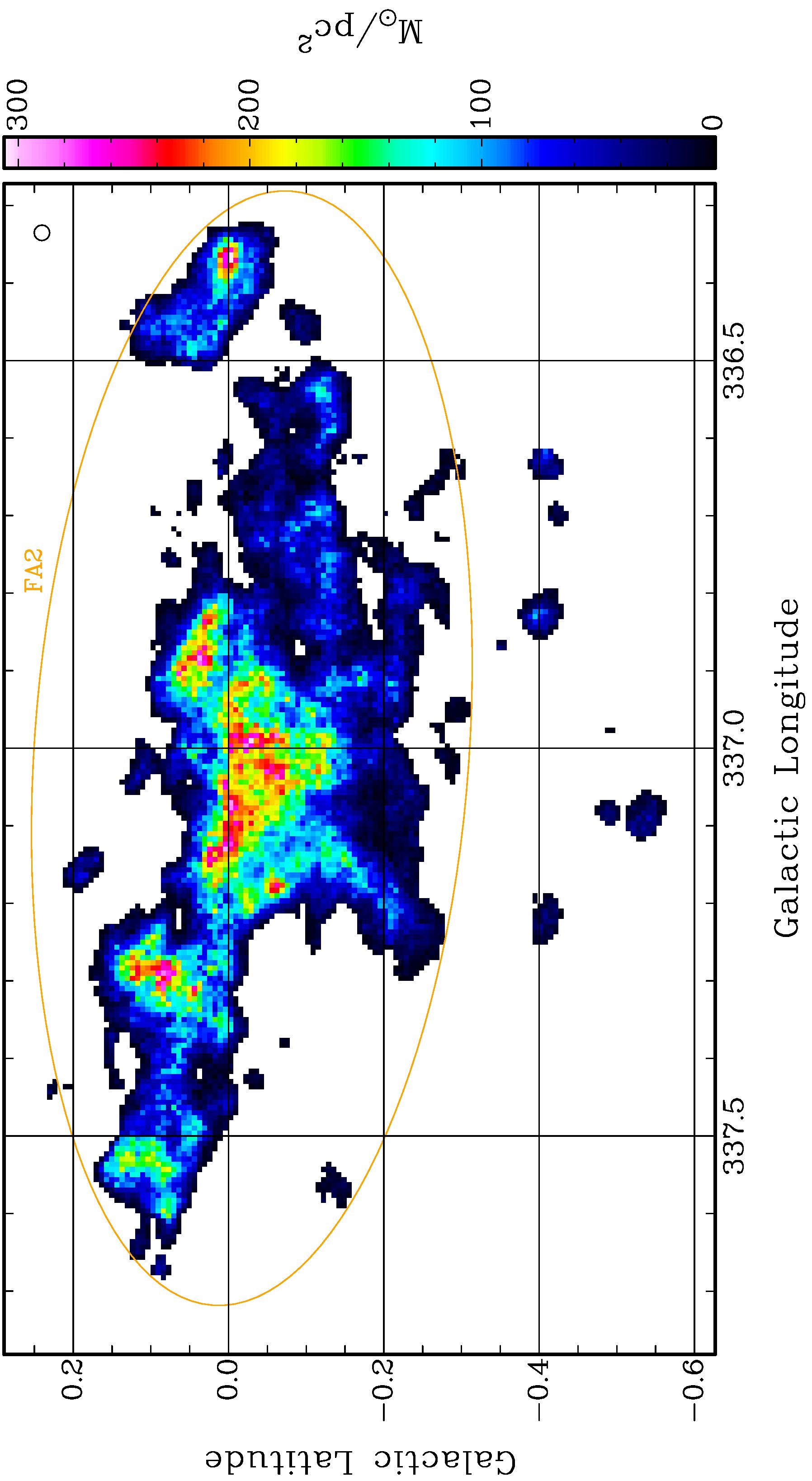}}

\vspace{-4mm}
\centerline{\includegraphics[angle=-90,scale=0.30]{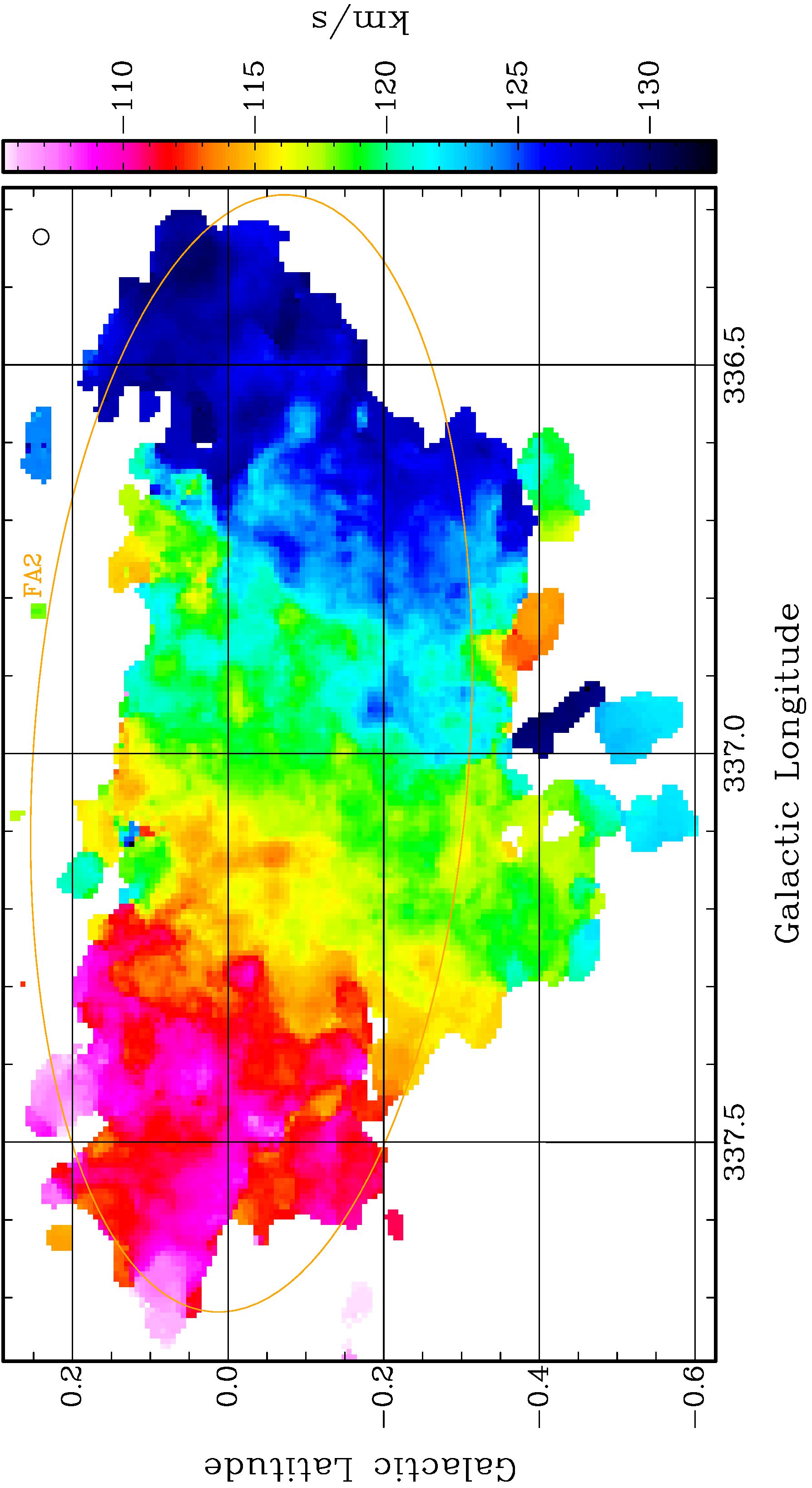}~~\includegraphics[angle=-90,scale=0.30]{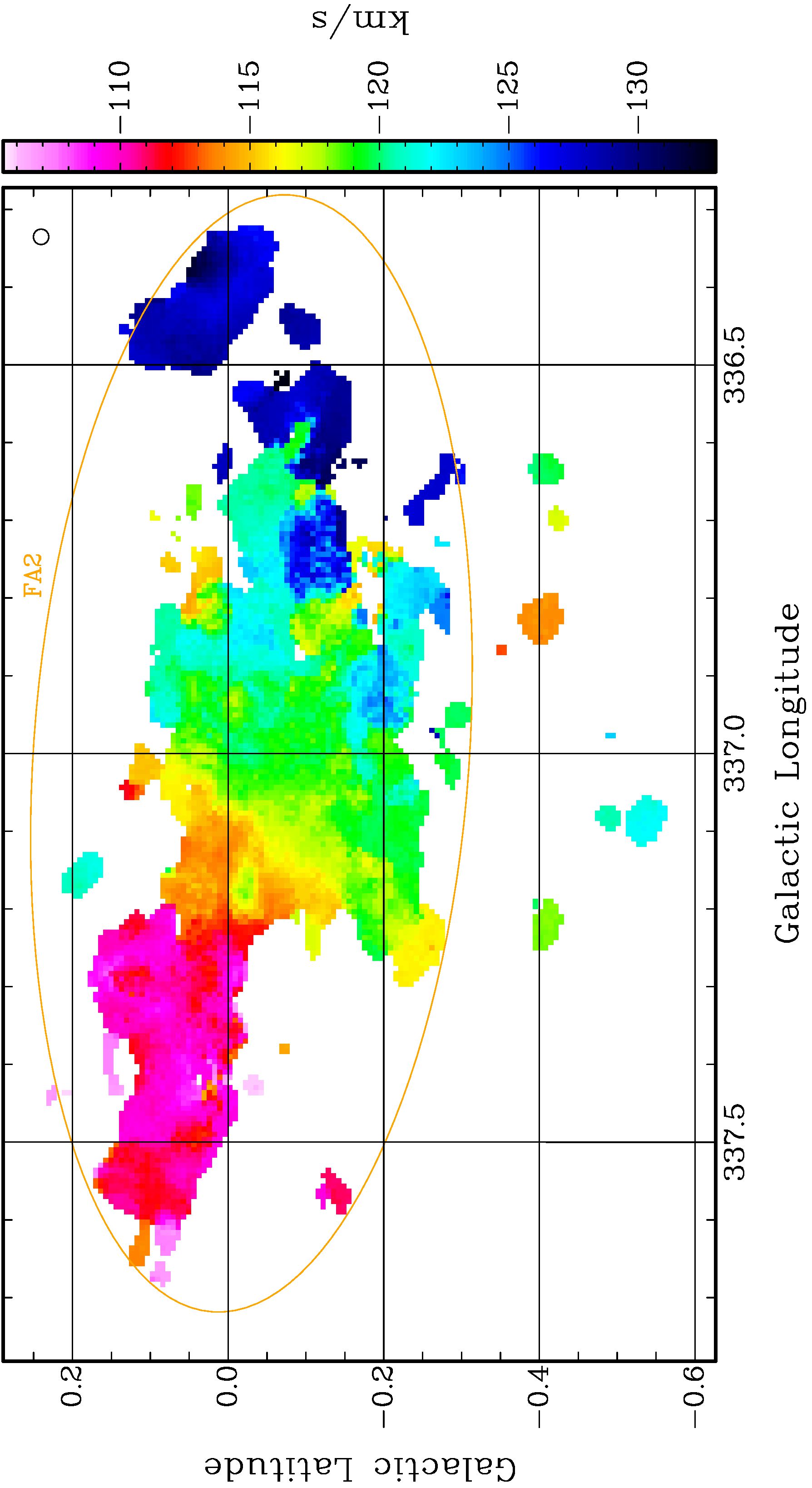}}

\vspace{-4mm}
\centerline{\includegraphics[angle=-90,scale=0.30]{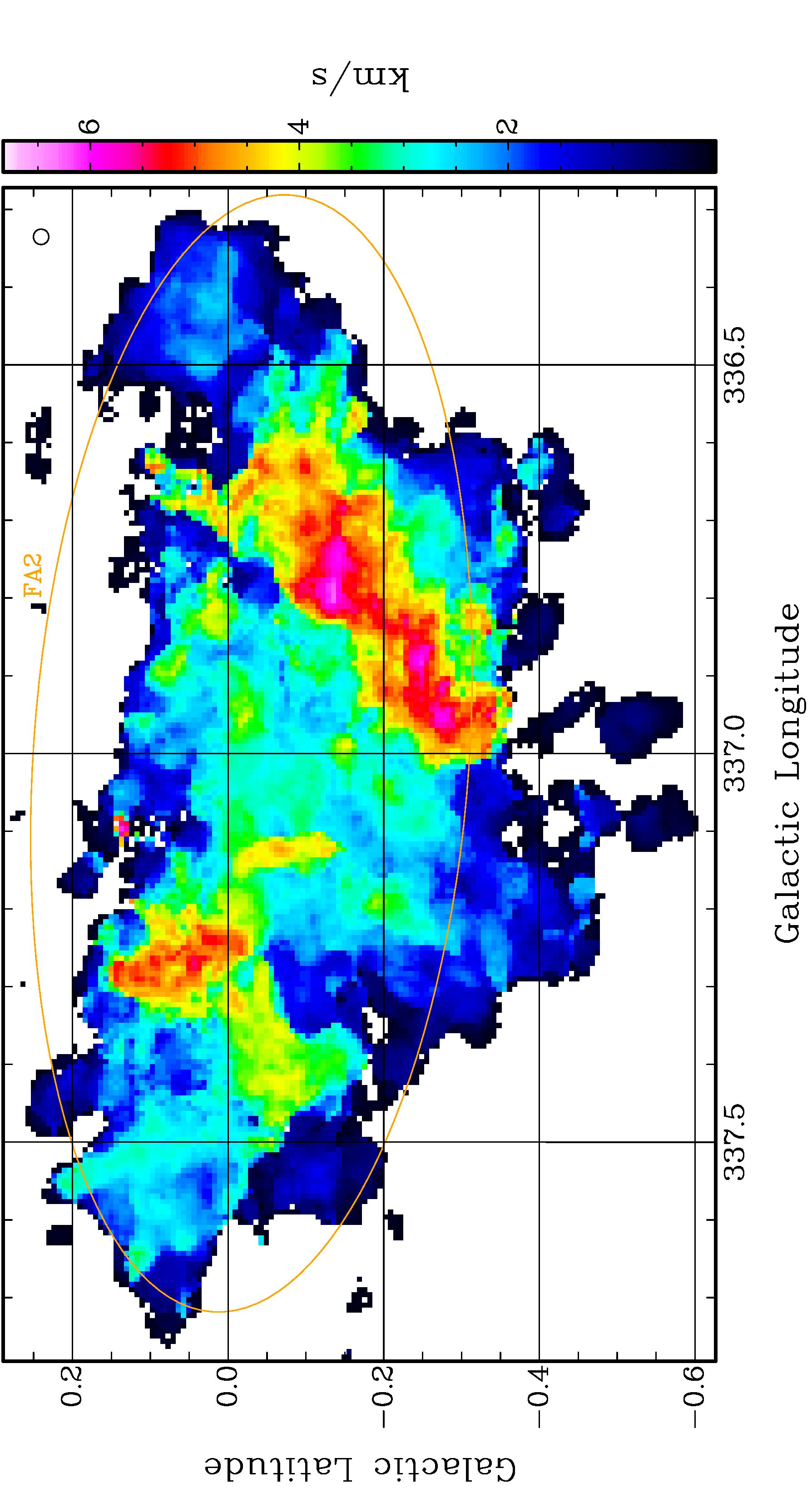}~~\includegraphics[angle=-90,scale=0.30]{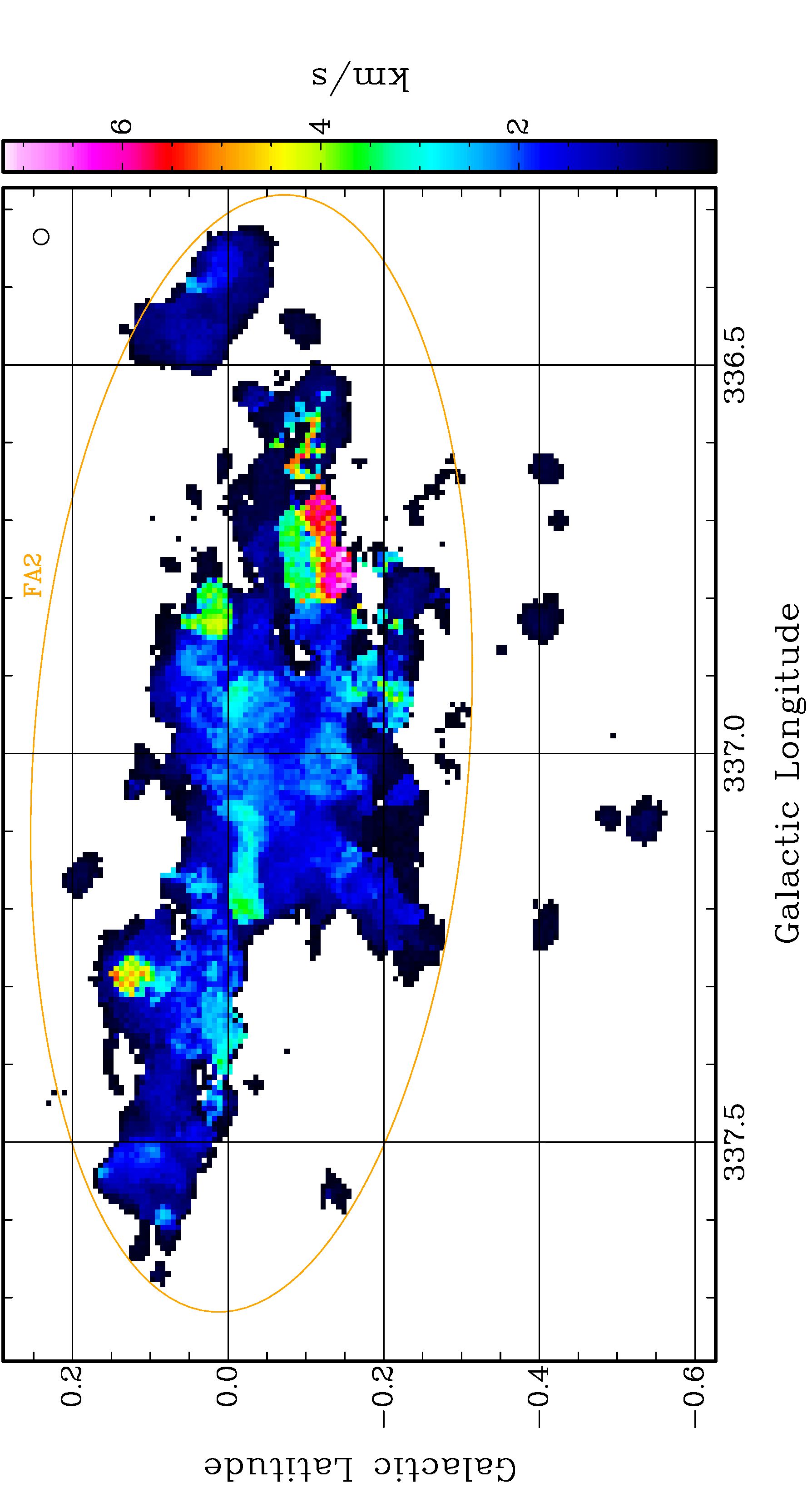}}

\vspace{-1mm}
\caption{\footnotesize Separate moments of FA2, masked to the orange \lv\ polygon cutout shown in Fig.\,\ref{ara}, of ({\em left column}) the \tco\ and ({\em right column}) \nco\ cubes: respectively, the 0th, 1st, and 2nd moments in rows 1--3.  The nominal outline of FA2 is shown approximately by a labelled orange ellipse in each panel. $$ $$
\label{FA2moms}}
\vspace{-7mm}
\end{figure*}

\vspace{1mm}Alternatively, FA1's extreme velocity gradient is so unique, one is tempted to consider an even more radical option: that it is a completely separate entity to normal Galactic molecular cloud complexes, perhaps akin to a gas-rich dwarf galaxy that happens to lie close to the nominal Galactic Plane; or, a very massive external gas cloud like a High-Velocity Cloud, but one possibly in the process of either falling into or being tidally stripped by the Milky Way.  Since the Sgr dwarf lies only $\sim$20\degr\ away on the sky from the Ara Clouds, it is at least conceivable that the gas in these clouds was stripped (in this case, by ram pressure) from the dwarf during its last pericentre passage, and is now infalling.  The latter scenario could readily explain its velocity gradient, but then its distance would be observationally indeterminate (although probably beyond the 12\,kpc discussed above), since its \vlsr\ would not be meaningfully connected to normal Galactic rotation.  Also, FA1's flat and thin aspect would be difficult to understand in a dynamical accretion scenario.  Still, its vaguely cometary shape in Figure \ref{FA1moms} lends some circumstantial support to this hypothesis, as does the inferred ratio of gas (if FA1 lies at 20\,kpc; see below) to total-Sgr-dwarf mass \citep[from the simulation of][]{a25} of 2.4$\times$10$^7$M\solar/4$\times$10$^9$M\solar\ $\approx$ 6\%, making the original Sgr galaxy a not-unreasonably gas-rich dwarf.

\vspace{1mm}This notion may also be relevant to a consideration of FA2, which we turn to next (see {\color{red}Fig.\,\ref{FA2moms}}).  Here again, the cloud is unremarkable compared to other Galactic GMCs, except for its very large velocity gradient (middle panels of Fig.\,\ref{FA2moms}), which together with FA's, stretches over $\sim$115\,\kms.  Its near- and far-kinematic distances are around 6.5 and 8.5\,kpc, respectively, meaning that by itself, it appears to be close to the tangent distance at this longitude, 7.5\,kpc.  However, if it is associated with FA1, median distances for them both would be around 4.5 or 11\,kpc.  The total mass of FA2 (from the top right panel of Fig.\,\ref{FA2moms}) is not as high as (only $\sim$10\%) that of FA1, but still significant at any of these distances, 6.0$\times$10$^5$\,M\solar\,($d$/10\,kpc)$^2$.  %mean Nco = 66.819908 Msun/pc2, 6662 pixels, sum = 445,154 Msun/pc2, converted to scaling at d = 1 kpc: 6026.744 Msun.  At nominal d=12 kpc, this = 867,851.2 Msun

\vspace{1mm}Given all this information, how do we discern the nature of these clouds?  Are any of the hypothesis offered, that of a fairly massive molecular cloud falling into the Galaxy (whether originally from the Sgr dwarf, or not), or that of a separate, neighbouring, gas-rich, dwarf galaxy in the Galactic Plane but some distance beyond, viable?  The dwarf neighbour hypothesis is perhaps a little more far-fetched, but by combining the information on FA1 \& 2, can we at least rule that idea out?

\vspace{1mm}For a dwarf galaxy, assumed here for simplicity not to be associated with the Sgr dwarf, we compare two masses in order to get some constraint on the distance.  The first mass is the total molecular mass as evaluated above for the various kinematic distances, but which in principle scales as $d^2$ from the \nco\ moment-0 maps,
\begin{eqnarray}
	M_{\rm mol} &=& [a (\pi/648,000) d]^2~\sum (N_{\rm CO,FA1}+N_{\rm CO,FA2})~ \\								%% EQ.C5
	&=& 5.93\times10^4\,{\rm M}_{\odot}~\left(d/{\rm kpc}\right)^2~. \nonumber %53,268.50+6026.744 = 59,295.244
\end{eqnarray}
Here $M$ is the integrated molecular mass of the clouds in units of M\solar, $a$=24$''$ is the pixel size which is converted to pc at distance $d$ via the small-angle formula, and the sum is over all \nco\ pixels (for both FA1 and FA2) in units of M\solar\,pc$^{-2}$.  For the two clouds, at mean pixel values of 137 and 66\,M\solar\,pc$^{-2}$, the 28,694 and 6,662 pixels (respectively) sum to the value in the second line, at a distance of 1 kpc. %3.93$\times$10$^6$\,M\solar\,pc$^{-2}$, while the pixels scale to 0.116\,pc per kpc of distance. %, and the integration is over Galactic coordinates $l$,$b$ in radians.  In the second line the integration has been converted to a sum over pixels in the \nco\ map:. %$a$=1.496$\times$10$^{11}$m, $S$ is the pixel scale in pc,  .  
%have no pre-ordained distance measure, so we posit a heliocentric $d$ = 50\,kpc by analogy with the Magellanic Clouds (and implying a Galactocentric distance $d_G$ $\approx$ 42\,kpc), with the understanding that any physical constraints on the hypothesis will be scalable with $d$.  With a $\sim$4$\times$ larger distance than the far-kinematic mean, the projected diameter of FA1+2 is now about 2\,kpc %
%which is somewhat smaller than that of the SMC, $\sim$6\,kpc. %but the molecular mass is well under the SMC's estimated 4$\times$10$^7$\,M\solar.  

\vspace{1mm}The other mass comes from interpreting the total velocity extent, conservatively put at 110\,\kms\ and independent of $d$, as part of a rotation curve with amplitude $V_{\rm rot}$ $\approx$ 55\,\kms.  This gives a gravitating mass\footnote{The dwarf's systemic velocity is then either \vlsr\ $\approx$ --73 or --93\,\kms, depending on whether we take the mean \vlsr\ of FA1+2 within the observed range, or use a kinematic midpoint based on the \vlsr\ profile seen in Fig.\,\ref{ara}.  If the latter is more correct, the rotation curve would be rather lopsided, with a redshifted amplitude $V_{\rm rot}$ $\sim$ 75\,\kms\ and a blueshifted amplitude $V_{\rm rot}$ $\sim$ 35\,\kms, rescaling the gravitating mass on each side as $V_{\rm rot}^2$.} (i.e., baryonic for a self-gravitating gas cloud, or presumably mostly stellar for an SMC-like dwarf galaxy) for the assumed orbital scale,
\begin{eqnarray}
	M_{\rm grav} &=& RV_{\rm rot}^2/G = d \Delta l V_{\rm rot}^2/2G~ \\					%% EQ.C6
	&=& 18.4\times10^6\,{\rm M}_{\odot}~\left(d/{\rm kpc}\right)~, \nonumber
\end{eqnarray}
where the postulated rotating disk of radius $R$ has been derived from half the total longitude extent $\Delta$$l$ $\approx$ 3\degree\ of FA1+2, and so scales in proportion to $d$. %  1000pc x 3.0857215e16 m x (1.5deg x pi/180) x 55000^2 m2s-2 / 6.673x10-11 kg-1m3s-2 = 3.662e37 kg = 18.4102e6 Msun
%OLD: $\sim$ 7$\times10^8$\,M\solar. % = 1000pc x 206,000 AU x 1.5e11 m x 55000^2 m2s-2 / 6.673x10-11 kg-1m3s-2 = 7.0e8 Msun
For the Milky Way, LMC, and SMC, the molecular mass is $\sim$1--3\% of the stellar mass, so parametrising the gas fraction as $M_{\rm grav}$ = $f$$M_{\rm mol}$, we can set $f$$\approx$50 and solve for the distance $d$ where the mass scales match. %now compare the gravitationally-implied molecular mass, say 1.4$\times10^7$\,M\solar, with the observed, rescaled molecular mass for FA1+2 of $\sim$1.3$\times10^6$\,M\solar. % total?  =972,000 Msun for FA1 only; NO!!
% Then 59295 d2 x 50 = 18.4102e6 d, so d = 18.4102e6/(50x59295) = 6.209697 kpc; Mmol = 2.286,452e6 Msun
Under these assumptions, we obtain $d$ $\approx$ 6.2\,kpc and $M_{\rm mol}$ $\approx$ 2.3$\times$10$^6$\,M\solar.

\vspace{1mm}Within this distance, any normal dwarf with that velocity gradient would have insufficient mass to bind the molecular gas, unless it were gas-poor and $f$$>$50.  But then we are already well within the disk of the Milky Way, so any resemblance to an external galaxy would be lost anyway, besides being physically contradictory to conditions within the disk.

\vspace{1mm}At the other extreme, beyond 6.2\,kpc a normally-constituted dwarf galaxy would have too much gravitating mass to produce the observed ``rotation curve,'' unless it were more gas-rich and $f$$<$50, bringing the implied gravitating mass into consistency.  Again, we are unlikely to be looking at a dwarf galaxy anyway, unless $d$ \gapp 15\,kpc or more, since FA1+2 would still be within the Milky Way's disk.  But the further away these clouds are, the less is the likelihood of a normal dwarf galaxy --- at a distance of $\sim$310\,kpc, FA1+2 would need to be {\bf \em all gas} to have the observed rotation curve, if still self-gravitating.  Beyond that distance, the implied molecular mass would be gravitationally unstable by itself, with that size and velocity gradient.

% Figure C36: FA3 moments
\begin{figure*}[t]
{\includegraphics[angle=-90,scale=0.325]{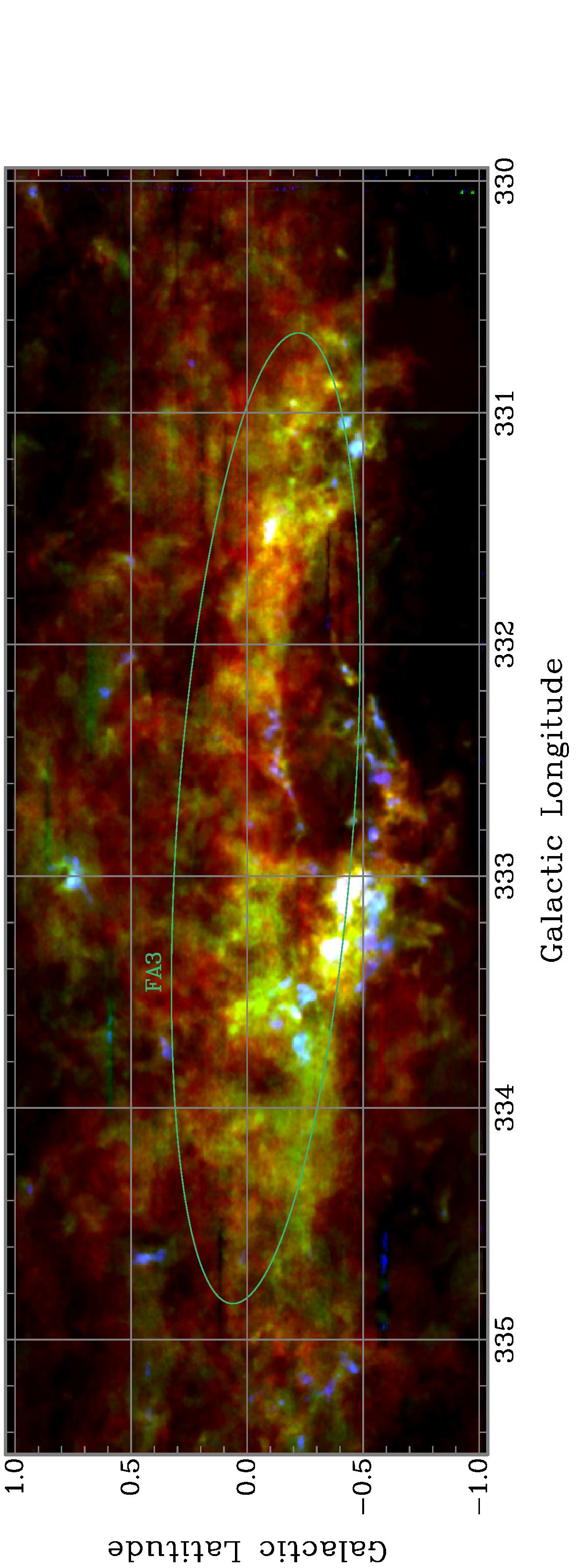} \includegraphics[angle=-90,scale=0.325]{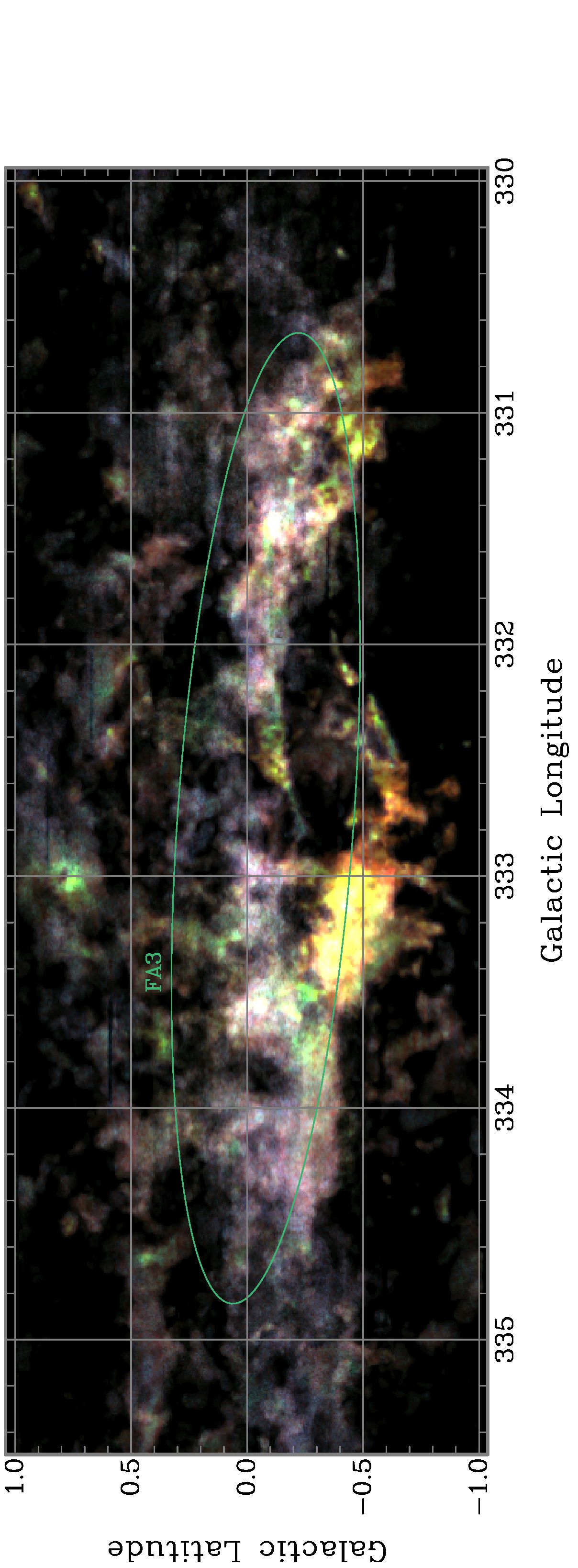}}

\vspace{-4mm}
{\includegraphics[angle=-90,scale=0.325]{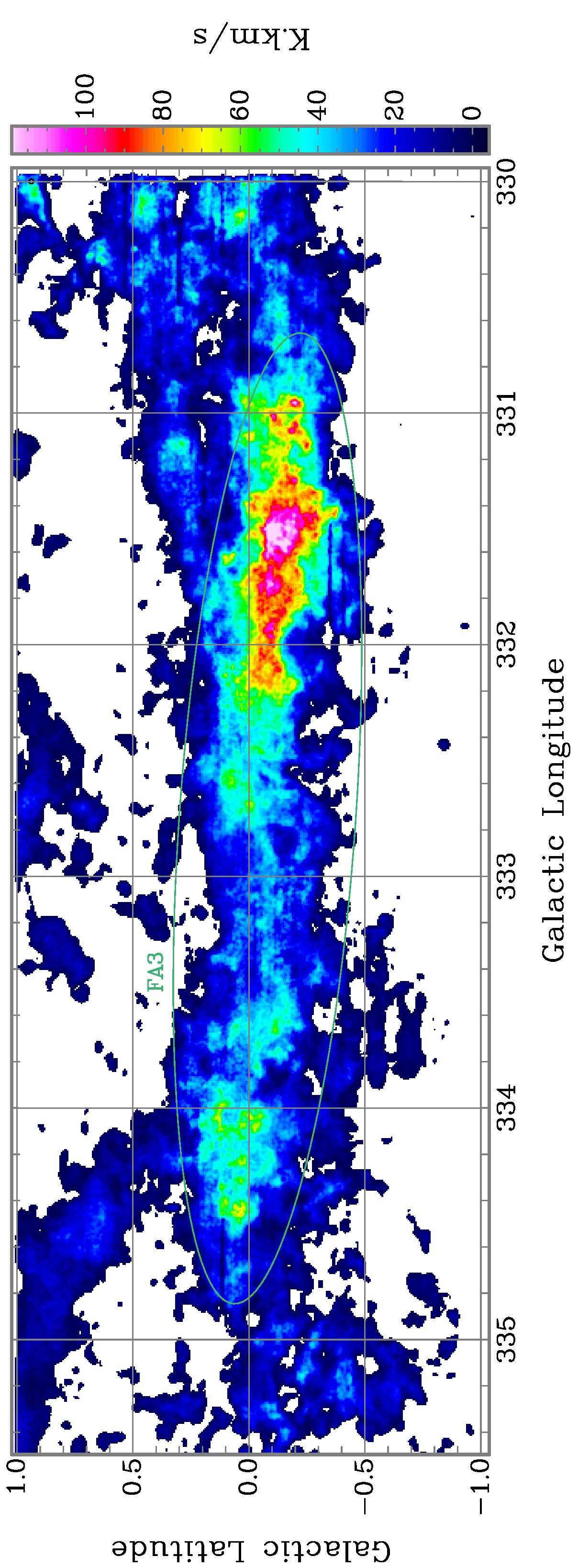} \includegraphics[angle=-90,scale=0.325]{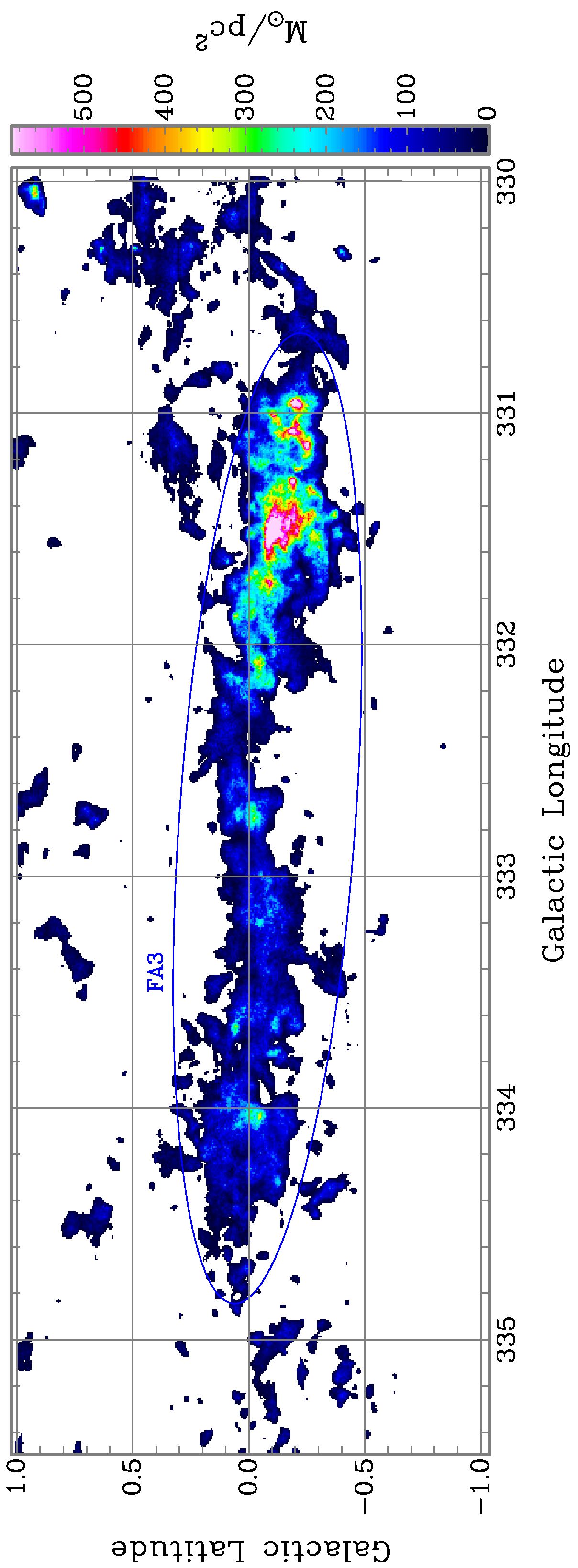}}

\vspace{-4mm}
{\includegraphics[angle=-90,scale=0.325]{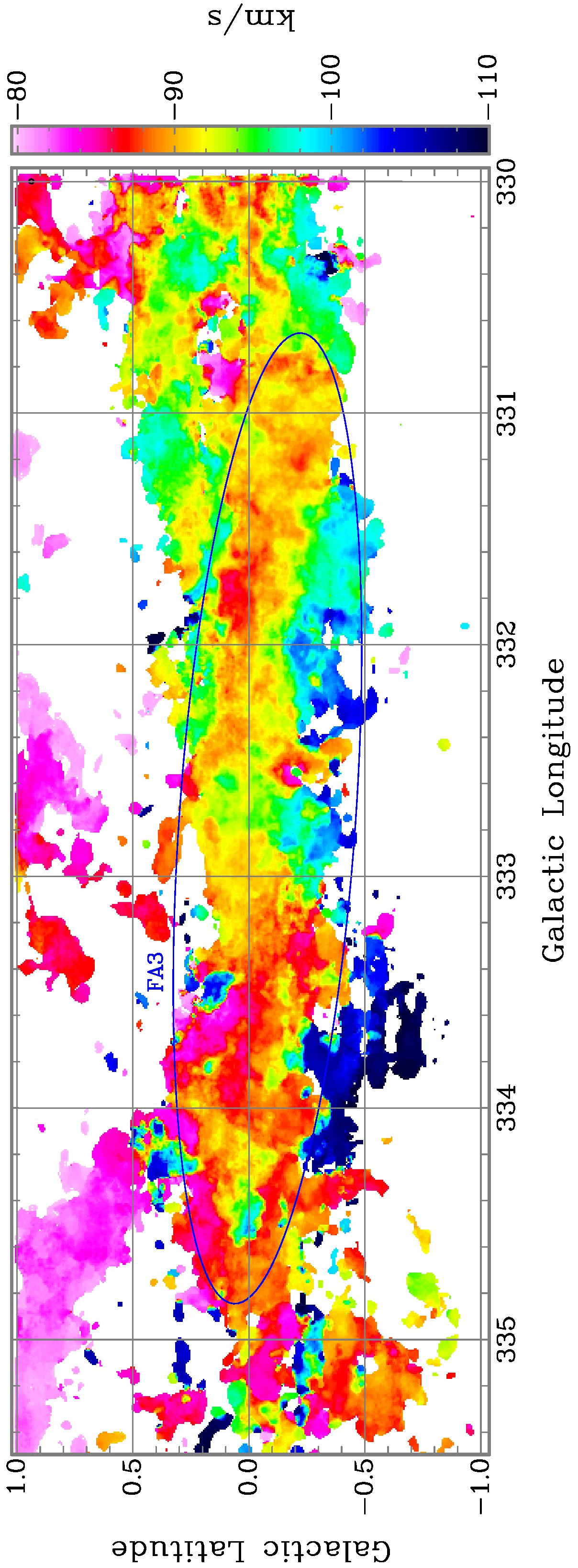} \includegraphics[angle=-90,scale=0.325]{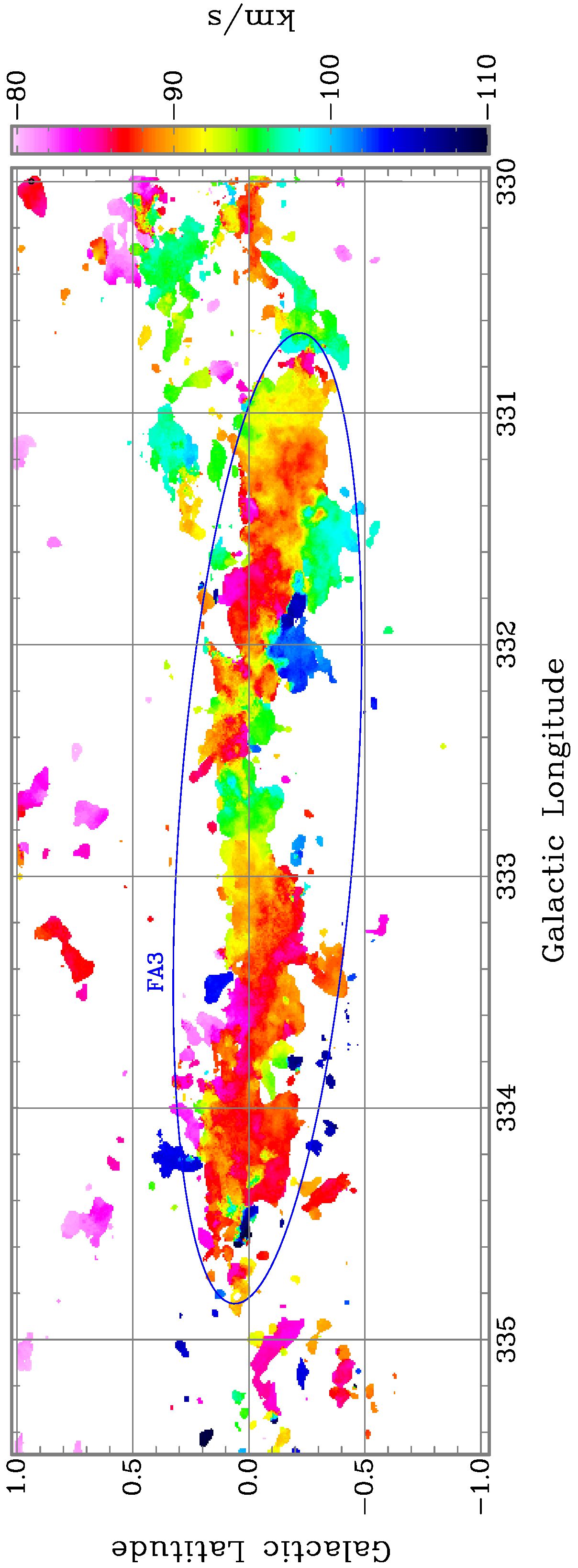}}

\vspace{-4mm}
{\includegraphics[angle=-90,scale=0.325]{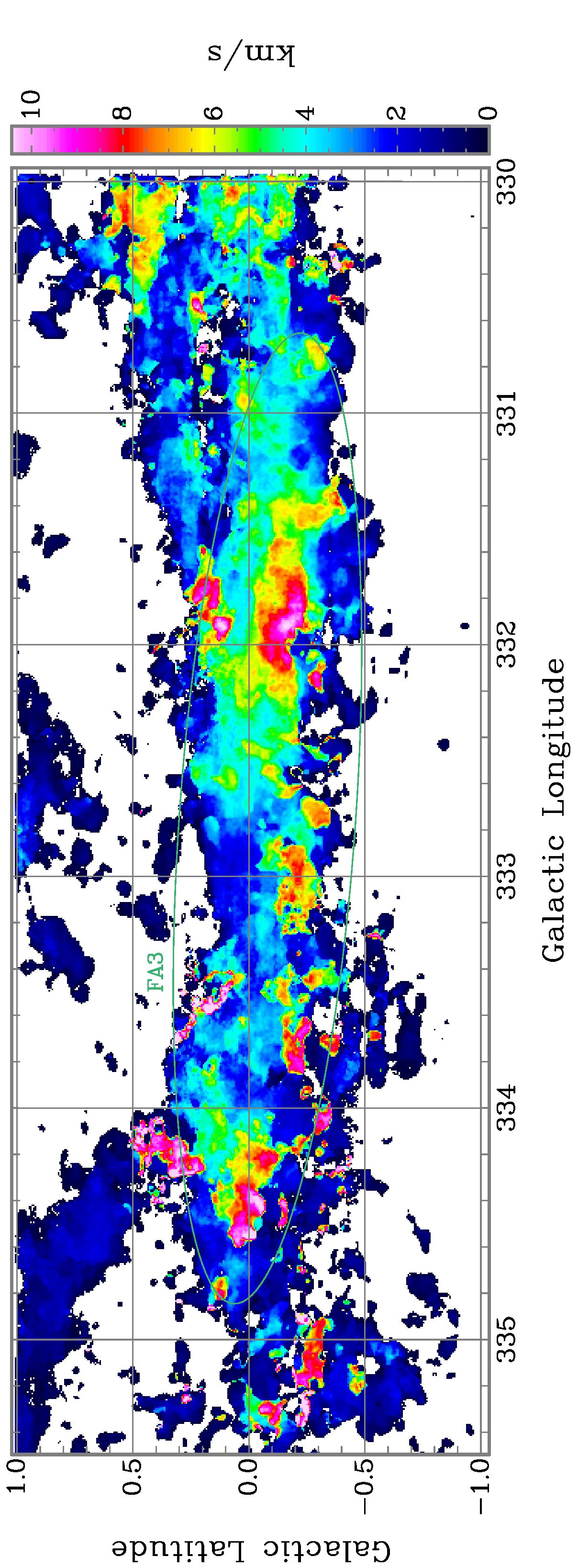} \includegraphics[angle=-90,scale=0.325]{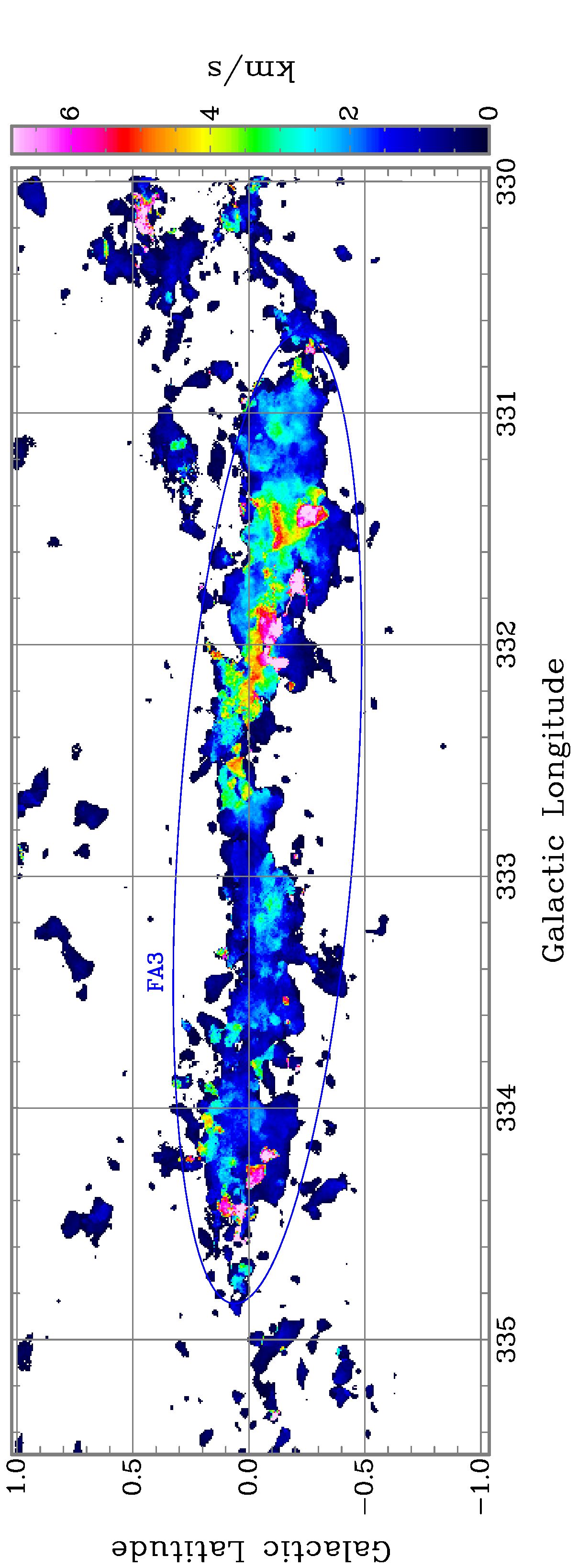}}

\vspace{-2mm}
\caption{\footnotesize ({\em Top row}) RGB composites integrated over all \vlsr\ of ({\em left}) the \tco, \ttco, and \ceto\ intensities as in Fig.\,\ref{full121318-mom0}, and ({\em right}) the mean-\tex, total \nco, and mean-$\tau$ as in Fig.\,\ref{fullTexZMtau-mom0}.  ({\em Rows 2--4}) Similar multi-moment displays for FA3 as for FA1+2 (Figs.\,\ref{FA1moms}, \ref{FA2moms}), but masked to the blue \lv\ square in Fig.\,\ref{ara}.  FA3 (approximately outlined by a labelled blue or green ellipse) separates quite cleanly in \lbv\ space from any surrounding features in the \tex, \nco, and $\tau$ cubes, even while its presence in the all-\vlsr\ RGB composites (top row) is difficult to make out, due mainly to the very bright emission from the foreground G333 complex \citep{q15}. $$ $$
\label{FA3moms}}
\vspace{-8mm}
\end{figure*}

\vspace{1mm}Despite these somewhat awkward numbers, we can salvage the scenario of a newly-recognised dwarf galaxy neighbour to the Milky Way under the following circumstances.  It could conceivably lie within a range of distances 20\,kpc \lapp\ $d$ \lapp\ 300\,kpc if it were relatively gas-rich, respectively about 7\%--100\% gas at this range of distances, with a corresponding size and mass over this range of roughly 1--16\,kpc and 2$\times$10$^7$--6$\times$10$^9$\,M\solar.  At the lower end of the range, such a new neighbour would be closer to us than the Magellanic Clouds (50--60\,kpc), and smaller ($\sim$13\%) \& less massive ($\sim$1\%) than the Small Magellanic Cloud, consistent with a somewhat gas-rich dwarf galaxy.  At the furthest distances, it would be up to 5$\times$ further away than the Magellanics, 2--3$\times$ larger than either, and about as massive as the SMC, but very gas-rich.

\vspace{1mm}Alternatively, the more plausible scenarios for the nature of FA1+2 include a very massive farside feather, or a cloud undergoing some kind of shear, possibly tidally-induced, to produce the velocity gradient in gas that may be falling into the disk.  In the latter scenario, however, we have no obvious distance scale upon which to place the clouds if neither kinematic distance can be relied upon, unless FA1+2 are stripped remnants of the Sgr dwarf, or lie in the far end of the Near 3\,kpc Arm, just downstream from the far end of the Milky Way's bar.  In this case, the kinematic distances shown in Figures \ref{12co-bgt-YX0fl} and \ref{ZM-bgt-YX0fl} would be close to those expected for such features, and the large masses understandable in that context.  But if outside the disk of the Galaxy, the molecular mass would be even larger, \gapp 10$^7$\,M\solar.  Clearly, some additional observational data (deep infrared imaging?) are needed to clear up this mystery.

\vspace{1mm}Finally, we briefly discuss FA3 ({\color{red}Fig.\,\ref{FA3moms}}), the third major far-kinematic molecular complex identified from our $\zeta^+$ filtering.  Figures \ref{both-ld0-b1p3m}--\ref{both-ld2h-b1p3m} put FA3 at $d$ = 9.7$\pm$0.5\,kpc as a large GMC, due to its extreme flatness in $b$ and $z$.  In contrast, we took a near-kinematic distance of 5\,kpc in the analysis of Paper II \citep{q15}, placing it in the Norma spiral arm.  With a more careful excision of FA3 from the foreground RCW\,106/G333 complex in \lv\ space, the current BGT model would actually place its nearside position $\sim$0.5\,kpc inside \cite{r19}'s Norma arm, at 4.8$\pm$0.2\,kpc.  The nearer distance was preferred in Paper II from a combination of the latitude distribution of the molecular emission, plus HI absorption.

\vspace{1mm}However, the moment maps in Figure \ref{FA3moms} weaken the argument for a near distance based on latitude, so it may be instructive to compare these alternatives in more detail.  With our new cutout (about half the area used in Paper II) we have a projected size in $l$$\times$$b$ = 2\fdeg0$\times${0\fdeg1}; at the two distances this corresponds to 168$\times$7\,pc or 339$\times$15\,pc;  %2.1 deg diameter x 3600'' x (4800 or 9700 AU/arcsec) / (206265 AU/pc) = 175.929 or 355.524 pc; at 2.0 deg, sizes are 168 or 339 pc
%height dispersions $\sigma_z$ $\sim$ 7 or 15\,pc, 
maximal height excursions from the mean $\pm$0\fdeg2 $\approx$ 17 or 35\,pc; and total masses $\sim$ 1.0 or 4.2$\times$10$^6$\,M\solar\,pc$^{-2}$ % 102.4174 Msun/pc2 x 32,465 pix x 24pi/648000 = 1,037,155 Msun or 4,235,501 Msun
from a mean surface density $\Sigma_{\rm H_2}$ = 102\,M\solar\,pc$^{-2}$ over 32,465 pixels.
The inferred projected mass from Paper II was 2.8$\times$10$^6$\,M\solar\ with an average surface density 130\,M\solar\,pc$^{-2}$, but this was based on just the \tco\ ThrUMMS data over the larger area.

\vspace{1mm}None of these values strongly rule out one distance or the other, but overall, the flatness argues for a farside location, while the HI absorption and convenient placement near the Norma Arm argues for a nearside location.  In balance, we favour the near-kinematic distance of 4.8\,kpc, subject to possible revision with more detailed HI data from (e.g.) the GASKAP survey.

\vspace{1mm}More generally, the examples of FA1--3 show that our $\zeta^+$ filtering is quite conservative, in the sense that it places most clouds at their near-kinematic distance, as might be expected based purely on sensitivity arguments.  That even a GMC as flat as FA3 might nevertheless lie at its near distance makes the configuration of FA1+2 even more striking.  In other words, among all molecular clouds seen in the ThrUMMS data, whether in individual species such as \tco\ or through parameters of the radiative transfer analysis such as \nco, their very large velocity gradients (whether taken together or individually) are unique.  In particular, the \nco\ \lv, \ld, or \xy\ maps show that this gradient is not just an artifact of one species' data.  Any model of where the clouds lie and what they represent must take these gradients into account.

\end{document}